\documentclass[
12pt, 
english, 
singlespacing 
draft, 
nolistspacing, 
liststotoc, 
parskip, 
headsepline, 
consistentlayout, 
]{MastersDoctoralThesis} 

\usepackage[utf8]{inputenc} 
\usepackage[T1]{fontenc} 

\usepackage{newtxtext}   

\usepackage[backend=bibtex,style=ieee,natbib=true]{biblatex} 

\usepackage[autostyle=true]{csquotes} 
\usepackage{CJKutf8}  
\usepackage[switch,columnwise]{lineno}
\usepackage[para,online,flushleft]{threeparttable}
\usepackage{xcolor}
\usepackage{flushend}
\definecolor{MYCOLOR}{RGB}{102,0,51}
\usepackage{hyperref,comment}
\hypersetup{colorlinks=true,%
citecolor=blue,%
filecolor=black,%
linkcolor=MYCOLOR,%
urlcolor=blue}

\usepackage{amsmath, bm, booktabs,amssymb,amsfonts}
\usepackage{array,enumitem}
\usepackage{multirow}
\usepackage{filecontents}
\usepackage{algorithm, algorithmic}
\usepackage{graphicx}
\usepackage{textcomp}

\thesistitle{Resilient-Economic Coordinated Robust Operation of Integrated Electric-Gas Systems} 
\supervisor{Prof. Tianshu Bi} 
\examiner{} 
\degree{Doctor of Philosophy} 
\author{Ahmed Rabee Kamel Sayed} 
\addresses{Cairo University Street, Giza, Egypt} 

\subject{Electrical Engineering} 
\university{\href{http://https://english.ncepu.edu.cn}{North China Electric Power University}} 
\department{\href{http://https://english.ncepu.edu.cn}{School of Electrical and Electronic Engineering}} 
\group{\href{https://laps.ncepu.edu.cn/}{Integrated Energy System}} 
\faculty{\href{http://faculty.university.com}{Faculty Name}} 

\AtBeginDocument{
\hypersetup{pdftitle=\ttitle} 
\hypersetup{pdfauthor=\authorname} 
\hypersetup{pdfkeywords=\keywordnames} 
}
\usepackage{caption}   
\usepackage{mathtools}   
\newcommand{\Lagr}{\mathcal{L}}

\begin{document}

\frontmatter 

\pagestyle{plain} 


\begin{titlepage}
\vspace*{.01\textheight}
\begin{center}

\begin{figure}[th]
    \centering
    \includegraphics[width=10cm]{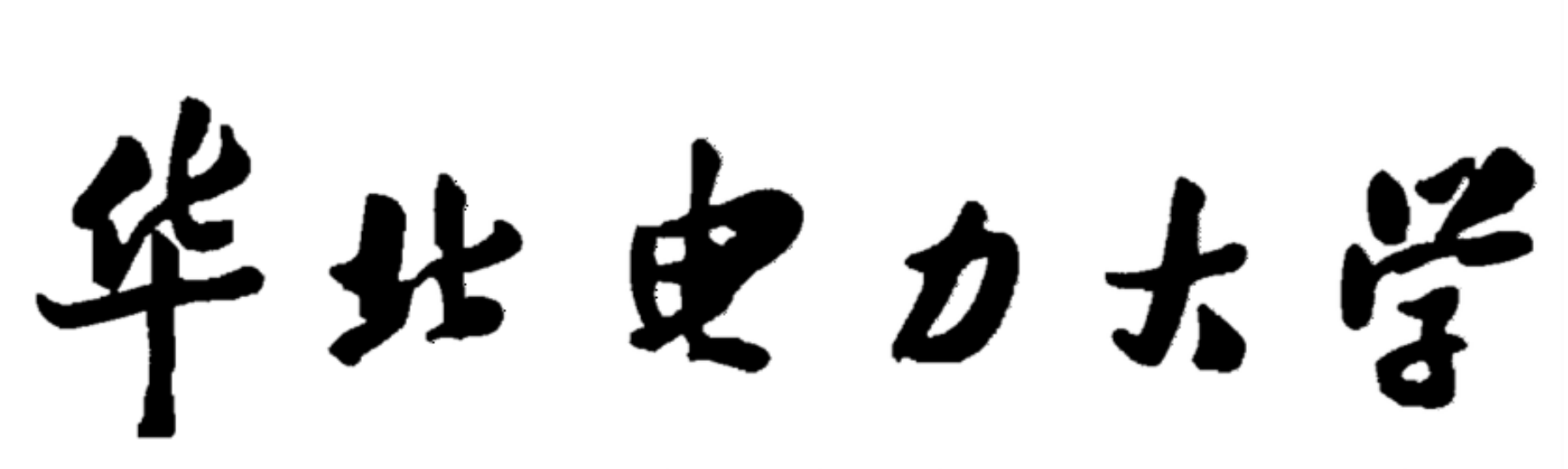}
    \label{fig:NCEPU_Chinese}
\end{figure}\vspace{-0.5cm}
{\scshape\LARGE \univname\par}\vspace{2cm}

\begin{CJK*}{UTF8}{gbsn}
\LARGE 博士学位论文   \vspace{0.5cm} 

{\huge \bfseries 韧性与经济性协调的电-气综合能源系统鲁棒运行方法研究  \par}\vspace{0.5cm}
\end{CJK*}

{\huge \bfseries \ttitle\par}\vspace{2cm} 
\vfill

\href{https://scholar.google.co.in/citations?user=oOzlgR8AAAAJ&hl=en}{ \LARGE\textcolor[rgb]{0.00,0.00,0.00}{\authorname}} 
\vspace{2cm}

\vfill

{\large \today}\\[4cm] 

\vfill
\end{center}
\end{titlepage}

\begin{titlepage}

\begin{CJK*}{UTF8}{gbsn}
  \begin{tabular}{l r}
  国内图书分类号：\textcolor[rgb]{1.00,0.00,0.00}{TM73}&\hspace{14em} 学校代码：\textcolor[rgb]{1.00,0.00,0.00}{10079} \\
  国际图书分类号：\textcolor[rgb]{1.00,0.00,0.00}{621.3} & 密级：公开
\end{tabular} \hspace{15em}
\end{CJK*}
\vfill
\vspace*{.06\textheight}

\begin{center}
\begin{CJK*}{UTF8}{gbsn}
\LARGE 工学博士学位论文 \vspace{2cm} 

{\huge \bfseries 韧性与经济性协调的电-气综合能源系统鲁棒运行方法研究  \par}\vspace{3cm}

\vfill

\Large
\begin{tabular}{l l}
  博士研究生 :& \authorname  \\
  导师 :& 毕天姝 教授 \\
  申请学位 :& 工学博士  \\
  学科 :& 电气工程  \\
  专业 :& 电力系统及其自动化  \\
  所 在 学 院 :& 电气与电子工程学院  \\
  答 辩 日 期 :& 2020年6月  \\
  授予学位单位 :& 华北电力大学
\end{tabular}
\end{CJK*}
\vfill

\end{center}
\end{titlepage}
%
%
%
%
%
%
%
\begin{titlepage}
\begin{tabular}{l r}
        Classified Index : \textcolor[rgb]{1.00,0.00,0.00}{TM73}&\hspace{14em} School Code : \textcolor[rgb]{1.00,0.00,0.00}{10079} \\
        ISBN : \textcolor[rgb]{1.00,0.00,0.00}{621.3} & Class : Public
    \end{tabular} \hspace{15em}
\begin{center}

\vspace*{.06\textheight}
{\scshape\LARGE \univname\par}\vspace{1.5cm} 
\textsc{\Large Doctoral Thesis}\\[0.5cm] 

{\huge \bfseries \ttitle\par}\vspace{2cm} 

\emph{By}\\
{\large \href{https://scholar.google.co.in/citations?user=oOzlgR8AAAAJ&hl=en}{\authorname}} 
\vfill
\emph{Under the Supervision of} \\
{\large \href{https://scholar.google.com/citations?user=9k3ltT8AAAAJ&hl=zh-CN}{\supname} }

\vfill

\large \textit{A thesis submitted in fulfillment of the requirements\\ for the degree of \degreename}\\[0.3cm] 
\textit{in the}\\[0.4cm]
\deptname\\[2cm] 

\vfill


\end{center}
\end{titlepage}

\begin{CJK*}{UTF8}{gbsn}
\begin{declarationC}
\vspace{-1cm}
\begin{center}
{\Large \textbf{华北电力大学博士学位论文原创性声明}}\\
\end{center} \vspace{1cm} \large
\noindent 本人郑重声明：此处所提交的博士学位论文《韧性与经济性协调的电-气综合能源系统鲁棒运行方法研究》，是本人在导师指导下，在华北电力大学攻读博士学位期间独立进行研究工作所取得的成果。据本人所知，论文中除已注明部分外不包含他人已发表或撰写过的研究成果。对本文的研究工作做出重要贡献的个人和集体，均已在文中以明确方式注明。本声明的法律结果将完全由本人承担。

作者签名：        \hspace{16em}               日期： ~~~    年  ~~~ 月  ~~~ 日
\vfill
\begin{center}
{\Large \textbf{华北电力大学博士学位论文使用授权书}} \\
\end{center} \vspace{1cm} \large
《韧性与经济性协调的电-气综合能源系统鲁棒运行方法研究》系本人在华北电力大学攻读博士学位期间在导师指导下完成的博士学位论文。本论文的研究成果归华北电力大学所有，本论文的研究内容不得以其它单位的名义发表。本人完全了解华北电力大学关于保存、使用学位论文的规定，同意学校保留并向有关部门送交论文的复印件和电子版本，允许论文被查阅和借阅，学校可以为存在馆际合作关系的兄弟高校用户提供文献传递服务和交换服务。本人授权华北电力大学，可以采用影印、缩印或其他复制手段保存论文，可以公布论文的全部或部分内容。

本学位论文属于（请在以上相应方框内打“√”）：

保密   □，在      年解密后适用本授权书

不保密  □

作者签名：           \hspace{16em}                 日期：  ~~~    年   ~~~ 月   ~~~ 日

导师签名：                 \hspace{16em}               日期：    ~~~  年    ~~~ 月    ~~~日
\vfill
\end{declarationC}\end{CJK*}


\begin{declaration}
\noindent I, \authorname, declare that this thesis titled, \enquote{\ttitle} and the work presented in it are my own. I confirm that:

\begin{itemize}
\item This work was done wholly or mainly while in candidature for a research degree at \univname.
\item Where any part of this thesis has previously been submitted for a degree or any other qualification at \univname ~or any other institution, this has been clearly stated.
\item Where I have consulted the published work of others, this is always clearly attributed.
\item Where I have quoted from the work of others, the source is always given. With the exception of such quotations, this thesis is entirely my own work.
\item I have acknowledged all main sources of help.
\item Where the thesis is based on work done by myself jointly with others, I have made clear exactly what was done by others and what I have contributed myself.\\
\end{itemize}

\noindent Signed:\\

\noindent Date:\\
\end{declaration}

\cleardoublepage


%
%


\begin{abstract}
  \addchaptertocentry{\abstractname} 

    Interactions between power and gas systems, which are both large and complex, have been gradually intensified during the last decades, predominantly due to the propagation of large fleet natural gas-fired power units  (GPUs)  and the technology developments of power-to-gas (P2G) facilities. These interactions not only bring significant economic benefits to the society but also provide additional operating flexibilities, which are essential to handle fluctuations of the large-scale renewable power generation (RPG) and power system contingencies. Moreover, neglecting these interactions in power system operation may not only result in infeasible operation status in the gas systems but also increase the decision-making operation costs of both systems. Previous studies suffered from two major drawbacks, namely (1) they assumed the existence of only one utility that has full control authority over the power system and gas system; (2) the economic interactions between power systems and gas systems have been neglected, which goes against the current industrial practice.  This research revisits the day-ahead resilient and economic operations of power systems considering the economic and physical interactions with gas systems, which are characterized by the modeling of bilateral energy purchase contracts and operational constraints of gas systems, respectively. The main work of the thesis is as follows:
     \begin{enumerate}
       \item Propose a tri-level resilient operational framework to optimize the operational performances of power systems under the worst-case $N-k$ contingencies. The proposed model considers gas contracts with gas systems, where firm gas supply contracts and gas reserve contracts are formulated in the pre- and the post-contingency stages, respectively.
       \item Emerging P2G facilities to mitigate the surplus RPG outputs, bidirectional gas contracts are inevitable. A two-stage robust model of the energy management problem for the power distribution networks (PDNs) is proposed. According to the current gas contracting mechanism, flexible real-time contracts may still be signed for the low-probability utilized reserved GPU outputs in practice. To balance the robustness and the conservativeness of the operation strategy, a two-stage distributionally robust contracting model is proposed.
       \item A robust operational equilibrium solution method for the interactive markets of power and gas systems is proposed, where the bidirectional interactions include energy contracts, and the impacts of the uncertainties of wind generation outputs on the two markets are characterized. To guarantee the robustness of market equilibrium against uncertainties, the power and gas market-clearing models become two-stage robust ones.
     \end{enumerate}

         In brief, this thesis provides a novel perspective and solution for the resilient and economic coordinated robust operation for the integrated electric-gas systems (IEGSs) under uncertainties. The proposed robust scheduling decision frameworks are practically compatible with the existing industrial operations of the IEGSs, and they are expected to be employed in the operation of IEGSs with large-scale integration of RPG, extreme weather and operating failures, to provide technical and optimal energy management for improving the resilient and economic operation of IEGSs, and to realize the secure and economic operation of the integrated systems against uncertainties.


\end{abstract}
\begin{CJK*}{UTF8}{gbsn}
\begin{abstractC}


{\huge \textbf{摘要}} \\

    近年来，由于燃气发电比例上升及电转气技术的发展，电力系统与天然气系统之间的耦合显著增强，逐步形成了新的能源产供消系统，称为电-气综合能源系统。电-气综合能源系统不仅提高了全社会用能经济性，还为电力和天然气网络提供了额外的运行灵活性，对于应对电力网络中风光发电出力不确定性与突发故障事件至关重要。然而，目前电力与天然气系统的调控由不同机构负责，因此在任一能源系统调控决策中均须合理建模其他能源系统运行工况对本系统运行可行性与经济性产生的影响。为实现不确定环境下电力系统安全经济运行，本文提出了一套韧性与经济性协调的电-气综合能源系统鲁棒运行方法。论文主要工作如下：
    \begin{enumerate}
      \item 提出了$N-k$ 故障条件下电-气综合能源系统韧性增强鲁棒调度运行方法，通过建模电力系统在故障前与故障后两个阶段对天然气系统的燃气需求，刻画了电力与天然气系统在运行可行性与经济性两方面的强耦合关系，仿真算例验证了所提模型在增强电力系统韧性方面的有效性。
      \item 提出了考虑风光发电出力不确定性的电-气综合能源系统鲁棒调度运行方法，通过在电力系统调度模型中考虑天然气系统运行约束与电力系统两阶段购气合同建模，反映了天然气系统对电力系统鲁棒运行策略可行性与经济性方面的影响，仿真结果表明了在电力系统鲁棒调度决策中考虑天然气系统的必要性。
      \item 提出了基于实际供能量与供能设备预留备用空间的节点能价机制，设计了基于节点能价的电-气耦合能源市场双边结算框架，提出了电-气耦合能源市场鲁棒均衡分析方法，保障了不确定环境下电-气综合能源系统用能经济性与安全性，算例结果验证了鲁棒均衡策略的有效性。
    \end{enumerate}

    总之，本文为不确定环境下电-气综合能源系统韧性与经济性协调的鲁棒运行决策提供了崭新视角和解决思路，所设计的鲁棒调度决策框架与现阶段电-气综合能源系统运行实际相容，有望应用于含高比例风光发电、极端天气与运行故障工况下电-气综合能源系统运行问题，为提升电-气综合能源系统韧性与运行经济性提供理论指导，实现不确定性环境下电-气综合能源系统安全经济运行。

\textbf{关键词:} 经济调度，韧性调度，电-气综合能源系统，不确定性，能量管理，能源市场.

\vfill
\end{abstractC}
\end{CJK*}
\renewcommand{\contentsname}{Table of Contents}
\tableofcontents 

\listoffigures 

\listoftables 
\begin{Acronyms}{lll}
AC     & Alternating Current. &\\
BRD     & Best-Response Decomposition Algorithm. &\\
CPF   & Conventional Power Flow. &\\
DC    & Direct Current. &\\
DCP      & Difference-of-Convex Programming. & \\
DRO     & Distributional Robust Optimization. & \\
DP      & Dynamic Programming. &\\
ED      & Economic Dispatch. &\\
EIA      & Energy Information Administration. &\\
EM       & Energy Management. &\\
ECs       & Electricity Contracts. &\\
EMO       & Electricity Market Operator. &\\
GCs      & Gas Contracts. &\\
GDN     & Gas Distribution Network. &\\
GMO     & Gas Market Operator. &\\
GFC     &Gas Flow Correction Method. &\\
GPU      & Gas-fired Power Units. & \\
GSO     & Gas System Operator. &\\
G2P      & Gas-to-Power. &\\
IEGS    & Integrated Electric-Gas System. &\\
IES     & Integrated Energy System. &\\
IPS   & Independent Power System. &\\
KKT      & Karush-Kuhn-Tucker. &\\
LMEPs         & Locational Marginal Electricity Prices. &\\
LMFGPs      & Locational Marginal Firm Gas Prices. &\\
LMRGPs      & Locational Marginal Reserved Gas Prices. &\\
LP       & Linear Programming. &\\
MILP   & Mixed Integer Linear Programming. &\\
MISOCP & Mixed-Integer Second-Order Cone Programming. &\\
NCUC     &Network Constraints Unit Commitment. &\\
NLP   & Non-Linear Programming. &\\
OECD   & Organization for Economic Cooperation and Development (it is to promote &\\
& the economic welfare of its members, which include 37 nations in Europe, &\\
& the Americas, and the Pacific). &\\
OGF & Optimal Gas Flow. &\\
OPGF   &Optimal Power-Gas Flow. &\\
OPF    & Optimal Power Flow. &\\
PLA    &Piecewise Linear Approximation. &\\
PSO   & Power System Operator. &\\
PDN   & Power Distributed Network. &\\
P2G   & Power-to-gas facilities. &\\
P-CCP & Penalty Convex-Concave Procedure. &\\
RD     & Resilient Dispatch. &\\
RO      & Robust Optimization. &\\
RPG   & Renewable Power Generation. &\\
RCV   & Relative Constraint Violation. &\\
SCP     & Sequential Convex Procedure. &\\
SOCP   & Second-Order-Cone Programming. &\\
SO   & Stochastic Optimization. &\\
SQP     & Successive Quadratic Programming. &\\
SOC      & Second-Order Cone. & \\
S-MISOCP & Sequential-Mixed-Integer Second-Order Cone Programming Algorithm. &\\
UC    & Unit Commitment. &\\
WPLs      & Wind Penetration Levels. &\\
\end{Acronyms}
%
%
%


\begin{Nomenclature}{lll} 
%
%
\multicolumn{3}{l}{\emph{A. Sets and indices}}   \\ \addlinespace
$c \in \mathcal{C}$ &              Gas compressors. & \\
$s \in \mathcal{S}$ &              Gas storages. & \\
$d \in \mathcal{D}_p / \mathcal{D}_g$ &   Electricity/Gas demands. & \\
$h \in \mathcal{H}$ &              Gas-to-power contracts. & \\
$i,o \in \mathcal{I}$ &            Gas network nodes. & \\
$w \in \mathcal{W}$ &              Gas wells or gas sources. & \\
$l \in \mathcal{L}$ &              Power transmission or distribution lines. & \\
$n,m \in \mathcal{N}$ &            Power transmission buses or distribution nodes. & \\
$p \in \mathcal{P}$ &              Gas passive pipelines, bidirectional set is $\mathcal{P}^\pm$. & \\
$r \in 1...R$ &                    Iteration index of the C\&CG Algorithm. & \\
$t \in \mathcal{T}$ &              Time periods. & \\
$u \in \mathcal{U}_g/\mathcal{U}_n$ &  Gas-fired power units (GPUs) /Non-GPUs, $\mathcal{U} = \mathcal{U}_g \cup \mathcal{U}_n$. & \\
$z \in \mathcal{Z}$ &              P2G facilities. & \\
$e \in \mathcal{E}$ &              Wind farms. & \\
$j \in \mathcal{J}$ &              Power-to-gas contracts. & \\
\addlinespace \addlinespace 
\multicolumn{3}{l}{\emph{B. Parameters}}   \\ \addlinespace
$C_u(.)$ &                             Quadratic cost function of power units. & \\
$\mu_h$ &                              Day-ahead or pre-contingency gas prices for firm gas. & \\
$\mu_h^+ / \mu_h^-$ &                  Day-ahead or pre-contingency gas prices for reserved gas. & \\
$\mu_h^{2+} / \mu_h^{2-}$ &            Real-time gas prices for reserved gas. & \\
$C_j^+/ C_j^-$ &                       Penalties of P2G contract avoidance. & \\
$C_j^{2-}/ C_j^{2+}$ &                 Penalties/revenue of P2G production adjustments in two-stage contracting. & \\
$C_j$ &                                Day-ahead gas prices for P2G gas contract. & \\
$C_w$ &                                Cost of gas production from wells. & \\
$C_w^+/C_w^-$ &                        Real-time up/down reserve production costs. & \\
$C_u^+ / C_u^-$ &                      Real-time regulation costs for non-GPUs. & \\
$C_d/C_n$ &                            Penalties of non-served electric demands $d$/connected to bus $n$. & \\
$C_e$ &                                Penalties of wind power curtailment. & \\
$C_i^u/C_i^y$ &                        Attacker/Defense costs. & \\
$C_h$ &                                Penalties of unserved gas demands in G2P contracts. & \\
$\beta_n$ &                            Electricity prices at power buses. & \\
$C_i^+/C_i^-$ &                        Prices of upward/downward gas reserves provided by end-users. & \\
$\overline{{C}}_i$ &                   Penalties of unbalanced gas nodes. & \\
$C_i$ &                                Penalties of gas load shedding. & \\
$\overline{P}_u/ \underline{P}_u$ &    Upper/lower active power limits of power units. & \\
$\overline{Q}_u/ \underline{Q}_u$ &    Upper/lower reactive power limits of power units. & \\
$\overline{P}_z/\underline{P}_z$ &      Upper/lower power limits of P2G units. & \\
$\overline{R}_{u}^+/ \overline{R}_{u}^-$ &  Ramping up/down of power units. & \\
$r_l /x_l $ &                          Series resistance/reactance of power lines. & \\
$G_n /B_n$ &                           Shunt condunctance/susceptance of power nodes. & \\
$P_{d}/Q_{d}$ &                        Active/reactive power demands. & \\
$\hat{W}_{e,t}$ &            Forecasted power outputs of wind farms. & \\
$\overline{P}_{e,t} / \underline{P}_{e,t}$ &   Maximum/minimum forecasting power outputs from wind farms. & \\
$\overline{I}_l$ &                     Ampacity limit of power lines. & \\
$\overline{V}_n / \underline{V}_n$ &   Voltage limits of power nodes. & \\
$\overline{F}_w/\underline{F}_w$ &      Upper/lower gas production limits of well or source $w$. & \\
$\overline{\Pi}_i/\underline{\Pi}_i$ &     Pressure limits of gas nodes. & \\
$\overline{F}_p/\underline{F}_p$ &         Gas flow limits of a pipeline. & \\
$\overline{L}_s/\underline{L}_s$ &         Upper/Lower limits of working volume of storage $s$ at time $t$. & \\
$\overline{f}_s^{in}/\overline{f}_s^{out}$ &  Maximum capacities of the injection/withdraw rates of storage $s$. & \\
$\alpha_{c}$ &                         Gas consumption factor of compressors. & \\
$\gamma_c$  &                          Maximum compression factor of compressors. &\\
$F_{d}$ &                           Gas demands. & \\
$\eta_z/ \eta_u$ &                     Efficiency of P2G units/GPUs. & \\
$\Phi$ &                               Electricity-to-gas conversion factor. & \\
$\chi_p^m/\chi_p^f$ &                  Line pack/Weymouth equation constants. & \\
$\underline{\Theta}_n/\overline{\Theta}_n$ & minimum/maximum limits bus angles\\
$\tilde{\pi}$ & The mathematical constant $\tilde{\pi}\approx3.1416$. &\\
$\overline{P}_l$ &                     Power flow capacity of power lines. & \\
$\Gamma^e/\Gamma^t$ &                  Wind budget based on number of wind farms/time periods. & \\
$\varepsilon / \xi, \varrho$ &         Convergence parameters of . & \\
$\tau$ &                                Penalty coefficients for penalized problems in S-MISOCP algorithm. & \\
$\overline{\mu},\;\underline{\mu},\sigma$ & penalty growth rate coefficients. & \\

\addlinespace \addlinespace 
\multicolumn{3}{l}{\emph{C. Decision variables}}   \\ \addlinespace
$c_{u,t}$ & Power units generator statues (UC). & \\
$f_{w,t}$ & Gas production of gas wells or sources. & \\
$p_{u,t}/ q_{u,t}$ &                   Active/reactive power of power units. & \\
$p_{l,t}/ q_{l,t}$ &                   Active/reactive power flow of power lines. & \\
$p_{j,t}/\varrho_{j,t}$ &              Utilized power by/produced gas from P2G units. & \\
$p_{e,t}$ &                        Real-time production power of wind farms. & \\
$\rho_{h,t}$ &                        Firm gas in G2P contracts. & \\
$\rho^+_{h,t} / \rho^-_{h,t}$ &        Upward/downward reserved gas in G2P contracts. & \\
$\rho^{2+}_{h,t} / \rho^{2-}_{h,t}$ &    Upward/downward real-time gas amounts in the two-stage contracting. & \\
$g_{j,t}$ &                        Scheduled gas in P2G contracts. & \\
$u_{l}/y_l$ &                  Attacker/defender decisions for power line $l$. & \\
$h_l$ &                  Availability statues of power line $l$. & \\
$\triangle p^+_{u,t} /\triangle p^-_{u,t}$ &     Upward/downward regulation power from non-GPUs. & \\
$\triangle p_{d,t}/\triangle q_{d,t}$ &      Active/reactive power load shedding. & \\
$w_{e,t}$ &               Real-time wind power outputs.& \\
$\triangle w_{e,t}$ &     Wind power curtailment. & \\
$\triangle g^+_{h,t} /\triangle g^-_{h,t}$ & Gas deviations in P2G contracts. & \\
$i_{l,t}$ &                        Squared current of power lines. & \\
$v_{n,t}$ &                        Squared voltage of power buses. & \\
$\theta_{n,t}$ &                   Voltage angle at power buses. & \\
$\pi_{i,t}$ &                      Pressure of gas nodes. & \\
$\pi_{i,t}^+/\pi_{i,t}^-$ &      Pressures of sending/receiving nodes of a pipeline. & \\
$f_{p,t}$ &                 Average gas flow of passive pipelines. & \\
$f_{c,t}^{in}/f_{c,t}^{out}$ &               Inlet/Outlet gas flow of compressors. & \\
$f_{p,t}^{in} / f_{p,t}^{out}$ & Inlet/Outlet flows of gas pipelines. & \\
$f_{s,t}^{in} / f_{s,t}^{out}$ & Inlet/Outletflows of gas storage. & \\
$m_{p,t}$ &                        Line pack. & \\
$l_{s,t}$ &                        Working volume of storages. & \\
$u_{w,t}^u, u_{w,t}^l$ &              Wind uncertainty binaries, they can be written as $\xi_{w,t}^u, \xi_{w,t}^l$  & \\
$UB / LB $ &                   Upper/lower bound of C\&CG algorithms. & \\
$Gap$ &                     Optimality gap of C\&CG algorithms, $(UB-LB)/LB$. & \\
$s$ &                                Auxiliary variables for the penalized problems in S-MISOCP algorithms. & \\

\end{Nomenclature}

\begin{ListOfPubliscations}
  \addchaptertocentry{List of Publications} 

   This document is based on the work of 6 articles (2 in journals with JCR, 2 under review, and 2 in conferences). These are:

\begin{enumerate}[label=\textbf{(Paper \Alph*)}]
      \item
            Ahmed R. Sayed, Cheng Wang, and Tianshu Bi. "Resilient operational strategies for power systems considering the interactions with natural gas systems." Applied energy, vol. 241, no. 1, pp. 548-66, May 2019.
      \item
            Ahmed R. Sayed, Cheng Wang, Junbo Zhao, and Tianshu Bi. "Distribution-level Robust Energy Management of Power Systems Considering Bidirectional Interactions with Gas Systems." IEEE Transactions on Smart Grid, vol. 11, no. 3, pp. 2092-2105, May 2020.
      \item
            Ahmed R. Sayed, Cheng Wang, Tianshu Bi, and Arsalan Masood. "A Tight MISOCP Formulation for the Integrated Electric-Gas System Scheduling Problem." In 2018 2nd IEEE Conference on Energy Internet and Energy System Integration (EI2), IEEE, 2018,  pp. 1-6.

      \item
            Ahmed R. Sayed, Cheng Wang, Tianshu Bi, Mohamed Abdelkarim Abdelbaky, and Arsalan Masood. "Optimal Power-Gas Flow of Integrated Electricity and Natural Gas System: A Sequential MISOCP Approach." In 2019 3rd IEEE Conference on Energy Internet and Energy System Integration (EI2), IEEE, 2019,  pp. 283-288.

      \item
            Ahmed R. Sayed, Cheng Wang, Sheng Chen, Ce Shang, and Tianshu Bi. "Two-stage Distributionally Robust Gas Contracting for Power System Operation." Submitted for publication to IEEE Transactions on Smart Grid.
      \item
            Ahmed R. Sayed, Cheng Wang,  Wei Wei, Tianshu Bi, and Mohammad Shahidehpour. "Robust Operational Equilibrium for Electricity and Gas Markets Considering Bilateral Energy and Reserve Contracts." Submitted for publication to IEEE Transactions on Power Systems.

            \vspace{4mm}

   \hspace{-15mm} During the course of the Ph.D. study, the following publications have been prepared, but they
   \vspace{-1mm}
   \hspace{-15mm} are omitted from the thesis document because they are not related to the main objective.

      \item Masood, Arsalan, Junjie Hu, Ai Xin, Ahmed R. Sayed, and Guangya Yang. "Transactive Energy for Aggregated Electric Vehicles to Reduce System Peak Load Considering Network Constraints." IEEE Access 8, 2020, pp. 31519-31529.
      \item Masood, Arsalan, Ai Xin, Junjie Hu, Salman Salman, Ahmed R. Sayed, and Mishkat Ullah Jan. "FLECH Services to Solve Grid Congestion." In 2018 2nd IEEE Conference on Energy Internet and Energy System Integration (EI2), IEEE, 2018, pp. 1-5.
      \item Sayed A. Zaki, Honglu Zhu, Jianxi Yao, Ahmed R. Sayed, and Mohamed Abdelkarim Abdelbaky "Detection and Localization the Open and Short Circuit Faults in PV Systems: A MILP Approach." Accepted for publication in 2020 The 2nd Asia Energy and Electrical Engineering Symposium (AEEES 2020).
    \end{enumerate}

\end{ListOfPubliscations}




\mainmatter 

\pagestyle{thesis} 



\chapter{Introduction} 

\label{Chapter1} 


\newcommand{\keyword}[1]{\textbf{#1}}
\newcommand{\tabhead}[1]{\textbf{#1}}
\newcommand{\code}[1]{\texttt{#1}}
\newcommand{\file}[1]{\texttt{\bfseries#1}}
\newcommand{\option}[1]{\texttt{\itshape#1}}

\section{Research Background \& Motivations}
The reliable, resilient and economic operation of critical infrastructures, such as electricity, water, natural gas, cooling, transport and telecommunication, is important to strengthen and support economic and social activities in modern society. The electric power system is the most critical infrastructure system because electricity plays an important role in the secure and continuous operation of these systems.  However, existing electric power grids experience different forms of vulnerabilities and random failures, such as extreme weather, terrorism, component aging/failure, unexpected generator or power line outages, and human errors, which may result in widespread economic and social contingencies. For example, extreme weather has caused power outages with damages ranging from \$$20$ to \$$55$ billion in the USA \cite{gordon2008protection}, blackouts such as Hurricane Katrina in $2005$ \cite{kwasinski2009telecommunications}, the Japan Earthquake in $2011$ \cite{adachi2011restoration}, Hurricane Sandy in $2012$ ($N-90$ event) \cite{henry2016impacts}, and transmission line contingencies in South Australia \cite{operator2017black}. Natural disasters, such as extreme weather are expected to increase in the future due to climate change \cite{estrada2015economic}. In addition, vulnerabilities to terrorist attacks could cause more severe system disruptions than natural disasters \cite{national2013resilience}. From $1999$ to $2002$, more than $150$ terrorist attacks on power networks worldwide have been reported \cite{national2012terrorism}. These vulnerabilities make it crucial to evaluate the performance and facilitate decision-making with regard to the power grid under contingencies by analyzing the power system vulnerability.

\begin{figure}[!ht]
        \centering
            \includegraphics[width=11cm]{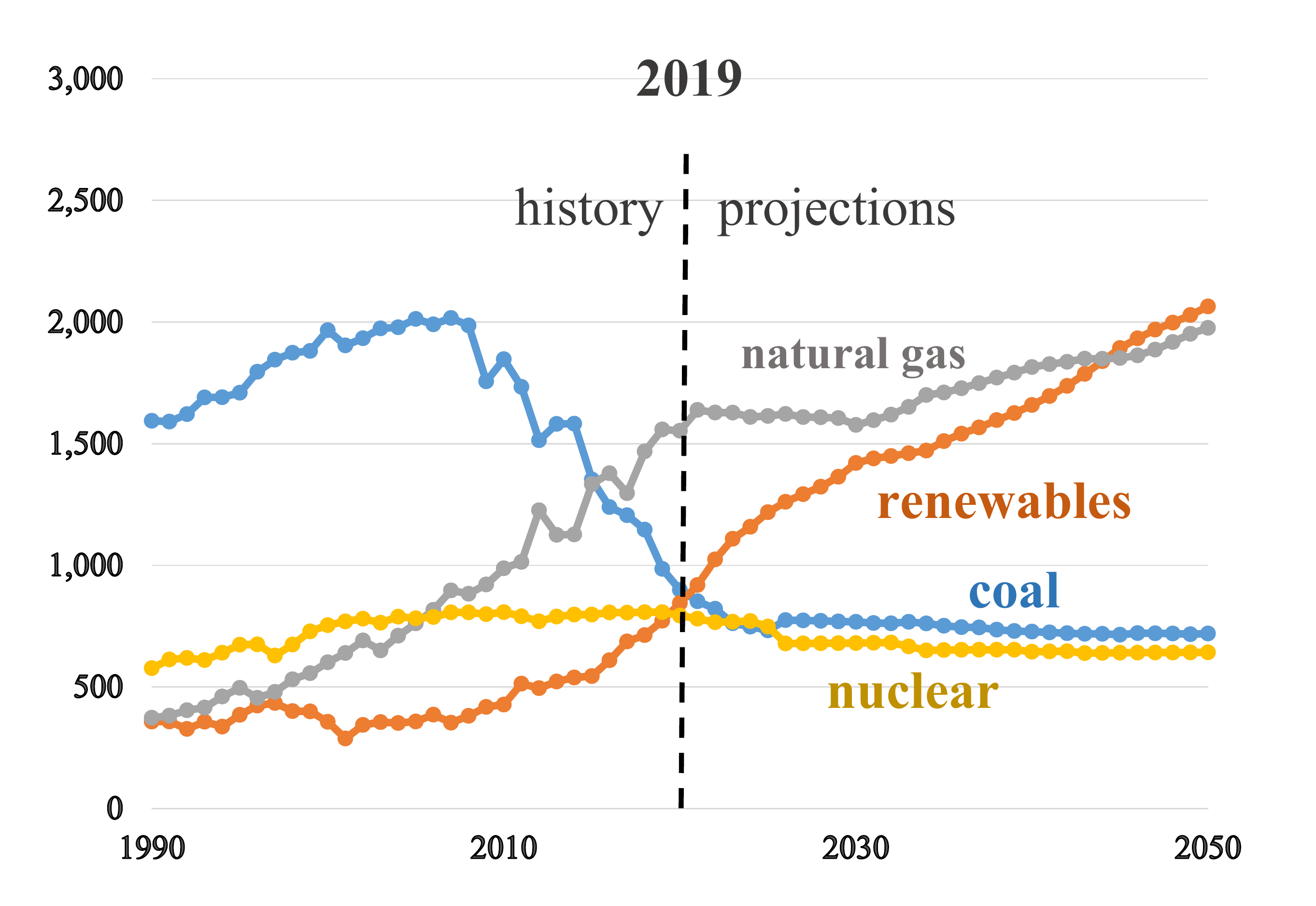} 
            \caption{Electricity generation in the US from selected fuels.\\ {\small Source: Annual Energy Outlook 2020 http://www.eia.gov}}
   \label{fig:Ch1WindGas}
    \end{figure}

Besides, climate change and environmental concerns have been major driven forces for the utilization of renewable energy resources, such as wind and solar power generation, around globe \cite{ibrahim2000renewable, zhang2016climate}.  In this regard, the top two CO\text{${}_2$} emitters, China and the US, pledged to increase their wind energy utilization to 20\% by $2030$ \cite{lu2016challenges}. According to the US energy information administration (EIA), renewable share in the US grid continue to increase from about $800$ billion MWh to $2100$ billion MWh by $2050$, as shown in Figure~\ref{fig:Ch1WindGas}. Therefore, renewable resources displace the conventional thermal power units, which today provide many services, including frequency and voltage control, generation reserves and stability services, to the reliable power system operation. However, the utilization of wind power at a large scale brings new challenges for energy management in power systems because of the variable and uncertain output features of renewables.

Power system operator (PSO) is responsible for providing reliable and economical energy management (EM) strategies, while satisfying power services to end-users, under power system uncertainties, such as fluctuations of renewables and demands, as well as contingencies. Under such uncertainties, other interconnected systems may be competent to offer some of operational flexibility services, and energy transactions between electricity and other systems are widely used to equilibrium their operation. However, these transactions are only valuable if interconnected systems have different characteristics. Each infrastructure have a specific inherent energy storage capability, therefore it can offer different levels of operational flexibility for the electricity network, and this level can be increased with proper mechanisms of energy market and communication. Besides the competition and complex energy transactions, larger transport networks and deregulation have driven forces to motivate the consideration of synergies between these networks.

Consequently, the concept of integrated energy system (IES) has gradually been realized to manage multiple subsystems together in numerous applications starting from planning to operation, breaking up the existing behavior of separate management. Traditionally, each energy system is planned, designed and operated without the responsive awareness of entanglements in other subsystems. Technical and economic knowledge as well as production and operation models developed independently may not nowadays be suitable for an efficient and secure overall operation. In the IES, production, storage, transmission, distribution and utilization of energy are coordinated and co-optimized \cite{jia2015thought}. Closer integration of different energy infrastructures might solve some of the challenges of experiencing different forms of contingences as well as integrating large-scale high-penetrating renewable energy sources, and suggest potential economic and environmental benefits for modern society. Closer integration can also enhance the efficiency, flexibility, stability and sustainability of the overall energy system.

Due to their direct and physical connection with energy consumers, IESs have become the most effective and essential part of energy supply, and IESs performance would have a great impact on the social activities and industrial practice. Therefore, in the recent economic development, ensuring the reliability of IES is a crucial need \cite{liu2016modelling, li2016reliability}.  In \cite{hughes2004developing}, IES structure is enhanced by integrating the communication systems, where the basic steps of the incorporation are discussed. A robust optimization model for the IES operation is proposed in \cite{martinez2013robust} to consider wind uncertainty in an electricity-coal-gas system. A general energy flow model is presented, while modeling hydrogen system in the IES \cite{hajimiragha2007optimal}.  Chemical energy utilization is studied with combined heating and power system in \cite{jin2007integrated}, where the utilization efficiency is improved.  Based on a cost-benefit theory, an evaluation method is established on urban IES to increase its utilization efficiency \cite{zheng2010incremental}.  An operation model for a new type of community-level IES is proposed in \cite{cartes2007novel}. Interested reader can refer to \cite{collins2017integrating, connolly2010review, wang2018review, lin2016study, mirakyan2013integrated} for the state-of-the-art reviews on the structures, models, analysis and solution methodologies for the IES.

\begin{figure}[!ht]
        \centering
            \includegraphics[width=15.5cm]{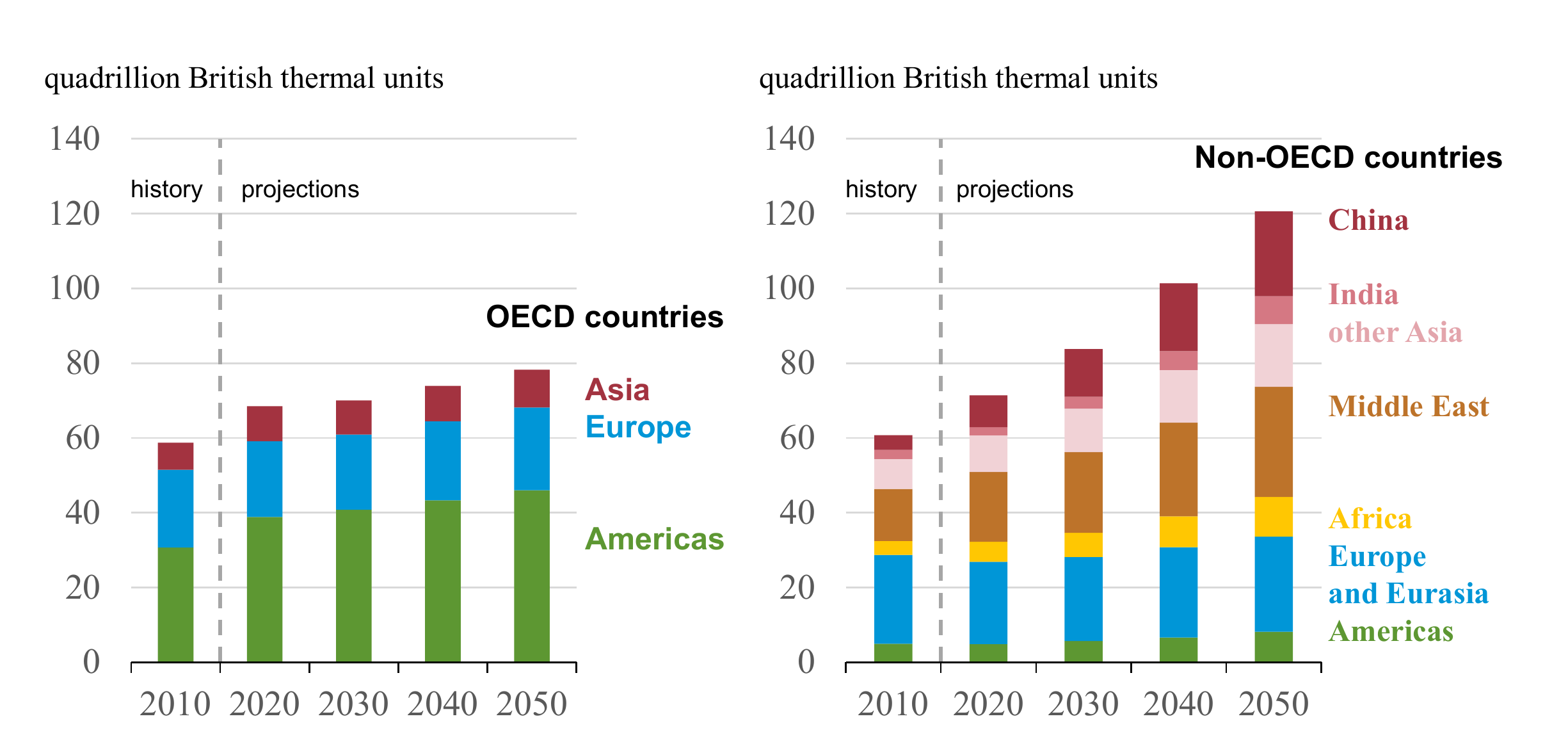} 
            \caption{World gas system consumption for OECD (left) and non-OECD (right) countries.\\ {\small Source: Annual Energy Outlook 2020 http://www.eia.gov}}
   \label{fig:Ch1Gas1}
    \end{figure}

\begin{figure}[!ht]
        \centering
            \includegraphics[width=12.5cm]{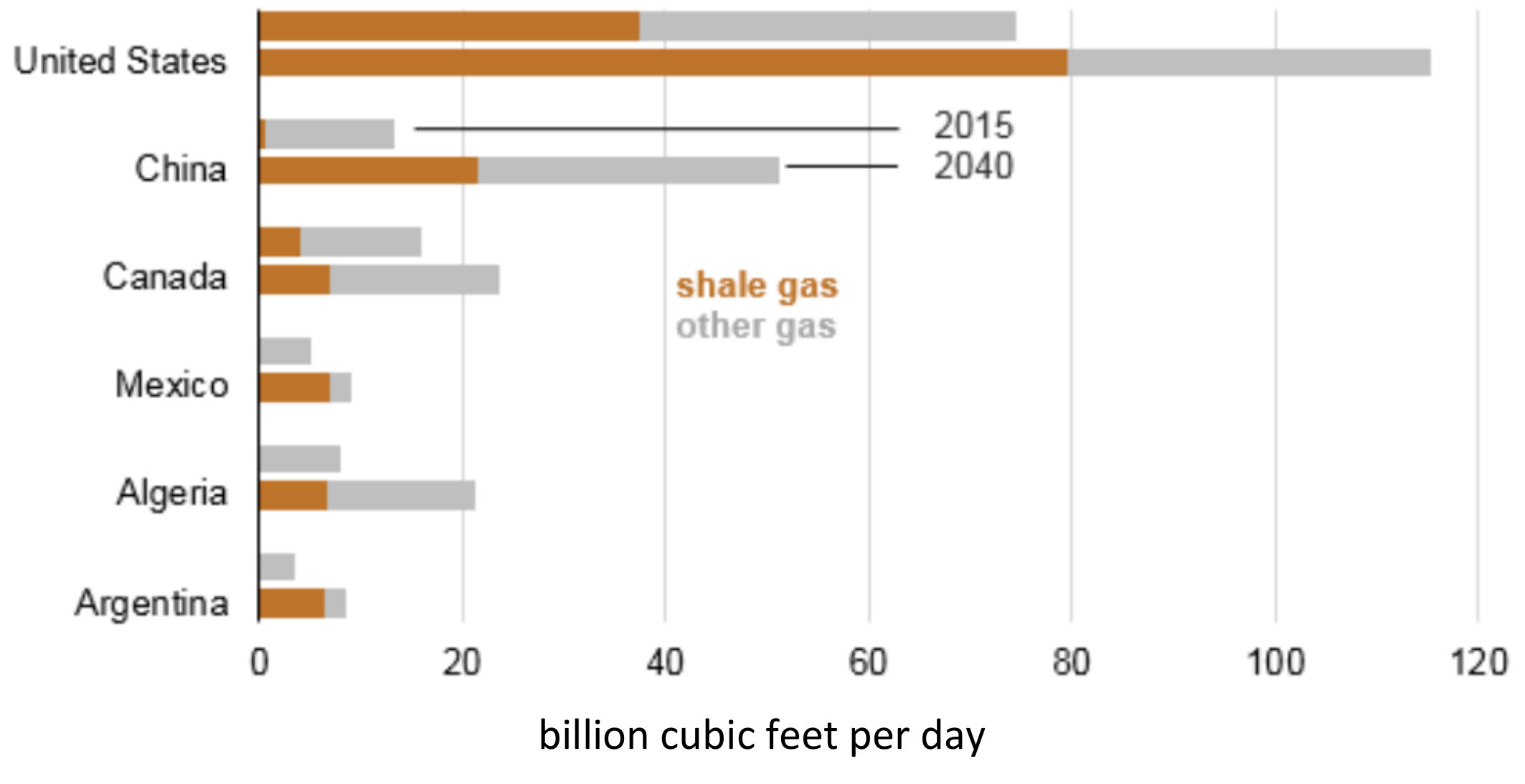} 
            \caption{World gas system production by gas types.\\ {\small Source: https://www.eia.gov}}
   \label{fig:Ch1Gas2}
    \end{figure}

\begin{figure}[!ht]
        \centering
            \includegraphics[width=15.5cm]{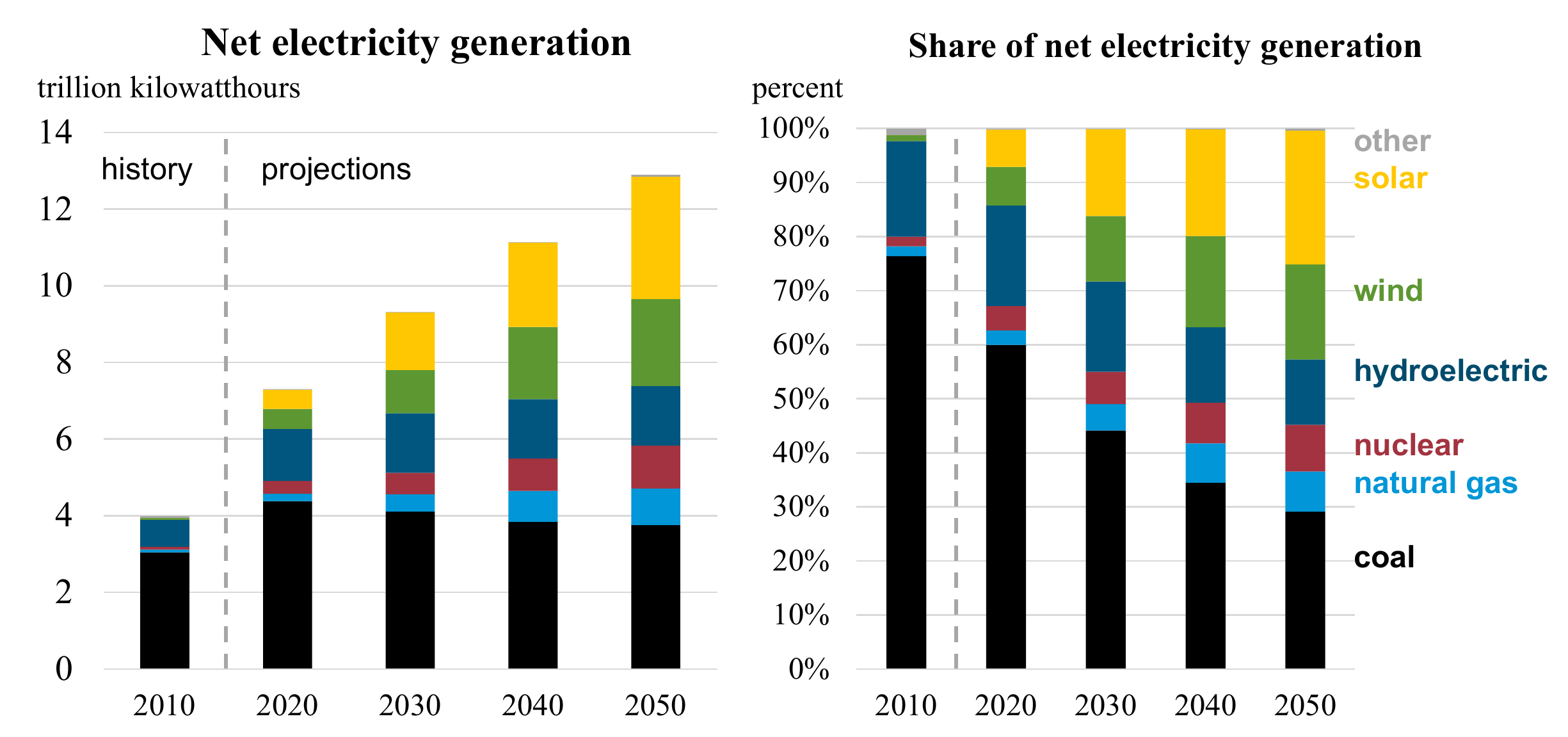} 
            \caption{Projected growth rate in the electricity generation in China.\\ {\small Source: Annual Energy Outlook 2020 http://www.eia.gov}}
   \label{fig:Ch1Gas3}
    \end{figure}

Natural gas has provided a strong importance in the global energy balance as the most energy cost-effective fossil fuel, and it act as a bridge in the transition to a near-zero emission IES system \cite{kerr2010natural, EIA, myhrvold2012greenhouse}. This progress was resulted from $1980$s due to new concerns from a potential global warming. Recently, because of advanced technologies in gas cracking and extraction, natural gas reserves have encouraged the growth of this energy worldwide. Besides, among other fuels, natural gas has a great position predominantly due to robust production, low carbon emission and abundant accessibility of gas sources. Moreover, thanks to the shale gas revolution, which is enhanced by the development in horizontal drilling and hydraulic cracking technologies, the gas prices are decreasing significantly \cite{stevens2012shale}. Therefore, high-efficiency gas is promoted as the second largest energy source/consumption over the world \cite{EIA}. Figure~\ref{fig:Ch1Gas1} displays the growth in gas consumption from $2010$ to $2014$ for OECD countries (left) and non-OECD countries (right), indicating that the worldwide consumption may be raised from $113$ trillion cubic feet (Tcf) to $185$ Tcf. Shale gas is projected to increase by $30\%$ of world gas production in $2040$. Figure~\ref{fig:Ch1Gas2} exhibits the shale gas production, which grows from $42$ billion cubic feet per day (Bcf/d) to $168$ Bcf/d in $2015-2040$ for the six countries that have shale resources.
Although coal and nuclear are expected to remain as the main fuels used in electricity generation, global environmental concerns and energy prices have motivated the developments to produce electricity from renewable and natural gas energies, respectively, as shown in Figure~\ref{fig:Ch1WindGas}. According to EIA, China will increase renewable energy utilization from about $1$ TWh to $5$ TWh, along with the growth deployment of natural gas to $10\%$ of electricity generation. This situation is illustrated in Figure~\ref{fig:Ch1Gas3}.

Beside the significant development in the energy market mechanisms, which offers a free competition for private investors, the technologies of gas-fired power units (GPUs), due to their environmental benefits, operational flexibility, high efficiency and fast response, led to increase natural gas share in the electricity sector. Compared with the traditional coal-fired power units, GPUs can be constructed and operated in less than two years rather than ten years, and they could facilitate greenhouse gas emissions \cite{myhrvold2012greenhouse, qadrdan2015impact, zhang2016climate}. An explanation of the quick response of GPUs to handle power demand fluctuations is introduced in \cite{watson2001constructing}, where global efficiencies of GPUs reach $60\%$, unlike coal-fired units, which have $33\%$ efficiency at most. As a result, it is expected that $60\%$ of new electric power units will be fueled by natural gas by $2035$ \cite{EIA}. The increase in GPUs deployment lead to a growing interdependence between the top two complex systems, electricity and gas that poses systemic challenges for resilient and economic operation of the interacted systems.

This interdependency is intensified due to not only the wide deployment of GPUs but also the advanced technologies of power-to-gas (P2G) facilities, which are the most well-qualified solution for the long-term energy storage in the existing bulk power system integrated large-scale renewable energy. Introducing more flexibility to mitigate renewable energy curtailment is one of the main challenges in power system operation \cite{johnson2014assessment, lund2015review, kondziella2016flexibility}. Because the natural gas can be stored with large capacities in a cost-effective manner, P2G facilities are recently employed to effectively convert electricity into gas, which further is stored, transported and reutilized by gas networks. P2G facilities allow power system to be interacted with other energy systems, such as transport and heating systems. The idea to convert electricity into gas for storage was firstly developed in $1978$ \cite{long1980method}. Several pilot sites are constructed throughout the world, indicating the strong importance in this technologies \cite{bailera2017power}. The existing researches \cite{yang2018modeling, chuan2017robust, belderbos2019storage} agree that P2G facilities can help the power system operation in mitigating the fluctuations of energy loads, the surplus renewable energy, recycling CO\textsubscript{2}  and offering ancillary services.

Neglecting the physical interactions between the two systems may not provide the optimal decision for power system operators and it may cause physical violations, such as under/over nodal pressure and/or gas well production capacity violations, in gas systems \cite{zlotnik2016coordinated}. Any inadequacy in the coordination between the two systems may result a sequence of shutdown of electric generators, and may lead to a blackout. However, this increased interaction, which is also referred to as an integrated electric-gas systems (IEGSs) in the literature, encounters issues related to the secure, reliable, and resilient operation of the IEGS.

\section{Literature Review}
The short-term operation of power and gas systems has been influenced with the increasing interdependence on each other.  For instance, power system operating decisions are influenced by natural gas prices, which are received from gas system operator (GSO), i.e., consumed gas by GPUs as well as the electricity prices will be directly affected. Also the production schedule of gas system depends on the injected gas to/from GPUs/P2G units, and further will impact on the operational costs. This situation may be more challengeable when power and gas demands are simultaneously peak. Furthermore, gas supplied to GPUs has high priority to be curtailed under gas system congestion because this amount of gas is usually signed as interruptible contracts \cite{liu2011coordinated}.  Therefore, PSO are progressively tending to appeal more flexibility from gas systems to assure continuous supply. On the other hand, facing volatile energy demands, which are difficult to forecast, represents security issues in the operation of IEGS because guarantee gas supply continuity under varying loads is not straightforward task for GSO.  Natural gas infrastructure take long time to response due to slow dynamics of gas, therefore, its operating decisions must be early and timely planned. The travelling velocity of the gas is much slower than electricity, its maximum value is about $50$ km/h \cite{gallagher2013natural}.

To this end, there are many economical and physical interactions, which may impact the operation of one system corresponding to the other, and appropriate planning and operation are necessary for accurate coordination in production, delivery and utilization, considering uncertainties of the integrated renewable energies as well as system contingencies.  Although different research works have been conducted in the pertinent literature in the past $20$ years to address the issues of interdependency and resilient-economic operation of IEGSs \cite{he2018coordination}, there are significant research gaps between the existing works and the practical application, that need to be fulfilled.
\begin{enumerate}
  \item Most studies on the coordinated operation of power and gas systems share one underlying assumption, namely, the existences of one operator or utility that has full control and operation authority over the both systems. This operator minimizes all costs associated with energy production and provides optimal decisions for the combined system. However, in industrial practice, there are significant institutional and administrative barriers to operate the two systems in a holistic manner \cite{chen2019operational}. Power system(s) and gas system(s) are operated by different utilities and they are unsynchronized in most countries and regions as in European countries \cite{bertoldi2006energy} and in China \cite{deng2017study}. This lack of synchronization indicates that the total fuel cost minimization determined by the IEGS models might not be a realistic operational objective for autonomous sub-systems and, therefore, bilateral energy trading is inevitable.
  \item	For the power system resilient operation, the operational mode of the electric power system in post-contingency conditions might be significantly different from that of the pre-contingency stage, such as sudden start-up or shut-down of fast-response generators and rapid increases or decreases in generator outputs to minimize operating losses; a similar trend is observed for the gas demands of GPUs. Moreover, GPUs usually provide interruptible gas supply services according to current gas industrial practices \cite{liu2009security} and the gas contracts are usually determined by considering day-ahead contracts because real-time contracting would be costly and inconvenient \cite{bouras2016using}. In other words, GPUs cannot execute the planned regulations without appropriate gas contracting. However, this economic interactions between the two systems, including reserved gas contracts, have been neglected in the existing resilient operation models.
  \item	In power system integrating large-scale renewable energy sources, the real-time operation of power system might be changed from the day-ahead dispatch largely owing to the renewable uncertainties. This means the outputs of GPUs and P2G facilities might deviate from their day-ahead schedules, to mitigate the operation losses or the surplus wind generation. In this regard, modeling bidirectional energy contracts is necessary. Moreover, the purchased gas can be divided into two parts, including the firm part for the dispatched outputs day-ahead and the variable one for the real-time utilized reserves, respectively, which suggests the size of the contract is directly related to the operation strategies of power systems. The recent studies only consider the modeling of firm gas contracts in power system operation \cite{wang2017strategic, liu2009security}, where the reserved gas contracts and the impacts of uncertainty on the contracts as well as P2G gas contracts are missing.
  \item	No attempt has been found in the literature that considers power system uncertainties in a robust optimization (RO) approach to analyze the equilibrium between the electricity and gas markets, where the main difficulty is how to reflect the impacts of power system uncertainties on the gas system, and vice versa.

  \item	Beside the drawbacks in decision-making framework modeling, there are also computational difficulties to identify the optimal and feasible decisions for the interacted power and gas systems. Finding the optimal gas flow (OGF) has drawn attention from researchers due to its non-convexity, as it is originated from the nonlinear partial differential equations, which are commonly reformulated by piecewise linear approximation (PLA) methods, and second-order cone (SOC) relaxation. However, these reformulations either introduce high computational burden due to large number of integer variables or infeasible OGF due to inexact relaxation. It should be noted that the steady-state gas flow model is widely adopted in the literature, neglecting the mass flow rate inside pipelines and slow gas dynamics that may provide suboptimal solutions. Moreover, most of the recent studies mainly concentrate on transmission level, however, stronger interactions in the IEGS are observed in the distribution-level \cite{evans2013age, wang2018convex}, in which the active and reactive power are coupled as the bus voltages are notably influenced by active power variations. Furthermore, additional computational challenges in the coupled power and gas system operation, especially under uncertainties, would be introduced.
\end{enumerate}

\section{Research Objectives and Challenges}
    The main purpose of this thesis is to revisit the resilient and economic operation of power systems against contingencies as well as renewable energy uncertainties in terms of decision-making, considering the interactions of power systems with gas systems. With this objective, this work seeks to satisfy the aforementioned gaps between the existing researches and the industrial application concerning the lack of neglecting physical and/or economic interactions with gas systems. The proposed operation models must be able to provide a level of reliability and flexibility, which is required for PSO, and secure and feasible optimal decisions for both systems. To achieve that, this thesis first introduces how to model and optimize accurately, in both transmission and distribution levels, the coupled power and gas system. Afterward, different power system dispatch models are developed and efficiently optimized against $N-k$ contingencies and volatile wind power uncertainties, where energy contracts are modeled. Finally, pool-based market mechanism for electricity and gas markets is proposed to separately clear the interacted and independent markets under uncertainties. This purpose is achieved by realizing the following explicit objectives and challenges:
    \begin{enumerate}
      \item Introducing an accurate and efficient detailed formulations of both natural gas system and power system that can be incorporated in the optimization models, and providing the state-of-the-art of modeling the interdependent power and gas systems.

            Compared with power flow, the gas flow travels with limited velocities due to the slow dynamics of gas system. Therefore, gas flow needs a response time to be delivered. Additionally, natural gas is compressible and it can be stored in pipelines, known as line pack. Such circumstance should be considered in modeling the short-term operation. The DC power flow is usually employed to formulate the power network constraints in the coupled power and gas system optimization problems. However, in the distribution level, the active and reactive power are coupled as the bus voltages are notably influenced by active power variations, and AC power flow is the only choice to be adopted.

      \item Proposing accurate, applicable and efficient solution methodologies for the coupled power and gas system optimization problems, in both transmission and distribution levels, considering the gas dynamics.

            Different reformulation and solution methods are developed in the literature to handle the general gas flow and/or power flow equations. The proposed methods must outperform the existing ones in terms of accuracy, optimality and computational burden. The feasibility and quick convergence must be guaranteed. This is not easy task, not only because power and gas systems are NP-hard optimization problems, but also they deliver their energies over the time frame,

    \item	Establishing a resilient operational framework for power systems that considers physical and economic interactions with gas systems.

        The physical interaction can be achieved by adding the security and feasibility constraints of the gas system into the proposed framework, while the economic interaction can be completed by modeling firm and reserved gas contracts to be utilized in the pre-contingency and the post-contingency stages, respectively. The proposed model must be robust against $N-k$ contingencies and be able to find the optimal gas contracts and defensive strategy. In fact, framework cannot be directly solved by decomposition algorithms, and a nested one is required.

    \item Develop a robust economic dispatch model for power distribution systems that considers bidirectional day-ahead gas contracting.

        To consider bidirectional energy trade with the gas system, two types of gas contracts must be modeled, namely, gas-to-power (G2P) contracts for GPUs and P2G contracts for P2G facilities. The non-convex nonlinear power and gas flow equations must be incorporated in the optimization model. Finding the optimal and feasible solution of the two-stage model is challengeable, because of the presence of non-convex equations in the real-time and day-ahead stages, therefore a novel method is needed that encourages the fifth objective.

    \item  Develop an efficient approach for solving two-stage robust optimization (RO) models with the non-convex nonlinear power and gas flow equations.

        Convex relaxation methods have been implemented to find the optimal power-gas flow (OPGF) owing to their computational benefits. However, convex relaxation is exact under mild conditions, therefore, they are not tight enough, and the solution exactness cannot be guaranteed. The non-convex equations can be reformulated as difference-of-convex programming (DCP) functions, and a sequential convex procedure (SCP) can be adopted to find a more accurate and feasible solution. Consequently, a quadrable-loop procedure based on the SCP and the nested column and constraint (NC\&CG) algorithm can be adopted to tackle the two-stage RO model.

    \item Establishing a distributionally robust economic dispatch model for power systems that considers bidirectional two-stage gas contracting.

        Incorporating both the day-ahead and real-time gas contracts in power system operation is necessary to consider the costly real-time contracts for the low-probability utilized reserved GPU outputs in practice. The proposed model can be insensitive on the exact probability distribution of renewable generation, and the decisions robustness and conservativeness can be adjusted.

    \item	Characterizing a robust operational equilibrium for the interactive markets of power and gas systems, considering the impacts of the uncertainties of wind generation outputs on the two markets.

        The proposed framework considers that the two markets must be independently operated and allow limited information exchange, including only the prices and demands of both systems for contract agreements. Besides, under equilibrium, impacts of power system uncertainties must be reflected on the gas system, and vice versa. The superiority of the robust operational equilibrium over the deterministic one and its effectiveness under limited data exchange must be confirmed.

    \item Validate proposed frameworks and solution methodologies with different numerical simulations and case studies.

        Data and test systems employed in case studies must be able to illustrate the applicability of the proposed frameworks and suggested approaches in the industrial practice. Unfortunately, during the thesis working period it was not easy to find real system data. However, the proposals are validated on test systems used in the literature, or similar to them.
    \end{enumerate}
\section{Contributions and Publications}
    The major contribution of this research is to develop optimization models for the coordinated power and gas systems that able to provide optimal resilient and economic operational strategies for the PSO considering the bidirectional physical and economic interactions with gas systems. System operators, planners and utilities can employ the proposed models and algorithms to optimally and fast identify the decision variables for both electric power and natural gas systems.

    Initially, Chapter~\ref{Chapter2} shows how to model and coordinate the interacted power and gas systems. Then, Chapter~\ref{Chapter3} proposes two solution methodologies to handle the non-convexity of the power and gas flow quadratic equations. The main contributions of this chapter are:    (1) A novel SOC relaxation method is adopted to convexify the Weymouth equation considering both the gas flow dynamics and bidirectional of gas flow, resulting in a MISOCP framework; (2) The multi-slack-node method with the Levenberg-Marquardt algorithm is derived as GFC method to obtain a feasible, near to optimal, energy flow solution for the IEGS; (3) Feasibility and accuracy guarantee. A S-MISCOP algorithm is proposed to find the OPGF for IEGS. Based on DCP, the nonconvex branch power flow and Weymouth gas flow equalities are decomposed as MISOCP constraints, which are easier to be solved than the original nonlinear problem. The proposed algorithm is a sequence of solving penalized MISOCP problems, and its feasibility is guaranteed by controlling its penalties;  (4)  Fast and reliable convergence. Because S-MISOCP algorithm is a local heuristic approach, it is influenced by the initial point. Therefore, a high-quality initial point is suggested and an adaptive penalty growth rate is developed to adjust the main objective weight in the penalized problem; (5)  Numerical simulations have been conducted to validate the proposed GFC method and the S-MISOCP algorithm performances, the recommended initial point effectiveness, and the adaptive penalty rate adoption.

    These two solution methodologies have been published as:
    \begin{itemize}
      \item
            Ahmed R. Sayed, Cheng Wang, Tianshu Bi, and Arsalan Masood. "A Tight MISOCP Formulation for the Integrated Electric-Gas System Scheduling Problem." In 2018 2nd IEEE Conference on Energy Internet and Energy System Integration (EI2), IEEE, 2018,  pp. 1-6. DOI:  https://doi.org/10.1109/EI2.2018.8582239
      \item
            Ahmed R. Sayed, Cheng Wang, Tianshu Bi, Mohamed Abdelkarim Abdelbaky, and Arsalan Masood. "Optimal Power-Gas Flow of Integrated Electricity and Natural Gas System: A Sequential MISOCP Approach." In 2019 3rd IEEE Conference on Energy Internet and Energy System Integration (EI2), IEEE, 2019,  pp. 283-288.
            \vspace{-2mm}

            DOI:  https://doi.org/10.1109/EI247390.2019.9062250
    \end{itemize}

    Throughout Chapter~\ref{Chapter4}, this thesis contributes with formulating a robust day-ahead dispatch model for electric power system against $N-k$ contingencies. The main contributions in this chapter are as follows; (1) A tri-level resilient operational framework of power systems that considers contracts with gas systems is established, where firm gas supply contracts and gas reserve contracts are considered in the pre-contingency and the post-contingency stages, respectively. To the best of our knowledge, this is the first attempt to consider gas reserve contracts in a robust model. The operational constraints of the gas systems are considered, which guarantees their operational feasibility and security in both the pre- and post-contingency stages. Compared with IEGS robust models presented in the literature, the proposed model considers the dynamic state of the gas system. Moreover, as a result of considering contracts, a new kind of attack strategy emerges, i.e., the consumption of gas below/above the reserved (contracted) values; (2) Unlike the most tri-level models where the lower level decision variables are continuous, there are binary variables in the lower level optimization problem in the proposed model. The additional binaries originate from the linearization of the nonlinear non-convex Weymouth equation, as well as the on/off control of the generators in the post-contingency stage, and they are used to determine the potential attack region \cite{wang2016robust}. Therefore, the NC\&CG algorithm proposed by \cite{zeng2013solving} is applied to solve the proposed tri-level model after adjusting its stopping criteria.  This model has been published as:
    \begin{itemize}
      \item
            Ahmed R. Sayed, Cheng Wang, and Tianshu Bi. "Resilient operational strategies for power systems considering the interactions with natural gas systems." Applied energy, 2019 May, 1;241:548-66, DOI: https://doi.org/10.1016/j.apenergy.2019.03.053
    \end{itemize}

    Then, in Chapter~\ref{Chapter5}, two operational models for optimal power system operation with bidirectional gas contracts are proposed. The first is a robust EM model for the PDN with RPG uncertainties. The main contributions of this study are twofold; (1) A tri-level robust dispatch model is established for the PDN considering both physical and economic interactions with the gas systems. Specifically, the physical interaction is achieved by adding the security and feasibility constraints of the gas system into the EM problem of the PDN, while the economic interaction is completed by modeling firm and reserved gas contracts for both G2P and P2G; (2)  A quadruple-loop algorithm for the proposed robust EM problem of the PDN is devised, where the second and forth loops are S-MISOCP algorithms to enhance the solution feasibility in the day-ahead and real-time dispatch stages, respectively, and the first and third loops are column-and-constraint (C\&CG) algorithms to tackle the tri-level decision-making structure with binary recourse. This work has been published as:
    \begin{itemize}
      \item
            Ahmed R. Sayed, Cheng Wang, Junbo Zhao, and Tianshu Bi. "Distribution-level Robust Energy Management of Power Systems Considering Bidirectional Interactions with Gas Systems." IEEE Transactions on Smart Grid, vol. 11, no. 3, pp. 2092-2105, May 2020. DOI: https://doi.org/10.1109/TSG.2019.2947219
    \end{itemize}

    The second model is a distributionally robust two-stage contracting model. Compared with the literature, the salient feature of this study is that a two-stage distributionally robust model is proposed for signing bidirectional energy contracts with gas systems from the perspective of power system operator (PSO). To the best knowledge of the authors, this work is the first attempt to incorporate both the day-ahead and real-time gas contracts in power system operation. This model is submitted for publication as
    \begin{itemize}
      \item
            Ahmed R. Sayed, Cheng Wang, Sheng Chen, Ce Shang, and Tianshu Bi. "Two-stage Distributionally Robust Gas Contracting for Power System Operation." Submitted for publication to IEEE Transactions on Smart Grid.
    \end{itemize}

    Finally, a method for the robust operational equilibrium seeking of the coupled electricity and gas markets is proposed. The main innovations are multi-fold: (1) A robust operational equilibrium for the coupled electricity and gas markets is characterized considering the uncertainties of RPG as well as bidirectional energy and reserve contracts; (2)   Inspired by \cite{ye2016uncertainty} and     \cite{fang2019introducing}, the marginal energy and reserve prices for electricity and gas markets are derived based on the cost causation principle to reflect the impacts of uncertainties; (3) The BRD algorithm is proposed to identify the characterized operational equilibrium, where the electricity and gas markets are separately cleared by the C\&CG and NC\&CG algorithms, respectively; (4) The superiority of the robust operational equilibrium over the deterministic one, its effectiveness under limited data exchange, the importance of considering the gas dynamics, and solution procedure performance have been verified by numerical results. This work is submitted for publication as
    \begin{itemize}
      \item
            Ahmed R. Sayed, Cheng Wang,  Wei Wei, Tianshu Bi, and Mohammad Shahidehpour. "Robust Operational Equilibrium for Electricity and Gas Markets Considering Bilateral Energy and Reserve Contracts." Submitted for publication to IEEE Transactions on Power Systems.
    \end{itemize}

\vspace{5mm}

    The abovementioned models and algorithms are intended to be employed by energy companies, planners and operators, interested in resilient and economic operations for power system with the physical and economic interactions with gas systems. We hope that this thesis gets some value to the “State Grid Corporation of China” and the "Egyptian Electricity Holding Company and Subsidiaries" in the near future, to their operation of Chinese and Egyptian electricity networks with a high level of security, economy and resiliency.

\section{Thesis Outline}
\begin{figure}[!ht]
        \centering
            \includegraphics[width=15.5cm]{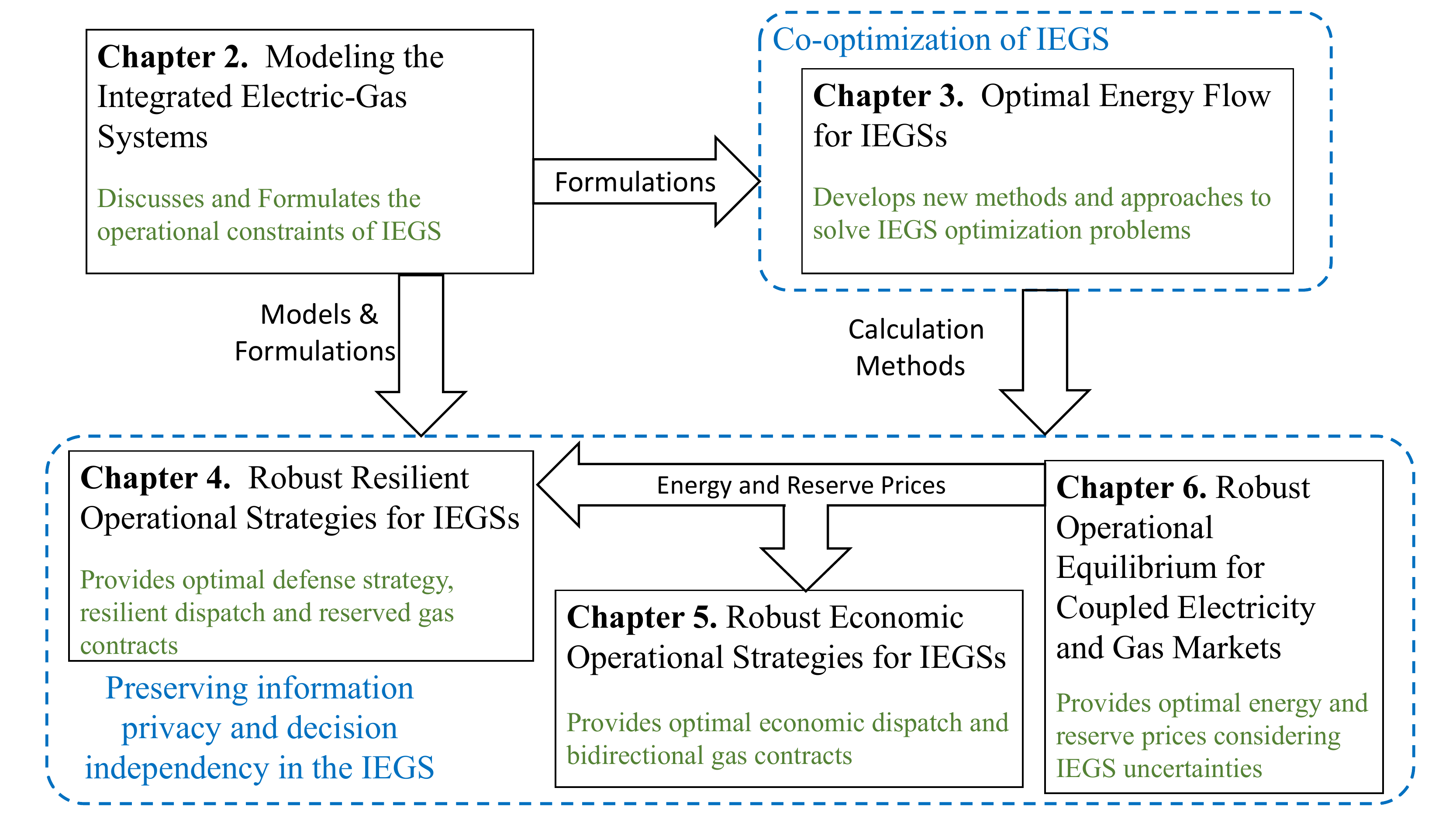} 
            \caption{Thesis structure. }
   \label{fig:Ch1Outline}
    \end{figure}

    The remainder of this document is organized into six chapters, each divided into sections and subsections. Figure~\ref{fig:Ch1Outline} displays the overall thesis structure to show the connections among chapters, indicating the main motivations of each one.  Chapters begin with a brief introduction to describe their motivations and contents. Each mathematical formulation or advised algorithms are validated with different numerical examples, including decision-making strategies and computational comparisons. The scalability tests have been conducted for all proposed models and approaches. It should be noted that all models in the thesis consider the gas dynamics. Finally, each chapter terminates with its conclusions. The proposed models and the thesis' contributions introduced in each chapter are listed in Table~\ref{tab:Ch1Chapters}.

    \textbf{Chapter~\ref{Chapter2}} starts by a description of the physical structure of gas system, indicating the mathematical model of the principal components. The set of PDE, which represents the gas system dynamics is listed. Then, the approximated dynamic- and steady-state gas flow models are depicted. Chapter 2 introduces the electric power system modeling, focusing on formulations, applications and versions of the optimal power flow (OPF). The similarities and differences between electricity and gas system and the coupling components are illustrated. Finally, a survey on the available coordination strategies between the two systems is presented.

   \textbf{Chapter~\ref{Chapter3}} focuses on identifying the optimal power-gas flow (OPGF) in the coupled power and gas system. Initially, different solution methodologies for solving gas system and power system optimization problems are illustrated. Then, based on convex relaxation methods, a gas flow correction methods is proposed to guarantee the exactness and feasibility from the relaxed formulation. Case studies are conducted to illustrate the effectiveness and features of the proposed two methodology, and to compare with the widely adopted PLA methods. Then, a sequential-MISOCP algorithm is designed to solve OPGF, considering the AC-OPF at the distribution level. Case studies validate the accuracy and feasibility of the proposed algorithm as well as its performance and convergence are discussed.

    \textbf{Chapter~\ref{Chapter4}} starts with a detailed introduction of power system resilience models, IEGS resilience models and the proposed model, indicating the main contributions of the work in this chapter. The pre- and post- contingency operational constraints are illustrated and gas contracts are modeled considering its two subcontracts, namely firm and reserved gas contracts. The proposed model is solved by NC\&CG algorithm. Finally, the necessity of considering economic and physical interactions between power systems and natural gas systems and the effectiveness of the proposed model and algorithm are verified by numerical simulations of two test systems.

    \textbf{Chapter~\ref{Chapter5}} focuses on the economic operation of power system against wind power uncertainties. Two optimization models are proposed, namely robust day-ahead operation with bidirectional gas contracting and two-stage distributionally robust gas contracting models, respectively. The main features, problem formulation and solution methodology and case studies of each model are separately presented.

    \textbf{Chapter~\ref{Chapter6}} beginnings with a detailed introduction of the existing studies from market perspectives, indicating the main contributions of the presented work. A pool-based market mechanism is proposed and the electricity and gas markets are modeled. Each market is individually cleared and the operational equilibrium is identified by the best-response decomposition (BRD) algorithm. Finally, case studies are conducted to verify the superiority of the robust operational equilibrium over the deterministic one and its effectiveness under limited data exchange.

    This document finishes with the main conclusions and future works in \textbf{Chapter~\ref{Chapter7}}.

 \begin{table}[!hbtp]
        \caption{The proposed models and contributions in the thesis} \label{tab:Ch1Chapters}
        \centering
        \includegraphics[width=15cm]{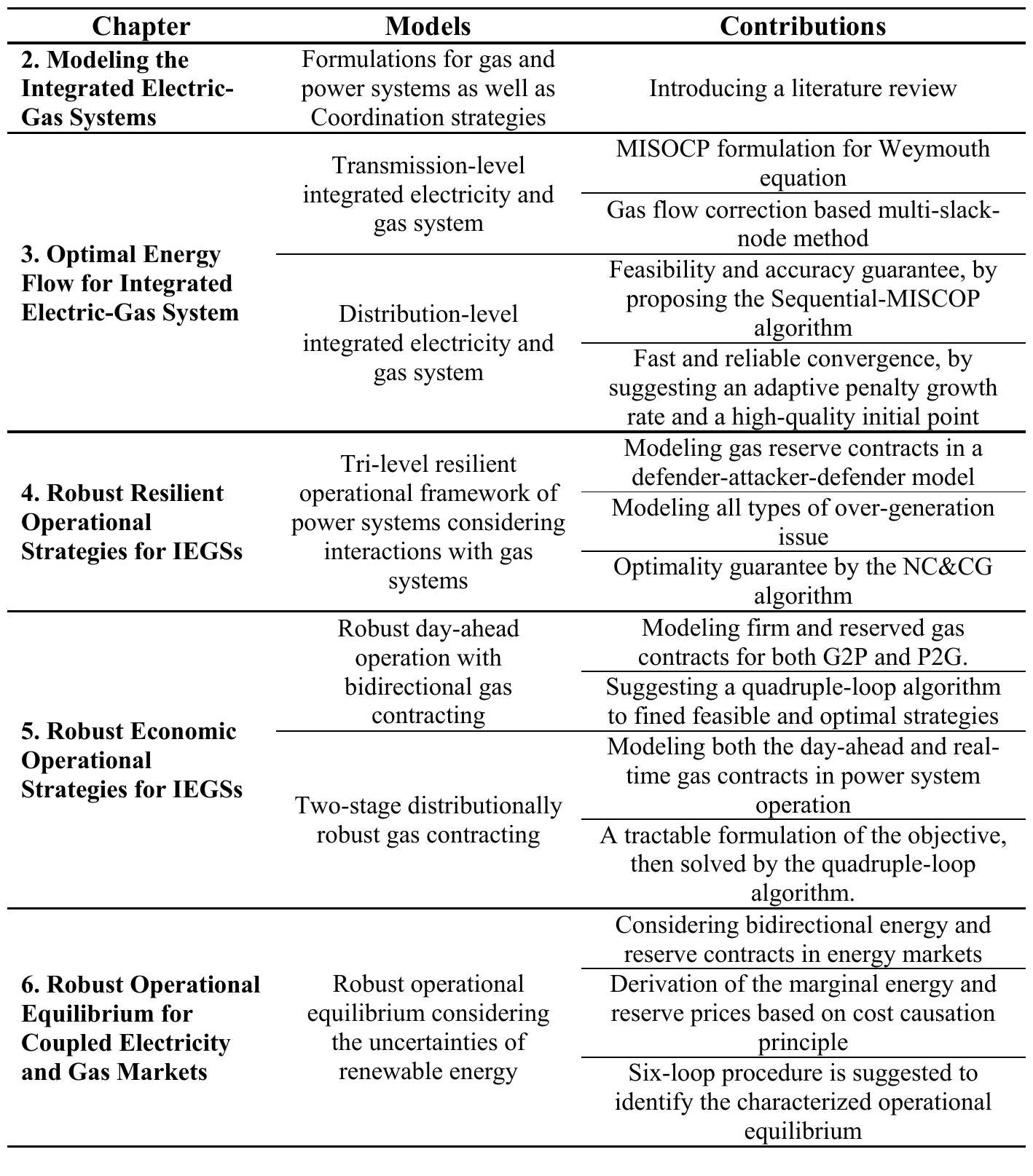}
    \end{table}


\chapter{Modeling the Integrated Electric-Gas Systems} 

\label{Chapter2} 



     Predominantly due to the propagation of large fleet natural gas power units (GPUs) and the technology developments of power-to-gas (P2G) facilities, interactions between power and gas systems have been noticeably enhanced in transmission \cite{yang2018modeling} and distribution \cite{wang2019convex, saldarriaga2013holistic} levels. These interactions not only bring significant economic and environmental benefits to the society but also provide additional operating flexibilities, which are essential to handle fluctuations of renewable energy and demands, as well as contingencies \cite{zlotnik2016coordinated, he2018coordination}. Moreover, neglecting these interactions in power system operation may not only result in infeasible operation status in the gas systems \cite{zlotnik2016coordinated}, but also increase the decision-making operation costs of both systems \cite{wang2017strategic}. This intensified interaction gradually brings quite a few research interests on modeling, simulation and analyzing the coupled power and natural gas systems \cite{he2017robustb, liu2018operational}. This chapter focuses on the advancements of modeling and coordinating the gas and power systems that could be adopted to analyze the resilient and economic operations of the coupled system.

     Section \ref{sec:Ch2Gas} focuses on modeling the gas systems. It starts by discussing the physical structure of gas system to provide a brief description of the principle components and their mathematical models. A set of partial differential equations (PDE) are defined, and, consequently, the approximated dynamic- and steady-sate gas flow models are obtained. Section \ref{sec:Ch2Gas} mainly focuses on modeling the optimal power flow (OPF) and its applications and versions. The AC-OPF and its approximated DC-OPF are arranged at the end. Finally, an analogy between power and gas systems, indicating the similarities and differences between the two energy systems as well as modeling the physical interactions are provided in Section \ref{sec:Ch2Interdep}. Finally, the existing coordination scenarios in operation are discussed.

\section{Natural Gas System Modeling} \label{sec:Ch2Gas}
    There are different gas models have been found in the literature that mathematically formulate the relationships between the physical quantities of the gas infrastructure. The existing models can be classified into four categories \cite{Midthun2007Optimization, correa2015optimal}: i) investment models, which are used to provide the planning decisions for site investments; ii) value chain models, in which, the complete system stages, including production, transportation, storage and marketing, are simultaneously optimized. iii) transportation models, which are adopted for studying the gas industrial. A tradeoff between the accuracy of decision-making strategies and the complexity of the gas transportation model is important to solve the model. A steady-state model is solved by the simplex algorithm in \cite{de2000gas}. A multi-period flow model is presented in \cite{martin2006mixed} to minimize the operational costs of compressors; iv) equilibrium models, which are used in gas markets, they are usually formulated as complementarity problems \cite{zhuang2008complement}. This thesis focuses on the transportation model, which is applicable to be incorporated in the power system optimization problems.

\subsection{Physical Structure of Gas System}
\begin{figure}[th]
\centering
\includegraphics[width=9.5cm]{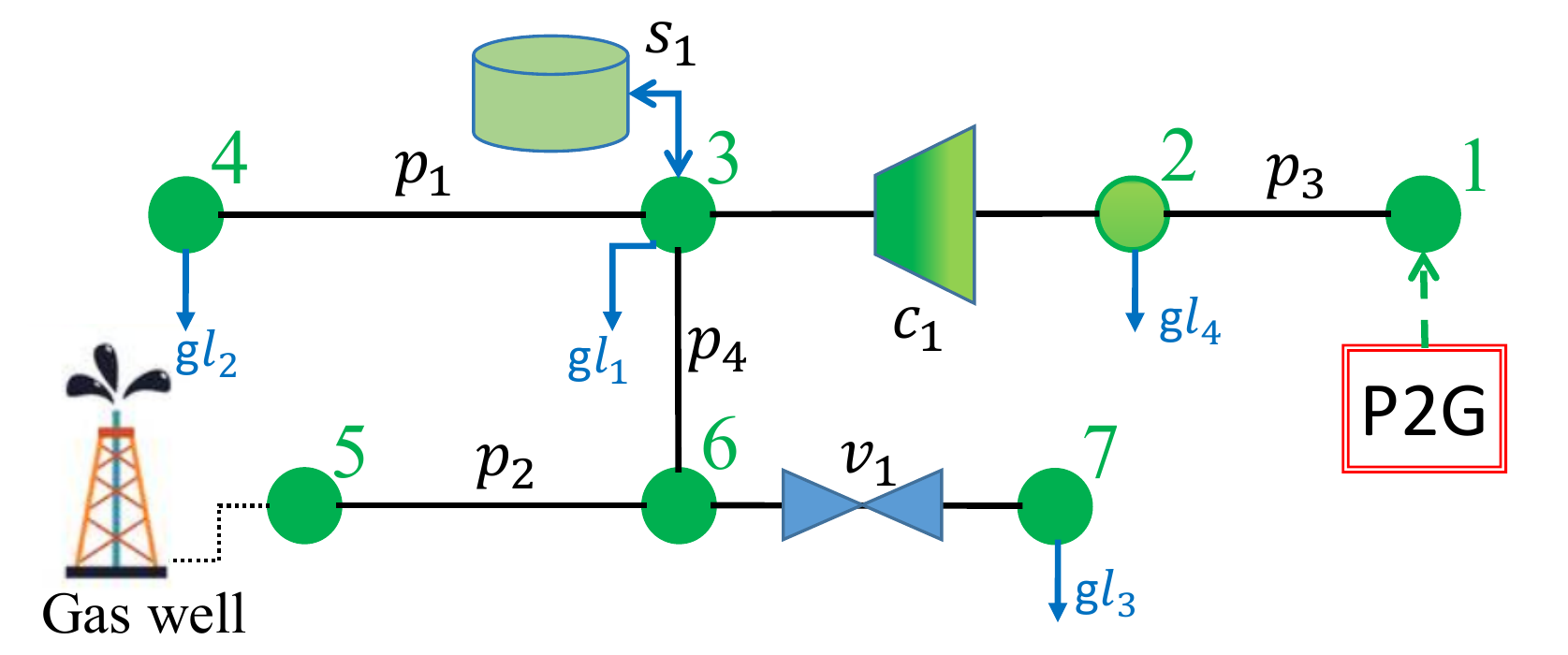} 
\caption{Gas System Topology}
\label{fig:Ch2GasTopology}
\end{figure}

    The gas system comprises several components, which are serving the delivery process, starting from production stage, moving with transportation stage, reaching to the consumption stage. All components are connected together by gas nodes, similar to the buses in power system. Figure~\ref{fig:Ch2GasTopology} displays a simple gas system topology, which depicts the main components of gas systems, including one gas well, four pipelines $p_1$-$p_4$, one valve $v_1$, one compressor station $c_1$, and one gas storage $s_1$. A brief description of the gas system components and their mathematical models are provided as follows.

\subsubsection{Gas Sources}
    In this study, natural gas is not only produced from gas wells but also from the power-to-gas (P2G) facilities, which are adopted to convert the excessive wind energy into gas by methanation process. The first gas source is discussed in this subsection, while the latter is comprehensively illustrated in Section \ref{sec:Ch2P2G}.

\begin{figure}[th]
\centering
\includegraphics[width=11cm]{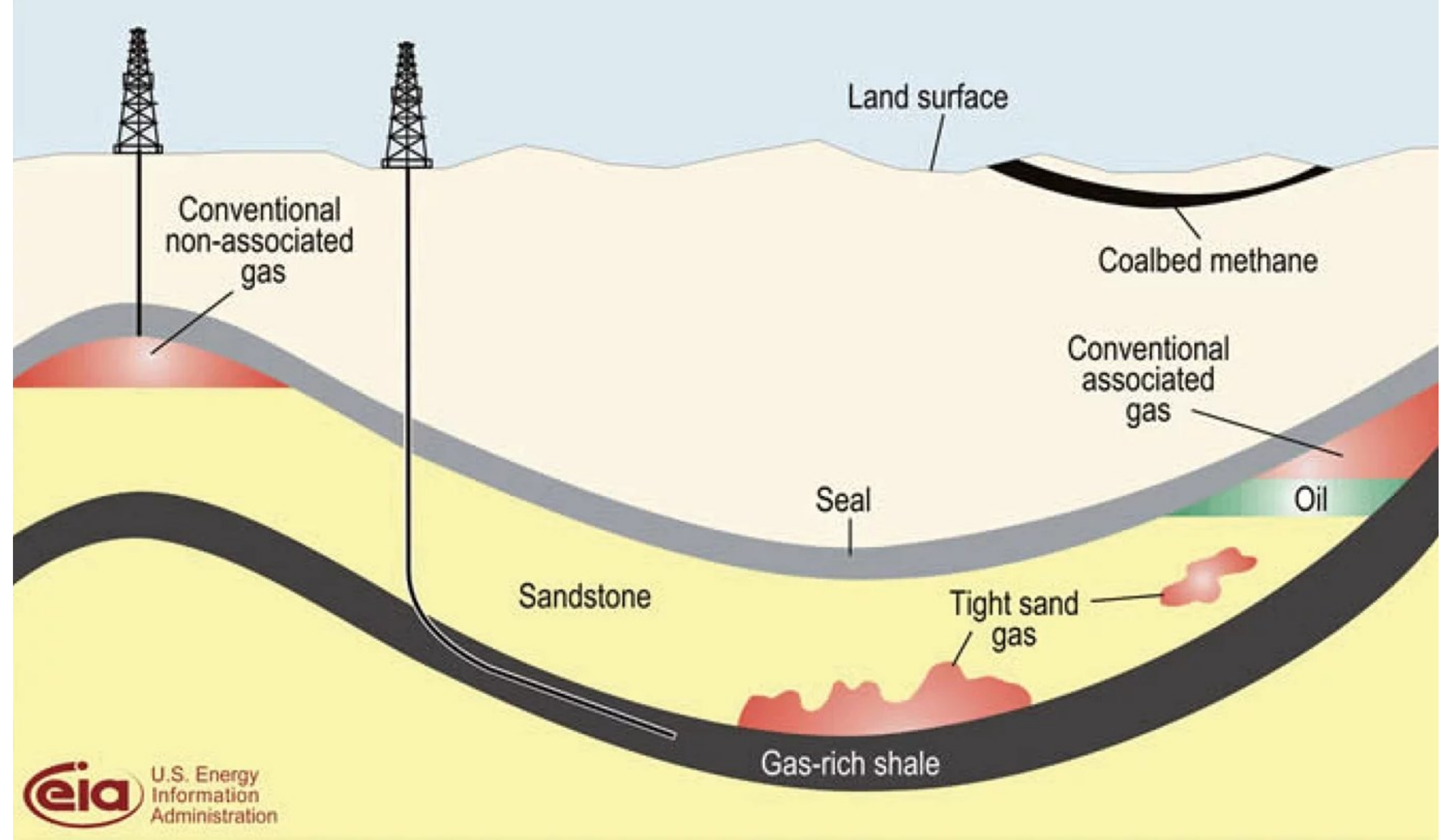} 
\caption[Schematic geology for the four types of gas resources]{Schematic geology for the four types of gas resources \\ {\small Source: https://www.eia.gov/energyexplained/natural-gas}}
\label{fig:Ch1Wells}
\end{figure}

    Natural gas has four different types, namely, conventional gas, unconventional gas, associated gas and coalbed gas, based on the formations of its land, where the surrounded area has large cracks and layer spaces, pore spaces like shale and sandstone, deposits of crude oil, and coal deposits, respectively, as displayed in Figure~\ref{fig:Ch1Wells}. To find natural gas, first the geologists locate the most likely formations that could contain gas deposits using seismic surveys. Second, if the survey introduces a positive indication, an exploratory well is examined to guarantee an acceptable quality and quantity of the available sored gas. Third, more wells are drilled vertically and/or horizontally in case of good information achieved by the examined well \cite{zou2017geological}.  For the unconventional gas production, natural gas is extracted by a new technological method, which is called hydraulic fracturing or fracking \cite{stevens2012shale}. In this method, the gas is forced by water, sand and chemicals with a high pressure down the well, and the formations are frittered, to release the gas from rocks up to the surface. The produced gas is called wet natural gas, because it has impurities, such as water vapor and propane. Therefore, it needs to be processed in the processing plants before sending it to the pipelines.

In fact, the injected gas volumes are subjected to technical or contractual ranges. Technically, the gas reservoir and the installed equipment have their industrial capacities, such as pressure range and maximum gas flow.  Contractually, the production fields might be controlled by several owners, who has rights to establish the maximum and minimum levels of production. These levels are usually signed in take-or-pay gas contacts with a predefined time intervals, and the producers have to deliver the contracted values \cite{Midthun2007Optimization}. Therefore, the gas wells production is modeled as
\begin{align}
    \underline{F}_w \le {f}_{w,t} \le \overline{F}_w,\; \forall w,t, \label{eq:Ch2Well}
\end{align}
where, ${f}_{w,t}$ is the gas flow rate injected from gas well $w$ at time $t$. $\underline{F}_w$ and $\overline{F}_w$ are the minimum and maximum production capacities.

It should be noted that, for sake of simplicity, the system operational constraints include only the processed gas, therefore, the whole production process is neglected and the gas is considered to be heterogamous \cite{correa2015optimal, Correa2015Integrated}, i.e., the gas quality is same in all situations. Interested readers can refer to \cite{borraz2010optimiz} for modeling the pooling problem considering the gas quality issues.

\subsubsection{Gas Storage}
Unlike power system, natural gas can be stored with large volumes. Gas storage provides additional operating flexibilities for the gas system, particularly in the short-term operation, to mitigate any congestion introduced by sudden increase/decrease of gas demands, fluctuations in gas prices, or system contingencies. Therefore, they are considered as an effective technique that is able to increase the gas system reliability in case of insufficient gas production. The peak-load storage are installed close to the gas demands to mitigate unserved gas loads at peak intervals, because the gas storages are in the injection (charging) state at off-peak load periods, and they are in the withdraw (discharging) state during the peak-loads or low gas prices. Gas storage can be classified as: i) underground storage, such as aquifers, salt caverns, and  abandoned gas reservoirs, which might have a large capacity so it is cost-effective; ii) aboveground storage for liquefied natural gas. In industrial, there are stored mass inside pipelines, called line pack, which is implicit natural gas storage. The line pack is discussed in Section \ref{sec:Ch2SGasDynamic}.

The gas volume inside gas storage can be expressed as
\begin{align}
    l_{s,t}  = l_{s,t-1} + {f}_{s,t}^{in} - {f}_{s,t}^{out}, \; \forall s,t, \label{eq:Ch2StorageL}
\end{align}
where, $l_{s,t}$ is the working volume of storage $s$ at time $t$, ${f}_{s,t}^{in}$ and ${f}_{s,t}^{out}$ are the \mbox{in-/outlet} gas flow of storage $s$.  The working volume should be limited by the storage minimum and maximum capacities ($\underline{L}_s$ and $\overline{L}_s$) as
\begin{align}
        \underline{L}_s \le  {l}_{s,t} \le \overline{L}_s,\; \forall s,t. \label{eq:Ch2StorageLimit}
\end{align}

The injection and withdraw rates are non-convex functions with the stored gas volume. These functions are linearized by piecewise approximation in \cite{keyaerts2012gas}. A basic representation of injection and withdraw rates can be expressed as
\begin{align}
         {f}_{s,t}^{in} \le  \overline{{f}}_{s}^{in}, \;\; {f}_{s,t}^{out} \le  \overline{{f}}_{s}^{out},\; \forall s,t. \label{eq:Ch2StorageRate}
\end{align}
where, $\overline{{f}}_{s}^{in}$ and $\overline{{f}}_{s}^{out}$ are the maximum capacities of the injection and withdraw rates, respectively.

In this thesis, most of studies assume that all gas storages are nonstrategic elements, i.e., they are in closed state, to highlight the effectiveness of considering the gas dynamics and gas line pack inside pipelines.

\subsubsection{Gas Valves}
    Gas valve is a controllable device that can regulate and/or direct the gas flow by opening, closing and partially opening pathways, and other functions. It is a strategic element in the gas system, i.e., it is employed to manage the gas flow rates, such as sectionalizing for maintenance, isolating under contingencies, preventing the excessive pressure, and forcing the flow in a certain direction. The existing literature presents five types of gas valves according to their functions as summarized in \cite{correa2015optimal}:
\begin{itemize}
  \item
	\textit{Check valve}: it fixes the gas flow direction, i.e., the gas is allowed to move in a specified direction and other directions are prohibited.
  \item
	\textit{Ordinary valve}: it is used to connect the pipelines to the gas nodes, if their end terminals have equal pressures.
  \item
	\textit{Bypass valve}: it is connected in parallel with the gas compressor to allow the reverse gas flow. Hence, the gas compressor is protected during its turning off, i.e., the valve acts as a flywheel.
  \item
	\textit{Block valve}: it is used to block the gas flow for isolating and sectionalizing a part of gas system for maintenance or operational reasons.
  \item
	\textit{Control valve}: it is employed to reduce the gas pressure at sink nodes or to regulate the gas flow inside pipelines.
\end{itemize}

Detailed models for the above gas valves can be found in \cite{moritz2007mixed}. In this study, the gas nodal pressures are considered uniform, therefore, the gas valves are nonstrategic elements in the gas system, in other words, the gas system optimization problem adopts the valves in static states. This assumption is widely adopted and common treatment in the recent coupled power and gas models.

\subsubsection{Gas Compressor Stations}\label{sec:Ch2SComp}
To overcome the pressure drop caused by the gas pipeline friction, gas compressor stations are employed to increase the pressure to its desired level, similar to the step-up transformer in power system. They are commonly installed close to underground storage at intervals of $50-100$ miles \cite{shahidehpour2005impact}. Gas compressors can be classified as:
\begin{itemize}
  \item
    \textit{Gas-driven compressor}: it is powered by gas turbines by consuming the required gas from pipeline flow. It is a traditional type and widely installed in the existing networks.

  \item
    \textit{Electricity-driven compressor}: it is equipped with electric motors. Although it increases the interdependency between power and system, and additional electricity contracts are needed, it introduces environmental benefits to the modern society so it is employed in the new gas networks.
\end{itemize}

 A detailed compressor model is given in Appendix~\ref{App:Comp}. Unfortunately, this model is non-convex and it is difficult to solve and poses tractable challenges to the gas system. Many attempts are found in the literature to propose a simplified models \cite{he2017robust, Correa2015Integrated, correa2015optimal, Midthun2007Optimization, liu2009security} that optimize the consumed energy, assume a constant pressure ratio, adopt a constant loss factor, linearize the consumption function, or adopt the Newton-Raphson method. Because the objective of this study is finding the optimal economic and resilient operation for the coupled power and gas system, the commonly adopted and simplified compressor model is employed in the thesis work. In simplified model, the gas flow direction inside the compressor is predetermined, therefore the in- and outlet nodes are known. Besides, according to \cite{borraz2009improving}, the consumed gas is usually in range $3\%-5\%$ of the pipeline gas flow. Therefore, the simplified compressor model is defined as
    \begin{align}
      {\pi}_{i,t} \le {\pi}_{o,t} \le \gamma_c {\pi}_{i,t}, \forall c,t, (i,o) \in c   \label{eq:Ch2Comp1} \\
      0 \le {f}_{c,t}^{out} = (1-\alpha_c) {f}_{c,t}^{in}, \; \forall c\in \mathcal{C},t,     \label{eq:Ch2Comp2}
    \end{align}
    where ${\pi}_{i,t} / {\pi}_{o,t}$ and ${f}_{c,t}^{in}/{f}_{c,t}^{out}$ are the inlet/outlet pressures and gas flows, $\gamma_c$ is the maximum compression factor with the specified direction, and $\alpha_c$ is the gas consumption factor, which equal zero for the electricity-driven compressor and $0.03-0.05$ for gas-driven compressor.

    Note that these simplifications have not a great influence on the solution accuracy for the coupled power and gas system, however, it provide a more tractable model to be optimally solved.

\subsubsection{Gas Nodes}
    Gas node is the connection point between gas system elements, including gas sources, gas loads, compressors and valves. It also connects the gas system with the power system through the coupled components, such as GPUs and P2G facilities. The gas pressure is same for all sections in the node, i.e., the node has a constant gas flow per period and there is no gas dynamics in the node. The gas nodal balancing equation of gas node $i$ at time $t$ is accomplished by
    \begin{align}
     \sum_{w \in \mathcal{W}(i)} {f}_{w,t} + \sum_{p \in \mathcal{P}_1(i)} {f}_{p,t}^{out} - & \sum_{p \in \mathcal{P}_2(i)} {f}_{p,t}^{in} + \sum_{s \in \mathcal{S}(i)} ({f}_{s,t}^{out} - {f}_{s,t}^{in})+ \sum_{c \in \mathcal{C}_1(i)} {f}_{c,t}^{out} \nonumber  \\
     - \sum_{c \in \mathcal{C}_2(i)} {f}_{c,t}^{in} + \sum_{z \in \mathcal{Z}(i)}  \varrho_{z,t} & = \sum_{u \in \mathcal{U}_g(i)} \rho_{u,t} + \sum_{d \in \mathcal{D}_g(i)} F_{d,t},\; \forall i,t,\label{eq:Ch2Node}
    \end{align}
    where ${f}_{w,t}$ is the gas flow injected from gas well $w$;  ${f}_{p,t}^{in}/{f}_{p,t}^{out},\; {f}_{s,t}^{in}/{f}_{s,t}^{out}$ and ${f}_{c,t}^{in}/{f}_{c,t}^{out}$ are the \mbox{in-/outlet} gas flow of pipeline $p$, storage $s$ and compressor $c$, respectively; $\varrho_{z,t}$ is the produced gas from P2G facility $z$; $\rho_{u,t}$ is the utilized gas by GPU $u$; and $F_{i,t}$ is the total gas load at node $i$ and time $t$; $\mathcal{W}(i),\;\mathcal{S}(i),\;\mathcal{Z}(i)$, and $\mathcal{D}_g(i)$ are subsets of gas wells, gas storage, P2G units and gas demands connected with node $i$; and $\mathcal{P}_1(i)/\mathcal{P}_2(i)$ and $\mathcal{C}_1(i)/\mathcal{C}_2(i)$ are subsets of pipelines and compressors, whose ending/beginning terminals are connected with node $i$, respectively.

    Additionally, the gas pressures should be in a specified range according to the consumer requirements, technical restrictions or signed contracts. Therefore, the gas pressure $\pi_{i,t}$ of node $i$ and time $t$ is limited by the nodal upper and lower boundaries, i.e., $\underline{\Pi}_i$ and $\overline{\Pi}_i$, as follows.
      \begin{align}
        \underline{\Pi}_i \le {\pi}_{i,t} \le \overline{\Pi}_i,\; \forall i,t, \label{eq:Ch2Pressure}
      \end{align}
\subsubsection{Gas Loads}
    Natural gas is consumed by final users at any gas node. The gas pressure is reduced to a contractual value for distribution. Gas loads are signed with different levels and priorities that should be handled by system operator in an effective way. In some situations, gas systems cannot satisfy all demands, then higher priority clients are served first, by adjusting the penalties of unserved gas loads ($\triangle f_{d,t}$) in the objective function og the optimization problem, and relaxing the gas nodal balancing equation as
    \begin{align}
     \sum_{w \in \mathcal{W}(i)} {f}_{w,t} + \sum_{p \in \mathcal{P}_1(i)} {f}_{p,t}^{out} - & \sum_{p \in \mathcal{P}_2(i)} {f}_{p,t}^{in} + \sum_{s \in \mathcal{S}(i)} ({f}_{s,t}^{out} - {f}_{s,t}^{in})+ \sum_{c \in \mathcal{C}_1(i)} {f}_{c,t}^{out} \nonumber  \\
     - \sum_{c \in \mathcal{C}_2(i)} {f}_{c,t}^{in} + \sum_{z \in \mathcal{Z}(i)}  \varrho_{z,t} & = \sum_{u \in \mathcal{U}_g(i)} \rho_{u,t} + \sum_{d \in \mathcal{D}_g(i)} (F_{d,t}-\triangle f_{d,t}),\; \forall i,t,\label{eq:Ch2NodeShed}
    \end{align}

\subsubsection{Gas Pipelines} \label{sec:Ch2Pipes}
    The most important and critical element in the gas system is the pipelines because of the pressure gradients and physical characteristics of natural gas and they have large quantity of gas stored within them. Pipelines can be categorized into three major types according to their location: i) gathering pipelines, which are installed in the production plants to collect the gas from wellheads to the refining stations; ii) transmission pipelines, which deliver the refined gas with large quantities to the market area, i.e., distribution systems. They have wide diameters and long lengths; iii) distribution pipelines, which are used to distribute the transmitted gas to the final users. Because the refining process is not included in the gas model, as discussed above, the gathering pipelines are not modeled in this study. The typical values of working pressures/diamters of pipelines in the transmission and distribution systems are \SI{1.5e5}{}$\sim$\SI{8.5e5}{\pascal}/\SI{0.15}{}$\sim$\SI{1.22}{\metre} and \SI{0.4e5}{}$\sim$\SI{1.5e5}{\pascal}/\SI{0.025}{}$\sim$\SI{0.61}{\metre}, respectively \footnote{https://naturalgas.org/}~\footnote{https://blog.miragemachines.com/types-of-pipeline-every-oil-and-gas-engineer-should-know-about}.

    Compared with power flow, the gas flow travels with limited velocities due to the slow dynamics of gas system. Therefore, gas flow needs a response time to be delivered, and such circumstance should be considered in modeling the short-term operation. The optimization of gas dynamics are found in the existing literature with different denotations, such as transient optimization, time-dependent optimization, partial differential equations (PDE) gas problem, nonlinear mixed-integer optimization and gas dynamics optimization. For the sake of clarity, in the thesis, the gas system is denoted as dynamic-state or steady-state gas models, where the latter neglects the line pack. These two models are introduced in the following sections.

    To represent the gas system dynamics, a set of PDE are derived from the physics of gas particles. This set guarantees that the system is affected by transportation process only, and there is no lost/gained energy or gas mass. According to \cite{correa2015optimal}, this set can be defined and summarized in Appendix~\ref{APP:Dynamics}. The PDE of gas model dynamics can not be incorporated with power system optimization problem. In what follows, the dynamic- and steady-state gas flow models are presented in a tractable algebraic formulations.

\subsection{Dynamic-State Gas Flow Model} \label{sec:Ch2SGasDynamic}
    Equations in Appendix~\ref{APP:Dynamics}, namely continuity equation,  momentum equation, energy equation and state equation, express the gas dynamics in detail in the time and space terms. In order to derive a tractable formulation of the gas optimization model, some simplifications, which are previously employed by researchers to achieve acceptable results, and satisfy the gas industry requirements, are adopted. These simplifications are as follows:
    \begin{enumerate}
        \item
            The gas temperature is assumed to be equal to that of the surroundings \cite{herran2009modeling}, considering that gas pipelines are close the ground, or as a result of slow dynamics of gas. According to \cite{osiadacz2001comparison}, this assumption may introduces results with an error up to $2\%$. Note that, with this assumption, the energy equation holds, i.e., it can be dropped from PDE set.

        \item
            All pipelines are installed horizontally \cite{keyaerts2012gas, liu2011coordinated, Correa2015Integrated}. Therefore, the second term of \eqref{eq:Ch2MomEqn} is constant, and the terms in \eqref{eq:Ch2EneEqn} including the pipeline height $h$ are less challengeable. One can refer to \cite{xue2019optimizing, geissler2018solving, han2019operations} for modeling and solving the gas dynamic models with inclined pipelines.

        \item
            The accelerating forces in the momentum equation, i.e., kinetic energy $\upsilon \partial\lambda/\partial t$ and gravity force $\partial(\lambda\upsilon^2)/\partial x$, introduce less than $1\%$ of the friction force $G\lambda\partial h/\partial x$ \cite{herran2009modeling}. Therefore, the third and fourth terms of this equation can be dropped and the gas pressure-flow relationship depends only on the friction of pipelines \cite{Correa2015Integrated, liu2009security, keyaerts2012gas}.

        \item
            Based on the first assumption, the compressibility function in the state equation can be linearized in terms of pressure only, or it can be assumed as a fixed value based on the pressure range \cite{Midthun2007Optimization, keyaerts2012gas, correa2015optimal}.

        \item
            The widely used Dercy friction factor is adopted. Therefore, the friction factor $F$, in the right hand side of the  momentum equation, can be calculated by Colebrook-White formula or its modified version (see equations (2.16)-(2.17) in \cite{correa2015optimal}).

    \end{enumerate}

     By adopting the above simplifications, and replacing the density and velocity with the mass flow rate, i.e., $f=\frac{\tilde{\pi} D^2}{4} \frac{\lambda\upsilon}{\rho_0}$, the continuity and momentum equations are expressed in a simplified forms. However these forms are still PDE, which can not be incorporated in the optimization problem. Therefore, they need to be converted into ordinary algebraic equations, by discretizing the PDE in time and space. The discretization methods for a transient equations can be explicit and implicit methods. The explicit method uses the recent variables to calculate the next ones, it provides restrictions on the time step \cite{liu2011coordinated, herran2009modeling}. However, the implicit method is based on finite-difference approximations (i.e., several short pipelines) or finite-volume schemes (i.e., several finite volumes in a meshed geometry), which offer numerical stabilities for large gas networks \cite{liu2011coordinated, moritz2007mixed, Midthun2007Optimization, keyaerts2012gas}. Due to the simplicity, applicability and stability of the finite-difference approximations, a novel set of algebraic equations that represent the gas flow dynamics are formulated in \cite{correa2015optimal}.
    \begin{figure}[th]
        \centering
        \includegraphics[width=8cm]{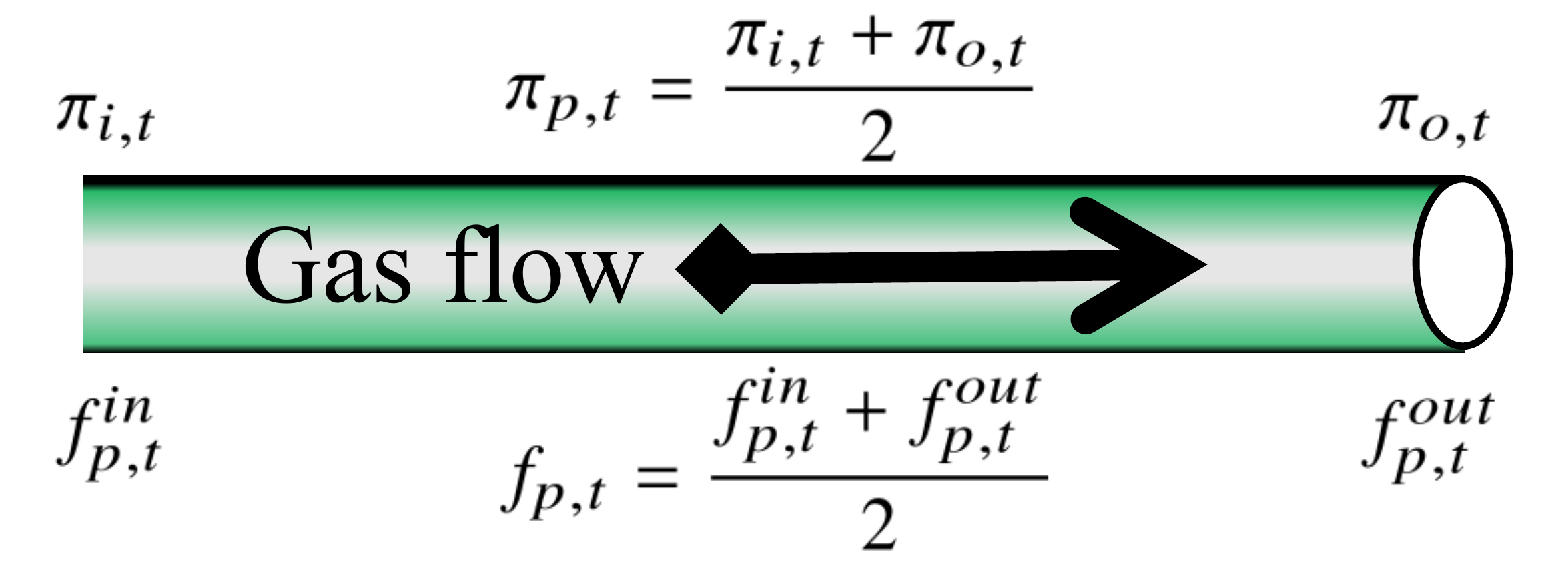} 
        \caption{Discretization of the PDEs, indicating the terminal and average values of pressures and gas flow.}
        \label{fig:Ch2Wey}
    \end{figure}

    Figure~\ref{fig:Ch2Wey} depicts the average and terminal values of pressures and gas flows for a discretized pipeline in space. The general flow equation, known as Weymouth equation, is derived from the momentum equation. Weymouth equation describes the relationship between the average gas flow and the terminal pressures of a pipeline. It is defined as
    \begin{align}
        {f}_{p,t} |{f}_{p,t}| & = \chi_p^f ({\pi}_{i,t}^2 - {\pi}_{o,t}^2), \; \forall p,t, (i,o) \in p \label{eq:Ch2Wey}\\
        {f}_{p,t} &= \frac{f_{p,t}^{in} + f_{p,t}^{out}}{2}, \; \forall p,t, \label{eq:Ch2AveFlow}
    \end{align}
    where ${f}_{p,t}$ is the average gas flow rate inside pipeline $p\in \mathcal{P}$ and time $t \in\mathcal{T}$, $\mathcal{P}$ and  $\mathcal{T}$ are the sets of gas pipelines and time intervals, ${\pi}_{i,t}$ and ${\pi}_{o,t}$ are the terminal pressures of a pipeline connected with gas nodes $i$ and $o$, respectively. $\chi_p^f$ is the Weymouth equation coefficient, which depends on the physical characteristics of the pipeline. This coefficient is calculated by
     \begin{align}
         \chi_p^f = \Phi^f \left(\frac{\tilde{\pi}}{4}\right)^2 \frac{T_0}{\pi_0 \rho_{air}} \frac{D_p^5}{L_p F_p T Z G_s}, \; \forall p \label{eq:Ch2WeyConst}
    \end{align}
    where $\tilde{\pi}\approx3.1416$ is the mathematical constant; the base temperature is $T_0=$\SI{273.15}{\kelvin}, the base pressure $\pi_0=$\SI{1.01325}{\bar}, base density of air is $\rho_{air}=$\SI{1.2922}{\kilogram\per\cubic\metre}, the gas temperature is $T=$\SI{281.15}{\kelvin}, the relative gravity of gas $G_s=$\num{0.6106}, and the compressibility factor adopted in the thesis models is $Z=$\num{0.8}. $D_p,\;L_p$ and $F_p$ are the physical parameters of pipeline $p$, namely diameter, length, and friction factor. Finally, $\Phi^f$ is the unit conversion factor, for example, if the pressure and the gas flow rate units are \si{\bar} and \si{\mega S \cubic\metre\per h}, respectively, then $\Phi^f=(3600)^2 \times 10^{-12}$.

    The approximated continuity equation depicts the relationship between the gas flow difference of a pipeline in space, i.e., the difference between in- and outlet gas flow, and the average pressure difference in time, i.e., the difference in gas pressure between the recent and previous time intervals. Therefore, it originates a time-dependent optimization model, which may introduces computational challenges. However, it formulates the industrial stored gas in pipelines, known as the line pack, by considering the gas flow difference in space. Line pack provides additional operating flexibilities for gas system operator to instantaneously balance the gas system in case of contingencies, peak demands or varying loads, under insufficient gas production schedules, which are based on longer time intervals (\num{15}-\num{120} minuets). Subsequently, the continuity equation can be expressed in terms of the line pack as
     \begin{align}
        {m}_{p,t} = \chi_p^m ( {\pi}_{i,t} + {\pi}_{o,t} ),\; \forall p,t, (i,o) \in p   \label{eq:Ch2GMMass1}\\
        {f}_{p,t}^{in} - {f}_{p,t}^{out} =  {m}_{p,t} - {m}_{p,t-1}, \; \forall p,t,   \label{eq:Ch2GMMass2}
    \end{align}
    where ${m}_{p,t}$ is the mass of gas stored in pipeline $p$ at time $t$. $\chi_p^m$ is the line pack coefficient, which is calculated by
    \begin{align}
         \chi_p^m = \Phi^m \frac{\tilde{\pi}}{8} \frac{T_0}{\pi_0} \frac{D_p^2 L_p}{T Z}, \; \forall p \label{eq:Ch2WeyConst}
    \end{align}
    where $\Phi^m$ is a unit conversion factor, for example, if the pressure and the gas flow rate units are \si{\bar} and \si{\mega S \cubic\metre\per h}, respectively, then $\Phi^m= 3600 \times 10^{-6}$.

    Based on the above discussion, the dynamic-state model is summarized as following:
    \begin{itemize}
        \item
            \textit{Gas production capacities}: \eqref{eq:Ch2Well}.
        \item
            \textit{Gas storage constraints}: working volume limits \eqref{eq:Ch2StorageL}-\eqref{eq:Ch2StorageLimit} and in-/outlet flow capacity \eqref{eq:Ch2StorageRate}.
        \item
            \textit{Gas compressors constraints}: compression and consumption constraints \eqref{eq:Ch2Comp1}-\eqref{eq:Ch2Comp2}.
        \item
            \textit{Nodal balancing equation}: \eqref{eq:Ch2Node} or with gas load shedding \eqref{eq:Ch2NodeShed}.
        \item
            \textit{Nodal pressure bounds}: \eqref{eq:Ch2Pressure}.
        \item
            \textit{Weymouth equation}: \eqref{eq:Ch2Wey} and average flow rate \eqref{eq:Ch2AveFlow}.
        \item
            \textit{Line pack}: \eqref{eq:Ch2GMMass1}.
        \item
            \textit{Continuity equation}: \eqref{eq:Ch2GMMass2}.

    \end{itemize}

\subsection{Steady-State Gas Flow Model} \label{sec:CH2SSS}
    When the transient fluctuations are neglected, the resulted formulation is the steady-state gas flow model. Compared with the dynamic-state gas model, the steady-state one does not consider the time-dependent equations, i.e., the continuity equation \eqref{eq:Ch2GMMass2}. Therefore, the line pack is neglected or assumed to be fixed, consequently, the inlet and outlet gas flows are the same ($f_{p,t}^{in}=f_{p,t}^{out}$), according to \eqref{eq:Ch2GMMass2}. And the model depends on the $\pi_{i,t}^2$ instead of $\pi_{i,t}$, as \eqref{eq:Ch2GMMass1} is dropped. Therefore, as suggested in \cite{de2000gas}, a simple substitutions can be adopted to simplify the steady-state model, where the nonlinear variables $\pi_{i,t}^2 ,\; \forall i,t$ are replaced with a linear one $\bar{\pi}_{i,t},\; \forall i,t$. Then the resultant formulation will be as the following:
    \begin{itemize}
        \item
            \textit{Gas production capacities}: \eqref{eq:Ch2Well}.
        \item
            \textit{Gas storage constraints}: working volume limits \eqref{eq:Ch2StorageL}--\eqref{eq:Ch2StorageLimit} and in-/outlet flow capacity \eqref{eq:Ch2StorageRate}.
        \item
            \textit{Gas compressors constraints}: compression constraints will be as
            \begin{align}
                {\bar{\pi}}_{i,t} \le {\bar{\pi}}_{o,t} \le \gamma_c^2 {\bar{\pi}}_{i,t}, \forall c,t, (i,o) \in c, \label{eq:Ch2Comp1SS}
            \end{align}
            and the consumption constraints are defined in \eqref{eq:Ch2Comp2}.
        \item
            \textit{Nodal balancing equation}: considering the gas load shedding
            \begin{align}
                \sum_{w \in \mathcal{W}(i)} {f}_{w,t} + \sum_{p \in \mathcal{P}_1(i)} {f}_{p,t} - & \sum_{p \in \mathcal{P}_2(i)} {f}_{p,t} + \sum_{s \in \mathcal{S}(i)} ({f}_{s,t}^{out} - {f}_{s,t}^{in})+ \sum_{c \in \mathcal{C}_1(i)} {f}_{c,t}^{out} \nonumber  \\
                - \sum_{c \in \mathcal{C}_2(i)} {f}_{c,t}^{in} + \sum_{z \in \mathcal{Z}(i)}  \varrho_{z,t} & = \sum_{u \in \mathcal{U}_g(i)} \rho_{u,t} + \sum_{d \in \mathcal{D}_g(i)} (F_{d,t}-\triangle f_{d,t}),\; \forall i,t,\label{eq:Ch2NodeShed2}
            \end{align}
        \item
            \textit{Nodal pressure bounds}:
            \begin{align}
                \underline{\Pi}_i^2 \le {\bar{\pi}}_{i,t} \le \overline{\Pi}_i^2,\; \forall i,t, \label{eq:Ch2PressureSS}
            \end{align}
        \item
            \textit{Weymouth equation}:
             \begin{align}
                {f}_{p,t} |{f}_{p,t}|  = \chi_p^f ({\bar{\pi}}_{i,t} - {\bar{\pi}}_{o,t}), \; \forall p,t, (i,o) \in p, \label{eq:Ch2WeySS}
             \end{align}

    \end{itemize}

     Equations \eqref{eq:Ch2Comp1SS} and \eqref{eq:Ch2PressureSS} are obtained by squaring each term in \eqref{eq:Ch2Comp1} and \eqref{eq:Ch2Pressure}, respectively. In the nodal balance equation \eqref{eq:Ch2NodeShed}, the inlet and outlet gas flow of the pipeline are replaced with the average one, and  the average flow rate equation \eqref{eq:Ch2AveFlow} is dropped, as $f_{p,t} = f_{p,t}^{in}=f_{p,t}^{out}$.

     Although the steady-state gas flow model introduces inaccurate decisions for gas system operator because of neglecting the slow dynamics of gas flow and disregarding the line pack, they merit to be studied due to some reasons, such as
     \begin{enumerate}
        \item
            The simplicity of the mathematical model that can be easily incorporated with power system optimization problems. This model can be adopted to find a quick solution for the main variables of the gas system. Therefore, it is commonly employed in the existing coupled power and gas studies.
        \item
            In planning perspectives, This model identifies acceptable solutions for large-scale systems within reasonable times. Because it neglects the line pack, which provides economic benefits to the system, its solutions could be low conservative as they consider the worst-line pack scenario, which is zero.
        \item
            In the distribution level, the line pack quantities are small, therefore, this model can fined a high-quality decision for system control and operation. Table~\ref{tab:Ch2TabTandD} presents a comparison between transmission and distribution levels in terms of the gas system parameters, including nodal pressures and pipeline dimensions. The range of stored gas within one unit length (\SI{1}{\metre}) is calculated by \eqref{eq:Ch2GMMass1} with the range of pressures and diameters. It is quit significant that the gas dynamics could be neglected in the distribution level under low pressures.
            \begin{table}[th]
                \caption{Typical values of gas system parameters}
                \label{tab:Ch2TabTandD}
                \centering
                \includegraphics[width=14cm]{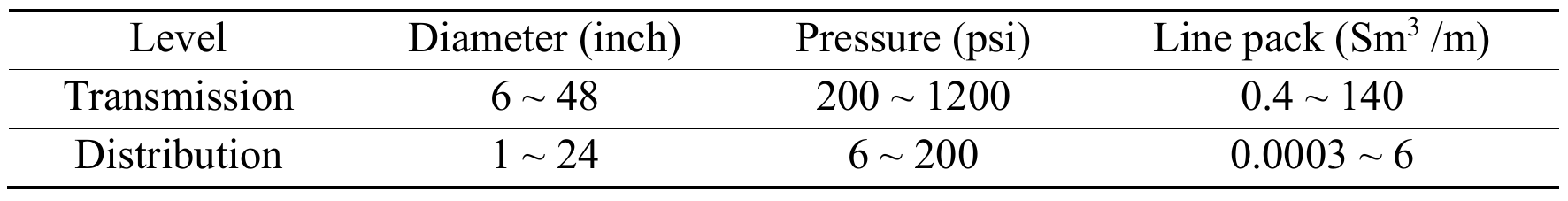} 
            \end{table}
        \item
            In this paper, many studies are presented to solve the dynamic-state gas flow model, and  numerical simulations are conducted to show the effectiveness of considering the gas flow dynamics. Therefore, the steady-state gas flow model is discussed as a reference.
     \end{enumerate}

\section{Electric Power Systems Modeling} \label{sec:Ch2Power}
    The power system is modeled as set of nodes (buses) interconnected by a set of branches, where the branches represent the power lines, transformers and cables, and the buses are the physical points for connecting the system components, including generators and loads. In the electric power system, identifying the optimal power flow (OPF) is one of the fundamental issues in power system planning, operation and markets. In this section, due to its importance in the coupled power and gas system modeling, the OPF is discussed, including its definitions, extensions, applications, and mathematical formulations.

\subsection{Optimal Power Flow}
\begin{figure}[!ht]
        \centering
            \includegraphics[width=14cm]{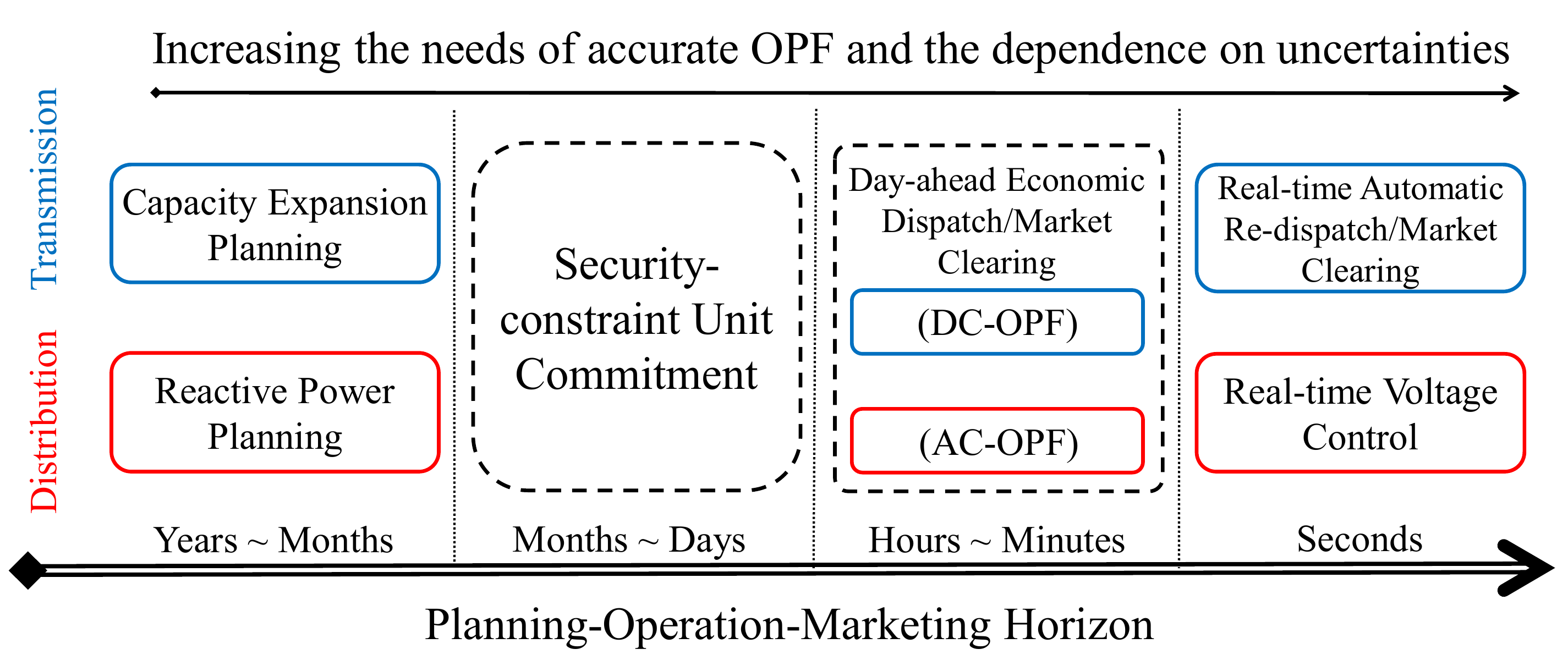} 
            \caption{Optimization and control procedures for power system planning, operation and market}   \label{fig:Ch2OPF}
    \end{figure}

    The OPF optimization problem was firstly introduced in $1962$ by Carpentier \cite{carpentier1962contribution}. It is well-studied in the literature to formulate the physical and economic constraints according to the electrical laws and engineering decisions (see the recent surveys \cite{frank2012optimal, frank2012optimal2, frank2016introduction, bienstock2013progress}). Figure.~\ref{fig:Ch2OPF} dramatically displays the optimization and control procedures for power system management, indicating that the accuracy of OPF is important with short time intervals. The OPF is applied for long-term transmission-level planning decisions, security-constrained unit commitment (SCUC), and day-ahead/real-time economic dispatch (ED).

    Before presenting the mathematical formulations of the OPF models, the conventional power flow (CPF) is firstly discussed.

\subsubsection{Conventional Power Flow}
     The CPF is to compute any feasible solution for the system equations, neglecting the operational costs or the objective function. Its compact form can be expressed as
    \begin{align}
          \sum_{u \in  \mathcal{U}(n)} p_{u,t} + \sum_{l \in  \mathcal{L}(n)} {f}_{l_p,t}(\bm{v},\bm{\theta}) = \sum_{d \in  \mathcal{D}_p(n)} P_{d,t}, \; \forall n,t, \label{eq:Ch2CPFP}\\
          \sum_{u \in  \mathcal{U}(n)} q_{u,t} + \sum_{l \in  \mathcal{L}(n)} {f}_{l_q,t}(\bm{v},\bm{\theta}) = \sum_{d \in  \mathcal{D}_p(n)} Q_{d,t}, \; \forall n,t. \label{eq:Ch2CPFQ}
    \end{align}
     where $p_{u,t}$ and $q_{u,t}$ are the active and reactive power generated from unit $u$ at time $t$;  $P_{d,t}$ and $Q_{{u,t}}$ are the active and reactive power of demand $d$; ${f}_{l_p,t}$ and ${f}_{l_q,t}$ are the functions of active and reactive power flow of line $l$, respectively. These functions depend on the vector of voltage magnitude $\bm{v}$ and phase angle $\bm{\theta}$ at system buses, and their detailed expressions are discussed in Section~\ref{sec:Ch2SOPF}. $\mathcal{U}(n),\;  \mathcal{L}(n)$ and $\mathcal{D}_p(n)$ are subsets of power units, power lines and power demands connected with bus $n$, respectively.

     For each bus, there are four variables, namely net active power $p_{n,t}$, net reactive power $q_{n,t}$, voltage magnitude $v_{n,t}$ and phase angle $\theta_{n,t}$. To find a deterministic solution for the CPF, two out of four variables are fixed by assigning the system buses to one of the following three bus types \cite{frank2016introduction}: i) slack bus, which sets a voltage reference for all system buses, i.e., $v_{n,t}=1$ p.u. and $\theta_{n,t}=0$; ii) load bus, at which the net power is fixed, i.e., $p_{n,t}=P_{n,t}$ and $q_{n,t}=Q_{n,t}$; and iii) voltage-controlled bus, at which the active power and voltage magnitude are fixed by a local reactive source as a voltage regulator, i.e., $p_{n,t}=p_{n,t}^*$, and $v_{n,t}=v_{n,t}^*$. As the real power injections at the slack bus is free to provide a feasible solution for the CPF, only one slack bus can be assumed in the system model. It should be noted that the CPF is a deterministic problem that is solved with a number of equations equal the number of unknowns, however, it may provide an impractical solution, such as negative voltage magnitude and large power generation.

\subsubsection{Optimal Power Flow Versions}
     Compared with the CPF, the OPF is an optimization problem that combines the CPF with an objective function and a set of technical constraints to avoid any physical or technical violation \cite{frank2012optimal}. Several OPF versions have been found in the existing works, including i) static OPF, which handles the problem with single time interval \cite{baydar2019resilient}; ii) dynamic OPF, which handles the problem with multi-time periods, i.e., multi-period optimization problem \cite{ma2019multi}; iii) transient stability-constrained OPF, which simultaneously systemizes the static and dynamic OPF models in the same problem \cite{xu2016robust}; iv) security-constrained OPF, which considers the system constraints under contingencies \cite{venzke2018convex}; v) deterministic OPF, which neglects any uncertainty in the power system parameters; vi) stochastic or robust OPF, which considers the uncertainties of the power system \cite{bazrafshan2016decentralized}; vii) AC-OPF, which accurately formulates the system power flow equations, considering the reactive power injections, transmission losses and voltage constraints \cite{gupta2019optimal}; viii) DC OPF, which simplifies the AC-OPF; ix) mixed AC/DC OPF, which adopted in integrated AC-DC grids \cite{bahrami2017security}; x) multi-phase OPF, which considers $n$-conductor in the optimization problem \cite{bernstein2018multiphase}; xi) unbalanced three-phase OPF, which are adopted for unbalanced distribution systems \cite{sanseverino2015optimal}.

        The above versions do not cover all versions of OPF. And other extensions can be obtained by combining two or more versions from the above list, for example dynamic stochastic OPF. In fact, the OPF version is selected based on the solution accuracy, reliability and optimality that is mainly influenced by the objective function. Different objective functions are reported to consider one or more sub-objectives, such as generation costs, power losses, voltage violations, reactive power costs, carbon emission, power shedding, energy reserves, and energy imports.

\subsubsection{Optimal Power Flow Applications}
     Most of the recent OPF models are based on the classic formulations presented in \cite{carpentier1962contribution}, where a classic ED is formulated. The objective of the classic ED model is to minimize the total operational costs of power production as
     \begin{align}
          \min_{p_{u,t}, q_{u,t}, v_{n,t}, \theta_{n,t}, p_{l,t}, q_{l,t}} &\sum_{t \in  \mathcal{T}} \sum_{u \in  \mathcal{U}} C_u(p_{u,t}) \label{eq:Ch2ED}\\
          s.t. \hspace{2em}& \eqref{eq:Ch2CPFP}-\eqref{eq:Ch2CPFQ},  \\
               & \underline{P}_u \le p_{u,t} \le \overline{P}_u, \; \forall u,t, \;\; \underline{Q}_u \le q_{u,t} \le \overline{Q}_u, \; \forall u,t, \label{eq:Ch2EDBound} \\
               & \underline{V}_n \le v_{n,t} \le \overline{V}_n, \; \forall n,t, \;\; \underline{\Theta}_n \le \theta_{n,t} \le \overline{\Theta}_n, \; \forall n,t.
    \end{align}
     where the production cost $C_u(.)$ is a quadratic convex function, and $\underline{P}_u/\overline{P}_u, \; \underline{Q}_u/\overline{Q}_u, \; \underline{V}_n/\overline{V}_n$ and $\underline{\Theta}_n/\overline{\Theta}_n$ are the maximum/minimum physical and technical constraints for power generation and bus voltages, respectively.

     Besides the classic ED model, there are many applications of OPF that can be employed to overcome the difficulties in operation, control, planning and marketing. These applications include i) optimal reactive power flow, known as VAR control, which include the effects of tap-changing and phase-shifting transformers \cite{frank2012optimal}; ii) reactive power planning, in which, new reactive power sources are optimally allocated \cite{frank2012optimal2}; iii) network constraints unit commitment (NCUC), which couples the unit commitment (UC) problem with the power flow equations \cite{chen2019modeling}. The UC problem refers to the optimal operating schedule (on-off statues) of all power units. Appendix~\ref{App:UC} presents the tight, compact and computational efficient UC model proposed by  Morales-Espana et al \cite{morales2012tight}. In the thesis, this model is employed to fined a predefined UC decisions to be used in the proposed operational models for the coupled power and gas systems; iv) security-constrained ED (SCED), which identifies an optimal ED considering the power system contingencies \cite{frank2015optimal}.

\subsection{AC Optimal Power Flow Model}\label{sec:Ch2SOPF}
    The AC-OPF is the transformation of the complex power flow equations into algebraic ones to be employed in a mathematical optimization problem. There are two different models for AC-OPF, namely bus injection model and branch flow model. The bus injection model is the most compact form, however, the branch power flow model is more convenient to be reformulated into a convex relaxation OPF problems. It should be noted that this thesis concentrates on solving the branch power flow model, whose formulation is provided in this subsection, and the possible formulations of the bus injection model is attached in Appendix~\ref{App:OPF}.

    The branch flow model is derived from the voltage-current relationship of a power line $l$ connected with two buses $m$ and $n$, which can be expressed as
    \begin{gather}
          \vec{i}_{l,t} = \vec{y}_l (\vec{v}_{m,t}-\vec{v}_{n,t}), \; \forall l, (m,n)\in l, t. \label{eq:Ch2BF1}
    \end{gather}
    where $\vec{i}_{l,t}, \;\vec{v}_{n,t}$ and $\vec{y}_l$ are the phasor vector of the branch current, bus voltage and branch admittance, respectively. Baran and Wu introduced the original branch flow model to optimize the capacitor allocation problem in $1989$ \cite{baran1989optimal}. The branch flow model is a set of three complex equations
    \begin{gather}
          \vec{v}_{n,t} - \vec{v}_{m,t} = \vec{z}_l \,\vec{i}_{l,t}, \; \forall l, (m,n)\in l,t, \label{eq:Ch2BFM1}\\
          \vec{s}_{l,t} = \vec{v}_{n,t} \, (\vec{i}_{l,t}) ^ {\star}, \; \forall l, (m,n)\in l,t, \label{eq:Ch2BFM2}\\
          \sum_{l \in  \mathcal{L}(n)} {f}_{l_p,t}(\bm{v},\bm{\theta}) +j \sum_{l \in  \mathcal{L}(n)} {f}_{l_p,t}(\bm{v},\bm{\theta}) =  \sum_{l \in  \mathcal{L}_1(n)} \vec{s}_{l,t} - \sum_{l \in  \mathcal{L}_2(n)} (\vec{s}_{l,t}-\vec{z}_l\, {i}_{l,t}^2) + \vec{y}_n^h\, {v}_{n,t}^2, \; \forall n,t. \label{eq:Ch2BFM3}
    \end{gather}
    where $\vec{z}_l=r_l + j \, x_l$ is the total series impedance of power line $l$; $\vec{y}_n^h=G_n + j \, B_n$ is the total shunt admittance at bus $n$. $\mathcal{L}_1(n)$ and $\mathcal{L}_2(n)$ are subsets of power lines whose final and initial terminals are connected with bus $n$, respectively; ${v}_{n,t}$ and ${i}_{l,t}$ are the magnitude value of vectors $\vec{v}_{n,t}$ and $\vec{i}_{l,t}$, respectively; $\vec{s}_{l,t}$ is the apparent power of branch $l$, and "${}^\star$" denotes complex conjugation

    The set \eqref{eq:Ch2BFM1}--\eqref{eq:Ch2BFM3} can be decomposed as pairs of real an imaginary terms as follows: i) using the squared voltage variable $\bar{v}_n=v_n^2$ instead of $v_n$; ii) using the squared current variable $\bar{i}_l=i_l^2$ instead of $i_l$; iii) incorporating the nodal balancing equations \eqref{eq:Ch2CPFP}--\eqref{eq:Ch2CPFQ} with the resultant forms. Therefore, the complete AC branch flow-based OPF model is
    \begin{gather}
        \sum_{u \in \mathcal{U}(n)}{p}_{u,t} + \sum_{l \in \mathcal{L}_1(n)}({p}_{l,t} - r_l \bar{i}_{l,t} ) -\sum_{l \in \mathcal{L}_2(n)}{p}_{l,t}- G_n \bar{v}_{n,t} = \sum_{d \in \mathcal{D}_p(n)}(P_{d,t} - \triangle p_{d,t}), \; \forall n,t,   \label{eq:Ch2BFOPF1}\\
        \sum_{u \in \mathcal{U}(n)}{q}_{u,t} + \sum_{l \in \mathcal{L}_1(n)}({q}_{l,t} - x_l \bar{i}_{l,t} )-\sum_{l \in \mathcal{L}_2(n)}{q}_{l,t} - B_n \bar{v}_{n,t} = \sum_{d \in \mathcal{D}_p(n)}(Q_{d,t} - \triangle q_{d,t}), \; \forall n,t, \label{eq:Ch2BFOPF2}\\
        \bar{v}_{n,t} = \bar{v}_{m,t} - 2(r_l \, {p}_{l,t} + x_l \, {q}_{l,t}) + (r_l^2+x_l^2) \,\bar{i}_{l,t},\; \forall l,t, \label{eq:Ch2BFOPF3}\\
        {p}_{l,t}^2 + {q}_{l,t}^2 = \bar{v}_{n,t} \, \bar{i}_{l,t},\; \forall l,t. \label{eq:Ch2BFOPF4}
    \end{gather}
    where the active power shedding $\triangle p_{d,t}$ and reactive power shedding $\triangle q_{d,t}$ are considered. For sake of simplicity, the squared voltage and current variables are directly denoted as $v_{n,t}$ and $i_{l,t}$ in the thesis studies, in case of adopting branch power flow model.

    In order to fully prepare the AC-OPF model to be included in an optimization problem, a set od boundary constraints are required to define the upper and lower limits of the decision variables. Bus voltages are bounded with the physical and engineering limits as
    \begin{gather}
         \underline{V}_{n}^2 \le {v}_{n,t} \le \overline{V}_n^2 ,\; \forall n,t, \label{eq:Ch2ACVlimi3}
    \end{gather}

     The branch current is limited with the power line capacity as
     \begin{gather}
         \underline{I}_{l}^2 \le {i}_{l,t} \le \overline{I}_l^2 ,\; \forall l,t, \label{eq:Ch2ACIlimi}
    \end{gather}

\subsection{DC Optimal Power Flow Model} \label{sec:Ch2SDC}
    The AC-OPF is nonlinear and non-convex model, which brings additional complexities for the coupled power and gas system optimization problems. Therefore, the DC-OPF model, which is a linear approximation for the AC-OPF model, is usually employed to formulate the power network constraints. It is named as DC-OPF as its equations resemble the power flows in a DC network, however, it is still working for the AC networks.

    There are some assumptions required to derive the DC-OPF: i) reactive power flow is neglected; ii) power system is lossless, i.e., all power line resistances are very small ($\approx0$); iii) there are reactive power sources that can regulate all the system buses to be equal $1$; and  iv) the difference between the voltage angles of two connected busses is very small, i.e., $\sin(\theta_n-\theta_m) \approx\theta_n-\theta_m$. Therefore, the DC-OPF model will be formulated as
    \begin{gather}
        {p}_{l,t} = \frac{{\theta}_{m,t}-{\theta}_{n,t}}{x_l} ,\; \forall l,t, (m,n) \in l. \label{eq:Ch2DCOPF1}\\
        \sum_{u \in \mathcal{U}(n)}{p}_{u,t} + \sum_{l \in \mathcal{L}_1(n)}{p}_{l,t} -\sum_{l \in \mathcal{L}_2(n)}{p}_{l,t}= \sum_{d \in \mathcal{D}_p(n)}(P_{d,t} - \triangle p_{d,t}), \; \forall n,t.  \label{eq:Ch2DCOPF2} \\
        -\tilde{\pi} \le \theta_{n,t} \le \tilde{\pi} ,\; \forall n,t, \label{eq:Ch2DCAngle} \\
        -\overline{p}_{l,t} \le {p}_{l,t}\le \overline{p}_{l,t},\; \forall l,t. \label{eq:Ch2DCFlow}
    \end{gather}

    It should be noted that, under normal operating conditions, the DC-OPF model provides quite accurate power flows with a low execution time due to is linearity. However, it may lead to insufficient solutions under stressed power systems, where the bus angle differences are large and overestimation of bus voltages. Besides, the active and reactive power are coupled in the distribution levels, and the voltage is significantly influenced by the reactive power flows, therefore, adopting the DC-OPF model in the distribution level provides considerable errors in the power system decisions. In this paper, the DC-OPF and AC-OPF models are employed to model the power network at transmission and distribution levels, respectively.

\section{Interdependent Power and Gas Systems} \label{sec:Ch2Interdep}
    As discussed earlier that the shale gas revolution is enhanced by the development in horizontal drilling and hydraulic cracking technologies, therefore, the gas prices are decreasing significantly \cite{stevens2012shale}. Besides its economical advantages, the new concerns from the potential global warming have driven forces to raise its importance in the global energy balance due to the low carbon emissions and high environmental benefits \cite{kerr2010natural, EIA, myhrvold2012greenhouse}. It is observed that natural gas is promoted as the second largest energy source over the world \cite{EIA}. Therefore, and as an outcome of the progress in the deregulation and competition, new investments have been enrolled in the electricity markets by installing GPUs due to their economical and quick construction, friendly to the environment, and high operational efficiency and flexibility, which are necessary to handle the power system uncertainties, such as demand response, renewable power generation (RPG) output fluctuations as well as contingencies. The installation of GPUs has been so high, for example, it is expected that $60\%$ of new electric power units will be fueled by natural gas by $2035$ \cite{EIA}. That discusses why researcher give some attentions on modeling and solving the interacted power and gas systems during the last $20$ years. Before that, the two systems were separately analyzed and optimized. Furthermore, due to the advanced technologies of P2G facilities, which are recently employed to effectively convert electricity into gas to be stored, transported and reutilized by gas networks. P2G facilities are the most well-qualified solution for providing more flexibility to mitigate RPG output \cite{lund2015review}. Therefore, the interactions between power and gas systems are enhanced, and neglecting them may not provide the optimal decision for power system operators and may cause physical violations.

    In this section, an analogy between power and gas systems is presented, indicating the similarities and differences between the two energy systems and modeling the physical interactions (coupling components). Finally, the existing coordination scenarios in operation and planning are provided. It should be noted that modeling the economic interactions in the coupled system depend on the coordination strategy, the considered coupling components and the type of optimization technique, i.e., stochastic optimization (SO), robust optimization (RO) and distributionally RO (DRO). These interactions are rarely introduced in the recent studies, and their models are systematically proposed in Chapters~\ref{Chapter4}--\ref{Chapter5}.

\subsection{Similarities and Differences}
    The two energy systems, power and gas, are network industries, where the energy production units are connected to energy utilisers through transmission or distribution sub-systems. These two networks share some similarities while also owing distinguished characteristics. Traditionally, electronic-hydraulic analogies are prepared to explain how the electricity works \cite{}, where the electric components are represented by hydraulic ones, as the electric current is invisible and it is hard to illustrate the electric operations. However, recently, the modeling and solution techniques in power systems are more developed than those in natural gas systems. Therefore, power-gas analogy is needed to understand the similarities and differences and to apply the previous knowledge of power analysis. Table~\ref{tab:Ch2Analogy} presents a list of equivalent variables, components and models that are employed in this study.

    In the energy production, electricity is generated by non-GFUs, RPG sources and GFUs, and natural gas is produced from gas wells and P2G facilities, where GFUs and P2G facilities consume natural gas and electricity from gas and power systems, respectively. In the energy transportation, power transformers (gas compressors or reducers) are required to adjust the voltage (pressure) levels, and circuit breakers (gas valves) are installed to control the directions of power (gas) flow. The electricity and natural gas are derived through power lines and pipelines, respectively, in both transmission and distribution levels. DC power flow model is the simplified version of AC power flow, as shown Section~\ref{sec:Ch2SDC}, by ignoring the reactive power flow. And the steady-state gas flow model is an approximation of the dynamic-state model, as shown Section~\ref{sec:CH2SSS}, by neglecting the gas line pack. Therefore, one might consider that the AC-OPF and DC-OPF models are analogous to the dynamic- and steady-state gas flow models, respectively. However, the more accurate AC-OPF model is required to be used in the distribution-level, while the dynamic-state gas model is more suitable in the transmission-level to consider the large volumes of gas line pack. The interested reader can refer to \cite{he2018coordination, correa2015optimal} for more details about power-gas analogies.

    Although the two networks have many similarities that would help to understand the one system's operation in terms of the other's, there are very important differences, which are necessary to be considered. i) \textit{Speed of energy flow}: electricity travels at ultra-high speeds reach the speed of light, while natural gas moves at a slower speed within $40$--$50$~mi/h \cite{he2018coordination}. Therefore, electric power generation and consumption are balanced instantaneously. And gas takes longer time to be derived to consumers due to the gas compressibility and low velocity. ii) \textit{Energy storage}: large-scale electric storage is uneconomical and inconvenient, while the gas storage is cost-effective as it can be stored in storage facilities as well as in transmission pipelines as line pack. Therefore, additional flexibilities are provided to balance the swings of gas demands and to handle the network contingencies. The major source of these swings is the GPUs demands, which are employed to mitigate power system uncertainties. As a result, economical and physical interactions  must be considered in the day-ahead operation or in the pre-contingency to identify the optimal, feasible, economic and resilient operations for power system.
    \begin{table}[!th]
        \caption{Electricity-Gas analogy}
        \label{tab:Ch2Analogy}
        \centering
        \includegraphics[width=14.5cm]{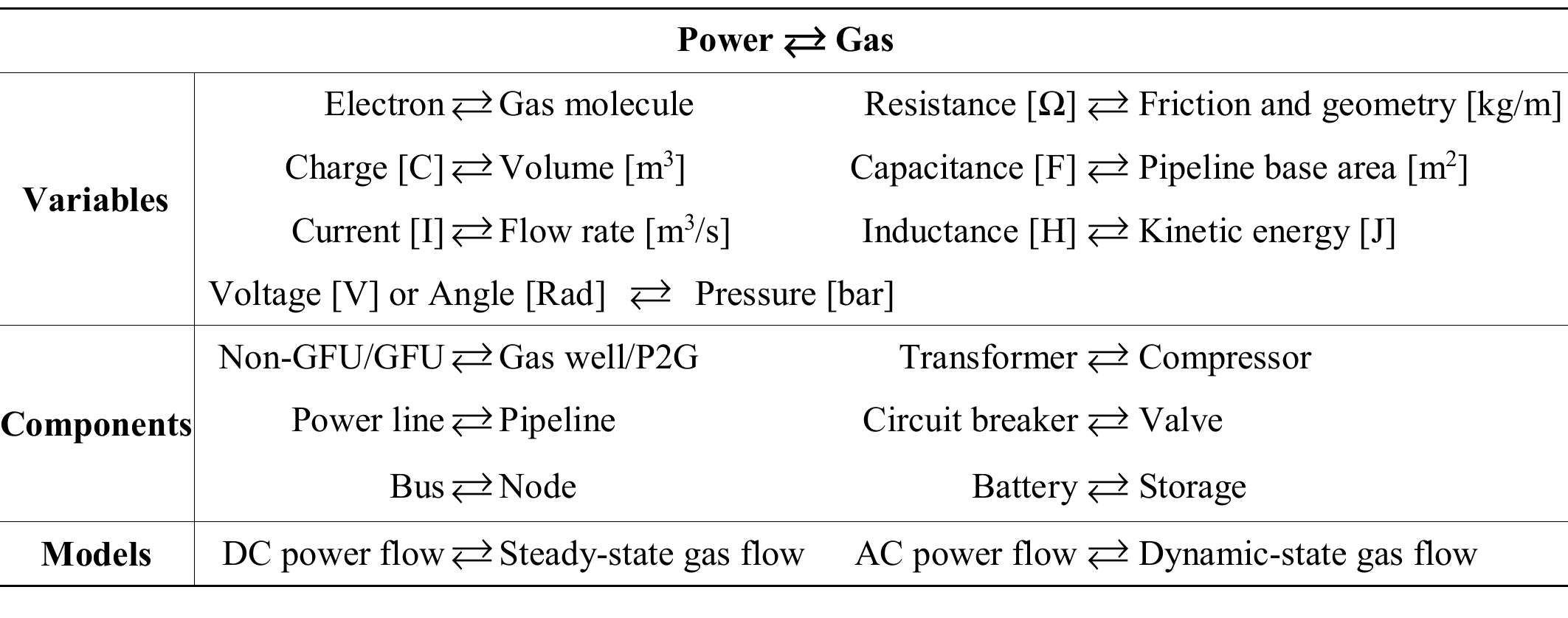} 
    \end{table}

\subsection{Coupling Components}
    GPUs, P2G facilities and electric-driven compressors represents the linkages between the power and gas systems. More specifically, the electricity network relies on the gas network for providing the gas fuel to GPUs and withdrawing the produced gas from P2G facilities. While the gas network relies on the power grid to supply the electric-driven compressor station for enhancing the gas transportation process.

\subsubsection{Gas-fired Power Units}
    GPUs are the most important and critical coupling components because of the following reasons:
    \begin{enumerate}
      \item
        Compared with other power units, such as coal- and oil-fueled units, natural gas is usually not stockpiled in the site. In other words, GPUs utilize just-on-time the delivered gas from gas network \cite{he2018coordination}. Therefore their operational flexibility directly depend on the gas network adequacy and capacity.
      \item
        In the industrial practice, there are two major types of gas delivery service, namely the firm and interruptible gas services, and GPUs usually get the latter due to their cost-effective \footnote{https://learn.pjm.com/three-priorities/keeping-the-lights-on/gas-electric-industry/natural-gas-electric-market.aspx} \cite{wang2019convex}. Moreover, the natural gas end-users have higher priorities than GPUs to be supplied with gas. Therefore, under insufficient gas supply or tight pipelines capacity, peak gas loads could impact on the interruptible gas delivery of GPUs, that introduces operational issues in the power system \cite{clegg2015integrated}.
      \item
         Due to the environmental concerns, large-scale renewable energy are promised to make a great contribution to future energy system. However, this contribution introduces additional challenges for the power system operation. In such conditions, GPUs are necessary to mitigate the renewable uncertainties by providing their flexible dispatchability and their high ramping capacity. In addition, the gas prices are decreasing significantly because of the shale gas revolution \cite{stevens2012shale}. Therefore, GPUs installation has been significantly increased, in the last decade around the world, to share the largest power generation capacity \cite{yang2018modeling, wang2019convex, zhang2015impact}. That is expected to continue increasing in the future as well. In turn, the operational flexibility of gas system is important to cope with volatile gas demands for GPUs.
    \end{enumerate}

    Three types of GFUs are introduced in the existing studies that are industrially employed, namely single-cycle gas-fired turbine, combined-cycle GFU, and dual-fuel units. The single-cycle turbine is the simplest GFU, where a consumption engine is employed to convert the gas into mechanical power, which further transformed to electric power. The combined-cycle GFU comprises a steam unit and multiple gas turbines, where the waste heat from turbines is utilized in the steam unit to increase the power plant efficiency \cite{guan2018unified,  pan2017data}. The dual-fuel power plant could switch from gas fuel to other types under insufficient gas delivery. Therefore, they have the ability to shave the peak gas demands and to provide the power system operational reliability and security \cite{defourny2017value}.

    GFU gas consumption can be calculated by
    \begin{gather}
           \frac{\Phi}{\eta_u} \left( G_u(c_{u,t} p_{u,t}) + G_u^+ y_{u,t} + G_u^- z_{u,t} \right), \; \forall u \in  \mathcal{U}_g,\;t  \label{eq:Ch2GUF}
    \end{gather}
    where $c_{u,t},\; y_{u,t}$ and $z_{u,t}$ are the UC decisions (please refer to Appendix \ref{App:UC}), $G_u(.)$ is a convex function for the energy in MWh required to produce one unit of electricity; $G_u^+$ and $G_u^-$ are the amount of energy in MWh required in starting up and shutting down the GPU, respectively; $\eta_u$ and $\Phi$ are the GFU efficiency and power to gas conversion factor.

\subsubsection{P2G Facilities} \label{sec:Ch2P2G}
    The excessive RPG outputs could be accommodated by electric storages such as pumped-storage hydro units \cite{yao2019optimal, jiang2011robust}, batteries \cite{strang2017feasibility}, and compressed air facilities \cite{cai2019optimal}. However, these techniques can offer limited capacities for energy storage because of their technical and economical aspects \cite{he2018coordination}. Because the natural gas can be stored with large capacities in a cost-effective manner, considering the gas line pack inside pipelines, P2G facilities are recently employed to effectively convert electricity into natural gas, which further is stored, transported and reutilized by gas networks. Therefore, dissipated energy through renewable energy curtailment could be significantly mitigated. The existing researches \cite{yang2018modeling, chuan2017robust, belderbos2019storage} agree that P2G facilities can help the power system operation in mitigating the fluctuations of energy loads, the surplus renewable energy, recycling CO\textsubscript{2}  and offering ancillary services.

    The P2G process has been proposed in $1980s$-$1990$ \cite{hashimoto1999global}, and the first P2G pilot plant for CO\textsubscript{2} recycling by sea water is built in $2003$ \cite{gotz2016renewable}. P2G process has two major steps: 1) water electrolysis, which produces hydrogen from electric power; 2) methanization, which produces methane from hydrogen using a suitable carbon source. There are three types of P2G facilities that are realized or under planning, namely i) alkaline electrolysis, which is the cheapest and well-understood type; ii) polymer electrolyte membranes, which is more newer with higher flexibility; iii) solid oxide electrolysis, which is the most recent one and still at the laboratory stage \cite{ghaib2018power}. Typically, the over all efficiency of P2G units is about $26\%$-$51\%$ \cite{gotz2016renewable}. It should be noted that the hydrogen storage capacities are restricted with technical and legislative considerations, while the methane can be injected to the gas system pipelines.

    In the coupled power and gas system operation, P2G facilities are addressed as power loads in the power system and gas sources in the gas system. According the recent literature \cite{chuan2017robust, wang2018equilibrium}, P2G units can be linearly modeled as
    \begin{gather}
           \varrho_{z,t} = \Phi\eta_z p_{z,t} , \; \forall z \in  \mathcal{Z},t,  \label{eq:Ch2P2G} \\
           \underline{P}_{z,t} \le  p_{z,t} \le \overline{P}_{z,t} , \; \forall z \in  \mathcal{Z},t  \label{eq:Ch2P2G2}.
    \end{gather}
    where $\varrho_{z,t}$ is the amount of gas production from unit $z$ at time $t$, $p_{z,t}$ is the consumed electric power, $\Phi$ and $\eta_z$ are the energy conversion factor and P2G efficiency, $\underline{P}_{z,t}/\overline{P}_{z,t}$ is the minimum and maximum consumption power, respectively.

\subsubsection{Electric-driven Compressor Stations}
    As discussed in Section~\ref{sec:Ch2SComp}, compressor stations are installed along the pipelines to facilitate the gas transportation. They keep the gas pressure in its technical and contractual levels that might be dropped due to the pipelines friction, long distance and elevation difference. The detailed compressor model and its simplified one, as well as the solution methods are presented in Appendix~\ref{App:Comp} and Section~\ref{sec:Ch2SComp}.

\subsection{Coordination Strategies}

    Due to the intensive strengthen interdependencies between power and gas systems\footnote{https://www.pjm.com/markets-and-operations.aspx}\textsuperscript{,}\footnote{https://www.eia.gov/todayinenergy/detail.php?id=34612}, it may not be practical or technical reasonable to separately model and optimize the two energy systems without considering their physical and economical interactions. In order to address this interdependency, three types of coordination strategies are discussed in the existing studies:
    \begin{enumerate}
      \item \textbf{Power system optimization models considering physical constraints of gas system}

            In this coordination, the impact of gas infrastructure capacities on the gas fuel delivering to GPUs is addressed. Due to the high priority of gas end-users, especially when the gas and power loads are peak together, and when the gas system operational constraints are not considered, power system operating decisions may be sub-optimal or infeasible. Therefore, quite a few researchers have incorporating the physical constraints of gas system in the power system optimization models, focusing on the security-constrained unit commitment (SCUC) problem. A simplified linear constraints of the gas system is included in the SCUC problem. \cite{sohrabi2019coordination}, while the transient behavior of gas system is considered in  \cite{badakhshan2019security}.  Gas marginal prices are defined to optimize the SCUC \cite{chen2019unit}. A SCUC problem is proposed to consider the AC-OPF and gas dynamics in a two-stage optimization model, and the reactive power dispatch is optimally obtained. Gas network awareness is analyzed in \cite{byeon2019unit} through a SCUC model.

      \item \textbf{Gas system optimization models considering gas consumptions of power system}

            In this coordination, the impact of gas demands of GPUs on the nodal pressure levels, line pack and gas system operational security has been investigated. Specifically, the time-varying gas consumptions are incorporated in the gas system optimization models to explore the risk of RPG fluctuations \cite{chertkov2015pressure}. In \cite{behrooz2017dynamic}, a gas network simulator with the stochastic optimization technique is established to find the optimal planning strategy under future uncertainties introduced by GPUs demands. PDE equations of gas dynamics have been discretized in \cite{zlotnik2015optimal} to fined the optimal control of transient gas flow inside pipelines. Interested readers can refer to the up-to-date models \cite{zlotnik2015model, sukharev2019mathematical, ghasemi2018importance}, which focus on this coordination.

      \item \textbf{Co-optimization models for power and gas systems}

          Unlike the above two coordination types, which optimize operational costs of one system considering the physical interactions of another, the two systems are coordinated and optimized together to improve energy efficiency, to reduce the overall planning as well as operating costs, and to decrease the dependence on fossil fuel. Numerous studies have been conducted to discuss the interdependency and optimal operation of integrated electric-gas system (IEGS) \cite{he2018coordination, shahidehpour2005impact}. The importance of considering natural gas infrastructure in the IEGS with high RPG penetrations is investigated in \cite{devlin2016importance, golari2014two}. With P2G technology, the excessive RPG can be absorbed to facilitate power/gas load leveling \cite{chuan2017robust} and to control the wholesales in the electricity and gas markets \cite{khani2017optimal}. A steady-state economic dispatch of the IEGS is established considering GPUs and P2G facilities in \cite{zeng2016steady} to enable bi-directional energy flow, and the impact of demand response is investigated in \cite{cui2016day}. To mitigate the uncertainties associated with wind generation, an interval optimization IEGS model is presented in \cite{bai2016interval}, and a stochastic UC for IEGSs is introduced in \cite{alabdulwahab2015stochastic} to solve the issues of random power outages and electricity load forecasting errors, based on which the impacts of P2G facilities are analyzed in \cite{guandalini2015power}. The $N-1$ contingency model for IEGSs was analyzed from economic and security-related aspects for a single outage in \cite{correa2014security}, and the model was improved by considering the spinning and regulation reserves in \cite{liu2018day}. Furthermore, researchers have studied different decentralized algorithms to derive high-quality decisions for both systems considering limited data exchange and preserving decision independency \cite{alkano2016distributed, nguyen2017distributed, liang2019generalized,he2016robust, wen2017synergistic}.

    \end{enumerate}

        The reliable, resilient and economic operation of the electric power system is essential to strengthen and support economic and social activities in modern society. The thesis work concentrates on the resilient and economic operations of the electric power system. Therefore, the second coordination strategy is not considered in the following chapters.

        In fact, different studies have presented numerous resilient and economic operation models to consider the power system vulnerabilities and uncertainties. Some of them identify the critical components of the power system \cite{crucitti2004model, ouyang2012time, ouyang2014multi, adachi2008serviceability, arroyo2010bilevel, aksoy2019generative, salmeron2009worst}.  However, the mere identification and protection of the vulnerable components do not assure an optimal defense plan in case of serious system disturbances. Therefore, other models were developed to determine the optimal protection strategies for such vulnerabilities \cite{hausken2009minmax, yuan2014optimal, yuan2016robust, huang2017integration, lai2019tri, fang2017optimizing, lin2018tri}. Others focus on how to model and solve the RPG output uncertainties, for example see the recent surveys \cite{kumar2017recent, deng2019power, wang2018review}. The aforementioned models can be used to determine the optimal operational strategies based on the requirements of the electricity utilities, however, they neglect the physical interactions between power systems and natural gas systems. Therefore, the “independent power system” (IPS)  model may not provide the optimal decision for power system operators (PSOs) and it may cause physical violations for the interacted gas systems, such as under/over nodal pressure and/or well production capacity violations \cite{zlotnik2016coordinated}.

        Consequently, the third coordination strategy, which overcome the above mentioned issue of neglecting physical interactions, provides a strong solution in terms of energy efficiency improvement and cost-effective perspectives.  However, in this coordination, the IEGS models share one underlying assumption, namely, the existences of one operator or utility that has full control and operation authority over the both systems. This operator minimizes all costs associated with energy production and provides optimal decisions for the combined system. However, in industrial practice, there are significant institutional and administrative barriers to operate the two systems in a holistic manner \cite{chen2019operational}. Power system(s) and gas system(s) are operated by different utilities and they are unsynchronized in most countries and regions as in European countries \cite{bertoldi2006energy} and in China \cite{deng2017study}. This lack of synchronization indicates that the total fuel cost minimization determined by the IEGS models might not be a realistic operational objective for autonomous sub-systems.

        Therefor, the first coordination strategy is the most suitable and practical one to be used in the recent operation mechanisms of the coupled power and gas system.  The existing studies, which characterize this coordination, suffered from three major drawbacks, namely (1) they concentrate on the physical interactions with gas system, neglecting the economical ones. Energy contracts, including firm and reserved gas contracts, are absent that may provide unrealistic decisions for power system operators; (2) they focus only on the UC problem, neglecting the resilient and economic dispatches (ED), which admit computational challenges due to the non-convexity of gas dynamics before and after uncertainty realization. In fact, ED against contingencies or RPG fluctuations are used to find the locational marginal energy and reserve prices in the energy markets; (3) The recent solution methodologies cannot guarantee the solution feasibility in case of adopting the dynamic-state gas flow model. It should be noted that chapters~\ref{Chapter4}--\ref{Chapter6} provide compatible operational strategies with the existing industrial practices for the power system and energy markets, based on the this coordination strategy along with improving its derelictions in the recent work. Moreover, Chapter~\ref{Chapter3}, which focuses on the optimal energy flow by proposing novel solution methodologies, neglecting the decision-making challenges, employs the third coordination strategy through deterministic models.

\section{Conclusions and Discussions}
    The interdependencies and interactions between the largest two energy systems, electric power system and natural gas systems, are intensified due to the wide deployment of coupling components, namely GPUs, P2G facilities and electric-driven compressors, as a result of their operational flexibilities, advanced technologies, and environmental benefits. This chapter provides the mathematical models and formulations, which represent the physical structure of the two energy systems,  and recent developments in the coupling and coordination strategies. The presented formulations are employed for the proposed models and solution methods in the next chapters.

    This chapter started with discussing the physical structure of the natural gas system and the mathematical formulations of the main components. The system components used in the thesis work are gas sources, compressors, gas nodes, pipelines, valves and storages. The dynamic-state gas flow model, which is rarely used in the recent works due to its complexity, is adopted in the thesis work to provide additional operating flexibility and practical system representation. The steady-state gas flow models is formulated to be compared with the dynamic-state model.

    Then, the electric power system modeling is presented. OPF, which is the fundamental issue in power system planning and operation, is formulated and its versions and applications are discussed. The AC-OPF and DC-OPF are mathematically modeled to be employed in the distribution and transmission levels electric systems, respectively. Finally, the similarities and differences between the electric power system and gas system are provided, indicating the physical interactions between them. Different types of coordinations between the two systems are listed, while demonstrating their applicability with recent industrial practice.

    Different simplifications are employed to obtain tractable formulations for power and gas systems to be incorporated in their operational optimization problems. However, more developments are required to control and systemize the final solution accuracy and the computational burden. For example, the simplified dynamic-state gas flow model is formulated under some assumptions applied on the PDEs, and the gas flow inside compressors and their consumed energy are usually approximated into linear constraints. In the IEGS optimization problems, it is generally to apply the DC-OPF, neglecting the reactive power flow, transformer models, voltage tolerances, and exact operations of power generators. These simplifications could provide errors in the integrated system operation. The questions are that 1) is the exactness of the final decisions acceptable for the interactive utilities; 2) is it possible to deduce the solution quality with such assumptions; 3) is it possible to better represent the gas system dynamics with low computational burden.  In fact, the IEGS research is in its first era, and it is very fast in modeling and solution methodology developments.  Modern modeling techniques are suggested to represent the energy systems, such as \cite{zhang2020transient,chevalier2020dynamic, garcia2020generalized}, which can be employed in IEGS decision-making frameworks.


\chapter{Optimal Energy Flow for Integrated Electric-Gas Systems} 

\label{Chapter3} 




    Optimal energy flow is the most fundamental problem in the energy system operation, and the enhanced interdependencies between power and natural gas systems provide additional challenges to this problem from transmission to distribution levels. For gas system operational constraints, the steady-state gas flow model is extensively adopted, considering the inlet and outlet gas flow of a pipeline are equal, neglecting the gas line pack. The non-convex Weymouth equation imposes a major complexity on seeking the feasible and optimal gas flow (OGF). For the power system operational constraints, the DC optimal power flow (OPF) model is commonly adopted for simplicity in the transmission level IEGS. The accurate AC-OPF must be adopted in the distribution level IEGS, however, it is also non-convex and presents more difficulties in the IEGS optimization problems. This chapter is about how to find the optimal power-gas flow (OPGF) in the IEGS, guaranteeing its feasibility and optimality. Two different efficient methods based on convex optimization approaches are proposed for transmission and distribution levels, respectively.

    A comprehensive study on the existing methods and approaches suggested to solve the OPGF problem is presented in Section~\ref{sec:Ch3OGF}.  These methods include the nonlinear and linear programming, dynamic programming, simulation, heuristics and convex relaxations. In Section~\ref{sec:Ch3MathForm}, day-ahead multi-period frameworks for economic dispatch are formulated for the transmission- and distribution-level IEGSs, respectively. In these models, the gas flow compressibility and slow traveling velocity as well as bidirectional gas flow are considered. Three different solution methodologies are provided to solve the OPGF problems in Section~\ref{sec:Ch3Meth}. The first method is to reformulate the transmission-level IEGS model into a mixed integer linear programming (MILP) framework using piecewise linear approximation (PLA) of the quadratic Weymouth equation. It is extensively adopted in the literature for IEGS optimization problems.

    In the second method, which is presented in Section~\ref{sec:Ch3GFCMethod}, the Weymouth equations are relaxed as second-order-cone programming (SOCP) constraints. As a result of considering bidirectional gas flow inside pipelines, the proposed model is converted into a mixed integer SOCP (MISOCP) framework. A gas flow correction (GFC) method is proposed based on the multi-slack-node method and the Levenberg-Marquardt algorithm to provide a tight relaxation and find exact production schedules for the IEGS. This work has been published as
    \begin{itemize}
      \item
            Ahmed R. Sayed, Cheng Wang, Tianshu Bi, and Arsalan Masood. "A Tight MISOCP Formulation for the Integrated Electric-Gas System Scheduling Problem." In 2018 2nd IEEE Conference on Energy Internet and Energy System Integration (EI2), IEEE, 2018,  pp. 1-6. DOI:  https://doi.org/10.1109/EI2.2018.8582239
    \end{itemize}

    Then, another contribution is provided to adopt the AC-OPF in the power system operational constraints. Section~\ref{sec:SCPMethod} focuses on finding the OPGF in the coupled system at distribution level, where AC-OPF is adopted. A sequential-MISOCP (S-MISOCP) algorithm is proposed to find the OPGF in the formulated IEGS optimization model. The non-convex power flow and gas flow equations are decomposed as difference-of-convex programming (DCP) functions, which are reformulated as MISOCP constraints. Starting with an initial point, a sequence of penalized MISOCP problems are solved to find a feasible OPGF close to, if not equal to, the optimal one. The feasibility and quick convergence are guaranteed by designing an adaptive penalty growth rate and suggesting a high-quality initial point, respectively. Moreover, bidirectional gas flow inside pipelines is considered to incorporate meshed gas networks. This work has been published as
    \begin{itemize}
      \item
            Ahmed R. Sayed, Cheng Wang, Tianshu Bi, Mohamed Abdelkarim Abdelbaky, and Arsalan Masood. "Optimal Power-Gas Flow of Integrated Electricity and Natural Gas System: A Sequential MISOCP Approach." In 2019 3rd IEEE Conference on Energy Internet and Energy System Integration (EI2), IEEE, 2019,  pp. 283-288.
            \vspace{-2mm}

            DOI:  https://doi.org/10.1109/EI247390.2019.9062250
    \end{itemize}

    Finally, in Section~\ref{sec:Ch3Simulations}, the reformulated MISOCP model with the GFC method is compared with the widely adopted analogous models from the literature, namely  MILP model and MISCOP relaxed model without the proposed GFC method. Case studies are conducted to highlight the line pack consideration in the integrated system operation. Other studies have been conducted on a distribution-level IEGS test system to validate the S-MISOCP algorithm performance and convergence with both the dynamic-state gas flow and the AC power flow models.

    In fact, the above solution methods are adopted in this chapter to optimize deterministic IEGS models, which disregard the system uncertainties. However, the S-MISOCP algorithm can be employed to guarantee the decisions feasibility in each stage of non-deterministic optimization models due to its tractability. This algorithm is modified to be applied in solving two-stage robust and distributionally robust optimizations models against renewable generation uncertainties in chapters~\ref{Chapter5}--\ref{Chapter6}.

\section{The State-of-the-Art Methods} \label{sec:Ch3OGF}

        The objective of system operators is minimizing the operating costs of energy transportation in their systems, fulfilling all physical, economical, technical, contractual, and legal constraints as well as any type of interactions with other systems. The optimization problems of both power and gas networks in planning and operation are challengeable. Nowadays, planners and operators manage larger and more complex transport grids with significant growth in production and consumption, with bilateral energy transactions, and with higher levels of interconnection between energy networks. Identifying an accurate, feasible and optimal solution of the mathematical formulations for the interconnected systems, either transmission or distribution for power systems or dynamic- or steady-state for gas systems, represents a major challenge. In this section, the existing solution methods that are adopted in solving the OGF and OPF are separately presented.

    For power system modeling, due to the simplicity and linearity of DC-OPF model, which can be efficiently solved by the most of commercial solvers, it is usually used to formulate the electric power systems in the IEGSs literature models (for example but not limited to \cite{correa2014security, wang2016robust, Correa2015Integrated, he2016robust, liu2018day, wang2019convex, chen2019operational, wang2018strategic, chen2019equilibria}). These studies mainly concentrate on the transmission level, where  the difference between the voltage angles of two connected busses is very small and the power lines are lossless. However, stronger interactions in the IEGS are observed in the distribution-level \cite{evans2013age, gao2017robust, qiu2018scheduling}. In the power distributed network (PDN), the AC-OPF is a fundamental issue of PDN operation. The AC-OPF problem, which can be formulated as a branch flow model \cite{baran1989network} or bus injection model \cite{dommel1968optimal}, is a non-convex framework with quadratic constraints, please refer to Section~\ref{sec:Ch2Power} for more details. Various studies have been conducted to find the OPF in power system operation. Because different review articles and surveys have addressed the applicable and efficient solution methods and approaches in the pertinent literature, there is no need to comprehensively discuss the recent methodologies.  An up-to-date comprehensive survey on the AC-OPF solution methodologies is presented in \cite{frank2012optimal} and \cite{frank2012optimal2} for deterministic and non-deterministic models with AC-OPF formulations, respectively. Convex relaxation methods, such as SOC relaxation \cite{li2012exact}, convex quadratic relaxation \cite{coffrin2015qc}, and semi-definite relaxation \cite{farivar2013branch}, could provide the optimal objective value if the relaxation is exact. In \cite{li2012exact}, it is proven that SOC relaxation is exact for PDN with the radial network under mild conditions. These conditions include the objective function that is convex, increasing with all power injecting sources, non-decreasing with power loads, and increasing with line losses. These conditions are sensitive to the objective function and system data, therefore SOC relaxation method may fail to find a feasible AC-OPF. In such circumstances, valid inequalities are suggested to enhance the relaxation tightness \cite{kocuk2015strong}, however, they could not provide a zero-optimality gap. Therefore, the local heuristic penalty convex-concave procedure (P-CCP) based on difference-of-convex programming (DCP) introduced in \cite{R35}, is utilized to locate a feasible and (local) optimal solution for AC-OPF problems \cite{wei2017optimal}.

    For natural gas system modeling, there are two major approaches for gas networks, namely numerical simulation and optimization frameworks. The simulation approaches are implemented to identify the actual response of gas network by a certain number of runs under different control variables \cite{osiadacz1996different}. The simulation results can describe the original nonlinear gas flow equations with a level of accuracy based on how discretized the PDEs is. However, the simulation approaches can not guarantee the solution optimality and they need the knowledge and experience of the operator to increase its performance \cite{borraz2010optimiz}. To simulate the gas network, some numerical methods can be found in \cite{woldeyohannes2011simulation, dorin2008modelling}. On the other hand, the mathematical optimization frameworks are modeled to find an optimal set of decision and control variables through a single optimization run. Besides being NP-hard, the optimization problem is nonlinear non-convex due to the presence of gas flow equations, therefore, simplifications and/or relaxation must be done, and the time and space discretization is not as fine as employed in simulation.

    The literature in the techniques of modeling and solving the gas system is vast. The interested reader can refer to \cite{rios2015optimization, corlu2020optimizing} for the state-of-the-art reviews. The main techniques are discussed below, demonstrating how the non-convex gas flow equations are addressed.

\subsection{Nonlinear Programming and Heuristic Techniques}
    Nonlinear techniques, which include nonlinear programming (NLP) and mixed-integer NLP (MINLP), are adopted to optimize the gas system operation problem, considering the gas flow nonlinearities in their original form. Three different approaches are proposed to solve this problem \cite{correa2015optimal}: i) algorithms for nonlinearly constrained optimization problems, for example interior point methods (IPMs) and successive quadratic programming (SQP); ii) global optimization algorithms, such as the branch-and-reduce algorithm; iii) iterative algorithms using linear and nonlinear solvers, where, in the first stage, the linear solver provides a good initial point to the nonlinear one in the next stages. The steady-state gas model is formulated as MINLP to be solved in its original nonlinear form to be solved by BARON in \cite{selot2008short}. In \cite{munoz2003natural}, the gas flow directions are determined by a MILP solver as an initial solution, then the CONOPT solver is used to optimize the gas network. With the same sequence, a primal-dual IPM with the Newton-Raphson approach are employed in \cite{qiu2018scheduling}, and the MINOS solver is utilized in \cite{guldmann1999optimizing}, considering the wind speed uncertainties \cite{martinez2013robust}. The dynamic-state gas model is solved by SQP method \cite{yan2016co2}, general purpose global optimization \cite{borraz2010optimiz} or iterative algorithms \cite{xue2019optimizing}, where the gas nonlinearities are relaxed into MILP to find a suitable integer variables, then NLP or SQP tackled the original formulation.

    Some studies have been conducted to optimize the gas system operation by applying non-classical (or heuristic) techniques. An heuristic algorithm is proposed to solve the steady-state gas flow problem in \cite{borraz2010optimiz}. A hybrid tabu search algorithm \cite{borraz2009improving} and a simulated annealing algorithm \cite{mahlke2007simulated} are developed to minimize the gas consumption by gas-driven compressors. Different works have adopted the genetic algorithm and particle swarm algorithm to find a high quality solution for the gas problem, such as \cite{panapakidis2017day, hu2016nsga} and \cite{wang2018scenario, yousif2018application}, respectively.

    Although these techniques consider the original nonlinear form, they cannot guarantee global optimality due to considering the non-convexity \cite{geissler2012using, moritz2007mixed}. For large-scale gas networks, nonlinear solvers are not able to handle the dynamic models due to their limitations in search space without expert information as well as in utilizing the underlying equations, and they could not guarantee global convergence, such as the iterative methods. Therefore, these techniques could complicate the integrated operations for multiple energy systems \cite{moritz2007mixed}.

\subsection{Dynamic Programming Techniques}
    Dynamic programming (DP) algorithms has proved to efficiently solve the gas problem with its nonlinearity and discontinuity constraints \cite{wong1968optimization}.  A multi-time-period optimization model is solved by a hybrid approximate DP algorithm in \cite{shuai2020real}, considering the advantages of model predictive control (MPC). A two-level model is suggested for transient gas flow model in \cite{zhang2016minimizing}, where the levels are pipeline and compressor levels, respectively, and the subproblems are handled by DP algorithm. In \cite{zhu2020adp}, a DP-based decentralized algorithm is advised to find the optimal energy flow of IEGS. In \cite{demissie2017multi}, a multi-objective optimization model is suggested to combine the power consumption minimization and the gas delivery maximization in one optimization problem for IEGS. Based on tree decomposition algorithm, the DP algorithm is applied for optimal gas flows considering compressor stations and the cost fuel minimization with a reduction technique \cite{borraz2004non}. An integrated gas and hydrothermal system is optimally operated by DP algorithm based on dual decomposition and Lagrangian relaxation \cite{unsihuay2007short}.

    However, the DP algorithms might have some limitations. Kelling et al. \cite{kelling2000practical} have introduced some concerns for the several partial solutions, which need variables monitoring and standard solution ranges. Furthermore, it is argued that DP algorithms are limited for radial gas networks and their execution time exponentially increase with the network size \cite{borraz2010optimiz}.

\subsection{Linear Programming Techniques}
    One way to overcome many of the aforementioned disadvantages of NLP techniques is to replace the nonlinear equations by approximated linear functions. The optimal solution would depend on the linearization accuracy, which can be measured and controlled. Therefore, with suitable and predefined tolerances, linear techniques guarantee that the global optimal solution can be found, that is not achieved by NLP or heuristic methods due to the presence of non-convexities. Due to the trustworthiness and straightforwardness of linear programming (LP) algorithms, linearization methods, including LP and mixed-integer LP (MILP), have been extensively applied in literature on gas flow optimization \cite{alabdulwahab2015stochastic, correa2014gas, shao2016milp}.

    The gas flow equations are approximated into a set of upper planes obtained from the first order Taylor expansion with fixed gas flow directions, and they are replaced by several linear inequality constraints in \cite{midthun2007optimiz}. However, the formulated convex feasible set may not provide the optimal objective in some cases, such as maximizing the line pack inside pipelines that is needed to mitigate the gas load variations \cite{pepper2012implementation}. Consequently, this approximation is not able to consider bidirectional gas flow and it forces the gas flow to be in a certain range \cite{correa2014security}. In \cite{pepper2012implementation}, a successive-LP (SLP) algorithm \cite{palacios1982nonlinear} is applied to better represent the flow equations, where the lower bound of the convex set obtained from \cite{midthun2007optimiz} is updated by penalizing some deviation variables in the objective function and the feasible region would be get narrow. This approach is employed to solve a multi-period optimization model for IEGS \cite{chaudry2008multi}, and to check the subproblem gas feasibility \cite{liu2011coordinated} for a coupled power and gas system. In \cite{de2000gas}, the steady-state gas model is iteratively solved by simplex algorithm, where two linear problems are formulated: the first one is to fined an initial point while neglecting compressor model; the second one considers all constraints. Another iterative algorithm is proposed in \cite{van2004math}, where the nonlinear function is approximated by a dynamically adjusted linear plane to mitigate the solution error. The Newton-Raphson method is adopted in \cite{liu2009security} with a starting point obtained from the projection method, which is well-addressed in \cite{correa2014security}.

    To overcome the iterative algorithms, which could introduce some difficulties in convergence, especially in the integrated energy system, the gas flow equation can be expressed as MILP formulation. In \cite{he2018co}, the sign function of Weymouth equations are relaxed by introducing binary directional indicator variables, and the resultant equalities are linearized by a Taylor series expansion method. Due to their fast, robust and applicability, piecewise linear approximation (PLA) methods are widely adopted to be solved by MILP techniques. A review of PLA methods is provided in \cite{lin2013review} to analyze their computational efficiencies. Martin et al. \cite{martin2006mixed} introduce a MILP formulations for the steady-state gas model. A multi-choice method is proposed in \cite{urbina2007combined} to approximate the squared variables of gas flow for the steady-state case in multi-energy systems. To consider gas dynamics, A novel special ordered set (SOS) constraints solved by a branch-and-cut algorithm \cite{mahlke2010mixed}, two-dimensional PLA \cite{geissler2011mixed}, one-dimensional PLA \cite{alabdulwahab2015stochastic} and a generalized incremental method \cite{domschke2011combination} are developed to formulate the gas flow nonlinearities into MILP forms. Valuable studies employ these formulations in the coupled power and gas optimization problems, e.g. \cite{he2016robust, Correa2015Integrated, shao2016milp, alabdulwahab2015stochastic, shao2016milp}.

    According to the above motivations, Correa and S{\'a}nchez \cite{correa2014gas} contribute theoretical and computational comparisons of MILP formulations for the gas network in both steady- and dynamic-state conditions, indicating the benefits and drawbacks of each PLA model. This study includes seven different PLA models, and it was shown that the incremental model outperformed the others in terms of computational time and accuracy. The incremental PLA model, which is the most widely adopted method to reformulate Weymouth equations into a tractable form to be employed in non-deterministic optimization models, is utilized in chapter~\ref{Chapter4} to identify the resilient operational strategies of power system with gas systems interactions. This model is theoretically and computationally compared with the proposed methods developed in this chapter. A detailed formulation of the incremental PLA model is presented in Appendix~\ref{App:PLA}.

\subsection{Convex Relaxation Techniques}

    Besides the LP techniques, other convex relaxations have drawn much attention in the pertinent literature to reformulate the non-convex general flow equations. A geometric programming approach is proposed for the fuel cost minimization for gas networks in \cite{misra2014optimal, hasle2007geometric}.  Different convex reformulations are introduced in \cite{humpola2015convex} for general nonlinear problems, where they are computationally analyzed. As an effective and efficient convexification method, second-order cone (SOC) relaxation are rapidly developed. In \cite{he2017robust}, the quadratic equality gas flow equations are relaxed into SOC programming (SOCP) for the distribution level IEGS with the assumption of fixed directions of gas flow. However, considering unidirectional gas flow is improper for the IEGS. Consequently, a SOC relaxation is proposed in \cite{borraz2016convex} to drive a high-quality solutions considering flow directions and on/off constraints. The resulting model is formulated as mixed-integer SOCP (MISOCP) framework, which is adopted on different practical large-scale gas systems. Further, this model is applied in IEGS, e.g., \cite{wen2017synergistic, he2017robustb, schwele2019coordination}.

Nevertheless, when the convex relaxations are not tight enough, the solution exactness cannot be guaranteed, yielding infeasible or suboptimal operating decisions. Therefore, a provable feasibility guarantee is non-trivial. To find a more accurate solution, there are two methods are suggested:
\begin{enumerate}
  \item Gas flow correction (GFC) method. In \cite{liu2018day}, a novel SOC relaxation is proposed to mathematically transform the non-convex programming problem into MISOCP model. Then, the Newton-Raphson algorithm is employed to correct the gas flow feasibility based on multi-slack-node gas flow calculation method. Although this study provide a high-quality decisions within suitable solution time, the line pack inside pipelines is neglected, and there is no attempt have been done to employ the GFC method in solving the IEGS optimization models in the dynamic-state conditions.

  \item Sequential convex programming (SCP) approach. In \cite{wang2018convex}, the Weymouth equations are reformulated as difference-of-convex programming (DCP) functions, and the SCP algorithm proposed in \cite{R35} is designed for the steady-state gas flow model in the IEGS, considering fixed gas flow directions. A decentralized operation model for IEGS is solved by an alternating direction method of multipliers (ADMM) in \cite{he2017decentralized}, where the gas flow feasibility is guaranteed by the SCP algorithm, where the DC-OPF is adopted.
\end{enumerate}

    Based on the above discussion, this chapter provides two novel methodologies to solve the OPGF problem considering both gas dynamics and bidirectional gas flow. The methodologies are based on the GFC and SCP methods to solve IEGS at transmission and distribution levels, respectively.
\section{Mathematical Formulations} \label{sec:Ch3MathForm}

    In this section, day-ahead multi-period models for economic dispatch are formulated for transmission- and distribution-level IEGSs, respectively. In these models, the gas flow compressibility and slow traveling velocity as well as bidirectional gas flow are considered. Before presenting the mathematical formulation of these IEGS models, some commonly prerequisite simplifications and assumptions, adopted in the literature, are stated as follows:
    \begin{enumerate}
    \item
        In general, (i) this study focuses on solving the OPGF with a deterministic model, and the uncertainties of power and gas systems are not considered \cite{wang2018convex} \cite{wei2017optimal}; (ii) the generated power prices and produced gas prices are known before optimization; (iii) there is one cooperator who has carte blanche to manage and control both power and gas systems.

    \item
        In power system modeling:

        \begin{itemize}
          \item In the transmission-level IEGS model, (i) the power system operates in a steady state, and the DC power flow model is adopted; (ii) All power units are fast response; (iii) Unit commitment (UC) is known. However, the proposed model can easily be extended to include traditional coal-fired units with minimum on/off times, and to include UC variables, please refer to Appendix~\ref{App:UC} for the UC problem.
          \item In the distribution-level IEGS model, (i) the PDN is radial with a balanced three-phase system; (ii) the branch flow model is employed, prohibiting bidirectional power flow \cite{wei2017optimal}; (iii) the required gas for GPUs are fully controlled by the generated active power only.
        \end{itemize}

    \item
        In gas system modeling, (i) the approximated Weymouth equation and dynamic-state gas flow model presented in Section~\ref{sec:Ch2SGasDynamic} are used; (ii) the linear model of P2G facilities and compressors \cite{wang2018strategic, correa2014security, correa2015optimal} are adopted.
    \end{enumerate}

\subsection{Transmission-level IEGS Model}
    Starting with the IEGS objective function definition, the power and gas operational constraints are listed, respectively. Finally, the holistic model is presented.

    \subsubsection{Objective Function}
        The objective function is to minimize the total operating costs associated with all energy suppliers by optimizing power system production costs along with the gas well production costs. In \eqref{eq:Ch3GFCObj}, the production costs of both systems are defined in the first two terms, and the non-served energy demands are penalized in the second two terms.
        \begin{align}
          \min_{\Omega}  & \sum_{\forall t} \big[ \sum_{\forall u \in \mathcal{U}_n} C_u (p_{u,t}) + \sum_{\forall w} C_w f_{w,t} + \sum_{\forall d \in \mathcal{D}_p} C_d \triangle p_{d,t} + \sum_{\forall d \in \mathcal{D}_g} C_f \triangle f_{d,t} \big] \label{eq:Ch3GFCObj}
        \end{align}
        where $\Omega$ is the set of all decision variables; $C_u(.)$ is the cost functions of non-GPUs; $C_w$ is the cost of gas production at gas wells; and $C_d/C_f$ is penalty of power/gas load shedding.

\subsubsection{Power System Operational Constraints} \label{sec:Ch3GFC_Power}
        The power system operational constraints are derived from Section~\ref{sec:Ch2SDC}, considering the power generation capacities. They are composed by:
        \begin{align}
           & \text{Power flow equation: \eqref{eq:Ch2DCOPF1},} \hspace{18em} \label{eq:Ch3DCOPF1}\\
           & \text{Bus angle limits: \eqref{eq:Ch2DCAngle},} \label{eq:Ch3DCAngle}\\
           & \text{Power flow limits: \eqref{eq:Ch2DCFlow},} \label{eq:Ch3DCFlow}\\
           & \text{Nodal balancing equation: \eqref{eq:Ch2DCOPF2},} \label{eq:Ch3DCOPF2}\\
           & \text{Generation capacities: } \nonumber \\
           &\hspace{10em} {c}_{u,t}\underline{P}_{u,t} \le {p}_{u,t} \le {c}_{u,t}\overline{P}_{u,t}, \; \forall u,t,   \ \label{eq:Ch3aPGa}\\
           & \text{Maximum ramping up and down limits: } \nonumber \\
           &\hspace{10em} -\overline{R}_{u}^- \le p_{u,t} - p_{u,t-1} \overline{R}_{u}^+, \; \forall u,t \label{eq:Ch3DCOPFRamp}        \end{align}
        where ${c}_{u,t}$ is a predetermined UC decision; $\underline{P}_{u,t} / \overline{P}_{u,t}$ is the minimum/maximum limit of power generation; and $\overline{R}_{u}^- / \overline{R}_{u}^+$ is the maximum  ramping down/up capacity.

        \subsubsection{Natural Gas Operational Constraints}  \label{sec:Ch3GFC_Gas}
        Gas system operational constraints are derived from Section \ref{sec:Ch2SGasDynamic} to consider the gas flow dynamics. They are composed by:
        \begin{align}
           & \text{Gas production capacities: \eqref{eq:Ch2Well},} \hspace{16em} \label{eq:Ch3aWell}\\
           & \text{Gas compressors constraints: \eqref{eq:Ch2Comp1}--\eqref{eq:Ch2Comp2},} \label{eq:Ch3aComp}\\
           & \text{Nodal pressure bounds: \eqref{eq:Ch2Pressure},} \label{eq:Ch3aPressure}\\
           & \text{Weymouth equation: \eqref{eq:Ch2Wey},} \label{eq:Ch3aWey}\\
           & \text{Average flow rate equation: \eqref{eq:Ch2AveFlow},} \label{eq:Ch3aAveFlow}\\
           & \text{Mass flow equation: \eqref{eq:Ch2GMMass1},} \label{eq:Ch3aGMMass1}\\
           & \text{Continuity equation: \eqref{eq:Ch2GMMass2},} \label{eq:Ch3aGMMass2}\\
           & \text{GFUs gas consumption: \eqref{eq:Ch2GUF} can be simplified as} \nonumber\\
           & \hspace{12em} \rho_{u,t} =  \frac{\Phi}{\eta_u} p_{u,t},\; \forall u \in \mathcal{U}_g,t, \label{eq:Ch3GFC-GUF}\\
           & \text{Nodal balancing equation:} \nonumber\\
           & \hspace{8em} \sum_{w \in \mathcal{W}(i)} {f}_{w,t} + \sum_{p \in \mathcal{P}_1(i)} {f}_{p,t}^{out} -  \sum_{p \in \mathcal{P}_2(i)} {f}_{p,t}^{in} + \sum_{c \in \mathcal{C}_1(i)} {f}_{c,t}^{out} \nonumber  \\
            & \hspace{8em}- \sum_{c \in \mathcal{C}_2(i)} {f}_{c,t}^{in} = \sum_{u \in \mathcal{U}_g(i)} \rho_{u,t} + \sum_{d \in \mathcal{D}_g(i)} (F_{d,t}-\triangle f_{d,t}),\; \forall i,t.\label{eq:Ch3GFCNodeShed}
            \end{align}
\subsubsection{The Holistic IEGS}
    The holistic non-convex IEGS model, at transmission level, can be cast as
    \begin{subequations} \label{eq:GFCModel}\begin{align}
          \min_{\Omega}  & \; \eqref{eq:Ch3GFCObj}\\
          s.t:\; & \text{Power system constraints: \eqref{eq:Ch3DCOPF1}--\eqref{eq:Ch3DCOPFRamp} }\\
          & \text{Gas system constraints: \eqref{eq:Ch3aWell}--\eqref{eq:Ch3GFCNodeShed} }\\
           & \Omega = \{\triangle p_{d,t}, \triangle \triangle f_{d,t}, {p}_{u,t}, {\rho}_{u,t}, {p}_{l,t} , {\theta}_{i,t},{f}_{w,t}, f_{p,t}, \pi_{i,t}, f_{p,t}^{in} , f_{p,t}^{out}, f_{c,t}^{in} , f_{c,t}^{out}, {m}_{p,t}\}
    \end{align} \end{subequations}

\subsection{Distribution-level IEGS Model}
    A stronger interactions in the IEGS have been observed in the distribution-level, where the interdependencies and interconnections between electricity and gas systems are intensified, due to the wide deployment of GPUs and P2G facilities for, respectively, their high operational flexibilities and advanced technologies. To highlight these interdependencies, the operational constraints of each coupled system and their coupled constraints are presented separately.

\subsubsection{PDN Operational Constraints} \label{sec:Ch3GFC_PDN}
        The PDN operational constraints are derived from Section \ref{sec:Ch2SOPF}, by adopting the branch flow model in the economic dispatch problem that can be summarized as
        \begin{gather}
        \underline{P}_u \le {p}_{u,t} \le \overline{P}_u;\:\;\: \underline{Q}_u \le {q}_{u,t} \le \overline{Q}_u,\forall u,t \label{eq:Ch3PLa}  \\
        \overline{R}_{u}^- \le {p}_{u,t} - {p}_{u,t-1} \le  \overline{R}_{u}^+,\; \forall u,t   \label{eq:Ch3PLb} \\
        0\le {p}_{z,t} \le \overline{P}_z,\forall z,t  \label{eq:Ch3PLc} \\
        0 \le p_{l,t};\;\; 0 \le q_{l,t},\; \forall l,t   \label{eq:Ch3PLd} \\
        0 \le i_{l,t} \le I_l^2,\; \forall l,t  \label{eq:Ch3PLe}\\
        \underline{V}_i^2 \le {v}_{i,t} \le \overline{V}_i^2,\; \forall i>1,t;\;\; {v}_{1,t}=1,\; \forall t  \label{eq:Ch3PLf}\\
        {v}_{m,t} = {v}_{n,t} -2(r_l p_{l,t} + x_l q_{l,t}) + (r_l^2 + x_l^2)i_{l,t}, \; \forall l,t  \label{eq:Ch3PLg}\\
        p_{l,t}^2 + q_{l,t}^2 = v_{i,t}i_{l,t}, \; \forall l,t \label{eq:Ch3PNL}
        \end{gather}

        In  \eqref{eq:Ch3PLa}, the generated active power and reactive power from all power units are restricted by their capacities. Ramping up and ramping down capacities of all power units are defined in \eqref{eq:Ch3PLb}. The active power consumption by P2G facilities is limited in \eqref{eq:Ch3PLc}. To fix the power flow direction, \eqref{eq:Ch3PLd} defines the lower boundary of active and reactive power. Squared line current and squared nodal voltage are restricted in \eqref{eq:Ch3PLe} and \eqref{eq:Ch3PLf} for power lines and power nodes, respectively. The line voltage drop is defined in \eqref{eq:Ch3PLg}, where $r_l, x_l, p_{l,t},$ and $q_{l,t}$ are the series resistance, series reactance, active and reactive power flows of line $l$, respectively. Finally, the non-convex power flow equation is expressed in \eqref{eq:Ch3PNL}.

\subsubsection{GDN Operational Constraints}  \label{sec:Ch3GFC_GDN}
        The GDN operational constraints are derived from Section \ref{sec:Ch2SGasDynamic} to consider the gas flow dynamics. They are composed by:
       \begin{align}
           & \text{Gas production capacities: \eqref{eq:Ch2Well},} \hspace{16em} \label{eq:Ch3bWell}\\
           & \text{Gas compressors constraints: \eqref{eq:Ch2Comp1}--\eqref{eq:Ch2Comp2},} \label{eq:Ch3bComp}\\
           & \text{Nodal pressure bounds: \eqref{eq:Ch2Pressure},} \label{eq:Ch3bPressure}\\
           & \text{Average flow rate equation: \eqref{eq:Ch2AveFlow},} \label{eq:Ch3bAveFlow}\\
           & \text{Mass flow equation: \eqref{eq:Ch2GMMass1},} \label{eq:Ch3bGMMass1}\\
           & \text{Continuity equation: \eqref{eq:Ch2GMMass2},} \label{eq:Ch3bGMMass2}\\
           & \text{GFUs gas consumption: \eqref{eq:Ch2GUF} can be simplified as} \nonumber\\
           & \hspace{12em} \rho_{u,t} =  \frac{\Phi}{\eta_u} p_{u,t},\; \forall u \in \mathcal{U}_g,t, \label{eq:Ch3SCP-GUF}\\
           & \text{P2G gas production: \eqref{eq:Ch2P2G}.} \label{eq:Ch3P2G}\\
           & \text{Weymouth equation: \eqref{eq:Ch2Wey},} \label{eq:Ch3bWey}
       \end{align}

\subsubsection{Coupling Constraints}
     All types of coupling components, namely GPUs, P2G facilities and electric-driven compressors, are considered. These components mainly interact in the nodal power and gas balancing equations, which are relaxed by adding the active and reactive power and gas loads shedding, as follows.
    \begin{gather}
        \sum_{u \in \mathcal{U}(n)}{p}_{u,t} + \sum_{e \in \mathcal{E}(n)} \hat{P}_{e,t} + \sum_{l \in \mathcal{L}_1(n)} (p_{l,t} - r_l i_{l,t})
        - \sum_{l \in \mathcal{L}_2(n)} p_{l,t} \nonumber \\
        = G_n v_{n,t} + \sum_{z \in \mathcal{Z}(n)} p_{z,t}
        + \sum_{c \in \mathcal{C}^e(n)} \alpha_c f_{c,t}^{in}/ \Phi + \sum_{d \in \mathcal{D}_p(n)} P_{d,t}(1-\delta_{d,t}), \; \forall n,t   \ \label{eq:Ch3PGa}\\
        \sum_{u \in \mathcal{U}(n)}{q}_{u,t} + \sum_{l \in \mathcal{L}_1(n)} (q_{l,t} - x_l i_{l,t}) - \sum_{l \in \mathcal{L}_2(n)} q_{l,t}  = B_n v_{n,t} + \sum_{d \in \mathcal{D}_p(n)} Q_{d,t}(1-\delta_{d,t}), \; \forall i,t \label{eq:Ch3PGb}\\
        \sum_{p \in \mathcal{P}_1(i)} {f}_{p,t}^{out} - \sum_{p \in \mathcal{P}_2(i)} {f}_{p,t}^{in} + \sum_{c \in \mathcal{C}_1(i)} {f}_{c,t}^{out} - \sum_{c \in \mathcal{C}_2(i)} {f}_{c,t}^{in} + \sum_{z \in \mathcal{Z}(i)} \varrho_{z,t} + \sum_{w \in \mathcal{W}(i)} {f}_{w,t}  \nonumber \\ = \sum_{u \in \mathcal{U}_g(i)} \rho_{u,t} + \sum_{d \in \mathcal{D}_g(i)} (F_{d,t}-\triangle f_{d,t}) ,\; \forall i,t \label{eq:Ch3PGc}\\
        0 \le \delta_{d,t} \le 1, \forall t,d \in \mathcal{D}_p;\;\; \triangle f_{d,t} \ge F_{d,t}, \forall t,d \in \mathcal{D}_g.  \label{eq:Ch3PGd}
    \end{gather}
    In the above expressions, the nodal active and reactive power balancing equations are defined in \eqref{eq:Ch3PGa} and \eqref{eq:Ch3PGb}, respectively.  $\mathcal{U}(n), \mathcal{E}(n), \mathcal{Z}(n),\mathcal{C}^e(n) $, and $\mathcal{D}_p(n)$ are subsets of power generators, wind farms, P2G facilities, electric-driven compressors, and power loads connected to node $n$, respectively. Subsets $\mathcal{L}_1(n)/\mathcal{L}_2(n)$ are the feeders whose initial/final node is $n$.  The balance equation for gas nodes is \eqref{eq:Ch3PGc}, where $\mathcal{W}(i), \mathcal{Z}(i), \mathcal{U}_g(i)$, and $\mathcal{D}_g(a)$ are subsets of gas sources, P2Gs, GPUs, and gas loads connected with node $i$. $\mathcal{C}_1(i)/\mathcal{C}_2(i)$ are subsets of compressors whose final/initial node is $i$. $\mathcal{P}_1(i)/\mathcal{P}_2(i)$ are subsets of pipelines whose terminal/starting node is $i$. The upper boundaries of active and reactive power load shedding, as well as gas load shedding, are defined in \eqref{eq:Ch3PGd}, where $\delta_{d,t}$ is the proportion of electric power load shedding.

\subsubsection{The Holistic IEGS Model}
    The objective of the IEGS model is to minimize the total operational costs for the integrated system, as defined in \eqref{eq:Ch3SCPObj}. Similar to \eqref{eq:Ch3GFCObj}, the objective of the proposed model includes the costs of power generated from non-GPUs, costs of the gas consumed from all gas sources, and costs of power and gas load shedding. Note, that the model objective is a convex quadratic function due to the presence of fuel consumption cost functions in the first term. The holistic model can be cast as
    \begin{subequations}\label{eq:Ch3SCPModel}\begin{align}
          \min_{\Omega}  & \;  \sum_{\forall t} \big[ \sum_{\forall u \in \mathcal{U}_n} C_u (p_{u,t}) + \sum_{\forall w} C_w f_{w,t} + \sum_{\forall d \in \mathcal{D}_p} C_d^p P_{d,t}\delta_{d,t} + \sum_{\forall d \in \mathcal{D}_g} C_d^f \triangle f_{d,t} \big] \label{eq:Ch3SCPObj}\\
          s.t:\; & \text{Power system constraints: \eqref{eq:Ch3PLa}--\eqref{eq:Ch3PNL}.} \label{eq:Ch3SCPObja}\\
          & \text{Gas system constraints: \eqref{eq:Ch3bWell}--\eqref{eq:Ch3P2G}.} \label{eq:Ch3SCPObjb}\\
          & \text{Coupling constraints: \eqref{eq:Ch3PGa}--\eqref{eq:Ch3PGd}.} \label{eq:Ch3SCPObjc}\\
          & \Omega = \{\delta_{d,t}, \triangle f_{d,t}, {p}_{u,t},{q}_{u,t}, {\rho}_{u,t}, {p}_{l,t} , {i}_{l,t} , {v}_{n,t},{f}_{w,t}, f_{p,t}, \pi_{i,t}, f_{p,t}^{in} , f_{p,t}^{out}, f_{c,t}^{in} , f_{c,t}^{out}, {m}_{p,t}\}
    \end{align} \end{subequations}

    In fact, because of the non-convexity of gas flow equations \eqref{eq:Ch3aWey} or \eqref{eq:Ch3bWey} and power flow equations \eqref{eq:Ch3PNL}, the above two models are not ready to be solved by commercial solvers. In what follows, the given problems are reformulated and solution methods are proposed to find a feasible decisions.

\section{Optimal Power-Gas Flow Calculation for IEGSs} \label{sec:Ch3Meth}

    In this section, three solution methodologies are provided to solve the above IEGS models.

\subsection{Piecewise Linear Approximation Method} \label{sec:Ch3MILP}
    This method can only be employed in the transmission level IEGS models \eqref{eq:GFCModel} because it can only handel the non-convexity of Weymouth equations, and PLA of the non-convex power flow equations provides unacceptable errors. It is to reformulate the IEGS model into a MILP framework using PLA of the quadratic Weymouth equation \eqref{eq:Ch3aWey}.

    Various models based on PLA are presented in \cite{correa2014gas}, and the incremental model outperforms the others according to its computational time and accuracy. In Appendix~\ref{App:PLA}, the incremental PLA model for a general nonlinear function $\Im(x)$, such as the squared nodal pressure, i.e., $\pi_{i,t}^2, \pi_{o,t}^2$ and the squared pipeline flow, i.e., $f_{p,t} |f_{p,t}|$, is introduced. Decreasing the linearization error can be accomplished by: (i) increasing number of segments $S$; (ii) selecting the breakpoint values $\left(x_i,\Im(x_i)\right)$. The optimal breakpoints are derived in \cite{correa2014security} to reduce the maximum approximation tolerances; (iii) using practical conditions to reduce the operating intervals of nodal pressures \cite{Correa2015Integrated, liu2018day}.

        Therefore, the MILP model for the IEGS dispatch problem is
        \begin{subequations}\label{eq:Ch3MILP}
        \begin{align}
          \min_{\Omega}  & \; \eqref{eq:Ch3GFCObj}\\
          s.t:\; & \text{Power system constraints: \eqref{eq:Ch3DCOPF1}--\eqref{eq:Ch3DCOPFRamp} }\\
          & \text{Gas system constraints: \eqref{eq:Ch3aWell}--\eqref{eq:Ch3aPressure}, \eqref{eq:Ch3aAveFlow}--\eqref{eq:Ch3GFCNodeShed} }\\
          & \text{PLA models for } \pi_{i,t}^2, \; \pi_{o,t}^2 \text{ and } f_{p,t}|f_{p,t}|. \\
           & \Omega = \{\triangle p_{d,t}, \triangle f_{d,t}, {p}_{u,t}, {\rho}_{u,t}, {p}_{l,t} , {\theta}_{i,t},{f}_{w,t}, f_{p,t}, \pi_{i,t}, f_{p,t}^{in} , f_{p,t}^{out}, f_{c,t}^{in} , f_{c,t}^{out}, {m}_{p,t}, Var_{PLA} \}
        \end{align} \end{subequations}
        where $Var_{PLA}$ is the linearization variables, which include all continuous and binary variables for the squared nodal pressures and squared pipelines average flow.

    \subsection{Gas Flow Correction Method} \label{sec:Ch3GFCMethod}
    This method is developed to be employed in the transmission level IEGS models \eqref{eq:GFCModel}.  The novel SOC relaxation method presented in \cite{liu2018day} is adopted to convexify the quadratic Weymouth equation  \eqref{eq:Ch3aWey} considering both the gas flow dynamics and bidirectional gas flow, resulting in an MISOCP framework mathematically. A GFC method, which is based on the multi-slack-node method and the Levenberg-Marquardt algorithm, is designed to calculate the optimal energy flow solution for the IEGS.

    \subsubsection{Formulating the MISOCP Model} \label{sec:MISOCPModel}
        Before convexifying the Weymouth equation, directional binary variables $z_{p,t}$ are introduced with the Big-M method to reformulate the equation as in \eqref{eq:Ch3Weytwo} without the sign function. If $\pi_{i,t}$ is greater/lower than $\pi_{o,t}$, then the direction variable $z_{p,t}$ would equal $1/0$ by \eqref{eq:Ch3Weya}, and $f_{p,t}$ is forced to be positive/negative value by \eqref{eq:Ch3Weyb}. The gas flow limits, i.e., $\underline{F}_p$ and$ \overline{F}_p$, can be obtained from \eqref{eq:Ch2Wey}, by substituting nodal pressure ranges.
        \begin{gather}
            (1-z_{p,t}) (\underline{\Pi}_i-\overline{\Pi}_o) \le \pi_{i,t} - \pi_{o,t} \le z_{p,t} (\overline{\Pi}_i-\underline{\Pi}_o), \label{eq:Ch3Weya}\\
            (1-z_{p,t}) \underline{F}_p \le f_{p,t} \le z_{p,t} \overline{F}_p  \label{eq:Ch3Weyb} \\
            f_{p,t}^2 = \begin{cases}
            \chi_p^f ({\pi}_{i,t}^2 - {\pi}_{o,t}^2), z_{p,t}=1, \\
            \chi_p^f ({\pi}_{o,t}^2 - {\pi}_{i,t}^2), z_{p,t}=0\end{cases}, \;\; \forall p,t, (i,o) \in p. \label{eq:Ch3Weytwo}
        \end{gather}

        The inlet and outlet pressures of a pipeline $p$, i.e., $\pi^+_{p,t}$ and $\pi^-_{p,t}$, are assigned in \eqref{eq:CH3Weyc1}--\eqref{eq:Ch3WeyNoDirc}, where \eqref{eq:CH3Weyc1}--\eqref{eq:CH3Weyc2} and \eqref{eq:CH3Weyc3}--\eqref{eq:CH3Weyc4} represent the pressures in case of positive and negative flow directions, respectively. In order to decrease binary variables, \eqref{eq:CH3Weyc1}--\eqref{eq:CH3Weyc4} are adopted only for the bidirectional pipelines $p\in \mathcal{P}^\pm$. The inlet and outlet pressures of unidirectional pipelines, which are connected with gas sources or gas loads at far terminals, are assigned directly by \eqref{eq:Ch3WeyNoDirc}. Therefore, Weymouth equations \eqref{eq:Ch3Weytwo} are replaced with \eqref{eq:CH3Weyc1}--\eqref{eq:Ch3WeyReform}, without the absolute function.
        \begin{gather}
             (1-z_{p,t}) ( \overline{\Pi}_o-\underline{\Pi}_i) \ge \pi^+_{p,t} - \pi_{i,t} \ge (1-z_{p,t}) (\underline{\Pi}_o- \overline{\Pi}_i) , \;\; \forall p\in \mathcal{P}^\pm,t, (i,o)\in p \label{eq:CH3Weyc1}\\
             (1-z_{p,t}) ( \overline{\Pi}_i-\underline{\Pi}_o) \ge \pi^-_{p,t} - \pi_{o,t} \ge (1-z_{p,t}) (\underline{\Pi}_i- \overline{\Pi}_o) , \;\; \forall p\in \mathcal{P}^\pm,t, (i,o)\in p \label{eq:CH3Weyc2}\\
             z_{p,t}     ( \overline{\Pi}_i-\underline{\Pi}_o) \ge \pi^+_{p,t} - \pi_{o,t} \ge z_{p,t}     (\underline{\Pi}_i- \overline{\Pi}_o) , \;\; \forall p\in \mathcal{P}^\pm,t, (i,o)\in p \label{eq:CH3Weyc3}\\
             z_{p,t}     ( \overline{\Pi}_o-\underline{\Pi}_i) \ge \pi^-_{p,t} - \pi_{i,t} \ge z_{p,t}     (\underline{\Pi}_o- \overline{\Pi}_i) , \;\; \forall p\in \mathcal{P}^\pm,t, (i,o)\in p \label{eq:CH3Weyc4}\\
             \pi^+_{p,t} =  \pi_{i,t} ,\;\; \pi^-_{p,t} = \pi_{o,t}    , \;\; \forall p\in \mathcal{P/P}^\pm,t, (i,o)\in p \label{eq:Ch3WeyNoDirc}   \\
            f_{p,t}^2 = \chi^f_p (\pi^{+2}_{p,t} - \pi^{-2}_{p,t}) , \;\; \forall p,t. \label{eq:Ch3WeyReform}.
        \end{gather}

        The quadratic equation \eqref{eq:Ch3WeyReform} can be relaxed as an inequality constraints as shown in \eqref{eq:Ch3Weyineq}. Note that \eqref{eq:Ch3Weyineq} is a proper cone and its canonical form is presented in \eqref{eq:Ch3Weycone}.

        \begin{gather}
            f_{p,t}^2 + \chi^f_p \pi_{p,t}^{+2} \le \chi^f_p \pi_{p,t}^{-2}  \label{eq:Ch3Weyineq} \\
            \begin{Vmatrix}
                 f_{p,t} \\ \sqrt{ \chi^f_p } \pi_{p,t}^-
            \end{Vmatrix}_2 \le  \sqrt{\chi^f_p} \pi_{p,t}^+  \label{eq:Ch3Weycone}
        \end{gather}

        Therefore, The MISOCP model for the IEGS dispatch problem is
         \begin{subequations}\label{eq:Ch3SOCP}
         \begin{align}
          \min_{\Omega}  & \; \eqref{eq:Ch3GFCObj}  \label{eq:Ch3MISOCPobj}\\
          s.t:\; & \text{Power system constraints: \eqref{eq:Ch3DCOPF1}--\eqref{eq:Ch3DCOPFRamp} }\\
          & \text{Gas system constraints: \eqref{eq:Ch3aWell}--\eqref{eq:Ch3aPressure}, \eqref{eq:Ch3aAveFlow}--\eqref{eq:Ch3GFCNodeShed} }\\
          & \text{Gas flow direction: \eqref{eq:Ch3Weya}--\eqref{eq:Ch3Weyb} \text{ and } \eqref{eq:CH3Weyc1}--\eqref{eq:Ch3WeyNoDirc} } \\
          & \text{SOC relaxation: \eqref{eq:Ch3Weycone}} \\
          \Omega = &\{\triangle p_{d,t}, \triangle f_{d,t}, {p}_{u,t}, {\rho}_{u,t}, {p}_{l,t} , {\theta}_{i,t},{f}_{w,t}, f_{p,t}, \pi_{i,t}, z_{p,t}, \pi_{p,t}^+, \pi_{p,t}^-, f_{p,t}^{in} , f_{p,t}^{out}, f_{c,t}^{in} , f_{c,t}^{out}, {m}_{p,t} \} \label{eq:Ch3MISOCPobj2}
        \end{align} \end{subequations}

\subsubsection{Multi-Slack-Node Based Newton-Raphson Algorithm}
        The MISOCP model \eqref{eq:Ch3SOCP} is provided to optimize the day-ahead dispatch for IEGS. It fails to find the exact gas flow as it is merely an approximation of the original model, especially for gas transmission networks. Accordingly, it is necessary to provide a correction method for gas flow after obtaining the solution of the MISOCP model. In \cite{yuan2017unified}, a multi-slack-node method with Newton-Raphson algorithm is suggested to calculate the steady-state OPGF, assuming unidirectional gas flow. This method is improved to consider bidirectional flows in an economic dispatch IEGS model under $K-1$ contingency criteria \cite{liu2018day}. However, the improved method neglects the line pack. In this paper, gas flow is corrected by Multi-slack-node model with Levenberg-Marquardt algorithm considering gas flow dynamics.

        Gas flow equations \eqref{eq:Ch3SOCNode}--\eqref{eq:Ch3SOCComp} are reformulated from the gas system operational constraints, namely the nodal balance equation \eqref{eq:Ch3GFCNodeShed}, Weymouth equation \eqref{eq:Ch3aWey}, mass flow equation, and the compressor flow constraints \eqref{eq:Ch3aComp}, which can be written as ${\pi}_{o,t} = \gamma_c {\pi}_{i,t},$ $ 1 \le \gamma_c \le \overline{\gamma}_c,$ $\forall c,t, (i,o) \in c$. $\overline{\gamma}_c$ is the maximum compression ration. $g_{i,t}^1$ is the nodal flow unbalance for node $i$ at time $t$, $g_{p,t}^2$ is the pipeline flow  unbalance for pipeline $p$ at time $t$, $g_{i,t}^3$ is the line pack unbalance for pipeline $p$ at time $t$, $g_{c,t}^4$ is the compressor pressure unbalance for compressor $c$ at time $t$. Each node connected with gas well is considered as slack node \cite{yuan2017unified}, and the amount of gas unbalance $\triangle g_{t}$ of all nodes at time $t$ is adjusted by multiple gas wells according to \eqref{eq:Ch3SOCdelta}. ${f}_{w,t}^0$ is the optimal gas flow from gas wells obtained from the MISOCP model \eqref{eq:Ch3SOCP}. $\beta_{w,t}$ is the participation factor of supplier $w$.
        \begin{gather}
            g_{i,t}^1 = \sum_{w \in \mathcal{W}(i)} {f}_{w,t} + \sum_{p \in \mathcal{P}_1(i)} {f}_{p,t}^{out} -  \sum_{p \in \mathcal{P}_2(i)} {f}_{p,t}^{in} + \sum_{c \in \mathcal{C}_1(i)} {f}_{c,t}^{out} \nonumber  \\
            - \sum_{c \in \mathcal{C}_2(i)} {f}_{c,t}^{in}  - \sum_{u \in \mathcal{U}_g(i)} \rho_{u,t} - \sum_{d \in \mathcal{D}_g(i)} (F_{d,t}-\triangle f_{d,t}) = 0,\; \forall \label{eq:Ch3SOCNode}
            \end{gather}
            \begin{gather}
            g_{p,t}^2 =  (f_{p,t}^{in} + f_{p,t}^{out} )|f_{p,t}^{in} + f_{p,t}^{out}|  - 4\chi_p^f ({\pi}_{i,t}^2 - {\pi}_{o,t}^2)=0, \; \forall p,t, (i,o) \in p, \label{eq:Ch3SOCWey}\\
            g_{p,t}^3 =  2f_{p,t}^{in} - 2 f_{p,t}^{out}  - \chi_m^f ({\pi}_{i,t} + {\pi}_{o,t}) + \chi_m^f ({\pi}_{i,t-1} + {\pi}_{o,t-1})=0, \; \forall p,t, (i,o) \in p, \label{eq:Ch3SOCMass}\\
            g_{p,t}^4 =  {\pi}_{o,t} -\gamma_c {\pi}_{i,t} = 0,\; \forall c,t, (i,o) \in c, \label{eq:Ch3SOCComp}\\
            f_{w,t} =  {f}_{w,t}^0 -\beta_{w,t} \triangle{g}_{t},\; \beta_{w,t}= \frac{{f}_{w,t}^0}{\sum_{\forall w}{f}_{w,t}^0}, \;\;\forall w,t,i \in w. \label{eq:Ch3SOCdelta}
        \end{gather}

    Let $Y_t$ and $X_t$ be the unbalance and state vectors at time $t$  as defined in \eqref{eq:Ch3GFCX}--\eqref{eq:Ch3GFCY}, respectively.
      \begin{gather}
        X_t = [f_t^{in},f_t^{out},f_c,\pi_t,\triangle{g}_{t} ],\;\; \forall t, \label{eq:Ch3GFCX}\\
            Y_t(X_t) = [g_t^{1},g_t^{2},g_t^{3},g_t^{4}]_{X_t},\;\; \forall t, \label{eq:Ch3GFCY}
      \end{gather}
      where $g_t^1,g_t^2,g_t^3,$ $g_t^4,f_t^{in},$ $f_t^{out},f_c,$ and $\pi_t$ are vectors for all $g_{i,t}^1,g_{p,t}^2,$ $g_{p,t}^3,g_{c,t}^4,g_{p,t}^{in}, $ $f_{p,t}^{out},f_{c,t},$ and $\pi_{i,t}$, respectively. The derivative matrix (Jacobian) between $Y_t$ and $X_t$ can be calculated by
      \begin{gather}
         A_t = \begin{bmatrix}
                \frac{\partial g_t^{1}}{\partial f_t^{in}}  & \frac{\partial g_t^{1}}{\partial f_t^{out}} & \frac{\partial g_t^{1}}{\partial f_t^{c}} & 0 & \frac{\partial g_t^{1}}{\partial \triangle g_t} \\
                \frac{\partial g_t^{2}}{\partial f_t^{in}}  & \frac{\partial g_t^{2}}{\partial f_t^{out}} & 0 & \frac{\partial g_t^{2}}{\partial \pi_t}& 0 \\
                \frac{\partial g_t^{3}}{\partial f_t^{in}}  & \frac{\partial g_t^{3}}{\partial f_t^{out}} & 0 & \frac{\partial g_t^{3}}{\partial \pi_t} & 0 \\
                0 & 0 & 0 & \frac{\partial g_t^{4}}{\partial \pi_t} & 0\\
            \end{bmatrix}, \; \forall t, \label{eq:Ch3GFCMatA}
      \end{gather}

      As a result of considering the continuity equation for line pack in \eqref{eq:Ch3SOCMass}, the derivative matrix between $Y_t$ and $X_{t-1}$ can be calculated by
      \begin{gather}
      B_{t,t-1} = \begin{bmatrix}
                 0 & 0 & 0 & 0 \\
                 0 & 0 & 0 & 0 \\
                 0 & 0 & \frac{\partial g_t^{3}}{\partial \pi_{t-1}} & 0 \\
                 0 & 0 & 0 & 0
            \end{bmatrix}_{X_{t-1}}, \; \forall t, \label{eq:Ch3GFCMatB}
            \end{gather}

      The initial line pack is considered to be equal the final line pack that could to be operated in the next day, i.e.,  $m_{p,1}= m_{p,T},\;\; \forall p$. Therefore, the overall Jacobian is presented in \eqref{eq:Ch3GFCJacob} and all derivatives for $X$ are obtained from \eqref{eq:Ch3GFCXall}--\eqref{eq:Ch3GFCYall}. Levenberg-Marquardt algorithm is presented to find a more exact solution for the proposed MISOCP model.
    \begin{gather}
             X = [X_1^{\top},X_2^{\top},X_3^{\top} \ldots X_T^{\top}]^{\top}, \label{eq:Ch3GFCXall}\\
             Y(X) = [Y_1(X_1)^{\top},Y_2(X_2)^{\top},Y_3(X_3)^{\top} \ldots Y_T(X_T)^{\top}]^{\top} \label{eq:Ch3GFCYall}\\
             J(X) = \begin{bmatrix}
                 A_1 & 0 & 0 & \ldots & 0 & B_{1,T} \\
                 B_{2,1} & A2 & 0 & \ldots & 0 & 0 \\
                 0 & B_{3,2} & A_3 & \ldots & 0 & 0 \\
                 \vdots & \vdots & \vdots & \ddots & 0 & 0\\
                 0 & 0 & 0 & \ldots & A_{T-1} & 0 \\
                 0 & 0 & 0 & \ldots &  B_{T,T-1} & A_T
            \end{bmatrix}_{X}  \label{eq:Ch3GFCJacob}
        \end{gather}

\begin{algorithm}[!htb]
\caption{Gas Flow Correction Algorithm}
\label{GFC}
\begin{algorithmic}[1]
    \STATE Set iteration index $k=0$, $\triangle g_t=0,\;\forall t$. Set convergence tolerance $\epsilon$, and Levenberg-Marquardt algorithm factors $\lambda^{(0)}$  and $0 < \mu_1 < 1 < \mu_2$. \\

    \STATE Solve MISOCP model \eqref{eq:Ch3SOCP} to find the optimal solution.
    \STATE Create $X^{(k)}$ from \eqref{eq:Ch3GFCXall}, then evaluate:  $Y^{(k)}=Y( X^{(k)} )$, and $J^{(k)}=J( X^{(k)} )$ \\

    \STATE If $|J^{(k)^{\top}} Y^{(k)}  - J^{(k-1)^{\top}} Y^{(k-1)}| \le \epsilon$, terminate with optimal $X^{(k)}$; else, go to Step 5. \\

    \STATE Compute: $X^{*}= X^{(k)} - [J^{(k)^{\top}}J^{(k)} + \lambda^{(k)} I]^{-1} J^{(k)} Y^{(k)} $. \\

    \STATE If $\| Y( X^{*}) \|^2 < \| Y( X^{(k)}) \|^2$, then $X^{(k+1)} = X^*, \;\; \lambda^{(k+1)} = \mu_1\lambda^{(k)}$; \\
    else, $X^{(k+1)} = X^{(k)}, \;\; \lambda^{(k+1)} = \mu_2\lambda^{(k)}$.

    \STATE $k=k+1$, and go to Step 3.
\end{algorithmic}
\end{algorithm}

\subsection{Sequential-MISOCP Algorithm} \label{sec:SCPMethod}

    Motivated by the discussion in Section~\ref{sec:Ch3OGF}, this study proposes a computational framework for distribution level multi-period OPGF problem based on DCP. Owing to the emerging P2G facilities and renewable energy outputs into the IEGS, bidirectional energy conversion is inevitable, and the conditions discussed in \cite{li2012exact} are not fulfilled. Therefore, SOC relaxation for the power flow generally provides inexact AC-OPF of the PDN. For the gas distribution network (GDN), in order to allow bidirectional gas flow, the sign function of the Weymouth equation is replaced with MILP constraints and quadratic equalities, similar to the treatment discussed in Section~\ref{sec:MISOCPModel}. These equalities and branch power flow quadratic equalities are reformulated as DCP functions. Following the algorithm proposed in \cite{wang2018convex}, The P-CCP proposed in \cite{R35} is designed to solve the OPGF for IEGS. Its convergence is proved in \cite{R35} for general DCP problems and another proof has been presented in \cite{wei2017optimal} for the branch flow OPF model. It should be noted that our work is an extension for \cite{wang2018convex} and the main differences are: (i) our work considers bidirectional gas flow pipelines to deal with meshed grid GDN; (ii) \cite{wang2018convex} directly adopts the SOC relaxation on power flow without an exactness guarantee. Moreover, three different types of coupling components are considered in this study: GPUs, P2G facilities, and electric-driven gas compressors. As well as, the dynamic gas model is adopted.

    The main contributions in this method are summarized as:
    \begin{enumerate}
    \item
        Feasibility and accuracy guarantee. A S-MISCOP algorithm is proposed to find the OPGF for IEGS. Based on DCP, the non-convex branch power flow and Weymouth gas flow equalities are decomposed as MISOCP constraints, which are easier to be solved than the original nonlinear problem. The proposed algorithm is a sequence of solving penalized MISOCP problems, and its feasibility is guaranteed by controlling its penalties.
    \item
        Fast and reliable convergence. Because S-MISOCP algorithm is a local heuristic approach, it is influenced by the initial point. Therefore, a high-quality initial point is suggested and an adaptive penalty growth rate is designed to adjust the main objective weight in the penalized problem.
    \end{enumerate}

\subsubsection{DCP Reformulation}
    Solving the above model requires much computation efforts, due to the presence of the nonlinear and nonconvex power flow and Weymouth equations. Fortunately, the power flow constraints and Weymouth equations can be formulated as DCP problem by expressing the proposed model constraints as difference of two convex functions. Note that the concave function $g(x)$ of a DCP constraint can be linearized as $\hat{g}(x,\overline x)$ at point $\overline x$ by \eqref{eq:Ch3CCP1}, which is the first-order Taylor expansion.
        \begin{gather}
            \hat{g}(x,\overline x) \cong g(\overline x) + \nabla g(\overline x)^{\top} (x-\overline x)  \label{eq:Ch3CCP1}
        \end{gather}

\begin{enumerate}
 \item {Power flow equation reformulation}:

        The quadratic power flow equation can be written as two inequality constraints as
        \begin{gather}
                4p_{l,t}^2 + 4q_{l,t}^2 + (v_{n,t}-i_{l,t})^2 \le (v_{n,t}+i_{l,t})^2, \;\; \forall l,t, \label{eq:Ch3simplePQ1}\\
                (v_{n,t}+i_{l,t})^2 \le 4p_{l,t}^2 + 4q_{l,t}^2 + (v_{n,t}-i_{l,t})^2, \;\; \forall l,t. \label{eq:Ch3simplePQ2}
        \end{gather}

        The first inequality \eqref{eq:Ch3simplePQ1} is an SOC constraint, and its canonical form is \eqref{eq:Ch3PQcone1}. Using \eqref{eq:Ch3CCP1}, the right-hand side of \eqref{eq:Ch3simplePQ2} can be replaced by its linear approximation. Given $\;[ \overline{p}_{l,t} \;\overline{q}_{l,t} \;\overline{v}_{n,t} \;\overline{i}_{l,t}]^{\top} $ as an initial point, the constraint \eqref{eq:Ch3simplePQ2} can be substituted with the approximated canonical form \eqref{eq:Ch3PQcone2}, where $\Gamma_{l,t}$ is an auxiliary variable.
    \begin{gather}
        \begin{Vmatrix}
            2p_{l,t} \\ 2q_{l,t} \\ (v_{n,t}-i_{l,t})
        \end{Vmatrix}_2 \le (v_{n,t}+i_{l,t}), \;\; \forall l,t,  \label{eq:Ch3PQcone1} \\
        \begin{Vmatrix}
            2 (v_{n,t}+i_{l,t}) \\ \Gamma_{l,t} -1
        \end{Vmatrix}_2   \le  \Gamma_{l,t} +1, \;\; \forall l,t,   \label{eq:Ch3PQcone2} \\
        \Gamma_{l,t} = 8\overline{p}_{l,t}p_{l,t} + 8\overline{q}_{l,t}q_{l,t} + 2 (\overline{v}_{n,t}-\overline{i}_{l,t})(v_{n,t}-i_{l,t}) \nonumber \\
        - 4\overline{p}_{l,t}^2 - 4\overline{q}_{l,t}^2 - (\overline{v}_{n,t}-\overline{i}_{l,t})^2, \;\; \forall l,t. \nonumber
    \end{gather}

    \item  {Gas flow equation reformulation:}

        Weymouth equations are firstly rearticulated without the sign function by using a directional binary variable $z_{p,t}$, as introduced in Section \ref{sec:MISOCPModel}, by using \eqref{eq:Ch3Weya}--\eqref{eq:Ch3Weyb} and \eqref{eq:CH3Weyc1}--\eqref{eq:Ch3WeyNoDirc}. Secondly, \eqref{eq:Ch3WeyReform} is converted into two opposite inequality constraints as
        \begin{gather}
            f_{p,t}^2 + \chi^f_p \pi^{-2}_{p,t} \le \chi^f_p \pi^{+2}_{p,t}, \;\; \forall p,t, \label{eq:Ch3Weyineq1} \\
            \chi^f_p \pi^{+2}_{p,t} - (f_{p,t}^2 + \chi^f_p \pi^{-2}_{p,t}) \le 0 , \;\; \forall p,t \label{eq:Ch3Weyineq2}
        \end{gather}

        The first inequality is an SOC constraint, and its canonical form is \eqref{eq:Ch3Weycone1}. Similar to \eqref{eq:Ch3simplePQ2}, given $[\overline{f}_{p,t}\;\overline{\pi}^+_{p,t}\;\overline{\pi}^-_{p,t}]^{\top}$ as an initial point, the second inequality \eqref{eq:Ch3Weyineq2} is substituted with the approximated canonical form \eqref{eq:Ch3Weycone2}, after linearizing the right-hand side by \eqref{eq:Ch3CCP1}, where $\Lambda_{p,t}$ is an auxiliary variable.
        \begin{gather}
            \begin{Vmatrix}
                f_{p,t} \\ \sqrt{\chi^f_p} {\pi}^-_{p,t}
            \end{Vmatrix}_2 \le  \sqrt{\chi^f_p} {\pi}^+_{p,t}   \label{eq:Ch3Weycone1}   , \;\; \forall p,t,\\
            \begin{Vmatrix}
                2 \sqrt{\chi^f_p} {\pi}^+_{p,t} \\ \Lambda_{p,t} -1
            \end{Vmatrix}_2 \le  \Lambda_{p,t} +1, \label{eq:Ch3Weycone2} \\
             \Lambda_{p,t} = 2\chi^f_p \overline{\pi}^-_{p,t} {\pi}^-_{p,t} + 2 \overline{f}_{p,t}f_{p,t} - \chi^f_p \overline{\pi}^-_{p,t} - \overline{f}_{p,t}^2 , \;\; \forall p,t \nonumber
        \end{gather}

    \end{enumerate}

\subsubsection{The Compact Form}
    The compact form of the proposed model, after above reformulations of nonlinear equations, is
   \begin{subequations}\label{eq:Ch3SCPModel}
   \begin{align}
        &\min_{\bm{x},\hat{\bm{x}}} f(\bm{x})   \label{eq:Ch3SCPIEGS}\\
        & s.t. \hspace{1mm} \bm{A}\bm{x} \le \bm{B}   \label{eq:Ch3SCPIEGSa}\\
         & \lVert \bm{D}_{h,t}\bm{x} \rVert_2  \le \bm{d}_{h,t}\bm{x}, \forall h,t,  \label{eq:Ch3SCPIEGSb}\\
        & \lVert \bm{E}_{h,t}(\hat{\bm{x}})\bm{x} + \bm{F}_{h,t}(\hat{\bm{x}}) \rVert_2 \le \bm{e}_{h,t}(\hat{\bm{x}}) \bm{x}+ \bm{f}_{h,t}(\hat{\bm{x}}), \forall h,t  \label{eq:Ch3SCPIEGSc}
    \end{align}\end{subequations}
    where $\bm{x}$ is the decision variables for both systems, including the continuous and binary variables. Due to the need of finding suitable linearization points used in the approximated cones \eqref{eq:Ch3PQcone2} and \eqref{eq:Ch3Weycone2}, $\hat{\bm{x}}$ is considered as a decision variable for the IEGS problem. $\bm{A}$ and $\bm{B}$ can be easily obtained from the MILP constraints \eqref{eq:Ch3Weya}--\eqref{eq:Ch3Weyb}, \eqref{eq:CH3Weyc1}--\eqref{eq:Ch3WeyNoDirc} and \eqref{eq:Ch3SCPObja}--\eqref{eq:Ch3SCPObjc}. The exact SOC constraints \eqref{eq:Ch3PQcone1} and \eqref{eq:Ch3Weycone1} are compressed in \eqref{eq:Ch3SCPIEGSb}, while the approximated SOC constraints \eqref{eq:Ch3PQcone2} and \eqref{eq:Ch3Weycone2} are defined in \eqref{eq:Ch3SCPIEGSc}.

   \subsubsection{The Proposed Algorithm Structure}
    The S-MISOCP algorithm starts with an initial infeasible linearization point $\hat{\bm{x}}$, a sequence of MISOCP problems, which penalize the constraints violations, are solved while updating the linearization point in each iteration. Therefore, with suitable algorithm parameters, a quick convergence can be achieved by shifting the infeasible solution to a feasible one very close to, or equal to, the optimum. The convergence proof is discussed in \cite{R35}. In fact, S-MISOCP algorithm is a local heuristic approach, and its performance and the solution quality are influenced by:
\begin{enumerate}
    \item {The Problem Infeasibility}:  the feasibility of the original problem, which suggests the algorithm would fail to converge if it  is infeasible. For this reason, the power and gas load shedding are added to relax the operational constraints. If load shedding occurs, the upgradation of system components is important to provide a secure operation for the IEGS.
    \item {The Initial Point Selection}: the proposed algorithm starts with a convexified counterpart of the original model, which needs to be parameterized by an initial point. According to \cite{wang2018convex} and \cite{wang2018risk}, the initial point has crucial impacts on the quality of the final solution, solver time and iterations number. Therefore, to find a high-quality initial point, we recommend using the relaxed MISOCP problem with penalizing the right-hand side of \eqref{eq:Ch3PQcone2} and \eqref{eq:Ch3Weycone2}, as follows.
        \begin{subequations}\label{eq:Ch3Relaxed}
        \begin{align}
            &\min_{\bm{x}} f(\bm{x}) + \lambda_p \sum_{\forall t} \sum_{\forall l} i_{l,t} + \lambda_g \sum_{\forall t} \sum_{\forall p} \pi^+_{p,t}  \label{eq:Ch3SCPIEGS0}\\
            &\hspace{1mm} s.t. \hspace{1mm} \eqref{eq:Ch3SCPIEGSa}-\eqref{eq:Ch3SCPIEGSb}
        \end{align}\end{subequations}
        where $\lambda_p$ and $\lambda_g$ are small values that control the focus of the model objective on the constraints violation.

    \item {The Algorithm Parameters}:  selecting suitable parameters and using adaptive penalty growth rate, which is suggested in this study, provide fast and feasible solutions.
\end{enumerate}

        Compared with the standard penalty growth rate introduced in \cite{R35}, where a global penalty coefficient $\tau$ is selected for all the convexified constraints, each convexified constraint is assigned with its own penalty coefficient, and  an adaptive rule is designed for updating it. This allows us to better capture the impact of slack variables on the objective and to facilitate convergence. The proposed rate depends on the relative constraint violation ($RCV$), which can be calculated by
        \begin{gather}
            RCV_{h,t} = \varphi_{h,t} /(\bm{e}_{h,t}(\hat{\bm{x}}) \bm{x}+ \bm{f}_{h,t}(\hat{\bm{x}})), \;\; \forall h,t \label{eq:Ch3Penalty}
        \end{gather}
        where, $\varphi_{h,t}$ is the value of the approximated cones violations. The adaptive penalty, which is used in step 5 of the proposed algorithm with iteration $k$, is achieved by
        \begin{align}
            &\text{If } RCV_{h,t} \le \varepsilon, \text{Then, } \tau_{h,t}^k = \mu\tau_{h,t}^{k-1};  \nonumber\\
            &\text{Else, }\tau_{h,t}^k = \tau_{h,t}^{k-1} \; \min\{\overline{\mu},\; \max[\underline{\mu},\; \sigma \, RCV_{h,t}]\}. \label{eq:Ch3Adaptive}
        \end{align}

        In the above formula, where $\overline{\mu},\underline{\mu}$ are limits of penalty rate coefficient, and $\sigma$ is a fixed constant that controls the rate. Selecting suitable parameters provides fast and reliable convergence with a high-efficiency solution. Note that the solution might be suboptimal at higher iteration number with high penalties because the weight of violations is greater than the main objective function. In order to decrease this weight, $\mu$ is added to decrease penalties of inviolated constraints. And its range should be $1 \ge \mu > 1/\underline{\mu}$ to avoid any fluctuations in penalties between iterations.

        Theoretically, the convergence of P-CCP only holds for continuous problems \cite{R35}, and it is not always guaranteed for MISOCP models due to their discontinuity. Based on our experiences, directional binaries obtained from the relaxed problem would remain fixed after the first few iterations, which is consistent with the observation in \cite{he2017decentralized}. Therefore, the binary variables can be fixed after the beginning iterations, which is tuned as $5$ in this work. Then, the original MISOCP model can be converted into an SOCP with fixed binary variables, which means the S-MISOCP algorithm would degenerate to a standard P-CCP and its convergence can be guaranteed.

\begin{algorithm}[!htb]
\caption{The S-MISOCP Algorithm}
\label{alg2}
\begin{algorithmic}[1]
    \STATE Set the convergence parameters $\varepsilon, \epsilon, \tau^{max}$, the maximum number of iterations $K^{max}$, initial penalties $\tau^{0}_{h,t}$, penalty growth rate limits $\overline{\mu},\underline{\mu}$, penalty growth rate coefficient $\sigma$, and iteration index $k=0$ \\

    \STATE Solve the relaxed problem \eqref{eq:Ch3Relaxed} to update $\hat{\bm{x}}$.\\

    \STATE Solve the penalized problem \eqref{eq:Ch3Penalized} to update ${\bm{x}^k}$, $\varphi_{h,t}^k$. \begin{subequations}\label{eq:Ch3Penalized}\begin{align}
         &obj^k = \min_{\bm{x},\varphi_{h,t}} f(\bm{x}) + \sum_{\forall t} \sum_{\forall h} \tau^k_{h,t} \varphi_{h,t} \label{eq:SMISOCPA}\\
        &s.t. \hspace{1mm} \eqref{eq:Ch3SCPIEGSa}-\eqref{eq:Ch3SCPIEGSb}, \nonumber\\
        &\hspace{2em}  \lVert \bm{E}_{h,t}(\hat{\bm{x}})\bm{x} + \bm{F}_{h,t}(\hat{\bm{x}}) \rVert_2 \le \bm{e}_{h,t}(\hat{\bm{x}}) \bm{x}+ \bm{f}_{h,t}(\hat{\bm{x}})+\varphi_{h,t}, \forall h,t.  \nonumber
    \end{align}\end{subequations}\\

    \STATE If \eqref{eq:SMISOCPB}--\eqref{eq:SMISOCPC} or $k > K^{max}$ is satisfied, terminate;
    Else, go to step 5. \begin{align}
         \varphi_{h,t}^k & \le \varepsilon(\bm{e}_{h,t}(\hat{\bm{x}}) \bm{x}^k+ \bm{f}_{h,t}(\hat{\bm{x}}) ),\; \forall h,t, \label{eq:SMISOCPB}\\
                obj^{k-1} - obj^{k} &\le \epsilon \label{eq:SMISOCPC}
         \end{align}

    \STATE Update $\tau^k_{h,t}$ by \eqref{eq:Ch3Penalty}, $\hat{\bm{x}}= \bm{x}^k,\; k=k+1$, go to step 3.
\end{algorithmic}
\end{algorithm}

\section{Simulation Results} \label{sec:Ch3Simulations}
\subsection{Case Studies with Transmission-level IEGS}
\begin{figure}[!ht]
        \centering
            \includegraphics[width=8cm]{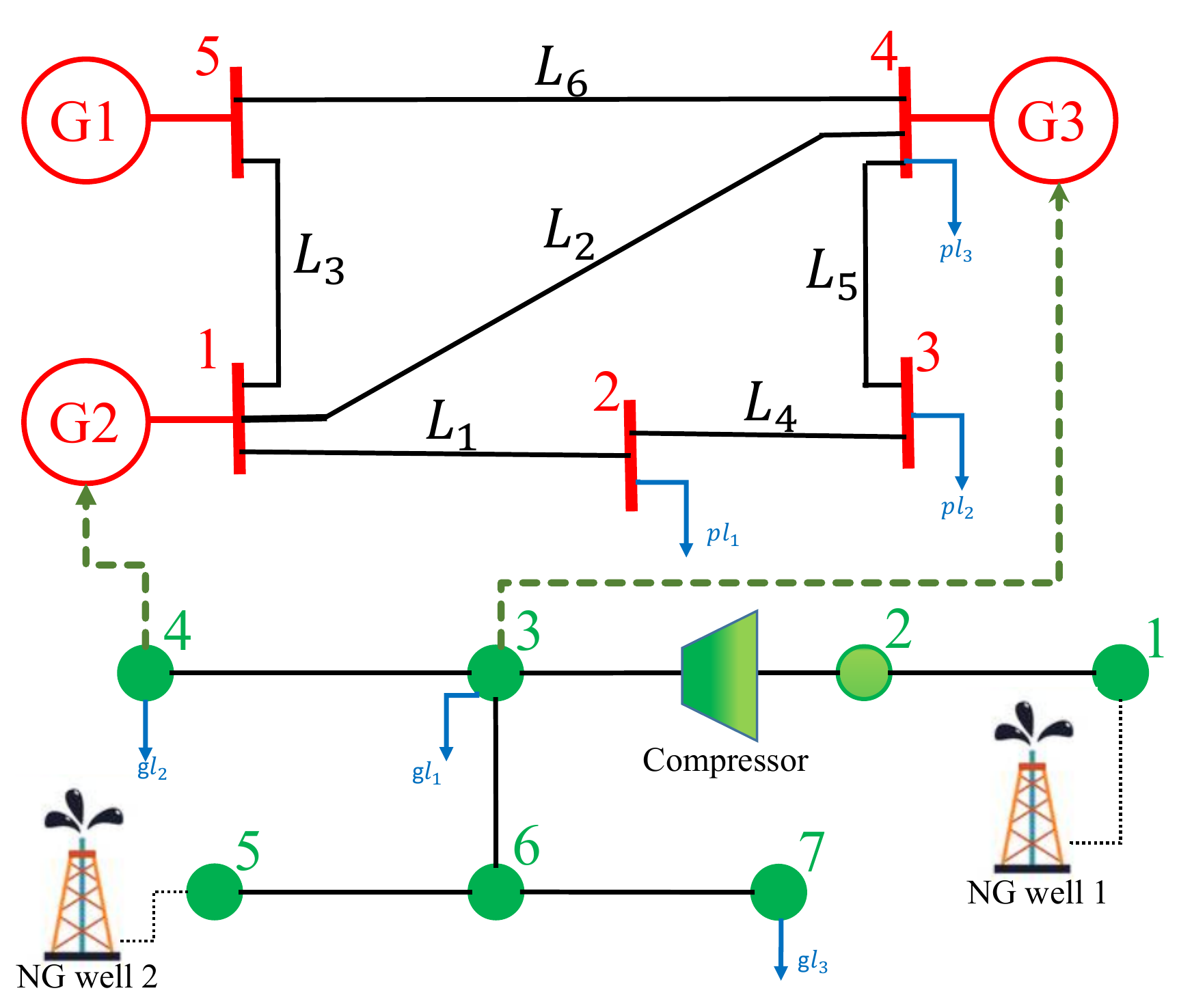} 
            \caption{Topology of the test system}   \label{fig:Ch3Topology1}
    \end{figure}

    A transmission-level IEGS test system, containing a $5$-bus power system and a $7$-bus gas system, is examined to illustrate the effectiveness and features of the proposed GFC method as well as the incremental PLA method. Figure~\ref{fig:Ch3Topology1} shows the topology of the integrated system infrastructure. The details of all parameters of the integrated system and unit commitment are found in Appendix~\ref{App:5Bus} and Appendix~\ref{App:sevenNode}. In the figure, $B, G, L, pl, W, C,$ and $gl$ are used with subscripts to denote the power buses, generators, power lines, power loads, gas wells, compressors, and gas loads, respectively.

    \subsubsection{Effectiveness of Break-Points on the MILP Model}
        MILP model \eqref{eq:Ch3MILP} is applied on the test system to demonstrate the effect of number of segments and selection of breakpoints used in PLA of Weymouth equation. To be clear results, the Weymouth equation contains three nonlinear elements, namely $\pi_{i,t}^2, \pi_{o,t}^2$ and $f_{p,t} |f_{p,t}|$, the pressure squared is linearized by $S$ segments, while the gas flow is linearized by $2*S$ due to the absolute function. It should be noted that these nonlinear terms are replaced by piecewise linearization variables. As a result, they may not be considered as a decision variable in the optimization problem. The Weymouth equation error is more important than the interpolation error for each variable. Correa et al. \cite{correa2015optimal} provide an analysis on the interpolation error, which affects the Weymouth error. Weymouth equation error ($error\%$) is the maximum difference occurred between the two sides of Weymouth equation for all pipelines at any time, as defined in \eqref{eq:Ch3WeyError}. Table~\ref{tab:Ch3MILP} presents the results of different number of segments. By increasing $S$, $error\%$ decreases while CPU time increases.
        \begin{gather}
            error\% = \max \left( \bigg| \frac{f_{p,t}^2 - \chi^f_p |\pi_{i,t}^2 - \pi_{i,t}^2|}{ f_{p,t}^2}  \bigg| \times 100 \right), \;\; \forall p,t \label{eq:Ch3WeyError}
        \end{gather}

    \begin{table}[!th]
        \caption{The effect of segments number and breakpoints selection on the Weymouth error and CPU time}
        \label{tab:Ch3MILP}
        \centering
        \includegraphics[width=13cm]{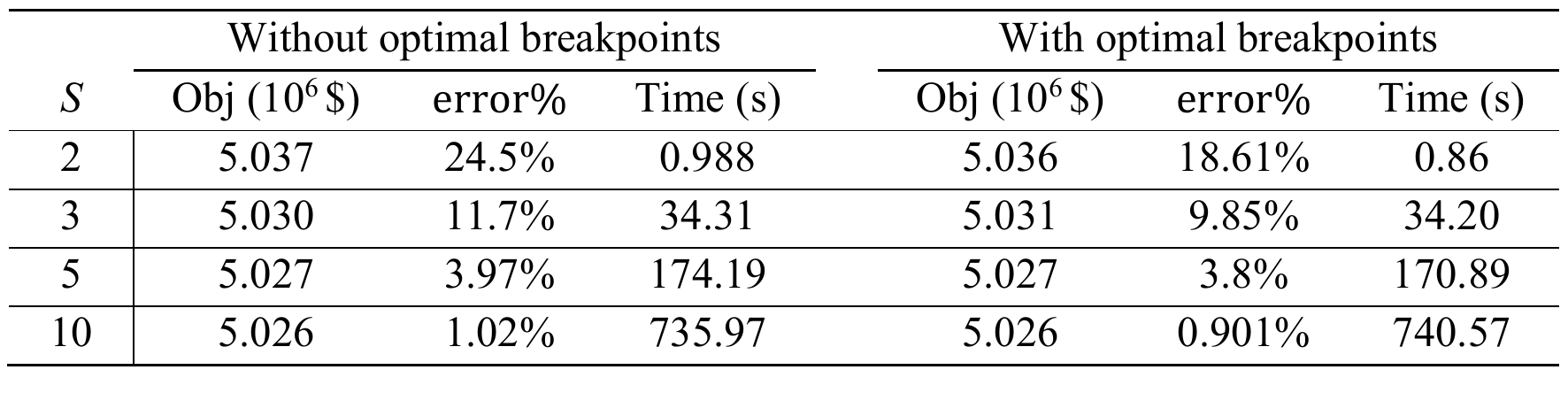} 
    \end{table}

        Using optimal breakpoints improves the incremental model without increasing $S$ as shown in Table~\ref{tab:Ch3MILP}. We use the formulation used in \cite{correa2014gas} to decrease the squared linearization error, and the maximum error is forced to be less than a tolerance value. The optimal breakpoints clearly decrease the Weymouth error with almost same CPU time. The objective value is slightly increased/decreased by breakpoints selection.

\subsubsection{Performance of the MISOCP Model}
     MISOCP model \eqref{eq:Ch3SOCP} is applied on the test system to demonstrate its effectiveness and the computational efficiency. Table~\ref{tab:Ch3MISOCP1} presents different stress levels (loading) on the gas and electricity infrastructure, which are denoted as G and E respectively. It shows the maximum error of Weymouth equation obtained by \eqref{eq:Ch3WeyError} before and after using the proposed GFC method, denoted as Error1 and Error2, respectively. With increasing the gas stress, the gas production cost is subsequently increased to feed the additional gas load, therefore the total IEGS cost increases. It is notable that using GFC method able to decrease the maximum error of the Weymouth equation.
    \begin{table}[!th]
        \caption{GFC method effectiveness under different stress levels on IEGS}
        \label{tab:Ch3MISOCP1}
        \centering
        \includegraphics[width=13cm]{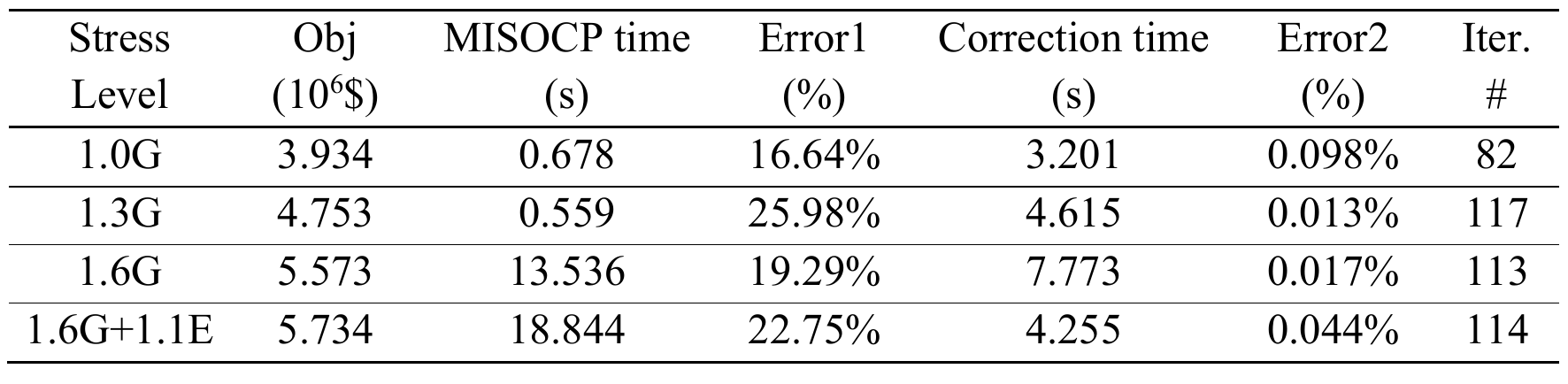} 
    \end{table}

         MISOCP model with the proposed GFC method is compared with the MILP model in Table~\ref{tab:Ch3MISOCPMILP} under different stresses on the IEGS. Moreover, as the MILP solution exactness depends on the number of segments $S$ used in the PLA model, two MISOCP models are formulated with different $S$, namely $S=10$ and  $S=20$. The proposed GFC method outperforms the MILP model, especially for high load stress. In the first two cases, the MILP model reaches the optimal objective value but with longer execution time and greater maximum error compared with the proposed MISOCP model. With $20$ segments, although the MILP model needs very long time, it cannot provide the optimal objective with accurate decisions compared with the proposed method, as shown in the last row of Table~\ref{tab:Ch3MISOCPMILP}.

    \begin{table}[!th]
        \caption{Comparison between MISOCP and MILP models under stress levels on IEGS}
        \label{tab:Ch3MISOCPMILP}
        \centering
        \includegraphics[width=13cm]{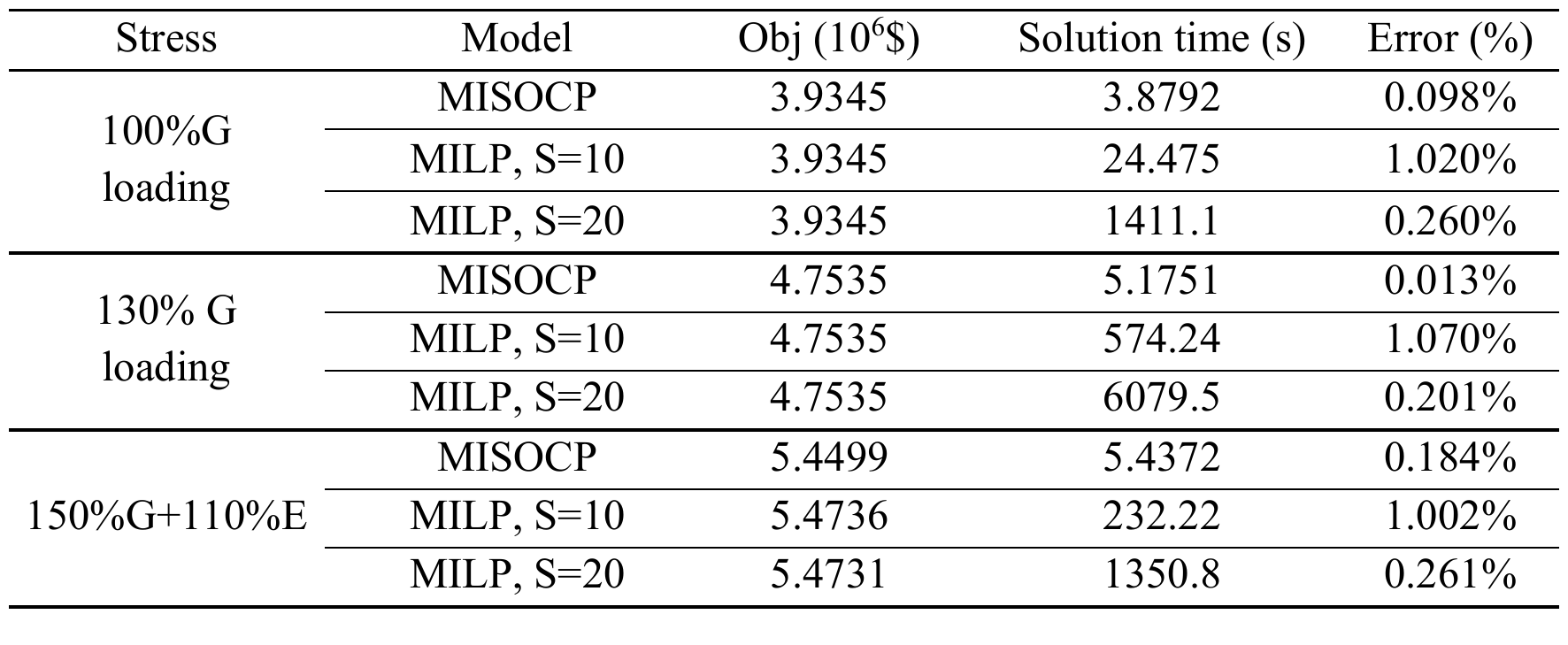} 
    \end{table}

\subsubsection{Comparison with the Steady-State Gas Model}
\begin{figure}[!ht]
        \centering
            \includegraphics[width=11cm]{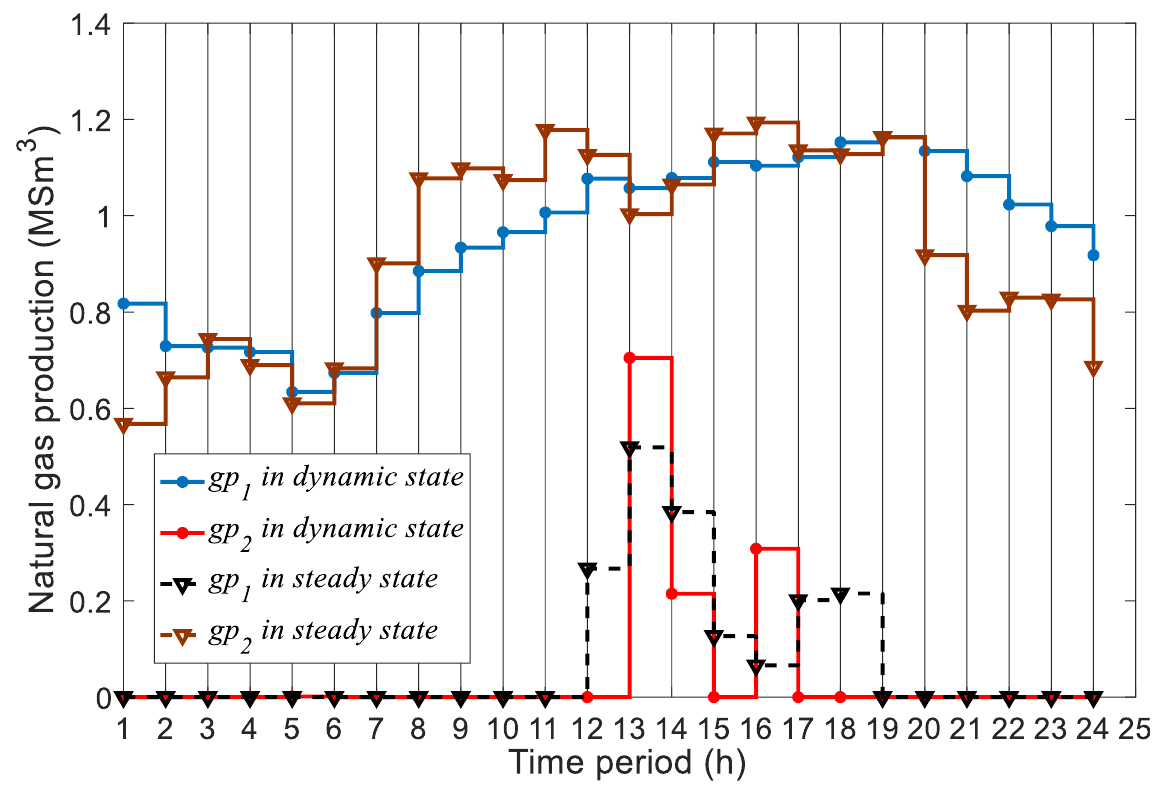} 
            \caption{Production scheduling of gas wells in both the dynamic- and steady-state conditions}   \label{fig:Ch3SteadyState}
    \end{figure}

        The main purpose of the proposed model is to find the optimal day-ahead schedule for both power units and gas suppliers. It is better to consider any assumption which may provide suboptimal decisions. With the consideration of the line pack and traveling velocity of gas, the proposed model has a significant effect on the production scheduling. The following two patterns are presented to discuss this effectiveness. Pattern $1$ corresponds to steady-state gas flow model, please refer to Section~\ref{sec:CH2SSS} for the model formulations. In pattern $2$, the dynamic-state gas flow model considers the line pack as formulated in the proposed model. Figure~\ref{fig:Ch3SteadyState} plots the day-ahead gas production for the two patterns in case of 150\% gas stress level. Considering the line pack inside pipelines provides more operational flexibility in pattern $2$ compared with pattern $1$, therefore, the schedules are different. Additionally, the objective cost would be dissimilar, it equal \$$5317.2\times10^3$ for pattern $1$, while it is \$$5311.7\times10^3$ for pattern $2$.

\subsection{Case Studies with Distribution-level IEGS}
    In this subsection, the S-MISOCP algorithm performance is evaluated, and the impact of algorithm parameters and initial point selection on the solution quality is discussed. Moreover, computational comparisons between the proposed algorithm and MISOCP relaxation method is presented. All the below results are conducted on personal PC with $8$GB memory and Intel(R) Core(TM) i$5-3320$M CPU, using the MATLAB environment with YALMIP toolbox \cite{YALMIP} and Gurobi solver.

\begin{table}[!th]
        \caption{S-MISOCP algorithm parameters}
        \label{tab:Ch3Parameter}
        \centering
        \includegraphics[width=12cm]{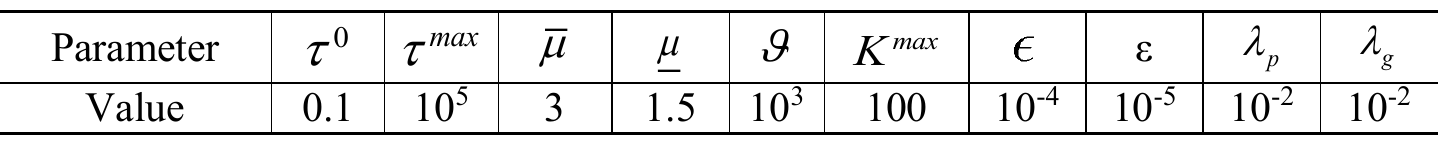} 
    \end{table}

\subsubsection{Test System Description}
\begin{figure}[!hbtp]
            \centering
            \includegraphics[width=12cm]{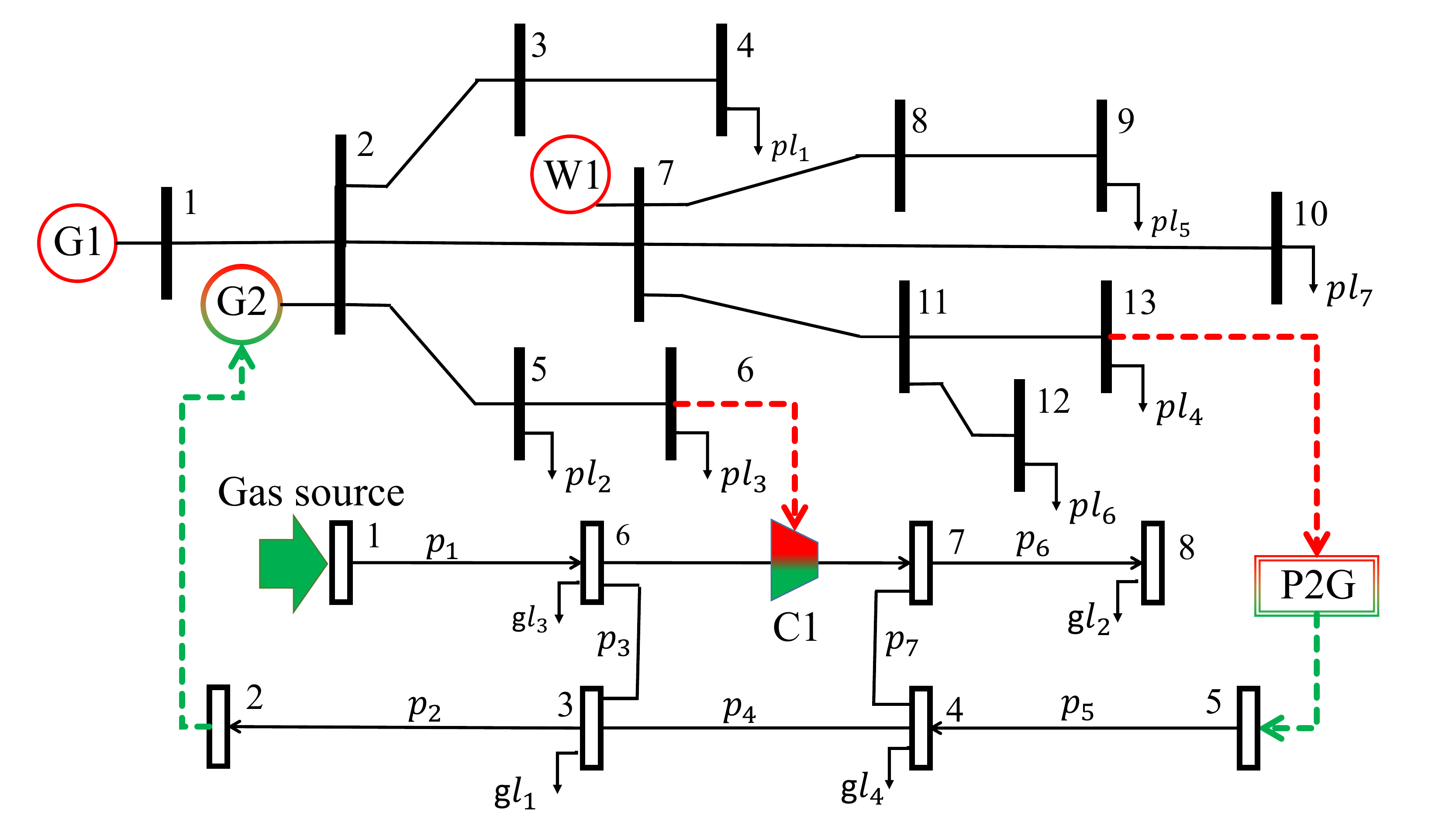}
            \caption{The test system topology.}
            \label{fig:Ch3SCPF1}
    \end{figure}

    The test system is a $13$-bus PDN integrated with $8$-node meshed gas network, and its topology is shown in Figure~\ref{fig:Ch3SCPF1}. The PDN has one non-GFU (G1), one wind farm (W1), seven power demands ($pl_1-pl_7$), and $11$ power lines, whereas, the gas system has one gas source, four gas demands ($gl_1-gl_4$), one electric-driven compressor (C1), one P2G facility (P2G), and seven pipelines ($p_1-p_7$). Note that the computational burden mainly depends on the number of power lines and gas pipelines due to their two pair cones. This burden is also influenced by the number of bidirectional pipelines, which are three pipelines, namely $p_3, p_4$ and $p_7$. System details, including wind power forecasting, are found in Appendix~\ref{App:13Bus} and Appendix~\ref{App:8Node}. Algorithm parameters are listed in Table~\ref{tab:Ch3Parameter}.

\subsubsection{Impact of Initial Point}
    The impact of the initial point on S-MISOCP algorithm is presented in this section. To better demonstrate this impact, the final objective values and execution time are reported with three different initial points, which are: (i) zero-initial point; (ii) relaxed MISOCP point, which is the solution vector of problem \eqref{eq:Ch3SCPModel}, excluding the approximated cones \eqref{eq:Ch3SCPIEGSc}; (iii) proposed initial point, which is obtained by \eqref{eq:Ch3Relaxed}. The effect of $\lambda_p$ and $\lambda_g$ is also presented in this study. Note that, the algorithm convergence parameters are kept same for all cases as given in Table~\ref{tab:Ch3Parameter}. Zero vector introduces the worst objective value with a longer computational time, while relaxed MISOCP vector is suitable to obtain a better solution within a time equals \SI{30}{\percent} of the zero vector time. The proposed vector, which outperforms the other two methods in the computational burden, is affected by the penalties used in \eqref{eq:Ch3SCPIEGS0}. With low values of $\lambda_p$ and $\lambda_g$, the solution acts as a relaxed MISOCP vector solution due to their low impact on the objective in \eqref{eq:Ch3SCPIEGS0}. While increasing these penalties, the solution time decreases as shown in the first three rows of Table~\ref{tab:Ch3SCPInitialPoint}. However, high values of $\lambda_p$ and $\lambda_g$ may provide bad initial vector, so the algorithm takes longer time to converge, as shown in the last row of the table. The reason is suboptimal objective, provided by initial vector, which needs more iterations to be recovered. We conclude that using proper values of penalties $\lambda_p$ and $\lambda_g$, fast, accurate and optimal solutions can be identified.

    \begin{table}[!htp]
        \caption{Computational comparisons between different initial vectors} \label{tab:Ch3SCPInitialPoint}
        \centering
        \includegraphics[width=13cm]{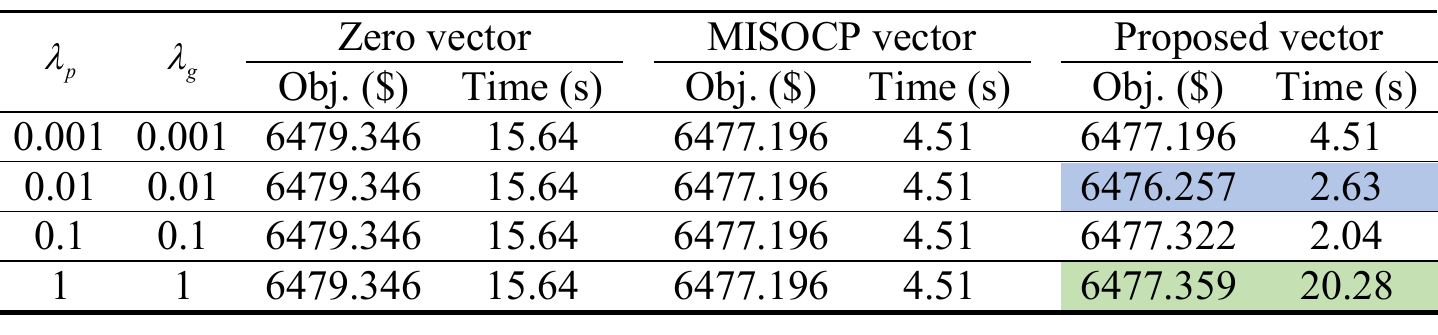}
    \end{table}

\subsubsection{Effectiveness of Adaptive Penalty}
    Compared with the fixed penalty growth rate (see e.g. \cite{wang2018convex}, \cite{R35}, \cite{wei2017optimal}), the proposed adaptive penalty growth rate has computational benefits, which are discussed in this section. Because the adaptive rate depends on the $RCV$, it provides less focus on low $RCV$ and concentrates more on high $RCV$ and the main objective. In order to show the effectiveness of the adaptive rate, numerical cases are conducted based on a different combination of power and gas demands. Table~\ref{tab:Ch3SCPAdaptive} shows a numerical comparison between the fixed and adaptive rates under four cases. Note that the power and gas load shedding are added to the power and gas nodal balancing equations, therefore, for any load stress a feasible solution can be found. For the fixed penalty rate, coefficients of \eqref{eq:Ch3Penalty} are set at $\mu=\underline{\mu}=\overline{\mu}=$\num{2}. While parameters listed in Table~\ref{tab:Ch3Parameter}  are used for adaptive penalty rate. In the first three cases, the fixed penalty rate provides the same results of an adaptive one because both of them starts with the same initial vector, which has small values of $RCV$, obtained by \eqref{eq:Ch3Relaxed}. Therefore, the algorithm converges quickly. In the last two cases, increasing the gas demands introduces more stressed IEGS and OPGF cannot be identified easily, therefore, the algorithm executes a large number of iterations.

    \begin{table}[!hbtp]
        \caption{Numerical comparisons between fixed and adaptive penalty rates} \label{tab:Ch3SCPAdaptive}
        \centering
        \includegraphics[width=12.5cm]{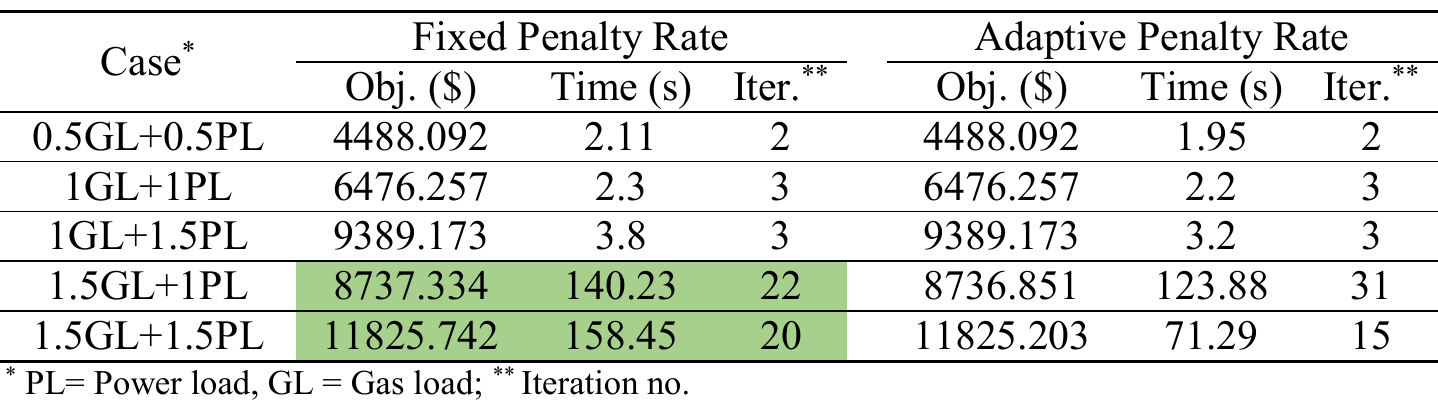}
    \end{table}

\begin{figure}[!hbp]
            \centering
            \includegraphics[width=11cm]{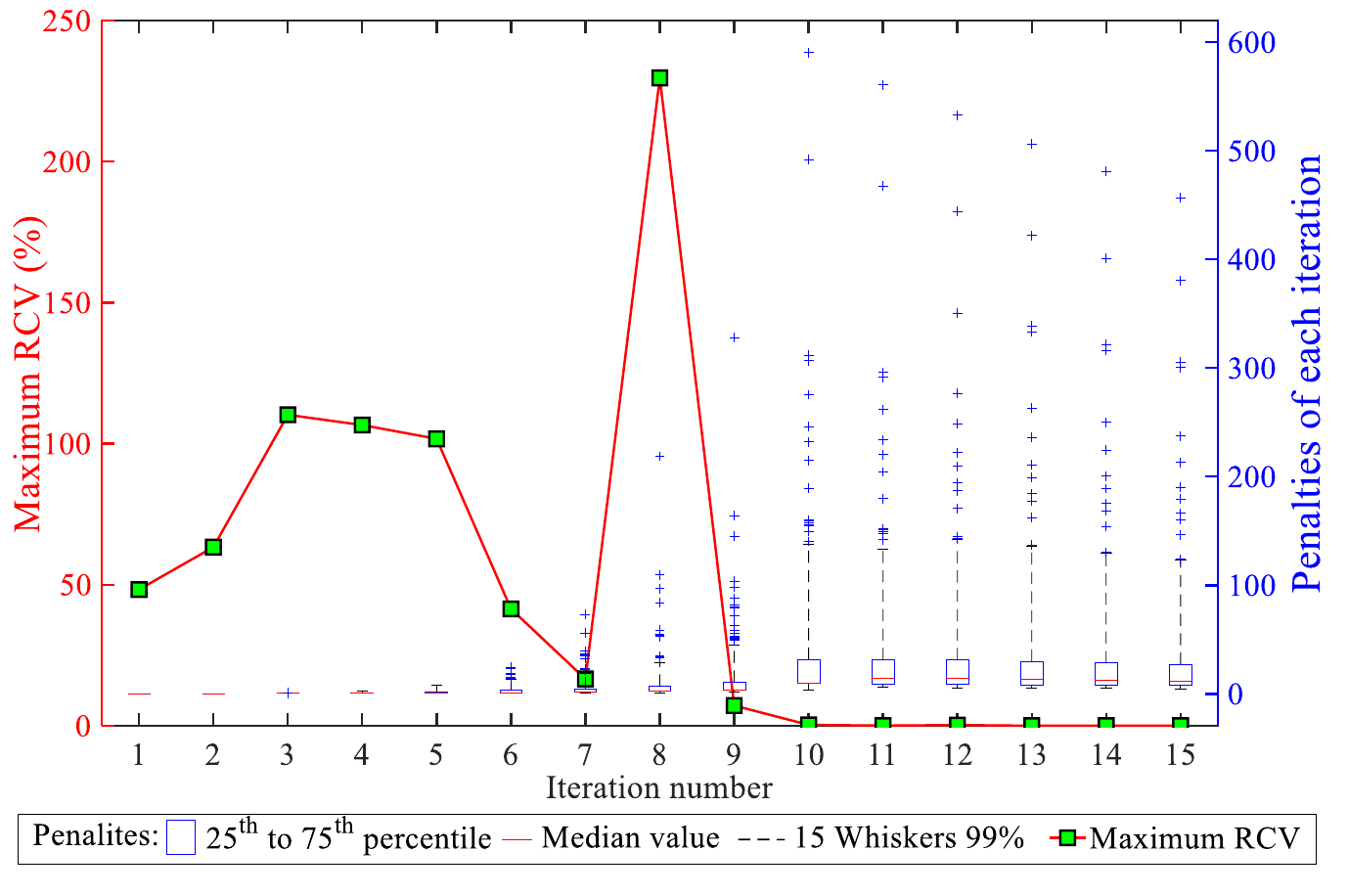}
            \caption{Maximum $RCV$ and penalties values for case 1.5GL+1.5PL.}
            \label{fig:Ch3SCPF2}
    \end{figure}

    To better present the effect of adaptive rate on the penalty values, Figure~\ref{fig:Ch3SCPF2}. displays all penalties at each iteration for the last case of Table~\ref{tab:Ch3SCPAdaptive}. Note that the number of penalties is $|\mathcal{T}|\times |\mathcal{P}\cup\mathcal{L}|$ at each iteration. It can be seen that the median value of penalties is very small as compared with the largest penalty value in each iteration. Due to the adoption of $\mu=$\num{0.95}, penalties start to decrease after the 9\textsuperscript{th} iteration to provide a high weight for the main objective $f(\bm{x})$.

\subsubsection{Comparison with MISOCP Relaxation Method}
\begin{figure}[!htbp]
            \centering
            \includegraphics[width=11cm]{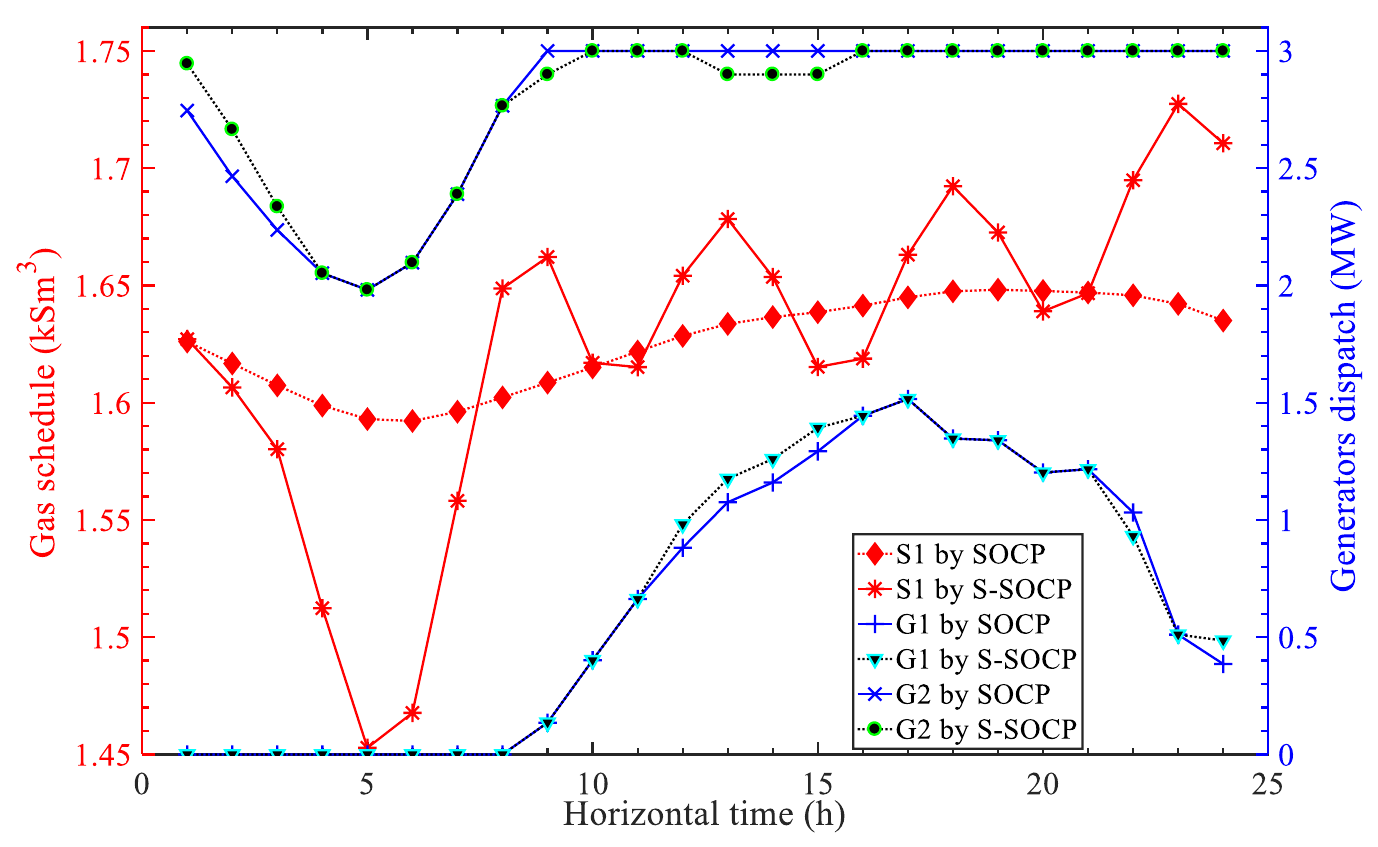}
            \caption{Energy production schedules obtained by MISOCP relaxation method and S-MISOCP algorithm.}
            \label{fig:Ch3SCPF3}
    \end{figure}

    MISOCP relaxation (MISOCPR) method, which is adopted in many studies (see e.g. \cite{borraz2016convex}, \cite{liu2019day}), is compared with the S-MISOCP algorithm. The MISOCPR method is obtained by solving \eqref{eq:Ch3Relaxed} without constraints \eqref{eq:Ch3SCPIEGSb}. It is clear that MISOCPR method takes smaller solution time, however, it may introduce infeasible solutions. Therefore, the objective values and maximum $RCV$ are reported from each method with different time periods and the results are listed in Table~\ref{tab:Ch3SCPMISOCPR}. The proposed algorithm provides total objective very close to, if not equal to, the optimal one obtained by the MISOCPR method. The maximum $RCV$ of power lines (MRCV\_P) and that of gas pipelines (MRCV\_G) provided by the MISOCPR method are larger than that of the proposed S-MISOCP algorithm. It is because the S-MISOCP terminates after checking the solution feasibility. Therefore, both MRCV\_P and MRCV\_G are below \num{e-5}. We can conclude that the decisions obtained by MISOCPR method are infeasible and may introduce insecure operations in IEGS. Figure~\ref{fig:Ch3SCPF3}. displays the dramatic difference in energy production schedules between the two methods. For power generation dispatch, active power of each generator from both methods is almost similar because of the low MCRV\_P in case of MISOCPR method. However, for gas schedules, the two methods provide completely different gas production decisions.

    \begin{table}[!htb]
        \caption{Effectiveness of the proposed algorithm compared with MISOCP relaxation method} \label{tab:Ch3SCPMISOCPR}
        \centering
        \includegraphics[width=12.5cm]{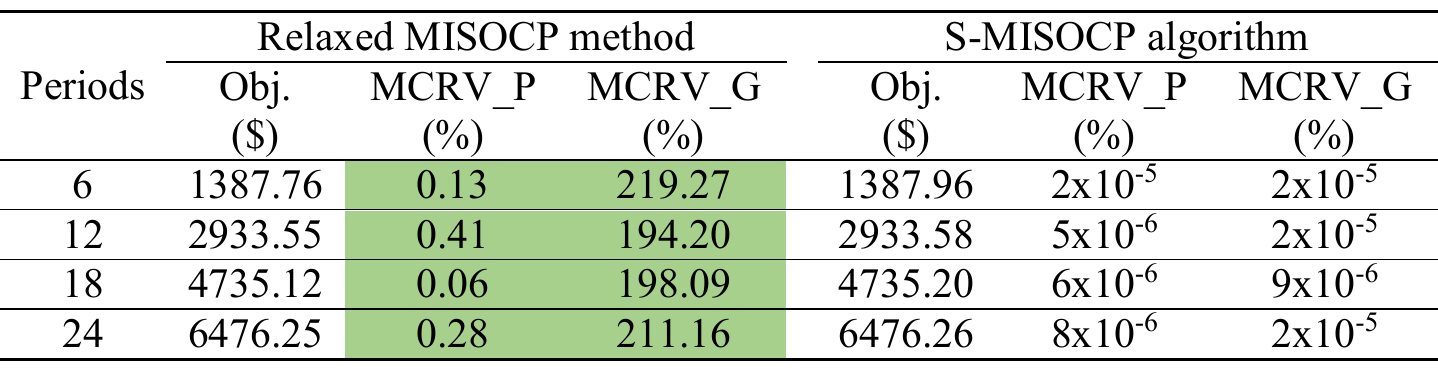}
    \end{table}

\section{Conclusions and Discussions}
        The optimal power-gas flow (OPGF) is the most fundamental problem in the interdependent power and natural gas systems. Initially, this chapter introduces a state-of-the-art techniques that are employed to solve OPGF problem, indicating the major advantages and drawbacks these techniques. Then, two different efficient methods based on convex optimization approaches have been proposed for transmission and distribution levels, respectively. Moreover, the commonly adopted PLA method, which reformulate the IEGS model into a MILP framework have been presented to be compared with the proposed ones.

        The first proposed method is the gas flow correction (GFC) method, in which the multi-slack-node method and the Levenberg-Marquardt algorithm are designed to consider the gas dynamics and bidirectional gas flow. Different case studies are conducted to show the effectiveness and the computational performance the proposed method, and the main conclusions are as follows.
        \begin{enumerate}
          \item The maximum Weymouth error decreases by GFC method from $25\%$ to $0.013\%$ as a worst-case.
          \item Increasing number of segments of PLA in the MILP model decreases the Weymouth error but it requires so long time. Using optimal breakpoints provide lower error with little increase of CPU time.
          \item GFC method is compared with 1) the MILP model based PLA; 2) MISOCP model based SOCP relaxation, and it is concluded that the proposed method provides a more optimal solution with shorter time.
        \end{enumerate}

            The second method is the S-MISOCP algorithm that finds the OPGF for IEGS at distribution-level, considering bidirectional energy conversions via GPUs, electric-driven compressors, and P2G facilities. In order to incorporate meshed gas networks, the sign function of Weymouth equations, for unfixed gas flow pipelines, are relaxed as MILP and quadratic constraints. The later and power flow quadratic equations are decomposed as MISCOP using DCP. The proposed algorithm is enhanced by (i) suggesting high-quality initial point instead of traditional or random ones, and (ii) adopting an adaptive penalty growth rate to control the main objective and violations weights in the penalized MISCOP problems. Finally, numerical results are conducted to evaluate the algorithm performance, showing the effectiveness of the suggested initial point and the adaptive penalty growth rate.

            In fact, the S-MISOCP algorithm can be employed to solve not only the distribution level IEGS with radial power networks but also the transmission level IEGS with DC-OPF model because it respects the gas flow directions inside pipelines and the DC-OPF is a linear set of constraints. In other words, transmission level IEGS can be solved by the S-MISOCP algorithm after removing the two opposite cones of AC-OPF for each branch. The main superiority of the proposed S-MISOCP algorithm is that it can be adopted to solve two-stage robust optimization (RO) models with the non-convex power and gas flow equations in both stages. This algorithm is called twice in the quadruple-loop procedure, which is proposed and employed in chapters~\ref{Chapter5}--\ref{Chapter6}.

            Although future work can include employing the proposed algorithm to solve a two-stage RO model, this work has been conducted in the thesis to consider power and gas system uncertainties, such as renewable outputs (see Chapter~\ref{Chapter5}), and to be integrated with bilateral gas-electricity marketing model (see Chapter~\ref{Chapter6}).  Future work also include solving the IEGS operational model under contingencies and demand response uncertainties. Enhancing its performance open quite a few new research directions, such as proposing better adaptive penalty rate with adjustable parameters or with additional factors, which could be controlled and updated at each iteration. Moreover, considering the AC-OPF model with bidirectional power flow and comparing with up-to-date studies, such as adjustable breakpoints for the PLA methods, are our subsequent works.


\chapter{Robust Resilient Operational Strategies for IEGSs} 

\label{Chapter4} 



Threatened by natural disasters and man-made attacks, the resilient operation of power systems has drawn increased attention, which gives rise to a greater demand for power generation assets with high operational flexibility, such as natural gas-fired power units (GPUs). This, in turn, results in a greater proportion of GPUs and greater interdependence between power and gas systems. As a consequence, the modeling of the interactions between power systems and natural gas systems to achieve operational resilience in power systems becomes extremely vital. This topic has been discussed by quite a few researchers; however, previous studies suffered from two major drawbacks, namely (1) they assumed the existence of only one utility that has full control authority over the power system and gas system; (2) the economic interactions between power systems and gas systems have been neglected, which goes against current industrial practice. In this study, the power system and gas system are regarded as two individual utilities and their physical and economic interactions are modeled by considering the fuel consumption of the GPUs and gas contracts and by guaranteeing the fuel availability of GPUs in the pre- and post-contingency stages, respectively. The proposed model is developed based on a two-stage robust decision-making framework to optimize the operational performances of power systems under the worst-case $N-k$ contingencies. To deal with the binary variables introduced by the linearization of the Weymouth equation and the on/off grid operation of generators, the nested column-and-constraint generation (NC\&CG) algorithm is adopted. The necessity of considering economic and physical interactions between power systems and natural gas systems and the effectiveness of the proposed model and algorithm are verified by numerical simulations of two test systems. This work has been published as
    \begin{itemize}
      \item
            Ahmed R. Sayed, Cheng Wang, and Tianshu Bi. "Resilient operational strategies for power systems considering the interactions with natural gas systems." Applied energy, 2019 May, 1;241:548-66, DOI: https://doi.org/10.1016/j.apenergy.2019.03.053
    \end{itemize}

This chapter is organized as follows. Section~\ref{sec:Ch4Intro} provides a state-of-the-art of the existing resilience model for the independent power system (IPS) resilience models, which disregard the interactions with gas infrastructures, and the integrated electric-gas system (IEGS) resilience models, which co-optimize the two infrastructures from economic and security perspectives, as well as it provides the main contributions of the work in this chapter. Section~\ref{sec:Ch4Prob} presents the mathematical formulations, including operational constraints of power and gas systems in both pre- and post-contingency stages, defender and attacker behavior models, firm and reserved gas contracts, and over-generation considerations. Section~\ref{sec:Ch4Methodol} describes the solution methodology by formulating the proposed non-convex model into mixed-integer linear programming (MILP) framework, and designing the NC\&CG algorithm with some recommendations to increase its performance. Numerical examples are provided in Section~\ref{sec:Ch4Res} to evaluate the performance and effectiveness of the proposed model. A thorough comparison is provided between the proposed model and (i) IPS resilience models and (2) IEGS resilience models. The effectiveness of the consideration of over-generation issues and gas line pack is discussed. The performance of the proposed solution methodology is validated with two large-scale IEGS test systems. Finally, the main conclusions and a brief discussion are drawn in Section~\ref{sec:Ch4Conc}.

\section{Introduction} \label{sec:Ch4Intro}
    The reliable and resilient operation of critical infrastructures, such as electricity, water, gas, and telecommunication, is important to strengthen and support economic and social activities in modern society. The electric power system is the most critical infrastructure system because electricity plays an important role in the secure and continuous operation of these systems. However, existing electric power grids experience different forms of vulnerabilities and random failures, such as extreme weather, terrorism, component aging/failure, unexpected generator or power line outages, and human errors, which may result in widespread economic and social contingencies. For example, extreme weather has caused power outages with damages ranging from \$$20$ to \$$55$ billion in the USA \cite{gordon2008protection}, blackouts such as Hurricane Katrina in $2005$ \cite{kwasinski2009telecommunications}, the Japan Earthquake in $2011$ \cite{adachi2011restoration}, Hurricane Sandy in $2012$ ($N-90$ event) \cite{henry2016impacts}, and transmission line contingencies in South Australia \cite{operator2017black}. Natural disasters, such as extreme weather are expected to increase in the future due to climate change \cite{estrada2015economic}. In addition, vulnerabilities to terrorist attacks could cause more severe system disruptions than natural disasters \cite{national2013resilience}. From $1999$ to $2002$, more than $150$ terrorist attacks on power networks worldwide have been reported \cite{national2012terrorism}. These vulnerabilities make it crucial to evaluate the performance and facilitate decision-making with regard to the power grid under contingencies by analyzing the power system vulnerability.

    Vulnerability analysis of power systems has drawn much attention in the pertinent literature with the focus on boosting the power system resilience and reliability. Quantitative and qualitative evaluations of the system resilience were summarized in \cite{lin2016study} and weather-affected resilience measures were discussed in \cite{panteli2015influence}. Different studies have presented numerous vulnerability analysis models, which can be grouped into two categories. In the first category, the model identifies the critical components of the power system \cite{crucitti2004model, ouyang2012time, ouyang2014multi, adachi2008serviceability, arroyo2010bilevel, aksoy2019generative, salmeron2009worst}. In \cite{crucitti2004model, ouyang2012time}, different models for the analysis of the vulnerability under random failures were presented. Reference \cite{crucitti2004model} formulates a simple model based on cascading failures by eliminating a number of system components randomly. In \cite{ouyang2012time}, the resilience management is improved based on the component failure probability and hazard frequencies. Likewise, natural disasters, which can result in the failure of multiple power system components simultaneously in a certain area, are usually simulated based on the probability of the damage states, such as hurricanes \cite{ouyang2014multi}, and earthquake \cite{adachi2008serviceability}. The vulnerability of critical components may not be the worst-case scenario in terms of disruption; therefore, mixed integer bi-level max-min models \cite{arroyo2010bilevel}, topology (graph) models \cite{aksoy2019generative}, and worst-case interdiction models \cite{salmeron2009worst} have been used to determine the worst attack strategies. However, the mere identification and protection of the vulnerable components do not assure an optimal defense plan in case of serious system disturbances. Therefore, models in the second category were developed to determine the optimal protection strategies for such vulnerabilities \cite{hausken2009minmax, yuan2014optimal, yuan2016robust, huang2017integration, lai2019tri, fang2017optimizing, lin2018tri}. In \cite{hausken2009minmax}, a min-max model was used to identify the optimal defense strategy and the worst attack scenarios under the optimum defense. To reduce the decision-making costs of the bi-level min-max model, a min-max-min model, which performs correction actions after the attack and considers the interaction between the power system defender and attacker, was employed in \cite{yuan2014optimal}. Under the same tri-level decision-making framework as \cite{yuan2014optimal}, quite a few variations and extensions have been reported by the literature, such as the consideration of multi-zone uncertainty in \cite{yuan2016robust}, the integration of preventive and emergency responses in \cite{huang2017integration}, the consideration of cyber-physical attacks in the communication network in \cite{lai2019tri}, the combination between system expansion (long-term planning) and switching operations (short-term operation) in \cite{fang2017optimizing}, and the distribution-level topology reconfiguration and DG islanding formulation in \cite{lin2018tri}.

    The aforementioned models can be used to determine the optimal protection plan and economic re-dispatch schedules based on the requirements of the electricity utilities, however, the models neglect the physical interactions between power systems and other energy systems such as natural gas systems, which provide the fuel for GPUs. Therefore, the “independent power system” (IPS) resilience model may not provide the optimal decision for power system operators (PSOs) and it may cause physical violations for other interacting systems, such as under/over nodal pressure and/or well production capacity violations in gas systems \cite{zlotnik2016coordinated}. It should be noted that interactions between power systems and gas systems have been significantly enhanced due to the increasing proportion of GPUs \cite{zhang2016climate}. In fact, the wide deployment of GPUs could also mitigate the fluctuations and uncertainties in power systems in a cost-effective manner \cite{devlin2016importance}, such as the outputs from renewable energy as well as contingencies, due to their high operational flexibility and efficiency \cite{correa2014security}. However, this increased interaction, which is also referred to as an integrated electricity and gas system (IEGS) in the literature, encounters issues related to the secure, reliable, and resilient operation of the power plants.

    In fact, quite a few studies have been conducted to address the issues of interdependency and the economic and resilient operation of IEGSs \cite{he2018coordination}. The solvability of the IEGS energy flow is discussed in \cite{chen2016identifying}. The gas system optimization problems are illustrated in detail in \cite{rios2015optimization}. A steady-state economic dispatch of the IEGS is established considering GPUs and power-to-gas (P2G) facilities in \cite{zeng2016steady} to enable bi-directional energy flow, and the impact of demand response is investigated in \cite{cui2016day}. The importance of considering the gas infrastructure in the IEGS with high penetration of wind power generation was investigated in \cite{golari2014two}. To mitigate the uncertainties associated with wind generation, an interval optimization IEGS model is presented in \cite{bai2016interval}, and a stochastic unit commitment (UC) for IEGSs is introduced in \cite{alabdulwahab2015stochastic} to solve the issues of random power outages and electricity load forecasting errors, based on which the impacts of P2G facilities are analyzed in \cite{guandalini2015power}. The $N-1$ contingency model for IEGSs was analyzed from economic and security-related aspects for a single outage in \cite{correa2014security}, and the model was improved by considering the spinning and regulation reserves in \cite{liu2018day}.

    It should be noted that the aforementioned power and gas co-optimization resilience models, namely IEGS models, share one underlying assumption, namely, the existences of one operator or utility that has full authority over both the power system and the gas system. This operator minimizes all costs associated with energy production and provides optimal decisions for the combined system. However, in most cases, the power system(s) and gas system(s) are operated by different utilities and they are unsynchronized in most countries and regions as in European countries \cite{bertoldi2006energy} and in China \cite{deng2017study}. This lack of synchronization indicates that the total fuel cost minimization determined by the IEGS model might not be a realistic operational objective for autonomous sub-systems and, therefore, bilateral energy trading is inevitable. In fact, the operational mode of the electric power system in post-contingency conditions might be significantly different from that of the pre-contingency stage, such as sudden start-up or shut-down of fast-response generators and rapid increases or decreases in generator outputs to minimize operating losses; a similar trend is observed for the gas demands of GPUs. Moreover, GPUs usually provide interruptible gas supply services according to current gas industrial practices \cite{liu2009security} and the gas contracts are usually determined by considering day-ahead contracts because real-time contracting would be costly and inconvenient \cite{bouras2016using}. In other words, GPUs cannot execute the planned regulations without appropriate gas contracting. In addition, both systems act in conjunction during contingencies and gas systems are ready to assist power utility according to real-time contracts.

    In this chapter, the resiliency of power systems against contingencies in terms of decision-making is revisited considering the interactions of power systems with gas systems. A robust day-ahead dispatch model for electric power system against $N-k$ contingencies is proposed. The model detects the worst-case attack against power systems and identifies the optimal gas contracts with preventive and corrective actions; this is accomplished by optimizing the economic generation dispatch in both the pre-contingency and post-contingency stages. In addition, the steady-state constraints of the gas system \cite{correa2014security} and the approximated flow dynamics are also considered, which increase the operational flexibility to satisfy the power system prerequisites. This model does not only provide optimal decisions in the pre-contingency stage, including the gas contracts, defensive strategy, and generation outputs but also provides the optimal adjustable strategy of the generator output in the post-contingency stage. In this study, the gas contract is considered as a combination of two sub-contracts, namely, the firm gas contract and the reserved gas contract. The firm gas contract determines the scheduled amount of gas required from the power utility to supply GPUs listed in the gas contract during the pre-contingency stage. In the other words, if there is no contingency, the gas demands from the power utility will be satisfied by the amount of gas scheduled in this contract. While, in post-contingency stage, as discussed above, the gas demands of GPUs may be changed. This change should obey the reserved gas contract which describes the scheduled positive (above the firm value) and negative (below the firm value) reserved gas. Alternatively, the power utility could sign a costly real-time contract. The interdependency and interactions between the power and gas systems are considered during normal and abnormal conditions, thus the gas network security can be guaranteed by suitable contracts and is not affected by the utilization of gas reserves. The operational goal of the power system operator is to achieve economic and secure operation with optimal gas contracts, while the operating costs of the gas systems are neglected. The main contributions in this chapter are as follows.
    \begin{enumerate}
    \item
        A tri-level resilient operational framework of power systems that considers contracts with gas systems is established, where firm gas supply contracts and gas reserve contracts are considered in the pre-contingency and the post-contingency stages, respectively. To the best of our knowledge, this is the first attempt to consider gas reserve contracts in a robust model. The operational constraints of the gas systems are considered, which guarantees their operational feasibility and security in both the pre- and post-contingency stages. Compared with IEGS robust models presented in the literature, the proposed model considers the dynamic state of the gas system. Moreover, as a result of considering contracts, a new kind of attack strategy emerges, i.e., the consumption of gas below/above the reserved (contracted) values.
    \item
        Unlike the most tri-level models where the lower level decision variables are continuous, there are binary variables in the lower level optimization problem in the proposed model. The additional binaries originate from the linearization of the nonlinear non-convex Weymouth equation, as well as the on/off control of the generators in the post-contingency stage, and they are used to determine the potential attack region \cite{wang2016robust}. Therefore, the NC\&CG algorithm proposed by \cite{zeng2013solving} is applied to solve the proposed tri-level model after adjusting its stopping criteria.
    \end{enumerate}

\section{Problem Formulation} \label{sec:Ch4Prob}
\begin{figure}[!htbp]
            \centering
            \includegraphics[width=14cm]{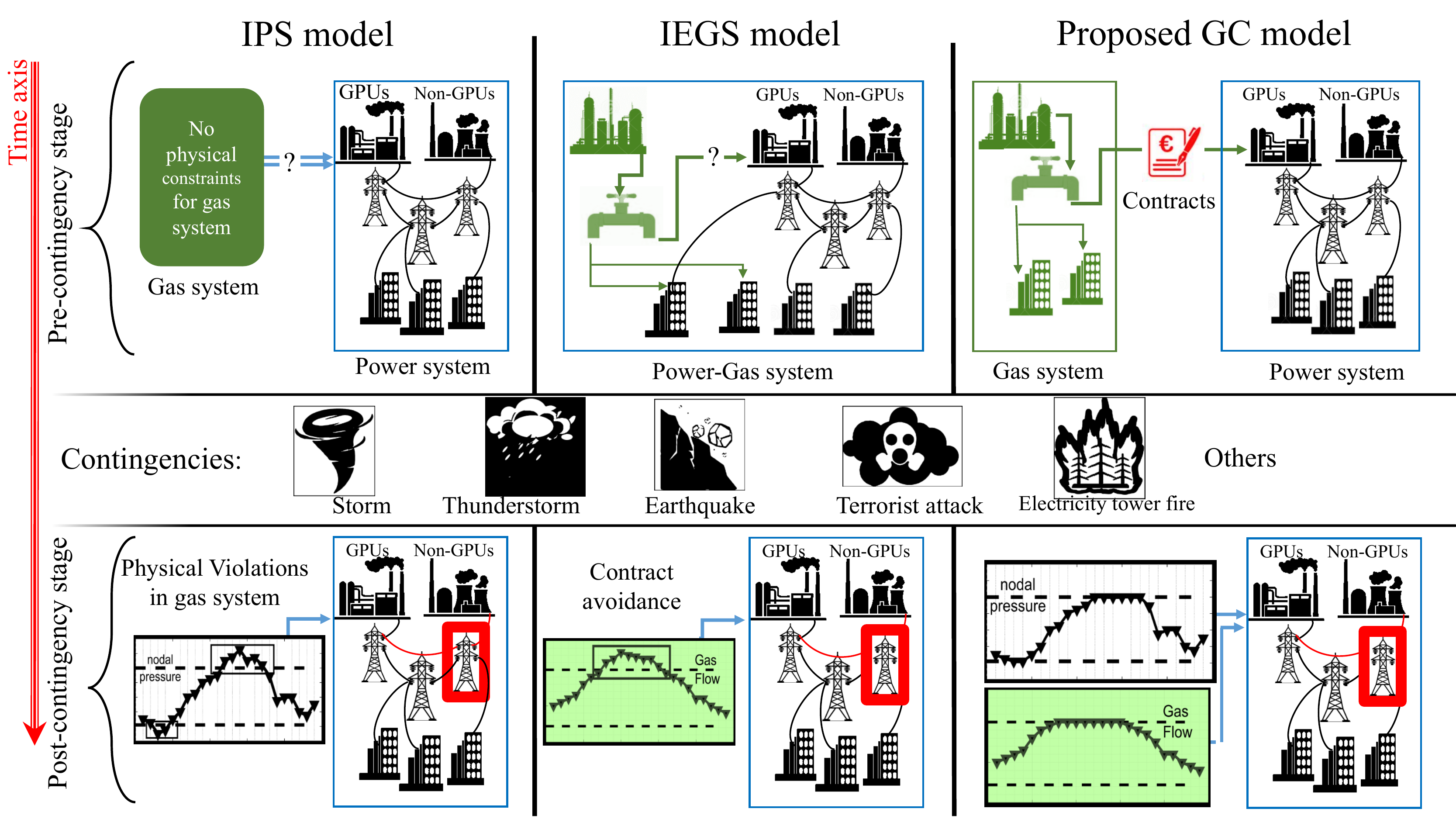}
            \caption{Operational layout in the pre- and post-contingency stages for the IPS model, IEGS model and the proposed GC model.}
            \label{fig:Ch4Description}
    \end{figure}

    For the ease of illustration, a schematic diagram of the aforementioned power resilience models, i.e., the IPS model, the IEGS model, and the proposed gas contracting (GC) model, is presented in Figure~\ref{fig:Ch4Description}. The salient features of the aforementioned models are as follows.
    \begin{enumerate}
    \item
        IPS model: this model treats gas systems as black boxes, where no operational constraints of the gas system are considered. This treatment may lead to violations of the gas constraints, particularly in the post-contingency stage, such as over-/under-pressure of gas nodes and might trigger cascading failures in gas systems.
    \item
         IEGS model: this model treats the power system and the gas system as one system and considers all operational constraints for both systems while neglecting the shared energy contracts. Because there are different utilities for the two systems, this model may result in over-optimistic solutions and cause contract avoidance in practice.
    \item
        GC model: this model minimizes the operational costs of the power system while considering the operational constraints of both the power system and gas system. The physical and economic interactions between the power system and the gas system are modeled by adding the fuel consumed by the GPUs to the nodal gas balancing constraint and gas contract costs in the power system objective function.
    \end{enumerate}

    The mathematical formulation of the proposed GC model is presented in this section after describing the prerequisite assumptions and simplifications commonly adopted in the literature:
    \begin{enumerate}
      \item
            In power system modeling, (i) the power system operates in a steady state and the transient state after the attack is ignored; (ii) the DC power flow model  presented in Section~\ref{sec:Ch2SDC} is adopted; (iii) all power generators are fast-response generator, therefore, the cut-off/connection process is completed without any delay, (iv) the UC is predetermined, and the interested reader can refer to Appendix~\ref{App:UC} for the UC problem formulation. These simplifications are commonly used in power system planning \cite{yuan2014optimal, yuan2016robust, huang2017integration,wang2016robust}. However, the proposed model can easily be extended to include the start-up and shut-down decisions of traditional coal-fired units with minimum on/off time constraints.
      \item
        In gas system modeling, (i) all gas storages are considered closed (not modeled) to highlight the significance of the gas dynamics; (ii) the simplified compressor model \cite{he2018coordination, correa2014security, wang2016robust}, which is presented in Section~\ref{sec:Ch2SComp}, is adopted.
      \item
        In contract modeling, the prices of both the firm gas in the pre-contingency stage and the reserved gas below/above the firm gas that is remained/consumed during the post-contingency stage are driven from the gas system operator. And the power system operator has received these prices before identifying the optimal gas contacts.
      \item
        Defense and attack strategies modeling: (i) a deterministic malicious attack analysis is implemented; (ii) any component is unavailable only if it is attacked and is not defended; (iii) in the proposed model, it is considered that only power lines are attacked but different component types such as generators and power-gas connection lines can also be included in the model.
    \end{enumerate}

    The overall objective function of the proposed model is presented in \eqref{eq:Ch4Obj1}. It consists of two parts, which are expressed by \eqref{eq:Ch4Obj2} and \eqref{eq:Ch4Obj3}, respectively. \eqref{eq:Ch4Obj2} depicts the operational costs $\Gamma_{pre}$ in the pre-contingency stage, including the power generation costs from all generators and costs for the reserved gas. Therefore, the day-ahead gas contracts are optimized by considering its two parameters in the objective function. The first parameter, which is the firm gas contact to be consumed by GPUs in the pre-contingency stage, is relevant to the power generation costs of GPUs. And the second parameter is the reserved gas contract. \eqref{eq:Ch4Obj3} provides the regulation costs $\Gamma_{post}$ in the post-contingency stage, namely, the re-dispatch costs of the non-GPUs, denoted by the first two terms and the penalties for the non-served electricity load, described by the third term. Note that there is no need to add the re-dispatch costs of GPUs to the regulation costs because they are optimized by considering the reserved gas in the operational costs. In this model, the best defense strategy and available resources in the pre-contingency stage such as the firm/reserved gas contracts and generation outputs are identified by the upper-level problem. The virtual attack strives to maximize worst-case disruption (regulation costs). In the lower level decision-making, the feasible resources are deployed, and generator re-dispatch is optimized to mitigate this disruption.
    \begin{gather}
      \min \; \Gamma_{pre} + \max \; \min \;\Gamma_{post}, \label{eq:Ch4Obj1} \\
      \Gamma_{pre} = \sum_{\forall t} \Big[ \sum_{u \in \mathcal{U}_n} {C}_u ( {p}_{u,t}^0) + \sum_{h \in \mathcal{H}} (\mu_{h,t} \rho_{h,t} +\mu_{h,t}^+ \rho_{h,t}^+ +\mu_{h,t}^- \rho_{h,t}^-) \Big] \label{eq:Ch4Obj2} \\
      \Gamma_{post} = \sum_{\forall t} \Big[ \sum_{u \in \mathcal{U}_n} ({C}_u^+ \triangle p_{u,t}^+ + {C}_u^- \triangle p_{u,t}^-) + \sum_{d \in \mathcal{D}_p(n)} {C}_d  \triangle p_{d,t} \Big]  \label{eq:Ch4Obj3}
    \end{gather}

    In this study, the decision-maker of the proposed GC model is the PSO, though the operational constraints of the gas system are involved. Different from existing works, which assumes the power systems and gas systems are controlled by one utility, economic behaviors, such as gas purchase contracts, are modeled in the formulation to reflect the multiple-decision-maker reality. The modeling of gas operational constraints in the power system decision-making problem is to guarantee the feasibility of the signed contracts with respect to the gas system operation. In other words, the gas operational constraints reflect the influence of signed gas contracts on the gas system from the feasibility perspective, and the gas system can always have a feasible operation status as long as the proposed tri-level model is successfully solved. The economic influences of the purchased gas from the power systems on the gas system have been considered in the prices of the contracts.

    \subsection{Pre-contingency constraints}
    To highlight the interdependency between the power system and the gas system, their operational constraints are introduced separately.

        \subsubsection{Power System Operational Constraints} \label{sec:Ch4PowerPre}
        The power system operational constraints are derived from Section~\ref{sec:Ch2SDC}, by using the pre-contingency decision variables. They are composed by
        \begin{gather}
         \underline{P}_{u} c_{u,t} \le p_{u,t}^0 \le \overline{P}_{u} c_{u,t}, \; \forall u,t, \label{eq:Ch4PowerPre1} \\
         -\overline{R}_{u}^- \le p_{u,t}^0 - p_{u,t-1}^0 \le \overline{R}_{u}^+, \; \forall u,t, \label{eq:Ch4PowerPre2} \\
        \sum_{u \in \mathcal{U}(n)}{p}_{u,t}^0 + \sum_{l \in \mathcal{L}_1(n)}{p}_{l,t}^0 -\sum_{l \in \mathcal{L}_2(n)}{p}_{l,t}^0= \sum_{d \in \mathcal{D}_p(n)} P_{d,t}, \; \forall n,t.  \label{eq:Ch4PreNode} \\
        -\tilde{\pi} \le \theta_{n,t}^0 \le \tilde{\pi},\; \forall n\in \mathcal{N}-1,t, \;\;\theta_{1,t}^0=0,\; \forall t, \;\;  \label{eq:Ch4PreAngle} \\
        -\overline{P}_{l,t} \le {p}_{l,t}^0 \le \overline{P}_{l,t},\; \forall l,t.   \label{eq:Ch4PreFlow}\\
         {p}_{l,t} = \frac{{\theta}_{m,t}^0-{\theta}_{n,t}^0}{x_l} ,\; \forall l,t, (m,n) \in l. \label{eq:Ch4PreOPF1}
        \end{gather}
        where the superscript symbol $({}^0)$ indicates the pre-contingency variables; ${c}_{u,t}$ is a predetermined UC decision; $\underline{P}_{u,t} / \overline{P}_{u,t}$ is the minimum/maximum limit of power generation; and $\overline{R}_{u}^- / \overline{R}_{u}^+$ is the maximum  ramping down/up capacity; $\tilde{\pi}\approx3.1416$ is the mathematical constant.

        \subsubsection{Natural Gas Operational Constraints}  \label{sec:Ch4GasPre}
        Considering gas consumption of GPUs in the nodal balancing equation, the dynamic-state gas flow model presented in Section \ref{sec:Ch2SGasDynamic}, is composed by
        \begin{gather}
            \underline{F}_w \le {f}_{w,t}^0 \le \overline{F}_w,\; \forall w,t, \label{eq:Ch4Wellpre} \\
            \underline{\Pi}_i \le {\pi}_{i,t}^0 \le \overline{\Pi}_i,\; \forall i,t, \label{eq:Ch4PressurePre}\\
            {\pi}_{i,t}^0 \le {\pi}_{o,t}^0 \le \gamma_c {\pi}_{i,t}^0, \forall c,t, (i,o) \in c   \label{eq:Ch4Comp1Pre} \\
            0 \le {f}_{c,t}^{out,0} = (1-\alpha_c) {f}_{c,t}^{in,0}, \; \forall c\in \mathcal{C},t,     \label{eq:Ch4Comp2Pre}
            \end{gather}
            \begin{gather}
            \sum_{w \in \mathcal{W}(i)} {f}_{w,t}^0 + \sum_{p \in \mathcal{P}_1(i)} {f}_{p,t}^{out,0} - \sum_{p \in \mathcal{P}_2(i)} {f}_{p,t}^{in,0} + \sum_{c \in \mathcal{C}_1(i)} {f}_{c,t}^{out,0} \nonumber  \\
            - \sum_{c \in \mathcal{C}_2(i)} {f}_{c,t}^{in,0}  = \sum_{h \in \mathcal{H}(i)} \rho_{h,t} + \sum_{d \in \mathcal{D}_g(i)} F_{d,t},\; \forall i,t,\label{eq:Ch4GFCNodePre}\\
            {m}_{p,t}^0 = \chi_p^m ( {\pi}_{i,t}^0 + {\pi}_{o,t}^0 ),\; \forall p,t, (i,o) \in p   \label{eq:Ch4GMMass1Pre}\\
            {f}_{p,t}^{in,0} - {f}_{p,t}^{out,0} =  {m}_{p,t}^0 - {m}_{p,t-1}^0, \; \forall p,t,   \label{eq:Ch4GMMass2Pre}\\
            {f}_{p,t}^0 = \frac{f_{p,t}^{in,0} + f_{p,t}^{out,0}}{2}, \; \forall p,t, \label{eq:Ch4AveFlowPre}\\
            {f}_{p,t}^0 |{f}_{p,t}^0|  = \chi_p^f ({\pi}_{i,t}^{0^2} - {\pi}_{o,t}^{0^2}), \; \forall p,t, (i,o) \in p. \label{eq:Ch4WeyPre}
        \end{gather}
        where $\mathcal{H}(i)$ is a subset of gas contracts, whose GPUs are supplied from node $i$; the firm gas amounts $\rho_{h,t}$ are defined in Section~\ref{sec:Ch4GasContr}.

\subsection{Behaviors of the Attacker and the Defender}
    It has been demonstrated that transmission lines are one of the most vulnerable assets in power systems \cite{arroyo2010bilevel,yuan2014optimal,motto2005mixed}. Therefore, we assume that there is a virtual attacker, who attacks transmission lines and has a limited attack budget, which is expressed as
    \begin{gather}
        \sum_{\forall l} C_l^u u_l \le k,\; u_l\in \{0,1\}, \; \forall l, \label{eq:Ch4AttBudg}
    \end{gather}
    where $k$ is the attack budget and $C_l^u$ is the cost of attack $u_l$. To mitigate the impacts of the attack, the power system operator can deploy defensive resources, such as hardening the lines or sending out line patrol crews aside from dispatching recourse resources in the post-contingency stage. Similarly, we assume that the defense budget of the PSO is limited \cite{arroyo2010bilevel,hausken2009minmax, yuan2014optimal, yuan2016robust, huang2017integration, lai2019tri, fang2017optimizing,wang2016robust}, as defined by
    \begin{gather}
        \sum_{\forall l} C_l^y y_l \le a,\; y_l\in \{0,1\}, \; \forall l,  \label{eq:Ch4DefBudg}
    \end{gather}
    where $a$ is the defense budget and $C_l^y$ is the cost of defender $y_l$. Combining the defense resource deployment and attack strategies, the availability of the lines are defined by \eqref{eq:Ch4AttDef}. The line is off-line only if it is attacked and is not defended. In other words, the line cannot be off-line if it is defended.
    \begin{gather}
        h_l = 1 - u_l + u_l y_l, h_l\in \{0,1\}, \; \forall l.  \label{eq:Ch4AttDef}
    \end{gather}

    \subsection{Gas Contracts} \label{sec:Ch4GasContr}
    The gas consumed by the GPU is subject to the pipeline capacity and contracts. The two terms of gas contract are firm gas and reserved gas. The firm gas contract specifications are mathematically presented in \cite{liu2009security} with UC, and in \cite{bouras2016using} as take-or-pay contracts. The proposed model provides the mathematical equations of the contracts signed between the gas utility and the electric utility. The interested reader is referred to \cite{liu2009security}. The firm gas amount can be defined as \eqref{eq:Ch4Firm}. In the proposed model, the gas contract costs include the cost of the firm gas in the pre-contingency stage and the cost of the reserved gas below/above the firm gas, which may remain or be consumed in the post-contingency stage. The common contract types are take-or-pay contracts and interruptible contracts \cite{bouras2016using}. The gas consumed by generator $u$ after an attack ($\frac{\Phi}{\eta_u} p_{u,t} $) is subjected to \eqref{eq:Ch4Reserve}. $\mathcal{U}_g(h)$ is a subset of the GPUs listed in contract $h$. The reserved gas below/above the firm gas is limited by the lower boundaries defined in \eqref{eq:Ch4ReserveBound}.
    \begin{gather}
        \rho_{h,t} =  \sum_{\forall u\in \mathcal{U}_g(h)} \frac{\Phi}{\eta_u} p_{u,t}^0,\; \forall h,t \label{eq:Ch4Firm}\\
        -\rho_{h,t}^- \le  \sum_{\forall u\in \mathcal{U}_g(h)} \frac{\Phi}{\eta_u} (p_{u,t} - p_{u,t}^0) \le \rho_{h,t}^+,\; \forall h,t \label{eq:Ch4Reserve}\\
        \rho_{h,t}^+ , \; \rho_{h,t}^- \ge 0,\; \forall h,t. \label{eq:Ch4ReserveBound}
    \end{gather}

    \subsection{Post-contingency Constraints}
    In the post-contingency stage, the model constraints include the transmission line availability \eqref{eq:Ch4AttDef} and the GPUs’ gas consumption constraints \eqref{eq:Ch4Reserve}. The constraints of the power and gas systems are described in the next section.
        \subsubsection{Power System Constraints}
\begin{figure}[!htbp]
            \centering
            \includegraphics[width=6cm]{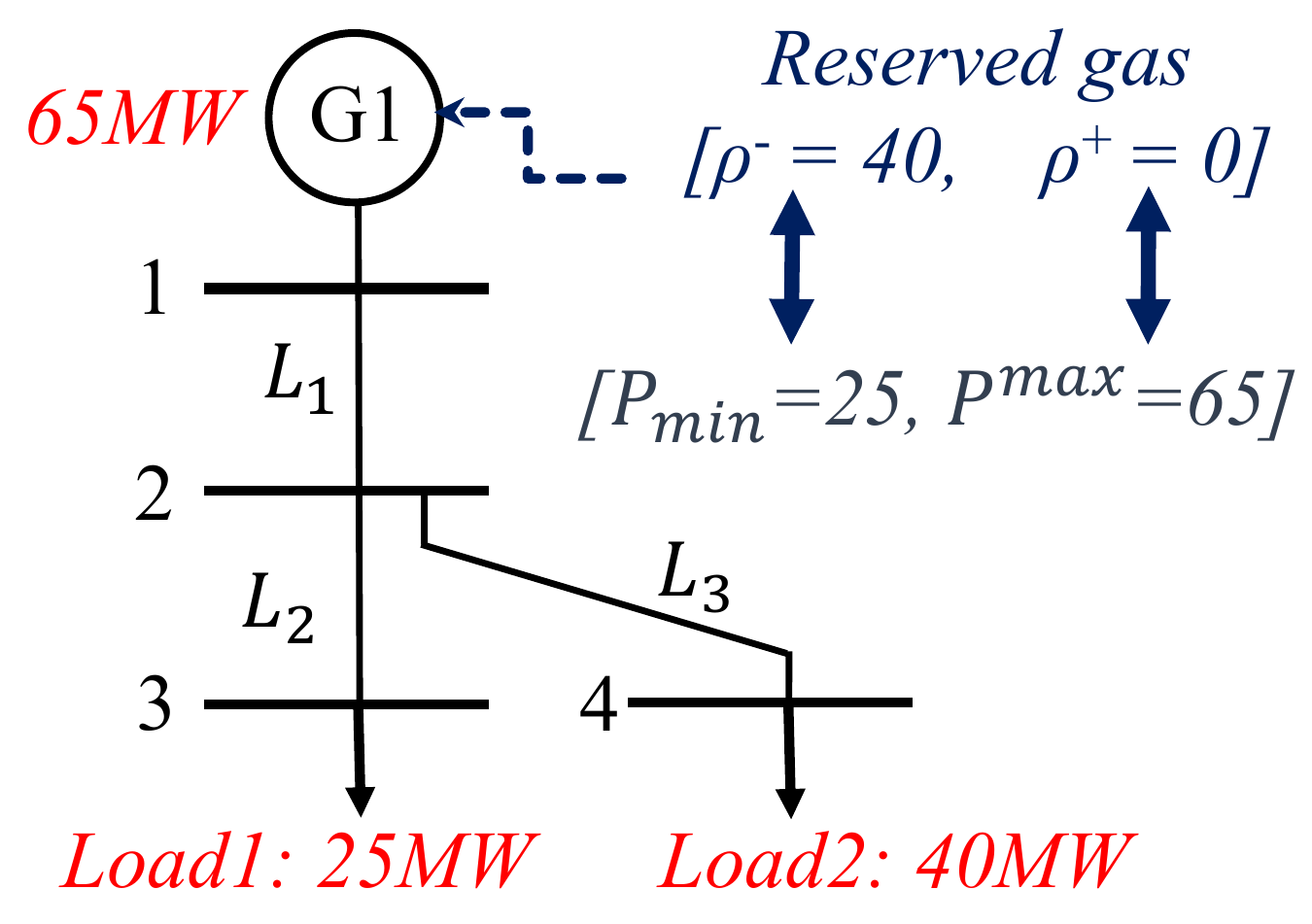}
            \caption{Illustration example of minimum output capacity constraint violation.}
            \label{fig:Ch4OG}
        \end{figure}

         Equations \eqref{eq:Ch4PowerPost1}--\eqref{eq:Ch4PowerPost2} and \eqref{eq:Ch4PowerPost4} define the generation capacities of the GPUs and non-GPUs, respectively. The ramping up and down limits for the generated power from all units are defined in \eqref{eq:Ch4PowerPost3}. Over-generation (OG), which is defined as a larger energy supply than demand, was first modeled by \cite{wang2016robust}. It was introduced to detect two attack strategies that may cause more damage than non-served power loads and reserved generated power. The first attack strategy is the violation of ramping down of a generator and the second attack strategy is the violation of the minimum output of a generator. The third attack strategy is introduced in the proposed model after considering the gas contracts. It is the violation of consuming gas by the GPUs below the reserved values. For example, if the reserved gas of contract $h$ is optimized as $\rho_{h,t}^+=\rho_{h,t}^-=0$, the outgoing feeders of the contracted GPUs cannot be attacked and these power plants have to generate the same amount of power in the post-contingency stage as in pre-contingency stage according to the constraints defined by \eqref{eq:Ch4Reserve}.
        \begin{gather}
          p_{u,t} = \delta_{u,t} {p}_{u,t}^*, \; \forall u\in \mathcal{U}_g,t, \label{eq:Ch4PowerPost1} \\
         \underline{P}_{u} c_{u,t} \le p_{u,t}^* \le \overline{P}_{u} c_{u,t}, \; \forall u,t, \label{eq:Ch4PowerPost2} \\
         -\overline{R}_{u}^- - (1-\delta_{u,t}) \overline{P}_{u}\le p_{u,t} - p_{u,t-1} \le \overline{R}_{u}^+ + (1-\delta_{u,t}) \overline{P}_{u}, \; \forall u,t, \label{eq:Ch4PowerPost3} \\
         \underline{P}_{u} \delta_{u,t} \le p_{u,t} \le \overline{P}_{u} \delta_{u,t}, \; \forall u,t, \label{eq:Ch4PowerPost4}
        \end{gather}
        Besides the existing two kinds of OG attack strategies reported by \cite{wang2016robust},  the following example is introduced to illustrate the third kind of attack strategy.  In Figure~\ref{fig:Ch4OG}, the test system has $4$ buses, one GPU $G1$, and two loads $Load1$, $Load2$. The firm and reserved gas contracts have been signed in the pre-contingency stage, say the firm value is $65$MW, and $40$ and $0$ are the reserved values below and above the firm, respectively. Therefore, the minimum and maximum outputs of $G1$ are $25$MW and $65$MW, respectively. According to the assumption that only power lines could be attacked, therefore, $L1$, $L2$ and, $L3$ are the attack decision. For simplicity, the attack and defender budgets are set as $1$ and $0$ respectively, and all cost coefficients in \eqref{eq:Ch4AttBudg}--\eqref{eq:Ch4DefBudg} are ones. In general, the worst attack scenario should be $L1$, as the non-served load will reach $65$MW. However, without considering OG, it is not allowed as the reserved gas lower bound would be violated. In this regard, the only feasible attack scenario is $L2$ and the corresponding unserved load is $25$MW, which is over-optimistic.

\begin{figure}[!htbp]
            \centering
            \includegraphics[width=9cm]{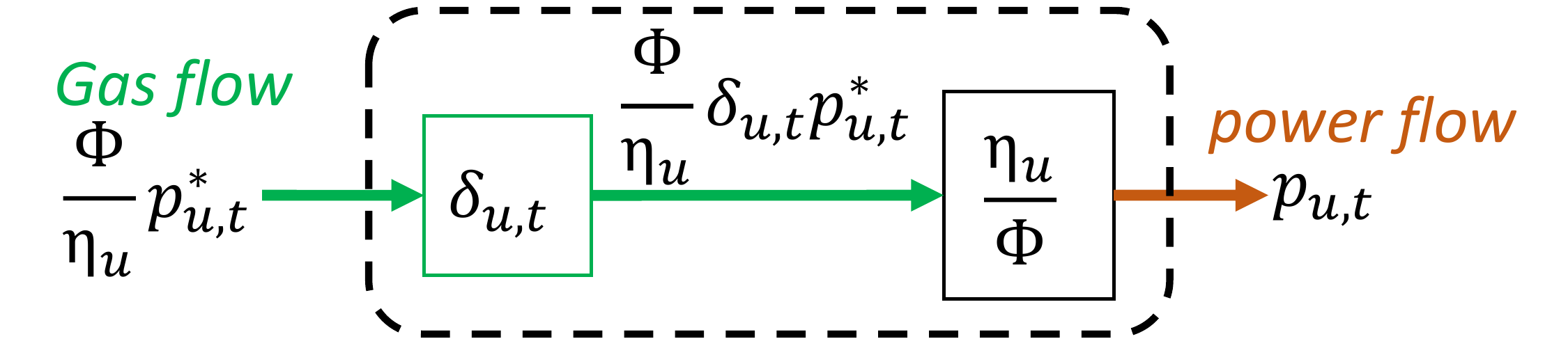}
            \caption{A simple block diagram of a GPU.}
            \label{fig:Ch4OG2}
        \end{figure}

         Therefore, a binary variable $\delta_{u,t}$ is added to consider OG during the post-contingency stage. As shown in Figure~\ref{fig:Ch4OG2}, the GPU needs gas with the value $\Phi p_{u,t}/\eta_u$ to generate power of $p_{u,t}$. On the other hand, the allowable gas $\Phi p_{u,t}^*/\eta_u$  is always restricted by the reserve constraints defined in \eqref{eq:Ch4Reserve}. Therefore, if the required gas $\Phi p_{u,t}/\eta_u$ fulfilled the contracts, then  $\delta_{u,t}$ equals $1$, otherwise,  $\delta_{u,t}$ equals $0$ and the GPU will be suddenly shut down. For non-GPUs, the first two attack strategies are detected using \eqref{eq:Ch4PowerPost3}and \eqref{eq:Ch4PowerPost4}, respectively. For GPUs, the first strategy is detected using \eqref{eq:Ch4PowerPost3}, whereas the last two strategies are detected using \eqref{eq:Ch4PowerPost1}--\eqref{eq:Ch4PowerPost2}.

        To provide supply and attack relaxations during the post-contingency stage, load shedding is added to the power nodal balance equation \eqref{eq:Ch4PostNode}. The range of load shedding is defined in  \eqref{eq:Ch4PostShedd}. The bus angle is defined by \eqref{eq:Ch4PostAngle}. If the line is off-line ($h_l=0$), \eqref{eq:Ch4PostFlow} forces the power flow of that line to be equal to zero; otherwise, it can be calculated from the phase angle difference defined in \eqref{eq:Ch4PostOPF1}. The reserved power generated by the non-GPUs is defined by \eqref{eq:Ch4PostResr1}--\eqref{eq:Ch4PostResr2}.
        \begin{gather}
            \sum_{u \in \mathcal{U}(n)}{p}_{u,t} + \sum_{l \in \mathcal{L}_1(n)}{p}_{l,t} -\sum_{l \in \mathcal{L}_2(n)}{p}_{l,t}= \sum_{d \in \mathcal{D}_p(n)} (P_{d,t}-\triangle p_{d,t}), \; \forall n,t, \label{eq:Ch4PostNode} \\
            0 \le \triangle p_{d,t} \le P_{d,t}, \; \forall d \in \mathcal{D}_p,t,  \label{eq:Ch4PostShedd} \end{gather}
            \begin{gather}
            -\tilde{\pi} \le \theta_{n,t} \le \tilde{\pi},\; \forall n\in \mathcal{N}-1,t, \;\;\theta_{1,t}=0,\; \forall t, \;\;  \label{eq:Ch4PostAngle} \\
            -\overline{p}_{l,t} h_l \le {p}_{l,t} \le \overline{p}_{l,t} h_l,\; \forall l,t,   \label{eq:Ch4PostFlow}\\
            -2\tilde{\pi} (1-h_l) \le x_l {p}_{l,t} - ({\theta}_{m,t}-{\theta}_{n,t} ) \le 2\tilde{\pi} (1-h_l) ,\; \forall l,t, (m,n) \in l, \label{eq:Ch4PostOPF1} \\
            \triangle{p}_{u,t}^+,\; \triangle{p}_{u,t}^- \ge 0,\; \forall u\in\mathcal{U}_n,t,\label{eq:Ch4PostResr1} \\
            -\triangle{p}_{u,t}^- \le {p}_{u,t} - {p}_{u,t}^0 \le \triangle{p}_{u,t}^+ ,\; \forall u\in\mathcal{U}_n,t. \label{eq:Ch4PostResr2}
        \end{gather}

    \subsubsection{Natural Gas System Constraints}
         The operating constraints of gas system in the post-contingency stage can be obtained by replacing the pre-contingency decision variables with post-contingency ones in \eqref{eq:Ch4Wellpre}--\eqref{eq:Ch4WeyPre}, namely removing the (${}^0$) symbols of the decision variables in those constraints. Such overlapped constraints are not listed.

    \subsection{Linearization of Nonlinear Terms}
    Note that there are products of a binary variable and a continuous variable in \eqref{eq:Ch4PowerPost1} as well as the nonlinearities in the Weymouth equation. Therefore, the upper level and lower level problems (pre- and post-contingency stages) represent mixed-integer nonlinear programming (MINLP) mathematically, which is not readily solvable by commercial solvers.

    The nonlinear constraints \eqref{eq:Ch4PowerPost1} is replaced by the following linear ones.
    \begin{gather}
         \underline{P}_{u} \delta_{u,t} \le p_{u,t} \le \overline{P}_{u} \delta_{u,t}, \; \forall u\in \mathcal{U}_g,t, \label{eq:Ch4PowerLP1} \\
         \underline{P}_{u}(1- \delta_{u,t}) \le p_{u,t} - p_{u,t}^* \le \overline{P}_{u} (1- \delta_{u,t}), \; \forall u\in \mathcal{U}_g,t, \label{eq:Ch4PowerLP2}
        \end{gather}

    In Appendix~\ref{App:PLA}, the incremental model for a general nonlinear function $\Im(x)$, such as the squared nodal pressure, i.e., $\pi_{i,t}^2, \pi_{o,t}^2$ and the squared pipeline flow, i.e., $f_{p,t} |f_{p,t}|$, is introduced.

    \subsection{The proposed GC Model Levels}

    Figure~\ref{fig:Ch4TriLevel} illustrates the purpose and decision variables of each level. In other words, it displays the roles of the system operator in both the pre-contingency and post-contingency stages and the purpose of the virtual disruptive agent. In the following subsections, each level will be discussed in detail.

    \subsubsection{Upper Level Problem}
\begin{figure}[!htbp]
            \centering
            \includegraphics[width=12cm]{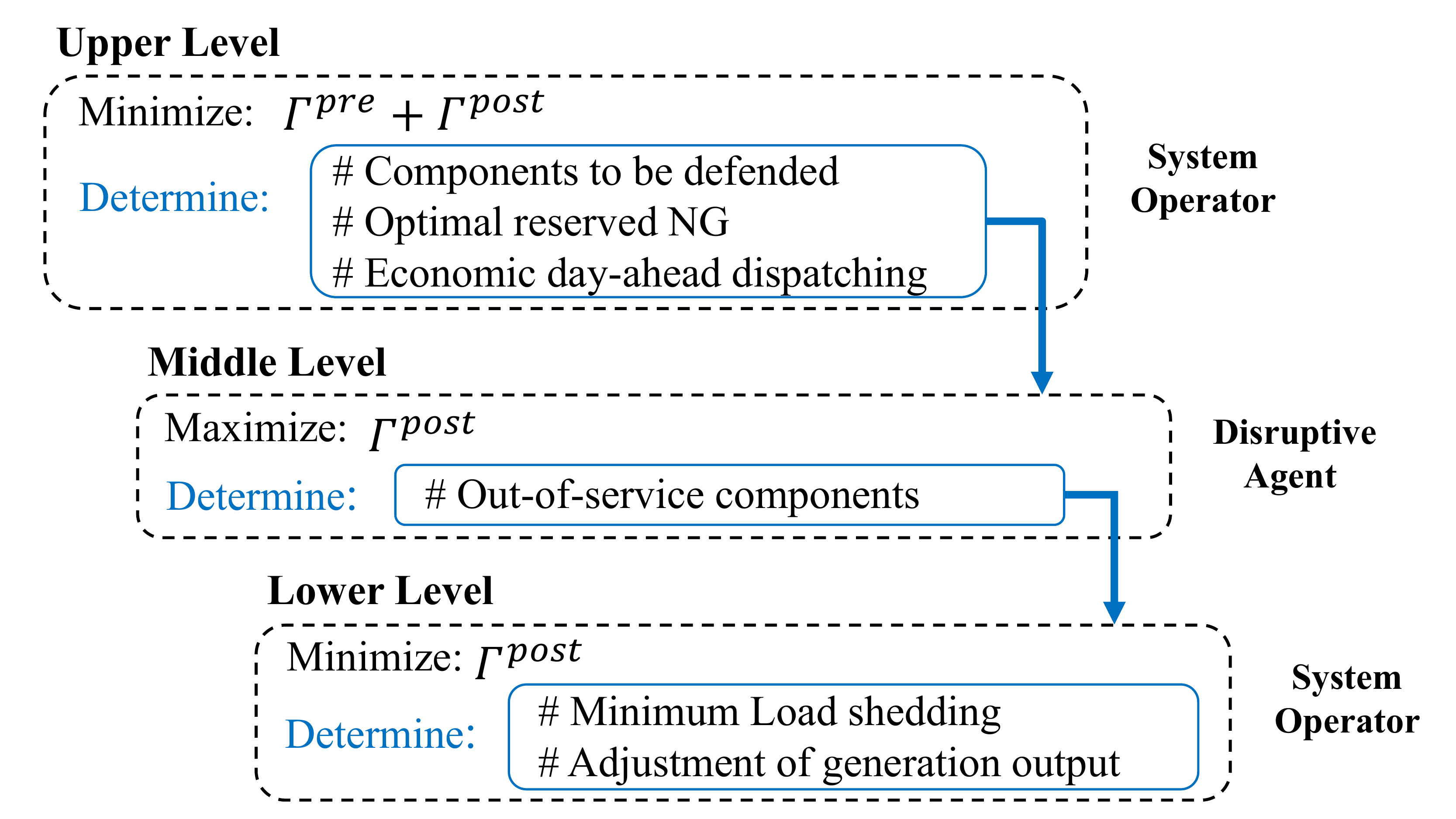}
            \caption{The three levels of the proposed model.}
            \label{fig:Ch4TriLevel}
        \end{figure}

    In this level, the decision variables are the defender action, the gas contracts, and the variables of the power and gas systems in the pre-contingency stage, including generator dispatch. The system operator seeks to minimize the overall cost in the pre-contingency (operation) and post-contingency (regulation) stages as shown in Figure~\ref{fig:Ch4TriLevel}.

    \begin{subequations}\label{eq:Ch4Upperobj} \begin{align}
          \min_{\bm{y},\; \bm{w},\; \bm{\alpha}}  & \Gamma^{pre} \; + \;  \Gamma^{post} \\
          s.t:\; & \text{Power system constraints: \eqref{eq:Ch4PowerPre1}--\eqref{eq:Ch4PreOPF1}.} \\
          & \text{Gas system constraints: \eqref{eq:Ch4Wellpre}--\eqref{eq:Ch4AveFlowPre} and PLA models for: \eqref{eq:Ch4WeyPre}.}     \\
          & \text{Defense budget: \eqref{eq:Ch4DefBudg}}. \\
          & \text{Firm gas and reserved boundaries in contracting: \eqref{eq:Ch4Firm} and \eqref{eq:Ch4ReserveBound}.}  \\
          & \bm{y} = \{y_{l,t}\}, \;\;  \bm{w} = \{ {\rho}_{u,t}, {\rho}_{u,t}^+, {\rho}_{u,t}^-\}, \;\; \bm{y}, \bm{w} \in arg\{\text{Middle-level problem \eqref{eq:Ch4Middleobj}}  \}. \\
          & \bm{\alpha} = \{{p}_{u,t}^0, {p}_{l,t}^0 , {\theta}_{n,t}^0, {f}_{w,t}^0, f_{p,t}^0, \pi_{i,t}^0, f_{p,t}^{in,0} , f_{p,t}^{out,0}, f_{c,t}^{in,0} , f_{c,t}^{out,0}, {m}_{p,t}^0, PLA_{Pre}\}
    \end{align}\end{subequations}
        where $PLA_{Pre}$ is the linearization variables, which include all continuous and binary variables for the squared nodal pressures and squared pipelines average flow of Weymouth equation \eqref{eq:Ch4WeyPre}. Because some variables in the upper level will influence the middle/lower level decision-making, the upper-level decision variables are divided into three vectors: $\bm{y}$, which is the defender action for all power lines, $\bm{w}$, which is the generator dispatch in the pre-contingency stage and the contracting decisions, and $\bm{\alpha}$, which represents the remaining variables of the power and gas systems in the pre-contingency stage. It should be noted that $f_{p,t},\pi_{i,t},\;f_{p,t}|f_{p,t}|$ and $\pi_{i,t}^2$ are represented by piecewise linearization variables, as discussed in Appendix~\ref{App:PLA}. As a result, they may not be considered as decision variables.

\subsubsection{Middle Level Problem}
    The objective of the middle-level problem is to maximize the regulation costs by finding the optimal (worst) attack strategy $\bm{u}$, which is the only decision variable in this level. Thus, the problem is defined as
     \begin{subequations}\label{eq:Ch4Middleobj} \begin{align}
          \max_{\bm{u}}  & \;\;  \Gamma^{post} \\
          s.t:\; & \text{Attack budget: } \eqref{eq:Ch4AttBudg}, \text{ and availability statues: } \eqref{eq:Ch4AttDef} \\
          & \bm{u} = \{u_{l,t}\}, \;\; \bm{y}, \bm{w}, \bm{u} \in arg\{\text{Lower-level problem \eqref{eq:Ch4Lowerobj}} \}.
    \end{align}\end{subequations}

\subsubsection{Lower Level Problem}
    Under to the worst attack strategy, the lower level problem aims to find the minimum load shedding with minimum power adjustments from the non-GPUs by re-dispatching all power units. The power adjustments from the GPUs are already optimized in the upper level by optimizing the gas contracts ($\rho_{h,t},\; \rho_{h,t}^+,\; \rho_{h,t}^-$). The lower decision variables are divided into $\bm{x}$ for continuous variables and $\bm{z}$ for binary variables.
    \begin{subequations}\label{eq:Ch4Lowerobj}\begin{align}
          \min_{\bm{x},\; \bm{z}}  & \;\;  \Gamma^{post} \\
          s.t:\; & \text{Power system constraints: \eqref{eq:Ch4PowerPost2}--\eqref{eq:Ch4PowerPost4} and \eqref{eq:Ch4PowerLP1}--\eqref{eq:Ch4PowerLP2}.} \\
          & \text{Gas system constraints: \eqref{eq:Ch4Wellpre}--\eqref{eq:Ch4AveFlowPre} with removing (${}^0$) symbol.}     \\
          & \text{PLA models for: \eqref{eq:Ch4WeyPre} and availability statues: \eqref{eq:Ch4AttDef}.} \\
          & \text{Gas contracting:  \eqref{eq:Ch4Reserve}.}  \\
          & \bm{z} = \{\delta_{u,t}, PLA_{Post}^b\}, \;\; \bm{x} = \{{p}_{u,t}, {p}_{u,t}^*, {p}_{l,t} , {\theta}_{n,t}, \triangle{p}_{u,t}^+, \triangle{p}_{u,t}^-, {f}_{w,t},  \\
          & f_{p,t}, \pi_{i,t}, f_{p,t}^{in} , f_{p,t}^{out}, f_{c,t}^{in} , f_{c,t}^{out}, {m}_{p,t}, PLA_{Post}^c\}.
    \end{align}\end{subequations}
    where $PLA_{Post}^c $ and $PLA_{Post}^b$ are the continuous and binary linearization variables for the for the nonlinear terms in the gas flow equation, respectively.

\section{Solution Methodology} \label{sec:Ch4Methodol}
Tri-level models are generally solved by decomposition methods such as Benders decomposition \cite{bertsimas2012adaptive} and the column-and-constraint generation (C\&CG) algorithm \cite{zeng2013solving}. In the proposed model, the presence of binary variables resulting from the linearization and OG prevents dualizing the lower level problem. Therefore, the C\&CG algorithm cannot be used directly to solve the model, and the NC\&CG algorithm presented in \cite{zhao2012exact} is adopted and is modified by editing its stopping criteria. The compact form of the proposed model is presented before discussing the implementation of the NC\&CG algorithm.

    \subsection{The Compact Form of the Proposed Model}
    For ease of analysis, the compact form of the proposed GC model is given below:
    \begin{subequations}\label{eq:Ch4Compact}\begin{align}
          \min_{\bm{y},\; \bm{w},\; \bm{\alpha}}& \;\;  f(\bm{w}) \max_{\bm{u}} \;\; \min_{\bm{x},\; \bm{z}} \bm{d}\bm{x} \label{eq:Ch4Compact1}\\
          s.t:\; &\bm{R}\bm{y} + \bm{A}\bm{w} + \bm{M}\bm{\alpha} \ge \bm{J},  \label{eq:Ch4Compact2} \\
          & \bm{h} = \bm{1} - \bm{u} + \bm{u}\cdot\bm{y}, \label{eq:Ch4Compact3} \\
          & \bm{B}\bm{u} \le k ,  \label{eq:Ch4Compact4} \\
          &\bm{E}\bm{x} + \bm{G}\bm{z} \ge \bm{F} - \bm{Q} \bm{h} - \bm{D} \bm{w},  \label{eq:Ch4Compact5}.
    \end{align}\end{subequations}

    where, $ f(\bm{w})$ denotes the operational costs in the pre-contingency state ($\Gamma^{pre}$) and $\bm{d}\bm{x}$ denotes the regulation costs in the post-contingency state ($\Gamma^{post}$). $\bm{R}, \bm{A}, \bm{M}, \bm{J}, \bm{B}, \bm{E}, \bm{G}, \bm{F},$ $\bm{Q}$ and $\bm{D}$ are the coefficient matrices of the model constraints. $ f(\bm{w})$ and constraints \eqref{eq:Ch4Compact2} can be derived from the upper-level problem. The constraint of the middle-level problem (attack budget) is defined in \eqref{eq:Ch4Compact4}. $\bm{d}\bm{x}$ and the constraints defined in \eqref{eq:Ch4Compact5}  can be determined from the lower level problem. Finally, constraints \eqref{eq:Ch4Compact3}  can be derived from the availability statues presented in \eqref{eq:Ch4AttDef}, where $\bm{u}\cdot\bm{y}$ denotes the element product. There is no need to linearize this product because the two variables are used at different levels.

    \subsection{The NC\&CG Algorithm}
\begin{figure}[!tb]
            \centering
            \includegraphics[width=13.5cm]{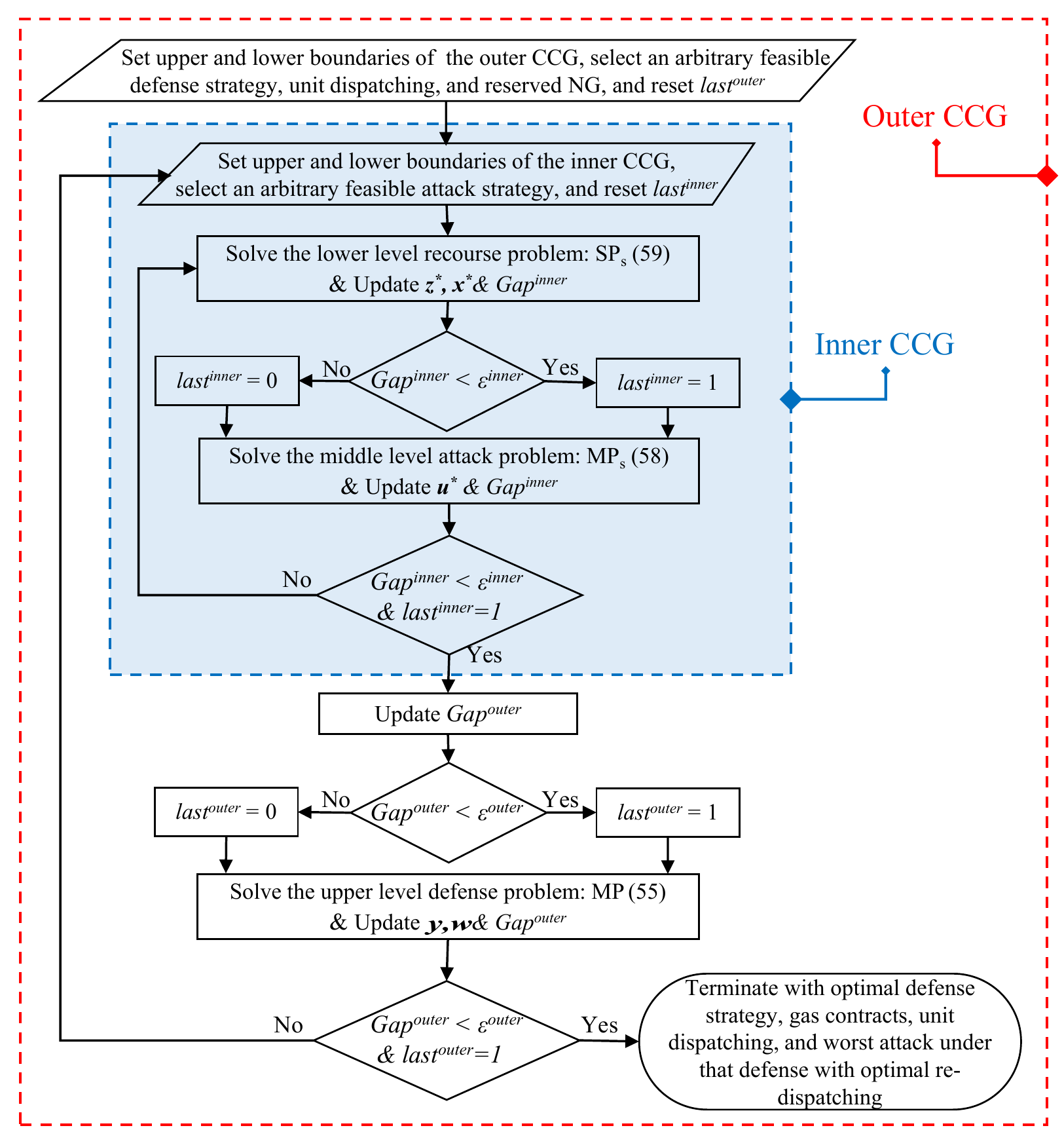}
            \caption{The NC\&CG algorithm layout.}
            \label{fig:Ch4NCCH}
    \end{figure}

    As shown in Figure~\ref{fig:Ch4NCCH}, once an arbitrary feasible decision has been derived from the upper and middle-level problems (i.e., $\bm{y}, \bm{w}$  and $\bm{u}$), then the inner C\&CG deals with the recourse problem to provide optimal binaries as a primal cut solution to the middle-level attack problem. The inner gap, which equals the relative difference between the middle and lower level objectives, is verified twice at each iteration to guarantee that the optimal decision vectors $\bm{u}^*, \bm{x}^*$, and $\bm{z}^*$ are achieved. The inner C\&CG provides the worst attack strategies as a primal cut solution to the upper-level defense problem in the outer C\&CG algorithm. Similar to the inner gap, the outer gap is checked twice at each iteration to find the optimal preventive actions $\hat{\bm{y}}$ and $\hat{\bm{w}}$ in the pre-contingency stage (upper level).

    \subsubsection{Outer C\&CG}
        We assume that the inner C\&CG algorithm can solve the sub-problem defined in \eqref{eq:Ch4SP} and derives an optimal attack scenario $\bm{u}^*$ and the worst-case value of the sub-problem $\Psi^*(\hat{\bm{y}},\hat{\bm{w}})$; subsequently, the outer C\&CG is implemented to solve the bi-level min-max (first-stage) problem. The details are presented in Algorithm~\ref{Ch3alg1}.

    \subsubsection{Inner C\&CG}
        The inner C\&CG is used to solve sub-problem \eqref{eq:Ch4SP} by expanding it into the tri-level form defined in \eqref{eq:Ch4SPTri}. A strong duality reformulation for the last level (linear problem) outperforms the use of the Karush-Kuhn-Tucker (KKT) conditions \cite{crucitti2004model} because of a large number of binary variables used to linearize the KKT conditions; this number equals the summation of total variables and constraints in the last level. In contrast, in strong duality, the variables used in the linearization are continuous and there is a small number of variables that equals the number of components that might be attacked or have to be defended, such as power lines in this study. The model \eqref{eq:Ch4SPTri} represents the sub-problem in the tri-level formulation based on strong duality. The vectors $\hat{\bm{y}}$ and  $\hat{\bm{w}}$ are derived from the outer C\&CG and become parameters in the inner C\&CG;  ${\bm{\lambda}}$ denotes the dual variables of the last level. In the objective, the only nonlinear product is  ${\bm{\lambda}}^{\top} \bm{Q} \bm{h}$ , which is linearized by  ${\bm{\gamma}}$ with additional constraints defined in \eqref{eq:Ch4MPs}, where $\sum{\bm{\gamma}}^r$ means the summation of the members inside vector ${\bm{\gamma}}^r$, and $\overline{M}$ is a large sufficient number.
        \begin{subequations}\label{eq:Ch4SPTri}\begin{align}
            \max_{\bm{u}} & \;\; \min_{\bm{z}}  \;\; \min_{\bm{\lambda}} \; (\bm{F} - \bm{Q} \bm{h} - \bm{D}\hat{\bm{w}} - \bm{G}\bm{z})^{\top}  \bm{\lambda} \\
          s.t:\; & \bm{h} = \bm{1} - \bm{u} + \bm{u}\cdot\hat{\bm{y}}, \\
          & \bm{B}\bm{u} \le k ,  \\
          &\bm{E}^{\top} \bm{\lambda} = \bm{d}^{\top},\;\; \bm{\lambda} \ge 0.
        \end{align}\end{subequations}

\begin{algorithm}[!ht]
\caption{Outer C\&CG Algorithm}
\label{Ch3alg1}
\begin{algorithmic}[1]
    \STATE Set boundary parameters $UB^{outer}=\infty, LB^{inner}=-\infty, R=0$, convergence parameters $\varepsilon, \;last^{outer}=0$ and select an arbitrary feasible $\hat{\bm{y}}, \hat{\bm{w}}$, then go to Step $4$.

    \STATE Solve problem \textbf{MP}  \eqref{eq:Ch4MP}  to update $\hat{\bm{y}},\hat{\bm{w}},\hat{\bm{\alpha}},{\varphi}^*, LB^{outer}=\textbf{MP}^*$ and $Gap^{outer}=(UB^{outer}-LB^{outer})/UB^{outer}$. \begin{subequations}\label{eq:Ch4MP}\begin{align}
          \textbf{MP}:\;\; & \min_{\bm{y},\; \bm{w},\; \bm{\alpha},\; \varphi,\;\bm{x}^r,\;\bm{z}^r} \;\;  f(\bm{w}) + \varphi \\
          s.t:\; &\bm{R}\bm{y} + \bm{A}\bm{w} + \bm{M}\bm{\alpha} \ge \bm{J},  \\
          & \varphi \ge \bm{d} \bm{x}^r, \;\; \forall r=\{1\ldots R\}, \\
          & \bm{h}^r= \bm{1} - \bm{u}^{*,r} + \bm{u}^{*,r} \bm{y}, \;\; \forall r=\{1\ldots R\},  \\
          &\bm{E}\bm{x}^r + \bm{G}\bm{z}^r \ge \bm{F} - \bm{Q} \bm{h}^r - \bm{D} \bm{w}, \;\; \forall r=\{1\ldots R\} .
    \end{align}\end{subequations} \\

    \STATE If $Gap^{outer} \le \varepsilon \;\&\; last^{outer}$, terminate; else, $R=R+1$, $\bm{u}^{*,R}=\bm{u}^*, last^{outer}=0$.

    \STATE Solve problem \textbf{SP} \eqref{eq:Ch4SP} to update $\bm{u}^*,\bm{x}^*,\bm{z}^*, UB^{outer}=\min\{UB^{outer}, f(\hat{\bm{w}})+\textbf{SP}^*\}$ and $Gap^{outer}=(UB^{outer}-LB^{outer})/UB^{outer}$.  \begin{subequations}\label{eq:Ch4SP}\begin{align}
           \textbf{SP}:\;\;& \max_{\bm{u}} \;\; \min_{\bm{x},\; \bm{z}} \bm{d}\bm{x} \hspace{13em} \\
          s.t:\; & \bm{h} = \bm{1} - \bm{u} + \bm{u}\cdot\hat{\bm{y}}, \\
          & \bm{B}\bm{u} \le k ,  \\
          &\bm{E}\bm{x} + \bm{G}\bm{z} \ge \bm{F} - \bm{Q} \bm{h} - \bm{D}\hat{\bm{w}},  .
    \end{align} \end{subequations}\\

    \STATE  If $Gap^{outer} \le \varepsilon$, set $last^{outer}=1$; else, $last^{outer}=0$, and go Step $2$.
\end{algorithmic}
\end{algorithm}

\begin{algorithm}[!ht]
\caption{Inner C\&CG Algorithm}
\label{Ch4alg2}
\begin{algorithmic}[1]
    \STATE Set boundary parameters $UB^{inner}=\infty, LB^{inner}=-\infty, R=0$, convergence parameters $\varepsilon,\; last^{inner}=0$ and select an arbitrary feasible ${\bm{u}}^*$, then go to Step $4$.

    \STATE Solve problem \textbf{MPs}  \eqref{eq:Ch4MPs}  to update ${\bm{u}}^*,\; UB^{inner}=\textbf{MPs}^*$ and $Gap^{inner}=(UB^{inner}-LB^{inner})/UB^{inner}$.  \begin{subequations}\label{eq:Ch4MPs}\begin{align}
          \textbf{MPs}:\;\; &\max_{\bm{u},\;\bm{h},\;\bm{\lambda}^r,\;\bm{\gamma}^r,\;\vartheta} \vartheta \\
          s.t:\; & \bm{h} = \bm{1} - \bm{u} + \bm{u}\cdot\hat{\bm{y}}, \\
          & \bm{B}\bm{u} \le k ,  \\
          & \vartheta \le {\bm{\lambda}^r}^{\top}(\bm{F} - \bm{Q} \bm{h}^r - \bm{D} \hat{\bm{w}}- \bm{G}\bm{z}^r) - \sum \bm{\gamma}^r, \;\; \forall r=\{1\ldots R\},  \\
          &\bm{E}^{\top} \bm{\lambda}  = \bm{d}^{\top}, \;\; \bm{\lambda}\ge 0, \;\; \forall r=\{1\ldots R\},  \\
          &\overline{M}\bm{h}\le \bm{\gamma}^r  \le \overline{M}\bm{h}, \;\; \forall r=\{1\ldots R\},  \\
          &\overline{M}(1-\bm{h})\le \bm{Q}^{\top} \bm{\lambda}^r - \bm{\gamma}^r  \le \overline{M}(1-\bm{h}), \;\; \forall r=\{1\ldots R\}.
    \end{align} \end{subequations}\\

    \STATE If $Gap^{inner} \le \varepsilon \;\&\; last^{inner}$, terminate; else, $R=R+1$, $\bm{z}^{*,R}=\bm{z}^*, last^{inner}=0$.

    \STATE Solve problem \textbf{SPs} \eqref{eq:Ch4SPs} to update $\bm{x}^*,\bm{z}^*, LB^{inner}=\max\{LB^{inner}, \; \bm{dx}\}$ and $Gap^{inner}=(UB^{inner}-LB^{inner})/UB^{inner}$. \begin{subequations}\label{eq:Ch4SPs}\begin{align}
           \textbf{SPs}:\;\;& \min_{\bm{x},\; \bm{z},\; \bm{h}} \bm{d}\bm{x} \hspace{20em} \\
          s.t:\; & \bm{h} = \bm{1} - \bm{u}^* + \bm{u}^*\cdot\hat{\bm{y}}, \\
          &\bm{E}\bm{x} + \bm{G}\bm{z} \ge \bm{F} - \bm{Q} \bm{h} - \bm{D}\hat{\bm{w}}.
    \end{align} \end{subequations}\\

    \STATE  If $Gap^{inner} \le \varepsilon$, set $last^{inner}=1$; else, $last^{outer}=0$, and go Step $2$.
\end{algorithmic}
\end{algorithm}

        In the traditional NC\&CG proposed in \cite{zhao2012exact}, if the duality gap is below the convergence tolerance value $\varepsilon^{outer}/\varepsilon^{inner}$, the stopping criteria are fulfilled for the outer/inner C\&CG algorithm, where the duality gap is measured only after the master problem. This means if the master problem achieves its optimal objective, which means it converges with previous sub-problem objectives (minimal duality gap), the algorithm terminates even if the master decision is not optimal. To illustrate that, in the inner C\&CG algorithm, for each iteration, the \textbf{MPs} in step $2$ represents the worst attack strategy based on the experience gained from previous iterations. For example, at iteration $R$, we have $\bm{u}_1$ and  $\bm{u}_2$, which provide the same objective value $UB_{inner}$ based on the experience of $R-1$ iterations. If the \textbf{MPs} decision is  $\bm{u}_1$, the \textbf{SPs} in step $4$ provides $LB_{inner}\simeq UB_{inner}$. For the next iteration, $R+1$, the \textbf{MPs} decision is either $\bm{u}_1$  or $\bm{u}_2$, which results in the same $UB_{inner}$. But if the \textbf{MPs} decision is $\bm{u}_2$ and the \textbf{SPs} decision results in deploying available resources to minimize the damage resulting from $\bm{u}_2$, the inner C\&CG will terminate due to a sub-optimal attack strategy. This stopping criterion provides the optimal objective value but it may fail to find the optimal attack strategy or optimal values of the lower level variables. This problem may perturb the outer C\&CG if it is not considered in the inner C\&CG because the optimal attack strategies cannot be found and, consequently, the outer C\&CG may not converge. Therefore, the concept of $last^{outer}/last^{inner}$  is introduced in the outer/inner C\&CG algorithms and their values are updated before and after the execution of the sub-problem \textbf{SP}/\textbf{SPs}, respectively.

\section{Simulation Results} \label{sec:Ch4Res}
    In this section, a $5$-Bus-$7$-Node electricity-gas integrated energy system is examined to illustrate the effectiveness and features of the proposed model and algorithm; the IEEE $39$-Bus-$20$-Node and the IEEE $118$-Bus-$20$-Node test systems are employed to study the scalability of the implemented algorithm. The load-shedding penalty price is set at \$$1000$/MWh and $\overline{M}$ used for the linearization in the inner C\&CG algorithm is set at $10^5$. In fact, this value affects the computational time and depends on the expected value of the inner C\&CG objective \cite{wang2016robust}. The convergence tolerance value $\varepsilon$ is set at $0.1\%$. The number of segments used in the linearization of the Weymouth equation is set at $6$ for both the gas flow and gas pressure. The numerical experiments are performed using MATLAB R$2020$a with the YALMIP toolbox \cite{YALMIP} and Gurobi $8.1$ on a personal laptop with Intel(R) Core(TM) i$5-3320$M CPU and $8.00$ GB RAM.

    Figure~\ref{fig:Ch4Topology1} shows the topology of the test system. The details and UC of the system are described in Appendix~\ref{App:5Bus} and Appendix~\ref{App:sevenNode}. The system has $6$ power lines, $2$ GPUs, $1$ non-GPUs, $3$ power loads, $2$ gas wells, $1$ compressor, $5$ passive pipelines, and $3$ gas loads. In the figure, $G, L, pl, W, C$, and $gl$ are used with subscripts to denote the power generators, power lines, power loads, gas wells, compressors, and gas loads, respectively. The targets of the defender and attacker are the $6$ power lines and any other components of the system cannot be attacked. We consider that the attack and defense cost coefficients $C^u_l$ and $C^y_l$ are ones, respectively. There are $2$ contracts $h_1$ and $h_2$ for generator $G2$ and $G3$, respectively. To demonstrate the effectiveness and practicability of the proposed model, various tests are performed under different defense and attack budgets where the defense budget range is $[1, 5]$ and the attack range is $[1, 5$-defense budget]. To make it easy, D and A are paired with numbers to denote the defense and attack budgets. For example, “D1A2” means the defense budget is $1$ and the attack budget is $2$.

    \begin{figure}[!ht]
        \centering
            \includegraphics[width=8cm]{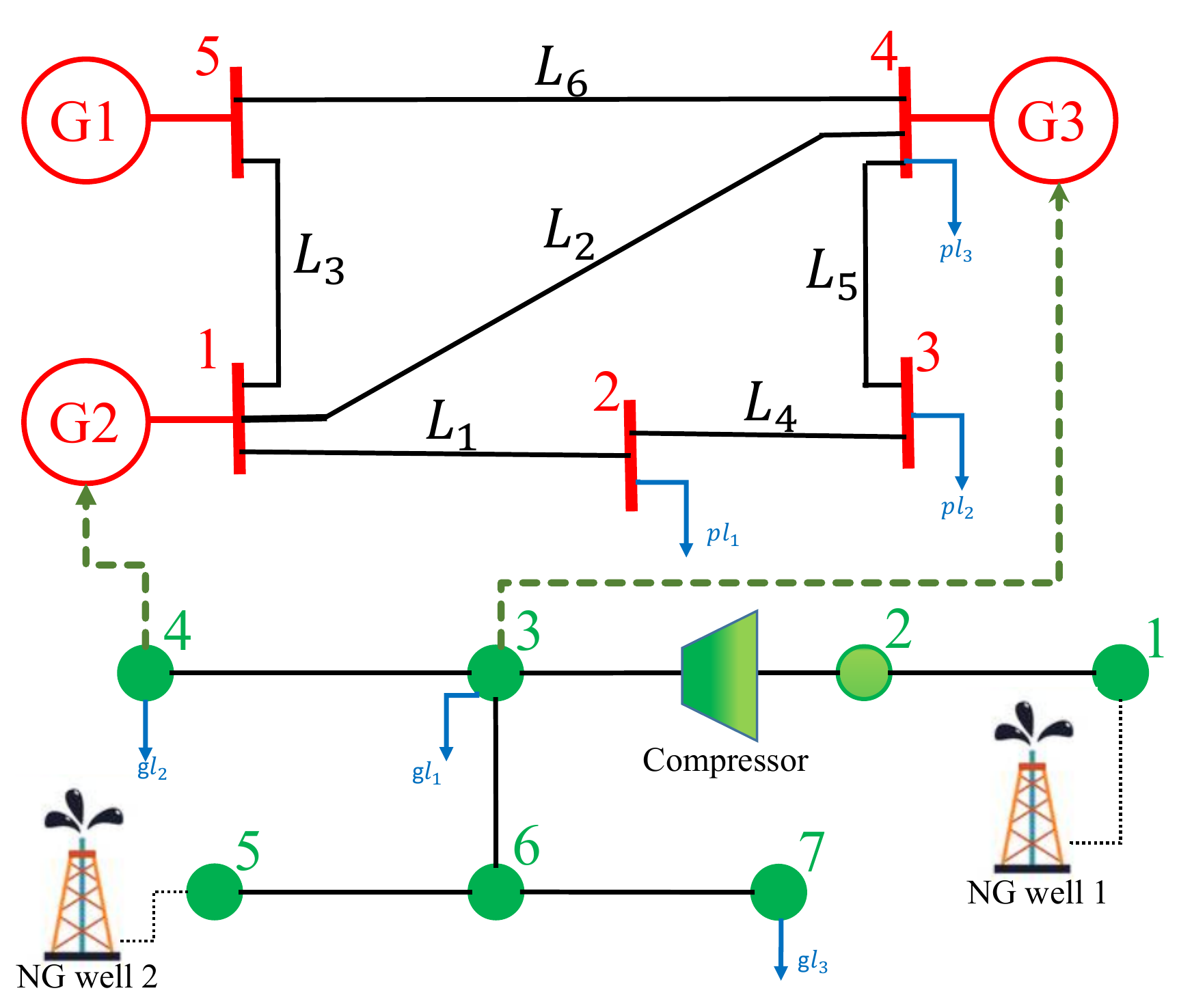} 
            \caption{Topology of the test system}   \label{fig:Ch4Topology1}
    \end{figure}

\subsection{Comparison with the Co-optimization Models } \label{sec:Ch4ResPGCO}
    To demonstrate the importance of optimizing the gas contracts for improving the power system resilience, a comparison in terms of economics and security is performed with IEGS models, which do not consider contracts between utilities and gas suppliers. Different cases are evaluated based on different defense and attack budget combinations. The following two models are compared;
    \begin{enumerate}
      \item The IEGS model represents the tri-level IEGS resilience models, which disregard the contracts during the IEGS resilient operation. In this study, to provide a fair comparison, the objective of this model is to minimize the operation and regulation costs only for the power system, while the cost of the reserved gas (day-ahead contracts) in the upper-level problem is not considered. The required gas, which would be consumed if the attack occurred, is calculated using \eqref{eq:Ch4reserv1}--\eqref{eq:Ch4reserv2}, where $p_{u,t}^*$  is the generated power from the GPU $u$ in the post-contingency stage under the worst-case attack. This value will be used for the real-time contracts, which are more expensive than day-ahead contracts.
          \begin{gather}
            \rho_{h,t}^+  = \max\Big\{0, \; \sum_{u\in \mathcal{U}_g(h)} \frac{\Phi}{\eta_u} (p_{u,t}^*-p_{u,t}^0)\Big\}, \;\; \forall h,t, \label{eq:Ch4reserv1}\\
            \rho_{h,t}^-  = \max\Big\{0, \; \sum_{u\in \mathcal{U}_g(h)} \frac{\Phi}{\eta_u} (p_{u,t}^0-p_{u,t}^*)\Big\}, \;\; \forall h,t.\label{eq:Ch4reserv2}
          \end{gather}

      \item Proposed GC model that considers the reserved gas in the upper-level problem.
    \end{enumerate}

        Each model provides an optimal defense strategy to minimize the power system disruption ($\Gamma^{post}$) under any malicious attack. The optimal protection strategy may be the same for both models because the cost of the reserved gas (if it is considered in the GC model) is small compared to the non-served power load penalty. Table~\ref{tab:Ch4PGCO} displays the economic performance of the proposed model for $3$ different cases. The numerical results for the two stages are shown, i.e., the pre-contingency and post-contingency stages. In the pre-contingency stage, each model provides the optimal operation cost ($\Gamma^{pre}$), which includes the cost of the day-ahead contracts for the reserved gas in the GC model. Therefore, the $\Gamma^{pre}$ is higher for the GC model than for the IEGS model, as shown in the table. Under the optimal defense and for generator dispatch, the inner C\&CG is applied to determine the worst attack strategy (limited by the attack budget of the case), which results in the re-dispatch of all generators to mitigate the power load shedding during the post-contingency stage. This re-dispatch indicates the changes in the gas consumed by the GPUs during the post-contingency stage. Therefore, in the GC model, the reserved gas is contracted and optimized as a day-ahead contract, consequently, the GPUs operate according to the contracted values and there is no additional gas cost, whereas, in the IEGS model, real-time contracts may be commissioned to adjust to the changes in the consumed gas after the generators have been re-dispatched. The total amount of firm gas used in the pre-contingency stage daily and the total reserved gas below/above the firm gas ($\rho_{h,t}^+ ,\; \rho_{h,t}^- $) for each contract are listed in the table. Based on the total cost, the proposed GC model provides a more economical and resilient operation than the IEGS model.

    \begin{table}[!htb]
        \caption{Comparison of the GC and IEGS models for three cases} \label{tab:Ch4PGCO}
        \centering
        \includegraphics[width=14.5cm]{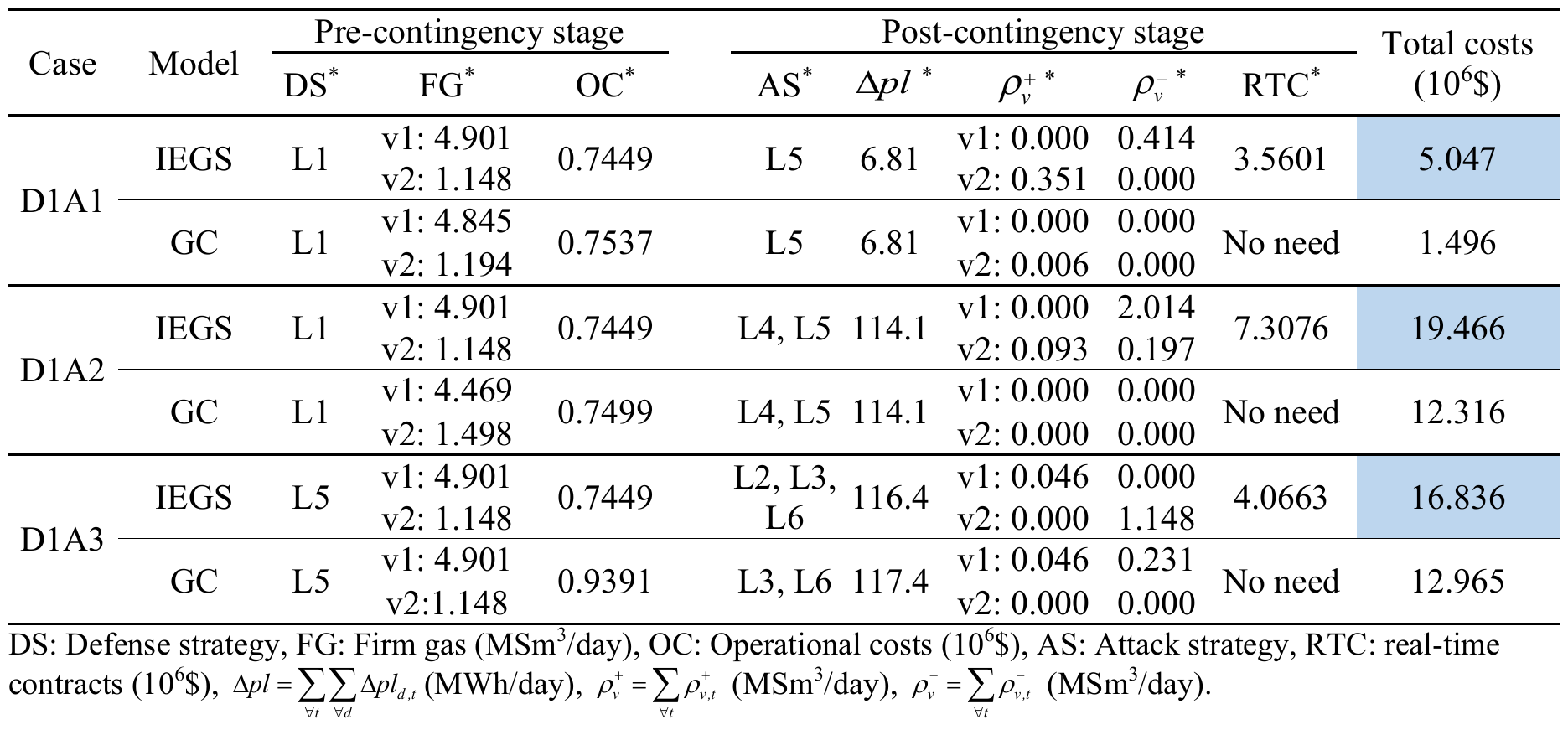}
    \end{table}

\begin{figure}[!ht]
        \centering
            \includegraphics[width=13cm]{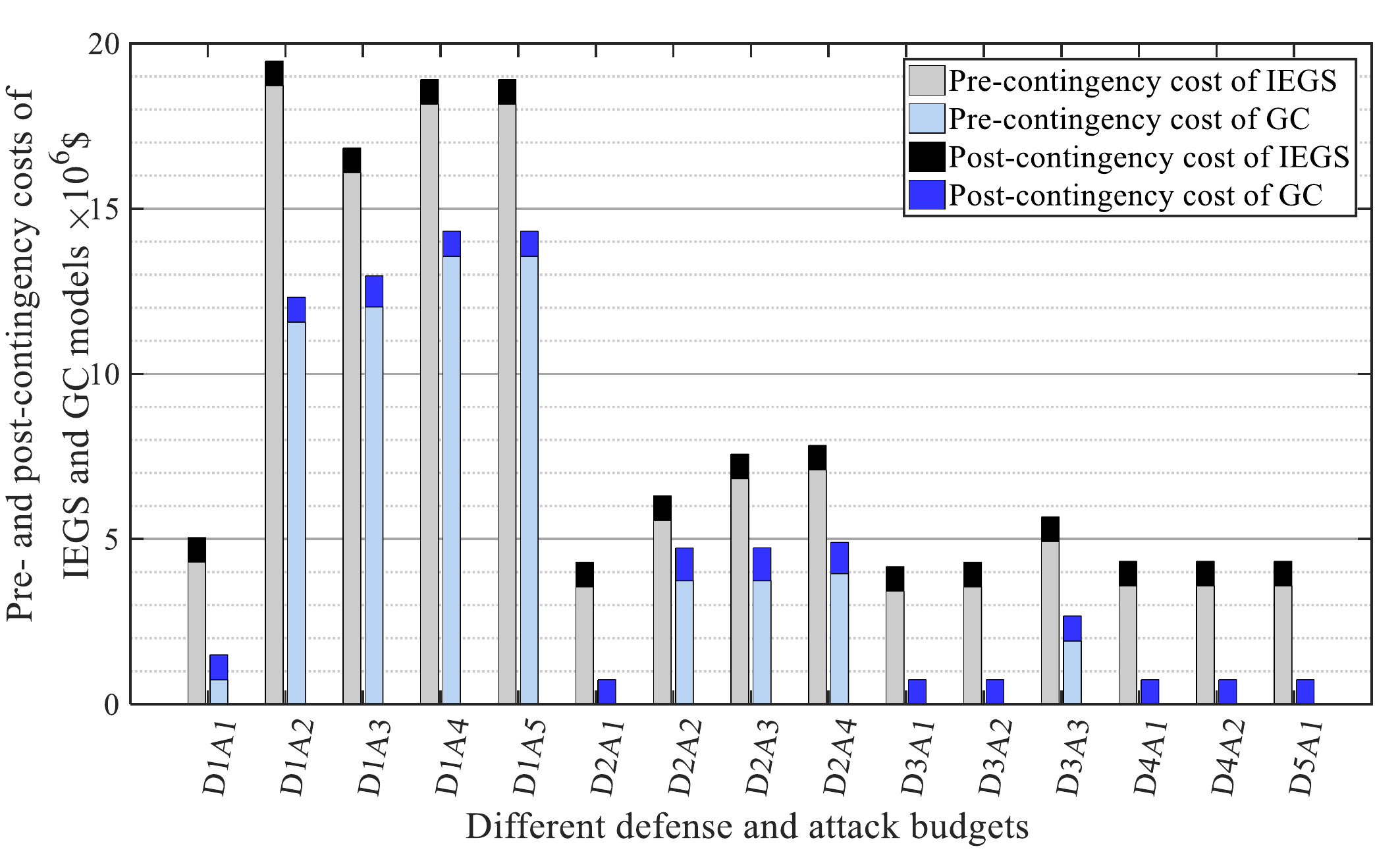} 
            \caption{Pre- and post-contingency costs of the IEGS and proposed GC models for $15$ cases.}   \label{fig:Ch4PGCO15Case}
    \end{figure}

        To generalize the recent conclusion, both models are applied to all $15$ cases with different attack and defense budgets. In Figure~\ref{fig:Ch4PGCO15Case}, the post-contingency cost includes the non-served power load penalty, the cost of the reserved power generated from non-GPUs, and the cost of real-time contracts. The pre-contingency cost includes the operational cost and the cost of the day-ahead contracts. The post-contingency cost and total cost of the IEGS model are much greater than those of the GC model for all $15$ cases.

\begin{figure}[!ht]
        \centering
            \includegraphics[width=13cm]{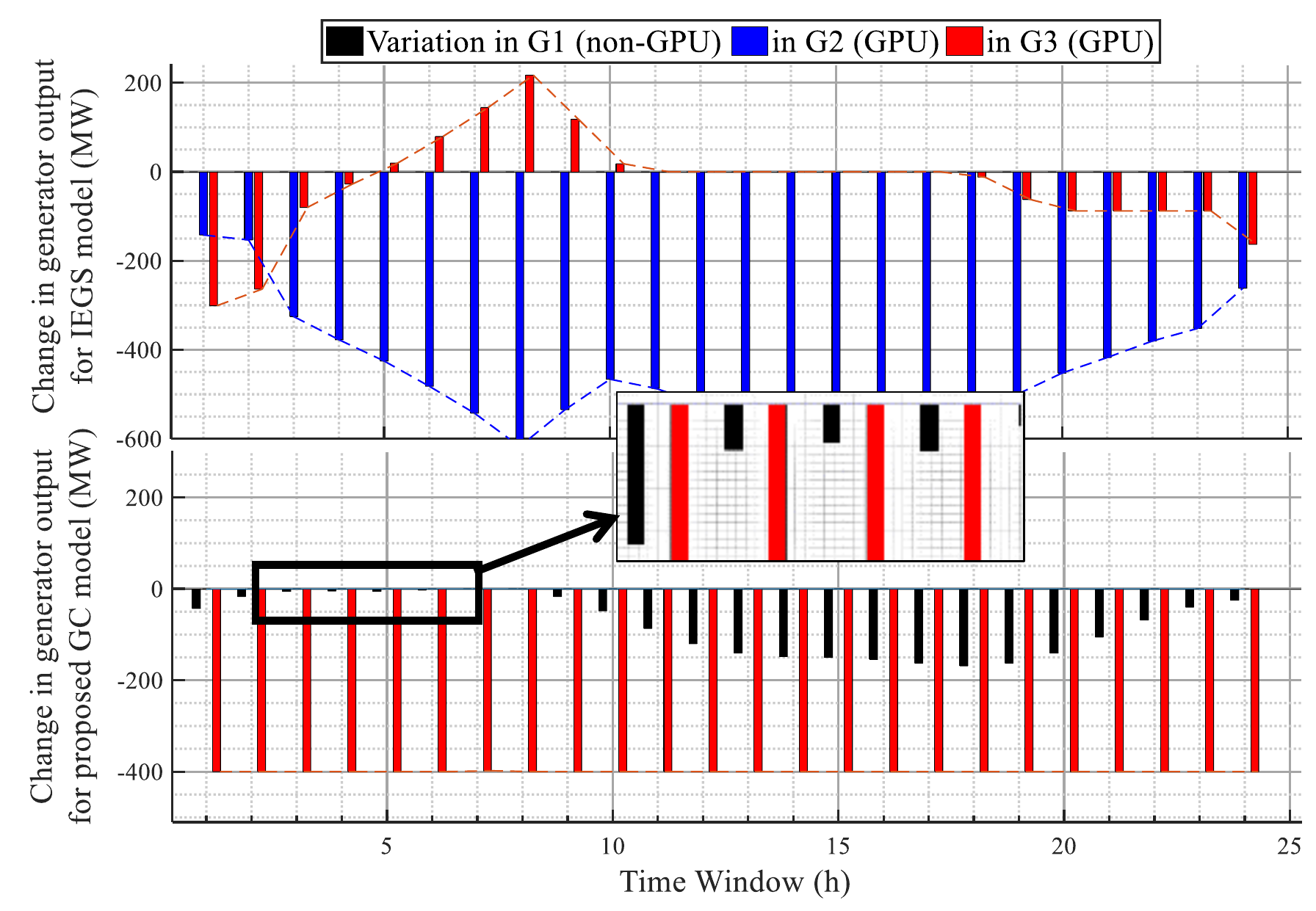} 
            \caption{Generator output adjustment before and after the contingency.}   \label{fig:Ch4PGCOConting}
    \end{figure}

        The adjustment of the generator output during the post-contingency stage is displayed in Figure~\ref{fig:Ch4PGCOConting} for the case “D1A2”. For the same defense and attack strategies (see Table~\ref{tab:Ch4PGCO} for this case), the IEGS model seeks to minimize the reserved power from non-GPUs and aims for minimum power shedding. Therefore, in the post-contingency stage, the $G1$ (non-GPU) schedule time is the same as in the pre-contingency schedule provided there are no reserves in the $G1$. The GPUs share the remaining power load (i.e., high gas reserves). In contrast, the GC model protects the outgoing feeder $L1$ of the $G2$ (GPU), which has the largest power capacity, to ensure that $pl_1$ and $pl_3$ are satisfied. Therefore, $G2$ will produce the same power and will not use any reserved gas during the contingency as shown in Figure~\ref{fig:Ch4PGCOConting}.

\subsection{Comparison with the Independent Operation Models} \label{sec:Ch4ResPSO}
        In this subsection, case studies are provided to demonstrate the effectiveness of considering the gas system operational constraints in the proposed model. Physical-based and economic-based comparisons are conducted between the proposed GC model and the IPS model, which disregards the interactions with the gas system in both the pre- and post-contingency stages. The two models are
        \begin{enumerate}
          \item The IPS model, which optimizes the resilient operation of the power system without considering the effect on the gas system. Therefore, \eqref{eq:Ch4Wellpre}--\eqref{eq:Ch4WeyPre} for pre- and post-contingency stages are not used in this model. To be fair in this comparison, the objective includes the day-ahead contracts. To check the gas system feasibility and security, the power system requirements, which are firm gas in the pre-contingency stage and reserved gas in the post-contingency stage, are considered as a gas load in the gas model.

          \item The proposed GC model.
        \end{enumerate}

        In this comparison, case “D1A2” is selected as a benchmark with different levels of gas load stress. It is shown in Table~\ref{tab:Ch4PSOTab} that each model identifies the best defense strategy and required firm gas during the pre-contingency operation and detects the worst attack strategy which affects the system with high regulation cost. The table displays the numerical results of the two models for three cases based on gas stress. In Case $\#1$ (there is no additional stress), the gas system is able to supply gas according to the power system requirements, therefore, the two models are feasible and results in the same total cost. In Case $\#2$, increasing $gl_1$ results in a stressed gas system, especially if this load is connected to node $3$, which supplies the largest GPU $G2$. The proposed model considers all physical constraints of the gas system, therefore, it optimizes the required gas in the pre- and post-contingency stages according to the gas system’s ability and feasibility unlike the IPS model, which seeks to provide the minimum total costs and disregards the interactions. Similarly, in Case $\#3$, although the gas load is redistributed without increasing the gas load, the IPS model fails to find the suitable gas production scheduling based on the requirements of the power system. In addition, the IPS model may provide an incorrect protection strategy against $N-k$ contingencies, particularly in large power systems.

    \begin{table}[!htb]
        \caption{Computational results of the Case “D1A2” for the proposed GC model and the IPS model} \label{tab:Ch4PSOTab}
        \centering
        \includegraphics[width=14.5cm]{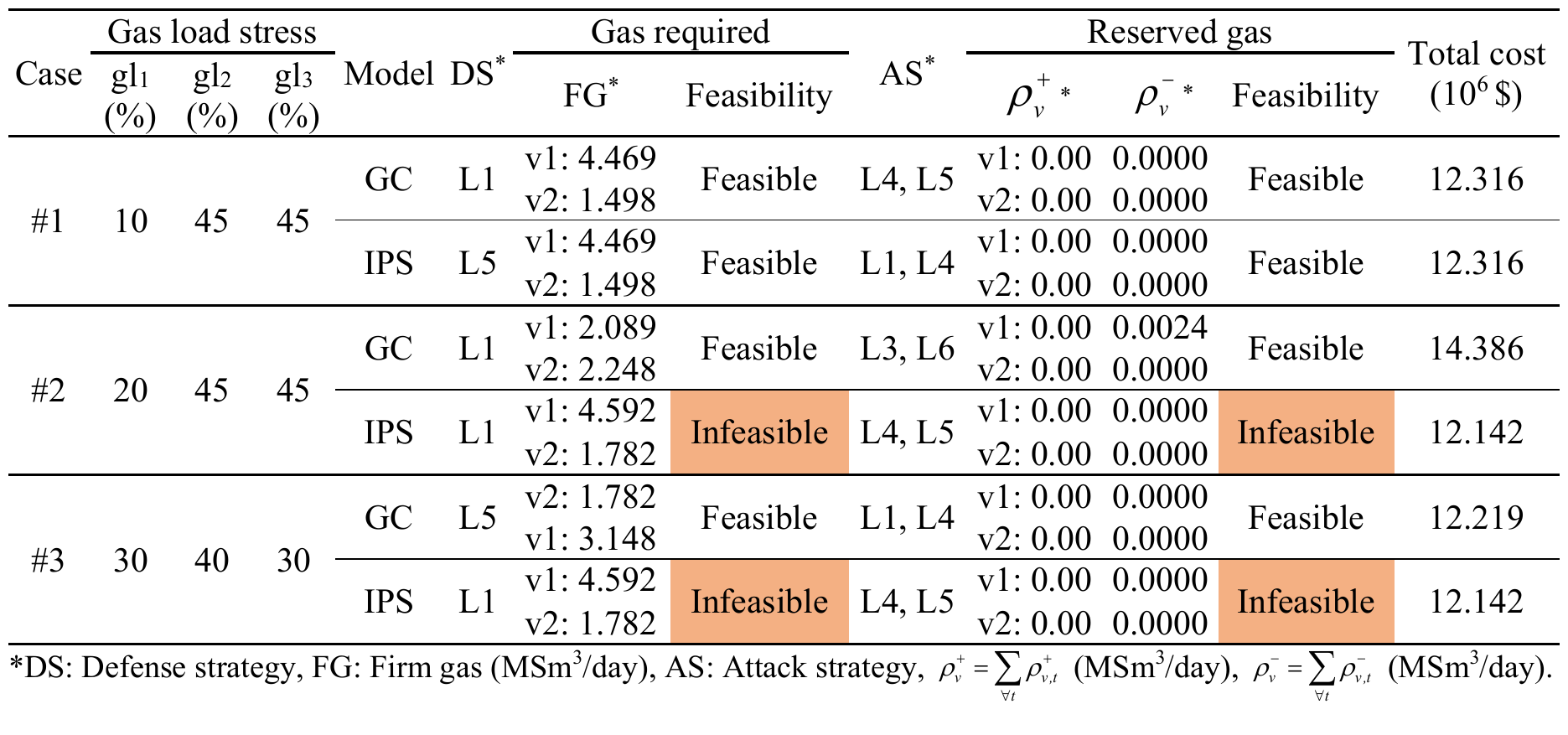}
    \end{table}

\begin{figure}[!ht]
        \centering
            \includegraphics[width=13cm]{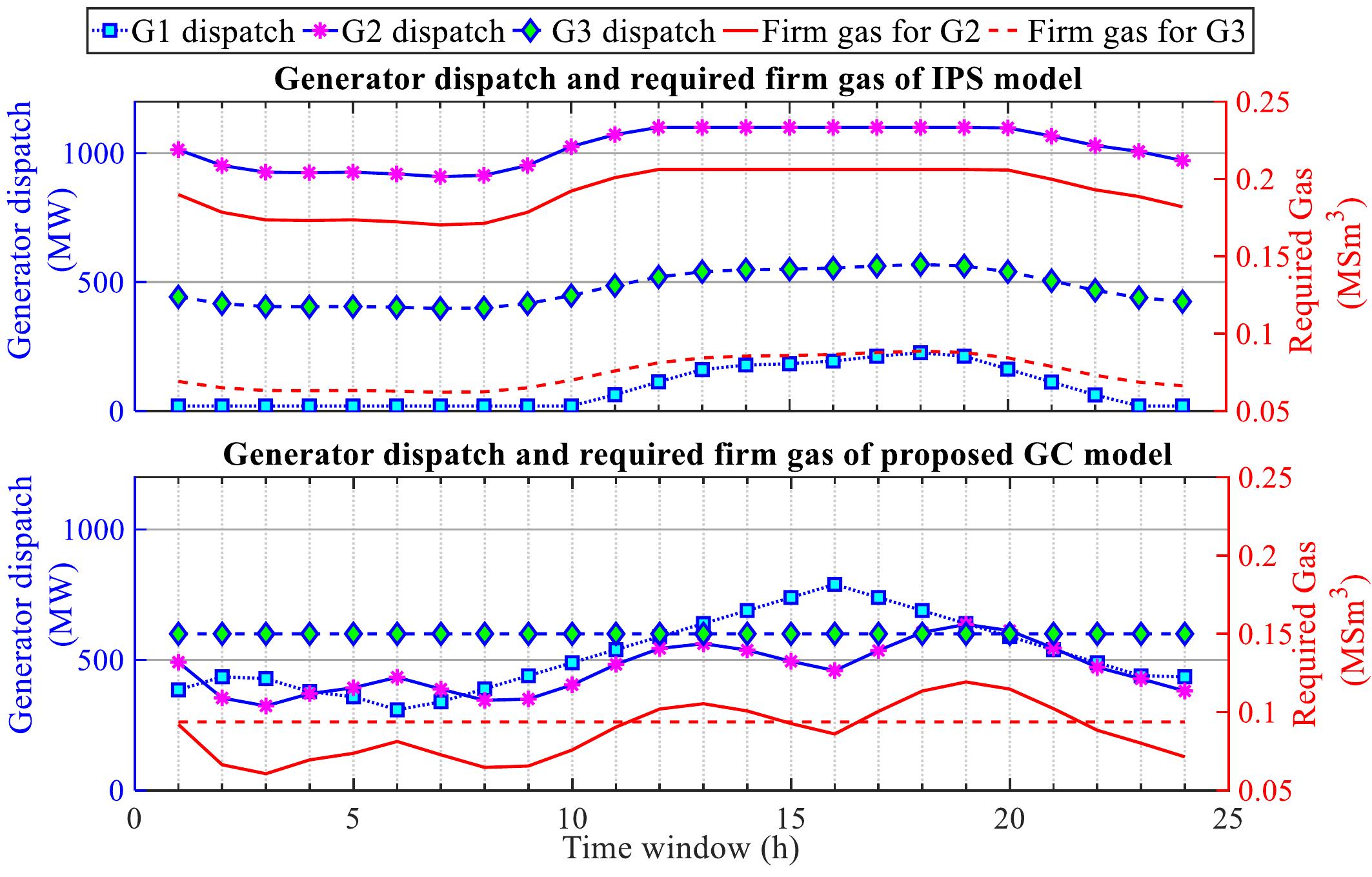} 
            \caption{Generator output and the required firm gas in the pre-contingency stage for the IPS model and the proposed GC model for the Case “D1A2” with gas load stress $\#2$.}   \label{fig:Ch4PSODispatch}
    \end{figure}

\begin{figure}[!ht]
        \centering
            \includegraphics[width=15.5cm]{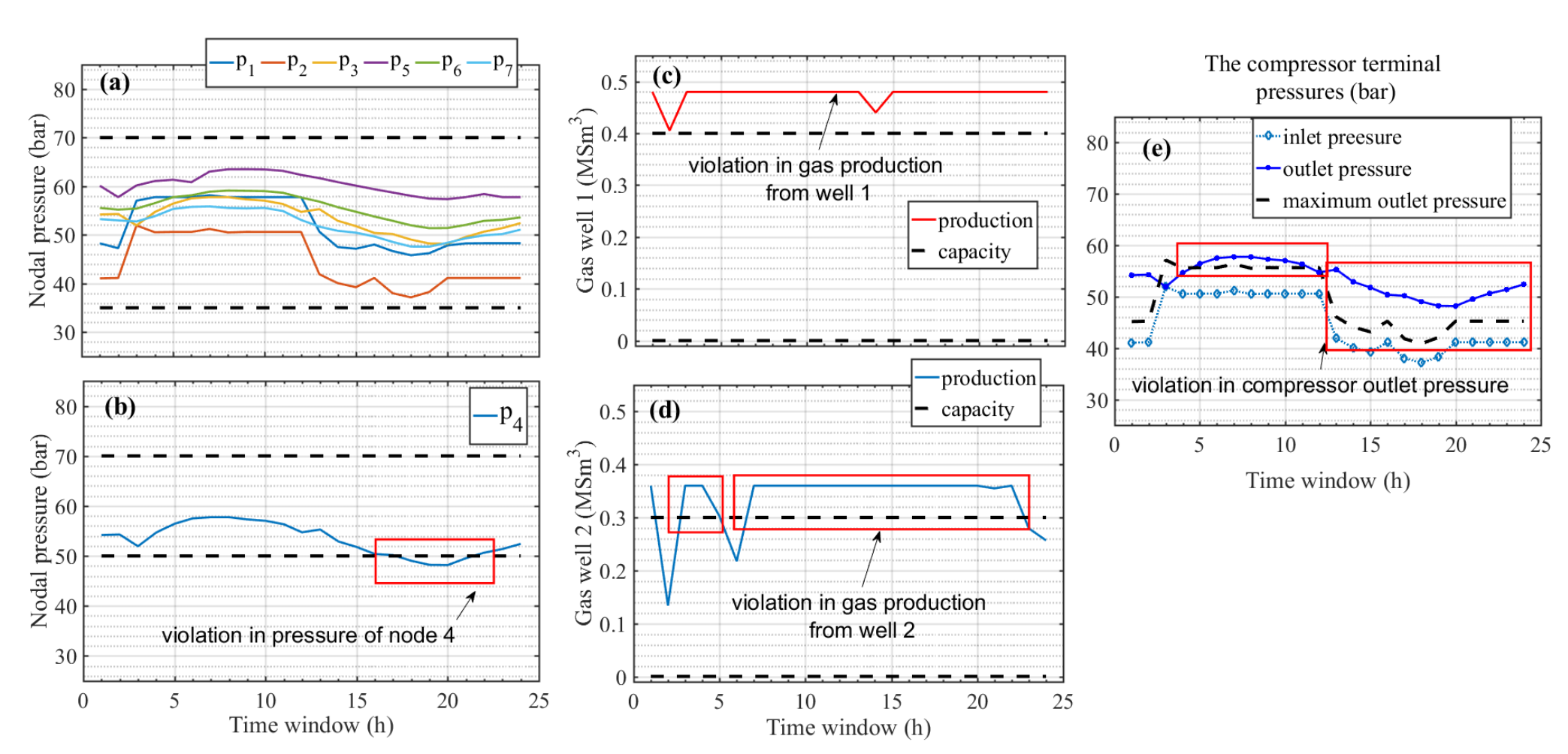} 
            \caption{Physical violations in the IPS model for Case “D1A2” and gas stress $\#2$; (a) gas pressure of all nodes (except node $4$) and the boundaries, (b) gas pressure at node $4$ and the boundaries, (c) gas production from well $1$ and the capacities, (d) gas production from well $2$ and the capacities, (e) inlet/outlet pressures of the compressor.}   \label{fig:Ch4PSOPhys}
    \end{figure}

        Figure~\ref{fig:Ch4PSODispatch} shows the generator dispatch in the pre-contingency stage and the required firm gas for Case $\#2$. The power system requirements in the IPS model exceed \SI{0.2}{\mega S \cubic\metre}, which results in an infeasible gas system. In contrast, the proposed GC model results in a feasible and economic generator dispatch of \SI{0.12}{\mega S \cubic\metre} as a maximum of the firm gas by increasing the power generated from the non-GPU $G1$. Based on the power system requirements for the gas stress in this case, which results in an infeasible gas system, the gas well capacities, nodal pressure boundaries, and compressor pressure capacity are relaxed by replacing \eqref{eq:Ch4Wellpre}--\eqref{eq:Ch4Comp1Pre} with
        \begin{gather}
            \underline{F}_w (1-\triangle{f}_{w,t})\le {f}_{w,t}^0 \le \overline{F}_w (1+\triangle{f}_{w,t}),\; \forall w,t, \label{eq:Ch4WellpreR} \\
            \underline{\Pi}_i (1-\triangle{\pi}_{i,t})\le {\pi}_{i,t}^0 \le \overline{\Pi}_i(1+\triangle{\pi}_{i,t}),\; \forall i,t, \label{eq:Ch4PressurePreR}\\
            {\pi}_{i,t}^0 \le {\pi}_{o,t}^0 \le (\triangle\gamma_c+1) {\pi}_{i,t}^0, \forall c,t, (i,o) \in c   \label{eq:Ch4Comp1PreR}
        \end{gather}
        where $\triangle{f}_{w,t},\; \triangle{\pi}_{i,t}$ and $\triangle\gamma_c$ are solution tolerances for gas production, nodal pressure and compressor pressure ratio, respectively. In fact, the optimal gas flow and production scheduling is strongly affected by these tolerances. We applied many tests based on the values of these parameters which are gradually increased by $5\%$ step from zero to achieve the gas system feasibility. The minimum values are $20\%, 5\%$, and $25\%$ for $\triangle{f}_{w,t},\; \triangle{\pi}_{i,t}$ and $\triangle\gamma_c$, respectively, to provide a feasible gas flow. In Figure~\ref{fig:Ch4PSOPhys}, the physical violations are plotted for the selected case.

\subsection{Significance of Considering Over-generation}
        The consideration of OG in the proposed model is important for the deployment of the defense resources and to optimize the reserved gas of all GPUs to defend against any possible attack strategy. As mentioned before, the optimal protection, gas contracts, and generator dispatch are affected by all attack strategies, which are generated in previous iterations of the NC\&CG algorithm. If some of these attack strategies are not detected, the algorithm provides suboptimal decisions. In the proposed methodology, the inner C\&CG uses strong duality, which creates a new variable $\bm{\lambda}^r$ in each iteration that is independent of $\hat{\bm{w}},\;\hat{\bm{y}},\;\bm{u}$ and $\bm{z}$. This situation always provides a feasible solution for the middle-level problem and ignores the decision from the upper-level problem (i.e., $\hat{\bm{w}},\;\hat{\bm{y}}$), whereas $\bm{u}$ is restricted only by the attacker budget and it is not restricted by $\hat{\bm{w}},\;\hat{\bm{y}},\;\bm{\lambda}$ and $\bm{z}$  (i.e., it is not affected by the gas contracts). As a result, the \textbf{MPs} may provide $\bm{u}$, which leads to the shutdown of power plants or forces a GPU to consume below/above the allowable reserved gas. Therefore, if the OG is not considered, the \textbf{SPs} will be infeasible. Therefore, \eqref{eq:Ch4OGR} is added to the \textbf{MPs} \eqref{eq:Ch4MPs} to provide a feasible attack strategy only if OG is not considered.
        \begin{gather}\label{eq:Ch4OGR}
          \bm{E}\bm{x} + \bm{G}\bm{z} \ge \bm{F} - \bm{Q} \bm{h} - \bm{D}\hat{\bm{w}}.
        \end{gather}

        Table~\ref{tab:Ch4OGTab1} lists the results for four cases based on different defense and attack budgets with and without OG. In case “D1A1”, the OG has no effect because the worst attack strategies are detected and defended. Therefore, the optimal defense strategy and reserved gas are same with and without OG. In case “D2A2”, however, the defense plans are the same with and without OG and the optimal reserves of gas are not the same because using the OG constraints provides the possibility of new attacks that violate the gas contracts. Consequently, the upper-level problem detects these attacks and adjusts the reserve gas. As shown in Table~\ref{tab:Ch4OGTab1}, for the last three cases, the models that disregard the OG issues identify the optimal defense, gas contracts, and unit dispatch when there are no system disruptions (zero regulation cost $\Gamma^{pre}$, zero load shedding) under any malicious attacks.

    \begin{table}[!htb]
        \caption{Computational results for four cases with/without the consideration of over-generation} \label{tab:Ch4OGTab1}
        \centering
        \includegraphics[width=14cm]{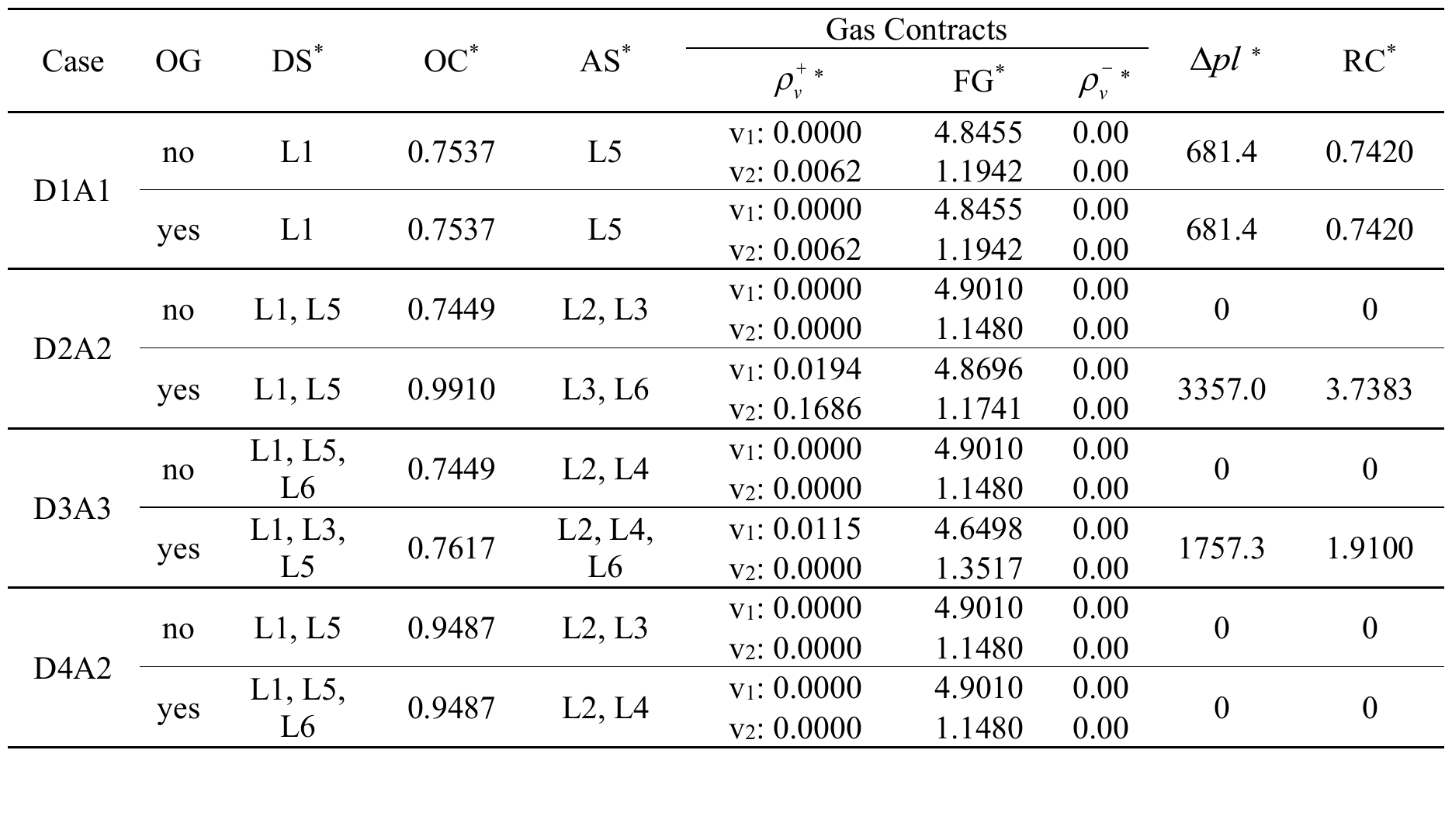}
    \end{table}

        Table~\ref{tab:Ch4OGTab2} lists the OG states for the three generators and three cases. In the first Case “D2A1”, no OG is detected (all binaries are ones); therefore the optimal decision-making is achieved without considering OG, as shown in Table~\ref{tab:Ch4OGTab1}. In Case “D2A2”, the worst attack strategy is detected, i.e., the violation of the $G1$ output below the minimum capacity. Therefore, OG only occurs for this generator. In the third Case “D2A3”, a larger attack budget provides different violations for the three generators.

    \begin{table}[!htb]
        \caption{Hourly over-generation states for the cases “D2A1”, “D2A2”, and “D2A3”} \label{tab:Ch4OGTab2}
        \centering
        \includegraphics[width=9.5cm]{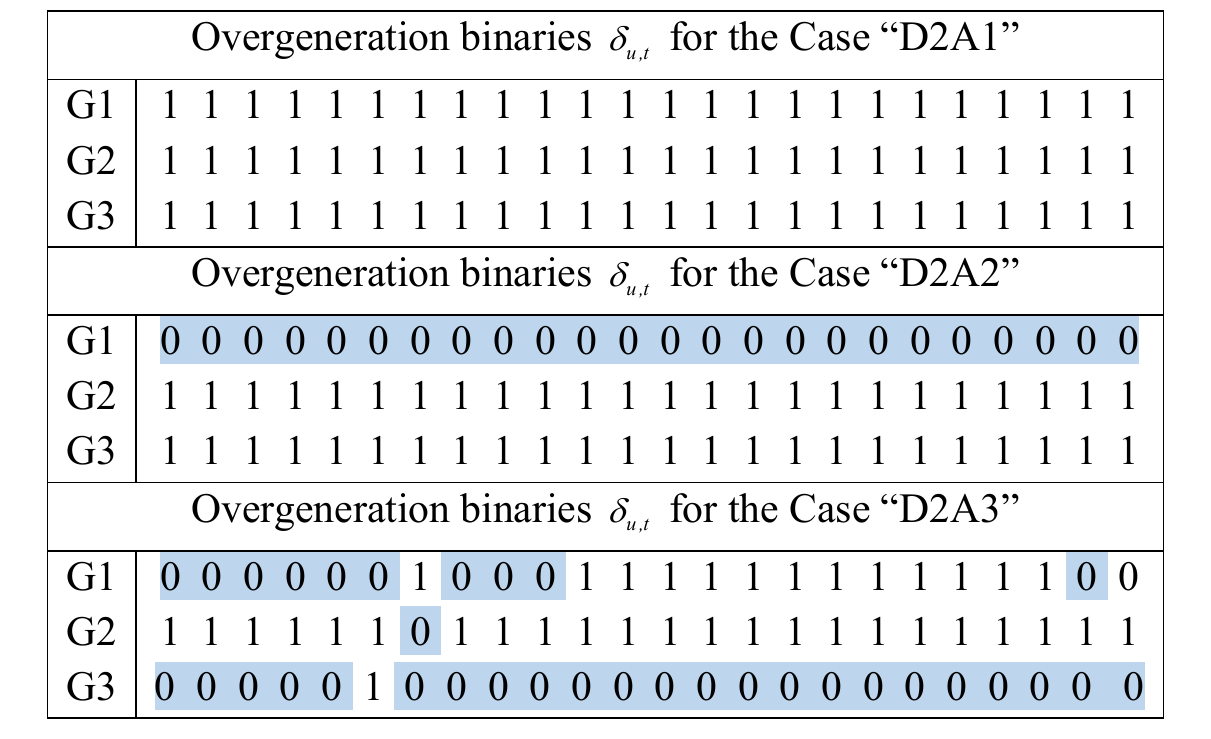}
    \end{table}

\begin{figure}[!ht]
        \centering
            \includegraphics[width=14cm]{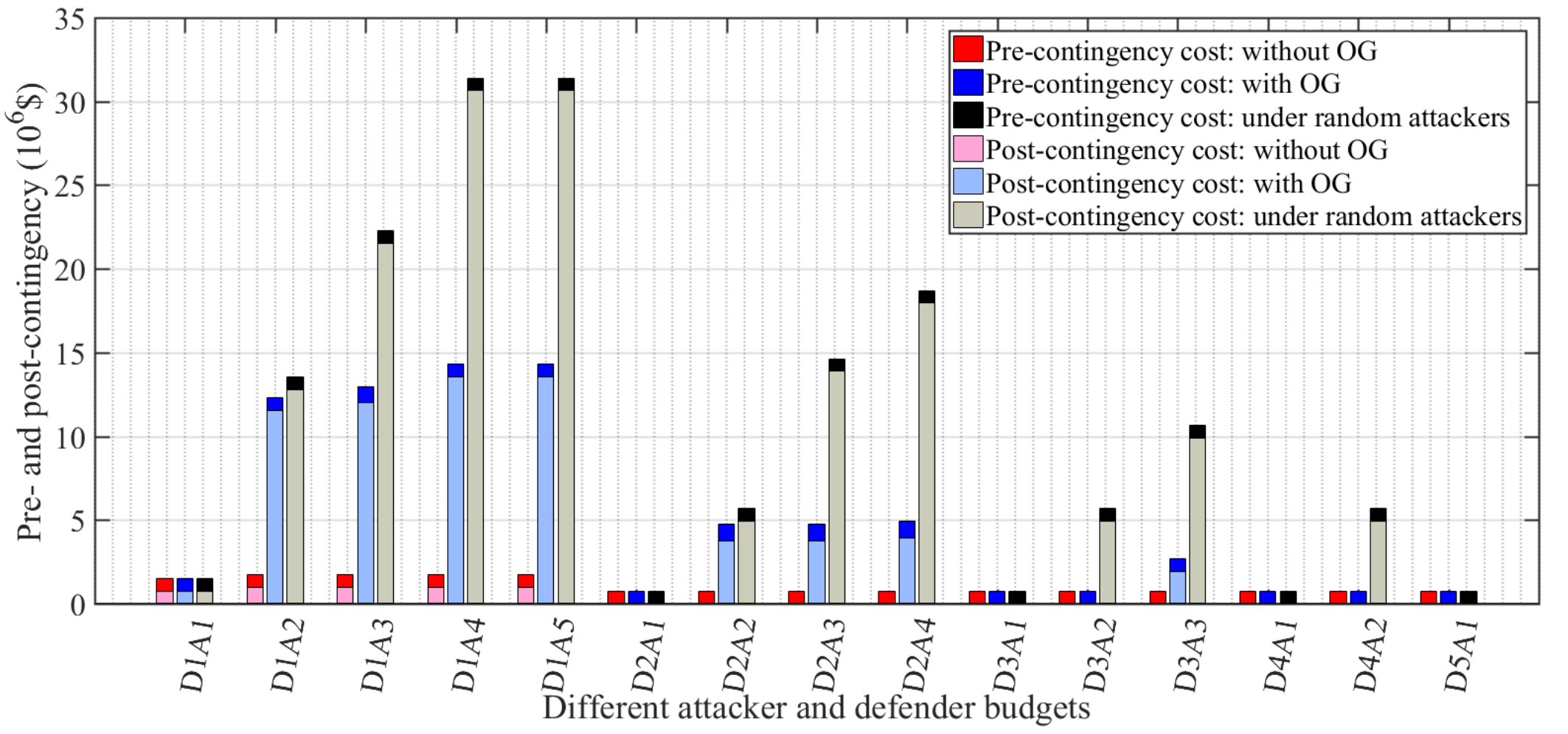} 
            \caption{Economic performance of the proposed model with and without considering over-generation for different cases.}   \label{fig:Ch4OGFig}
    \end{figure}

        To fully demonstrate the benefits, Figure~\ref{fig:Ch4OGFig} displays the operating costs in the pre-contingency stage $\Gamma^{pre}$ and the regulation costs in the post-contingency stage $\Gamma^{post}$ for the $15$ possible cases with and without OG. It is evident that the operating costs of the cases with OG are not always less than those of the cases without OG. However, when random attack strategies are included in the lower level problem to determine the worst attack using the same decision ($\hat{\bm{w}},\;\hat{\bm{y}},\;\hat{\bm{\alpha}}$) obtained without OG, the regulation costs are much greater than for the OG cases. In the aforementioned situation of a random attack strategy, the lower level problem is always feasible by forcing to shut down a generator if (i) the required power is lower than its minimum capacity, (ii) ramping up/down of the required power exceeds its limits, (iii) for GPUs, the consumed gas exceeds the reserved gas.

\subsection{Significance of Considering the Dynamic Gas Model}
        As shown in subsections \ref{sec:Ch4ResPGCO} and \ref{sec:Ch4ResPSO}, it is very important to determine the optimal contracts while considering the gas interactions. The main purpose of the proposed GC model is to find the best protection plan with the optimal reserved gs under malicious attacks. Therefore, any assumption in the gas system model that may provide suboptimal or infeasible decisions for power system operators has to be known. The consideration of the line pack and traveling velocity of the gas has an effect on the proposed model. Therefore, we have
        \begin{enumerate}
          \item Steady-state model: The gas model in the pre- and post-contingency stages is in a steady state without considering the line pack. Please, refer to section~\ref{sec:CH2SSS} for more details.
          \item Dynamic-state model: the dynamic gas system is modeled by considering the line pack as formulated in the proposed model.
        \end{enumerate}

        Table~\ref{tab:Ch4SSModel} lists the defense strategies, optimal gas contracts, worst attack strategy, and the associated total cost for four cases with different gas loads. Under normal loading, the steady-state model cannot provide the power system with the required gas during the pre-contingency stage (there is no power load-shedding) and it provides an infeasible solution, whereas the dynamic model does provide the required amount of gas. Similarly, in the second case, even though the gas load is lower, the steady-state model results in an infeasible solution. Therefore, we present a feasible gas load distribution with different defense and attack budgets “D1A3” and “D2A3”. If the line pack is not considered as in the steady-state model, the model may fail to find the best defense strategy, as is the case for the dynamic model in case “D1A3”. Another drawback occurs if the steady-state model is adopted, i.e., the contract parameter values. This model provides different amounts of firm and reserved gas for both the pre- and post-contingency stages, resulting in costly contracts, as shown for the last two cases. It is clear that the dynamic model offers more flexibility because it handles the bidirectional gas flows and gas line pack. Therefore, it is important to consider the dynamic gas model during the optimization of the power system’s resilient operation.

    \begin{table}[!htb]
        \caption{Computational results for four cases with/without the consideration of over-generation} \label{tab:Ch4SSModel}
        \centering
        \includegraphics[width=15cm]{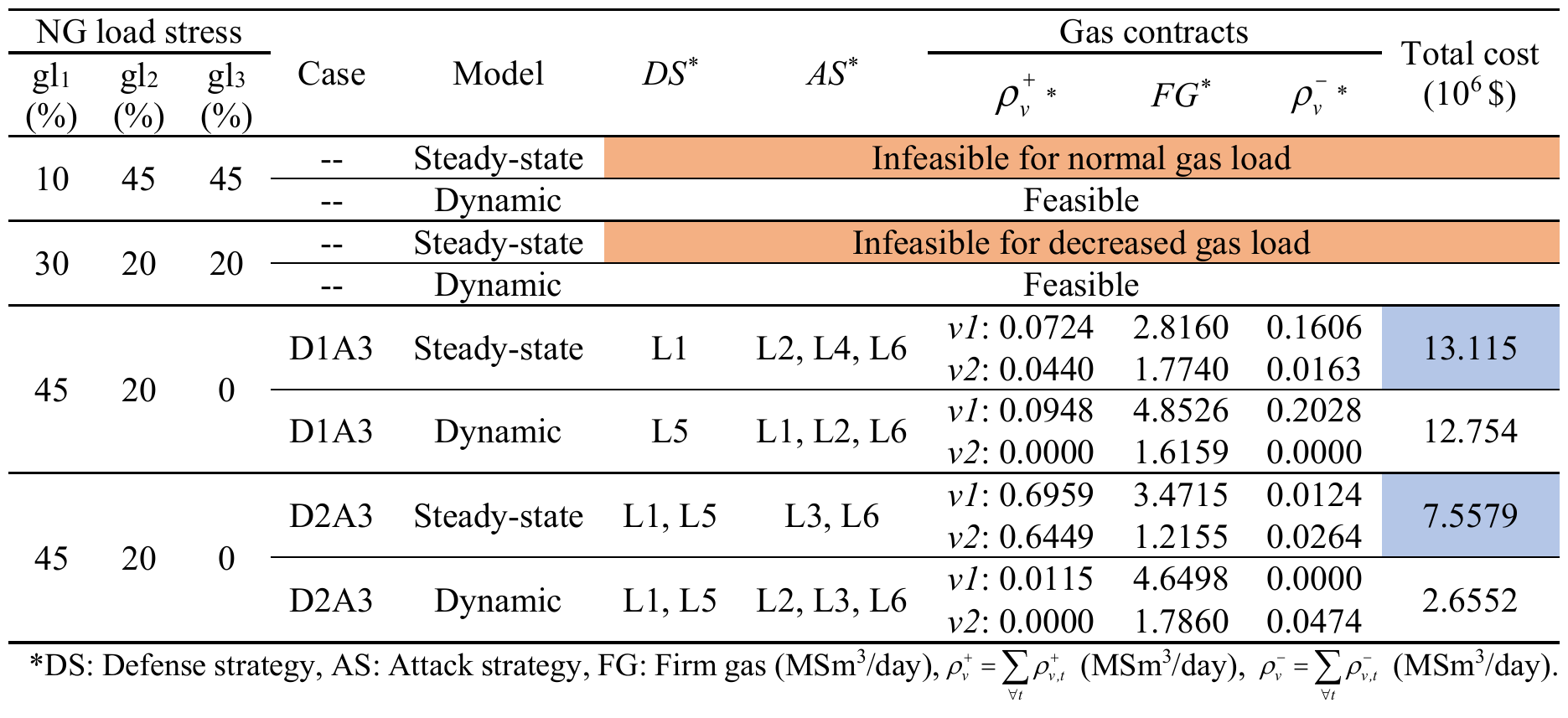}
    \end{table}

\subsection{Impacts of Defense and Attack Budgets}
        In general, the planning and protection of power lines are always beneficial \cite{yuan2014optimal}. Table~\ref{tab:Ch4DefAtt} lists the total cost for all possible cases ($15$ cases). Using only one defender reduces the total cost by about half if this defender is not used. The results show that the power system operator can choose a suitable protection design based on the expected attack and the trade-off between benefits and costs. For example, if the attack budget is one, protecting one power line reduces the resilient operation cost to below half and there is no need to defend more than two power lines. Similarly, defending $3$ power lines is sufficient to provide full protection if the attack budget equals $2$. To provide full protection against any attack, it is not necessary to protect all power lines because full protection is achieved by hardening only $4$ power lines. The results demonstrate that we can identify the suitable defense budget based on the expected attack budgets, especially for large power systems.

    \begin{table}[!htb]
        \caption{Total cost during the pre- and post-contingency stages for all cases based on different defense and attack budget combinations} \label{tab:Ch4DefAtt}
        \centering
        \includegraphics[width=10cm]{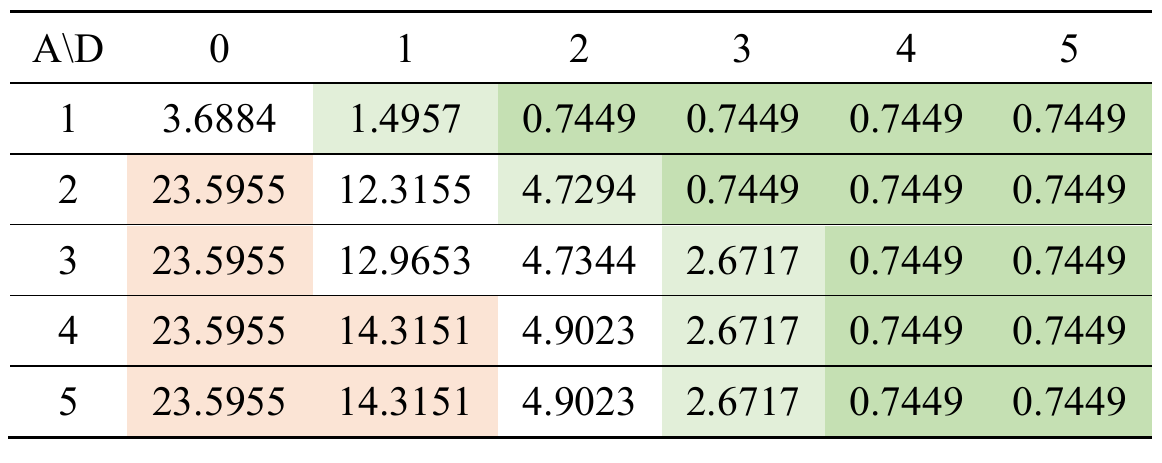}
    \end{table}

\begin{figure}[!ht]
        \centering
            \includegraphics[width=13.5cm]{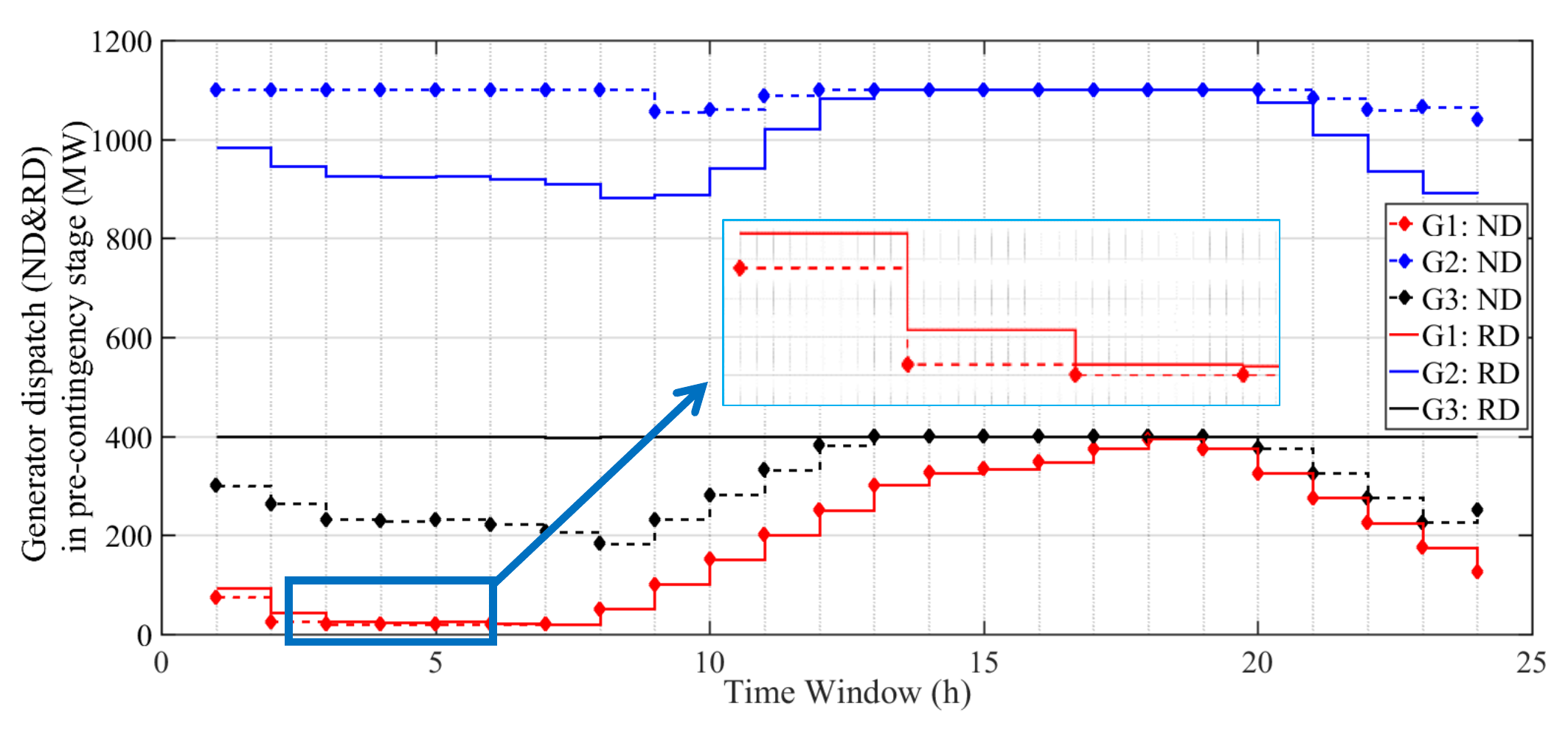} 
            \caption{Time schedules in the pre-contingency stage for normal and resilient dispatch for the Case “D1A2” in \textbf{TS-I}.}   \label{fig:Ch4DefED}
    \end{figure}

        Adding the UC to the upper level provides very important benefits during optimization \cite{chen2016identifying}. However, the proposed model uses only economic dispatch in the upper level and it considers that the UC is already known. In the upper curves in Figure~\ref{fig:Ch4DefED}, the dashed lines represent a normal dispatch (ND), which is achieved by optimizing the UC without considering the power system uncertainties (deterministic model). The solid lines represent resilient dispatch (RD), which is optimized in the upper level of the proposed model (robust model) for the Case “D1A2”. Unit re-dispatch under the worst attack scenario ($L4, L5$) is used in the lower curves. The change in the time schedules during the normal state is to decrease the value of the reserved gas from all units so that it can be used under any attack strategy, even if RD is more costly than ND. The normal operating costs for resilient dispatch (\$$749928.8$) are higher than those for normal dispatch (\$$744889.8$) by only $0.6\%$.

\subsection{Scalability Tests of the Proposed Algorithm} \label{sec:Ch4LargeScale}
    In this section, we select an IEEE $39$-Bus power system coupled with a $20$-Node gas system (modified high-calorific gas network), denoted as \textbf{TS-I}, as well as IEEE $118$-Bus power system coupled with the same gas system, denoted as \textbf{TS-II}, as large-scale power-gas test systems. The proposed model and algorithm is used to determine their scalability. The test systems topology, parameters, and UC are described in further detail in Appendix~\ref{App:5Bus} and Appendix~\ref{App:sevenNode}. The load-shedding penalty price is set at \$$1000$/MWh and $\overline{M}$ used for the linearization in the inner C\&CG algorithm is set at $10^4$. The convergence tolerance values and  are $0.1\%$. The number of segments used in the linearization of the Weymouth equation is $4$ for both the gas flow and gas pressure. We consider a problem with $2$ periods from $t = 2$ to $t = 3$ as the target slot \cite{wang2016robust}.


    \begin{figure}[!ht]
        \centering
            \includegraphics[width=15.5cm]{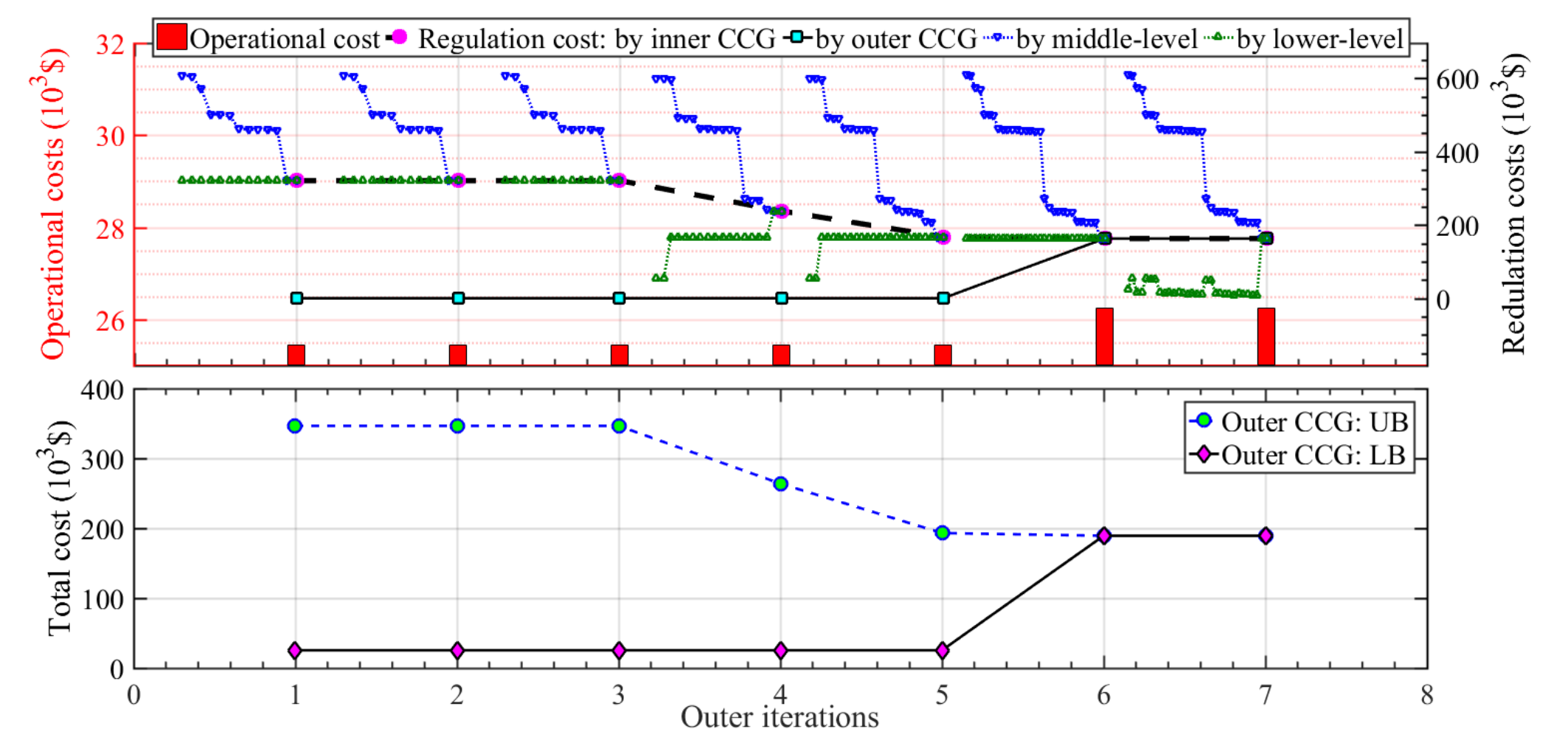} 
            \caption{Tracing the operational, regulation, and total costs during the inner and outer iterations for case “D3A2” in  \textbf{TS-I}. }   \label{fig:Ch4D3A2}
    \end{figure}

    Figure~\ref{fig:Ch4D3A2} displays the upper and lower boundaries of the inner and outer C\&CG algorithms, as well as the operational and regulation costs in the NC\&CG iterations for case “D3A2” in the \textbf{TS-I}. In the first outer iteration, the recourse action (lower-level problem) can alone reduce the system disruption from about \$$680,000$ to \$$320,000$ without the optimal decision-making in the upper level. Defending the most vulnerable components obtained from the first iteration of the outer C\&CG does not provide any advantages as shown in the second iteration (same regulation cost as the inner C\&CG). In the first $6$ outer iterations, the outer C\&CG seeks to maintain the operational cost at \$$25,460$ but the outer C\&CG gap is still high (see the lower graph in Figure~\ref{fig:Ch4D3A2}). Therefore, the operational cost increase to $26,250$ in the $7$\textsuperscript{th} outer iteration. The last outer iteration is executed to confirm that the optimal solution is achieved.

    Table~\ref{tab:Ch4ReTimees} lists the total computation time of the NC\&CG algorithm and the inner and outer C\&CG iteration numbers for different four Cases “D1A2”, “D1A2”, “D3A2”, and “D1A4” in the two test systems, namely \textbf{TS-I} for the IEEE $39$-Bus-$20$-Node  and \textbf{TS-II} for the IEEE $118$-Bus-$20$-Node systems. The computation time increases as the defender and attacker budgets increase. The outer C\&CG iteration number is influenced more by the defender budget than the attacker budget, whereas the inner C\&CG iteration number is affected only by the attacker budget.

    \begin{table}[!htb]
        \caption{Computation times for the selected four cases} \label{tab:Ch4ReTimees}
        \centering
        \includegraphics[width=12.5cm]{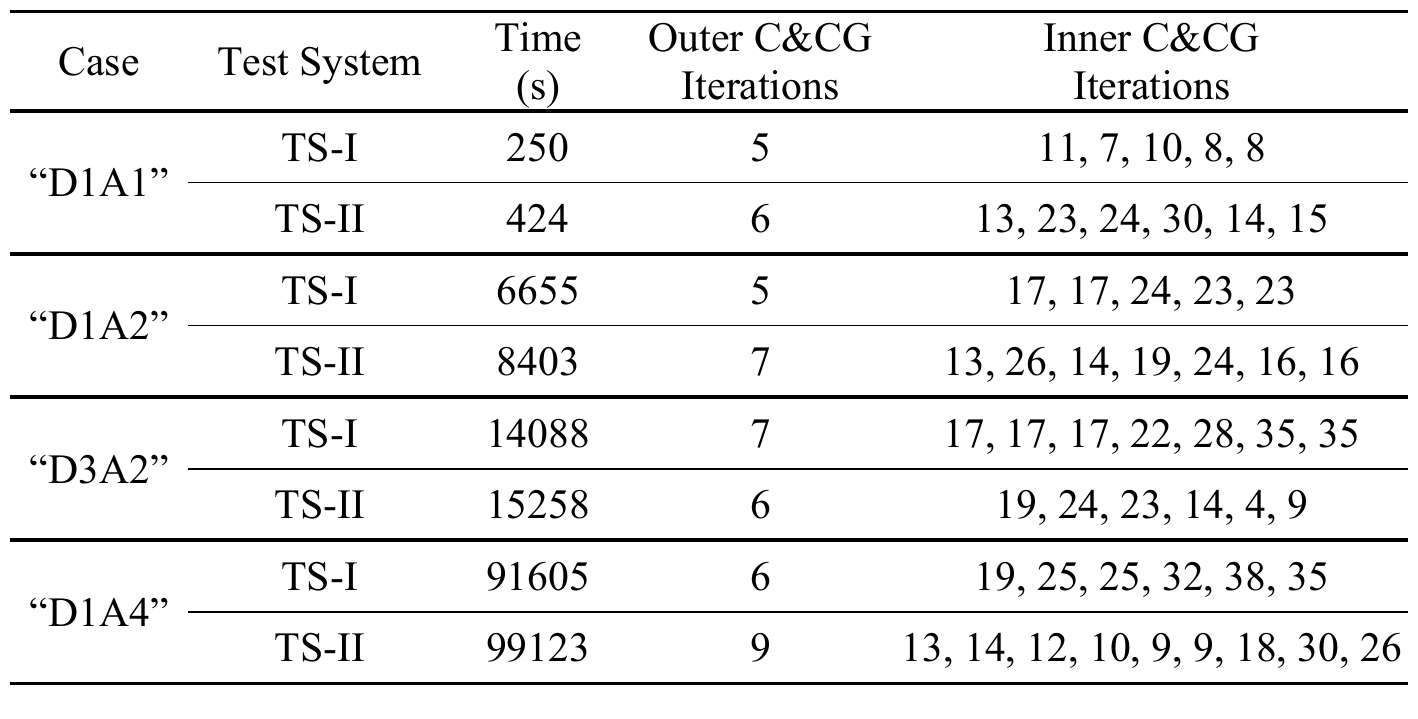}
    \end{table}

\begin{figure}[!ht]
        \centering
            \includegraphics[width=15cm]{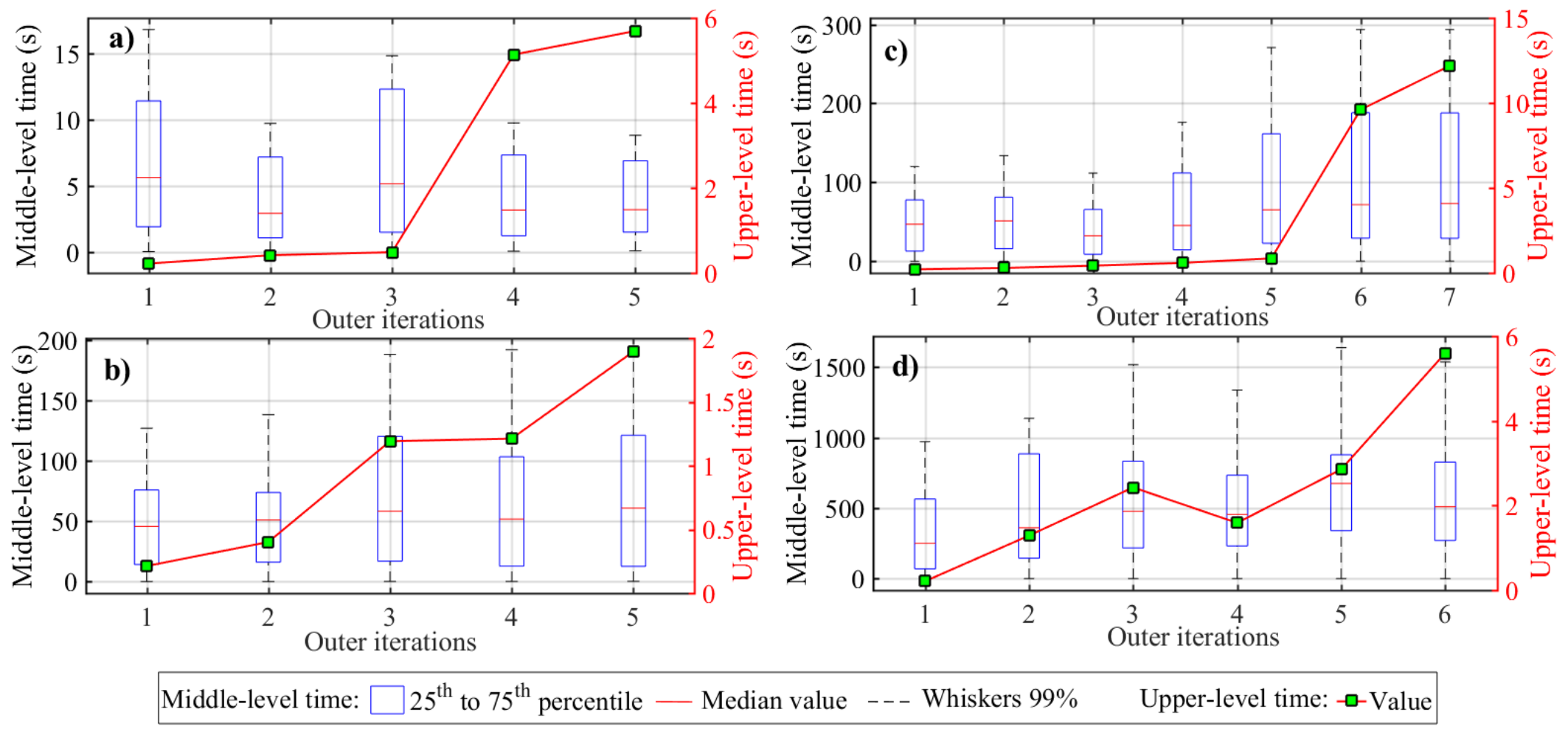} 
            \caption{Middle-and upper-level computational times for \textbf{TS-I}; (a) Case “D1A1”, (b) Case “D1A2”, (c) Case “D3A2”, (c) Case “D1A4”.}   \label{fig:Ch4AllTime}
    \end{figure}

    Figure~\ref{fig:Ch4AllTime} displays the computational time for middle- and upper-level problems for each outer iteration in case of \textbf{TS-I}. In case “D1A1”, the middle-level time is less than $18$s for all inner iterations whereas this time reaches $200, 300$, and $1500$s in Cases “D1A2”, “D3A2”, and “D1A4”, respectively, due to the increase in the defender budget. The upper-level time ranges from $0.2$s to $6$s with defender budget equals $1$ (Figure~\ref{fig:Ch4AllTime}.a, Figure~\ref{fig:Ch4AllTime}.b, Figure~\ref{fig:Ch4AllTime}.d), while it is doubled with defender budget equals $3$ (Figure~\ref{fig:Ch4AllTime}.c). The solver time of lower-level problem is small (less than $0.02$s for any iteration) compared with middle-and upper-level times, therefore it is not plotted in Figure~\ref{fig:Ch4AllTime}.

    Before providing our recommendations to address the scalability issue, we discuss the reasons for the high computation cost. The main reason for the long computation time is not the upper-level calculation (see Figure~\ref{fig:Ch4AllTime}) but rather the presence of the $\overline{M}$ constraints and the binary variables ($h_l, \;\forall l$) of the power line availability in the middle-level problem, which suffers from the additional primal cut constraints and variables. Therefore, the attack budget has a strong effect on the inner C\&CG execution time and consequently on the overall computation time. The binary variables resulting from the linearization of the Weymouth equation and the OG have little effect on the computation time because they are used as parameters in the middle-level problem. Therefore, the selection of a suitable value of $\overline{M}$ will reduce the computational cost \cite{wang2016robust}. This value depends on the maximum regulation cost under any possible attack strategy (limited by its budget). Table~\ref{tab:Ch4ReBigM} presents the computation time for Case “D1A1” with different $\overline{M}$ values. It should be kept in mind that very low $\overline{M}$ values do not result in a realistic evaluation of the attack decision in the middle-level problem as shown in Table~\ref{tab:Ch4ReBigM}. Based on the aforementioned discussion, we recommend the following assumptions and suggestions to reduce the computational burden:
    \begin{enumerate}
      \item Reduce the defender and attacker budget; this can be achieved by selecting the most frequent and vulnerable components based on the power system operators’ experience and by ensuring that healthy and protectable components cannot be attacked.
      \item Provide suitable defender and attacker strategies based on the operators’ experience rather than using arbitrary strategies.
      \item Select a suitable $\overline{M}$ value for the middle-level problem based on the expected regulation cost.
      \item Increase the inner and upper convergence tolerance values ($\varepsilon$) to obtain a balance between the accuracy and the computational burden.
      \item Reduce the number of binaries used in the linearization of the Weymouth equation. Or use a convexification method for Weymouth equation such as convex-concave procedure \cite{R35}.
    \end{enumerate}

    \begin{table}[!htb]
        \caption{Computation times for case “D2A1” with different $\overline{M}$ values} \label{tab:Ch4ReBigM}
        \centering
        \includegraphics[width=11cm]{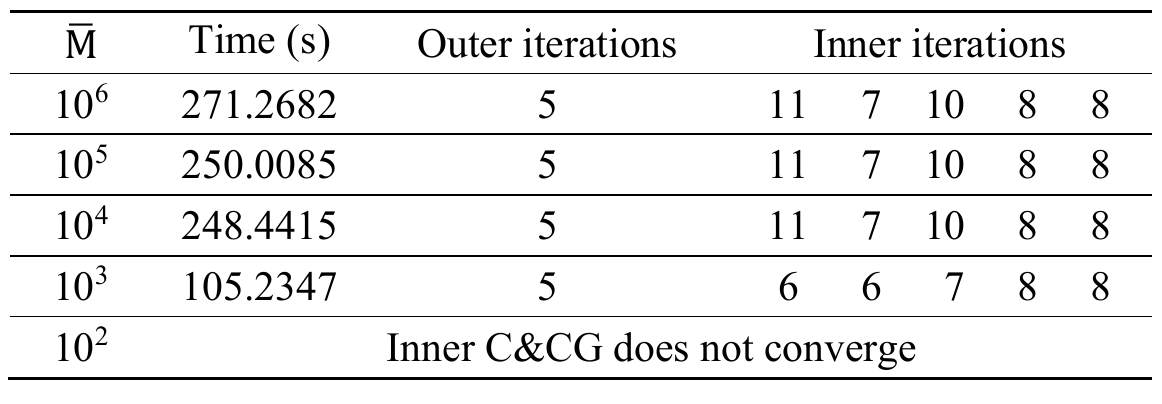}
    \end{table}

\section{Conclusions and Discussions} \label{sec:Ch4Conc}

    Natural disasters and malicious attacks can cause serious threats for electric infrastructures and affect the normal operation of power systems, especially under heavy loads. Due to their fast response, good regulation capacity, relatively high efficiency, and low generation costs, GPUs have been playing increasingly larger roles in the resilient operation of power system, such as quick power flow distribution adjustments in the pre-contingency stage and picking up important loads in the post-contingency stage. These actions have significantly improved the physical interdependency between power systems and gas systems. In addition, in most cases, power systems and gas systems are operated by different utilities, suggesting inevitable economic behaviors between the two energy systems, such as gas purchase contracts. Due to the fact that the utilization of the superior regulation capabilities of GPUs relies on a reliable gas supply, it is essential to model the physical and economic interactions between power systems and gas systems for resilient decision-making.

    In this chapter, the interactions between power systems and gas system during resilient operations are considered from both the physical perspective, i.e., the consideration of the operational and security constraints of the gas system, and the economic perspective, which is addressed by modeling the gas contracts including the here-and-now gas demands in the pre-contingency stage and the wait-and-see fuel consumption in the post-contingency stage. To improve the operational performance of the power system in the worst-case contingency scenario as well as to consider practical decision-making, a two-stage robust optimization and defender-attacker-defender model are used in the mathematical model. Due to the linearization of the non-convex Weymouth equation used in the gas network and the modeling of the on/off grid operation of the generators, binary variables are used for the post-contingency stage decision-making. In this regard, an additional decomposition algorithm is needed to coordinate the results of the middle- and lower-level problems aside from the regular decomposition algorithm for the upper- and middle- and-lower-level problems, this results in a nested problem, which is solved by the nested column-and-constraint algorithm. Various numerical simulations of the two-test systems demonstrate the necessity of considering both physical and economic interactions with the gas system and avoid over-optimistic results driven by integrated modeling.

    The results of this study open quite a few new research directions.
    As this study presents the first attempt that considers the reserved gas contracts along with firm gas contracts in a robust optimization (RO) model, it is upgraded to 1) consider bidirectional gas contracts; 2) consider a more practical two-stage (day-ahead and real-time) contracting mechanism; 3) incorporating more uncertainty sources such as renewable power generation. These improvements have been addressed in Chapter~\ref{Chapter5}. Other future works are to decrease the conservativeness of the resilient robust strategy of IEGS using data-driven decision-making theories, and to consider different types of contingencies \cite{wang2018risk}, such as power unit outages and gas-power pipeline malfunctions \cite{wang2016robust}, as well as to consider regional and timely behavior of attacker \cite{shao2017integrated}.


\chapter{Robust Economic Operational Strategies for IEGSs} 

\label{Chapter5} 




Due to increasing penetration of variable and uncertain renewable power generation (RPG), as well as stronger interdependency with gas systems
facilitated by the deployment of gas-fired power units (GPUs) and
power-to-gas (P2G) facilities, the secure and economic energy management (EM)
of the power systems has become more challenging in the decision-making and computational burden perspectives. Most existing works on the
EM problem of the power systems (1) neglect the bidirectional interactions of the power system with other energy systems or (2) assume that the power and gas systems are run by one utility while neglecting the fact that there are different regulation authorities for them in most occasions. This chapter revisits the day-ahead operation of large-scale RPG integrated power systems considering the physical and economic interactions with gas systems. Specifically, the physical interaction is simulated by incorporating the gas system operation constraints, and the economic interaction is realized by modeling the gas contracting mechanism.

The previous chapter proposed a two-stage robust decision-making framework to optimize the operational performances of power systems under the worst-case $N-k$ contingencies, where the day-ahead gas contracts are modeled. Emerging P2G facilities to mitigate the surplus RPG outputs, bidirectional gas contracts are inevitable. This chapter develops two operational models for optimal EM in the power system with bidirectional gas contracts.

The first model is to find the economic and reliable EM decisions for power distribution networks (PDNs) with RPG integration. A tri-level robust dispatch model is established for the PDN considering bidirectional interactions with the gas distribution networks (GDNs). Two types of gas contracts are modeled, namely, gas-to-power (G2P) contracts for GPUs and P2G contracts for P2G facilities. A quadruple-loop solution procedure is developed for
the proposed tri-level EM model, including two column-and-constraint (C\&CG)
loops for the two-stage decision-making framework, and two sequential mixed
integer second-order cone program (S-MISOCP) loops to enhance the solution
feasibility with respect to the non-convex power flow and Weymouth equations. This model is illustrated in Section~\ref{sec:Ch5RO}, indicating the mathematical formulation and the proposed quadruple-loop procedure. This model has been published as
\begin{itemize}
      \item
            Ahmed R. Sayed, Cheng Wang, Junbo Zhao, and Tianshu Bi. "Distribution-level Robust Energy Management of Power Systems Considering Bidirectional Interactions with Gas Systems." IEEE Transactions on Smart Grid, vol. 11, no. 3, pp. 2092-2105, May 2020.  DOI: https://doi.org/10.1109/TSG.2019.2947219
    \end{itemize}

The second model incorporates both the day-ahead and real-time gas contracts in power system operation. To balance the robustness and the conservativeness of the operation strategy, a distributionally robust optimization (DRO) based decision-making framework is derived, and two-stage contracting mechanism is proposed. The quadruple-loop solution procedure is designed to tackle the computation burden brought by the DRO as well as the non-convexities in gas system modeling. This model is illustrated in Section~\ref{sec:Ch5DRO}, indicating the mathematical formulation and the solution procedure. This model is submitted for publication as
\begin{itemize}
      \item
            Ahmed R. Sayed, Cheng Wang, Sheng Chen, Ce Shang, and Tianshu Bi. "Two-stage Distributionally Robust Gas Contracting for Power System Operation." Submitted for publication to IEEE Transactions on Smart Grid.
    \end{itemize}

 The effectiveness of the proposed models and solution methodologies are verified by simulation results of several moderate test systems from distribution to transmission levels in Section~\ref{sec:Ch5Res}, where the significance of the proposed models compared with the literature analogues, the DRO-based modelling compared with SO- and RO-based ones, the two-stage contracting compared with one-stage contracting, and the solution method scalability are verified by the simulation results. Finally, the main conclusions and discussions are drawn in Section~\ref{sec:Ch5Con}.

\section{Introduction}
    Climate change and environmental concerns have been major driven forces for the utilization of renewable energy resources, such as wind and solar power generation, around globe \cite{ibrahim2000renewable,zhang2016climate}. In this regard, the top two CO\text{${}_2$} emitters, China and the US, pledged to increase their wind energy utilization to 20\% by 2030 \cite{lu2016challenges}. However, the integration of wind power at a large scale brings new challenges for EM in power systems, especially in the distribution level. This is because of the variable and uncertain output features of RPG as well as the complex operation conditions in PDN. Therefore, it becomes essential to make economic and reliable EM decisions for PDNs with RPG integration.

    To this end, many efforts have been made on the EM of PDNs with a concentration on optimal power flow (OPF) problem, which is one of the basis of PDN operational analysis. An up-to-date survey on OPF optimization methods is provided in \cite{frank2012optimal, frank2012optimal2}. In \cite{bazrafshan2016decentralized}, reactive power management is developed and solved with the decentralized algorithm. A stochastic EM problem is formulated in \cite{lopez2017two} considering slow- and fast-timescale controls. Unlike stochastic approaches, robust optimization approaches consider all the possible realizations of renewable uncertainties within a prescribed uncertainty set irrespective of their probabilities. An adjustable robust OPF is proposed in \cite{jabr2013adjustable}, where controllable generators are adjusted affinely according to the wind generation outputs. Meanwhile, quite a few inspiring works have been carried out on coordinating the active and reactive power of PDNs in a robust manner with additional modeling efforts on the continuous and discrete reactive power compensators \cite{ding2015two}, the on-load tap changer ratios and energy storage systems \cite{gao2017robust}, the price-based demand response \cite{zhang2017robust} and the photovoltaic output uncertainties \cite{ding2016two}.

    The aforementioned works can provide economic and robust EM decision for PDNs in the view of power systems. However, they neglect economic and physical interactions with other energy systems, such as GDNs, which feed GPUs and could damp the surplus RPG through P2G facilities. Therefore, these models, which are referred as the independent power system (IPS) model in this study, may cause physical violations such as under/over- pressure in gas systems, as discussed in Chapter~\ref{Chapter4}.

    Owing to the rapid development of the P2G technology and wide deployment of GPUs, interactions between power and gas systems have been noticeably enhanced in transmission \cite{yang2018modeling} and distribution \cite{wang2019convex, saldarriaga2013holistic} levels. This intensified interaction gradually appreciates the concept of integrated electricity and gas system (IEGS) and brings quite a few research interests on the IEGS model \cite{he2017robustb, liu2018operational}, in which power and gas systems are coordinated and co-optimized for planning, economic dispatch, and resilient operation. To utilize the gas system flexibility and reliability, the IEGS dispatch problem is formulated in \cite{Correa2015Integrated}, where the gas dynamics are modeled. The mutual dependence in IEGS and importance of considering gas systems, especially, under high wind penetrations, are investigated in \cite{devlin2016importance}. To consider wind uncertainties in the IEGS, different models are proposed in the literature, such as stochastic optimization models \cite{alabdulwahab2015stochastic}, the interval optimization model \cite{bai2016interval}, the adjustable robust dispatch model \cite{wang2019convex}, and the two-stage robust model \cite{he2016robust}.

    Note that the aforementioned IEGS dispatch models share one core assumption, that is, power and gas systems are supervised and controlled by one system operator. This operator has full authority to control and optimize all energy resources. However, in most cases, the two systems are operated by different utilities and information synchronization policy may be restricted~\cite{bertoldi2006energy}. Therefore, the IEGS model might provide suboptimal, or even infeasible decisions for subsystems. Under the multi-party decision-making reality as well as the necessity of bidirectional energy conversion, signing energy purchase contracts become inevitable. In industrial practice, GPUs are usually supplied with interruptible gas services under day-ahead G2P contracts, which are more economical and convenient than real-time contracts \cite{zlotnik2016coordinated, bouras2016using}. Moreover, the gas produced from P2G facilities is injected into the gas system under P2G contracts. Nevertheless, the real-time operation of PDNs might be changed from the day-ahead dispatch largely owing to the renewable uncertainties. This means the outputs of GPUs and P2G facilities might deviate from their day-ahead schedules, to mitigate the operation losses or the surplus wind generation. In this regard, there should also be contracts for the reserved gas besides the contracts for firm energy.

\section{Robust Day-ahead Operation with Bidirectional Gas Contracting} \label{sec:Ch5RO}

    To overcome the drawbacks of the IPS and IEGS models, the EM problem of PDNs is revisited considering interactions with gas systems. In this section, a tri-level robust dispatch model of the PDN is presented, considering energy contracts with gas systems. The proposed model not only identifies the optimal gas contracts and generation dispatch in the day-ahead stage, but also provides the optimal re-dispatch strategy in the real-time stage. In addition, the approximated gas dynamics are adopted in both day-ahead and real-time stages to offer additional operational flexibility for the gas system. To consider bidirectional energy trade with the gas system, two types of gas contracts are modeled, namely, gas-to-power (G2P) contracts for GPUs and P2G contracts for P2G facilities. The main objective of the PDN operator is to manage available energy resources economically and to identify the optimal energy contracts with interacted systems irrespective of their operational cost (OC).

    Beside the challenge in decision-making framework modeling, there is also computational difficulty in the EM of the interacted power and gas systems. For the PDN, the active and reactive power are coupled as the bus voltages are notably influenced by active power variations. Recently, convex relaxation methods have been implemented to solve OPF owing to their computational benefits. As discussed in Section~\ref{sec:Ch3OGF},  that SOC relaxation is exact under mild conditions for radial networks \cite{li2012exact}, however, the exactness depends on the objective function. In the proposed model, the objective function is not strictly increasing with all injected active power such as wind power (zero-cost). Based on penalty convex-concave procedure (P-CCP), the S-MISOCP algorithm proposed in Section~\ref{sec:SCPMethod} is employed to guarantee the power and gas feasibility and the accuracy of SOC relaxation.  Based on the above discussion, this work is the first attempt to embed the S-MISOCP algorithm into the nested column-and-constraint (NC\&CG) algorithm to solve the tri-level model, whose upper and lower levels are non-convex.
    Compared with recent works, the main contributions of this study are twofold.
    \begin{enumerate}
    \item
        A tri-level robust dispatch model is established for the PDN considering both physical and economic interactions with the gas systems. Specifically, the physical interaction is achieved by adding the security and feasibility constraints of the gas system into the EM problem of the PDN, while the economic interaction is completed by modeling firm and reserved gas contracts for both G2P and P2G.
    \item
        A quadruple-loop algorithm for the proposed robust EM problem of the PDN is devised, where the second and forth loops are S-MISOCP algorithms to enhance the solution feasibility in the day-ahead and real-time dispatch stages, respectively, and the first and third loops are column-and-constraint (C\&CG) algorithms to tackle the tri-level decision-making structure with binary recourse.
    \end{enumerate}

    \subsection{Mathematical Formulation}

    \subsubsection{Model Descriptions}
        \par
\begin{figure}[!htbp]
                \centering
                \includegraphics[width=10cm]{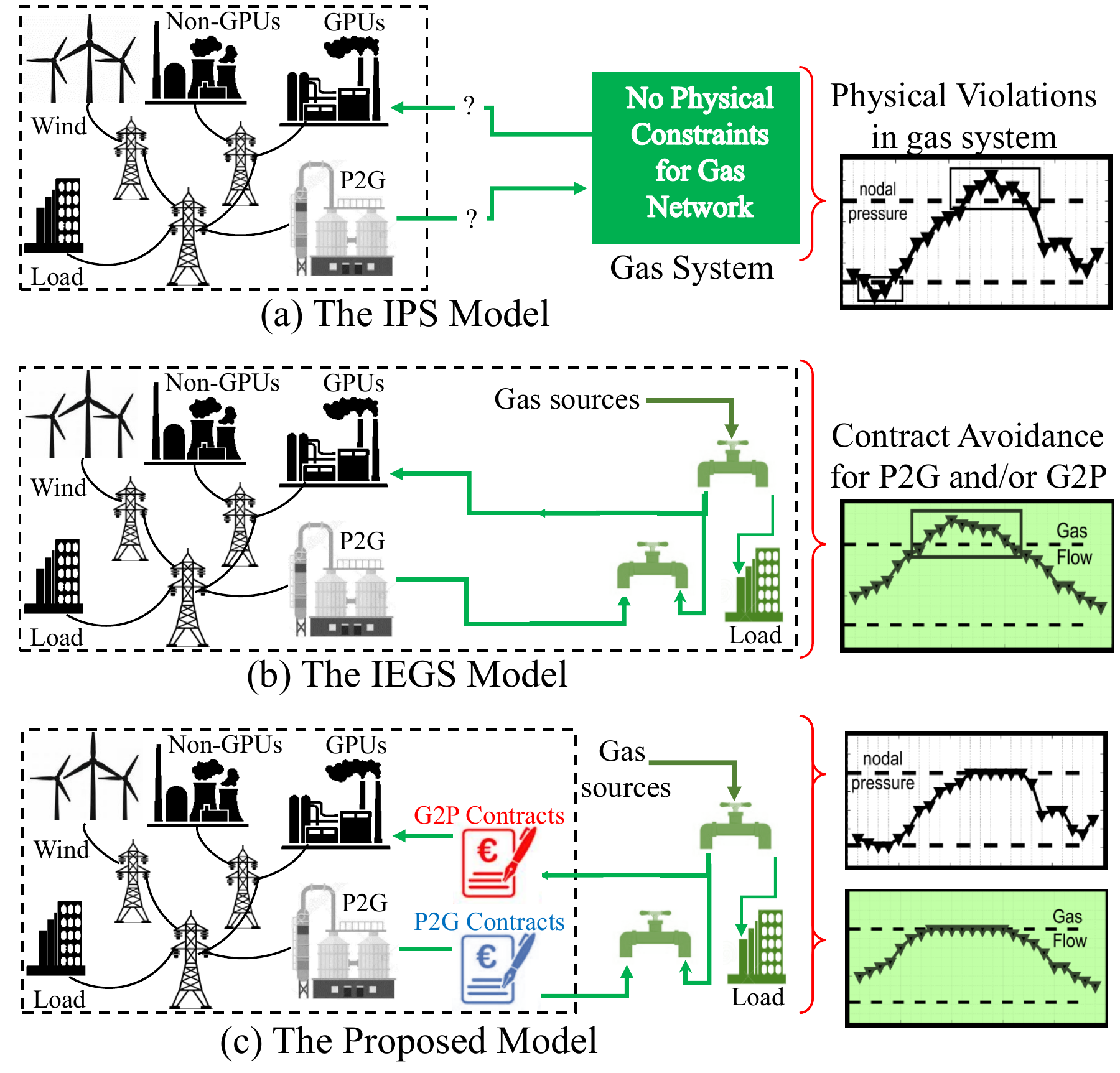}
                \caption{Schematic layout of the IPS, the IEGS, and the proposed models.}
                \label{fig:Ch5aF1}
        \end{figure}

        Based on the discussion above, three models are available to optimize and manage energy distributed for the PDN. Figure~\ref{fig:Ch5aF1} displays the schematic layout of these models and simplifies the salient features of each one: (i) the IPS model \cite{frank2012optimal2, bazrafshan2016decentralized, lopez2017two, jabr2013adjustable, ding2015two,gao2017robust, zhang2017robust, ding2016two}, which does not consider the operational and security constraints of the gas system, may cause physical violations, yielding a cascading failure in gas systems, particularly with high penetration of wind generation; (ii) the IEGS model \cite{yang2018modeling, wang2019convex, saldarriaga2013holistic, zlotnik2016coordinated, he2018coordination, he2017robustb, liu2018operational, Correa2015Integrated, devlin2016importance, alabdulwahab2015stochastic, bai2016interval, he2016robust}, which considers power and gas systems as one system and co-optimizes the total OCs, and neglects the energy transaction contracts, rendering a high probability of providing over-optimistic decisions and contract avoidance in practice; (iii) the proposed model, which finds the optimal decisions for the power system operator (PSO) considering the operational constraints of the gas system to avoid any physical violations and optimizes gas contracts to prevent any contract avoidance under the prescribed uncertainties in the day-ahead stage.

        Before presenting the detailed formulation of the proposed model, commonly used assumptions and simplifications are presented as follows
    \begin{enumerate}
    \item
        In general, (i) the distribution system may have some loops as well with the integration of renewable energy and storage. We may just say that this study only considers the scenario of radial structure \cite{ding2015two, li2012exact} and the mesh network will be addressed in our future work; (ii) cost coefficients of generated power from GPUs and non-GPUs are known; (iii) PDN and GDN are operated by different utilities.
    \item
        In PDN modeling, (i) the branch flow model in \cite{wei2017optimal, wang2018convex} is adopted, where the three-phase system is balanced and the power flow direction is fixed without any reverse power; (ii) for GPUs, the consumed gas and the OC depend only on the active power \cite{wang2018convex}; (iii) all P2G facilities are owned by the power system utility.
    \item
        In GDN modeling, (i) the approximated dynamic-state model for gas system presented in Section \ref{sec:Ch2SGasDynamic} is adopted.
    \item
        In contract modeling, P2G and G2P contracts are signed in the day-ahead stage and there are no real-time contracts. The gas prices of sale and purchase gas contracts are fixed. These prices and penalties of contract avoidance are obtained from the gas system operator in the day-ahead stage before determining the gas contracts.
        \end{enumerate}
    \par
        The proposed two-stage model aims to identify the optimal EM strategy with the best gas contracts by minimizing both OC and regulation cost (RC). The optimal OC is identified in the first (day-ahead) stage based on the forecasted outputs wind generation. While, in the second (real-time) stage, the model seeks to minimize the RC with the worst-case realization of the uncertainties. The overall objective function is presented in \eqref{eq:Ch5aObj1}, where $\Omega_1,\Omega_2$, and $\Omega_3$ are the day-ahead, wind uncertainty, and real-time decision variables, respectively. Please refer to \eqref{eq:Ch5aAllModel} for their detailed expressions. Equation \eqref{eq:Ch5aOC} depicts the OC in the day-ahead stage, including the power production costs from all generators, costs of reserved gas in the G2P contracts, and revenue from sale of gas in the P2G contracts. Equation \eqref{eq:Ch5aRC} provides the RC in the real-time stage, including the costs of adjustable power production from non-GPUs and the penalties for the non-served power load, wind curtailment, and P2Gs output deviation from their contracts.
    \begin{gather}
          \min_{\Omega_1} \; OC \; + \;  \max_{\Omega_2} \; \min_{\Omega_3} \; RC \label{eq:Ch5aObj1}\\
          OC = \sum_{t}\Big[ \sum_{u} {C}_u ( \hat{p}_{u,t}) + \sum_{y} ({\mu}_h \rho_{h,t} + \mu_h^+ \rho_{h,t}^+ + {\mu}_h^- \rho_{h,t}^-) - \sum_{j} {C}_j g_{j,t}  \Big]   \label{eq:Ch5aOC}\\
          RC = \sum_{t} \Big[ \sum_{u \in \mathcal{U}_n} ({C}_u^+ \triangle p_{u,t}^+ + {C}_u^- \triangle p_{u,t}^-) + \sum_{d} {C}_d  \triangle p_{d,t} \nonumber \\
          + \sum_{e} {C}_e  \triangle w_{e,t} + \sum_{j} ({C}_j^+ \triangle g_{j,t}^+ + {C}_j^- \triangle g_{j,t}^- )  \Big]   \label{eq:Ch5aRC}
    \end{gather}

\subsubsection{Day-ahead Operational Constraints}
        Day-ahead constraint set constitutes of two parts. The first one is the operational constraints of PDN defined in \eqref{eq:Ch5aDAL1}--\eqref{eq:Ch5aPGb}, which are derived from Section~\ref{sec:Ch3GFC_PDN}, using the day-ahead decision variables, i.e., adding the hat ( $\hat{}$ ) symbol on the decision variables for those constraints. They are composed by
        \begin{align}
           & \text{Active and reactive power capacities of all units: \eqref{eq:Ch3PLa},}  \hspace{10em} \label{eq:Ch5aDAL1}\\
           & \text{Maximum ramping up and down limits: \eqref{eq:Ch3PLb},} \label{eq:Ch5aDAL2}\\
           & \text{P2G power consumption capacities: \eqref{eq:Ch3PLc},} \label{eq:Ch5aDAL3}\\
           & \text{Active and reactive power flow direction: \eqref{eq:Ch3PLd},} \label{eq:Ch5aDAL4}\\
           & \text{Squared line current limits: \eqref{eq:Ch3PLe},} \label{eq:Ch5aDAL5}\\
           & \text{Squared node voltage limits: \eqref{eq:Ch3PLf},} \label{eq:Ch5aDAL6}\\
           & \text{Voltage drop equation: \eqref{eq:Ch3PLg},} \label{eq:Ch5aDAL8}\\
           & \text{Power flow equation: \eqref{eq:Ch3PNL},} \label{eq:Ch5aDAL9}\\
           & \text{Active power nodal balancing equation: } \nonumber \\
           &\hspace{4em} \sum_{u \in \mathcal{U}(n)}\hat{p}_{u,t} + \sum_{e \in \mathcal{E}(n)} \hat{W}_{e,t} + \sum_{l \in \mathcal{L}_1(n)} (\hat{p}_{l,t} - r_l \hat{i}_{l,t}) - \sum_{l \in \mathcal{L}_2(n)} \hat{p}_{l,t} \nonumber \\
           & \hspace{6em} = G_n \hat{v}_{n,t} + \sum_{z \in \mathcal{Z}(n)} \hat{p}_{z,t}
            + \sum_{d \in \mathcal{D}_p(n)} P_{d,t}, \; \forall n,t   \ \label{eq:Ch5aPGa}\\
            & \text{Reactive power nodal balancing equation:} \nonumber \\
            &\sum_{u \in \mathcal{U}(n)}\hat{q}_{u,t} + \sum_{l \in \mathcal{L}_1(n)} (\hat{q}_{l,t} - x_l\hat{i}_{l,t}) - \sum_{l \in \mathcal{L}_2(n)} \hat{q}_{l,t}  = B_n \hat{v}_{n,t} + \sum_{d \in \mathcal{D}_p(n)} Q_{d,t}, \; \forall i,t \label{eq:Ch5aPGb}
        \end{align}

        Similarly, the second part is GDN constraints are derived in \eqref{eq:Ch5aDAG1}--\eqref{eq:Ch5PGc}.
        \begin{align}
           & \text{Gas production capacities: \eqref{eq:Ch2Well},} \label{eq:Ch5aDAG1}\\
           & \text{Gas compressors constraints: \eqref{eq:Ch2Comp1}--\eqref{eq:Ch2Comp2},} \label{eq:Ch5aDAG2}\\
           & \text{Nodal pressure bounds: \eqref{eq:Ch2Pressure},} \label{eq:Ch5aDAG3}\\
           & \text{Average flow rate equation: \eqref{eq:Ch2AveFlow},} \label{eq:Ch5aDAG4}\\
           & \text{Mass flow equation: \eqref{eq:Ch2GMMass1},} \label{eq:Ch5aDAG5}\\
           & \text{Continuity equation: \eqref{eq:Ch2GMMass2},} \label{eq:Ch5aDAG6}\\
           & \text{GPU gas consumption: \eqref{eq:Ch3GFC-GUF},} \label{eq:Ch5aDAG8}\\
           & \text{P2G gas production: \eqref{eq:Ch2P2G},} \label{eq:Ch5aDAG9}\\
           & \text{Weymouth equation: \eqref{eq:Ch2Wey}.} \label{eq:Ch5aDAGF}\\
           & \text{Gas nodal balancing equation: } \nonumber \\
           & \hspace{3em}\sum_{p \in \mathcal{P}_1(i)} \hat{f}_{p,t}^{out} - \sum_{p \in \mathcal{P}_2(i)} \hat{f}_{p,t}^{in} + \sum_{c \in \mathcal{C}_1(i)} \hat{f}_{c,t}^{out} - \sum_{c \in \mathcal{C}_2(i)} \hat{f}_{c,t}^{in} + \sum_{z \in \mathcal{Z}(i)} \hat{\varrho}_{z,t} + \sum_{w \in \mathcal{W}(i)} \hat{f}_{w,t}  \nonumber \\
           & \hspace{8em}= \sum_{u \in \mathcal{U}_g(i)} \rho_{u,t} + \sum_{d \in \mathcal{D}_g(i)} F_{d,t}, \; \forall i,t \label{eq:Ch5PGc}
        \end{align}

        It should be emphasized that, to guarantee the exactness of SOC relaxation, the cost function should be strictly increasing in all injected active power \cite{wei2017optimal}, even for radial PDNs. In this study, the objective function includes zero-cost wind power in the day-ahead stage and indirect negative-cost for injected power from GPUs, non-GPUs, surplus wind, and P2G units in the real-time stage, i.e., $\rho_{h,t}^-,\triangle p_{u,t}^-,\triangle w_{e,t}$, and $\triangle g_{j,t}^+$. Therefore, relaxation of power flow equation presented in \cite{li2012exact} is generally inexact for the proposed model. Consequently, the S-MISOCP algorithm should be applied on the power flow equations besides the gas flow equations.

        \subsubsection{Gas Contracts Modeling}
            According to current electricity industrial practice, GPUs usually have the interruptible gas delivery service in terms of cost-saving \cite{liu2009security}, and the gas delivery contracts are usually determined day-ahead or even earlier because real-time contracting would be costly and inconvenient \cite{bouras2016using}. Gas contracting models proposed in Section \ref{sec:Ch4GasContr}, is adopted in the proposed model. The G2P contract is characterized by two subcontracts, namely, contracts for firm and reserved outputs of GPUs, which is consistent with the two-stage power system dispatch. The firm gas contract provides the scheduled gas amounts in the day-ahead stage, which can be calculated by  \eqref{eq:Ch5aFirm}. The gap between the actual gas consumption of GPUs in the real-time stage and the amount in the firm gas contract must obey the reserved gas contract, which is defined in \eqref{eq:Ch5aReserve}--\eqref{eq:Ch5aReserveBound}.
    \begin{gather}
        \rho_{h,t} \ge  \sum_{\forall u\in \mathcal{U}_g(h)} \frac{\Phi}{\eta_u} p_{u,t},\; \forall h,t \label{eq:Ch5aFirm}\\
        -\rho_{h,t}^- \le  \sum_{\forall u\in \mathcal{U}_g(h)} \frac{\Phi}{\eta_u} (p_{u,t} - \hat{p}_{u,t}) \le \rho_{h,t}^+,\; \forall h,t \label{eq:Ch5aReserve}\\
        \rho_{h,t}^+ , \; \rho_{h,t}^- \ge 0,\; \forall h,t. \label{eq:Ch5aReserveBound}
    \end{gather}

        There are two methods to cope with the surplus wind energy, which are curtailment by the wind sector management and methanation by P2G facilities, respectively. The proposed model systematizes the two manners based on the curtailment penalty, which is managed by the PDN operator, and the G2P contract avoidance penalty, which is assigned from the gas system operator. The P2G output gas, which is injected into the gas system pipelines, should obey the P2G contracts. The scheduled sale values can be calculated by \eqref{eq:Ch5aP2G1} in the day-ahead stage. However, in the real-time stage, any gas variation (upward and downward) detected by \eqref{eq:Ch5aP2G2}--\eqref{eq:Ch5aP2G3} will be penalized.
        \begin{gather}
            g_{j,t} \le \sum_{j\in \mathcal{Z}(j)} \Phi \eta_z \hat{p}_{z,t} , \forall j,t,  \label{eq:Ch5aP2G1}\\
            -\triangle g_{j,t}^- \le \sum_{z \in \mathcal{Z}(j)} \Phi \eta_z (p_{z,t}-\hat{p}_{z,t}) \le \triangle g_{j,t}^+, \forall j,t,  \label{eq:Ch5aP2G2}\\
            \triangle g_{j,t}^- ,\; \triangle g_{j,t}^+ \ge 0, \forall j,t. \label{eq:Ch5aP2G3}
        \end{gather}

\subsubsection{Wind Power Generation Uncertainty Modeling}
    Wind power generation uncertainties can be formulated with various modeling approaches, such as interval-based model and continuous probability distribution function approximation \cite{bai2016interval}. In \cite{guan2013uncertainty}, different uncertainty set approaches for robust models are discussed. In this study, the uncertainty set used in \cite{jiang2011robust} is adopted. The uncertainty budgets are defined in \eqref{eq:Ch5aWind1}--\eqref{eq:Ch5aWind3}. The real-time outputs under uncertainty is defined in \eqref{eq:Ch5aWindP} based on the average, maximum, and minimum forecasted values.
        \begin{gather}
            \sum_{e}(\xi_{e,t}^+ + \xi_{e,t}^-) \le \Gamma^{e}, \forall t \label{eq:Ch5aWind1}\\
            \sum_{t}(\xi_{e,t}^+ + \xi_{e,t}^-) \le \Gamma^{t}, \forall e  \label{eq:Ch5aWind2}\\
            \xi_{e,t}^+ + \xi_{e,t}^- \le 1, \xi_{e,t}^+ , \xi_{e,t}^- \in \{0,1\} , \forall e,t   \label{eq:Ch5aWind3}\\
            w_{e,t} = \hat{W}_{e,t}  + (\overline{P}_{e,t} -\hat{W}_{e,t})\xi_{e,t}^+ + (\underline{P}_{e,t}-\hat{W}_{e,t})\xi_{e,t}^-, \forall e,t  \label{eq:Ch5aWindP}
        \end{gather}

\subsubsection{Real-time Operational Constraints}
        Most of the operation constraints in the real-time stage can be obtained by replacing the day-ahead decision variables with real-time ones in   \eqref{eq:Ch5aDAL1}--\eqref{eq:Ch5aDAL9} and \eqref{eq:Ch5aDAG1}--\eqref{eq:Ch5aDAGF}, namely removing the hat symbols of the decision variables in those constraints. Additionally, in real-time operation of the power network, wind generation curtailment and electrical load shedding are also practical means to recover the power balancing condition, whose adjustment ranges are shown in \eqref{eq:Ch5aRTL11}--\eqref{eq:Ch5aRTL12}. Meanwhile, the nodal power balancing condition should be modified by adding the wind generation curtailment and electrical load shedding terms, resulted in \eqref{eq:Ch5aPGaRT}--\eqref{eq:Ch5aPGbRT}. To quantify the regulation costs of non-GPUs in the real-time stage, \eqref{eq:Ch5aRTL13} is added to describe the outputs adjustment of non-GPUs.
\begin{gather}
     0 \le \delta_{d,t} \le 1,  \; \forall d,t, \label{eq:Ch5aRTL11} \\
     0 \le \triangle w_{e,t} \le w_{e,t}, \; \forall e,t,  \label{eq:Ch5aRTL12} \\ \sum_{u \in \mathcal{U}(n)} {p}_{u,t} + \sum_{e \in \mathcal{E}(n)} ({w}_{e,t} - \triangle{w}_{e,t}) + \sum_{l \in \mathcal{L}_1(n)} ({p}_{l,t} - r_l {i}_{l,t}) - \sum_{l \in \mathcal{L}_2(n)} {p}_{l,t} \nonumber \\
     = G_n {v}_{n,t} + \sum_{z \in \mathcal{Z}(n)} {p}_{z,t}
     + \sum_{d \in \mathcal{D}_p(n)} P_{d,t}(1-\delta_{d,t}), \; \forall n,t,   \label{eq:Ch5aPGaRT}\\
     \sum_{u \in \mathcal{U}(n)} {q}_{u,t} + \sum_{l \in \mathcal{L}_1(n)} ({q}_{l,t} - x_l {i}_{l,t}) - \sum_{l \in \mathcal{L}_2(n)} {q}_{l,t}  = B_n {v}_{n,t} + \sum_{d \in \mathcal{D}_p(n)} Q_{d,t}(1-\delta_{d,t}), \; \forall i,t, \label{eq:Ch5aPGbRT}\\
     -\triangle p_{u,t}^- \le p_{u,t} -  \hat{p}_{u,t} \le \triangle p_{u,t}^+, \; \triangle p_{u,t}^-,\triangle p_{u,t}^+  \ge 0, \; \forall t, u \in \mathcal{U}_n. \label{eq:Ch5aRTL13}
\end{gather}

\subsubsection{The Overall Model}
The holistic model of power system EM problem can be cast as follows
        \begin{subequations}\label{eq:Ch5aAllModel}        \begin{align}
                \min_{\Omega_1} \;& OC \; + \;  \max_{\Omega_2} \; \min_{\Omega_3} \; RC     \label{eq:Ch5aAllModela}\\
                s.t. \hspace{1mm}& \text{Day-ahead constraints: \eqref{eq:Ch5aDAL1}--\eqref{eq:Ch5PGc}}, \\
                &  \text{Real-time constraints: \eqref{eq:Ch5aWindP}, \eqref{eq:Ch5aDAL1}--\eqref{eq:Ch5aDAL9},  \eqref{eq:Ch5aDAG1}--\eqref{eq:Ch5aDAGF}, \eqref{eq:Ch5aRTL11}--\eqref{eq:Ch5aRTL13}}, \\
                &  \text{Gas contracts: \eqref{eq:Ch5aFirm}--\eqref{eq:Ch5aReserveBound},  \eqref{eq:Ch5aP2G1}--\eqref{eq:Ch5aP2G3}}, \\
                &  \text{Uncertainty set: \eqref{eq:Ch5aWind1}--\eqref{eq:Ch5aWind3}}, \\
                & \Omega_1 = \{\rho_{h,t}^+, \rho_{h,t}^-,\rho_{h,t}, g_{j,t}, \hat{p}_{u,t}, \hat{q}_{u,t}, \hat{p}_{z,t}, \hat{p}_{l,t}, \hat{q}_{l,t}, \hat{i}_{l,t},  \hat{v}_{n,t}, \hat{f}_{w,t}, \hat{\pi}_{i,t} , \hat{f}_{c,t}^{in}, \nonumber\\
                 & \hat{f}_{c,t}^{out}, \hat{f}_{p,t}^{in}, \hat{f}_{p,t}^{out} , \hat{f}_{p,t}, \hat{m}_{p,t}\} \\
                &  \Omega_2 = \{u_{e,t}^+, u_{e,t}^- \} \\
                 & \Omega_3 = \{\triangle p_{u,t}^+, \triangle p_{u,t}^-, \triangle g_{j,t}^+ \triangle g_{j,t}^- ,\triangle w_{e,t} ,\delta_{d,t} , w_{e,t},{p}_{u,t}, {q}_{u,t}, {p}_{z,t}, {p}_{l,t}, {q}_{l,t}, {i}_{l,t},{v}_{n,t}, \nonumber \\
             & {f}_{p,t} , {\pi}_{i,t} , {f}_{c,t}^{in}, {f}_{c,t}^{out} , {f}_{p,t}^{in} , {f}_{p,t}^{out} , {f}_{p,t}, {m}_{p,t}     \}
        \end{align}  \end{subequations}
    which can be viewed as a two-stage or tri-level model. The model reformulation and the corresponding solution methodology are introduced in the following section.

\subsection{Solution Methodology} \label{sec:4LOOPS}
In addition to the relatively complicated structure of the overall model, solving its each level also require much computation efforts, due to the presence of the nonlinear and nonconvex power flow and Weymouth equations. Fortunately, the power flow constraints and Weymouth equations can be formulated as DCP problem by expressing the proposed model constraints as difference of two convex functions. Referring to Section \ref{sec:SCPMethod}, equations \eqref{eq:Ch5aDAL9} and \eqref{eq:Ch5aDAGF} are reformulated as MISOCP constraints, which are convenient with decomposition algorithms.

\subsubsection{Power Flow Equation Reformulation}

        For ease of analysis, the general form of day-ahead and real-time power flow equations is given as
    \begin{equation}  
            p^2 + q^2 = vi \label{eq:Ch5asimplePQ}
    \end{equation}
    which can be further divided into two opposite inequality constraints. The first inequality is an SOC constraint, and its canonical form is \eqref{eq:Ch5aPQcone1}. Using \eqref{eq:Ch3CCP1}, the second inequality is reformulated as \eqref{eq:Ch5aPQcone1}, where $\;[\overline{p} \;\overline{q} \;\overline{v} \;\overline{i}]^{\top}$ is used aa a linearization point.
    \begin{gather}
        \begin{Vmatrix}
            2p \\ 2q \\ (v-i)
        \end{Vmatrix}_2 \le (v + i ),  \label{eq:Ch5aPQcone1} \\
        \begin{Vmatrix}
            2 (v+i) \\ \varpi -1
        \end{Vmatrix}_2   \le  \varpi +1, \;\; \varpi_{l,t} = 8\overline{p}p + 8\overline{q}q + 2 (\overline{v}-\overline{i})(v-i)
        - 4\overline{p}^2 - 4\overline{q}^2 - (\overline{v}-\overline{i})^2. \label{eq:Ch5aPQcone2}
    \end{gather}

\subsubsection{Gas Flow Equation Reformulation}

          The general form of the Weymouth equations in the day-ahead and real-time stages is presented as
        \begin{equation}  
                f |f| = \chi^f (\pi_i^2 - \pi_o^2)  \label{eq:Ch5asimpleWey}
        \end{equation}

    \par
        The sign function in \eqref{eq:Ch5asimpleWey} is firstly removed by introducing set of MILP constraints \eqref{eq:CH5aWeyc1}--\eqref{eq:Ch5aWeyNoDirc}, where $z$ is the directional binary variable and $\pi^+/\pi^-$ is the inlet and outlet pressures of the pipeline, respectively. Then the resultant equality constraint is split into two opposite inequality constraints. The first inequality is an SOC constraint, and its canonical form is \eqref{eq:Ch5aWeycone1}. Similar to \eqref{eq:Ch5aPQcone2}, given $[\overline{f}\;\overline{\pi}^+\;\overline{\pi}^-]^{\top}$ as an initial point, the second inequality  is substituted with the approximated canonical form \eqref{eq:Ch5aWeycone2}.
        \begin{gather}
             (1-z) ( \overline{\Pi}_o-\underline{\Pi}_i) \ge \pi^+ - \pi_{i} \ge (1-z) (\underline{\Pi}_o- \overline{\Pi}_i) , \;\; \forall \mathcal{P}^\pm \label{eq:CH5aWeyc1}\\
             (1-z) ( \overline{\Pi}_i-\underline{\Pi}_o) \ge \pi^- - \pi_{o} \ge (1-z) (\underline{\Pi}_i- \overline{\Pi}_o) , \;\; \forall \mathcal{P}^\pm \label{eq:CH5aWeyc2}\\
             z     ( \overline{\Pi}_i-\underline{\Pi}_o) \ge \pi^+ - \pi_{o} \ge z    (\underline{\Pi}_i- \overline{\Pi}_o) , \;\; \forall  \mathcal{P}^\pm \label{eq:CH5aWeyc3}\\
             z     ( \overline{\Pi}_o-\underline{\Pi}_i) \ge \pi^- - \pi_{i} \ge z     (\underline{\Pi}_o- \overline{\Pi}_i) , \;\; \forall \mathcal{P}^\pm \label{eq:CH5aWeyc4}\\
             \pi^+ =  \pi_{i} ,\;\; \pi^- = \pi_{o}    , \;\; \forall \mathcal{P/P}^\pm \label{eq:Ch5aWeyNoDirc}   \\
            \begin{Vmatrix}
                f \\ \sqrt{\chi^f} {\pi}^-
            \end{Vmatrix}_2 \le  \sqrt{\chi^f} {\pi}^+,   \label{eq:Ch5aWeycone1}\\
            \begin{Vmatrix}
                2 \sqrt{\chi^f} {\pi}^+ \\ \Lambda -1
            \end{Vmatrix}_2 \le  \Lambda+1, \;\;   \Lambda = 2\chi^f \overline{\pi}^- {\pi}^- + 2 \overline{f}f - \chi^f \overline{\pi}^- - \overline{f}^2. \label{eq:Ch5aWeycone2}
        \end{gather}

        Therefore, the compact form of the proposed model after reformulations of nonlinear equations is
        \begin{subequations}\label{eq:Ch5aCompact}\begin{align}
        \min_{\bm{y},\bm{w},\overline{\bm{y}}} \; & f(\bm{w}) + \max_{\bm{u}} \min_{\bm{x},\bm{z},\overline{\bm{x}}} \bm{e}^{\top}\bm{x} \label{eq:Ch5aCompacta} \\
                s.t. \hspace{1mm} & \bm{I}\bm{y}+\bm{J}\bm{w} \le \bm{C}                   \label{eq:Ch5aUpperMIL}  \\
            & \lvert\lvert \bm{A}_{v,t}\bm{y} \rvert\rvert_2  \le \bm{a}_{v,t}\bm{y}, \forall v \in \mathcal{P\cup L},  t, \label{eq:Ch5aUppercone1}\\
        &\lvert\lvert \bm{B}_{v,t} (\overline{\bm{y}}) \bm{y}+\bm{D}_{v,t}(\overline{\bm{y}}) \rvert\rvert_2 \le\bm{b}_{v,t}(\overline{\bm{y}})\bm{y}+\bm{d}_{v,t}(\overline{\bm{y}}), \forall v\in \mathcal{P\cup L} ,t,  \label{eq:Ch5aUppercone2}\\
            & \bm{S}\bm{u}  \le \bm{K}        \label{eq:Ch5aMiddleMIL} \\
        & \bm{E}\bm{x}+\bm{G}\bm{z} \ge \bm{F}- \bm{Q}\bm{u}-\bm{P}\bm{w}     \label{eq:Ch5aLowerMIL}\\
        & \lvert\lvert \bm{H}_{v,t}\bm{x} \rvert\rvert_2  \le \bm{h}_{v,t}\bm{x}, \forall v \in \mathcal{P\cup L},  t,   \label{eq:Ch5aLowercone1}\\
            & \lvert\lvert \bm{M}_{v,t}(\overline{\bm{x}})\bm{x}+\bm{N}_{v,t}(\overline{\bm{x}}) \rvert\rvert_2  \le \bm{m}_{v,t}(\overline{\bm{x}})\bm{x}+\bm{n}_{v,t}(\overline{\bm{x}}), \forall v\in \mathcal{P\cup L},t.  \label{eq:Ch5aLowercone2}
        \end{align}\end{subequations}

    where $\bm{w}$ is the reserved gas in G2P contracts, scheduled gas in P2G sale contracts, and committed power from all generators; $\bm{y}$ is the remained variables from the upper-level problem; $\bm{u}$ is the middle-level decision; $\bm{x}$ is the continuous variables in the lower-level problem, and $\bm{z}$ is the remained binary variables. Because of the approximation in cones \eqref{eq:Ch5aPQcone2} and \eqref{eq:Ch5aWeycone2}, initial points are required; therefore, $\overline{\bm{y}}$ and $\overline{\bm{x}}$ are considered as decision variables for upper- and lower-level problems, respectively. $\bm{I,J}$, and $\bm{C}$ are coefficients, and they can be driven from the linear constraints in \eqref{eq:Ch5aDAL1}--\eqref{eq:Ch5PGc}, \eqref{eq:Ch5aFirm}, \eqref{eq:Ch5aReserveBound}, \eqref{eq:Ch5aP2G1}, \eqref{eq:Ch5aP2G3} and \eqref{eq:CH5aWeyc1}--\eqref{eq:Ch5aWeyNoDirc}. $\bm{S}$ and $\bm{K}$ are the coefficients of \eqref{eq:Ch5aWind1}--\eqref{eq:Ch5aWind3}. $\bm{E, G, F, Q}$, and $\bm{P}$ are the coefficients of linear constraints in \eqref{eq:Ch5aReserve}, \eqref{eq:Ch5aP2G2}, \eqref{eq:Ch5aDAL1}--\eqref{eq:Ch5aDAG9} (with real-time variables), \eqref{eq:Ch5aDAG1}--\eqref{eq:Ch5aDAG9} (with real-time variables), \eqref{eq:Ch5aWindP}, \eqref{eq:Ch5aRTL13}, and \eqref{eq:CH5aWeyc1}--\eqref{eq:Ch5aWeyNoDirc}.  Equation \eqref{eq:Ch5aUppercone1} (\eqref{eq:Ch5aLowercone1}) can be obtained from \eqref{eq:Ch5aPQcone1} and \eqref{eq:Ch5aWeycone1} for upper (lower) level constraints. Similarly, SOC constraints in \eqref{eq:Ch5aPQcone2} and \eqref{eq:Ch5aWeycone2} are compacted in \eqref{eq:Ch5aUppercone2} (\eqref{eq:Ch5aLowercone2}) for upper (lower) level.
\begin{algorithm}[!tb]
\caption{The S-MISOCP Algorithm for RO model}
\label{Ch5S-MISOCP}
\begin{algorithmic}[1]
    \STATE Initialize the penalties \textsuperscript{a} $\tau_{v,t}^0, \;\tau_{v,t}^r$ / \textsuperscript{b} $\tau_{v,t}$, limits of penalty growth rates $\overline{\mu},\;\underline{\mu}$, penalty growth rate coefficient $\sigma$, maximum number of iterations $K^{max}$, and convergence parameters $\varepsilon$ and $\epsilon$. Set $k=0$. \\

    \STATE \textsuperscript{a} Solve the relaxed $\textbf{P1}$ \eqref{eq:Ch5aMP0}, and update
                   $\bm{y}^{(k)}$ and $\bm{x}^{r\,(k)}$, go to Step $4$.        \\
           \textsuperscript{b} Solve the relaxed $\textbf{P4}$ \eqref{eq:Ch5aSPs0}, and update
                    $\bm{x}^{(k)}$, go to Step $4$.       \\

    \STATE \textsuperscript{a} Solve the penalized \textbf{P1} \eqref{eq:Ch5aMPK} and update
                    $\bm{y}^{(k)},\bm{x}^{r\,(k)},s_{v,t}^{0\,(k)}$ and $s_{v,t}^{r\,(k)}$.  \\
           \textsuperscript{a} Solve the penalized \textbf{P4} \eqref{eq:Ch5aSPsK} and update
                    $\bm{x}^{(k)}$ and $s_{j,t}^{(k)}$.

    \STATE If \textsuperscript{a} \eqref{eq:Ch5aAlg1}-\eqref{eq:Ch5aAlg3} / \textsuperscript{b} \eqref{eq:Ch5aAlg4}-\eqref{eq:Ch5aAlg5}  is satisfied, OR $ k > K^{max}$, terminate; else, go to Step $5$. \begin{gather}
            \textbf{P1}^{(k-1)}-\textbf{P1}^{(k)} \le \epsilon, \label{eq:Ch5aAlg1}   \\
            s_{v,t}^{0\,(k)} \le \varepsilon \left(\bm{b}_{v,t}(\overline{\bm{y}})\bm{y}+\bm{d}_{v,t}(\overline{\bm{y}}) \right), \; \forall v,t \label{eq:Ch5aAlg2} \\
            s_{v,t}^{r\,(k)} \le \varepsilon  \left( \bm{m}_{v,t}(\overline{\bm{x}}^r)\bm{x}^r+\bm{n}_{v,t}(\overline{\bm{x}}^r) \right),\;\forall v,t,r, \label{eq:Ch5aAlg3} \\
            \textbf{P4}^{(k-1)}-\textbf{P4}^{(k)} \le \epsilon,    \label{eq:Ch5aAlg4}  \\
            s_{v,t}^{(k)} \le \varepsilon  \left(\bm{m}_{v,t} (\overline{\bm{x}}) \bm{x}+\bm{n}_{v,t} (\overline{\bm{x}})\right) ,\; \forall v,t.    \label{eq:Ch5aAlg5}
            \end{gather}\\

    \STATE  By  \eqref{eq:Ch3Adaptive}, update \textsuperscript{a} $\tau_{v,t}^0, \tau_{v,t}^r$ / \textsuperscript{b} $\tau_{v,t}$, set \textsuperscript{a} $\overline{\bm{y}}=\bm{y}^{(k)}, \overline{\bm{x}}^r =\bm{x}^{r\,(k)}$ / \textsuperscript{b} $\overline{\bm{x}} =\bm{x}^{(k)}$, and $k=k+1$, go to Step $3$.\\

    ${}^a\,$For problem \textbf{P1} \eqref{eq:Ch5aMPK};  ${}^b\,$For problem \textbf{P4} \eqref{eq:Ch5aSPsK}
\end{algorithmic}\vspace{-5.8mm}
\end{algorithm}
\subsubsection{The Quadruple-loop Algorithm: Nested C\&CG + Two S-MISOCP}
    Tri-level models can be solved by different techniques based on decomposition methods, such as Benders decomposition \cite{bertsimas2012adaptive} and C\&CG algorithm \cite{zeng2013solving}. The presence of binary variables $\bm{z}$ and finding a suitable point $\bm{\overline{x}}$ for the approximated SOC prevent the lower-level problem to be dualized with zero-gap. Therefore, the NC\&CG algorithm \cite{zhao2012exact} with its improvement discussed in Chapter~\ref{Chapter4} (see Algorithm~\ref{Ch3alg1}) is adopted in this study. In addition,  the S-MISOCP algorithm proposed in Chapter~\ref{Chapter3} (see Algorithm~\ref{alg2}), is suggested to find the values of $\overline{\bm{y}}$ and $\overline{\bm{x}}$ in each C\&CG process, resulting in a quadruple-loop solution procedure.

\begin{figure}[!htbp]
                \centering
                \includegraphics[width=12cm]{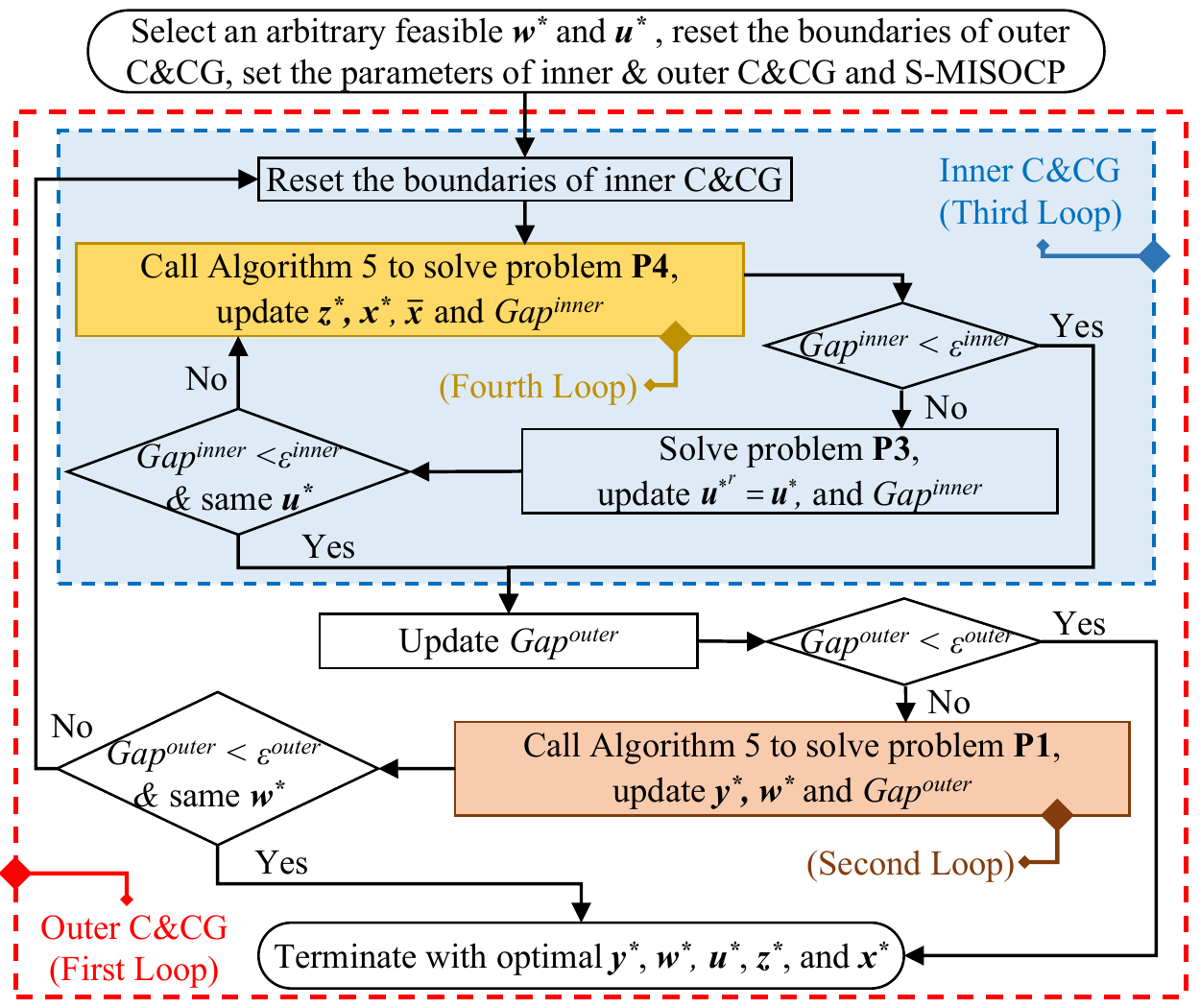}
                \caption{Flowchart of the proposed quadruple-loop algorithm.}
                \label{fig:Ch5aMethod}
        \end{figure}

    Figure~\ref{fig:Ch5aMethod} shows the proposed methodology and the interactions among different algorithmic loops. With an arbitrary feasible decision $\bm{w}^{*}$ and $\bm{u}^{*}$, the inner C\&CG algorithm, i.e., the third loop, starts to solve the lower-level problem using the S-MISOCP algorithm, namely the fourth loop, which provides a primal cut (i.e., $\bm{z}^{*}$, $\overline{\bm{x}}$) to the middle-level problem. The inner C\&CG stops when the inner gap is below a tolerance value. The inner C\&CG algorithm provides a primal cut (i.e., $\bm{u}^{*}$) to the upper-level problem, which is solved by the S-MISOCP algorithm (second loop) in the outer C\&CG algorithm (first loop). The outer C\&CG stops when the outer gap is below a tolerance value and the optimal decision $\bm{w}^{*}$ is achieved.

\begin{enumerate}[label=\roman*)]
 \item \emph{The first loop, i.e., outer C\&CG}: It is considered that the inner C\&CG algorithm can find the worst-case realization of wind generation uncertainties $\bm{u}^*$ by solving the max--min subproblem, denoted as \textbf{P2}, with a fixed value of $\bm{w}^*$.
        \begin{subequations}\label{eq:Ch5aP2}\begin{align}
                \textbf{P2:}\; &\max_{\bm{u}} \min_{\bm{z},\bm{x},\overline{\bm{x}}} \mathbf{e}^{\top} \bm{x} \\
                 s.t. \hspace{1mm}& \eqref{eq:Ch5aMiddleMIL}, \eqref{eq:Ch5aLowercone1}-\eqref{eq:Ch5aLowercone2}, \\
                 & \bm{E}\bm{x}+\bm{G}\bm{z} \ge \bm{F}- \bm{Q}\bm{u}-\bm{P}\bm{w}^{*}.
        \end{align}\end{subequations}

        The master problem of outer C\&CG, denoted as \textbf{P1} at R\textsuperscript{th} iteration is defined in \eqref{eq:Ch5aP1}, where $\eta$ is the maximum RC caused by $\bm{u}^{*^r}$, and indices $r \in \{1...R\}$, and $v \in \mathcal{P \cup L}$.
        \begin{subequations}\label{eq:Ch5aP1}\begin{align}
                \textbf{P1:}\; & \min_{\bm{y},\bm{w},\bm{\eta},\overline{\bm{y}},\bm{x}^r,\bm{z}^r,\overline{\bm{x}}^r}  f(\bm{w}) + \eta           \\
                s.t. \hspace{1mm} & \eqref{eq:Ch5aUpperMIL}-\eqref{eq:Ch5aUppercone2}, \label{eq:Ch5aMP1}\\
                & \eta \ge \bm{e}^{\top}\bm{x}^r, \; \forall r \label{eq:Ch5aMP2}\\
                & \bm{E}\bm{x}^r+\bm{G}\bm{z}^r \ge \bm{F}- \bm{Q} \bm{u}^{*^r} -\bm{P} \bm{w}, \; \forall r, \label{eq:Ch5aMP3}\\
                & \lvert\lvert \bm{H}_{v,t}\bm{x}^r \rvert\rvert_2  \le \bm{h}_{v,t}\bm{x}^r, \;\forall v,t,r,  \label{eq:Ch5aMP4}\\
                & \lvert\lvert \bm{M}_{v,t}(\overline{\bm{x}}^r)\bm{x}^r+\bm{N}_{v,t}(\overline{\bm{x}}^r) \rvert\rvert_2 \le
                \bm{m}_{v,t}(\overline{\bm{x}}^r)\bm{x}^r+\bm{n}_{v,t}(\overline{\bm{x}}^r), \; \forall v,t,r. \label{eq:Ch5aMP5}
        \end{align}\end{subequations}

   \item  \emph{The second loop, i.e., S-MISOCP algorithm for \textbf{P1}:}  Algorithm~\ref{Ch5S-MISOCP} starts with finding an initial vector $\overline{\bm{y}}$ and $\overline{\bm{x}}^r$, which can be obtained by solving the relaxed \textbf{P1} as follows
        \begin{subequations}\label{eq:Ch5aMP0}\begin{align}
                \min_{\bm{y},\bm{w},{\eta},\bm{x}^r,\bm{z}} & f(\bm{w}) + \eta  \\
                s.t. \hspace{1mm}& \eqref{eq:Ch5aUpperMIL}-\eqref{eq:Ch5aUppercone1}, \eqref{eq:Ch5aMP2}-\eqref{eq:Ch5aMP4}.
        \end{align}\end{subequations}

       Not that $\overline{\bm{y}}$ and $\overline{\bm{x}}^r$ may yield violations in \eqref{eq:Ch5aMP5} owing to the missing nonconvex half of each quadratic inequality pair when solving the relaxed \textbf{P1}. Therefore, auxiliary variables are added to convexify \eqref{eq:Ch5aMP5}, and their weighted sum is added in the objective function of the penalized \textbf{P1} \eqref{eq:Ch5aMPK} with a changeable penalty coefficient. Algorithm~\ref{Ch5S-MISOCP} is adopted to enhance the feasibility of the solution of the convexified problem with respect to the original problem.
        \begin{subequations}\label{eq:Ch5aMPK}\begin{align}
                \min_{\substack{\bm{y},\bm{w},\bm{\eta},\overline{\bm{y}} \\ \bm{x}^r,\bm{z}^r,\overline{\bm{x}}^r}} & f(\bm{w}) + \eta + \sum_{t} \sum_{v} \big[ \tau_{v,t}^0 s_{v,t}^0 + \sum_{r} \tau_{v,t}^r s_{v,t}^r \big]\;\; \\
                        s.t. \hspace{1mm} & \eqref{eq:Ch5aUpperMIL}-\eqref{eq:Ch5aUppercone1}, \eqref{eq:Ch5aMP2}-\eqref{eq:Ch5aMP4},  \\
                        & s_{v,t}^0 \ge 0, \forall v,t, \; \; s_{v,t}^r \ge 0,\forall v,t,r  \\
                        & \lvert\lvert \bm{B}_{v,t}(\overline{\bm{y}}) \bm{y}+\bm{D}_{v,t} (\overline{\bm{y}}) \rvert\rvert_2  \le \bm{b}_{v,t} (\overline{\bm{y}}) \bm{y}+ \bm{d}_{v,t}(\overline{\bm{y}}) +  s_{v,t}^0, \forall v,t \\
                        & \lvert\lvert \bm{M}_{v,t} (\overline{\bm{x}}^r)\bm{x}^r+ \bm{N}_{v,t} (\overline{\bm{x}}^r) \rvert\rvert_2  \le \bm{m}_{v,t} (\overline{\bm{x}}^r) \bm{x}^r +\bm{n}_{v,t} (\overline{\bm{x}}^r) + s_{v,t}^r, \forall v,t,r.
                \end{align}\end{subequations}

    Compared with the standard P-CCP introduced in \cite{R35}, where a global penalty coefficient $\tau$ is selected for all the convexified constraints, each convexified constraint is assigned with its own penalty coefficient, and  an adaptive rule designed for Algorithm~\ref{alg2} is employed to update penalty coefficients. This allows us to better capture the impact of slack variables on the objective and to facilitate convergence.  The relative constraint violation ($RCV$), which can be calculated by
        \begin{gather}
            RCV_{v,t}^0 = s_{v,t}^0 /\Big(\bm{b}_{v,t} (\overline{\bm{y}}) \bm{y}+ \bm{d}_{v,t}(\overline{\bm{y}})\Big), \;\; \forall v,t \label{eq:Ch5aTau1}\\
            RCV_{v,t}^r = s_{v,t}^r /\Big(\bm{m}_{v,t} (\overline{\bm{x}}^r) \bm{x}^r +\bm{n}_{v,t} (\overline{\bm{x}}^r)\Big), \;\; \forall v,t,r. \label{eq:Ch5aTau2}
        \end{gather}
        is assigned to the adaptive penalty growth rate equation defined in \eqref{eq:Ch3Adaptive} to update the Algorithm~\ref{Ch5S-MISOCP} penalties in Step $5$.

   \item \emph{The third and fourth loops, i.e., solving the \textbf{P2}}:  Problem \textbf{P2} indicates a bi-level programming with integer decision variables in the inner level, which needs to be decomposed into two subproblems and solved iteratively. With fixed values of $\bm{w}^*$ and $\bm{u}^*$, the inner-level problem of \textbf{P2}, denoted as \textbf{P4}, is as follows
        \begin{subequations}\label{eq:Ch5aSPs}\begin{align}
            \textbf{P4:} \; &\min_{\bm{x},\bm{z},\overline{\bm{x}}} \bm{e}^{\top} \bm{x}  \\
            s.t. \hspace{1mm} & \eqref{eq:Ch5aLowercone1}-\eqref{eq:Ch5aLowercone2}, \nonumber\\
             & \bm{E} \bm{x} +\bm{G}\bm{z} \ge \bm{F}- \bm{Q} \bm{u}^* -\bm{P} \bm{w}^*.
        \end{align}\end{subequations}

        Note that an initial vector is needed to formulate the approximated SOC constraint \eqref{eq:Ch5aLowercone2} in problem \textbf{P4}, which can be obtained by solving the relaxed \textbf{P4} \eqref{eq:Ch5aSPs0}.
         \begin{subequations}\label{eq:Ch5aSPs0}\begin{align}
                        \min_{\bm{x},\bm{z}} &  \; \bm{e}^{\top} \bm{x}  \\
                        \hspace{1mm} s.t. \hspace{1mm} & \eqref{eq:Ch5aLowercone1} , \\
                        & \bm{E}\bm{x}+\bm{G}\bm{z} \ge \bm{F}- \bm{Q}\bm{u}^*-\bm{P}\bm{w}^*.
                \end{align}\end{subequations}

            For the same reason of adding the slack variables in \eqref{eq:Ch5aMPK}, $s_{v,t}$ is added to the penalized \textbf{P4} \eqref{eq:Ch5aSPsK}.
                \begin{subequations}\label{eq:Ch5aSPsK}\begin{align}
                        \min_{\bm{x},\bm{z},\overline{\bm{x}}} & \; \bm{e}^{\top} \bm{x} + \sum_{t} \sum_{v} \tau_{v,t} s_{v,t}  \\
                        s.t. \hspace{1mm} & \eqref{eq:Ch5aLowercone1} , \nonumber\\
                         &\bm{E}\bm{x}+\bm{G}\bm{z} \ge \bm{F}- \bm{Q}\bm{u}^*-\bm{P}\bm{w}^*, \; \; s_{v,t} \ge 0, \forall v,t, \\
                         &\lvert\lvert \bm{M}_{v,t} (\overline{\bm{x}})\bm{x} +\bm{N}_{v,t} (\overline{\bm{x}}) \rvert\rvert_2  \le \bm{m}_{v,t} (\overline{\bm{x}}) \bm{x}+\bm{n}_{v,t} (\overline{\bm{x}}) + s_{v,t}, \forall v,t.
                \end{align}\end{subequations}
                where the $RCV$ can be calculated by
        \begin{gather}
            RCV_{v,t} = s_{v,t}^r /\Big(\bm{m}_{v,t} (\overline{\bm{x}}) \bm{x} +\bm{n}_{v,t} (\overline{\bm{x}})\Big), \; \forall v,t. \label{eq:Ch5aTau3}
        \end{gather}
        \par
        To express the objective of \eqref{eq:Ch5aSPsK} in a more compact manner, $[\bm{x}^\top \, \bm{s}^\top]^\top$ is aggregated as a new vector, denoted as $\bm{\alpha}$, rendering the tri-level form of \textbf{P2} as
                \begin{subequations}\label{eq:Ch5aMPs}\begin{align}
                        \max_{\bm{u}} \;  & \min_{\bm{z},\overline{\bm{x}}} \; \min_{\bm{\alpha}} \;  \widetilde{\bm{e}}^{\top} \bm{\alpha}    \\
                        s.t. \hspace{1mm}  & \eqref{eq:Ch5aMiddleMIL}   \\
                         & \widetilde{\bm{E}} \bm{\alpha} +\widetilde{\bm{G}}\bm{z} \ge \widetilde{\bm{F}} -\widetilde{\bm{Q}} \bm{u}^* -\widetilde{\bm{P}} \bm{w}^* \; : \bm{\lambda},  \\
                         & \lvert\lvert \widetilde{\bm{H}}_{j,t}\bm{\alpha} \rvert\rvert_2  \le \widetilde{\bm{h}}_{j,t} \bm{\alpha} \; : \bm{\beta}_{j,t},\bm{\omega}_{j,t}  , \; \forall j,t,  \\
                         & \lvert\lvert \widetilde{\bm{M}}_{j,t} \bm{\alpha} +\widetilde{\bm{N}}_{j,t} \rvert\rvert_2 \le  \widetilde{\bm{m}}_{j,t} \bm{\alpha} +\widetilde{\bm{n}}_{j,t} \; : \bm{\theta}_{j,t}, \bm{\sigma}_{j,t}, \forall j,t.
                \end{align}\end{subequations}

        With the results of \eqref{eq:Ch5aSPsK}, the inner-level minimization problem of \eqref{eq:Ch5aMPs} is ready to be dualized, and \eqref{eq:Ch5aMPs} will be reformulated as the single-level maximization problem \eqref{eq:Ch5aMPsFinal}, denoted as \textbf{P3}.
        \begin{subequations}\label{eq:Ch5aMPsFinal}\begin{align}
            \textbf{P3:} \; & \max_{\bm{u},\bm{\lambda},\bm{\beta},\bm{\omega}, \bm{\theta},\bm{\sigma},\bm{\gamma},\bm{\phi}} \bm{\phi}   \\
            s.t. \hspace{1mm} & \eqref{eq:Ch5aMiddleMIL}, \\
             & \widetilde{\bm{e}} = \widetilde{\bm{E}}^{\top} \bm{\lambda}^r + \sum_t \sum_v \Big[\widetilde{\bm{H}}_{v,t}^{\top} \bm{\beta}_{v,t}^r + \widetilde{\bm{h}}_{v,t}^{\top} \bm{\omega}_{v,t}^r + \widetilde{\bm{M}}_{v,t}^{r\top} \bm{\theta}_{v,t}^r + \widetilde{\bm{m}}_{v,t}^{r^{\top}} \bm{\sigma}_{v,t}^r \Big], \; \forall r, \\
              &\bm{\phi} \le \bm{\lambda}^{r^{\top}} \Big[ \widetilde{\bm{F}}-\widetilde{\bm{G}}\bm{z}^r -\widetilde{\bm{Q}}\bm{u}-\widetilde{\bm{P}}\bm{w}^* \Big] - \sum \bm{\gamma}^r - \nonumber\\
              & \hspace{14em} \sum_t \sum_v \Big[ \widetilde{\bm{N}}_{v,t}^{r^{\top}} \bm{\theta}_{v,t}^r + \widetilde{\bm{n}}_{v,t}^{r^{\top}} \bm{\sigma}_{v,t}^r \Big], \; \forall r, \\
              & \lVert \bm{\beta}_{v,t}^r \rVert_2 \le \bm{\omega}_{v,t}^r, \; \;\lVert \bm{\theta}_{v,t}^r \rVert_2 \le \bm{\sigma}_{v,t}^r, \forall v,t,r\;\; \bm{\lambda}^r \ge 0,\; \forall r,  \\
              & -\overline{M}\bm{u} \le \bm{\gamma}^r \le \overline{M}\bm{u}, \; \forall r,\\
               & -\overline{M}(1-\bm{u}) \le \widetilde{\bm{Q}}^{\top} \bm{\lambda}^r - \bm{\gamma}^r \le \overline{M}(1-\bm{u}), \; \forall r.
                \end{align}\end{subequations}
    where $\bm{\gamma}^r$ is the auxiliary variables to linearize the
    bilinear terms in \textbf{P3}, indices $r \in \{1...R\}$, and $v \in
    \mathcal{P \cup L}$..
\end{enumerate}

  \begin{algorithm}[!ht]
\caption{The NC\&CG Algorithm for RO model}
\label{Ch5NCCG}
\begin{algorithmic}[1]
    \STATE Select an arbitrary feasible \textsuperscript{a} $\bm{w}^{*}$ /  \textsuperscript{b} $\bm{u}^{*}$, set convergence parameters $\;\varepsilon, \; UB=\infty, \; LB=-\infty, \; R=0 $, go to Step $4$.\\

    \STATE      \textsuperscript{a} Call \textbf{Algorithm}~\ref{Ch5S-MISOCP} to solve problem \textbf{P1} \eqref{eq:Ch5aP1},  update $\bm{w}^*,\; LB=\textbf{P1}^*$, and $Gap=(UB-LB)/LB$. \\
        \textsuperscript{b} Solve problem \textbf{P3} \eqref{eq:Ch5aMPsFinal} to update $\bm{u}^*,\; UB=\textbf{P3}^*$ and $Gap=(UB-LB)/UB$.\\

    \STATE If $Gap \le \varepsilon \; \& \;$ same \textsuperscript{a} $\bm{w}^{*}$ / \textsuperscript{b} $\bm{u}^{*}$, terminate; else, go to Step $4$. \\

    \STATE  \textsuperscript{a} Call \textbf{Algorithm}~\ref{Ch5NCCG} to solve problem \textbf{P2} \eqref{eq:Ch5aP2}, and  update $\bm{u}^*,\; UB=\min\{UB, \textbf{P3}^*+f(\bm{w}^*)\}$.  \\
    \textsuperscript{b} Call \textbf{Algorithm}~\ref{Ch5S-MISOCP} to solve problem \textbf{P4} \eqref{eq:Ch5aSPs}, and update $\bm{z}^*,\overline{\bm{x}}, \; LB=\max\{LB, \mathbf{P4}^*\}$.\\

    \STATE  Calculate $Gap=(UB-LB)/UB$, If $Gap \le \varepsilon$, terminate; else, $R=R+1$, \textsuperscript{a} $\bm{u}^{*^r}=\bm{u}^{*}$ / \textsuperscript{b} $\bm{z}^{*^r}=\bm{z}^{*},\overline{\bm{x}}^r =\overline{\bm{x}}$, and create new matrices for \eqref{eq:Ch5aMPs}, go Step $2$.

    ${}^a\,$For the main Problem \eqref{eq:Ch5aCompact}; ${}^b\,$For Problem \textbf{P2} \eqref{eq:Ch5aP2}
\end{algorithmic}\vspace{-5.8mm}
\end{algorithm}

    As aforementioned, there are four loops in the developed algorithm, where the first and third loops are standard C\&CG procedures and their convergence property has been justified by  \cite{zhao2012exact}, and the rest two loops are P-CCP with binary variables, i.e. S-MISOCP algorithm, to identify feasible solutions for the optimal gas-power flow (OGPF) problem in different decision stages. Its convergence property has been discussed in Section \ref{sec:SCPMethod}. Directional binaries obtained from the relaxed problems \textbf{P1} and \textbf{P4} would remain fixed after the first few iterations, which is consistent with the observation in \cite{he2017decentralized}. Therefore, the binary variables can be fixed after the beginning iterations, which is tuned as $5$ in this work. Then, the original MISOCP model can be converted into an SOCP with fixed binary variables, and Algorithm~\ref{Ch5S-MISOCP} convergence can be guaranteed.

\section{Two-stage Distributionally Robust Gas Contracting} \label{sec:Ch5DRO}

    Most of the recent studies only consider the modeling of firm gas contracts in power system operation \cite{wang2017strategic, liu2009security}, where the reserved gas contracts and the impacts of uncertainty on the contracts are missing. In \cite{chen2017clearing}, a day-ahead market clearing model for IEGS is presented, where the reserved gas amounts are introduced and effectively priced. According to current gas contracting mechanism, though the gas fuel for the firm and part of the reserved outputs of GPUs follows the day-ahead contracts, which are much cheaper than the real-time ones \cite{bouras2016using}, flexible real-time contracts may still be signed for the low-probability utilized reserved GPU outputs in practice, as the corresponding costs are \textit{wait-and-see} rather than \textit{here-and-now}. The common treatment is to minimize the real-time gas adjustments in the IEGS optimization models, while they are not optimized with the day-ahead gas contracts \cite{he2017robustb,zhao2016unit}. In addition, the emerging P2G technologies not only provide an alternative way to accommodate the excessive wind power generation, but also create new opportunities in energy trading markets, as P2G contracts have to be signed for the injected gas from external gas sources.

    To determine the optimal power system operation strategy, various advanced optimization approach based decision-making frameworks have been proposed, including stochastic optimization (SO) \cite{zhang2016electricity,chen2017clearing}, robust optimization (RO) \cite{he2018co}, and distributionally robust optimization (DRO) \cite{he2019distributionally, wang2018risk} based ones, where the uncertainties in the first two approaches are described by deterministic distributions and uncertainty sets, respectively, and the ambiguity sets are constructed to represent the distribution of uncertainties in the last one. The SO based approaches can generate a high-quality solution as long as the prior distribution of the uncertainties is close enough to the actual one. However, they may not be suitable for determining gas contracts, as the high-loss rare events are hardly to be captured and purchase gas in real-time operation could be costly \cite{bouras2016using}. The solutions from RO based approaches could cover all the possible realizations of uncertainties within the prescribed uncertainty set, which is promising in security or reliability oriented applications, nevertheless, it may lead to over-conservative gas contracts as they care about the performance under the worst-case scenario.

Besides the modeling of gas contracts, the nonlinear Weymouth equations of the gas network also admit computational challenges, due to their non-convexity in the day-ahead and real-time stages. The quadrable-loop procedure proposed in Section~\ref{sec:4LOOPS} that is based on the S-MISOCP algorithm and the NC\&CG algorithm, is the most efficient method to tackle a two-stage RO model for power system operation with gas system constraints. Consequently, reformulating the two-stage DRO into RO model is main task in the proposed methodology of this section.

To bridge the gap between industrial practice and academic research on power-gas coordination, a DRO-based power system operation model is proposed, to determine the two-stage bidirectional contracting with gas systems. Compared with the literature, the salient feature of this work is that a two-stage distributionally robust model is proposed for signing bidirectional energy contracts with gas systems from the perspective of power system operator (PSO). To the best knowledge of the authors, this work is the first attempt to incorporate both the day-ahead and real-time gas contracts in power system operation.

\subsection{Mathematical Formulation}
    \subsubsection{The Two-stage Uncertainty Mitigation Procedure}

    In this work, a two-stage procedure, including the day-ahead stage and the real-time stage, is employed to mitigate the uncertainties originated from wind generation, where the uncertainty mitigation capability from controllable resources is committed and the committed capability in the day-ahead stage or additional regulation capability purchased in the real-time stage, usually costly, is utilized, respectively. It is a conventional practice in power industry as well as academic research \cite{he2017robust, chen2017clearing}.

    The operation goal of the PSO is to minimize the total costs and simultaneously meet its operation requirements. Meanwhile, the operation of power systems incorporates the bidirectional interactions with natural gas systems in both physical, say the gas demands of GPUs and generated gas by P2G units would influence the operating status of gas systems, and economic perspectives, usually through signing energy contracts. As fast-response controllable resources, the actual outputs (inputs) of GPUs (P2G units) cannot be known beforehand, considering the uncertainty of wind power generation, which indicates the gas demands (outputs) are also uncertain.

\begin{figure}[!htbp]
            \centering
            \includegraphics[width=11cm]{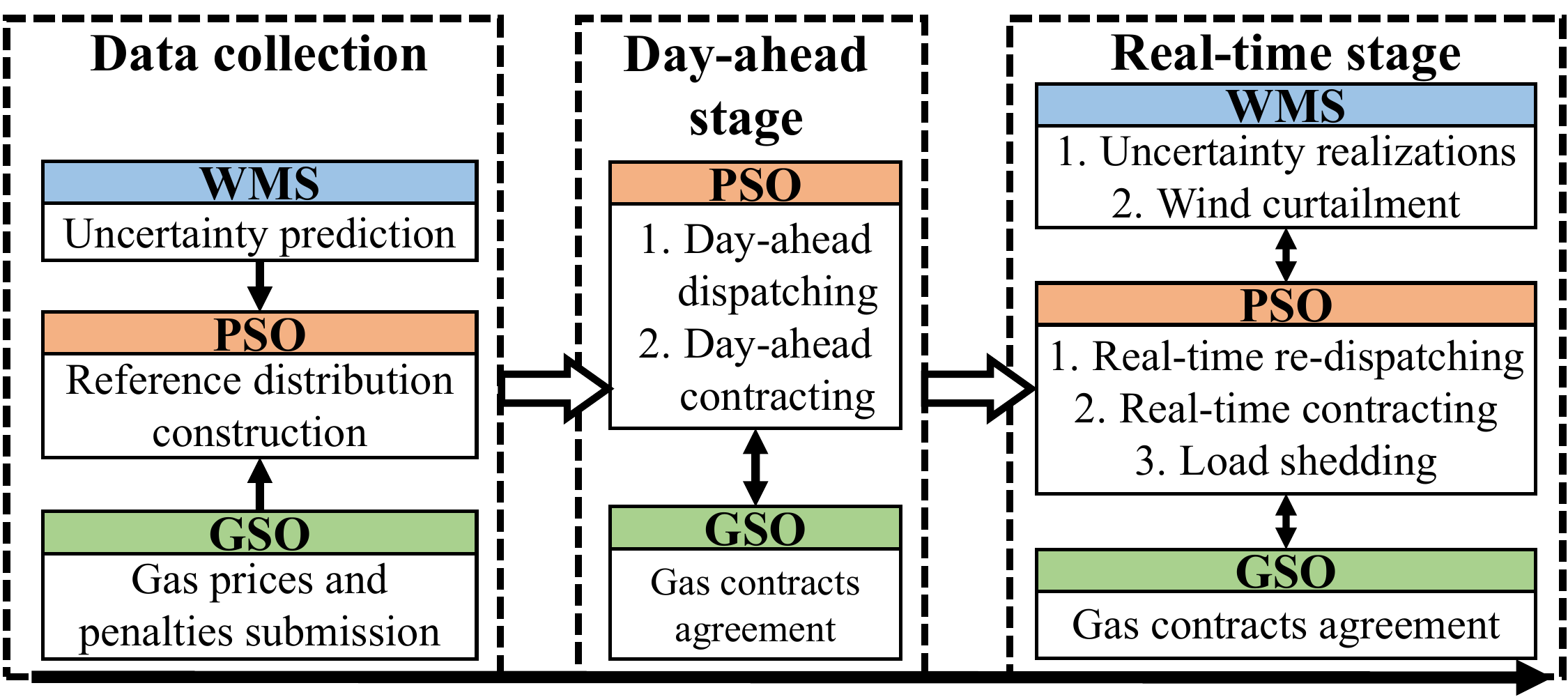}
            \caption{Decision-making process for the PSO.}
            \label{fig:Ch5F1}
    \end{figure}

    The two-stage decision-making process of the PSO considering the interactions with the gas system operator (GSO) is demonstrated in Figure~\ref{fig:Ch5F1}, which includes the following three steps: (i) the PSO constructs the reference distribution of wind power outputs based on historical data and the predicted outputs received from the wind management sector (WMS), and receives the gas prices and contract avoidance penalties from the GSO; (ii) in the day-ahead stage, the PSO commits the outputs of controllable resources according to the strategy obtained from the proposed DRO model, and communicates with the GSO for the day-ahead gas contracts agreement; (iii) in the real-time stage, the PSO adjusts the outputs of controllable resources, contacts with the GSO for real-time gas contracts agreement, and sends wind curtailment signals to the WMS.

    The overall objective of the proposed model is expressed in \eqref{eq:Ch5Obj1}, where the probability distribution $\bm{\mu}$ is uncertain and subject to a pre-defined ambiguous set $\mathcal{M}$. The operating costs (OC) in the day-ahead stage are defined in \eqref{eq:Ch5OC}, including the generation costs of non-GPUs, the day-ahead G2P contract costs, and the revenue from day-ahead P2G contracts. The regulation costs (RC) in the real-time stage is expressed by \eqref{eq:Ch5RC}, which incorporates the upward and downward adjustments costs of non-GPUs, penalties of non-served power loads, real-time G2P contract costs, penalties of wind curtailment, penalties/revenue of the shortage/surplus of real-time P2G contracts, respectively. From \eqref{eq:Ch5Obj1}, it can be observed that the goal of the PSO is to minimize the sum of the day-ahead dispatch costs and the expectation of the real-time regulation costs under the worst-case distribution. In what follows, the introductions of the constraints would be presented according to interaction types.
    \begin{gather}
        \min \;  OC \; + \; \max_{{\bm{\mu}} \in \mathcal{M}} \; {E}_{\bm{\mu}} [\min \; RC ] \label{eq:Ch5Obj1}\\
         OC = \sum_{\forall t} \Big[  \sum_{\forall u\in\mathcal{U}_n} C_u (\hat{p}_{u,t} )- \sum_{\forall j} C_j g_{j,t} + \sum_{\forall h} (\mu_h \rho_{h,t} +\mu_h^{+} \rho_{h,t}^{+} + \mu_h^{-} \rho_{h,t}^{-}) \Big]  \label{eq:Ch5OC}\\
        RC = \sum_{\forall t} \Big[ \sum_{\forall u \in \mathcal{U}_n}{} (C_u^+ \triangle p_{u,t}^+ +  C_u^- \triangle p_{u,t}^-) + \sum_{\forall d} C_d  \triangle p_{d,t} + \sum_{\forall h} (\mu_h^{2+} \rho_{h,t}^{2+} + \mu_h^{2-} \rho_{h,t}^{2-}) \nonumber\\
       + \sum_{\forall e} C_e  \triangle w_{e,t}  + \sum_{\forall j}{} C_j^{2-} \triangle g_{j,t}^- -\sum_{j} C_j^{2+} \triangle g_{j,t}^+   \Big]    \label{eq:Ch5RC}
    \end{gather}

\subsubsection{Economic Interactions: Two-stage Bidirectional Contracts Modeling}
    \begin{enumerate}
      \item {G2P Contracts Modeling}:

        As aforementioned, the G2P contracts are signed in two different time-scales, which are day-ahead and real-time, respectively. The firm gas fuel demands of GPUs, which can be calculated by \eqref{eq:Ch5G2PFirm}, are covered by the day-ahead G2P contracts. Meanwhile, the gas fuel for the outputs adjustments of GPUs in real-time operation are supported by both the day-ahead and the real-time G2P contracts. Specifically, \eqref{eq:Ch5G2PLimit1} and \eqref{eq:Ch5G2PLimit2} set non-negative limits for reserved gas contracts in day-ahead and real-time stages; \eqref{eq:Ch5G2PRes1} and \eqref{eq:Ch5G2PRes2} give upper and lower boundaries of the real-time bidirectional gas fuel variation of GPUs, respectively.
            \begin{gather}
                \rho_{h,t}^1 = \sum_{u\in\mathcal{U}_g(n)}\Phi \hat{p}_{u,t}/\eta_u, \ \forall h,t,  \label{eq:Ch5G2PFirm}\\
                \rho_{h,t}^{1-}, \rho_{h,t}^{1+} \geq 0, \forall h,t, \label{eq:Ch5G2PLimit1}\\
                \rho_{h,t}^{2-}, \rho_{h,t}^{2+} \geq 0, \forall h,t,  \label{eq:Ch5G2PLimit2} \\
                \sum_{u \in \mathcal{U}_g(n)} \Phi (\hat{p}_{u,t} - p_{u,t})/\eta_u \le \rho_{h,t}^{1-}+ \rho_{h,t}^{2-}, \forall h,t,  \label{eq:Ch5G2PRes1}\\
                \sum_{u \in \mathcal{U}_g(n)}  \Phi (p_{u,t} - \hat{p}_{u,t})/ \eta_u \le \rho_{n,t}^{1+} + \rho_{n,t}^{2+}, \forall h,t. \label{eq:Ch5G2PRes2}
        \end{gather}

\item {P2G Contracts Modeling}:

    Similar with GPUs, P2G facilities $z\in\mathcal{Z}$ are also controllable and can be adopted to mitigate the uncertainties of wind power generation. In the day-ahead stage, the operating points of P2G facilities are tuned according to the predicted value of wind generation outputs, and the amount of gas sold to the GSO can be calculated by \eqref{eq:Ch5P2G0}, which are the day-ahead P2G contracts. In the real-time stage, the gas shortage in the day-ahead P2G contracts due to inadequate wind power outputs would be penalized, and excessive power from wind farms can be converted into gas if the operation feasibility of the gas systems holds. Therefore, the real-time P2G contracts can be signed based on \eqref{eq:Ch5P2G2a}-\eqref{eq:Ch5P2G2c}.
    \begin{gather}
        g_{j,t} = \sum_{z\in \mathcal{Z}(j)} \Phi \eta_z \hat{p}_{z,t} , \forall j,t, \label{eq:Ch5P2G0}\\
        \triangle g_{j,t}^- \ge \sum_{z \in \mathcal{Z}(j)} \Phi \eta_z (\hat{p}_{z,t}-p_{z,t}), \forall z,t,  \label{eq:Ch5P2G2a} \\
        \triangle g_{j,t}^+ \le \sum_{z \in \mathcal{Z}(j)} \Phi \eta_z (p_{z,t}-\hat{p}_{z,t}), \forall z,t,  \label{eq:Ch5P2G2b}\\
        \triangle g_{j,t}^- , \triangle g_{j,t}^+ \ge 0, \forall j,t.  \label{eq:Ch5P2G2c}
    \end{gather}
\end{enumerate}

\subsubsection{Physical Interactions: Operation Constraints}
    In the power network, the operational constraints are derived from Sections~\ref{sec:Ch2SDC}, where UC of all the generators are assumed to be predetermined, the generation capacities and physical limits are considered and the hat symbol is used for day-ahead decision variables. They are composed by
        \begin{gather}
         \underline{P}_{u} c_{u,t} \le \hat{p}_{u,t} \le \overline{P}_{u} c_{u,t}, \; \forall u,t, \label{eq:Ch5PowerPre1} \\
         -\overline{R}_{u}^- \le \hat{p}_{u,t} -\hat{p}_{u,t-1} \le \overline{R}_{u}^+, \; \forall u,t, \label{eq:Ch5PowerPre2} \\
        \sum_{u \in \mathcal{U}(n)}\hat{p}_{u,t} + \sum_{e \in \mathcal{E}(n)}\hat{W}_{e,t} + \sum_{l \in \mathcal{L}_1(n)}\hat{p}_{l,t} -\sum_{l \in \mathcal{L}_2(n)}\hat{p}_{l,t} =  \sum_{z \in \mathcal{Z}(n)}\hat{p}_{z,t} + \sum_{d \in \mathcal{D}_p(n)} P_{d,t}, \; \forall n,t.  \label{eq:Ch5PreNode} \\
        -\tilde{\pi} \le \hat{\theta}_{n,t} \le \tilde{\pi},\; \forall n\in \mathcal{N}-1,t, \;\;\hat{\theta}_{1,t}=0,\; \forall t, \;\;  \label{eq:Ch5PreAngle} \\
        -\overline{p}_{l,t} \le \hat{p}_{l,t} \le \overline{p}_{l,t},\; \forall l,t.   \label{eq:Ch5PreFlow}\\
         \hat{p}_{l,t} = \frac{{\hat{\theta}}_{m,t}-\hat{\theta}_{n,t}}{x_l} ,\; \forall l,t, (m,n) \in l. \label{eq:Ch5PreOPF1}
        \end{gather}

    The operation constraints of the gas network are represented by \eqref{eq:Ch5Wellpre}--\eqref{eq:Ch5WeyPre} that is derived from the dynamic-state gas flow model presented in Section \ref{sec:Ch2SGasDynamic}.

    \begin{align}
           & \text{Gas production capacities: \eqref{eq:Ch2Well},} \label{eq:Ch5Wellpre}\\
           & \text{Gas nodal balancing equation: } \nonumber \\
           & \hspace{3em}\sum_{p \in \mathcal{P}_1(i)} \hat{f}_{p,t}^{out} - \sum_{p \in \mathcal{P}_2(i)} \hat{f}_{p,t}^{in} + \sum_{c \in \mathcal{C}_1(i)} \hat{f}_{c,t}^{out} - \sum_{c \in \mathcal{C}_2(i)} \hat{f}_{c,t}^{in} + \sum_{z \in \mathcal{Z}(i)} \hat{\varrho}_{z,t} + \sum_{w \in \mathcal{W}(i)} \hat{f}_{w,t}  \nonumber \\
           & \hspace{8em}= \sum_{u \in \mathcal{U}_g(i)} \rho_{u,t} + \sum_{d \in \mathcal{D}_g(i)} F_{d,t}, \; \forall i,t \label{eq:Ch5PGc}\\
           & \text{Gas compressors constraints: \eqref{eq:Ch2Comp1}--\eqref{eq:Ch2Comp2},} \label{eq:Ch5DAG2}\\
           & \text{Nodal pressure bounds: \eqref{eq:Ch2Pressure},} \label{eq:Ch5DAG3}\\
           & \text{Average flow rate equation: \eqref{eq:Ch2AveFlow},} \label{eq:Ch5DAG4}\\
           & \text{Mass flow equation: \eqref{eq:Ch2GMMass1},} \label{eq:Ch5DAG5}\\
           & \text{Continuity equation: \eqref{eq:Ch2GMMass2},} \label{eq:Ch5DAG6}\\
           & \text{GPU gas consumption: \eqref{eq:Ch3GFC-GUF},} \label{eq:Ch5DAG8}\\
           & \text{P2G gas production: \eqref{eq:Ch2P2G},} \label{eq:Ch5AveFlowPre}\\
           & \text{Weymouth equation: \eqref{eq:Ch2Wey}.} \hspace{15em} \label{eq:Ch5WeyPre}
        \end{align}
        where $\mathcal{Z}(i)$ is a subset of P2G units connected with node $i$, and $\mathcal{Z}(i)/\mathcal{H}(i)$ is a subset of P2G units/gas contracts, whose GPUs are supplied from node $i$.

    It should be noted that \eqref{eq:Ch5PowerPre1}--\eqref{eq:Ch5WeyPre} are day-ahead constraints for the coupled energy system. Further, most of the operation constraints in the real-time stage can be obtained by replacing the day-ahead decision variables with real-time ones in \eqref{eq:Ch5PowerPre1}--\eqref{eq:Ch5WeyPre} except \eqref{eq:Ch5PreNode}, namely removing the hat symbols of the decision variables in those constraints. Such overlapped constraints are not listed. In real-time operation of the power network, wind generation curtailment $\triangle w_{e,t}$ and electrical load shedding $\triangle p_{d,t}$ are also practical means to recover the power balancing condition, whose adjustment ranges are shown in \eqref{eq:Ch5RT1c}. Meanwhile, the nodal power balancing condition should be modified by adding the wind generation curtailment and electrical load shedding terms, resulted in \eqref{eq:Ch5RT1d}. To quantify the regulation costs of non-GPUs in the real-time stage, \eqref{eq:Ch5RT1g} is added to describe the outputs adjustment of non-GPUs.
        \begin{gather}
        0\le \triangle w_{e,t} \le W_{e,t},\forall e,t;\;\:\;\: 0 \le \triangle p_{d,t} \le P_{d,t},\forall d,t, \label{eq:Ch5RT1c} \\
        \sum_{u \in \mathcal{U}(n)}{p}_{u,t} + \sum_{e \in \mathcal{E}(n)}({W}_{e,t}-\triangle w_{e,t}) + \sum_{l \in \mathcal{L}_1(n)}{p}_{l,t} -\sum_{l \in \mathcal{L}_2(n)}{p}_{l,t} \nonumber \\
        =  \sum_{z \in \mathcal{Z}(n)}{p}_{z,t} + \sum_{d \in \mathcal{D}_p(n)} (P_{d,t}-\triangle p_{d,t}), \; \forall n,t. \label{eq:Ch5RT1d}\\
        -\triangle p_{u,t}^-\le p_{u,t}- \hat{p}_{u,t} \le \triangle p_{u,t}^+, \triangle p_{u,t}^- \ge 0, \triangle p_{u,t}^+ \ge 0,\forall t,u \in \mathcal{U}_n \label{eq:Ch5RT1g}
    \end{gather}

\subsubsection{Ambiguity Set Construction} \label{sec:Ambiguity}
    Given a set of historical data of wind generation outputs, they can be clustered as $K$ scenarios $\{\bm{W}_1,\bm{W}_2,...,\bm{W}_K\}$ by sample clustering or scenario reduction methods, and the probability weight coefficient $\mu^0_k$ for scenario $\bm{W}_k$ can be obtained as well. Then, the empirical distribution can be established as $\bm{\mu}^0=\{\mu_1^0,\mu_2^0,...,\mu_k^0\}$. However, the true distribution $\bm{\mu}=\{\mu_1,\mu_2,...,\mu_K\}$ may be different from $\bm{\mu}^0$, due to the lack of historical data in practical situations. Therefore, the ambiguity set is constructed using $L_\infty$ norm \cite{zhao2015data}, as shown in \eqref{eq:Ch5set}, which suggests the statistical distance between the true distribution and the empirical one would always be smaller than the tolerance along with change of the size of the data sample set adaptively.
    \begin{align}
        \mathcal{M} & = \left\{ \bm{\mu} \in \mathbb{R}_+^K | \;\lVert \bm{\mu}-\bm{\mu}^0 \rVert_\infty \le \sigma,\; \sum_{\forall k} \mu_k =1 \right\} \nonumber\\
                & = \left\{ \mu_k | \; \mu_k \ge 0, \; \max_{1 \le k \le K}\vert \mu_k - \mu_k^0 \vert \le \sigma, \; \sum_{\forall k} \mu_k =1 \right\} \nonumber\\
                & = \left\{ \mu_k | \; \mu_k \ge 0, \; - \sigma \le \mu_k - \mu_k^0 \le \sigma, \; \sum_{\forall k} \mu_k =1 \right\}  \label{eq:Ch5set}
    \end{align}

In \eqref{eq:Ch5set}, the tolerance value $\sigma$ depends on the amount of historical data $S$ and confidence level $\beta$. According to proposition $2$ in \cite{zhao2015data}, $\sigma$ can be calculated by

\begin{gather}  
        \sigma = \frac{1}{2S} \log \frac{2K}{1-\beta} \label{eq:sigma}
\end{gather}

As a matter of fact, the scenario set used in SO based works \cite{zhang2016electricity,alabdulwahab2015stochastic} can be viewed as a special case of \eqref{eq:Ch5set} with $\sigma=0$. In addition, \eqref{eq:Ch5set} incorporates the finite uncertainty set employed in RO by tuning $\sigma=1$.

\subsubsection{The Holistic Model}
    The completed form of the proposed model is shown as \eqref{eq:Ch5holistic}, where $\Omega_1$ and $\Omega_2$ define the sets of decision variables for the day-ahead and real-time stages, respectively.
    \begin{subequations}\label{eq:Ch5holistic}\begin{align}
        \min_{\Omega_1} \;   &  OC \; + \; \max_{\bm{\mu} \in \mathcal{M}} \; {E}_{\bm{\mu}} [\min_{\Omega_2} \; RC ]\label{eq:Ch5holistica}\\
        s.t. \hspace{1mm}  & \text{Day-ahead constraints: \eqref{eq:Ch5PowerPre1}--\eqref{eq:Ch5WeyPre}}, \\
        &  \text{Real-time constraints: \eqref{eq:Ch5PowerPre1}--\eqref{eq:Ch5PowerPre2}, \eqref{eq:Ch5PreAngle}--\eqref{eq:Ch5RT1g}}, \\
        &  \text{Gas contracts: \eqref{eq:Ch5G2PFirm}--\eqref{eq:Ch5P2G2c}}, \\
        &  \text{Ambiguity set: \eqref{eq:Ch5set}}, \\
            & \Omega_1 = \{\rho_{h,t}^1,\rho_{h,t}^{1+},\rho_{h,t}^{1-}, g_{j,t}, \hat{p}_{u,t}, \hat{p}_{z,t}, \hat{f}_{l,t}, \hat{\theta}_{n,t}, \hat{f}_{w,t} , \hat{\pi}_{i,t} , \hat{f}_{c,t}^{in}, \hat{f}_{c,t}^{out}, \hat{f}_{p,t}^{in} , \hat{f}_{p,t}^{out} , \hat{f}_{p,t}, \hat{m}_{p,t}  \}, \\
        & \Omega_2 = \{\triangle p_{u,t}^+, \triangle p_{u,t}^-, \rho_{h,t}^{2+}, \rho_{h,t}^{2-}, \triangle g_{j,t}^+ \triangle g_{j,t}^-, \triangle w_{e,t}, \triangle p_{d,t},  {p}_{u,t},{p}_{z,t}, {f}_{l,t}, {\theta}_{n,t}, {f}_{w,t} , {\pi}_{i,t}, \nonumber\\
         & \hspace{3em} {f}_{c,t}^{in} , {f}_{c,t}^{out} , {f}_{p,t}^{in} , {f}_{p,t}^{out} , {f}_{p,t}, {m}_{p,t}  \}.
 \end{align}\end{subequations}

    \par
    The above model is not readily solvable by commercial solvers due to the presence of nonconvex constraints caused by Weymouth equations \eqref{eq:Ch5WeyPre} and intractable objective function \eqref{eq:Ch5holistica}. Therefore, tractable reformulations for the Weymouth equations and the model objective are derived, before introducing the solution approach for the proposed model.

\subsection{Solution Methodology}
\subsubsection{Tractable Reformulations of the Proposed Model}
\begin{enumerate}
  \item {MISOCP-based Approximation for the Weymouth Equations}:

    Weymouth equations can be formulated as DCP constraints, which further are reformulated as MISOCP constraints.  By introducing a binary variable $\hat{z}_{p,t}$ and a pair of auxiliary variables $\hat{\pi}^+_{p,t},\hat{\pi}^-_{p,t}$, the sign function in \eqref{eq:Ch5WeyPre} can be removed, resulting the equivalent form as follows.
    \begin{gather}
        \left\{\forall t, p\in \mathcal{P}^\pm \; | \;
          (1-\hat{z}_{p,t}) \overline{M} \le (\hat{\pi}_{i,t} - \hat{\pi}_{o,t} )\| \hat{f}_{p,t} \le \hat{z}_{p,t} \overline{M}, \right. \label{eq:Ch5WeyDirc} \\
        (1-\hat{z}_{p,t}) \overline{M} \le (\hat{\pi}^+_{p,t} - \hat{\pi}_{i,t}) \| (\hat{\pi}^-_{p,t} - \hat{\pi}_{o,t})\le (1-\hat{z}_{p,t}) \overline{M}, \\
        \left. \hat{z}_{p,t} \overline{M} \le (\hat{\pi}^+_{p,t} - \hat{\pi}_{o,t}) \| (\hat{\pi}^-_{p,t} - \hat{\pi}_{i,t})\le \hat{z}_{p,t} \overline{M} \right\}, \label{eq:Ch5WeyDircEnd}\\
        \hat{\pi}^+_{p,t} = \hat{\pi}_{i,t}, \;\; \hat{\pi}^-_{p,t} = \hat{\pi}_{o,t},\;\; \forall t, p\in \mathcal{P}/ \mathcal{P}^\pm,     \label{eq:Ch5WeyNoDirc}   \\
        \hat{f}_{p,t}^2 = \chi_p^f (\hat{\pi}^{+^2}_{p,t} - \hat{\pi}^{-^2}_{p,t}), \; \forall p,t. \label{eq:Ch5WeyReform}
    \end{gather}

    In \eqref{eq:Ch5WeyDirc}, $\mathcal{P}^\pm$ is the subset of pipelines that have bidirectional gas flow, $\overline{M}$ is a sufficient large positive number, please refer to Section~\ref{sec:Ch3GFCMethod} for a detailed formulation, notation $\underline{C} \le a \| b \le \overline{C}$ represents that $a$ and $b$ have the same boundaries, i.e., $\underline{C} \le a\le\overline{C}$ and $\underline{C}\le b \le\overline{C}$.  For fixed flow pipelines, $\hat{\pi}^+_{p,t}$ and $\hat{\pi}^-_{p,t}$ are assigned directly by \eqref{eq:Ch5WeyNoDirc} to decrease binary variables $\hat{z}_{p,t}$. Further, \eqref{eq:Ch5WeyReform} can be converted into two opposite inequalities, where the canonical form of the first inequality, namely $\hat{f}_{p,t}^2 + \chi_p^f \hat{\pi}^{-^2}_{p,t} \le \chi_p^f \hat{\pi}^{+^2}_{p,t}$, is
    \begin{gather}
        \begin{Vmatrix}
            \hat{f}_{p,t} \\ \sqrt{\chi_p^f} \hat{\pi}^{-}_{p,t}
        \end{Vmatrix}_2 \le \sqrt{\chi_p^f} \hat{\pi}^{+}_{p,t}, \; \forall p,t \label{eq:Ch5Weycone1}
    \end{gather}

    Given an initial vector $[\bar{\hat{f}}_{p,t} \;\; \bar{\hat{\pi}}^{-}_{p,t}]^{\top}$, the right-hand side of the second inequality, namely $\chi_p^f \hat{\pi}^{+^2}_{p,t} \le \hat{f}_{p,t}^2 + \chi_p^f \hat{\pi}^{-^2}_{p,t}$, can be approximated as $\hat{\Lambda}_{p,t}$, hence its canonical form would be
    \begin{gather}
        \hat{\Lambda}_{p,t} = 2 \bar{\hat{f}}_{p,t}\hat{f}_{p,t} + 2\chi_p^f \bar{\hat{\pi}}^{-}_{p,t}\hat{\pi}^{-}_{p,t} - \bar{\hat{f}}_{p,t}^2 - \chi_p^f \bar{\hat{\pi}}^{-2}_{p,t}, \;\;  \begin{Vmatrix}
            2 \sqrt{\chi_p^f} \hat{\pi}^{+}_{p,t} \\ \hat{\Lambda}_{p,t} -1
        \end{Vmatrix}_2 \le  \hat{\Lambda}_{p,t} +1, \; \forall p,t. \label{eq:Ch5Weycone2}
    \end{gather}

    Till now, an MISOCP based approximation for \eqref{eq:Ch5WeyPre} is derived, which consists of \eqref{eq:Ch5WeyDirc}--\eqref{eq:Ch5WeyReform} and \eqref{eq:Ch5Weycone1}--\eqref{eq:Ch5Weycone2}.

  \item {Equivalent Reformulation for the Objective Function}:

    As discussed in Section \ref{sec:Ambiguity}, the distribution of wind generation outputs can be approximated by $K$ clustered scenarios, therefore, the objective of the proposed model can be written as
    \begin{gather}
        \min_{\Omega_1} \;  OC \; + \; \max_{\bm{\mu} \in \mathcal{M}} \sum_{ k} \mu_k \min_{\Omega_2} \; RC(\bm{W}_k) \label{eq:Ch5Obj2}
    \end{gather}
    where $RC(\bm{W}_k)$ denotes the regulation costs under scenario $\bm{W}_k$. As all the wind generation output scenarios are independent, the summation and minimization operators can be interchanged as
    \begin{gather}
        \min_{\Omega_1} \;  OC \; + \; \max_{\bm{\mu} \in \mathcal{M}} \min_{\Omega_2} \; \sum_{k} \mu_k RC(\bm{W}_k)
    \end{gather}
        making the proposed model a standard two-stage robust program.

\end{enumerate}

\subsubsection{The Compact Form}
    \par
    For ease of exposition, the compact form of the proposed model after reformulation is written as follows.
    \begin{subequations}\label{eq:Ch5Compact} \begin{align}
        \min_{\bm{u},\bm{y},\overline{\bm{u}}} \; & f(\bm{u},\bm{y}) + \max_{\bm{\mu}\in\mathcal{M}}  \;\min_{\bm{x}_k,\bm{z}_k,\overline{\bm{x}}_k} \sum_{k} \mu_k \bm{c}^{\top}\bm{x}_k   \\
            s.t. \hspace{1mm} &\bm{Au}+\bm{By} \le \bm{C},                    \label{eq:Ch5UpperMIL}  \\
            & \lVert \bm{D}_{p,t}\bm{y} \rVert_2  \le \bm{d}_{p,t}\bm{u}, \forall p,t,   \label{eq:Ch5Uppercone1}\\
            &\lVert \bm{E}_{p,t}^{\overline{\bm{u}}}\bm{u} + \bm{F}_{p,t}^{\overline{\bm{u}}} \rVert_2 \le \bm{e}_{p,t}^{\overline{\bm{u}}} \bm{u}+ \bm{f}_{p,t}^{\overline{\bm{u}}}, \forall p,t,  \label{eq:Ch5Uppercone2}\\
            & \bm{G}\bm{x}_k + \bm{Hz}_k \ge \bm{I}- \bm{J}\bm{W}_k - \bm{K}\bm{y}, \forall k,   \label{eq:Ch5LowerMIL}\\
             & \lVert \bm{L}_{p,t}\bm{x}_k \rVert_2  \le \bm{l}_{p,t}\bm{x}_k, \forall p,t,k,  \label{eq:Ch5Lowercone1}\\
         & \lVert \bm{P}_{p,t}^{\overline{\bm{x}}_k}\bm{x}_k + \bm{Q}_{p,t}^{\overline{\bm{x}}_k} \rVert_2  \le \bm{p}_{p,t}^{\overline{\bm{x}}_k} \bm{x}_k + \bm{q}_{p,t}^{\overline{\bm{x}}_k}, \forall p,t,k,    \label{eq:Ch5Lowercone2}
    \end{align}\end{subequations}
    where, $\bm{y}= \{\rho_{h,t},\rho_{h,t}^{1+},\rho_{h,t}^{1-}, \hat{p}_{u,t}, \hat{p}_{z,t}\}$ ; $\bm{u}$ collects the rest of day-ahead decision variables in $\Omega_1$; $\bm{x}_k$ and $\bm{z}_k$ are the continuous and binary variables in the real-time stage at wind output scenario $k$, respectively; to find the optimal initial vector for the approximated cone \eqref{eq:Ch5Weycone2}, $\overline{\bm{u}}$ and $\overline{\bm{x}}_k$ are considered as decision variables in day-ahead  and real-time stages, respectively; $\bm{A}, \bm{B}$, and $\bm{C}$ are coefficients of the first stage linear constraints \eqref{eq:Ch5G2PFirm}--\eqref{eq:Ch5G2PLimit1}, \eqref{eq:Ch5P2G0}, \eqref{eq:Ch5PowerPre1}--\eqref{eq:Ch5AveFlowPre} and \eqref{eq:Ch5WeyDirc}-\eqref{eq:Ch5WeyNoDirc}. $\bm{G}, \bm{H}, \bm{I}, \bm{J}$, and $\bm{K}$ are coefficients of the second stage linear constraints \eqref{eq:Ch5G2PLimit2}--\eqref{eq:Ch5G2PRes2}, \eqref{eq:Ch5P2G2a}--\eqref{eq:Ch5P2G2c}, \eqref{eq:Ch5PowerPre1}--\eqref{eq:Ch5PowerPre2}, \eqref{eq:Ch5PreAngle}--\eqref{eq:Ch5AveFlowPre}, \eqref{eq:Ch5RT1c}--\eqref{eq:Ch5RT1g} and \eqref{eq:Ch5WeyDirc}--\eqref{eq:Ch5WeyNoDirc}; SOC constraints \eqref{eq:Ch5Uppercone1} and \eqref{eq:Ch5Uppercone2} (\eqref{eq:Ch5Lowercone1} and \eqref{eq:Ch5Lowercone2}) are driven from the proper cones \eqref{eq:Ch5Weycone1} and \eqref{eq:Ch5Weycone2} for the day-ahead (real-time) Weymouth equations, respectively.

\subsubsection{The Overall Solution Procedure}
    The quadruple-loop procedure developed in Section \ref{sec:4LOOPS} is employed to tackle the proposed DRO-based model with $K$ clusters of wind power outputs. Tough the compact form of the proposed model admits a standard two-stage robust program with binary variables and SOC constraints in both stages, which can be solved by the NC\&CG algorithm. However, the quality of the MISOCP-based approximation for the Weymouth equation, which appears in both decision-making stages, depends on the linearization point, i.e. $\overline{\bm{u}}$ and $\overline{\bm{x}}_k$, not only influencing the optimality and feasibility of the day-ahead operation strategy, but also affecting the conservativeness of the strategy through the robust counterpart of the real-time decision-making model. Thus, the subsection begins with developing a method for a high-quality linearization point.
\begin{enumerate}
  \item {Two loops based S-MISOCP Algorithm}:

        From \eqref{eq:Ch5holistic}, the proposed model is a two-stage program, which can be solved by firstly decomposing it into a master-slave structure and then calling the cutting plane based iteration methods. If the binary variables in \eqref{eq:Ch5holistic} are fixed, the major obstacle that hinders the solution efficiency is the quadratic equalities, namely the simplified Weymouth equation \eqref{eq:Ch5WeyReform}, which means both the master and slave subproblems of the proposed model would be degenerated into DCP functions \cite{R35}. An efficient algorithm, which is called P-CCP, is devised in \cite{R35} to find a high-quality local optimum for DCPs. However, due to the existence of the binary variables, the convergence of the P-CCP in the proposed model cannot be guaranteed, as the convergence proof in \cite{R35} is merely valid for continuous problems. Therefore, the S-MISOCP algorithm presented in Chapter~\ref{Chapter3} (see Algorithm~\ref{alg2}), is employed to generate a high-quality initial point for the subproblems of the DRO-based model. Before introducing the details of the S-MISOCP algorithm, the tractable approximations for the master and slave subproblems of \eqref{eq:Ch5holistic} are given as \eqref{eq:Ch5P1Penalized}--\eqref{eq:Ch5P1b}, denoted as \textbf{F1}, and \eqref{eq:Ch5P2Penalized}--\eqref{eq:Ch5P2e}, denoted as \textbf{F2}, respectively.
    \begin{subequations} \label{eq:Ch5P1Penalized}    \begin{align}
        \textbf{F1:}\;\; & \min_{\bm{x}_k,\bm{z}_k,\bm{s}\ge\vec{0}} \; \sum_{k} (\mu_k^* \bm{c}^{\top}\bm{x}_k + \sum_{\forall t} \sum_{\forall p} \tau_{p,t,k} s_{p,t,k} )  \\
   s.t. \hspace{1mm} & \eqref{eq:Ch5Lowercone1}, \\
         & \bm{G}\bm{x}_k + \bm{Hz}_k \ge \bm{I}- \bm{J}\bm{W}_k - \bm{K}\bm{y}^*, \forall k,  \label{eq:Ch5P1a}  \\
        & \lVert \bm{P}_{p,t}^{\overline{\bm{x}}_k}\bm{x}_k + \bm{Q}_{p,t}^{\overline{\bm{x}}_k} +s_{p,t,k} \rVert_2  \le \bm{p}_{p,t}^{\overline{\bm{x}}_k} \bm{x}_k + \bm{q}_{p,t}^{\overline{\bm{x}}_k} +s_{p,t,k}, \forall p,t,k,  \label{eq:Ch5P1b}
    \end{align}\end{subequations}

    \begin{subequations} \label{eq:Ch5P2Penalized}  \begin{align}
      \textbf{F2:} \;\; & \min_{\substack{\bm{u}, \bm{y}, \rho,\bm{x}_k^r, \\ \bm{z}_k^r, \bm{\xi}^0, \bm{\xi}^r}}\; f(\bm{u},\bm{y})+ \vartheta + \sum_{t} \sum_{p} (\tau_{p,t}^0 s_{p,t}^0 + \sum_{r} \sum_{k} \tau_{p,t,k}^r s_{p,t,k}^r ) \\
        \hspace{0mm} s.t. \hspace{1mm} &\eqref{eq:Ch5UpperMIL}-\eqref{eq:Ch5Uppercone1}, \\
         & \bm{G}\bm{x}_k^r + \bm{Hz}_k^r \ge \bm{I}- \bm{J}\bm{W}_k- \bm{K}\bm{y},\; \forall k,r, \label{eq:Ch5P2a} \\
        & \vartheta \ge \sum_{k} \mu_k^{r*} \bm{c}^{\top} \bm{x}_k^r,\; \forall r, \;\; s_{p,t}^0 \ge 0, \forall p,t,  \label{eq:Ch5P2b} \\
        & \lVert \bm{L}_{p,t}\bm{x}_k^r \rVert_2  \le \bm{l}_{p,t}\bm{x}_k^r,\; \forall p,t,k,r, \;\; s_{p,t,k}^r \ge 0, \forall p,t,k,r,  \label{eq:Ch5P2c} \\
         & \lVert \bm{E}_{p,t}^{\overline{\bm{u}}}\bm{u} + \bm{F}_{p,t}^{\overline{\bm{u}}} + s_{p,t}^0 \rVert_2 \le \bm{e}_{p,t}^{\overline{\bm{u}}} \bm{u}+ \bm{f}_{p,t}^{\overline{\bm{u}}} + s_{p,t}^0 , \forall p,t, \label{eq:Ch5P2d} \\
         & \lVert \bm{P}_{p,t}^{\overline{\bm{x}}_k^r}\bm{x}_k^r + \bm{Q}_{p,t}^{\overline{\bm{x}}_k^r} + s_{p,t,k}^r\rVert_2  \le \bm{p}_{p,t}^{\overline{\bm{x}}_k^r} \bm{x}_k^r +  \bm{q}_{p,t}^{\overline{\bm{x}}_k^r} + s_{p,t,k}^r, \forall p,t,k,r.  \label{eq:Ch5P2e}
    \end{align}\end{subequations}

        In \textbf{F1}, $s_{p,t,k}$ is the non-negative slack variable and $\tau_{p,t,k}$ is the penalty coefficient; the objective is to minimize the sum of the expected real-time RC with a given distribution ($\mu_k^*,\ \forall k$) and the penalized violation for \eqref{eq:Ch5Lowercone2};  \eqref{eq:Ch5P1a} is the real-time operation constraints with fixed day-ahead stage variables;  the last set of constraints is the relaxed counterpart of \eqref{eq:Ch5Lowercone2} to detect the constraint violation. Similarly, in \textbf{F2}, $s_{p,t}^0$ and $s^r_{p,t,k}$ are the non-negative slack variable added to \eqref{eq:Ch5Uppercone2} and \eqref{eq:Ch5Lowercone2}, which are parameterized with the candidate worst-case distribution set $\{\bm{\mu}^r,\ \forall r\}$, respectively, and $\tau_{p,t}^0$ and $\tau_{p,t,k}^{r}$ are the corresponding penalty coefficients; $\vartheta$ is the auxiliary variable estimating the lower bound of the real-time RC \eqref{eq:Ch5RC}, which is constrained by \eqref{eq:Ch5P2b}; $\bm{x}_k^r$ and $\bm{z}_k^r$ are the real-time decision variables under distribution $\bm{\mu}^r$; \eqref{eq:Ch5P2d} and \eqref{eq:Ch5P2e} are the relaxed counterparts of \eqref{eq:Ch5Uppercone2} and \eqref{eq:Ch5Lowercone2}, respectively; the real-time linear constraints \eqref{eq:Ch5LowerMIL} and \eqref{eq:Ch5Lowercone1} under distribution $\bm{\mu}^r$ are incorporated in \eqref{eq:Ch5P2a} and \eqref{eq:Ch5P2c}, respectively.

        The details of the proposed S-MISOCP algorithm are as follows. Compared with the P-CCP, the parameter $J^{max}_{int}$ controlling the iteration process is added in the proposed S-MISOCP algorithm, beyond which the binary variables would be fixed to their solutions in iteration $J^{max}_{int}$, to enhance the algorithmic convergence. In other words, \textbf{F1} and \textbf{F2} would degenerate to standard DCPs after iteration $J^{max}_{int}$, and the S-MISOCP algorithm would become P-CCP accordingly, indicating the guaranteed convergence of the proposed algorithm.
        \begin{algorithm}[!tbh]
\caption{The S-MISOCP Algorithm for DRO Model}
\label{Ch5bS-MISOCP}
\begin{algorithmic}[1]
    \STATE Set $\overline{\mu}, \underline{\mu}, \sigma, J^{max}, J^{max}_{int}, \epsilon, \varepsilon, j=1$ and ${}^a$$\tau_{p,t,k}$ / ${}^b$$\tau_{p,t}^0, \tau_{p,t,k}^r$.

    \STATE Solve ${}^a\,$\textbf{F1} \eqref{eq:Ch5P1Penalized} without \eqref{eq:Ch5P1b} / ${}^b\,$\textbf{F2} \eqref{eq:Ch5P2Penalized} without \eqref{eq:Ch5P2d}--\eqref{eq:Ch5P2e}.

    \STATE Set ${}^a\,$$\overline{\bm{x}}_k=\bm{x}_k$/ ${}^b\,$$\overline{\bm{u}}=\bm{u}, \overline{\bm{x}}_k^r=\bm{x}_k^r$.

    \STATE If $j> J^{max}_{int}$, parameterize the binary variables in ${}^a\,$\textbf{F1} \eqref{eq:Ch5P1Penalized}/ ${}^b\,$\textbf{F2} \eqref{eq:Ch5P2Penalized} with the solutions in iteration $J^{max}_{int}$.

    \STATE Solve ${}^a\,$\textbf{F1}/ ${}^b\,$\textbf{F2} to update: ${}^a\,$$\bm{x}_k,\bm{z}_k,\bm{s}$ / ${}^b\,$$\bm{u},\bm{x}_k^r, \bm{z}_k^r,\bm{s}^0,\bm{s}^r$.

    \STATE If ${}^a\,$\eqref{alg:A2} / ${}^b\,$\eqref{alg:B2}\; is satisfied, or $j > J^{max}$, terminate; \\ Else, go to Step 7. \begin{gather}
        \textbf{F1}^{(j-1)}-\textbf{F1}^{(j)} \le \epsilon,\;\;
        s_{p,t,k} \le \varepsilon, \;\; \forall p,t,k. \label{alg:A2}\\
        \textbf{P2}^{(j-1)}-\textbf{P2}^{(j)} \le \epsilon,\;\;
        s_{p,t}^0 \le \varepsilon, \; \forall p,t, \;\; s_{p,t,k}^r \le \varepsilon, \; \forall p,t,k,r. \label{alg:B2}
     \end{gather}\\

    \STATE Update ${}^a\,$$\tau_{p,t,k}$ / ${}^b\,$$\tau_{p,t}^0, \tau_{p,t,k}^r$  using the adaptive penalty rate equation \eqref{eq:Ch3Adaptive}, then go to Step 3.\\

     ${}^a\,$For problem \textbf{F1} \eqref{eq:Ch5P1Penalized};  ${}^b\,$For problem \textbf{F2} \eqref{eq:Ch5P2Penalized}.
\end{algorithmic}\vspace{-5.6mm}
\end{algorithm}

    \item{Two loops based NC\&CG Algorithm}:

        Prior to calling the NC\&CG algorithm, the master and slave problems in each C\&CG loop need to be identified. The inner-loop C\&CG is to solve the max-min problem \textbf{F3}:
     \begin{subequations} \label{eq:Ch5P3}\begin{align}
        \text{\textbf{F3}}: \; & \max_{\bm{\mu}\in\mathcal{M}} \;\min_{\bm{x}_k,\bm{z}_k,\overline{\bm{x}}_k} \sum_{k} \mu_k \bm{c}^{\top}\bm{x}_k \\
        \hspace{1mm} s.t. \hspace{1mm} & \eqref{eq:Ch5Lowercone1}-\eqref{eq:Ch5Lowercone2}, \\
        &\bm{G}\bm{x}_k + \bm{Hz}_k \ge \bm{I}- \bm{J}\bm{W}_k - \bm{K}\bm{y}^*, \;\forall k.
    \end{align}\end{subequations}

    To drive the master problem for \textbf{F3}, (i) slack variables $s_{p,t,k}$ are included to relax the approximated cones \eqref{eq:Ch5Lowercone2}, and the inner level of \textbf{F3} becomes \eqref{eq:Ch5P3Tri}--\eqref{eq:P3triF}, which can be more compact by introducing a new vector, denoted as $\bm{\alpha}_k = [\bm{x}_k^{\top} \,\bm{s}_k^{\top}]^{\top}$, then (ii) the problem \textbf{F3} is expressed in its tri-level form, after creating the new matrices, as follows
    \begin{subequations} \label{eq:Ch5P3Tri}\begin{align}
        \max_{\bm{\mu}\in\mathcal{M}} \;  & \min_{\bm{z}_k,\overline{\bm{x}}_k} \; \min_{\bm{\alpha}_k} \; \sum_{\forall k} \mu_k \widetilde{\bm{c}}_k^{\top} \bm{\alpha}_k  \\
        \hspace{1mm} s.t. & \widetilde{\bm{G}} \bm{\alpha}_k +\widetilde{\bm{H}}\bm{z}_k \ge \widetilde{\bm{I}}_k \; : \bm{\lambda}_k, \; \forall k, \\
         & \lVert \widetilde{\bm{L}}_{p,t}\bm{\alpha}_k \rVert_2  \le \widetilde{\bm{l}}_{p,t} \bm{\alpha}_k \; : \bm{\beta}_{p,t,k}, \bm{\gamma}_{p,t,k}, \;\forall p,t,k, \\
         & \lVert \widetilde{\bm{P}}_{p,t}^{k}\bm{\alpha}_k + \widetilde{\bm{Q}}_{p,t}^{k} \rVert_2  \le \widetilde{\bm{p}}_{p,t}^{k} \bm{\alpha}_k + \widetilde{\bm{q}}_{p,t}^{k}  \; : \bm{\vartheta}_{p,t,k}, \bm{\omega}_{p,t,k} , \forall p,t,k. \label{eq:P3triF}
    \end{align}\end{subequations}

        Consequently, the inner-level problem of  \eqref{eq:Ch5P3Tri} can be directly dualized with the primal cut ($\overline{\bm{x}}_k^{r*}$ and $\bm{z}_k^{r*}$), rendering the master problem of inner C\&CG at $\text{R}^{\text{th}}$ iteration, denoted as \textbf{F4}.
        \begin{subequations} \label{eq:Ch5P4} \begin{align}
            \text{\textbf{F4}}: \; & \max_{\bm{\mu}, \varphi, \bm{\lambda}, \bm{\beta}, \bm{\gamma}, \bm{\vartheta}, \bm{\omega}} \; \varphi   \\
        \hspace{1mm} s.t. \hspace{1mm} & \mu_k \widetilde{\bm{c}}_k^{r{\top}} = \widetilde{\bm{G}}^{\top} \bm{\lambda}_k^r +\\
        & \sum_t \sum_p \begin{pmatrix}
        \widetilde{\bm{L}}_{p,t}^{\top} \bm{\beta}_{p,t,k}^r + \widetilde{\bm{l}}_{p,t}^{\top} \bm{\gamma}_{p,t,k}^r +\widetilde{\bm{P}}_{p,t}^{k,r\top} \bm{\vartheta}_{p,t,k}^r + \widetilde{\bm{p}}_{p,t}^{k,r\top} \bm{\omega}_{p,t,k}^r
        \end{pmatrix},\; \forall k,r, \\
        &  \varphi \le \bm{\lambda}_k^{r{\top}} \left( \widetilde{\bm{I}}_k^r - \widetilde{\bm{H}} \bm{z}_k^{r^*}\right) -\sum_t \sum_p \left( \widetilde{\bm{Q}}_{p,t}^{k,r\top} \bm{\vartheta}_{p,t,k}^r + \widetilde{\bm{q}}_{p,t}^{k,r\top} \bm{\omega}_{p,t,k}^r \right),\; \forall k,r, \\
       &  \lVert \bm{\beta}_{p,t,k}^r \rVert_2 \le \bm{\gamma}_{p,t,k}^r, \; \;\lVert \bm{\vartheta}_{p,t,k}^r \rVert_2 \le \bm{\omega}_{p,t,k}^r,\; \forall p,t,k,r, \\
        & \bm{\lambda}_k^r \ge 0\;, \forall k,r, \; \;\; \bm{\mu}\in\mathcal{M}.
    \end{align}\end{subequations}

    It should be noted that the master problem of the inner-loop C\&CG, which is \textbf{F4}, also serves as the slave problem of the outer-loop C\&CG. Meanwhile, the master problem of the outer-loop C\&CG has already been given in \textbf{F2}. By far, \eqref{eq:Ch5Compact} is readily solvable by calling the C\&CG algorithm twice.
\end{enumerate}

\begin{algorithm}[!tbh]
\caption{The C\&CG Algorithm for DRO Model}
\label{Ch5bNCCG}
\begin{algorithmic}[1]
    \STATE Set parameters $\varepsilon, UB=\infty, LB=-\infty, R=0$, and select an arbitrary feasible ${}^a\,$$\bm{\mu}^*$ / ${}^b\,$$(\bm{u}^*,\bm{y}^*)$, then go to step $4$.

    \STATE ${}^a\,$Solve problem \textbf{F1} \eqref{eq:Ch5P1Penalized} using Algorithm~\ref{Ch5bS-MISOCP}, update $(\bm{x}_k^*,\bm{z}_k^*)$ and $LB=\max\{LB,\; \textbf{F1}^*\}$. \\

    ${}^b\,$Solve problem \textbf{F3} \eqref{eq:Ch5P3Tri} using the inner C\&CG algorithm, update $\bm{\mu}^*$, and $UB=\min\{UB,\; \textbf{F3}^*+f(\bm{u}^*,\bm{y}^*)\}$.

    \STATE If $(UB-LB)/LB \le \varepsilon$, terminate; else, $R=R+1$, ${}^a\,$$(\bm{z}_k^{R*}=\bm{z}_k^*$, $\overline{\bm{x}}_k^{R*} =\bm{x}_k^*)$, and create new  matrices for \eqref{eq:Ch5P3Tri} / ${}^b\,$$\bm{\mu}^{R*}=\bm{\mu}^*$, go to Step 4.

    \STATE ${}^a\,$Solve problem \textbf{F4} \eqref{eq:Ch5P4}, update $\bm{\mu}^*,UB$.
    \\${}^b\,$Solve problem \textbf{F2} \eqref{eq:Ch5P2Penalized} using algorithm~\ref{Ch5bS-MISOCP}, update $(\bm{u}^*,\bm{y}^*),UB$.

    \STATE If $(UB-LB)/LB \le \varepsilon$ \&  same ${}^a\,$$\bm{\mu}^*$/ \ ${}^b\,$$(\bm{u}^*,\bm{y}^*)$, terminate; else, go Step $2$. \\

    ${}^a\,$For the inner C\&CG;  ${}^b\,$For the outer C\&CG.
\end{algorithmic}\vspace{-5.8mm}
\end{algorithm}

\begin{figure}[ht]
            \centering
            \includegraphics[width=12cm]{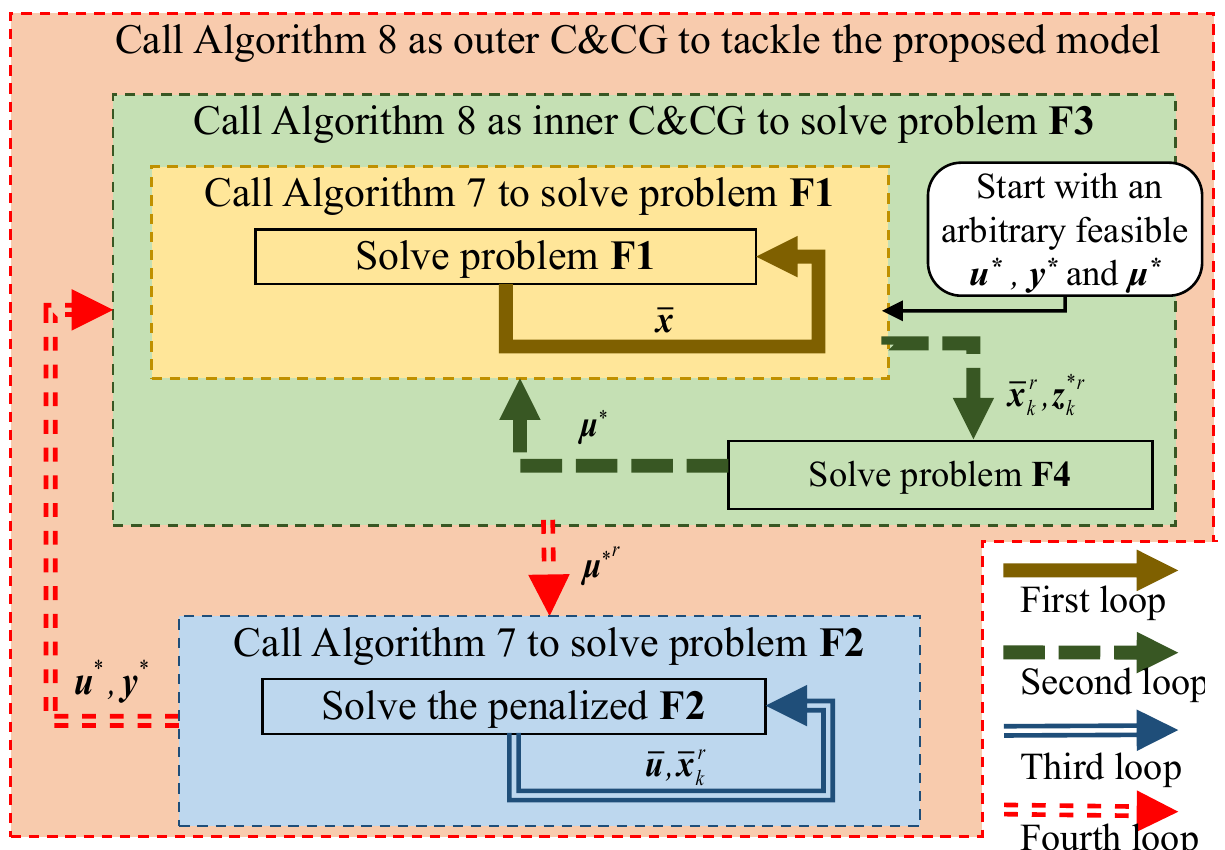}
            \caption{The schematic diagram of the overall solution procedure.}
            \label{fig:Ch5F2}
    \end{figure}

    The flowchart of the overall solution procedure is presented in Figure~\ref{fig:Ch5F2}, which contains four iteration loops. In the first loop, starting with an arbitrary feasible decision ($\bm{u}^*,\bm{y}^*$ and $\bm{\mu}^*$), Algorithm~\ref{Ch5bNCCG} is called to solve the lower-level problem \textbf{F1} and it provides a primal cut ($\bm{x}_k^{r*}$ and $\bm{z}_k^{r*}$) to the master problem of the inner C\&CG, i.e., the second loop. Then, the second loop is executed also by Algorithm~\ref{Ch5bNCCG}, which provides a primal cut ($\bm{\mu}^{r*}$) to the outer C\&CG. The master problem of outer C\&CG is solved by Algorithm~\ref{Ch5bS-MISOCP} to find optimal $\overline{\bm{u}}$ and $\overline{\bm{x}}_k^r$ in the third loop. Finally, outer C\&CG, i.e., the fourth loop, is terminated with optimal day-ahead decision $(\bm{u}^*,\bm{y}^*)$, and real-time decisions $(\bm{x}_k^*,\bm{z}_k^*,\, \forall k)$ under the worst-case distribution $\bm{\mu}^*$.

\section{Simulation Results} \label{sec:Ch5Res}
    In this section, the effectiveness of the proposed two models and the performance of the quadruple-loop algorithm are illustrated by examining four different test systems. Two of them are distribution level and the remaining are transmission level. It should be noted that the robust day-ahead operation model with bidirectional gas contracting is applied for the first two systems in subsections~\ref{sec:RO1}--\ref{sec:RO2}, while the last two are employed by the two-stage distributionally robust gas contracting in subsections~\ref{sec:DRO1}--\ref{sec:DRO2}.
\subsection{Test Systems Description}
        The tests systems are as follows.
         \begin{enumerate}
         \item A $13$-Node PDN interacted with an $8$-Node gas system, demoted as \textbf{TS-I}, is employed to study the robust day-ahead operation model with bidirectional gas contracting. Figure~\ref{fig:Ch5aF3} displays the test system topology. It has eleven power lines, one wind farm, one non-GPU, one GPU, seven power loads, four gas loads, one compressor, one P2G facility, and seven passive pipelines. We have one G2P contract for $G2$, and one P2G contract for the P2G facility. From this topology, we have four fixed direction pipelines \textit{p1}, \textit{p2}, \textit{p5}, and \textit{p6}, and three bidirectional pipelines \textit{p3}, \textit{p4}, and \textit{p7}. More details of the system, wind generation curves, parameters for the two algorithms, prices of gas, and penalties of non-served power loads, wind curtailment, and avoidance of P2G contracts can be found in Appendix~\ref{App:13Bus} and \ref{App:8Node}. We consider that the time window is $6$h; therefore, the wind budget ranges from $1$ to $6$. In the following results, we consider the wind budget to be $4$, which provides a feasible solution with a probability of greater than $95\%$  against uncertainties \cite{jiang2011robust}. Moreover, we represent cases based on the wind variation levels (WVLs), which represents the limits of the maximum relative deviations of wind generation uncertainties w.r.t. their predicted values.

        \begin{figure}[!htbp]
                \centering
                \includegraphics[width=9cm]{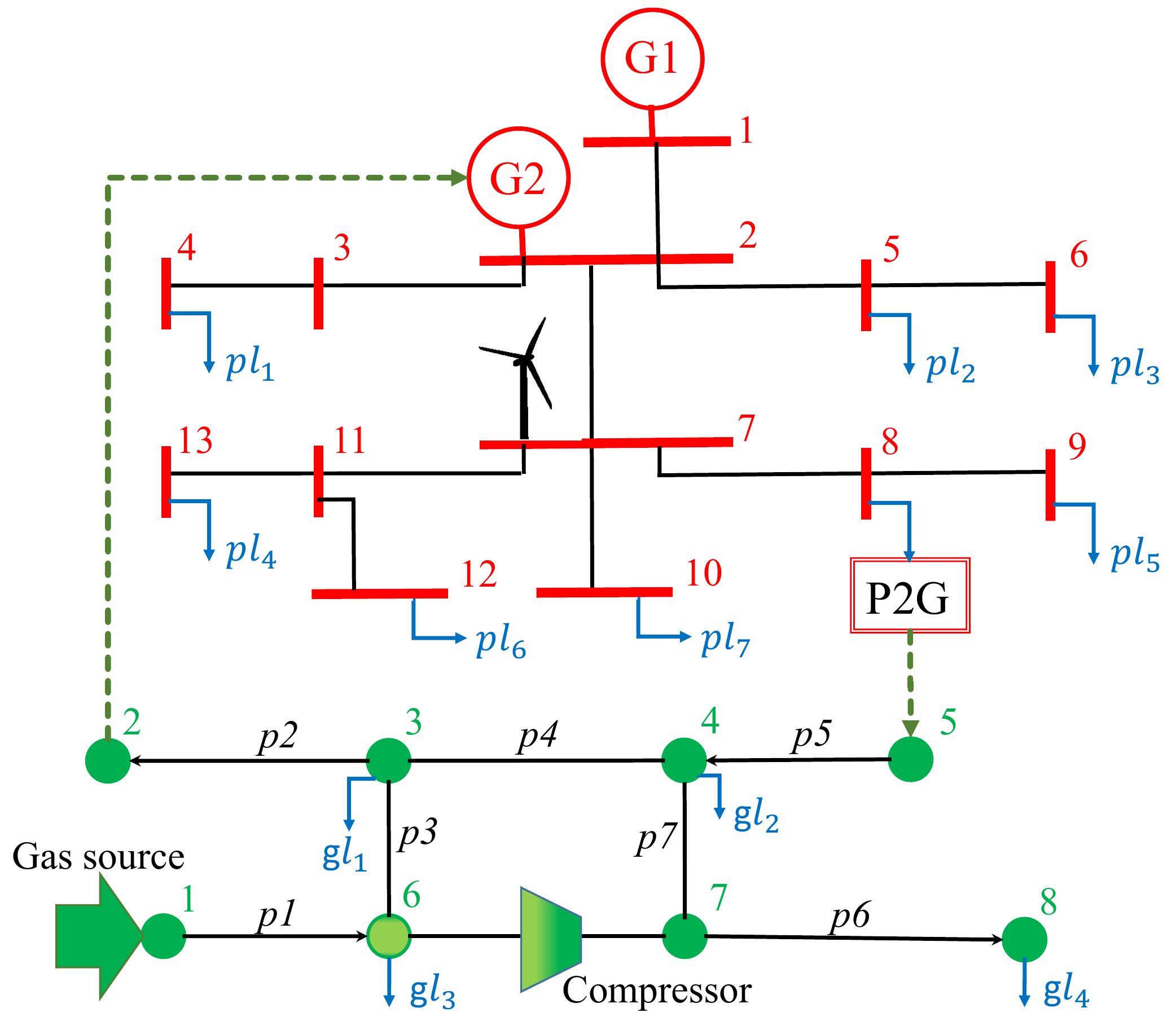}
                \caption{Topology of the test system.}
                \label{fig:Ch5aF3}
        \end{figure}

         \item A large-scale test system ($123$-Node PDN interacted with $20$-Node gas system), denoted as \textbf{TS-II}, is employed to study the scalability of the proposed algorithm at distribution level.
         \item A $5$-Bus-$7$-Node system, denoted as \textbf{TS-III}, is employed to study the two-stage distributionally robust gas contracting. The topology of \textbf{TS-III} is shown in Figure~\ref{fig:Ch5F3}. To construct the reference distribution $\bm{\mu}^0$, wind outputs are assumed to follow a multivariate Gaussian distribution \cite{zhao2015data,wang2018risk}, where their mean values can be found in \ref{App:5Bus} and the standard deviations equal half of their means. The distribution is used to generate a set of historical data samples, which are consequently utilized to create a histogram with $5$ pins \cite{zhao2015data}. Moreover, all data samples are checked to satisfy the wind farm power capacity, i.e.
                \begin{equation}
                    0 \le W_{e,t}^s \le \overline{W}_e, \; \forall e,t,s
                \end{equation}

                \begin{figure}[!htbp]
                \centering
                 \includegraphics[width=8cm]{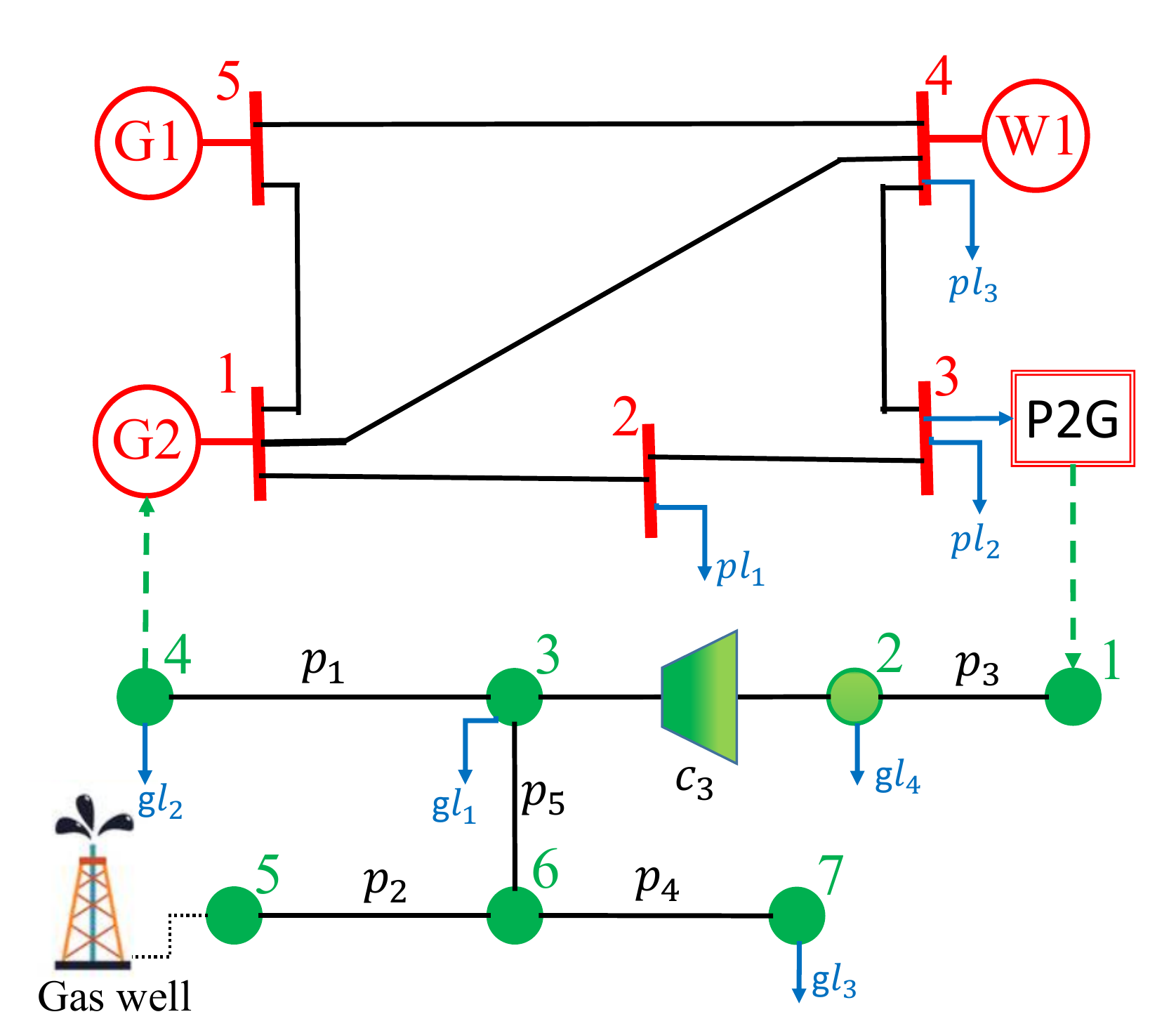}
                \caption{The test system topology.}
                \label{fig:Ch5F3}
                \end{figure}

         \item A $118$-Bus-$20$-Node system, denoted as \textbf{TS-IV}, is employed to study the scalability of the proposed algorithm at transmission level.
         \end{enumerate}
             Due to space limitation, please refer to Appendix~\ref{AppendixB} for detailed description of the selected test systems as well as the cost and algorithmic parameters.
             The numerical results are performed using MATLAB R$2018$a with Gurobi $8.1.0$ and YALMIP toolbox \cite{YALMIP} on a personal laptop with Intel(R) Core(TM) i$5-3320$M CPU and $8.00$GB RAM.

\subsection{Comparison with the IPS Model} \label{sec:RO1}
    Physical and economic-based comparisons are performed between the IPS model and the proposed model to reveal the effectiveness of considering the GDN constraints in the EM of the PDN. \textbf{TS-I} is selected to be the coupled system. The IPS model optimizes the economic operation of the PDN without taking into account the effect on the gas system feasibility. Therefore, the gas system constraints in \eqref{eq:Ch5aDAG1}--\eqref{eq:Ch5PGc} for day-ahead and real-time stages are dropped in this model. The objective function is the same for both models by including all gas contracts to provide a fair comparison. The feasibility of the gas system is checked after identifying the day-ahead contracts by considering the firm gas required in the G2P contract and the scheduled gas in the P2G contract as a gas load and gas source in the gas system, respectively.

    In this comparison, we present four cases; Case $1$ (normal WVL, normal gas load); Case $2$ ($\pm$10\% WVL, normal gas load); Case $3$ ($\pm$20\% WVL, normal gas load); and Case $4$ (normal WPL, high gas load). Table~\ref{tab:Ch5aPSO} displays the firm and reserved gas for the G2P contracts and scheduled gas for the P2G contract as cumulative hourly values in the time window. The gas system feasibility is denoted by F, which could be feasible (Y) or infeasible (N). Based on the signed contracts, the gas system should be feasible under any wind uncertainty. In Case $1$, the gas system is competent to supply/sink gas according to the requirements; therefore, the two models are feasible and provide the same total cost. The results from the two models are very close, in Case $2$, however, the gas system is infeasible with the contracts generated from the IPS model. In Case $3$, the increase of the WVL, the IPS model flunks to identify the suitable gas contracts to fulfill the requirements of the PDN. In Case $4$, increasing the gas load leads to a stressed gas system, which has a priority for supplying gas load \cite{wang2019convex}. Therefore, the proposed model provides a high operation cost. Unlike the IPS model, the proposed model selects the best contracts considering the gas system feasibility and priority. Moreover, multilevel pricing or bidding structure could be applied to consider the G2P contracts in the gas system priority.

\begin{table}[!htb]
        \caption{Physical comparison with the PSO model} \label{tab:Ch5aPSO}
        \centering
        \includegraphics[width=13cm]{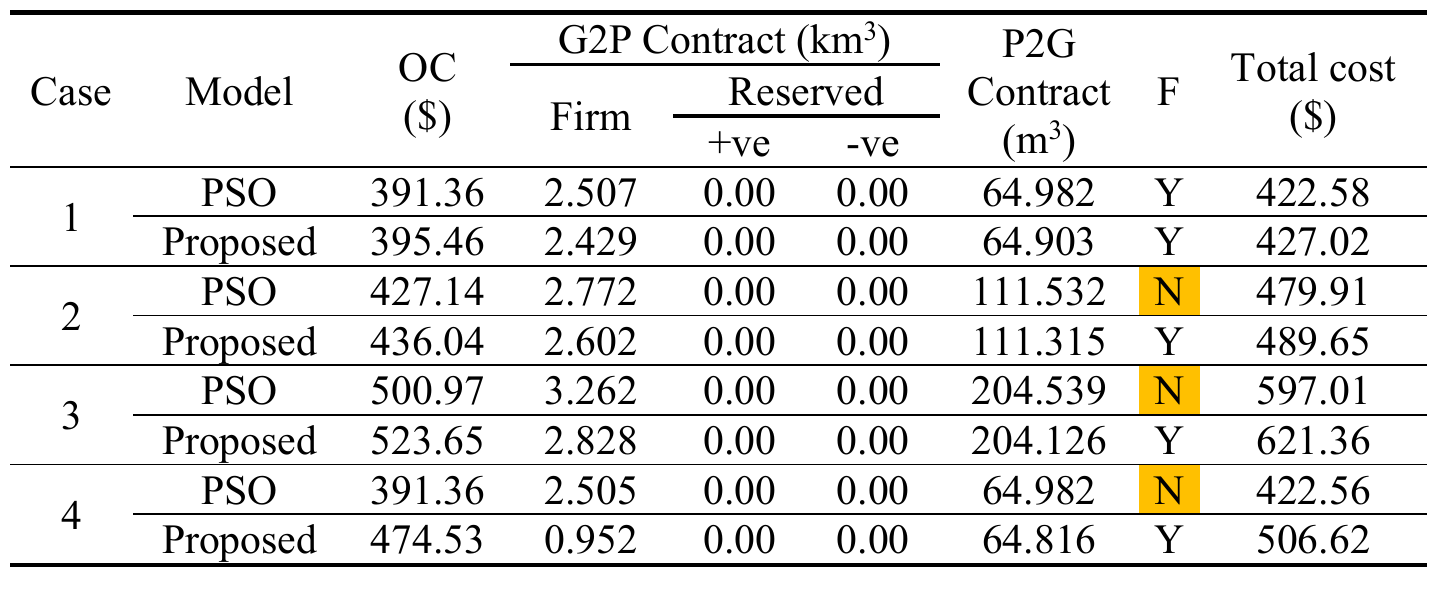}
    \end{table}

\subsection{Comparison Between the One-stage Contracting and IEGS Models}
    The importance of considering gas contracts in EM is discussed in this section. An economic comparison with the IEGS models, co-optimizing power and gas systems irrespectively of gas contracts, is performed for different cases based on WVLs. To achieve a fair comparison, the objective of the IEGS model is focused on the power system operation only while neglecting the gas system production cost. Its objective also does not include the cost of day-ahead gas contracts. Therefore, \eqref{eq:Ch5aFirm}--\eqref{eq:Ch5aP2G3} are dropped for this model. However, the PDN operator can sign costly real-time contracts based on the GPUs adjustment and P2G output deviations under uncertainties.

    Table~\ref{tab:CH5aIEGS} presents three different cases based on the WVLs to analyze the two models in a cost-effective manner. Each model provides the optimal dispatch to minimize the total operating cost. It is clear that the cost of day-ahead contracts in the IEGS model is zero because it does not consider them. Therefore, its total cost is lower. However, any change in the firm gas in G2P contracts is considered as a real-time contract. In addition, any variations in the P2G outputs will be penalized to mitigate any disturbance in the interacted gas systems. The table shows the real-time contracts under the worst uncertainty set, which is obtained by applying the inner C\&CG algorithm with the optimal solution $\bm{w}^*$. In contrast, the proposed model considers all the aforementioned problems in the day-ahead stage, so it provides a more economical operation than the IEGS model.

    \begin{table}[!htb]
        \caption{Economic comparison with the IEGS model under different WPLs} \label{tab:CH5aIEGS}
        \centering
        \includegraphics[width=13cm]{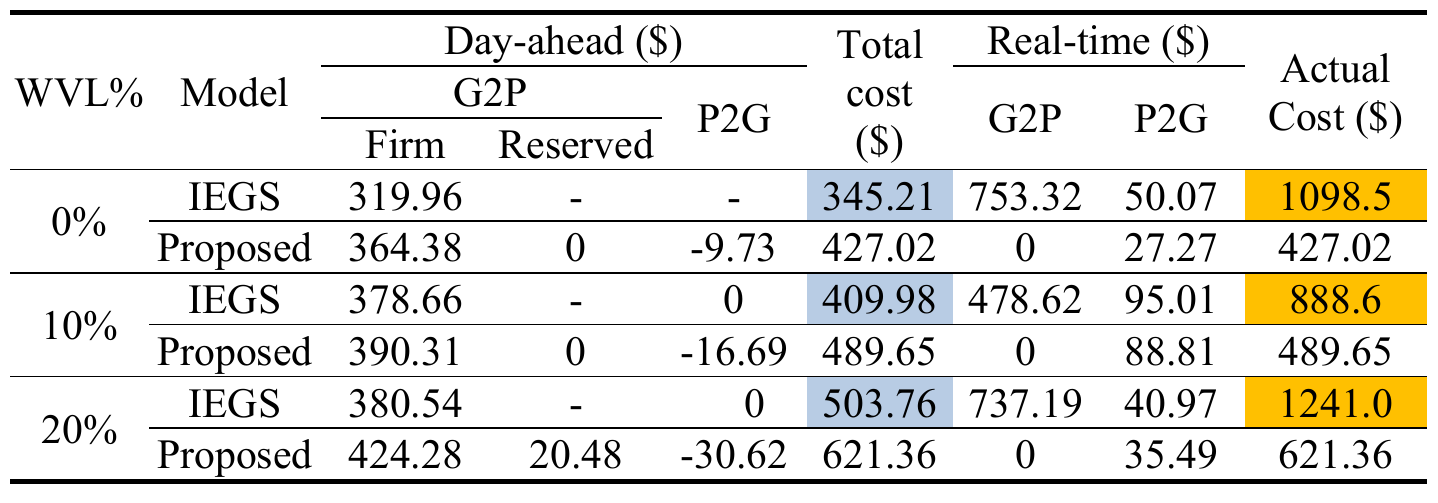}
    \end{table}

\subsection{Impacts of the Penalty Coefficients}
     The proposed model provides the ability for the PDN operator to control and identify the optimal scenario for wind generation management, i.e., curtailment or conversion to gas. In practice, gas prices and penalties of contract avoidance are driven from the gas system operator, and other penalties, in the proposed model, can be adjusted by the PDN operator. In this subsection, we present the effect of wind curtailment penalty $C_e$ on the gas contracts and OCs. Whereas the non-served load penalty $C_d$, prices of the adjustment in non-GPUs re-dispatch $C_u^+,C_u^-$, and gas prices $C_h,C_h^+,C_h^-,C_j,C_j^+,C_j^-$ are not changed in this comparison. Table~\ref{tab:Ch5aPricing} displays the influence of $C_e$ on the P2G contracts and curtailed wind energy. In Case I, $C_e$ is the same as the results above, i.e., \$$100$/MW, which is greater than the penalties of the P2G contract (\$$0.4/0.8$/\si{\cubic\metre} for up/down deviations). With decreasing $C_e$, the wind curtailment increases, whereas the P2G contract cost decreases, as shown in the table. The cost variation is not high (small system with the $6$ h operation); however, it should be considered for a large PDN. Therefore, by the PDN operator experience, $C_e$ can be adjusted to optimize and utilize the surplus wind energy.

\begin{table}[!htb]
        \caption{Impact of wind curtailment penalty \& ${C_e}$\ on the surplus wind energy} \label{tab:Ch5aPricing}
        \centering
        \includegraphics[width=14cm]{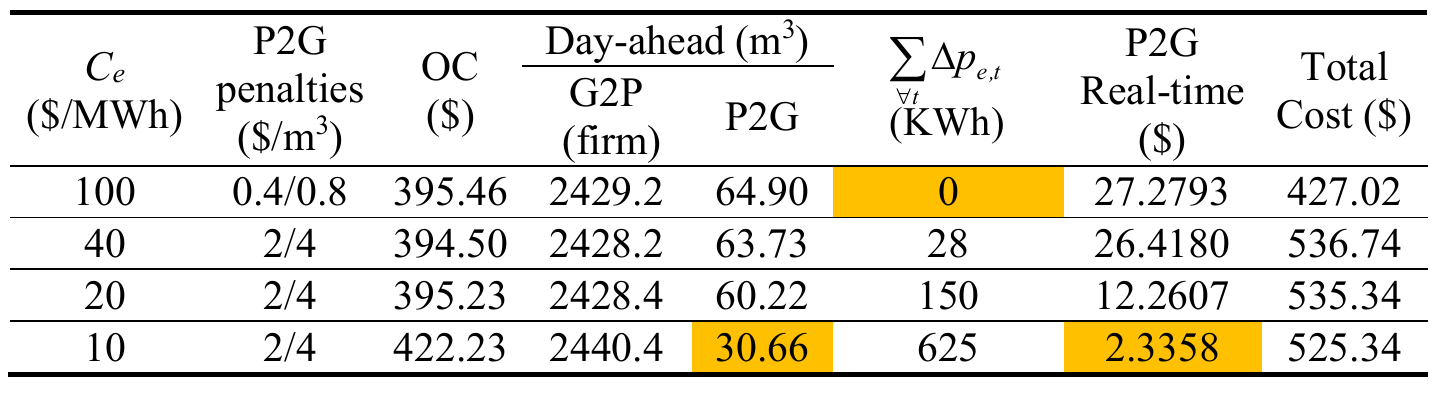}
    \end{table}

\subsection{Performance of the S-MISOCP Algorithm in RO Model}
    The S-MISOCP algorithm is proposed in Chapter~\ref{Chapter3}, where a detailed discussion is introduced, indicating its convergence and solution quality for deterministic IEGS optimization problems. In this subsection, it is compared with two widely used methods in the literature, namely, mixed-integer linear programming (MILP) formulation \cite{correa2014gas, Correa2015Integrated}  and MISOCP relaxation \cite{ding2015two, zlotnik2016coordinated, he2018coordination, kocuk2017new, borraz2016convex}. In the MILP model, the piece-wise linear approximation method (PLA) with optimal breakpoints is adopted for gas flow equation \eqref{eq:Ch5asimpleWey}, i.e., $f|f|$, and $\pi^2$. In \eqref{eq:Ch5asimplePQ}, the product $vi$ is equivalent to $0.25(U^2 - L^2)$, $U=v+i$, and $L=v-i$, so the PLA is adopted for $p^2, q^2, U^2$, and $L^2$. This method is highly influenced by the number of segments; therefore, two MILP models are applied with $20$ (MILP\_20) and $40$ (MILP\_40) segments. In the MISOCP relaxation method, quadratic equality constraints (power and gas flow equations) are reformulated as SOC constraints, and the resulted model will be as \eqref{eq:Ch5aMP0} for \textbf{P1} and \eqref{eq:Ch5aSPs0} for \textbf{P4}, without using the proposed S-MISOCP algorithm.

    Table~\ref{tab:Ch5aPerform} displays the results for the discussed models under different four cases based on solving problems \textbf{P1} \eqref{eq:Ch5aP1} and \textbf{P4}  \eqref{eq:Ch5aSPs}. With the optimal solution $\bm{w}^*$, problem \textbf{P4} is optimized under the worst-case uncertainty set ($\bm{u} = \bm{u}^*$) and zero uncertainty sets ($\bm{u}=\textbf{0}$) in Cases I and II, respectively. In Case I, compared with other models, the proposed algorithm provides the lowest maximum constraint violation (MCV) with the optimal objective cost. Note that MCV is calculated by \eqref{eq:MCV1} and \eqref{eq:MCV4} for problems \textbf{P1} and \textbf{P4}, respectively, and these values represent the relaxation gaps, please refer to the compact model \eqref{eq:Ch5aCompact} to drive MCV expressions. The relaxed MISOCP model usually introduces high maximum MCV in the power and gas flow equations. The high MCV from the latter might be originated from the ignorance of the gas system operation costs. With the worst $\bm{u}^*$ in Case II, MILP\_20 model presents an infeasible solution; however, MILP\_40 introduces a suitable cost with $0.3$\% MCV. Problem \textbf{P1} is solved with a deterministic uncertainty set $\bm{u}^*$ in Cases III and IV, where $\bm{u}^*$ equals zero and worst set, respectively.
        \begin{gather}
             \max \big( \frac{\bm{a}_{v,t}\bm{y}}{\parallel\bm{A}_{v,t}\bm{y} \parallel_2}-1, \forall t,v, \;\;\; \frac{\bm{h}_{v,t}\bm{x}^r}{\parallel \bm{H}_{v,t}\bm{x}^r \parallel_2}-1, \forall v,t,r \big) \;\; \label{eq:MCV1}\\
             \max \big(\frac{\bm{h}_{v,t}\bm{x}}{\parallel \bm{H}_{v,t}\bm{x} \parallel_2}-1, \forall v,t \big) \;\; \label{eq:MCV4}
        \end{gather}

    \begin{table}[!htb]
        \caption{Objective and computational performance of the S-MISOCP algorithm compared with other methods in RO models} \label{tab:Ch5aPerform}
        \centering
        \includegraphics[width=15.3cm]{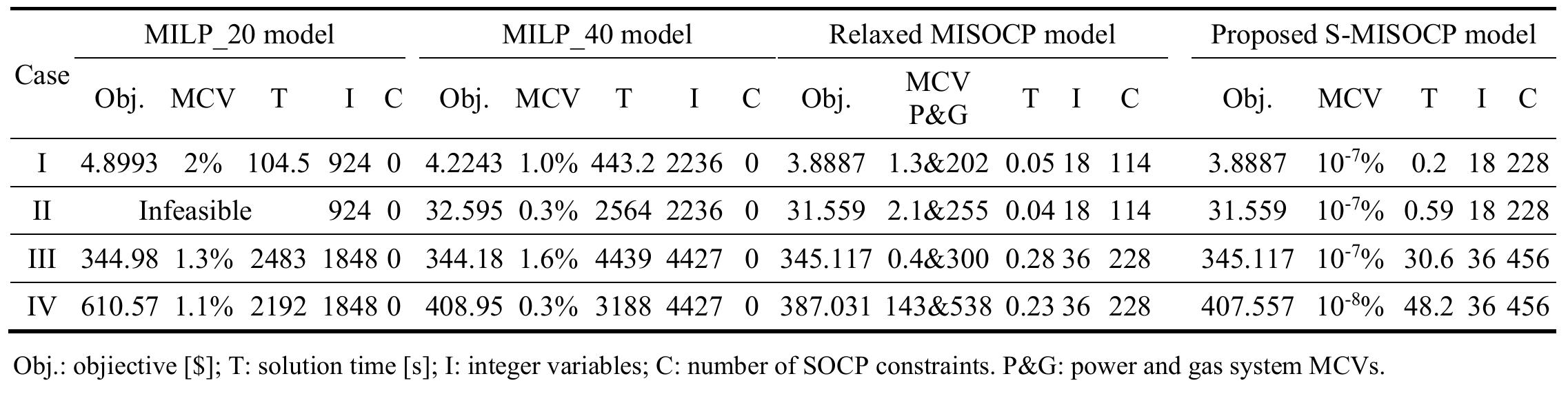}
    \end{table}

        The adaptive penalty rate method is applied to the simulations of Table~\ref{tab:Ch5aPerform}. To show the effectiveness of the proposed adaptive penalty rate method, computational comparisons with the traditional fixed penalty rate are conducted. Note that the adaptive rate ranges from $1.1$ to $2$ based on the associated constraint violation \eqref{eq:Ch3Adaptive}. The traditional rate is fixed at $1.5$, which is the average value of the range for the adaptive rate method. Penalties of Algorithm~\ref{Ch5S-MISOCP} are updated by \eqref{eq:Ch3Adaptive} and \eqref{eq:Tau1Re} for the adaptive and fixed rates, respectively. Algorithm~\ref{Ch5S-MISOCP} is applied to solve problem \textbf{P1} with the worst uncertainty set ($\bm{u} = \bm{u}^*$), which is identified under different uncertainty budgets, similar to Case~IV of Table~\ref{tab:Ch5aPerform}. Table~\ref{tab:Ch5aAdaptive} shows the performance of Algorithm~\ref{Ch5S-MISOCP} with adaptive and fixed penalty rates under different uncertainty budgets. From Table~\ref{tab:Ch5aAdaptive}, it can be observed that the adaptive rate method outperforms the traditional one in all the cases in terms of both solution quality and convergence speed as it controls the penalties according to the violation of each constraint individually. In the traditional fixed penalty rate method, the penalty coefficient grows equally for each constraint in each iteration of Algorithm~\ref{Ch5S-MISOCP}, which means the coefficient would become relatively large in a couple of iterations and may make the algorithm overemphasizing the weights of penalty terms, resulting in poorer solution quality.  From the last two columns of Table~\ref{tab:Ch5aAdaptive}, the penalty coefficients in the adaptive penalty rate method are much lower than the ones in the traditional method, the major reason is that the penalty coefficient of one constraint would stop increasing after its violation is beneath the given threshold.
        \begin{gather}
             \tau = min\Big[ \; \mu \tau, \;\; \tau^{max} \Big] \label{eq:Tau1Re}
        \end{gather}
        where $\mu$ is the fixed penalty growth rate.

 \begin{table}[!hbtp]
    \caption{Computational comparisons between the traditional and adaptive penalty rate methods} \label{tab:Ch5aAdaptive}
    \centering
    \includegraphics[width=13cm]{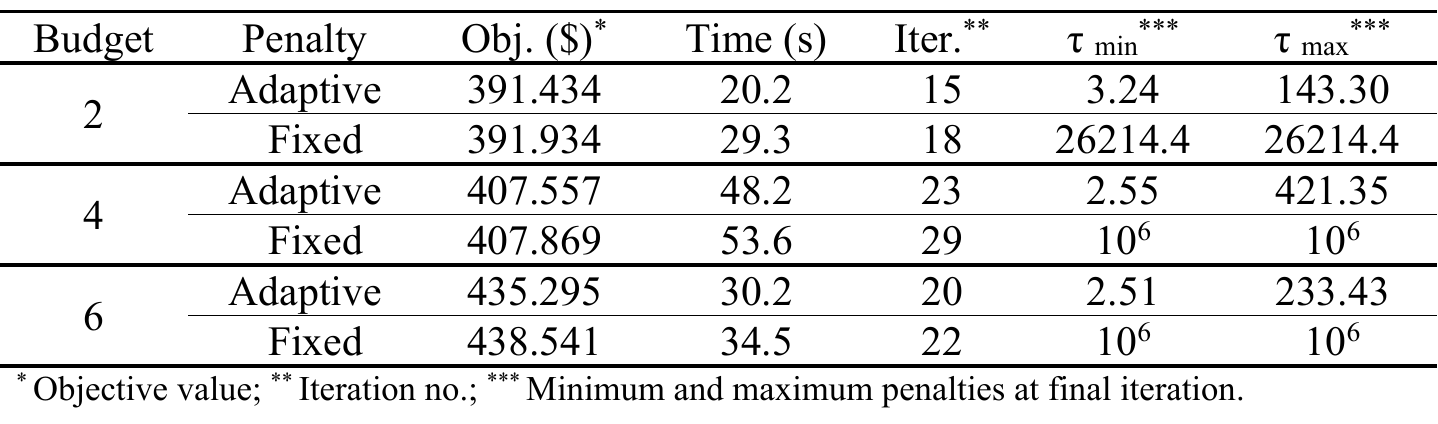}
\end{table}

\subsection{Scalability Tests of the Procedure with RO Models}\label{sec:RO2}
    In this section, we select a $123$-Node PDN coupled with a modified high-calorific gas network ($20$-Node gas system) as a large-scale IEGS test system to test the scalability of the proposed model and algorithms. The test system has $124$ power lines, $3$ wind farms, $5$ non-GPUs, $5$ GPUs, $85$ power loads, $9$ gas loads, $3$ compressors, $3$ P2G facilities, and $11$ passive pipelines. We have four G2P contracts for G2+G4, G6, G8, and G10, respectively. There are three P2G contracts for each P2G facility. Note that, although the gas network admits a meshed topology, only four passive pipelines are bidirectional and the gas flow directions of the rest seven can be fixed according to the topology of the gas network. Rest parameters of the system, wind generation forecast curves, algorithms parameters, and contracts prices can be found in Appendix~\ref{AppendixB}. We consider a problem with $6$ periods, where $t = 1$ to $t = 6$ are selected as the target slot.

\begin{figure}[h]
    \centering
    \includegraphics[width=13cm]{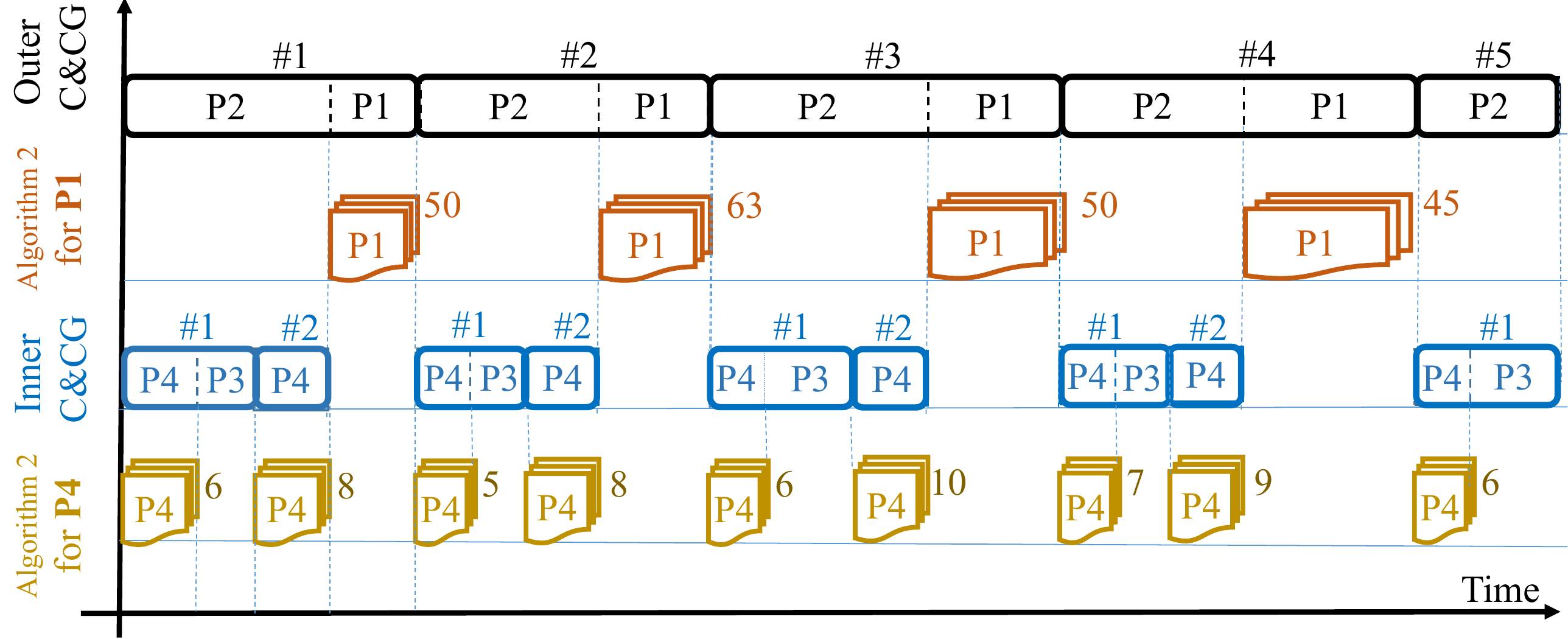}
    \caption{Sequences of solving problems and algorithms iterations in the proposed quadruple-loop algorithm for the large test system with wind uncertainty budget being $4$.}
                \label{fig:Ch5aLarge}
\end{figure}

    Figure~\ref{fig:Ch5aLarge} displays the iterations of algorithms and sequence of problems solved in the proposed quadruple-loop algorithm for the large-scale test system with uncertainty budget being $4$. Starting with solving \textbf{P4}  by Algorithm~\ref{Ch5S-MISOCP}, which converges after $6$ iterations, the inner C\&CG algorithm takes $2$ iterations to solve \textbf{P2}. The outer C\&CG algorithm, which calls Algorithm~\ref{Ch5S-MISOCP} and inner C\&CG algorithm four and five times, respectively, terminates after $5$ iterations. Meanwhile, the execution time of problem \textbf{P1} increases along with the iteration index of outer C\&CG due to the additional primal cuts of columns and constraints generated from previous iterations, which can be observed in Figure~\ref{fig:Ch5aLarge} as the execution time of each iteration of the second loop is approximated by the length of the red blocks. It should be noted that inner C\&CG algorithm usually terminates after a few iterations (no more than $2$) in these simulations, owing to the relatively small number of binary variables used in the lower-level recourse problem (directional gas flow). Table~\ref{tab:Ch5aLarge} summarizes the simulation results after applying the proposed algorithm on the above system with different wind uncertainty budgets, namely $0, 2, 4$, and $6$. It can be observed that the quadruple-loop algorithm always converges in a relatively reasonable number of iterations for each loop, and the total execution time is agreeable, where the simulation platform is a personal laptop rather than a high-performance work-station.

\begin{table}[!htbp]
        \caption{Computation times for the large test system under four different uncertainty budgets} \label{tab:Ch5aLarge}
        \centering
        \includegraphics[width=13cm]{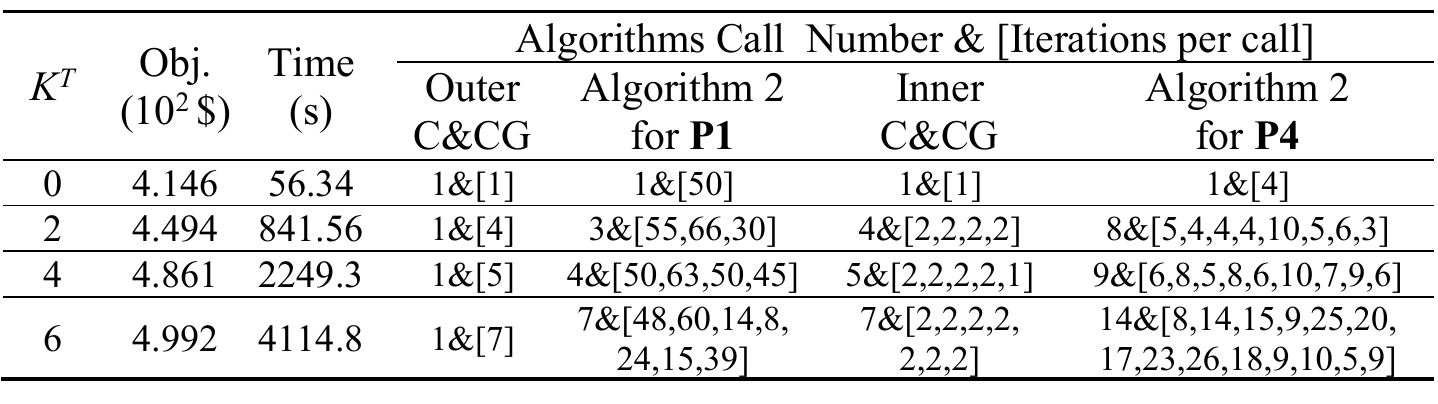}
    \end{table}

    Besides, based on the aforementioned analysis and simulation results, the following suggestions are recommended to reduce the computational costs:
    \begin{enumerate}
    \item
        For Algorithm~\ref{Ch5S-MISOCP}: as the violations from power or gas flow equations may still be relatively large after quite a few iterations due to the infeasibility of the initial point, we suggest (i) a high-quality initial point by adding the right hand side of the exact MISOCP constraints, i.e., $\lambda_p \sum_t \sum_l \hat{i}_{l,t} + \lambda_g \sum_t \max_p \hat{a}_{p,t}$, to the objective of the relaxed \textbf{P4} \eqref{eq:Ch5aSPs0} and $ \lambda_p \sum_t \sum_l \left(\hat{i}_{l,t}+\sum_r i_{l,t}\right) + \lambda_g \sum_t \max_p \left(\hat{\pi}^+_{p,t}+\sum_r {\pi}^+_{p,t}\right)$ to the objective of the relaxed  \textbf{P1}  \eqref{eq:Ch5aMP0}, where $\lambda_p$ and $\lambda_g$ are small factors. Same treatment have been done in Algorithm~\ref{alg2} using equation \eqref{eq:Ch3SCPIEGS0}; (ii) a relatively high initial penalty coefficient, such as $0.1$, to force the solution into a feasible region close to the one obtained from the relaxed problem; (iii) set lower solution quality requirement for the solver in the first few iterations with MISOCP problems (before fixing the binary variables), by increasing the relative and absolute optimality gaps, and selecting a suitable time limit. The reason is that, at the first iterations, the initial point needs to be adjusted and it is not necessary to find the optimal solutions. A worm-start is also recommended at these iterations by providing initial guess obtained from previous iterations to the solver.
    \item
        For nested C\&CG: (i) reduce the uncertainty budget to decrease the algorithm iterations (e.g. cases of Table~\ref{tab:Ch5aLarge}), by analyzing the periods when the outputs of wind generation are more likely to deviate significantly from their forecasted values based on historical data. (ii) select a suitable value of $\overline{M}$ used in problem \textbf{P3} \eqref{eq:Ch5aMPsFinal} as the algorithm is strongly influenced by this value, as discussed in Section \ref{sec:Ch4LargeScale}, which have additional suggestions to improve the NC\&CG algorithm performance.
    \end{enumerate}

\subsection{Comparison with SO and RO Models}\label{sec:DRO1}
    In this subsection, $2000$ samples are generated to construct the reference distribution $\bm{\mu}^0$. Then, a total of six sets of strategies (day-ahead decisions $\bm{u}^*$ and $\bm{y}^*$) can be obtained from the SO ($\sigma=0$), the RO ($\sigma=1$), and the proposed DRO ($\sigma=0.2, 0.4, 0.6, 0.8$) models. The total costs of the strategies from the SO and the RO models are $\$1.113\times{10^6}$ and $\$1.310\times{10^6}$, respectively. $100$ random distributions are created under each ambiguity set ($\sigma=0.2, 0.4, 0.6, 0.8$) to serve as the validation data. The performances of the strategies under the four validation distribution sets are summarized in Table~\ref{tab:Ch5SORO}, where the maximum and the average total costs under each validation set are listed. The proposed DRO outperforms both the SO and RO models under all sets of validation distributions. Compared with the SO model, the proposed model able to see all distributions in the ambiguity set, therefore it provides the optimal OC with best $\bm{y}^*$ to tackle any distributions of wind uncertainty. In addition, compared with the RO model, which identifies a high OC to satisfy the $100\%$ confidence level, the proposed model provides a less conservative decision that is optimal for each ambiguity set. Based on the above discussion, the proposed model has better performance on balancing the robustness and conservativeness of the dispatch strategy than the SO and RO models.

    \begin{table}[!hbtp]
        \caption{Comparison with the RO and SO based models.} \label{tab:Ch5SORO}
        \centering
        \includegraphics[width=13cm]{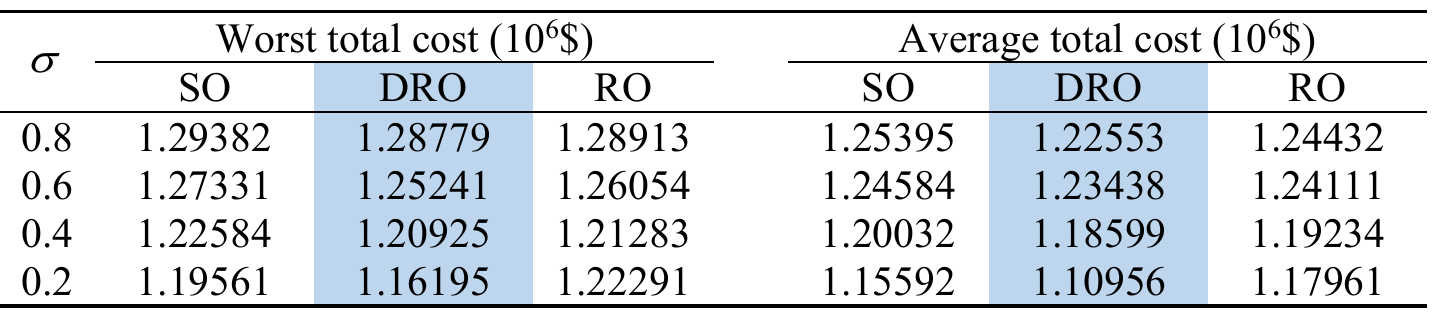}
    \end{table}

\subsection{Comparison Between the Two-stage Contracting and IEGS Models}
To highlight the necessity and effectiveness of gas contracts modeling, economic comparisons between the proposed model and the IEGS model are conducted. Specially, the DRO based IEGS model can be obtained by removing the contract related terms in the objective and the constraint set from the proposed model, where the production costs of the gas system are not included to provide a fair comparison by focusing only on the power system operational costs. It should be noted that the PSO is still able to sign a costly real-time purchase G2P gas contracts as well as a cheap real-time sale P2G gas contracts with the operation strategy from the IEGS model to control the regulation costs under the worst-case distribution. The costs of real-time G2P contracts can be calculated according to \eqref{eq:Ch5IERGSG2P}, while the income of real-time P2G contracts is defined in \eqref{eq:Ch5IERGSP2G}.
    \begin{gather}
        \sum_{h} \sum_{t} \sum_{u \in \mathcal{U}_n(h)} \frac{\Phi}{\eta_u} \begin{pmatrix}
            C_h^{2+}\max\{{p}_{u,t} - \hat{p}_{u,t},0\}
            + C_h^{2-} \max\{\hat{p}_{u,t} -{p}_{u,t},0\}
        \end{pmatrix},  \label{eq:Ch5IERGSG2P}\\
        - \sum_{j}\sum_{t} \sum_{z \in \mathcal{Z}(j)}C_j^{2+}\Phi\eta_z p_{z,t}. \label{eq:Ch5IERGSP2G}
    \end{gather}

    Table~\ref{tab:Ch5IEGS} lists the results of four cases under different confidence levels and P2G capacities, where $50$ samples are generated to construct the $5$-pin reference distribution. From Table \ref{tab:Ch5IEGS}, it can be observed that the proposed model outperforms the IEGS model in all four cases in terms of the out-of-pocket costs, as the PSO have to sign more expensive real-time G2P contracts, resulting in the increment of total costs, as well as cheaper real-time P2G contracts, leading to the decrement of the revenue, to mitigate the uncertainty of wind generation outputs, while it can have more reasonable contracts in the proposed model. In addition, the larger $\beta$ or the small capacity of P2G facilities leads to higher total costs. From \eqref{eq:sigma}, the $\sigma$ would increase along with $\beta$, which suggests the ambiguity set would become more conservative if $\beta$ increases, giving rise to higher total costs. Meanwhile, the impacts of confidence level $\beta$ on the total costs are more significant than the capacity of P2G facilities, due to the fact that the PSO has an alternative means to deal with the excessive wind generation, which is to curtail it and pay the fine.

    \begin{table}[!hbtp]
        \caption{Comparison between the proposed two-stage contracting model and the IEGS model}  \label{tab:Ch5IEGS}
        \centering
        \includegraphics[width=13cm]{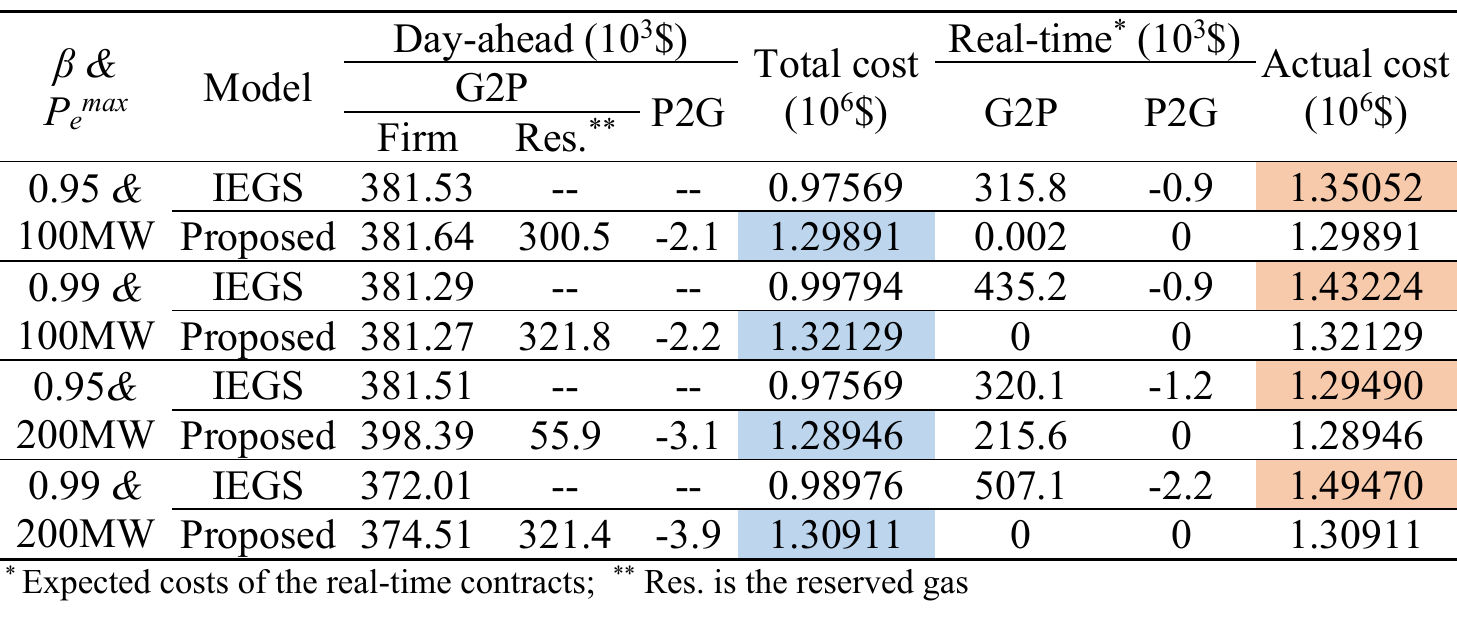}
    \end{table}

\subsection{Comparison Between Two-stage and One-stage Contracting Mechanisms}
In the sequel, the performance of the proposed two-stage contracting mechanism is compared with the one-stage contracting modeling (e.g. see \cite{chen2017clearing}). In the one-stage mechanism, the PSO can only sign day-ahead gas contracts, which means the real-time contracts related-costs are removed from the objective and $\rho_{h,t}^{2-}, \rho_{h,t}^{2+}, \triangle g_{j,t}^{+}$ and $\triangle g_{j,t}^{-}$ are forced to zero. $50$ samples are generated to construct the reference distribution $\bm{\mu}^0$ and the confidence level $\beta$ is set as $0.9$. The results are gathered in Table \ref{tab:OneStep}, where different wind generation curtailment penalty coefficients $C_e$ are tuned. The total wind curtailment over the day ($\triangle w_{e}$), the costs of gas contracts in the day-ahead (DA) and real-time (RT) stages and the expected total costs for the two mechanisms are listed in the table. Note that penalties $C_v$, which are regulated by the PSO, control the excessive amount of wind power outputs to be curtailed or converted into gas. The curtailed wind generation would increase if the $C_e$ decreases, as signing gas contracts would be less cost-effective than curtailing the wind generation. Moreover, the expected contracted values of gas in both day-ahead and real-time stages are decreasing along with $C_e$ due to the fact that low penalty value would weaken the influence of wind uncertainties. Consequently, the expected total costs decline for both mechanisms with the penalty reduction. The one-stage mechanism, which signs only day-ahead contracts and the real-time contracts are prohibited, provides larger expected total costs compared with the proposed two-stage mechanism as the latter has more flexibility to sign both day-ahead and real-time gas contacts and identifies the optimal decisions for the PSO.

    \begin{table}[!hbtp]
        \caption{Comparisons with between the proposed tow-stage and one-stage contracting mechanisms.} \label{tab:OneStep}
        \centering
        \includegraphics[width=13cm]{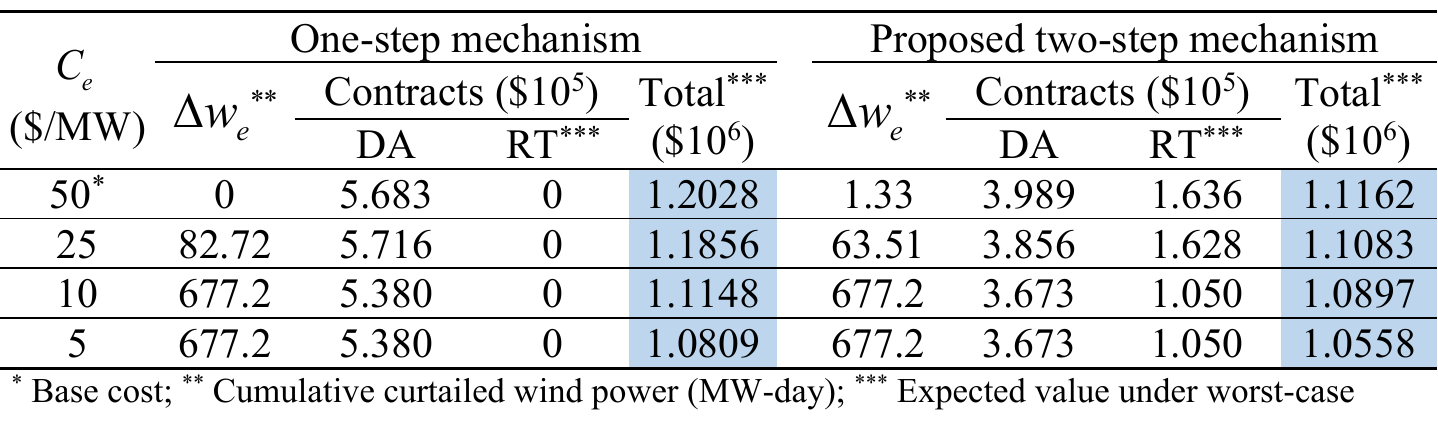}
    \end{table}

\subsection{Scalability Tests of the Procedure with DRO Models}\label{sec:DRO2}
\begin{figure}[!tbh]
            \centering
            \includegraphics[width=14cm]{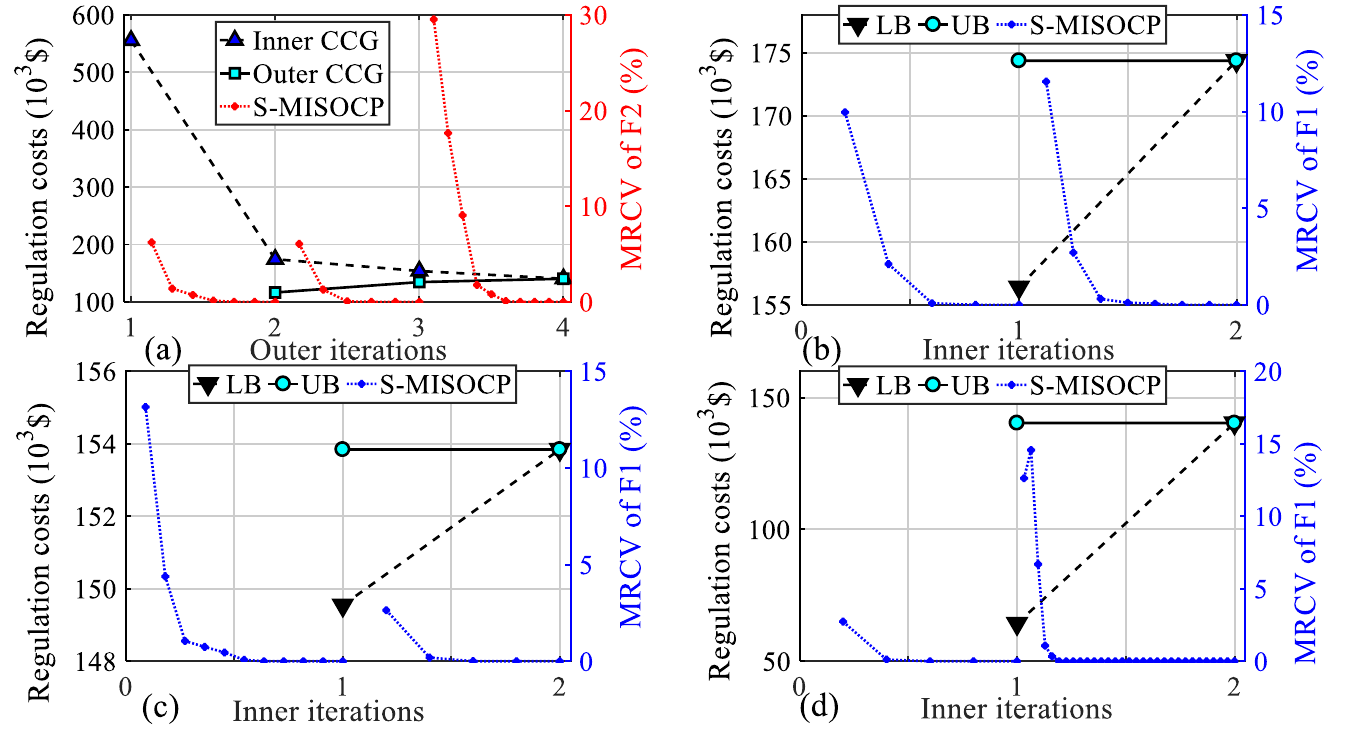}
            \caption{(a) RC obtained by the inner and outer C\&CG and the MRCV of \textbf{F2} by S-MISOCP; (b)-(d) RC obtained by the inner C\&CG (LB and UB) and the MRCV of \textbf{F1} by S-MISOCP for outer iterations (1)-(3), respectively.}
            \label{fig:Ch5Largef}
    \end{figure}

    In this subsection, the proposed methodology is applied on the large-scale test system \textbf{TS-IV} to evaluate its performance and scalability.  Economic interactions between power and gas systems are formulated as ten G2P and four P2G gas contracts for the $18$ GPUs and $4$ P2G units, respectively. $1000$ samples are generated to construct the reference distribution, and the confidence level is set at $0.95$. The problem is considered with $6$ periods and the target slots are from $1$ to $6$. Figure~\ref{fig:Ch5Largef} displays the optimality and feasibility of the proposed quadruple-loop procedure. In Figure~\ref{fig:Ch5Largef}(a), the outer C\&CG algorithm converges after only four iterations, where the inner C\&CG and the S-MISOCP algorithm are called four and three times to provide the UB and LB of RC, respectively. The \textbf{F2} feasibility is guaranteed by the S-MISOCP, which decreases the maximum relative constraints violation (MRCV) to $10^{-5}$. The first call of inner C\&CG is under an arbitrary day-ahead decision and the remaining three calls are organized in Figure~\ref{fig:Ch5Largef}(b)-(d), respectively, where each call takes two iterations to terminate as the number of binary variables (bidirectional gas flow) used in the recourse problem is relatively small (according to the topology of the gas network, we have only four pipelines with bidirectional gas flow and the remaining $17$ can be with fixed flow). The S-MISOCP is applied twice at each call to find a feasible solution of \textbf{F1}, and the first call is under an arbitrary $\bm{\mu}$, which explains the iteration curve of S-MISOCP before the first iteration of inner C\&CG in Figure~\ref{fig:Ch5Largef}(b)-(d). The MRCV of \textbf{F2} and \textbf{F1} are calculated by \eqref{eq:CH5MRCVP2} and \eqref{eq:Ch5MRCVP1}, respectively.
    \begin{gather}
        \max \big( \frac{\bm{d}_{p,t}\bm{u} }{ \lVert \bm{D}_{p,t}\bm{y} \rVert_2}\forall p,t, \frac{\bm{l}_{p,t}\bm{x}_k^r}{ \lVert \bm{L}_{p,t}\bm{x}_k^r \rVert_2} \forall p,t,k,r \big)-1,  \label{eq:CH5MRCVP2} \\
        \max \big( \frac{\bm{l}_{p,t}\bm{x}_k}{\lVert \bm{L}_{p,t}\bm{x}_k \rVert_2} \forall p,t,k \big)-1. \label{eq:Ch5MRCVP1}
    \end{gather}

       \begin{table}[!tbh]
        \caption{Computation times for the large-scale test system with different confidence levels.} \label{tab:Ch5Larget}
        \centering
        \includegraphics[width=13cm]{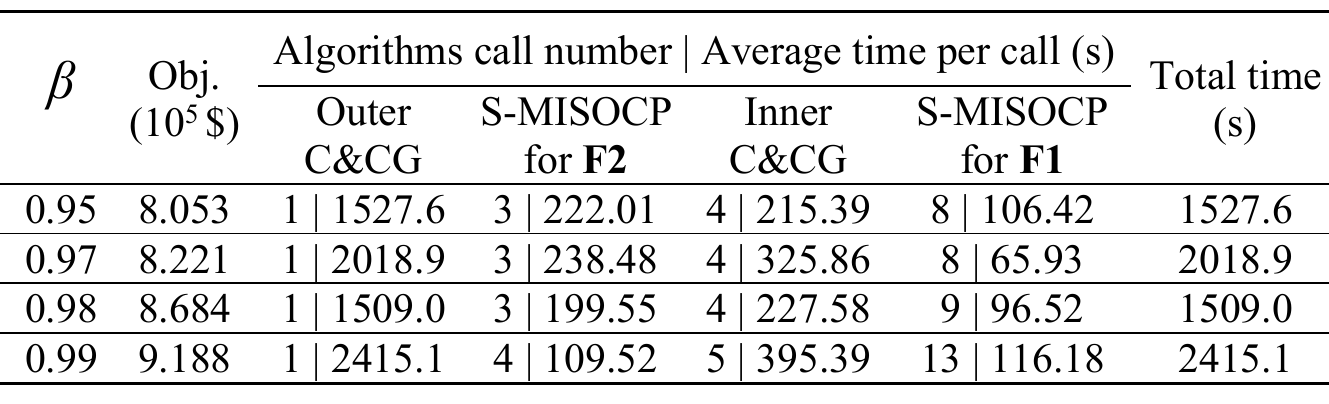}
    \end{table}

    Table~\ref{tab:Ch5Larget} lists the simulation results after applying the proposed procedure on the \textbf{TS-IV} with different values of confidence level $\beta$, namely $0.95, 0.97, 0.98,$ and $0.99$. It should note that NC\&CG algorithm usually terminates after a few iterations (no more than $5$ for outer C\&CG and $3$ for inner C\&CG), and the solution time for each algorithm is acceptable. It can be observed that the total execution time is mainly influenced by S-MISOCP algorithm, which appears in outer and inner iterations for \textbf{F2} and \textbf{F1}, respectively. Based on our experiences, the following recommendations, are pointed out to enhance the algorithm performance:
    \begin{enumerate}
    \item
        Select suitable coefficients in the penalty equation (\textcolor{MYCOLOR}{C}). As these coefficients may force the solver to focus on the violations rather than the main objective or drive the algorithm to execute more iterations to decrease MCRV, if their values are over-high or over-low, respectively. For example, see Table~\ref{tab:Ch5Recomm} in Cases \#1.
    \item
        Select a proper initial penalty. Poor choices of the initial penalty coefficient may lead to infeasibility issue or sub-optimal solutions.  For example, see Table~\ref{tab:Ch5Recomm} in Cases \#2.
    \item
        Divide \textbf{F1} into K sub-problems for each wind output scenario, as defined in \eqref{eq:Ch5P1new}, instead of solving the overall problem directly. Therefore, each sub-problem has fewer numbers of variables and constraints. Note that the solver will be called $K$ times to solve \eqref{eq:Ch5P1new}. In Cases \#3 of Table~\ref{tab:Ch5Recomm}, problem \textbf{F1} is solved with optimal values of $\bm{u}^*, \bm{y}^*$ and $\bm{\mu}^*$ two times: (i) one-shot using formula \eqref{eq:Ch5P1Penalized}, and (ii) k-shot using formula \eqref{eq:Ch5P1new}.
        \begin{gather}
        \sum_{\forall k} \mu_k^* \min_{\bm{x}_k,\bm{z}_k,\overline{\bm{x}}_k} \bm{c}^{\top}\bm{x}_k + \sum_{\forall t} \sum_{\forall p} \tau_{p,t,k} \xi_{p,t,k}. \label{eq:Ch5P1new}
    \end{gather}
    \end{enumerate}

    \begin{table}[!tbh]
        \caption{Computational performance of S-MISOCP algorithm under different parameters with $\beta=0.95$ and $S=1000$} \label{tab:Ch5Recomm}
        \centering
        \includegraphics[width=13cm]{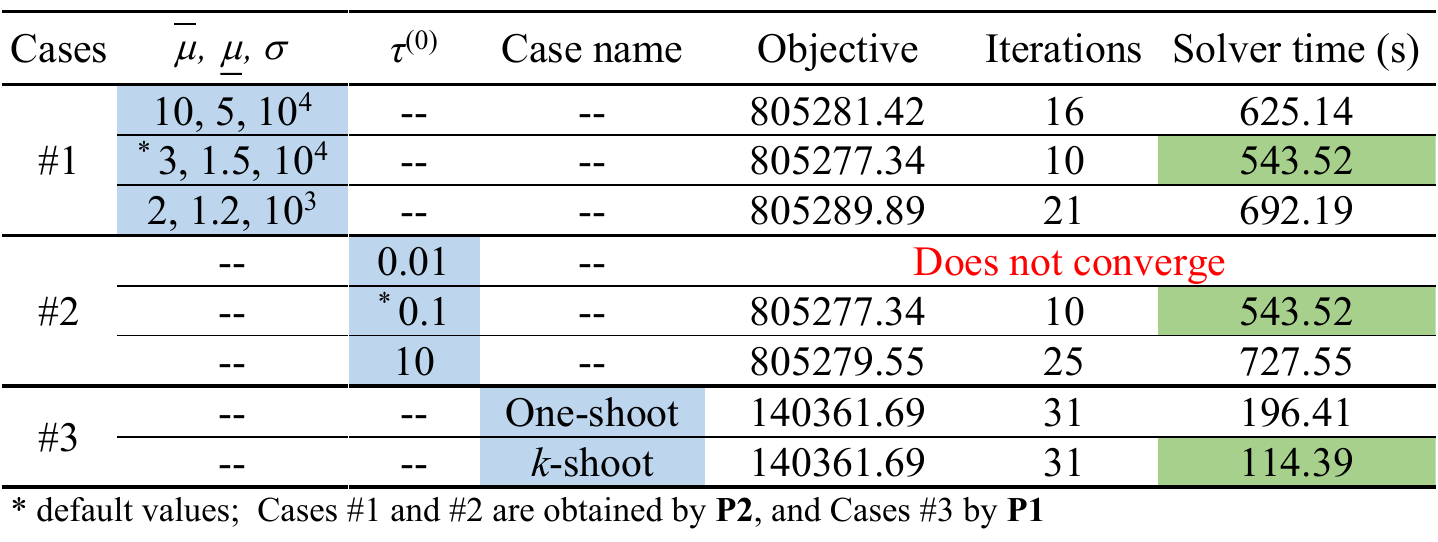}
    \end{table}

\section{Conclusions and Discussions} \label{sec:Ch5Con}
    With the increasing interactions between power and gas systems, energy contracts are desired to guarantee secure and reliable operations, as these systems are, in most cases, controlled by different operators. The integration of variable and uncertain wind power generation into power systems makes the contracting even more challenging. This chapter proposes two different operation models for the integrated electric-gas systems from the perspective of the PSO, where bidirectional contracts, including P2G and G2P, are mathematically formulated.

    The first model is a robust energy management (EM) model for the power distribution network against wind generation uncertainty, where both the physical, through the modeling of the gas system operation constraints, and the economic, by the modeling of bidirectional energy trading contracts, interactions with the gas system are considered. Mathematically, the proposed robust EM problem suggests a two-stage programming, where the summation of the day-ahead operation costs and the worst-case real-time regulation costs is minimized. To guarantee the robustness of the EM strategy, the contract for the reserved gas, which would be utilized for mitigating wind generation outputs deviation, is determined day-ahead along with the contract for firm energy.  To tackle the computational challenge brought by the nonconvex Weymouth equations in the two decision-making stages, a quadruple-loop solution procedure is devised, including two C\&CG loops and two S-MISOCP loops, through which a robust, feasible and nearly optimal solution can be obtained. Numerical simulations confirm that the proposed model outperforms both the IPS  and IEGS models. The impact of gas prices and contract avoidance penalties on the wind energy control (curtailment or methanation) are investigated. The effectiveness of the proposed methods is validated by the comparisons with other solution methods in terms of the computation performance as well as the solution quality.

    The second model is a distributionally robust two-stage contracting model, where bidirectional contracts, including P2G and G2P, can be signed in both day-ahead and real-time decision-making stages. The physical interactions between the power and gas systems, such as gas consumed by GPUs and electricity used by P2G facilities, would follow the signed contracts. The quadruple-loop solution procedure, which is used above for the RO model, is designed for this model to solve the DRO model with $K$ clusters of wind power outputs. Simulation results validate: i. the effectiveness of proposed DRO model over the SO and RO based ones; ii. the advantage of the two-stage contracting model over the one-stage one; iii. the scalability of the proposed solution methodology in transmission/distribution level IEGS.

    Future works include identifying the proper prices of gas contracts to guarantee the optimal revenue for gas systems instead of using predetermined ones, adopting extensive distributional uncertainty sets such as Kullback-Leibler divergence and Wasserstein distance based ones, and incorporating more types of uncertainties.


\chapter{Robust Operational Equilibrium for Coupled Electricity and Gas Markets} 

\label{Chapter6} 




Chapters~\ref{Chapter4}--\ref{Chapter5} proposed resilient-economic coordinated robust operation models for the interdependent power and gas systems, while formulating the physical and economic interactions between the two systems. These models focus on modeling the challenges in decision-making framework  as well as developing solution methodologies to overcome the  computational difficulty in the solving the non-convex nonlinear two-stage optimization problems, where energy and reserve prices are predetermined by system operators. Currently, this chapter seeks to derive the optimal values of these prices to complete the proposed models' compatibility to be applied in the existing industrial practice.

The increasing integration of uncertain and volatile renewable power generation (RPG) poses challenges not only to the operation of the interdependent electricity and gas systems but also their coupled markets. In this Chapter, a robust operational equilibrium solution method for the interactive markets of power and gas systems is proposed, where the bidirectional interactions include both firm energy and reserve contracting, and the impacts of the uncertainties of wind generation outputs on the two markets are characterized. The line pack as well as bidirectional flow characteristics of gas pipelines are depicted in the gas system model so as to improve its operational flexibility. To guarantee the robustness of market equilibrium against uncertainties, the power and gas market clearing models become two-stage robust ones. Column-and-constraint generation (C\&CG) based and the best response algorithms are devised to clear the two markets as well as to coordinate them, respectively. Simulation results validate the effectiveness of the proposed robust operational equilibrium model and the performance of the devised solution methodology. This work is submitted for publication as
    \begin{itemize}
      \item
            Ahmed R. Sayed, Cheng Wang,  Wei Wei, Tianshu Bi, and Mohammad Shahidehpour. "Robust Operational Equilibrium for Electricity and Gas Markets Considering Bilateral Energy and Reserve Contracts." Submitted for publication to IEEE Transactions on Power Systems.
    \end{itemize}

\section{Introduction}
    The extensively strengthened interdependencies between the power and gas systems not only suggest potential economic and environmental benefits for modern society but also can enhance the operational flexibility of both systems\footnote{https://www.pjm.com/markets-and-operations.aspx}\textsuperscript{,}\footnote{https://www.eia.gov/todayinenergy/detail.php?id=34612}, which is crucial to accommodate the uncertain renewable power generation  (RPG)\footnote{https://www.eia.gov/todayinenergy/detail.php?id=41533}. Despite these interdependencies, as discussed earlier, power and gas systems are operated independently in most occasions \cite{bertoldi2006energy}. Therefore, the co-optimization manner in the literature \cite{Correa2015Integrated,he2018co} may not be a realistic way to guide the operation of the two systems. Moreover, new challenges have appeared on pricing the resources which are used to mitigate the uncertainties of RPG in both electricity and gas markets \cite{ye2016uncertainty}. In this regard, the clearing mechanisms of power and gas markets in the uncertain and interactive environment need to be revisited.

    Recently, some inspiring studies investigated the interdependency of power and gas systems from the market perspective. Two market coordination methodologies for power and gas systems were proposed in \cite{gil2015electricity} to minimize the operational costs. A bi-level optimization model was proposed in \cite{cui2016day} to maximize the profits of the offering companies considering the demand response mechanism, where the steady-state model of the gas system was adopted. This model could be extended to identify the optimal strategies  of multiple offering companies, as shown in \cite{chen2019equilibria} and \cite{wang2017strategic}. In \cite{wang2018equilibrium} a coordinated market mechanism considering gas dynamics is proposed and solved by a sequential convex optimization. Ref. \cite{chen2019operational} proposed a methodology to identify the operational equilibrium of the integrated power-gas market with limited information exchange, where the Karush-Kuhn-Tucker (KKT) system of the model was derived. A price-based coordinated market mechanism was developed in \cite{zhao2018shadow} to clear the day-ahead markets of the two independent systems. It should be highlighted that the aforementioned works neglect the uncertainties of RPG during the decision-making process.

    To handle the uncertainties in the integrated power and gas systems, different optimization approaches have been adopted by the pertinent literature, such as stochastic optimization (SO) and robust optimization (RO), especially in operation and planning problems. However, limited work has been conducted to address the issue of operating the non-deterministic interactive markets, where SO is the common choice of the relevant ones \cite{ordoudis2019integrated, chen2017clearing, li2017security}.  Concretely, the focuses of \cite{ordoudis2019integrated, chen2017clearing, li2017security} are to minimize the total expected costs of the integrated market of power and gas systems, to allocate the reserves considering the gas system transmission capacities, and to demonstrate the value of P2G in practical cases, respectively. It should be noted that the market clearing results of those works depend on the predetermined scenarios of RPG, which might be over-optimistic if the scenarios are insufficient, yet a large number of scenarios would lead to huge computation burden. The greatest merit of the RO approach is that it guarantees the feasibility of the solution against all the realizations of uncertainties within the prescribed set. Therefore, it has driven more attention in electricity market problems with uncertainties  \cite{fang2019introducing, ye2016uncertainty, arroyo2005energy}.


    Based on the above discussion, there are significant research gaps between the existing works and the practical application, that need to be fulfilled. First, the core assumption in the non-deterministic market models is that one utility has full  control and operation authority over the power and gas systems. However, in industrial practice, there are significant institutional and administrative barriers to operate the two systems in a holistic manner \cite{chen2019operational}. Second, no attempt has been found in the literature that adopts the RO approach to analyze the equilibrium between the two markets, where the main difficulty is how to reflect the impacts of power system uncertainties on the gas system, and vice versa. Third, similar to the up-to-date researches in the electricity market \cite{fang2019introducing, ye2016uncertainty}, the gas system uncertainties introduced by GPUs demands need to be considered in its pricing scheme.

    In the following sections, a robust operational equilibrium seeking framework for electricity and gas markets considering bilateral energy trading is proposed. The proposed framework considers that the two markets are independently operated and allows limited information exchange, including only the prices and demands of both systems for contract agreements. In the electricity market, robust operation strategies including the gas consumption of GPUs during day-ahead and real-time stages against uncertainties of RPG as well as the electricity prices are determined. In the gas market, a robust production schedule is devised against the deviation of gas demands caused by the reserve utilization of GPUs, where the gas prices for firm energy and reserves can be obtained as well. The electricity and gas markets are cleared by column-and-constraint generation (C\&CG) and nested column-and-constraint generation (NC\&CG) algorithms, respectively. To deal with the nonconvex Weymouth equation in the gas market, the relaxation and the sequential penalty procedure are applied to guarantee the solution feasibility. Finally, the operational equilibrium between the two markets is tackled by the best-response decomposition (BRD) algorithm.

In light of this discussion, the innovations are multi-fold:
\begin{enumerate}
    \item
        A robust operational equilibrium for the coupled electricity and gas markets is characterized considering the uncertainties of RPG as well as bidirectional energy and reserve contracts.
    \item
        Inspired by \cite{ye2016uncertainty} and \cite{fang2019introducing}, the marginal energy and reserve prices for electricity and gas markets are derived based on the cost causation principle to reflect the impacts of uncertainties.
    \item
        The BRD algorithm is proposed to identify the characterized operational equilibrium, where the electricity and gas markets are separately cleared by the C\&CG and NC\&CG algorithms, respectively.
    \item
        The superiority of the robust operational equilibrium over the deterministic one, its effectiveness under limited data exchange, the importance of considering the gas dynamics, and solution procedure performance have been verified by numerical results.
\end{enumerate}

\section{Mathematical Formulation}
    \subsection{Pool-based Market Mechanism}
\begin{figure}[!hbtp]
            \centering
            \includegraphics[width=11cm]{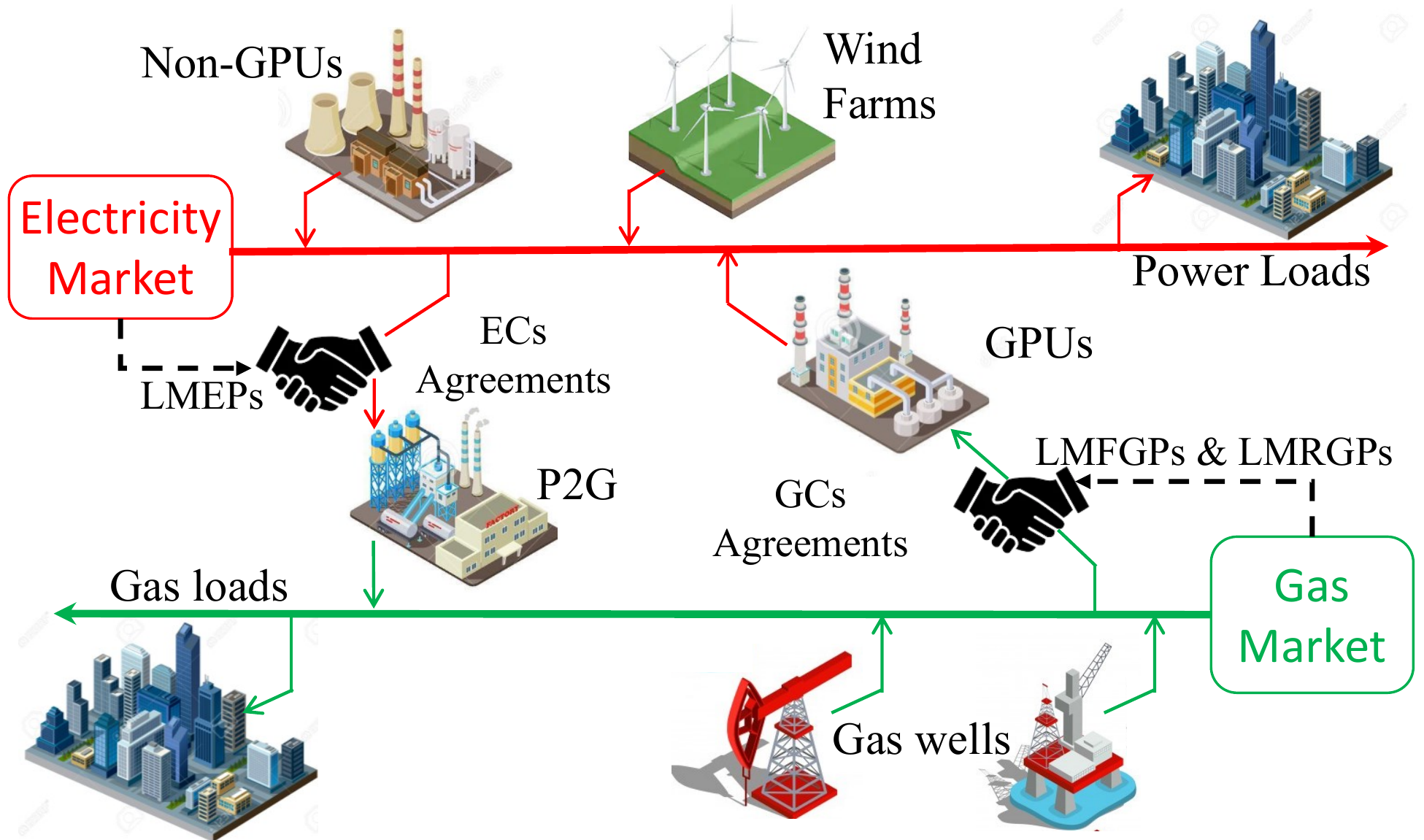}            \caption{Market mechanism for the coupled power and gas systems.}
            \label{fig:F1}
    \end{figure}

    Figure~\ref{fig:F1} displays the overall schematic diagram for the coupled electricity-gas operation. In the electricity market, the electricity market operator (EMO) decides the optimal robust dispatch strategy for all generators and signs the best gas contracts (GCs) for GPUs, considering the electricity consumed by P2G units  and the uncertainties of RPG. It should be noted that GCs are signed as two sub-contracts as discussed in Chapters $4$ and $5$: i) firm gas contract, which provides the required gas amounts for GPUs under the forecasted outputs of RPG in the day-ahead stage; ii) reserved gas contract, which defines the real-time gas consumption considering the utilization of upward and downward reserves of GPUs. The prices of GCs are obtained from the gas market clearing results, namely, the locational marginal firm gas prices (LMFGPs) and the locational marginal reserved gas prices (LMRGPs). In the gas market, the gas market operator (GMO) aims to find the robust gas production schedule against the uncertainties from the gas demands of GPUs. The GMO also identifies the best electricity contracts (ECs) to supply the P2G facilities according to locational marginal electricity prices (LMEPs), which can be received from the EMO.

    Some assumptions are listed as follows to simplify the mathematical formulation.
    \begin{enumerate}
        \item
        In general, (i) uncertainties only originates from RPG,, and energy demands are non-elastic; (ii) electricity and gas markets are cleared at the same time \cite{chen2019operational,chen2017clearing,wang2017strategic,wang2018equilibrium}.
        \item
        In the electricity market, (i) the lossless DC power flow model is adopted \cite{chen2019operational,bertoldi2006energy, ye2016uncertainty,gil2015electricity,cui2016day,zhao2018shadow}; (ii) unit commitment (UC) decisions are known \cite{wang2017strategic}.
        \item
        In the gas market, (i) the approximated gas line pack model is adopted; (ii) the simplified compressor and P2G models are employed \cite{wang2017strategic,wang2018equilibrium, chen2019operational,chen2019equilibria,bertoldi2006energy,gil2015electricity, cui2016day}; (iii) gas storages are non-strategic components (in a closed state) \cite{wang2017strategic,wang2018equilibrium}; (iv) P2G units are not reserve providers in the electricity market.
    \end{enumerate}

\subsection{Bilateral Energy and Reserve Contracting}
    Economical interactions between the two markets are modeled as bidirectional energy transactions. According to \cite{chen2017clearing}, day-ahead energy contracts are more convenient and cheaper than real-time ones, therefore, only day-ahead contracts are considered in this work.

    \subsubsection{Gas Contrcts modeling} In the electricity market, GCs are determined based on the gas prices received from the GMO. Under the forecasted outputs of RPG, the EMO signs firm GCs on the basis of base-case outputs of GPUs, which are defined in \eqref{eq:GCFrim}. Besides, GCs for reserved gas should satisfy the fluctuations of gas demands of GPUs in the real-time stage, as described by \eqref{eq:GCLimit} and \eqref{eq:GCReserve}.
        \begin{gather}
                 \rho_{h,t}=\sum_{u \in \mathcal{U}_g(h)} \Phi \hat{p}_{u,t} / \eta_u , \forall h,t, \label{eq:GCFrim}\\
                 0 \le \rho_{h,t}^- , \rho_{h,t}^+, \forall h,t, \label{eq:GCLimit}\\
                - \rho_{h,t}^-  \le \sum_{ u \in \mathcal{U}_g(h)} \Phi (p_{u,t} - \hat{p}_{u,t})/\eta_u  \le \rho_{h,t}^+, \forall h,t, \label{eq:GCReserve}
        \end{gather}
    where $u$, $h$ and $t$ are indices of generators, GCs and time periods, respectively; $\rho_{h,t}$ and $\rho_{h,t}^- / \rho_{h,t}^+ $ are the contracted firm and upward/downward reserved gas amounts of GC, respectively; $\Phi$ is the power-to-gas conversion factor; $\eta_u$ is the generation efficiency of GPU; $\mathcal{U}_g(h)$ is a subset of GPUs listed in contract $h$; $\hat{p}_{u,t}/p_{u,t}$ is the outputs of GPUs in day-ahead/real-time stages.

    \subsubsection{Electricity Contracts Modeling} In the gas market, ECs are optimized based on the electricity prices received from the EMO. In this study, two choices are provided for the excessive outputs of RPG: curtailed by the management sector of RPG; consumed by the P2G units. The contracted electricity consumed by P2G units, denoted as $p_{j,t}$, is defined in \eqref{eq:ECFrim}, where $j$ and $z$ are the indices for ECs and P2G units, respectively; $\mathcal{Z}(j)$ is a subset of P2G units; $\varrho_{z,t}$ denotes the produced gas from P2G; $\eta_z$ is the production efficiency. Gas production from P2G units is limited by their capacities in \eqref{eq:P2GLimit}, where $\underline{\varrho}_z/\overline{\varrho}_{z}$ is the lower/upper production capacity.
    \begin{gather}
            p_{j,t} = \sum_{z \in \mathcal{Z}(j)}{} \varrho_{z,t} / (\Phi \eta_z), \forall j,t,
              \label{eq:ECFrim}\\
            \underline{\varrho}_z \le \varrho_{z,t} \le \overline{\varrho}_{z} , \forall z,t  \label{eq:P2GLimit}
    \end{gather}

\subsection{Robust Clearing Model of the Electricity Market}
     The operation goal of the EMO is expressed in \eqref{eq:EMObj},  which is to minimize the sum of total day-ahead operational costs and the worst-case real-time regulation costs. In \eqref{eq:EMObj}, the first five terms are the generation cost of non-GPUs, the cost of GCs and penalties of power load shedding, respectively, and the last four terms express the costs of non-GPUs re-dispatching as well as penalties of wind curtailment and power load shedding. It should be noted that the real-time adjustment costs of GPUs are included in the day-ahead GC costs.
    \begin{gather}
            \min_{\bm{y}} \; \sum_{t}\Big[ \sum_{u \in \mathcal{U}_n} {C}_u ( \hat{p}_{u,t}) + \sum_{h \in \mathcal{H}} (\mu_{h,t} \rho_{h,t} +\mu_{h,t}^+ \rho_{h,t}^+ +\mu_{h,t}^- \rho_{h,t}^-) + \sum_{d \in \mathcal{D}_P} {C}_n  \triangle \hat{p}_{n,t} \Big] \nonumber \\
        + \;  \max_{\bm{\xi}\in\Upsilon} \; \min_{\bm{x}} \; \sum_{t} \Big[ \sum_{u \in \mathcal{U}_n} ({C}_u^+ \triangle p_{u,t}^+ + {C}_u^- \triangle p_{u,t}^-)+ \sum_{e \in \mathcal{E}} {C}_e  \triangle p_{e,t} + \sum_{d \in \mathcal{D}_p} {C}_n  \triangle p_{n,t} \Big]  \label{eq:EMObj}
    \end{gather}

    In \eqref{eq:EMObj}, $\bm{y}$ and $\bm{x}$ are the day-ahead and real-time decision vectors, respectively, and the uncertainty $\bm{\xi}$ follows a predefined uncertainty set $\Upsilon$; ${C}_u (.)$ is the convex cost function of non-GPUs. $\mu_{h,t}$ and $\mu_{h,t}^+/\mu_{h,t}^-$ are the contracted prices for firm and reserved gas, respectively; $\triangle p_{u,t}^-/\triangle p_{u,t}^+$ is the downward/upward adjustments of non-GPU outputs and its penalty is ${C}_u^-/{C}_u^+$; $\triangle \hat{p}_{n,t}/\triangle {p}_{n,t}$ is the load shedding in the day-ahead/real-time stage and its penalty is $ {C}_n$; $\triangle {p}_{e,t}$ is the curtailment of RPG and its penalty is $ {C}_e$.

    The operational constraints of the power system are defined in \eqref{eq:DAGen}--\eqref{eq:DABus}. Generation and ramping up/down capacities for both GPUs and non-GPUs are presented in \eqref{eq:DAGen}--\eqref{eq:DARamp2}. In \eqref{eq:DAAng}--\eqref{eq:DAshed}, bus angle $\hat{\theta}_{n,t}$, power flow $\hat{p}_{l,t}$, and load shedding limits are defined, respectively.  Considering the potential infeasibility caused by aggressive ECs, the bus balancing equation \eqref{eq:DABus} is relaxed by adding the load shedding term, where $\hat{W}_{e,t}$ is the predicted outputs of wind generation and $p_{n,t} = \sum_{d \in \mathcal{D}_p(n)}P_{d,t} + \sum_{j \in \mathcal{J}(n)}{p}_{j,t}, \; \forall n,t$ is the aggregated electricity demand.
    \begin{gather}
        U_{u,t}\underline{P}_u \le \hat{p}_{u,t} \le U_{u,t}\overline{P}_u,\forall t, u \in \mathcal{U}, \mathcal{U}=\mathcal{U}_n\cup\mathcal{U}_g, \label{eq:DAGen} \\
        \hat{p}_{u,t} - \hat{p}_{u,t-1} \le U_{u,t} R_u^+ + (1-U_{u,t+1}) \overline{P}_u ,\; \forall t,u \in \mathcal{U},   \label{eq:DARamp1} \\
        \hat{p}_{u,t-1} - \hat{p}_{u,t} \le U_{u,t+1} R_u^- + (1-U_{u,t}) \overline{P}_u ,\; \forall t,u \in \mathcal{U},   \label{eq:DARamp2} \\
        -\pi \le \hat{\theta}_{n,t} \le \pi,\; \forall t,n, \hat{\theta}_{1,t}=0,\; \forall t,   \label{eq:DAAng} \\
         |\hat{p}_{l,t} = (\hat{\theta}_{m,t}-\hat{\theta}_{n,t})/x_l | \le \overline{P}_l,\; \forall l,t, (m,n) \in l.   \label{eq:DAFlow}\\
         0 \le \triangle \hat{p}_{n,t} \le p_{n,t},\;\forall n,t,  \label{eq:DAshed}\\
        \sum_{u \in \mathcal{U}(n)}\hat{p}_{u,t} + \sum_{e \in \mathcal{E}(n)} \hat{P}_{e,t} + \sum_{l \in \mathcal{L}_1(n)} \hat{p}_{l,t}  - \sum_{l \in \mathcal{L}_2(n)} \hat{p}_{l,t} = P_{n,t} - \triangle \hat{p}_{n,t}, \; \forall n,t.
          \label{eq:DABus}
    \end{gather}

    In \eqref{eq:DAGen}--\eqref{eq:DABus}, $U_{u,t}$ is a predetermined unit commitment decision; $\underline{P}_u/\overline{P}_u$ and $R_u^-/R_u^+$ are the minimum/maximum generation limits and ramping down/up limits of generators, respectively; $\overline{P}_l$ and $x_l$ are the power flow capacity and the reactance of power line $l$; $n$ and $m$ are indices of power buses; $\mathcal{U}(n),  \mathcal{E}(n)$, and $\mathcal{D}_p(n)$ are subsets of generators, wind farms, power lines and power loads connected to bus $n$, and $\mathcal{J}(n)$ is a subset of ECs, in which P2G units are supplied from bus $n$; and $\mathcal{L}_1(n)/\mathcal{L}_2(n) $ is a subset of power lines connected with the end or start terminals.

    In this work, a box-like uncertainty set is employed \cite{ye2016uncertainty}, as shown in \eqref{eq:Uset}.
    \begin{gather}
           \Upsilon := \left\{ \begin{array}{l} \bm{\xi} =(\overline{\xi}_{e,t} , \underline{\xi}_{e,t})_{\forall e,t} \left| \sum_{e}(\overline{\xi}_{e,t} + \underline{\xi}_{e,t}) \le \Gamma^e, \forall t, \right.\\
           \hspace{5.5em} \left|\sum_{t}(\overline{\xi}_{e,t} + \underline{\xi}_{e,t}) \le \Gamma^t, \forall e, \right.\\
           \hspace{5.5em}\left|\overline{\xi}_{e,t} + \underline{\xi}_{e,t} \le 1, \right.\\
           \hspace{5.5em}\left|\overline{\xi}_{e,t} , \underline{\xi}_{e,t} \in \{0,1\} , \forall e,t \right.\end{array} \right\}  \label{eq:Uset}
        \end{gather}
        where $\overline{\xi}_{e,t}$ and $\underline{\xi}_{e,t}$ are uncertainty variables; $\Gamma_1$ and $\Gamma_2$ are the spatial and temporal uncertainty budgets, respectively. The actual outputs of wind generation is defined by \eqref{eq:WindP}, where $\overline{P}_{e,t}$ and $\underline{P}_{e,t}$ are its upper and lower bounds, respectively.
        \begin{gather}
            p_{e,t} = \hat{P}_{e,t}(1- \overline{\xi}_{e,t}- \underline{\xi}_{e,t})  + \overline{P}_{e,t} \overline{\xi}_{e,t} + \underline{P}_{e,t} \underline{\xi}_{e,t}, \forall e,t  \;\;\; \label{eq:WindP}
        \end{gather}

     In the real-time stage, the operational constraints are similar to those in the day-ahead stage. Specifically, some of them can be directly obtained by replacing the day-ahead variables in \eqref{eq:DAGen}--\eqref{eq:DABus} with the real-time ones, i.e., by removing the hat symbols. Here, the overlapped constraints are not shown. Further, the real-time power balancing equation allows wind generation curtailment, as shown in \eqref{eq:RTBus}. Accordingly, the upper and lower boundaries for wind generation curtailment is added as  \eqref{eq:RTWind}.  The outputs adjustment of non-GPUs can be measured by \eqref{eq:RTadj}. Note that the real-time GPUs outputs has been restricted in \eqref{eq:GCReserve}.
     \begin{gather}
        \sum_{u \in \mathcal{U}(n)}{p}_{u,t} + \sum_{e \in \mathcal{E}(n)} (p_{e,t} -\triangle p_{e,t})+ \sum_{l \in \mathcal{L}_1(n)} {p}_{l,t} - \sum_{l \in \mathcal{L}_2(n)} {p}_{l,t} = p_{d,t}-\triangle p_{d,t} , \; \forall n,t,  \label{eq:RTBus}\\
         0 \le \triangle p_{e,t} \le p_{e,t}, \forall e,t,   \label{eq:RTWind}\\
         - \triangle p_{u,t}^-\le  p_{u,t} - \hat{p}_{u,t}\le  \triangle p_{u,t}^+, \forall u,t.  \label{eq:RTadj}
    \end{gather}

\subsection{Robust Clearing Model of the Gas Market}
    Similarly, the goal of the GMO is to minimize the energy supply costs of the gas system, as shown in  \eqref{eq:objgas}, which includes the firm production and reserved gas costs, costs of ECs as well as penalties of GCs avoidance, gas load shedding and real-time gas imbalance. In this work, both the gas wells and gas loads can be gas reserve providers.
    \begin{gather}
        \min_{\Psi_1} \; \sum_{t}\Big[ \sum_{w \in \mathcal{W}} ({C}_w \hat{f}_{w,t}+{C}_w^+ R_{w,t}^+ + {C}_w^- R_{w,t}^-) + \sum_{j \in \mathcal{J}, n \in j} \beta_{n} p_{j,t} + \sum_{h \in \mathcal{H}} {C}_h \triangle\rho_{h,t}
        \nonumber\\
        + \sum_{i\in \mathcal{I}} ({C}_i \triangle{f}_{i,t} + {C}_i^+ f_{i,t}^+ + {C}_i^- f_{i,t}^- )\Big]+ \max_{\bm{g}} \min_{\Psi_2} \sum_{t}  \sum_{i \in \mathcal{I}} \overline{{C}}_i (V_{i,t}^+ + V_{i,t}^-)   \label{eq:objgas}
    \end{gather}

     In \eqref{eq:objgas}, $w$ and $i$ are the indices of gas wells and nodes, respectively; $\Psi_1$ and $\Psi_2$ are the day-ahead and real-time decision variable sets, respectively; $\bm{g} = \{g_{h,t}, \; \forall h,t\}$ represents the uncertain gas consumption of the power system and $g_{h,t}$  is limited by the GCs of reserved gas, as described by \eqref{eq:Usetg}; $\hat{f}_{w,t}$ and $R_{w,t}^+/R_{w,t}^-$ are the gas production and the up/down reserves of gas well, and their prices are ${C}_w$ and ${C}_w^+/{C}_w^-$, respectively; $\beta_{n}$ is the LMEP at bus $n$; $\triangle\rho_{h,t}$ denotes the violation of GCs and its penalty is ${C}_h$; $\triangle\hat{f}_{i,t}$ and $f_{i,t}^+/f_{i,t}^-$ are the gas load shedding and the provided up/down reserves, and their prices are ${C}_i$ and ${C}_i^+/{C}_i^-$, respectively; the real-time nodal gas imbalance is $V_{i,t}^+/V_{i,t}^-$, which is penalized with $\overline{{C}}_i$.
    \begin{gather}
        - \rho_{h,t}^- \le  g_{h,t} \le \rho_{h,t}^+  ,\; \forall h,t. \label{eq:Usetg}
    \end{gather}

    The operational constraints of the gas system in the day-ahead stage are defined in \eqref{eq:GMWell}--\eqref{eq:GMspin}. The gas production capacities and nodal pressure boundaries are described by \eqref{eq:GMWell} and \eqref{eq:GMPre}, respectively. Terminal pressures and gas flow of compressors are expressed in \eqref{eq:GMComp1}--\eqref{eq:GMComp2}. The line pack can be calculated by \eqref{eq:GMMass1} and its continuity equation is depicted by \eqref{eq:GMMass2}. Note that the gas nodal balancing equation \eqref{eq:GMNode} is relaxed by adding unserved gas amounts $\triangle\rho_{h,t}$ of the signed GCs and replacing the total gas loads $F_{i,t}$ with the bided one $\hat{f}_{i,t}$, to recover the operational feasibility. The served gas load is bounded by \eqref{eq:GMShed}. Weymouth equation is defined in \eqref{eq:GMWey1}--\eqref{eq:GMWey2}. Unserved gas amounts of GCs are limited in \eqref{eq:GMGC}, and lower boundaries of unserved gas loads and reserves are defined in \eqref{eq:GMLoad} and  \eqref{eq:GMspin}, respectively.
   \begin{gather}
        \underline{Q}_w \le \hat{f}_{w,t} \le \overline{F}_w,\; \forall w,t, \label{eq:GMWell} \\
        \underline{\Pi}_i \le \hat{\pi}_{i,t} \le \overline{\Pi}_i,\; \forall i,t, \label{eq:GMPre}\\
        \hat{\pi}_{i,t} \le \hat{\pi}_{o,t} \le \gamma_c \hat{\pi}_{i,t}, \forall c,t, (i,o) \in c,   \label{eq:GMComp1} \\
        0 \le \hat{q}_{c,t}^{out} = (1-\alpha_c) \hat{q}_{c,t}^{in}, \; \forall c\in \mathcal{C},t,     \label{eq:GMComp2}\\
        \hat{m}_{p,t} = K_p^m (\hat{\pi}_{i,t} + \hat{\pi}_{o,t} ),\; \forall p,t, (i,o) \in p , \label{eq:GMMass1}\\
        \hat{q}_{p,t}^{in} - \hat{q}_{p,t}^{out} =  \hat{m}_{p,t} - \hat{m}_{p,t-1}, \; \forall p,t,   \label{eq:GMMass2}\\
        \sum_{w \in \mathcal{W}(i)} \hat{f}_{w,t} + \sum_{p \in \mathcal{P}_1(i)} \hat{f}_{p,t}^{out} - \sum_{p \in \mathcal{P}_2(i)} \hat{f}_{p,t}^{in} + \sum_{c \in \mathcal{C}_1(i)} \hat{f}_{c,t}^{out} - \sum_{c \in \mathcal{C}_2(i)} \hat{f}_{c,t}^{in} \nonumber \\
        + \sum_{z \in \mathcal{Z}(i)}  \varrho_{z,t} = \sum_{h \in \mathcal{H}(i)} (\rho_{h,t}-\triangle\rho_{h,t}) + \hat{q}_{i,t},\; \forall i,t,   \label{eq:GMNode}\\
        0 \le \hat{f}_{i,t} \le F_{i,t} ,\; \forall i,t.   \label{eq:GMShed}\\
        \hat{f}_{p,t} = 0.5( \hat{f}_{p,t}^{in}+ \hat{f}_{p,t}^{out}), \; \forall p,t,   \label{eq:GMWey1}\\
        \hat{f}_{p,t} |\hat{f}_{p,t}| = \chi_p^f (\hat{\pi}_{i,t}^2 - \hat{\pi}_{o,t}^2), \; \forall p,t, (i,o) \in p,   \label{eq:GMWey2}\\
        0 \le \triangle\rho_{h,t} \le \rho_{h,t}, \; \forall h,t, \label{eq:GMGC} \\
        F_{hi,t}-\hat{f}_{i,t} \le \triangle f_{i,t}, \; \forall i,t, \label{eq:GMLoad} \\
        0 \le f_{i,t}^+, f_{i,t}^-, \; \forall i,t, \;\;  0 \le R_{w,t}^+, R_{w,t}^-, \; \forall w,t. \label{eq:GMspin}
   \end{gather}

     In \eqref{eq:GMWell}--\eqref{eq:GMspin}, $\underline{F}_w/\overline{F}_w$ and $\underline{\Pi}_i/\overline{\Pi}_i$ are the lower/upper production  and pressure limits, respectively; $\gamma_c$ and $\alpha_c$ are the compression and fuel consumption factors of the compressor; $\hat{f}_{c,t}^{out}/\hat{f}_{c,t}^{in}$ and $\hat{f}_{p,t}^{out}/\hat{f}_{p,t}^{in} /\hat{f}_{p,t}$ are the out-/in-flow of the compressor and out-/in-/average-flow of the pipeline, respectively; $K_p^m/K_p^f$ is mass flow/Weymouth equation coefficient; $\mathcal{W}(i),\mathcal{Z}(i)$, and $\mathcal{D}_g(i)$ are subsets of gas wells, P2G units and gas demands connected to node $i$, and $\mathcal{H}(i)$ is a subset of GCs, in which GPUs are supplied from node $i$; $\mathcal{P}_1(i)/\mathcal{P}_2(i)$ and $\mathcal{C}_1(i)/\mathcal{C}_2(i)$ are subsets of pipelines and compressors, whose ending/beginning terminals are node $i$, respectively.

        Similarly, most of the real-time operation constraints of the gas system can be derived from the day-ahead ones \eqref{eq:GMWell}-\eqref{eq:GMWey2}, by removing the hat symbols and including the reserved gas amounts of GCs $g_{h,t}$ as well as nodal violations $V_{i,t}^+/V_{i,t}^-$ in the nodal balancing equation \eqref{eq:GMNode}, as expressed in \eqref{eq:GMNodeRT}. Besides, gas well production and the gas loads follow \eqref{eq:GMadj1} and \eqref{eq:GMadj2}, respectively. \eqref{eq:GMLB} sets the non-negative restriction on the nodal gas imbalance variables.
   \begin{gather}
        \sum_{w \in \mathcal{W}(i)} {f}_{w,t} + \sum_{p \in \mathcal{P}_1(i)} {f}_{p,t}^{out} - \sum_{p \in \mathcal{P}_2(i)} {f}_{p,t}^{in} + \sum_{c \in \mathcal{C}_1(i)} {f}_{c,t}^{out} - \sum_{c \in \mathcal{C}_2(i)} {f}_{c,t}^{in}  \nonumber \\
        + \sum_{z \in \mathcal{Z}(i)}  \varrho_{z,t} = {f}_{i,t} + V_{i,t}^- + V_{i,t}^+ + \sum_{h \in \mathcal{H}(i)} (\rho_{h,t}-\triangle\rho_{h,t} + g_{h,t}),\; \forall i,t,   \label{eq:GMNodeRT}\\
       - R_{s,t}^- \le f_{s,t} - \hat{f}_{s,t}\le R_{s,t}^+ , \; \forall s,t, \label{eq:GMadj1}\\
       - R_{i,t}^- \le f_{i,t} - \hat{f}_{i,t}\le R_{i,t}^+ , \; \forall i,t, \label{eq:GMadj2}\\
        V_{i,t}^-, V_{i,t}^+ \ge 0  , \; \forall i,t.   \label{eq:GMLB}
\end{gather}

\section{Solution Methodology}
\subsection{Clearing the Electricity Market with Uncertainties}
    For ease of analysis, the electricity market robust clearing model is compacted as
    \begin{subequations}\label{eq:PCompact}\begin{align}
        \mathcal{E}(\bm{q}, \bm{\mu})= \min_{\bm{y}} &\mathbf{c}^{\top}\bm{y} + \bm{\mu}^{\top}\mathbf{C}\bm{y} + \max_{\bm{\xi}\in\Upsilon} \min_{\bm{x}} \mathbf{d}^{\top}\bm{x} \label{eq:PCompacta} \\
         s.t. \hspace{1mm}& \mathbf{A}_1\bm{y} \ge \mathbf{A}_2 - \mathbf{A}_3\bm{q},   \label{eq:PUpper1}  \\
        & \mathbf{D}_1\bm{y} + \mathbf{D}_2\bm{x} \ge \mathbf{D}_3- \mathbf{D}_4\bm{\xi}- \mathbf{D}_5\bm{q}. \label{eq:PLower1}
    \end{align}\end{subequations}
    where $\bm{y} = \{ \rho_{h,t},\rho_{h,t}^- , \rho_{h,t}^+, \hat{p}_{u,t}, \hat{p}_{l,t}, \triangle \hat{p}_{d,t}, \hat{\theta}_{i,t}\}$ and $\bm{x} = \{{p}_{u,t}, {p}_{l,t}, \triangle {p}_{d,t},\theta_{i,t},  \triangle p_{u,t}^-, \triangle p_{u,t}^+\}$ are the day-ahead and real-time decision vectors; $\bm{q}$ is the day-ahead decision vector of the gas system, and it aggregates the power demands listed in ECs, i.e., $p_{j,t}, \forall j,t$. The gas prices is $\bm{\mu}=\{ \mu_{h,t};\mu_{h,t}^+; \mu_{h,t}^-\}$, which is a coefficient vector. \eqref{eq:PUpper1} represents the first-stage constraints \eqref{eq:GCFrim}--\eqref{eq:GCLimit} and \eqref{eq:DAGen}--\eqref{eq:DABus}. \eqref{eq:PLower1} expresses the second-stage constraints \eqref{eq:GCReserve}, \eqref{eq:DAGen}--\eqref{eq:DAshed} (without the hat symbols) and  \eqref{eq:WindP}--\eqref{eq:RTadj}. Model \eqref{eq:PCompact}--\eqref{eq:PLower1} admits a standard two-stage robust program and can be solved by the C\&CG algorithm and its details are presented in Algorithm \ref{Ch6CCG}, where the electricity market subproblem (EM-SP) is defined as
    \begin{subequations}\label{eq:EM SP1}\begin{align}
        \text{EM-SP}: \max_{\bm{\xi}\in\Upsilon}  & \min_{\bm{x}} \mathbf{d}^{\top}\bm{x} \label{eq:EM SP1a} \\
        \hspace{1mm} s.t. \hspace{1mm} & \mathbf{D}_2\bm{x} \ge \mathbf{D}_3- \mathbf{D}_4\bm{\xi}-\mathbf{D}_5\bm{q}-\mathbf{D}_1\bm{y}^* \label{eq:SPLower}
    \end{align}\end{subequations}

     In \eqref{eq:EM SP1}, $\bm{y}^*$ is updated by the electricity market master problem (EM-MP), which can be expressed as a single-level optimization problem in
    \begin{subequations}\label{eq:SP2}\begin{align}
        \text{EM-SP}: & \max_{\bm{\xi}\in\Upsilon,\bm{\omega},\bm{u}} \bm{\omega}^{\top}(\mathbf{D}_3-\mathbf{D}_5\bm{q}-\mathbf{D}_1\bm{y}^*) - \sum \bm{u} \label{eq:SP2a} \\
        \hspace{1mm} s.t.  \;\;\; & \mathbf{D}_2^{\top}\bm{\omega} = \mathbf{d}, \;\; \bm{\omega} \ge0, \\
        & - \overline{M}(1-\bm{\xi}) \le \mathbf{D}_4^{\top}\bm{\omega}-\bm{u} \le \overline{M}(1-\bm{\xi}) \\
        & - \overline{M}\bm{\xi} \le \bm{u} \le \overline{M}\bm{\xi} \label{eq:SP2F}
    \end{align}\end{subequations}
    where $\bm{u}$ is an axillary variable equivalent to $\bm{\omega}^{\top}\mathbf{D}_4\bm{\xi}$ and $\overline{M}$ is a sufficient large positive number.

      The formulation of the electricity market-master problem EM-MP in the $r^{\text{th}}$ iteration is expressed in \eqref{eq:EM MP}, where the uncertainty vector ($\bm{\xi}^{*r},\forall r$) is dynamically generated by the EM-SP at each iteration, and $\varphi$ is the worst-case regulation costs.
    \begin{subequations}\label{eq:EM MP}\begin{align}
            \text{EM-MP}: &\min_{\bm{y},\varphi, \bm{x}^r} \mathbf{c}^{\top}\bm{y} + \bm{\mu}^{\top}\mathbf{C}\bm{y} + \varphi \label{eq:EM MPa} \\
        s.t. \hspace{1mm} &\mathbf{A}_1\bm{y} \ge \mathbf{A}_2 - \mathbf{A}_3\bm{q},  \label{eq:EMMPa}\\
        & \varphi \ge \mathbf{d}^{\top}\bm{x}^r, \forall r,  \label{eq:EMMPb}\\
        & \mathbf{D}_1\bm{y} + \mathbf{D}_2\bm{x}^r \ge \mathbf{D}_3- \mathbf{D}_4\bm{\xi}^{*r}- \mathbf{D}_5\bm{q}, \forall r.  \label{eq:EMMPc}
    \end{align}\end{subequations}

    Then, LMEP can be derived from the Lagrangian function $\Lagr(\bm{y}, \bm{x}^r, \varphi, {\bm{\lambda}},  {\bm{\pi}}^r, \bm{\alpha}^r)$ of EM-MP after identifying the worst uncertainty vector ($\bm{\xi}^{*r},\forall r$) by Algorithm~\ref{Ch6CCG}, where ${\bm{\lambda}},  {\bm{\pi}}^r$ and $\bm{\alpha}^r$ are lagrangian multipliers of \eqref{eq:EMMPa}, respectively. According to \cite{ye2016uncertainty} and \cite{schweppe2013spot}, the LMEP is calculated by
    \begin{align}
        \beta_{n,t} &= \frac{\partial\Lagr(\bm{y}, \bm{x}^r, \varphi, {\bm{\lambda}},  {\bm{\pi}}^r, \bm{\alpha}^r)}{\partial p_{n,t}} \label{eq:LMEP}, \;\; \forall n,t  \nonumber\\
        &= {\underline{\lambda}}_{n,t}-{\overline{\lambda}}_{n,t} - \overline{{\lambda}}_{n,t}^{\triangle} + \sum_{\forall r} ({\underline{\alpha}}_{n,t}^r-{\overline{\alpha}}_{n,t}^r - \overline{\alpha}_{n,t}^{\triangle,r}) , \forall n,t \hspace{5mm}
    \end{align}
    where ${\underline{\lambda}}_{n,t}/{\overline{\lambda}}_{n,t}$ and $\overline{\lambda}_{n,t}^{\triangle}$ are the dual variables of constraints \eqref{eq:DABus} and \eqref{eq:DAshed} (upper bound), respectively. ${\underline{\alpha}}_{n,t}^r/{\overline{\alpha}}_{n,t}^r$ and $\overline{\alpha}_{n,t}^{\triangle,r}$ are dual variables of real-time operation constraints \eqref{eq:RTBus} and \eqref{eq:DAshed} (upper bound) under the worst-case scenario at $r^{\text{th}}$ iteration. It should be noted that the impacts of uncertainties on the LMEP have been considered in \eqref{eq:LMEP}.

\begin{algorithm}[!ht]
\caption{The C\&CG Algorithm for Coordinated markets}
\label{Ch6CCG}
\begin{algorithmic}[1]
    \STATE Set parameters $\epsilon, LB=-\infty, UB=\infty, R=0$, select an arbitrary feasible \textsuperscript{a} $\bm{\xi}^*$, \textsuperscript{b} $\bm{g}^*$, \textsuperscript{c} $\bm{j}^*$.

    \STATE \textsuperscript{a} Solve EM-SP \eqref{eq:SP2}, update $\bm{\xi}^*,\bm{x}^*$ and $UB= \mathbf{d}^{\top}\bm{x}^* $. \\
    \textsuperscript{b} Call inner C\&CG algorithm  to solve GM-O-SP \eqref{eq:GMSP1}, update $\bm{g}^*$ and $UB= \bm{\vartheta}^* $. \\
    \textsuperscript{c} Solve GM-I-SP \eqref{eq:GMSP2}, update $\bm{j}^*,\bm{v}^*$ and $LB=  \mathbf{h}^{\top}\bm{v}^* $. \\

    \STATE If $(UB-LB)/LB \le \varepsilon$, terminate; else, $R=R+1$, \textsuperscript{a} $\bm{\xi}^{*R}=\bm{\xi}^*$,  \textsuperscript{b} $\bm{g}^{*R}=\bm{g}^*$, \textsuperscript{c} $\bm{j}^{*R}=\bm{j}^*$.

    \STATE  \textsuperscript{a} Solve EM-MP \eqref{eq:EM MP}, update $\bm{y}^*$ and $LB= \bm{\varphi}^* $. \\
    \textsuperscript{b} Solve GM-O-MP \eqref{eq:GMMP1}, update $\bm{q}^*$ and $LB= \bm{\psi}^* $. \\
    \textsuperscript{c} Solve GM-I-MP \eqref{eq:GMMP2}, update $\bm{g}^*$ and $UB= \bm{\vartheta}^* $. \\

    \STATE If $(UB-LB)/LB \le \varepsilon$ \& same \textsuperscript{a} $\bm{y}^*$, \textsuperscript{b} $\bm{q}^*$, \textsuperscript{c} $\bm{g}^*$, terminate; else, go Step 2.\\

    \textsuperscript{a} For electricity market; \textsuperscript{b}  For gas market; \textsuperscript{c}  For GM-O-SP \eqref{eq:GMSP1}
\end{algorithmic}\vspace{-6mm}
\end{algorithm}

\subsection{Clearing the Gas Market with Uncertainties}
    Besides the two-stage robust optimization based market clearing framework, the nonconvex Weymouth equations in both day-ahead and real-time stages increase the solution difficulty. Existing methods, which are adopted to approximate the Weymouth equation, such as second-order-cone (SOC) relaxation \cite{chen2019operational,chen2019equilibria} and linearization method \cite{correa2014gas}, can not guarantee the solution feasibility. In what follows, the solution procedure for the robust clearing of the gas market would be presented.

    The Weymouth equation is formulated as MISOCP constraints as discussed in Section~\ref{sec:Ch3GFCMethod}. That is achieved by writing sign-function-free form of Weymouth equation, as shown in \eqref{eq:WeyConvex}, with the indicator constraints \eqref{eq:Indicator1}--\eqref{eq:Indicator2}.
    \begin{gather}
        \hat{f}_{p,t}^2 = \chi_p^f (\hat{\pi}_{p,t}^{+,2} - \hat{\pi}_{p,t}^{-,2} ), \; \forall p,t.  \label{eq:WeyConvex}\\
        z_{p,t} =0 \; \Longleftrightarrow \; \hat{f}_{p,t} \ge 0, \hat{\pi}^+_{p,t}=\hat{\pi}_{i,t},\; \hat{\pi}^-_{p,t}=\hat{\pi}_{o,t}, \; \forall p,t, \label{eq:Indicator1}   \\
        z_{p,t} =1 \Longleftrightarrow \; \hat{f}_{p,t} \le 0, \hat{\pi}^+_{p,t}=\hat{\pi}_{o,t},\; \hat{\pi}^-_{p,t}=\hat{\pi}_{i,t}, \; \forall p,t.  \label{eq:Indicator2}
    \end{gather}
    where $z_{p,t}=\{0,1\}$ is the gas flow directional indicator. Note that \eqref{eq:Indicator1}--\eqref{eq:Indicator2} can be further represented as a logic equation using the big-M method, please refer to Section~\ref{sec:Ch3GFCMethod} for narrow boundaries. The quadratic constraint \eqref{eq:WeyConvex} can be written as two opposite inequalities
    \begin{gather}
        \hat{f}_{p,t}^2 + (\sqrt{\chi_p^f} \hat{\pi}^{-^2}_{p,t}) \le (\sqrt{\chi_p^f}\hat{\pi}^{+^2}_{p,t}), \; \forall p,t,   \label{eq:WeyIneq1}\\
        (\sqrt{\chi_p^q} \hat{\pi}^{+^2}_{p,t}) - [\hat{f}_{p,t}^2 + (\sqrt{\chi_p^q} \hat{\pi}^{-^2}_{p,t})] \le 0, \; \forall p,t. \label{eq:WeyIneq2}
    \end{gather}
    where the first inequality \eqref{eq:WeyIneq1} admits a convex cone constraint, and the latter one \eqref{eq:WeyIneq2} is nonconvex.

    The compact form of the gas market robust clearing model with relaxed Weymouth equation is presented as follows.
    \begin{subequations}\label{eq:GCompact}\begin{align}
            \mathcal{Q}(\bm{y}, \bm{\beta})= & \min_{\bm{q},\bm{z}} \;F(\bm{q}) + \max_{\bm{g}}\min_{\bm{j},\bm{v}} d(\bm{v})  \label{eq:GCompacta} \\
        \hspace{1mm} s.t. \hspace{1mm} & \eqref{eq:Usetg}, \;\; (\bm{q},\bm{z}) \in \mathcal{A}, \;\; (\bm{q},\bm{g},\bm{v},\bm{j}) \in \mathcal{B} \label{eq:GCompactA}  \\
        &   \hat{w}_{p,t}(\bm{q}) - \hat{g}_{p,t}(\bm{q}) \le 0, \forall p,t, \label{eq:GCompactB} \\
        &  {w}_{p,t}(\bm{v}) - {g}_{p,t}(\bm{v}) \le 0, \forall p,t, \label{eq:GCompactC}
    \end{align}\end{subequations}
    where $\bm{z}$ and $\bm{j}$ are directional binary variables in the day-ahead and real-time stages, respectively; $\bm{q} = \{p_{j,t}, \varrho_{z,t}, \triangle\rho_{h,t}, \hat{f}_{w,t},\hat{f}_{i,t}, \triangle{f}_{i,t}, R_{w,t}^+, R_{w,t}^-, f_{i,t}^+, f_{i,t}^-, \hat{\pi}_{i,t} ,\hat{f}_{c,t}^{out} , \hat{f}_{c,t}^{in}, \hat{f}_{p,t}^{out} , \hat{f}_{p,t}^{in},\hat{m}_{p,t}, $  $\hat{f}_{p,t}, \hat{\pi}^+_{p,t},\hat{\pi}^+_{p,t}\}$ and $\bm{v} = \{{f}_{s,t},{f}_{i,t}, {\pi}_{i,t} , {f}_{c,t}^{out},{f}_{c,t}^{in},{f}_{p,t}^{out},{f}_{p,t}^{in},{m}_{p,t}, {f}_{p,t}, {\pi}^+_{p,t}, {\pi}^-_{p,t} \}$ are the day-ahead and real-time continuous variable vectors, respectively; $\bm{y}$ is the day-ahead decision vector of the power system, and it aggregates the firm and reserved gas amounts listed in GCs, i.e., $\{\rho_{h,t}, \rho_{h,t}^+, \rho_{h,t}^-, \forall h,t\}$; $\bm{\beta}$ is the vector of LMEP; constraints \eqref{eq:GCompactB} and \eqref{eq:GCompactC} gather the nonconvex inequality \eqref{eq:WeyIneq2} in the day-ahead and real-time stages, respectively; $\mathcal{A}$ and $\mathcal{B}$ are constraint sets and their expressions are given as
    \begin{gather}
        \mathcal{A}= \{(\bm{q},\bm{z}) \; | \; \eqref{eq:GCReserve},\eqref{eq:ECFrim}\text{--}\eqref{eq:P2GLimit}, \eqref{eq:GMWell}\text{--}\eqref{eq:GMWey2}, \eqref{eq:Indicator1}\text{--}\eqref{eq:Indicator2},\eqref{eq:WeyIneq1}\} \label{eq:SetA} \\
        \mathcal{B}= \{(\bm{q},\bm{g},\bm{v},\bm{j}) \; | \; \eqref{eq:GMWell}\text{--}\eqref{eq:GMMass2}, \eqref{eq:GMShed}\text{--}\eqref{eq:GMWey2}, \eqref{eq:GMNodeRT}\text{--}\eqref{eq:GMLB}, \eqref{eq:Indicator1}\text{--}\eqref{eq:Indicator2}, \eqref{eq:WeyIneq1}\} \label{eq:SetA}
    \end{gather}

    Though model \eqref{eq:GCompact} is a two-stage robust problem, it is not readily solvable by the state-of-the-art decomposition algorithms, due to the existence of \eqref{eq:GCompactB} and \eqref{eq:GCompactC}. Therefore, mixed integer second-order cone (MISOC) approximations would be derived for the mixed integer nonlinear programs (MINLPs) in the day-ahead and real-time stages, and then the decomposition algorithms could be applied.

\subsubsection{Deriving MISOC Approximations for MINLPs}
    According to the results in Chapter~\ref{Chapter3} and Chapter~\ref{Chapter5}, where S-MISOCP algorithm is designed to solve deterministic, robust and distributionally robust optimization problems for coupled power and gas systems, the S-MISOCP algorithm is suggested to guarantee the solution feasibility for the gas market decisions.  The adaptive penalty growth rate is developed and high quality initial point is provided. The details are presented in Algorithm~\ref{Ch6S-MISOCP}, where the binary variables are fixed with their optimal values after iteration $I_1^{max}$, consequently the convergence can be guaranteed \cite{R35}.  Another non-trivial outcome of Algorithm~\ref{Ch6S-MISOCP} is the MISOC approximation of the original MINLP when the algorithm terminates, which is favorable for the dualization of the second-stage max-min problem of \eqref{eq:GCompact}.

   The core of the S-MISOCP algorithm is to derive SOC approximations for \eqref{eq:GCompactB} and \eqref{eq:GCompactC} which can be done by giving an initial point $(\bm{q}^0, \bm{v}^0)$ and replace the quadratic terms ${\hat{g}}_{p,t}$ and ${g}_{p,t}$ with their linear approximations  $\overline{\hat{g}}_{p,t}$ and $\overline{g}_{p,t}$, respectively, which are displayed in \eqref{eq:concave1}--\eqref{eq:concave2}.
    \begin{gather}
        \bar{\hat{g}}_{p,t}(\bm{q},\bm{q}^0) =  \hat{g}_{p,t}(\bm{q}^0) + \nabla \hat{g}_{p,t}(\bm{q}^0)^{\top} (\bm{q}-\bm{q}^0), \forall p,t, \label{eq:concave1} \\
        \bar{{g}}_{p,t}(\bm{v},\bm{v}^0) =  {g}_{p,t}(\bm{v}^0) + \nabla {g}_{p,t}(\bm{v}^0)^{\top} (\bm{v}-\bm{v}^0), \forall p,t \label{eq:concave2}
    \end{gather}

\subsubsection{The NC\&CG Algorithm}
    As there are binary variables, namely the gas flow direction variables, in the inner level of the second-stage problem of \eqref{eq:GCompact}, one more C\&CG is needed, making the overall decomposition algorithm for \eqref{eq:GCompact} a nested one \cite{wang2016robust, zhao2012exact}. The master and slave problems of the outer C\&CG algorithm are defined in \eqref{eq:GMMP1} and \eqref{eq:GMSP1}, respectively, which can be solved by Algorithm~\ref{Ch6CCG} and the inner C\&CG algorithm, respectively.
    \begin{subequations}\label{eq:GMMP1}\begin{align}
        \text{GM-O-MP}: & \min_{\bm{q},\bm{z},\bm{j}^r,\bm{v}^r} \;F(\bm{q}) + \psi + \sum_t \sum_p ( \hat{\tau}_{p,t} \hat{s}_{p,t} + \sum_r \tau_{p,t}^r s_{p,t}^r) \label{eq:GMMP1a} \\
        \hspace{1mm} s.t. \hspace{1mm}  & (\bm{q},\bm{z}) \in \mathcal{A};\; (\bm{q},\bm{g}^{*r},\bm{v}^r,\bm{j}^r) \in \mathcal{B}, \forall r; \; \psi \ge d(\bm{v}^r) , \forall r, \label{eq:GMMP1A}  \\
         &  \hat{w}_{p,t}(\bm{q}) - \bar{\hat{g}}_{p,t}(\bm{q},\bm{q}^0) \le \hat{s}_{p,t}, \; \forall p,t, \label{eq:GMMP1B}  \\
         &   {w}_{p,t}(\bm{v}^r) - \bar{{g}}_{p,t}(\bm{v}^r,\bm{v}^{0,r}) \le s_{p,t}^r, \; \forall p,t,r, \label{eq:GMMP1C}\\
         &  \hat{s}_{p,t}\ge 0, \forall p,t, \;\;\;  s_{p,t}^r\ge 0, \forall p,t,r. \label{eq:GMMP1D}
    \end{align}\end{subequations}
    \begin{gather}
        \text{GM-O-SP}: \max_{\bm{g}}\min_{\bm{j},\bm{v}} \; \{ \;d(\bm{v}) \; : \; \eqref{eq:Usetg},  \; (\bm{q}^*,\bm{g},\bm{v},\bm{j}) \in \mathcal{B}, \; \eqref{eq:GCompactC} \} \hspace{4mm} \label{eq:GMSP1}
    \end{gather}

     In \eqref{eq:GMMP1}, $\psi$ is the worst-case regulation costs under the contracted amounts of reserved gas $\bm{g}^{*r}$, which are dynamically generated by GM-O-SP and provided as a primal cut to the GM-O-MP; $\hat{s}_{p,t}/ s_{p,t}^r$ and $\hat{\tau}_{p,t}/ \tau_{p,t}^r$  are violations and the associated penalties, respectively. In \eqref{eq:GMSP1}, $\bm{q}^*$ is the optimal day-ahead decision from GM-O-MP.

    In the inner-loop C\&CG, the subproblem of GM-O-SP, defined in \eqref{eq:GMSP2}, can be solved by Algorithm~\ref{Ch6S-MISOCP}.
    \begin{subequations}\label{eq:GMSP2}\begin{align}
        \text{GM-I-SP}: & \min_{\bm{j}, \bm{v}}\; d(\bm{v}) + \sum_{t}  \sum_{p} \tau_{p,t} s_{p,t} \label{eq:GMSP2a} \\
        s.t. \hspace{3mm} & (\bm{q}^*,\bm{g}^*,\bm{v},\bm{j}) \in \mathcal{B}, \;\; s_{p,t}\ge 0, \forall p,t, \label{eq:GMSP2A} \\
        & {w}_{p,t}(\bm{v}) - \bar{{g}}_{p,t}(\bm{v},\bm{v}^{0}) \le s_{p,t}, \; \forall p,t.  \label{eq:GMSPC}
    \end{align}\end{subequations}
    where $\bm{g}^{*}$ is the worst-case uncertainty obtained from the master problem of GM-O-SP; $s_{p,t}$ denotes the constraint violation and has been penalized in the objective function with penalty $\tau_{p,t}$.

    The tractable formulation of the master problem of GM-O-SP is given as below.
    \begin{subequations}\label{eq:GMMP2}\begin{align}
        \text{GM-I-MP}: & \max_{\bm{g},\bm{\sigma}^r,\bm{\pi}_{k}^r, \bm{\theta}_{k}^r} \vartheta \label{eq:GMMP2a} \\
         s.t. \hspace{1mm} &  \eqref{eq:Usetg},  \;\; \bm{\sigma}^r \ge 0, \forall r,  \; \; \begin{Vmatrix}
            \bm{\pi}_{k}^r
        \end{Vmatrix}_2 \le  \bm{\theta}_{k}^r, \forall k,r, \label{eq:MPg1a}\\
            &\mathbf{H}_3^{\top}\bm{\sigma}^r + \sum_{k} (\mathbf{M}_{k}^{r\top} \bm{\pi}_{k}^r + \mathbf{m}_{k}^{r\top} \bm{\theta}_{k}^r )= \mathbf{D}^{\top}, \forall r, \label{eq:MPg1b}\\
            &\vartheta \le (\mathbf{H}_4- \mathbf{H}_1\bm{g} - \mathbf{H}_2\bm{j}^{*r})^{\top} \bm{\sigma}^r - \sum_{j} (\mathbf{N}_{k}^{r\top} \bm{\pi}_{k}^r + \mathbf{n}_{k}^{r\top} \bm{\theta}_{k}^r ), \forall r. \label{eq:MPg1c}
    \end{align}\end{subequations}
    where $\vartheta$ is the regulation costs under integer recourse actions $\bm{j}^{*r}$; $\bm{\sigma}^r/\bm{\pi}_{k}^r/ \bm{\theta}_{k}^r$ is the dual variables of \eqref{eq:GMMP2Low} at $r^{\text{th}}$ iteration;  $k=\{1,2...2|\mathcal{P}||\mathcal{T}|\}$ is the cone index. The bilinear terms in \eqref{eq:MPg1c}, namely $(\mathbf{H}_1\bm{g})^{\top} \bm{\sigma}^r$, can be linearized by the exact separation approach presented in Appendix~\ref{AP:GM-MP2}.
    \begin{subequations}\label{eq:GMMP2Low}\begin{align}
        \min_{\bm{w}} \; &D\bm{w} & \label{eq:GMMP2A} \\
        \hspace{1mm} s.t. \hspace{1mm}  & \mathbf{H}_1\bm{g}^* +\mathbf{H}_2\bm{j}^* +\mathbf{H}_3\bm{w} \ge \mathbf{H}_4: \;\;\; & \bm{\sigma}   \label{eq:GMMP2B}\\
         & \begin{Vmatrix}
            \mathbf{M}_{k}\bm{w}+\mathbf{N}_{k}
        \end{Vmatrix}_2 \le  \mathbf{m}_{k}\bm{w}+\mathbf{n}_{k}: & \bm{\pi}_{k}, \bm{\theta}_{k}, \; \forall k  \label{eq:GMMP2C}
    \end{align}\end{subequations}
    where $\bm{w} = [\bm{v}^\top, $ $ s_{p,t}, \forall p,t]^\top$; \eqref{eq:GMMP2C} represents the compact form of cones \eqref{eq:WeyIneq1} and \eqref{eq:GMSPC}; \eqref{eq:GMMP2B} is the remaining constraints in \eqref{eq:GMSP2A}.

    The gas market would be cleared if the NC\&CG algorithm converges, and the Lagrangian function $\Lagr(\bm{q},\psi, \bm{v}^r, \bm{\nu},\bm{\omega}^r, \bm{\kappa}^r)$ can be constructed based on the latest GM-O-MP, where $\bm{\nu}/\bm{\kappa}^r$ and $\bm{\omega}^r$ are the Lagrangian multipliers. Then, the LMFGP $\mu_{h,t}$ can be calculated by
    \begin{align}
        \mu_{h,t} &= \frac{\partial \Lagr(\bm{q},\psi, \bm{v}^r, \bm{\nu},\bm{\omega}^r, \bm{\kappa}^r)}{\partial \rho_{h,t}}, \; \forall h,t \nonumber \\
        &= {\underline{\nu}}_{i,t} -{\overline{\nu}}_{i,t} - \overline{{\nu}}^{\triangle}_{h,t} + \sum_{\forall r} ({\underline{\kappa}}_{i,t}^r -{\overline{\kappa}}_{i,t}^r ) , \; \forall i \in \mathcal{H}^{-1}(h) \label{eq:LMGP}
    \end{align}
    where ${\underline{\nu}}_{i,t}/{\overline{\nu}}_{i,t}$ and $\overline{{\nu}}^{\triangle}_{i,t}$ are the dual variables of constraints \eqref{eq:GMNode} and \eqref{eq:GMGC} (upper bound), respectively; ${\underline{\kappa}}_{i,t}^r/{\overline{\kappa}}_{i,t}^r$ is dual variable of the real-time operation constraint \eqref{eq:GMNodeRT} associated with the uncertainty scenario $\bm{g}^{*r}$;  $\mathcal{H}^{-1}(h)$ is a subset of gas nodes listed in contract $h$. Similarly, the impacts of the fuel consumption uncertainties of the GPPs on LMFGP have been considered, as the worst-case operation constraints of the gas system have been added in GM-O-MP.

\begin{algorithm}[!htb]
\caption{The S-MISOCP Algorithm for Gas Market Clearing}
\label{Ch6S-MISOCP}
\begin{algorithmic}[1]
    \STATE Set $I_1^{max}, I^{max}_2,\overline{\mu}, \underline{\mu},\sigma, \epsilon, \varepsilon, i=1$ and \textsuperscript{a} $\tau_{p,t}$ / \textsuperscript{b} $\tau_{p,t}, \tau_{p,t}^r$.

    \STATE Find the initial point: \textsuperscript{a}~$(\bm{v}^0)$ by solving GM-I-SP \eqref{eq:GMSP2} without \eqref{eq:GMSPC}/ \textsuperscript{b}~$(\bm{q}^0,\bm{v}^{0,r})$ by solving GM-O-MP \eqref{eq:GMMP1} without \eqref{eq:GMMP1B}--\eqref{eq:GMMP1C}.

    \STATE If $i> I^{max}_1$, fix gas flow directions with optimal ones in iteration $I^{max}_1$.

    \STATE Solve \textsuperscript{a}~GM-I-SP \eqref{eq:GMSP2} / \textsuperscript{b}~GM-O-MP \eqref{eq:GMMP1}.

    \STATE If  \textsuperscript{a}~\eqref{eq:SSPa} / \textsuperscript{b}~\eqref{eq:SPPb}, or $i > I^{max}_2$, terminate; else, go to Step 6. \vspace{-2mm}\begin{eqnarray}
        |\text{GM-I-SP}^{(i-1)}-\text{GM-I-SP}^{(i)} | \le \epsilon, \;\; s_{p,t} \le \varepsilon,\forall p,t \hspace{2em}\label{eq:SSPa} \\
        |\text{GM-O-MP}^{(i-1)}-\text{GM-O-MP}^{(i)} | \le \epsilon, \;\;\hat{ s}_{p,t},s_{p,t}^r \le \varepsilon,\forall p,t,r\label{eq:SPPb}
    \end{eqnarray}\\

    \STATE Apply the adaptive penalty rate \eqref{eq:Ch3Adaptive} to update \textsuperscript{a}~$\tau_{p,t}$/ \textsuperscript{b}~$\hat{\tau}_{p,t}, \tau_{p,t}^r$, $i=i+1$, update \textsuperscript{a}~$(\bm{v}^0)$ /\textsuperscript{b}~$(\bm{q}^0,\bm{v}^{0,r})$, then go to Step $3$.

     \textsuperscript{a}~For problem GM-I-SP \eqref{eq:GMSP2};  \textsuperscript{b}~For problem GM-O-MP \eqref{eq:GMMP1}
\end{algorithmic}\vspace{-5.8mm}
\end{algorithm}

    The LMRGP is calculated according to the cost causation principle \cite{chakraborty2017cost,ye2016uncertainty,fang2019introducing}. Therefore, LMRGP is equivalent to the costs caused by uncertainties $g_{h,t}^r$  from the view of the GMO. In other words, it is the marginal price of the additional unit of uncertainty, which equals
    \begin{align}
        {\mu}_{h,t}^r &= \frac{\partial \Lagr(\bm{q},\psi, \bm{v}^r, \bm{\nu},\bm{\omega}^r, \bm{\kappa}^r)}{\partial g_{h,t}^r} = {\underline{\kappa}}_{i,t}^r -{\overline{\kappa}}_{i,t}^r ,\;\; i \in \mathcal{H}^{-1}(h) \label{eq:UMGP}
    \end{align}

    Based on \emph{Lemma 1} in \cite{ye2016uncertainty}, the upward and downward LMRGPs can be aggregated as
    \begin{align}
        {\mu}_{h,t}^{+} = \sum_{{\forall r}}  \max \{ {\underline{\kappa}}_{i,t}^r -{\overline{\kappa}}_{i,t}^r , \; 0\} , \;\;i \in \mathcal{H}^{-1}(h) \label{eq:RMGP} \nonumber \\
        {\mu}_{h,t}^{-} = \sum_{{\forall r}} \max \{ {\overline{\kappa}}_{i,t}^r -{\underline{\kappa}}_{i,t}^r , \; 0\} , \;\;i \in \mathcal{H}^{-1}(h)
    \end{align}

\subsection{Seeking the Operational Equilibrium}
   To this end, seeking the equilibrium between the two markets boils down to the fixed point problem as follows.
    \begin{gather}
        [{\bm{q}}, \bm{\mu}] = \mathcal{E}^{-1}({\bm{y}}, \bm{\beta}) \;\; \& \;\; [{\bm{y}}, \bm{\beta}] = \mathcal{Q}^{-1}({\bm{q}}, \bm{\mu})
    \end{gather}

\begin{figure}[!hbtp]
            \centering
            \includegraphics[width=12cm]{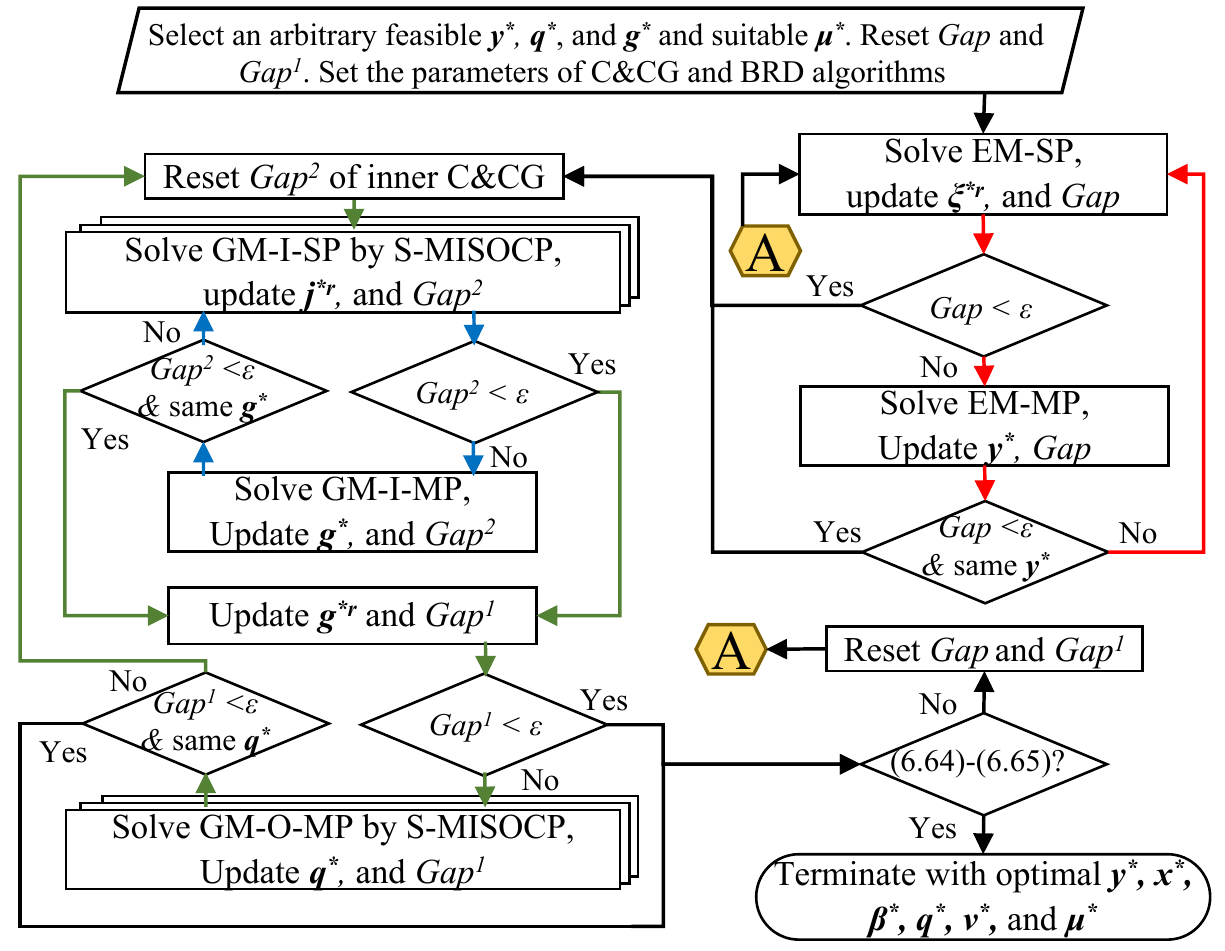}
            \caption{The proposed procedure for the interdependent market mechanism.}
            \label{fig:Ch6Flow}
    \end{figure}

     Similar with \cite{wang2018equilibrium, zhao2018shadow} and \cite{gabriel2012complementarity}, the BRD algorithm is devised to solve the aforementioned fixed point problem, where the algorithm starts with an initial guess of gas prices and power demands of ECs to solve the electricity market, and transmits the optimal GCs as well as electricity prices to the gas market, calculating its best ECs and gas prices, then a new iteration launches. A detailed flow chart of the overall solution procedure is given in Figure~\ref{fig:Ch6Flow} The details of the BRD algorithm are presented in Algorithm~\ref{alg3}. To enhance the convergence performance of the BRD algorithm, the following recommendations are made.
     \begin{enumerate}
        \item
            With fixed integer variables obtained from Steps $2$ and $4$ of Algorithm~\ref{alg3}, i.e., worst-case scenarios $(\bm{\xi}^{*r}, \bm{\delta}^{*r})$ and gas flow directions $(\bm{z}^*, \bm{j}^{*r})$, all continuous variables, including energy prices and demands, can be obtained by solving the KKT optimality conditions \cite{gabriel2012complementarity, chen2019operational} of both EM-MP and GM-O-MP, where uncertainties of reserved gas $\bm{g}^r$ would be replaced with $\mathbf{T}_1\bm{y}\mathbf{T}_2\bm{\delta}^{*r}$ as pointed out in Appendix~\ref{AP:GM-MP2}. Therefore, Step $5$ of Algorithm~\ref{alg3} can be replaced with solving the KKT conditions, and the BRD algorithm will only need to update the integer variables between the two markets.
        \item
            Passing a weighted combination of the recent iteration prices and the previous iteration ones, instead of the recent price only, to the electricity and gas market clearing models, which can be done before Step $2$ and Step $4$ of Algorithm~\ref{alg3}. Similar treatment has been found in the Cobweb algorithm applied in the national energy model system of the US \cite{murphy1998decomposition}.
     \end{enumerate}

\begin{algorithm}[!t]
\caption{The BRD Algorithm}
\label{alg3}
\begin{algorithmic}[1]
    \STATE Set $\varepsilon$, and $l=1$,  select suitable $\bm{\mu}^*$ and $\bm{q}^*$.

    \STATE Call Algorithm \ref{Ch6CCG} for the electricity market \eqref{eq:PCompact}, update ${\bm{y}}^*,\bm{\xi}^{*r}$.
    \STATE With fixed $\bm{\xi}^{*r}$, solve EM-MP \eqref{eq:EM MP} then update $\bm{\beta}$.\\

    \STATE Call Algorithm \ref{Ch6CCG} for the gas market \eqref{eq:GCompact}, update $\bm{z}^{*},\bm{\delta}^{*r},\bm{j}^{*r}$. \\

    \STATE With fixed $(\bm{z}^{*},\bm{g}^{*r},\bm{j}^{*r})$, solve GM-O-MP \eqref{eq:GMMP1} then update $\bm{\mu}$.\\

    \STATE If \eqref{eq:BIRDa}-\eqref{eq:BIRDb}, terminate; else, $l=l+1$, and go Step~$2$.  \begin{gather}
        |\mathcal{E}(\bm{q}^l, \bm{\mu}^l)-\mathcal{E}(\bm{q}^{l-1}, \bm{\mu}^{l-1})| \le \epsilon \label{eq:BIRDa} \\ |\mathcal{Q}(\bm{y}^l, \bm{\beta}^l)-\mathcal{Q}(\bm{y}^{l-1}, \bm{\beta}^{l-1})|\le \epsilon \label{eq:BIRDb}
    \end{gather}

\end{algorithmic}
\end{algorithm}

\section{Simulation Results}
\begin{figure}[!hbtp]
            \centering
            \includegraphics[width=13cm]{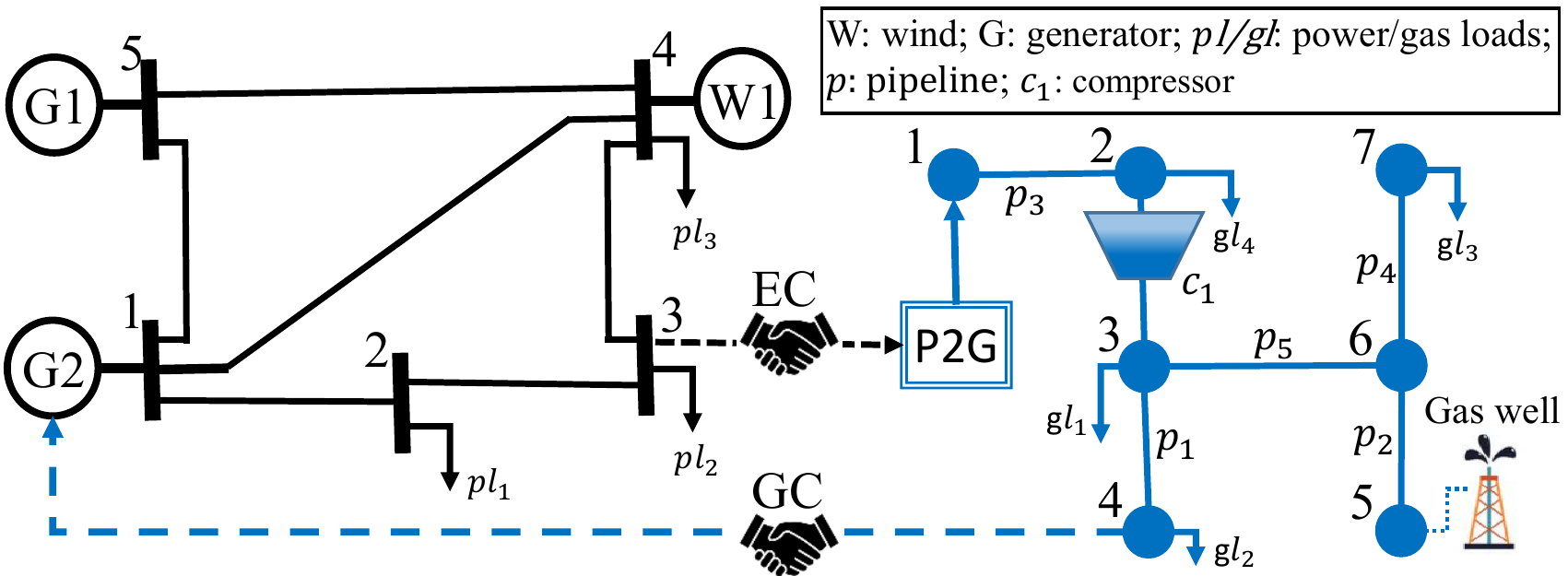}
            \caption{The topology of \textbf{TS-I}.}
            \label{fig:sys}
    \end{figure}

    In this section, the proposed model and solution methods are performed on two test systems, where the first one, denoted \textbf{TS-I}, includes a $5$-bus power network and a $7$-node gas network, and the second one, named \textbf{TS-II}, consists of a $118$-bus power network and a $20$-node gas network. The topology of \textbf{TS-I} is displayed in Figure~\ref{fig:sys}. Due to space limitations, a detailed description of the two test systems, parameters of the three algorithms and wind forecasted data are provided in Appendix~\ref{AppendixB}. The market models and algorithms are programmed using MATLAB with Gurobi solver and YALMIP toolbox \cite{YALMIP} on a computer with $8$~GB RAM and $2.6$~GHz.

\subsection{Base-case Analysis}
    In this sequel, \textbf{TS-I} is examined to provide the characterized equilibrium features. As shown in Figure~\ref{fig:sys}, we have one GC for the GPU and one EC for the P2G unit. The time periods are selected to be from $1$ to $6$, and according to Theorem $3$ in \cite{PriceRobustness}, the wind budget $\Gamma_2$ is set at $4$ to provide feasible decisions with probability more than $97\%$. The BRD algorithm starts with zero ${\bm{q}}$ and gas prices equal its production costs. All algorithms are terminated in a suitable number of iterations, where the BRD algorithm converges in three iterations, and the C\&CG algorithm for electricity (gas) markets and S-MISOCP algorithm converge with $6$ ($3$) and $5$ iterations in average, respectively.
\begin{figure}[!hbtp]
            \centering
            \includegraphics[width=13cm]{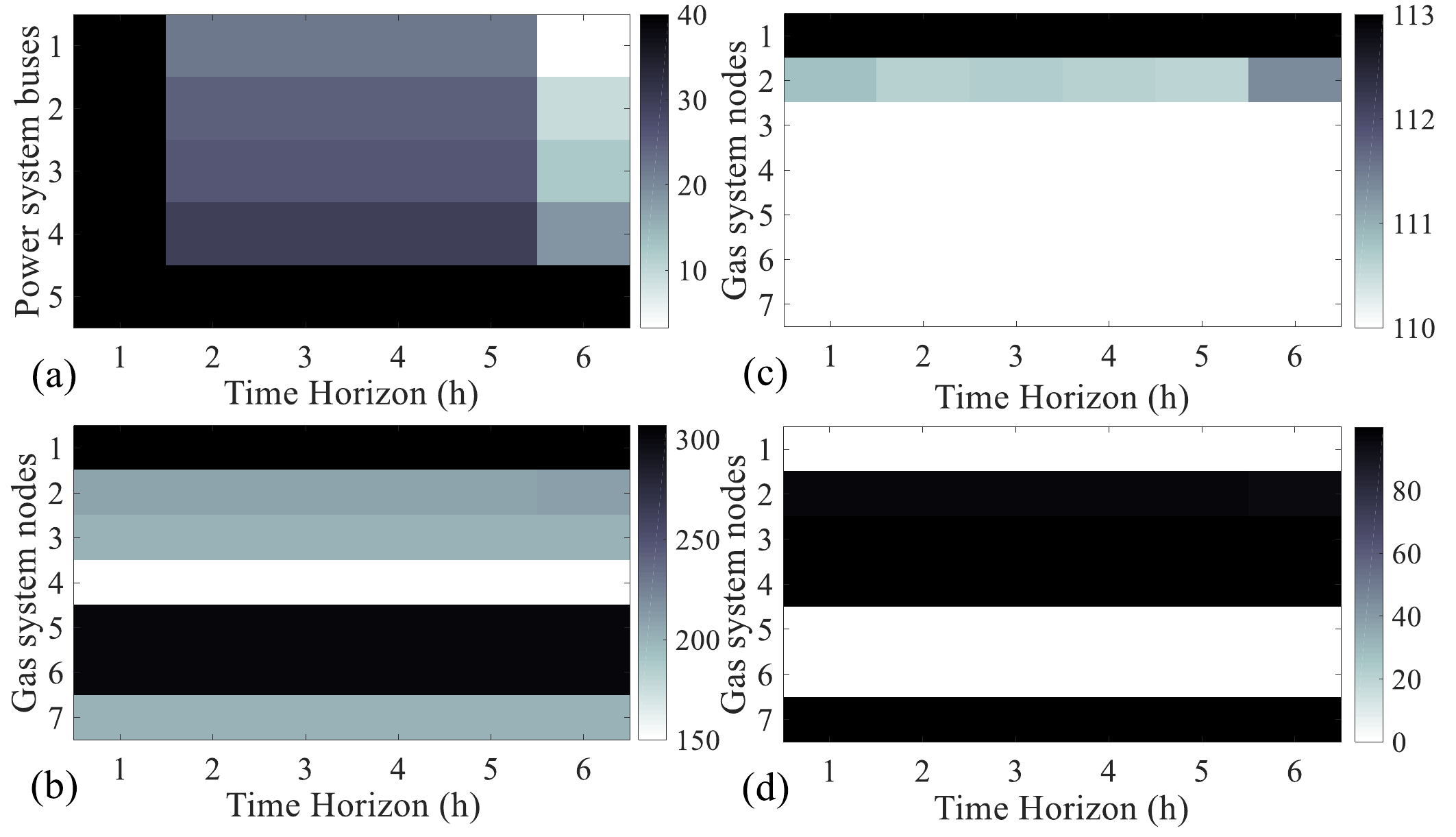}
            \caption{Bone Map for energy prices at equilibrium: (a) LMEP (\$/MWh);  (b) LMFGP (\$/kSm\textsuperscript{3}h);  (c) Upward LMRGP (\$/kSm\textsuperscript{3}h);  (d) Downward LMRGP (\$/kSm\textsuperscript{3}h) }
            \label{fig:Prices}
    \end{figure}
    The energy prices are depicted as a bone map in Figure.~\ref{fig:Prices}. The y-axes represent the power buses or gas nodes, and the x-axis displays the time intervals. The color bar of each sub-figure provides various colors for the energy prices range, for example, the LMEP range is $3.24-40$ \$/MWh. It can be observed from Figure.~\ref{fig:Prices}(a) that the LMEP remains unchanged during some intervals, such as hours $2-5$, while the first and last intervals have a large difference in LMEP due to the ramping generation constraints. The relatively high LMEPs at bus $5$ are caused by the expensive thermal generator $G1$ and the limited capacities of connected the neighboring power transmission lines. On the other hand, the LMFGPs and LMRGPs, as shown in Figure~\ref{fig:Prices}(b)-(d), are almost constant  in all intervals, reflecting the effectiveness of considering the line pack in the gas market model. Based on the simulation results, investments of flexible resources, such as energy storages, might be attracted to reduce energy prices at bus $5$ and nodes $1$, $5$ and $6$.

\subsection{Effectiveness of Modeling the Gas Dynamics}
    The effectiveness of modeling the gas system dynamics is studied by comparing with other gas system models, namely the steady-state \cite{chen2017clearing} and fixed gas flow direction \cite{chen2019operational,wang2018equilibrium} models. In the steady-state model, the stored mass of gas inside pipelines are neglected, and the inlet- and outlet-flow rates are equal. It can be modeled by dropping \eqref{eq:GMMass1}--\eqref{eq:GMMass2} and adding $q_{p,t}^{out}=q_{p,t}^{in}=q_{p,t}$ to the proposed gas market model. In the fixed direction model, the indicator binary variables $\bm{z}$ are predetermined, i.e., the day-ahead gas flow directions are known and these directions would not be changed in the intra-day operation, namely  $\bm{j}=\bm{z}$. The operating costs and energy prices are listed in Table~\ref{tab:LinePack} for the three gas models. Neglecting the line pack increases the operating costs of the electricity and gas markets by $18.74\%$ and $24.40\%$, respectively. Meanwhile, fixing the gas flow directions would bring these markets additional $\$2.723\times10^4$ and $\$5.533\times10^4$, respectively. It can be concluded that considering the gas line pack and bidirectional gas flow decreases the energy prices and operating costs.

    \begin{table}[!hbtp]
        \caption{Operating costs and energy prices at equilibrium with different gas system models} \label{tab:LinePack}
        \centering
        \includegraphics[width=13cm]{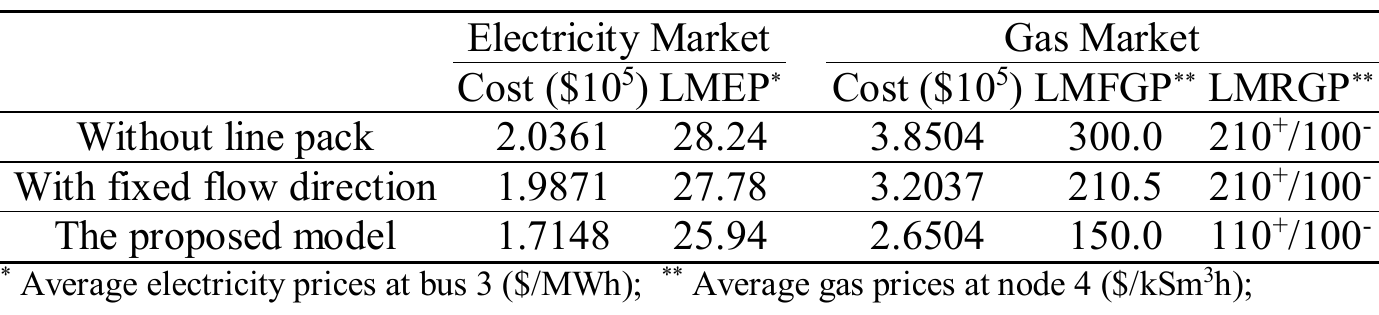}
    \end{table}

\subsection{Comparison with Deterministic Market Clearing Models}
    To reveal the effectiveness of considering wind generation uncertainties, the deterministic market clearing models \cite{gil2015electricity, cui2016day, chen2019equilibria, wang2017strategic, wang2018equilibrium, chen2019operational, zhao2018shadow} are compared with the proposed robust operational equilibrium under different wind penetration levels (WPLs). The operational equilibrium of the deterministic model can be obtained by setting the upper and lower boundaries of wind power outputs with the forecasted values, i.e., $\overline{P}_{e,t} = \underline{P}_{e,t} = \hat{W}_{e,t}, \forall e,t$. WPL represents the relative change in the boundary values of wind outputs, for example, when WPL$=+10\%$, the upper outputs are $\overline{P}_{e,t} = 1.1\times\overline{P}_{e,t}$, and the lower ones are $\underline{P}_{e,t}=0.9\times\underline{P}_{e,t}$. The simulation results are listed in Table~\ref{tab:Uncertainties}. In the deterministic model, there is no reserved gas requirement as the uncertainties are not considered. With the increment of WPL, the contracted reserved gas increases, while the firm amount decreases. The reason is that the LMFGPs would increase and the gas-fired generation is becoming less cost-effective than non-gas ones. However, when WPL is high, the need for the operational flexibility provided by GPU is significant. Therefore, the required gas demands in the GC would increase, from $632.9$ to $635.4$ $\text{kSm}^3\text{h}$ in this case. It is clear that the costs of the EC equal to zero, as the LMEPs are higher than both LMFGPs and LMRGPs considering energy conversion, and the operation of P2G is not cost-effective to the GMO.  Due to the ramping constraints, the LMEP may be non-positive value when the WPL is high, as shown in the last row of Table~\ref{tab:Uncertainties}. Therefore, the GMO would take advantage of the opportunity to sign EC as a revenue. In conclusion, it is important to make sure that the contracted gas amounts are deliverable  for GPUs to utilize  their flexibilities, and the proposed  market clearing framework can effectively reflect the impacts of wind outputs uncertainties.

    \begin{table}[!hbtp]
        \caption{Operational equilibria under different wind penetration levels} \label{tab:Uncertainties}
        \centering
        \includegraphics[width=13cm]{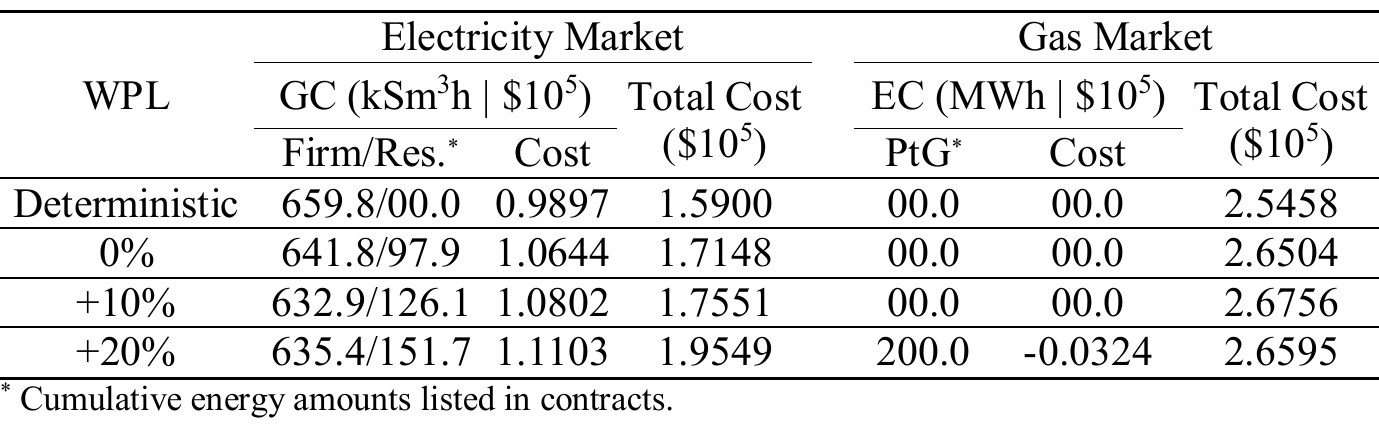}
    \end{table}

\subsection{Comparison with the Centralized Clearing Model}

    In this study, the independent operation (IO) mode of electricity and gas markets, which is in line with the industrial practice, is compared with the central operation (CO) mode of the integrated market adopted in \cite{ordoudis2019integrated, chen2017clearing, li2017security}.  In the CO mode, the objective is to minimize the total operational costs of the two systems in both day-ahead and real-time stages, and the operational constraints of the two systems are included. The robust clearing results of the CO mode operated market can also be obtained by calling Algorithm~\ref{Ch6S-MISOCP} and Algorithm~\ref{Ch6CCG}. The overall CO market model can be written as
    \begin{gather}
        \{\min \eqref{eq:EMObj}+\eqref{eq:objgas} : s.t.\; \eqref{eq:DAGen}\text{--}\eqref{eq:RTadj}, \eqref{eq:GMWell}\text{--}\eqref{eq:GMLB}, \eqref{eq:Indicator1}\text{--}\eqref{eq:WeyIneq2} \} \label{eq:CentOper} \nonumber
    \end{gather}
    where the energy contract terms are removed from the objective function. To provide a fair comparison with the CO mode, the net pocket-of-money (NPM), which equals the operational costs minus the revenue of contracts, for each market is calculated. The reserved gas volumes in the CO mode can be calculated by
    \begin{gather}
                \rho_{h,t}^+ = max \{\sum_{u \in \mathcal{U}_g(h)} \Phi (p_{u,t}^r - \hat{p}_{u,t})/\eta_u , \forall r, \; 0 \}, \forall h,t  \label{eq:CentRes1} \\
                \rho_{h,t}^- = max \{\sum_{u \in \mathcal{U}_g(h)} \Phi (\hat{p}_{u,t}-p_{u,t}^r)/\eta_u ,\forall r, \; 0 \}, \forall h,t \label{eq:CentRes2}
        \end{gather}

   The simulation results under different loading levels are summarized in Table~\ref{tab:IndepOperTab} with different energy demand levels. In the CO mode, due to the high operational flexibility of GPUs and excluding the cost of contracts from the objective, gas demands for the power system would increase. Consequently, the NPM of the electricity market increases as shown in the first two levels. When the loading level of the gas system increases, a high competition between the two systems would be introduced, and the gas market needs to pay additional money. Besides, in the IO mode, the increment of gas (power) demands provides economical benefits to the electricity (gas) market due to the extra energy listed in EC (GC), please see the $6$\textsuperscript{th} and $8$textsuperscript{th} columns of Table~\ref{tab:IndepOperTab}. Moreover, although the two markets are not cleared in the integrated mode, the total operational costs of the IO are very close to those of the CO. For example, in the first loading level, the total costs of the IO mode is \$$3.3\times10^5$, and that of CO is \$$3.299\times10^5$. Therefore, the social welfare impact (SWI), which is defined as the deviation of the total operating costs of the two markets from costs of the CO mode, is quite small, as shown in the last column of Table~\ref{tab:IndepOperTab}. The SWI can be viewed as the costs of preserving the data privacy of the two markets.

    \begin{table}[!hbtp]
        \caption{Economical comparisons between the independent and central market operations under different loading levels.} \label{tab:IndepOperTab}
        \centering
        \includegraphics[width=13cm]{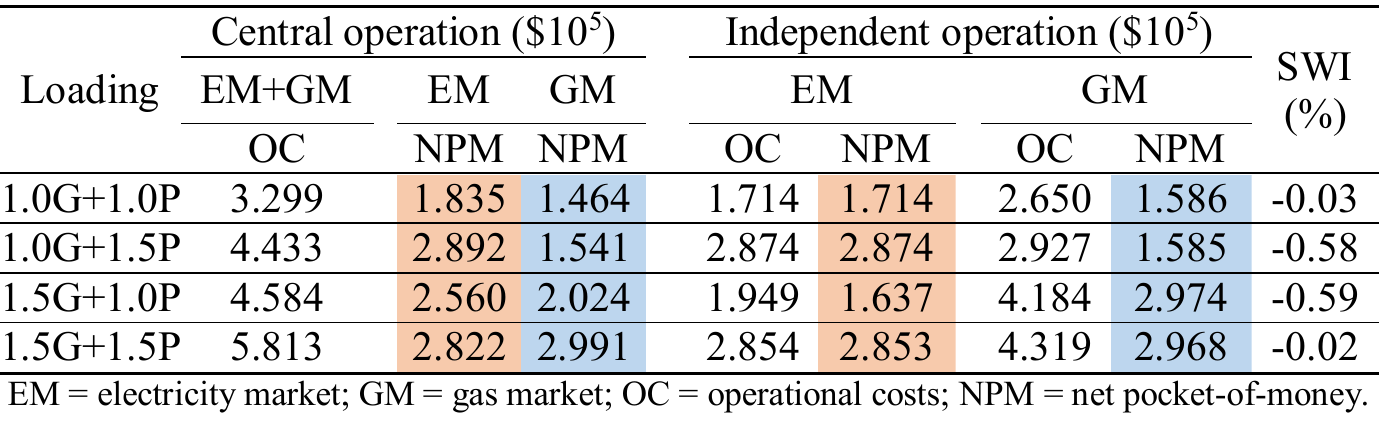}
    \end{table}

\subsection{Computational Efficiency Analysis}
\begin{figure}[!hbtp]
            \centering
            \includegraphics[width=13cm]{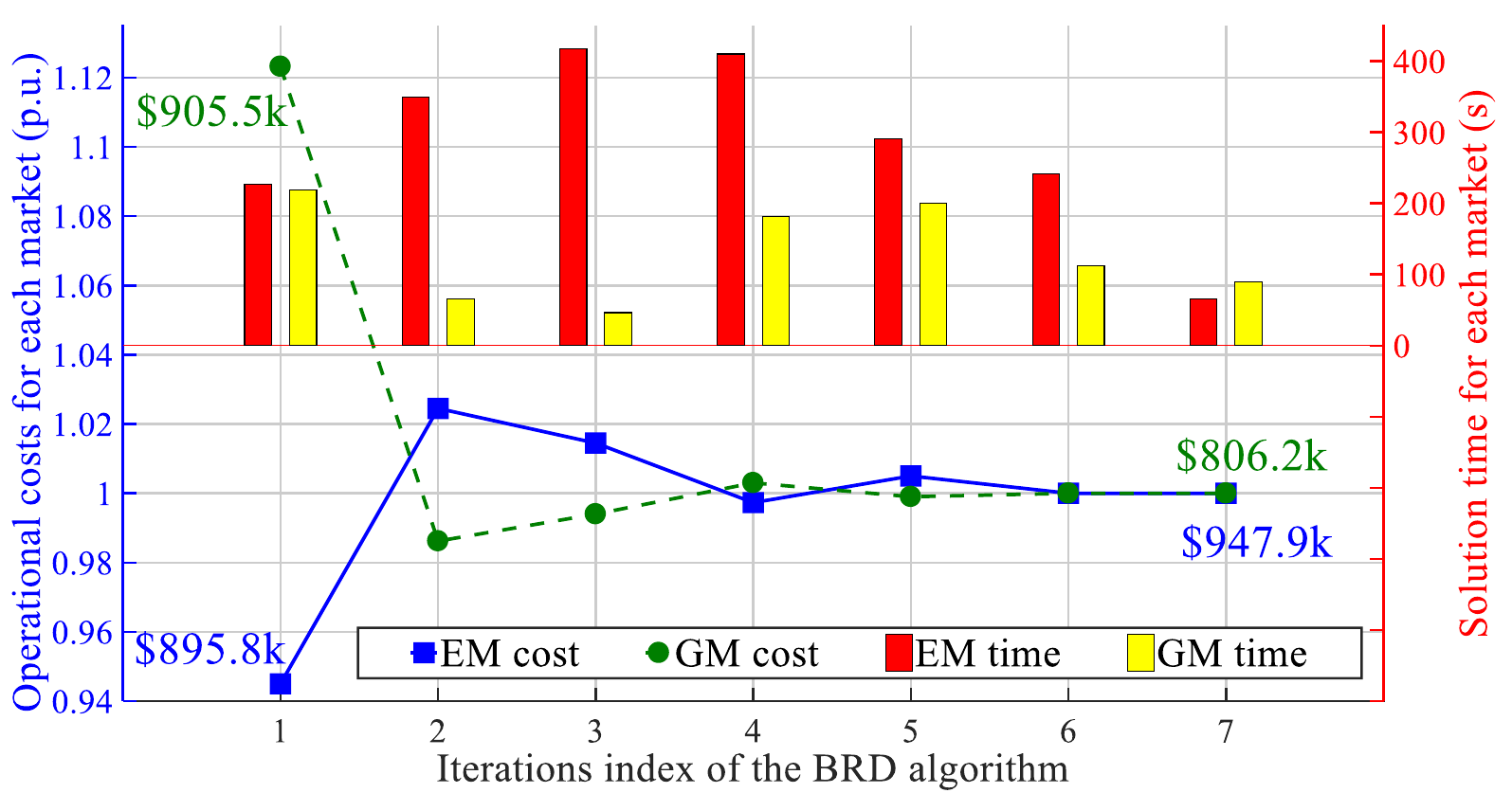}
            \caption{The performance of the BRD algorithm.}
            \label{fig:LargeFig}
    \end{figure}

    To demonstrate the computational efficiency of the proposed procedure, the devised algorithms are implemented on \textbf{TS-II}. The time intervals are considered from $1$ to $6$, and the uncertainty budgets are $\Gamma_1=2, \Gamma_2=3$.  The initial gas prices and EC in the BRD algorithm are set as the average production costs of gas wells and zero, respectively. The operational equilibrium is reached at the $7^\text{th}$ iteration  of the BRD algorithm. The iterative operational costs and execution time for electricity and gas market clearing are articulated in Figure~\ref{fig:LargeFig}. It is observed that the optimal costs of each market seems to iteratively move on a wave to a settled point, i.e., electricity and gas market clearing costs start from \$$895.8$k and \$$905.5$k and end at \$$947.9$k and \$$806.2$k, respectively. The total solution time is $2917.6$s, which is mainly spent on solving the EM-SP and GM-I-MP, as there are large numbers of binary variables. Therefore, we recommend adopting the suggestions proposed in Chapter~\ref{Chapter3}--Chapter~\ref{Chapter5} to enhance the performances of Algorithm~\ref{Ch6CCG} and Algorithm~\ref{Ch6S-MISOCP}. Moreover, the following suggestions are made to improve the computational efficiency of the BRD algorithm:
    \begin{enumerate}
        \item
            Assign all the scenarios obtained from the last iteration, i.e., $\bm{\zeta}^{*r}, \bm{\delta}^{*r}$, which probably are the worst-case scenarios, to decrease the solution time of EM-SP and GM-O-SP.
        \item
            Use an initial guess in solving each problem from the previous iteration, which can be done in  many commercial solvers.
    \end{enumerate}

     With the above suggestions, although the iteration number of the BRD algorithm remains unchanged, the solution time decreases to $1162.7$s ($\approx40\%$). That indicates the suitability of the proposed solution procedure for large-scale systems, considering it is programmed on a PC, rather than a high-performance workstation.

\section{Conclusions and Discussions}

This chapter proposes a method for the robust operational equilibrium seeking of the coupled electricity and gas markets, considering wind generation outputs uncertainties and bidirectional energy transactions. In the gas market clearing model, the gas flow dynamics is considered through the modeling of the line pack and the gas flow directions are allowed to change in both day-ahead and real-time operation stages, so as to maximize its operational flexibility. A six-loop solution procedure is devised to obtain the market equilibrium, including one C\&CG loop to clear the electricity market, two C\&CG loops for the robust clearing of the gas market, two S-MISOCP loops to recovery the solution feasibility in the day-ahead and real-time operation stages of the gas system, and one BRD loop to seek the robust equilibrium. Several suggestions and recommendations are made to enhance the algorithmic performance. Simulation results reveal the benefits of enabling the gas system with more operational flexibility, the superiority of the robust operational equilibrium over the deterministic one, and the gap between the proposed independent clearing framework and the centralized clearing one. Incorporating the contingencies into the uncertainty set as well as the construction of distributionally robust optimization based market clearing framework would be the future work.


\chapter{Conclusions and Future Works} 

\label{Chapter7} 


\section{Conclusions and Discussions}
    The resilient-economic robust operation of the most critical infrastructure energy system, i.e., the electric power system,  is important to strengthen and support economic and social activities in modern society because electricity plays an important role in the secure and continuous operation of other energy systems. However, existing electric power grids experience different forms of vulnerabilities and random failures, such as natural disaster and malicious attacks, which may result in widespread economic and social contingencies. On the other hand, climate change and environmental concerns have been major driven forces for the integration of renewable power generation (RPG) with the power systems around globe.  However, this integration at a large scale brings new challenges for power system operations because of the variable and uncertain output features of RPG. Therefore, it crucial to provide operation models for power systems against such uncertainties, i.e., contingencies and RPG fluctuations, to boost their resilience and reliability in a cost-effective manner.

    Moreover, due to their fast response, good regulation capacity, relatively high efficiency, and low generation costs, gas-fired power units (GPUs) have been playing increasingly larger roles in the resilient and economic operations of power system, such as quick power flow distribution adjustments in the pre-contingency stage, picking up important loads in the post-contingency stage, and mitigating the RPG penetrations in the real-time operation. These actions have significantly improved the physical interdependency between power systems and gas systems. This interdependency has been intensified due to not only the wide deployment of GPUs but also the advanced technologies of power-to-gas (P2G) facilities, which are the most well-qualified solution for the long-term energy storage in the existing bulk power system integrated large-scale RPG. Because the natural gas can be stored with large capacities in a cost-effective manner, P2G facilities are recently employed to effectively convert electricity into gas, which further is stored, transported and reutilized by gas networks.

    To this end, many efforts have been made on the power system resilient-economic operation that are categorized into two optimization models: (1) independent power system (IPS) optimization models, which determine the optimal operation strategies based on the requirements of the electricity utilities, however, they neglect the bidirectional physical interactions with natural gas systems. Therefore, they may not provide the optimal decisions for power system operators (PSOs) and it may cause physical violations for the interacted gas systems; (2) integrated electric-gas system (IEGS) co-optimization models, which overcome the above issue of neglecting physical interactions, and provide a strong solution in terms of energy efficiency improvement and cost-effective perspectives.  However, in most cases, power and gas systems are operated by different utilities, suggesting inevitable economic behaviors between the two energy systems. Due to the fact that the utilization of the superior regulation capabilities of GPUs relies on a reliable gas supply, it is essential to model the physical and economic interactions between power systems and gas systems for resilient and economic decision-making.

    Therefore, this research has developed different operation models for the integrated electric-gas systems from the perspective of the PSO, where the bidirectional interactions between power systems and gas systems are considered from both the physical perspective, i.e., the consideration of the operational and security constraints of the gas system, and the economic perspective, which is addressed by modeling the gas contracts including the here-and-now gas demands and the wait-and-see fuel consumption utilized before and after uncertainty realization, respectively. The developed operation models has proved their ability to provide a high level of reliability and flexibility, which is required for PSO, and secure and feasible optimal decisions for both systems.

    This research has fulfilled the existing gaps between the academic researches and the industrial application by concerning the lack of neglecting physical and/or economic interactions with gas systems. Initially, the question of how to model and solve the interdependent power and gas system has been addressed. Then, different power system dispatch models are developed and efficiently optimized against $N-k$ contingencies and volatile wind power uncertainties, where energy contracts are modeled. Finally, pool-based market mechanism is proposed to separately clear the interdependent electricity and gas markets under uncertainties.

    This study comprehensively discussed the accurate and efficient formulations of both natural gas system and power system that can be incorporated in the optimization models. Physical structures and operational constraints of each system, demonstrating the main components models, has been presented. The dynamic-state gas flow model, which is adopted in the thesis work to provide additional operating flexibility and practical system representation, has been formulated along with steady-state gas flow, AC power flow and DC power flow models. Different types of coordinations between the two systems are listed, while demonstrating their applicability with recent industrial practice.

    The work proposes two convex alternatives to solve the most fundamental problem in the IEGS operation, i.e., the optimal power-gas flow (OPGF). Considering the gas dynamics and bidirectional gas flow inside pipelines, which are adopted in all studies of the thesis, poses additional computational challenges to the OPGF problem. The first alternative, called gas flow correction (GFC) method, employs  the multi-slack-node method and designs the Levenberg-Marquardt algorithm to solve the IEGS at transmission level. The second alternative, named the S-MISOCP algorithm, finds the OPGF for IEGS at distribution-level, considering the non-convex power and gas flow equations. The proposed algorithm is enhanced by (i) suggesting high-quality initial point instead of traditional or random ones, and (ii) adopting an adaptive penalty growth rate to control the main objective and violations weights in the penalized MISCOP problems.

    Thereafter, the resiliency of power systems against contingencies in terms of decision-making is revisited considering the interactions of power systems with gas systems. A two-stage robust day-ahead dispatch model for electric power system against $N-k$ contingencies is proposed. The model detects the worst-case attack against power systems and identifies the optimal gas contracts with preventive and corrective actions; this is accomplished by optimizing the economic generation dispatch in both the pre-contingency and post-contingency stages. Due to the linearization of the non-convex Weymouth equation used in the gas network and the modeling of the on/off grid operation of the generators, binary variables are used for the post-contingency stage decision-making. Therefore, the resultant tri-level framework is solved by the nested column-and-constraint (NC\&CG) algorithm. This developments illustrates how:
    \begin{enumerate}
      \item The proposed model provides a more economical and resilient operation than the IEGS literature models for the PSO because of incorporating the reserved gas contracts in the robust optimization problem.
      \item The IPS literature models generally provide incorrect protection strategy against $N-k$ contingencies, particularly in large power systems, and infeasible decisions for the IEGS operation.
      \item Considering the over-generation issue is essential for the resilient optimization models to cover all possible malicious attacks.

      \item The NC\&CG algorithm can handle the proposed model, especially with the recommended suggestions, which increase its performances.

      \item Dynamic-state gas flow model offers more flexibility because it handles the bidirectional gas flows and gas line pack.
    \end{enumerate}

    In the above model, the day-ahead gas contracts are formulated as a combination of two sub-contracts, namely, the firm gas contract and the reserved gas contract. Emerging P2G facilities to mitigate the surplus RPG outputs, bidirectional gas contracts are inevitable.

    This research develops two operational models for optimal power system operation with bidirectional gas contracts, including P2G and gas-to-power (G2P). The first model is a robust energy management (EM) model for the power distribution network (PDN) against wind generation uncertainty, where both the gas system operation constraints and bidirectional energy trading contracts are considered. The second model is a distributionally robust two-stage contracting model, where bidirectional contracts can be signed in both day-ahead and real-time decision-making stages. To tackle the computational challenge brought by the nonconvex Weymouth equations in the two decision-making stages, a quadruple-loop solution procedure is devised for the first model, including two C\&CG loops and two S-MISOCP loops, through which a robust, feasible and nearly optimal solution can be obtained. The quadruple-loop solution procedure is also designed for the second model to solve the DRO model with $K$ clusters of wind power outputs. These robust economic formulations illustrate how:
    \begin{enumerate}
      \item The proposed economic frameworks outperform the existing IPS and IEGS models in the feasibility and economic perspectives, respectively.
      \item The ability of the proposed models to control and identify the optimal scenario for wind generation management, i.e., curtailment or conversion to gas.
      \item The proposed distributionally robust model has better performance on balancing the robustness and conservativeness of the dispatch strategy than the stochastic and robust optimization models.
      \item Incorporating both day-ahead and real-time gas contracts provides high performance strategies for the PSO.
      \item The proposed model and approaches are applicable to be adopted on the large-scale IEGS, particularly with the several suggestions and recommendations, which are made to enhance the algorithmic performance.
    \end{enumerate}

    Finally, the thesis work terminates with deriving the optimal values of energy and reserved prices to complete the proposed models' compatibility to be applied in the existing industrial practice. A method for the robust operational equilibrium seeking of the coupled electricity and gas markets is proposed, considering RPG uncertainties and bidirectional energy transactions. A six-loop solution procedure is devised to obtain the market equilibrium, including one C\&CG loop to clear the electricity market, two C\&CG loops for the robust clearing of the gas market, two S-MISOCP loops to recovery the solution feasibility in the day-ahead and real-time operation stages of the gas system, and one best-response-decomposition loop to seek the robust equilibrium. Several suggestions and recommendations are made to enhance the algorithmic performance. Simulation results reveal the superiority of the robust operational equilibrium over the deterministic one, and the gap between the proposed independent clearing framework and the centralized clearing one.

\section{Future Work Guidelines}
    To best of our knowledge, previous to this research there was no attempt to model the bidirectional economic interactions of power systems with gas systems in either resilient or economic operation models, and only three proposals consider the interactions between the two systems in the distribution level. Since coordinated operation for the interacted power and gas systems is still relatively young with different challenges, many studies are required to be conducted not only in developing more accurate and faster formulations for both power and gas systems but also in the developments of coordination mechanisms and modeling the coupling components. Next, a discussion on the new future directions, which can improve the current state-of-the-art operation models and solution methodologies, is presented.

\subsection{Modeling the Integrated Electric-gas Systems}

      For gas system modeling, different simplifications are employed to obtain tractable formulations to be incorporated in their operational optimization problems. However, more developments are required to control and systemize the final solution accuracy and the computational burden. For example, the simplified dynamic-state gas flow model is formulated under some assumptions applied on the PDEs, and the gas flow inside compressors and their consumed energy are usually approximated into linear constraints. These simplifications could provide errors in the integrated operation with power system. The unanswered questions are 1) is the exactness of the final decisions acceptable for the interactive utilities? 2) is it possible to deduce the solution quality with such assumptions? 3) is it possible to better represent the gas system dynamics with low computational burden? 4) what are the best lengthes and sizes of pipelines as well as time intervals of the optimization problem, at which the simplified dynamic-state model is still working?  In fact, the IEGS research is in its first era, and it is very fast in modeling and solution methodology developments.  Modern modeling techniques are suggested to represent the energy systems, such as \cite{zhang2020transient,chevalier2020dynamic, garcia2020generalized}, which can be employed in IEGS decision-making frameworks.

      For power system modeling, adopting the accurate AC-OPF, instead of the widely employed and simplified DC-OPF, would be an interesting future research direction to develop robust operational models that consider the reactive power flow, voltage stability and power line losses. Considering the unit commitment decisions along with the resilient and economic dispatch provide more reliable and economical benefits to the PSO. Some interesting studies that are provide tight, accurate and efficient UC problems are \cite{morales2012tight,morales2015tight}.

     For the IEGS coordination, it is observed that integrating the electricity and natural gas systems reduces energy costs, decreases environmental impacts and improves the overall stability and security. However, this integration needs to adopt new technologies for energy conversion, energy storage, efficient energy transportation systems and end-user flexibilities, and it also requires to remove a number of economic and regulatory barriers, which restrict the information exchange. New directions for the future work can be educed from these resent requirements, which enhance the IEGS operation. Moreover,  providing more accurate models for the coupling components, including GPUs, gas compressors and P2G facilities, as well as emerging the communication systems in the IEGS, introduce high performance for a reliable coordination mechanism. Furthermore, different uncertainties can be considered in the IEGS optimization models, such as demand response, local and sizes of the interacted gas systems.

\subsection{Solving the Integrated Electric-gas Systems}

    Future works in solving the optimization problems of IEGSs could be categorized into two main directions. The first direction is to improve the performance and the applicability of the proposed methods that could be accomplished by:
    \begin{enumerate}
      \item Employing the S-MISOCP algorithm in solving a two-stage RO model. This work has been conducted in the thesis to consider power and gas system uncertainties, such as renewable outputs (see Chapter~\ref{Chapter5}), and to be integrated with bilateral gas-electricity marketing model (see Chapter~\ref{Chapter6}). It is the first attempt to employ DCP algorithms in two-stage RO models. Additional work is opened to solve the IEGS operation models under contingencies and demand response uncertainties as well as with other optimization techniques, such as DRO and SO.
      \item GFC method is designed for transmission level IEGS with DC-OPF. New research is to drive novel energy flow correction methods that able to guarantee the solution feasibility of OPGF problems at distribution levels and to be incorporated with non-deterministic optimization models.
      \item Proposing better adaptive penalty rate for the S-MISOCP algorithm with adjustable parameters or with additional factors, which could be controlled and be updated at each iteration. Moreover, suggesting high-quality initial solution is challengeable, especially with large-scale practical systems.
      \item Considering the AC-OPF model with bidirectional power flow in the S-MISOCP algorithm for meshed power grids instead of the fixed directions in the radial networks.
    \end{enumerate}

    The second directions is to suggest new solution approaches. The future research could include
     \begin{enumerate}
       \item In fact, during this conducted research, novel approaches are proposed to tackle the nonlinearity and non-convexity of the IEGS problems. For example, an iterative PLA algorithm is designed in \cite{xue2019optimizing} to dynamically update the breakpoints instead of the fixed ones. Comparing and/or adopting with such up-to-date studies, are our subsequent works.
       \item Other advanced approaches and strategies, such as resilience-oriented information gap decision theory \cite{ma2019resilience} and geographical information system based models \cite{mehrtash2019graph}, have been implemented for IPS. New studies are stimulated to upgrade such studies for IEGS.
       \item The proposed quadruple-loop procedure calls the C\&CG algrithm two times to tackle the tri-level robust models with binary variables in the recourse problem. Other decomposition algorithms, such as \cite{mardan2019accelerated} and \cite{zetina2019exact}, are recently proposed to solve this type of problems. Future work could be to employ such algorithms in the quadruple-loop procedure.
     \end{enumerate}

    Finally, there are several research directions to investigate the data-driven models. Alternative ambiguity sets have been recently proposed in power system distributionally robust studies that include moment-based, Kullback-Leibler divergence-based and Wasserstein distance based ones. The required data, solution conservativeness and computational burden depend on the type of ambiguity set \cite{wang2018risk}. Tractable reformulations of the DRO models have a great interest from researchers to find high-quality conservative and robust decisions with low computational burden. It should be noted that, till finishing this research, there is no attempt has found to adopt the DRO-based models in the coupled electricity and gas markets.


\appendix 



\chapter{Reference Formulations}\label{AppendixA}
 \section{Nonlinear Gas Compressor Model} \label{App:Comp}

    The gas flow inside the compressor $f_{c,t}$ is defined as
    \begin{equation}\label{Ch2CompQ}
      f_{c,t} = sgn(\pi_{i,t}-\pi_{o,t}) \frac{H_{c,t}}{\bar{k}_c - \dot{k}_c \left(\frac{max\{\pi_{i,t},\pi_{o,t}\}}{min\{\pi_{i,t},\pi_{o,t}\}}\right)^{\hat{\alpha}_c} },\; \forall c,t, (i,o) \in c.
    \end{equation}
    where, the empirical parameters $\bar{k}_c, \; \dot{k}_c$ and $\hat{\alpha}_c$ depend on the compressor design and gas physical properties \cite{liu2009security, borraz2010optimiz}; $\pi_{i,t}$ and $\pi_{o,t}$ are terminal pressures of the compressor connected with nodes $i$ and $o$, respectively; $H_{c,t}$ is the controlled power of compressor $c$ at time $t$, it is restricted by a technical range as
    \begin{equation}\label{Ch2CompH}
      \underline{H}_{c} \le H_{c,t} \le \overline{H}_{c},\; \forall c,t.
    \end{equation}
    where $\underline{H}_{c}$ and $ \overline{H}_{c}$ are the minimum and maximum power for the compressor.
    The pressure ratio should be limited according to the compressor capacity as
    \begin{equation}\label{Ch2CompP}
      \underline{\Gamma}_{c} \le \frac{\max\{\pi_{i,t},\pi_{o,t}\}}{\min\{\pi_{i,t},\pi_{o,t}\}} \le \overline{\Gamma}_{c},\; \forall c,t, (i,o) \in c.
     \end{equation}
      where $\underline{\Gamma}_{c}$ and $ \overline{\Gamma}_{c}$ are the minimum and maximum pressure ratios for the compressor.

     For the gas-driven compressors, the equivalent gas flow consumed by the compressor is a convex function in the required power $H_{c,t}$ as
     \begin{equation}\label{Ch2CompGas}
      \Delta f_{c,t} = c_c + b_c H_{c,t} + a_c H_{c,t}^2,\; \forall c,t.
    \end{equation}
    where $c_c, \; b_c$ and $a_c$ are constant parameters.

\section{Formulation of the Exact Gas System Dynamics} \label{APP:Dynamics}

    To represent the gas system dynamics, a set of PDE are derived from the physics of gas particles. This set guarantees that the system is affected by transportation process only, and there is no lost/gained energy or gas mass. According to \cite{correa2015optimal}, this set can be defined and summarized as follows.
    \begin{itemize}
        \item
            \textit{Continuity equation}: it demonstrates the mass balance, and guarantees that the mass inside pipelines remains constant over the time unless a quantity of gas is withdrawn or injected into the pipeline. The continuity equation is expressed in \eqref{eq:Ch2ConEqn}, please refer to \cite{modisette2003physics} for the equation derivation.
             \begin{align}
                \partial(\lambda\upsilon)/\partial x + \partial\lambda/\partial t = 0 \label{eq:Ch2ConEqn}
             \end{align}
             where $\upsilon$ and $\lambda$ are the gas flow velocity and gas density, respectively.

        \item
            \textit{Momentum equation}: according to the Newton's second low, the momentum equation illustrates the rate of momentum of the gas particles and the resultant force on these particles. It is defined in \eqref{eq:Ch2MomEqn}, please refer to \cite{geissler2011mixed} for the equation derivation.
            \begin{align}
                \partial \pi/\partial x + G\lambda\partial h/\partial x + \upsilon \partial\lambda/\partial t + \partial(\lambda\upsilon^2)/\partial x  = \alpha \lambda\upsilon|\upsilon| / (2D) \label{eq:Ch2MomEqn}
             \end{align}
             where $\pi$ is the average gas pressure, and $G,\; h,\; \alpha$ and $D$ are the gravity force, pipeline height, friction and diameter, respectively. The terms represents the pressure gradient, friction force, gas flow rate, kinetic energy, and gravity force, respectively.

        \item
            \textit{Energy equation}: according to \cite{dorin2008modelling}, the law of conservative of energy is
            \begin{align}
                 \gamma\lambda = \frac{\partial}{\partial t} \left[ \lambda(CT+\upsilon^2/2+Gh)\right] + \frac{\partial}{\partial x} \left[ \lambda \upsilon(CT+ \pi/\lambda +\upsilon^2/2+Gh)\right] \label{eq:Ch2EneEqn}
             \end{align}
             where $\gamma$ is the heat transfer rate per unit time and mass, and $C$ and $T$ are the specific heat and temperature, respectively.

        \item
            \textit{State equation}: it expresses the relationship between the state variables of gas, including pressure, density and temperature. In \cite{correa2015optimal}, the thermodynamic equation is derived from universal gas law considering the gas compressibility $Z(\pi,T)$, as
            \begin{align}
                 \pi = \lambda RTZ(\pi,T) \label{eq:Ch2StaEqn}
             \end{align}
            where $R$ is the specific gas constant.

    \end{itemize}

\section{Unit Commitment Problem} \label{App:UC}
    The UC problem identifies the optimal generating schedule for a set of generators subjected to their technical constraints. This problem is a large-scale NP-hard problem that adopted to proceed a weekly or day-ahead dispatch by system operator. Due to its importance in power system operation, it attracts the researchers' attention during the last four decades. In this section, the tight, efficient and compact UC problem formulated in \cite{morales2012tight} is presented. This formulation considers the generation limits, ramping up/down rates and minimum up/down times.

    The objective function of system operator is to minimize the total operational costs. It is defined in \eqref{eq:Ch2UCObj} to respectively include the generation costs, startup and shutdown costs, and penalties of non-served power loads.
    \begin{align}
          \min_{\Omega} \sum_{t \in  \mathcal{T}} \left\{\sum_{u \in  \mathcal{U}} \left( C_u(c_{u,t}p_{u,t}) + C_u^U y_{u,t} + C_u^D z_{u,t} \right) + \sum_{d \in  \mathcal{D}_p} C_d \triangle p_{d,t} \right\} \label{eq:Ch2UCObj}
    \end{align}
    where $c_{u,t}$ is the commitment decision of unit $u$ at time interval $t$. $y_{u,t}$ and $z_{u,t}$ are the startup and shutdown variables and their associated costs are $C_u^U$ and $C_u^D$, respectively. $\triangle p_{d,t}$ is the power load shedding at time $t$ for demand $d$, and it is limited with a practical feasible limits as
    \begin{align}
          0 \le  \triangle p_{d,t} \le P_{d,t}, \; \forall d \in  \mathcal{D}_p,t. \label{eq:Ch2UCShedding}
    \end{align}

    The total generation power must equal the total served power at all times, that is guaranteed by
    \begin{align}
         \sum_{u \in  \mathcal{U}} p_{u,t} = \sum_{d \in  \mathcal{D}_p} (P_{d,t} - \triangle p_{d,t}), \; \forall t. \label{eq:Ch2UCNode}
    \end{align}

    Up and down spinning reserves ($r^+_{u,t},\;r^-_{u,t}$) are limited with a predefined operational constraints as
    \begin{align}
         r^+_{u,t} \ge \underline{R}^+_{u}, \;\; r^-_{u,t} \ge \underline{R}^-_{u}, \; \forall u,t. \label{eq:Ch2UCSpinning}
    \end{align}
    where $\underline{R}^+_{u}$ and $\underline{R}^-_{u}$ are the minimum limits for up and down spinning reserves, respectively. The generation limits are accomplished by
    \begin{gather}
         p_{u,t} + r^+_{u,t} \le \overline{P}_{u} c_{u,t} -  (\overline{P}_{u} -P^+_{u}) y_{u,t}- \max ({P}^+_{u}-P^-_{u}, \; 0) z_{u,t+1} , \; \forall u\in\mathcal{U}_1,t, \label{eq:Ch2UCGenlmt1} \\
         p_{u,t} + r^+_{u,t} \le \overline{P}_{u} c_{u,t} -  (\overline{P}_{u} -P^-_{u}) z_{u,t+1}- \max ({P}^-_{u}-P^+_{u}, \; 0) y_{u,t} , \; \forall u\in\mathcal{U}_1,t, \label{eq:Ch2UCGenlmt2} \\
         p_{u,t} + r^+_{u,t} \le \overline{P}_{u} c_{u,t} -  (\overline{P}_{u} -P^+_{u}) y_{u,t}- (\overline{P}_{u} -P^-_{u}) z_{u,t+1}, \; \forall u\in\mathcal{U}/\mathcal{U}_1,t, \label{eq:Ch2UCGenlmt3} \\
         p_{u,t} \ge \underline{P}_{u} c_{u,t} + r^-_{u,t}, \; \forall u,t. \label{eq:Ch2UCGenlmt4}
    \end{gather}
    where $\overline{P}_{u}$ and $\underline{P}_{u}$ are the maximum and minimum generation capacities, ${P}^+_{u}$ and $P^-_{u}$ are the startup and shutdown capabilities. $\mathcal{U}_1$ is a subset of power units that have minimum up time limit $T^+_u=1$. Ramping up and down constraints are defined as
    \begin{gather}
         p_{u,t} + r^+_{u,t} - p_{u,t-1} \le \overline{R}_{u}^+, \; \forall u,t, \label{eq:Ch2UCRU} \\
         p_{u,t-1} - p_{u,t-1} + r^-_{u,t} \le \overline{R}_{u}^-, \; \forall u,t, \label{eq:Ch2UCRD}
    \end{gather}
    where $\overline{R}^+_{u}$ and $\overline{R}^-_{u}$ are the maximum limits for ramping up and down generation capacities. The minimum up and down times are expressed as
    \begin{gather}
         \sum_{i=t-T^+_u+1}^t y_{u,i} \le c_{u,t}, \; \forall u,t\in[T^+_u,T], \label{eq:Ch2UCTU}\\
         \sum_{i=t-T^-_u+1}^t z_{u,i} \le 1-c_{u,t}, \; \forall u,t\in[T^-_u,T], \label{eq:Ch2UCTD}\\
          c_{u,t} - c_{u,t-1} =  y_{u,t} - z_{u,t}, \; \forall u,t. \label{eq:Ch2UCLogic}
    \end{gather}

    Note that \eqref{eq:Ch2UCLogic} is the startup and shutdown logical constraints. In order to guarantee the UC decisions are feasible and secure in practical operation, the network constraints, i.e., the CPF equations \eqref{eq:Ch2CPFP}-\eqref{eq:Ch2CPFQ}, are added  to the above UC problem to formulate the network-constrained UC (NCUC).

\section{Bus Injection Power Flow Model}\label{App:OPF}
Before presenting the bus injection model, the general power flow equation is firstly discussed. It expresses the relationship between buses voltage $\bm{v}$ and branches current $\bm{i}$, as follows.
    \begin{gather}
          \vec{\bm{i}}_t = \vec{\bm{Y}}_t \vec{\bm{v}}_t,\;\; \forall t. \label{eq:Ch2GFQ}
    \end{gather}
    where $\vec{\bm{v}}_t = \{\vec{v}_{1,t}, .... , \vec{v}_{|\mathcal{N}|},t\}$ is an  $|\mathcal{N}|$-dimensional phasor vector of the voltages at each bus at time $t$, $|\mathcal{N}|$ is the number of buses, $\vec{\bm{i}}_t = \{\vec{i}_{1,t}, .... , \vec{i}_{|\mathcal{N}|},t\}$ is an  $|\mathcal{N}|$-dimensional phasor vector of the current induced at each system bus. $\vec{\bm{Y}}_t$ is an $|\mathcal{N}|\times|\mathcal{N}|$-dimensional phasor matrix, known as bus admittance complex matrix. Please refer to the power system textbook \cite{zhu2015optimization} for the derivation and the final expressions for the bus admittance matrix considering the accurate models for transmission lines, cables and transformers. The power flow equation can be expressed in terms of power instead of currents as
    \begin{gather}
          \vec{\bm{s}}_t = \vec{\bm{v}}_t \bullet (\vec{\bm{Y}}\vec{\bm{v}}_t)^\star, \;\; \forall t. \label{eq:Ch2GFQ2}
    \end{gather}
    where $\vec{\bm{s}}_t=\bm{p}_t+j\bm{q}_t$ is an $|\mathcal{N}|$-dimensional phasor vector of the power at each bus. $\bm{p}_t$ and $\bm{q}_t$ are the active and reactive power injections. "$\bullet$" denotes element-wise multiplication, and "${}^\star$" denotes complex conjugation. Note that working with \eqref{eq:Ch2GFQ2} is more efficient and convenient than \eqref{eq:Ch2GFQ}, as it is independent on currents and directly computes the energy flows. Note that the following formulations provide the exact solution for the system power flows considering that the system is under a sinusoidal steady-state operation (no change in system frequency, phase shift and magnitude).

    Equation  \eqref{eq:Ch2GFQ2} can be written as two algebraic equations by equalizing the active power with the real terms and the reactive power with the imaginary terms.  The following equations, in order, are the most common formulations of the AC-OPF:
    \begin{itemize}
      \item Using the polar coordinates for voltage and rectangular coordinates for admittance, i.e., $\vec{v}_{n,t} = {v_{n,t}}\angle \theta_{n,t}, \; \vec{Y}_{mn}=G_{mn}+jB_{mn}$, then
          \begin{align}
            f_{l_p,t}(\bm{v}_t,\bm{\theta}_t) = v_{n,t}  v_{m,t} \big[ G_{mn} \cos(\theta_{n,t}-\theta_{m,t}) +  B_{mn} \sin(\theta_{n,t}-\theta_{m,t})\big], \; \forall n,t, \label{eq:Ch2ACOPF11}\\
             f_{l_q,t}(\bm{v}_t,\bm{\theta}_t) = v_{n,t} v_{m,t} \big[ G_{mn} \sin(\theta_{n,t}-\theta_{m,t}) +  B_{mn} \cos(\theta_{n,t}-\theta_{m,t})\big], \; \forall n,t. \label{eq:Ch2ACOPF12}
          \end{align}

      \item Using the polar coordinates for both voltage and admittance, i.e., $\vec{v}_{n,t} = {v_{n,t}}\angle \theta_{n,t}, \; \vec{Y}_{mn}= {Y_{mn}} \angle \delta_{mn}$, then
          \begin{align}
             f_{l_p,t}(\bm{v}_t,\bm{\theta}_t) = v_{n,t}  v_{m,t} Y_{mn} \cos(\theta_{n,t}-\theta_{m,t}-\delta_{mn}), \; \forall n,t, \label{eq:Ch2ACOPF21}\\
             f_{l_q,t}(\bm{v}_t,\bm{\theta}_t) = v_{n,t}  v_{m,t} Y_{mn} \sin(\theta_{n,t}-\theta_{m,t}-\delta_{mn}), \; \forall n,t. \label{eq:Ch2ACOPF22}
          \end{align}

      \item Using the rectangular coordinates for both voltage and admittance, i.e., $\vec{v}_{n,t} = a_{n,t}  + jb_{n,t}, \; \vec{Y}_{mn}=G_{mn}+jB_{mn}$, then
          \begin{align}
             f_{l_p,t}(\bm{v}_t,\bm{\theta}_t) =   G_{mn} (a_{n,t} a_{m,t} + b_{n,t} b_{m,t}) +  B_{mn} (b_{n,t} a_{m,t} - a_{n,t} b_{m,t}), \; \forall n,t, \label{eq:Ch2ACOPF31}\\
             f_{l_q,t}(\bm{v}_t,\bm{\theta}_t) =  G_{mn} (b_{n,t} a_{m,t} - a_{n,t} b_{m,t}) -  B_{mn} (a_{n,t} a_{m,t} + b_{n,t} b_{m,t}), \; \forall n,t. \label{eq:Ch2ACOPF32}
          \end{align}
    \end{itemize}

    Note that the fourth formulation, i.e., rectangular coordinates for the voltage  and polar coordinates for the admittance, has no computational benefits or practical use.

     A set of boundary constraints are required to define the upper and lower limits of the decision variables. They could be characterized as
    \begin{gather}
        -\tilde{\pi} \le \theta_{n,t} \le \tilde{\pi} ,\; \forall n,t,\;\;
        \underline{V}_{n} \le v_{n,t} \le \overline{V}_n ,\; \forall n,t \label{eq:Ch2ACVlimit1} \\
         \underline{A}_{n} \le a_{n,t} \le \overline{A}_n ,\;\; \underline{B}_{n} \le b_{n,t} \le \overline{B}_n ,\; \forall n,t \label{eq:Ch2ACVlimit2} \\
         f_{l_p,t}(\bm{v}_t,\bm{\theta}_t)^2 + f_{l_q,t}(\bm{v}_t,\bm{\theta}_t)^2 \le \overline{S}_{l,t}^2 ,\; \forall l,t. \label{eq:Ch2ACSlimit}
    \end{gather}
   In the above expressions, the range of bus voltage in polar and rectangular forms are defined in \eqref{eq:Ch2ACVlimit1} and \eqref{eq:Ch2ACVlimit2}, respectively; Finally, the active and reactive power flow is limited with the power line capacity by equation \eqref{eq:Ch2ACSlimit}.

\section{Exact Separation Approach for GM-I-MP}\label{AP:GM-MP2}
 Exact separation approach is adopted to linearize the GM-I-MP \eqref{eq:GMMP2}. Based on \textit{Lema 1} in \cite{zhao2018data}, there exists an optimal solution $\bm{g}^*$ such that $\bm{g}_{h,t}^*=\rho_{h,t}^+$, $\bm{g}_{h,t}^*=-\rho_{h,t}^-$, or $\bm{g}_{h,t}^*=0, \;\forall h,t$. Therefore, the following constraints satisfy the optimal solution.
    \begin{gather}
        {g}_{h,t} = \rho_{h,t}^+ \delta_{h,t}^+ -  \rho_{h,t}^- \delta_{h,t}^- ,\;\; \forall h,t, \label{eq:linear1} \\
         \delta_{h,t}^+ + \delta_{h,t}^- \le 1, \;\; \delta_{h,t}^+ , \delta_{h,t}^- =\{0,1\},\;\; \forall h,t.   \label{eq:linear2}
    \end{gather}

    Then, \eqref{eq:linear1} can be written as $\bm{g} = \mathbf{T}_1\bm{y}\mathbf{T}_2 \bm{\delta}$. And the nonlinear product $(\mathbf{H}_1\bm{g})^{\top} \bm{\sigma}^r$ can be replaced with $\bm{\varpi}^r$, which is linearized as
    \begin{gather}
        -\overline{M} (1- \bm{\delta}) \le (\mathbf{H}_1 \mathbf{T}_1\bm{y} \mathbf{T}_2)^{\top} \bm{\sigma}^{r} - \bm{\varpi}^r \le \overline{M}(1- \bm{\delta}) , \; \forall r,\;\;\;  \\
        -\overline{M} \bm{\delta} \le \bm{\varpi}^r \le \overline{M} \bm{\delta} , \; \forall r\;
    \end{gather}
    where $\overline{M}$ is a sufficient large positive number.

\section{Incremental Piecewise Linear Approximation Model} \label{App:PLA}
    Solving non-convex functions, such as Weymouth equation, poses difficulties in the NP-hard optimization problems. Piecewise linear approximation (PLA) methods provide a suitable solution to handel the nonlinearities of these functions. Different linearization models have been presented to solve the gas system problems. These models include special ordered set of type two \cite{moller2004mixed}, basic convex combination \cite{padberg2000approximating}, logarithmic \cite{vielma2011modeling}, disaggregated convex combination, disaggregated logarithmic, multiple choice \cite{balakrishnan1989composite}, and incremental \cite{markowitz1957solution} models. In \cite{correa2014gas}, all of these models are applied to the steady-state and dynamic-state gas  flow models and it was shown that the incremental model outperformed the others in terms of computational time and accuracy.

    The Weymouth equation has two nonlinear terms, which are the squared nodal pressure $\pi_{i,t}^2$ and the directional squared pipeline flow $f_{p,t}|f_{p,t}|$. These two terms are linearized individually by applying the incremental PLA model \eqref{eq:INC1}--\eqref{eq:INC4}, in which $\Im(x)$ is the nonlinear function of variable $x$ and it is defined by breakpoints $\{ \Im(x_1), \Im(x_2) ... \Im(x_{S+1}) \}$. A continuous variable $\lambda_k$ is introduced in \eqref{eq:INC3} to represent the portion of each segment $k$. Binary variables $\zeta_k$ are used in \eqref{eq:INC4} to select the active segment and to force the use of all continuous variables $\lambda_k$ of the lower segments. The linearization error decrement can be accomplished by increasing the number of segments $S$ and by changing the breakpoints values $\Im(x_k)$ \cite{correa2014security}.
    \begin{gather}
      \Im(x) \simeq \Im(x_1) + \sum_{k\in \{1,2...S\}} \left[ \Im(x_{k+1})-\Im(x_k)\right] \lambda_k  \label{eq:INC1}\\
      x = x_1 + \sum_{k\in \{1,2...S\}} \left[ x_{k+1}- x_k \right]  \label{eq:INC2}\\
      0 \le \lambda_k  \le 1, \;\; \forall k\in \{1,2...S\}, \label{eq:INC3} \\
      \lambda_{k+1}  \le \zeta_k \le \lambda_{k}, \;, \zeta_k \in \{0,1\}, \;\; \forall k\in \{1,2...S-1\} \label{eq:INC4}
    \end{gather}
    
\begin{figure}[!ht]
        \centering
            \includegraphics[width=12cm]{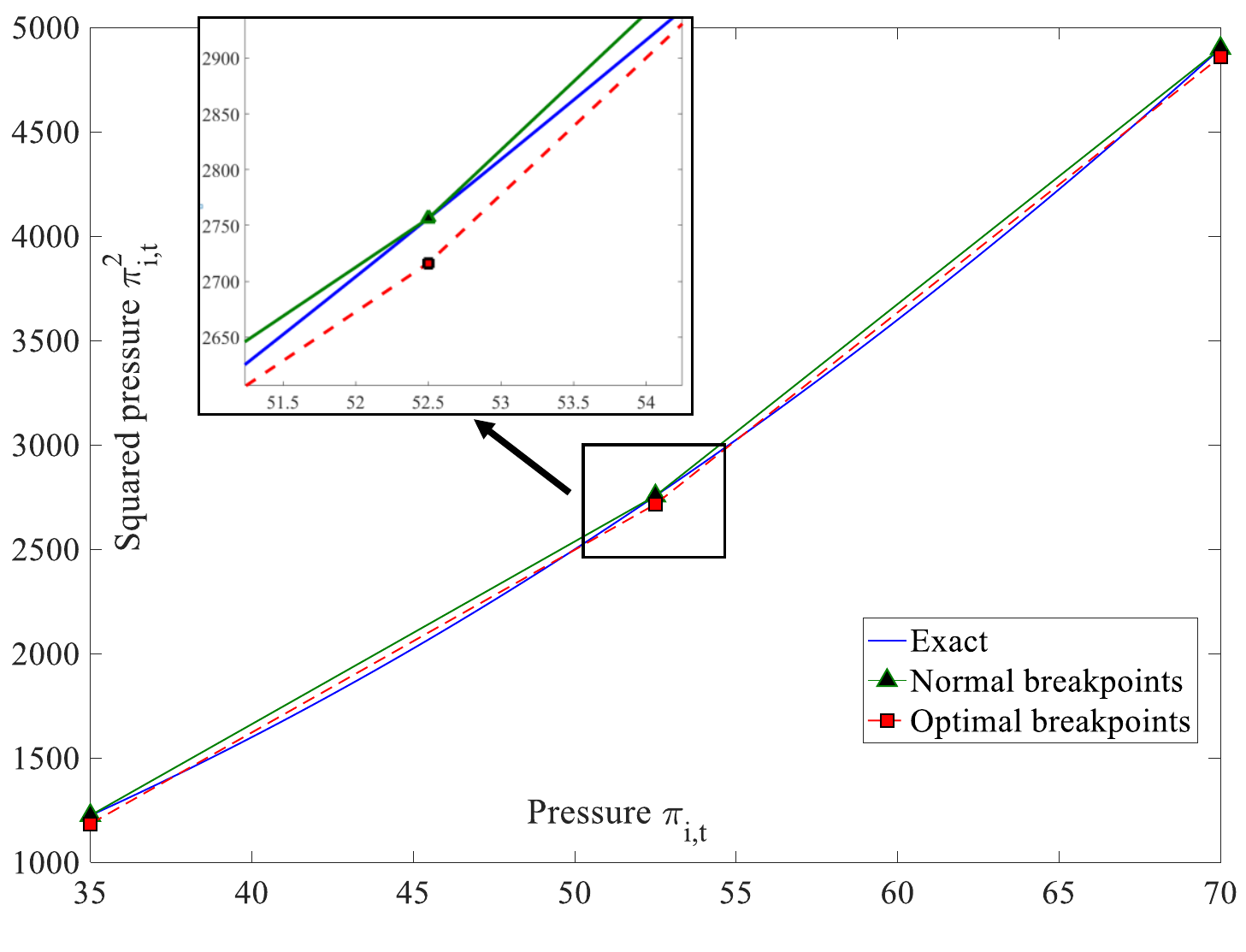} 
            \caption{Breakpoints of PLA model used to linearized the nodal pressure at node $1$ of the 7Nodes gas system}
   \label{fig:PLAPress}
    \end{figure}

For example, the breakpoints used for the incremental PLA model in case of 7Nodes gas system are displayed in Figure~~\ref{fig:PLAPress}. We select node $i=1$ to display its normal and optimal breakpoints with two segments. The pressure range is $35$bar to $70$bar. The squared pressure function $\Im(x)=\pi_{i,t}^2$ is depicted in the vertical axis.

\chapter{Energy Test Systems}\label{AppendixB}

\section{Power Transmission Systems}
\subsection{PJM--5Bus Power Transmission System} \label{App:5Bus}
   \begin{figure}[!ht]
        \centering
            \includegraphics[width=9cm]{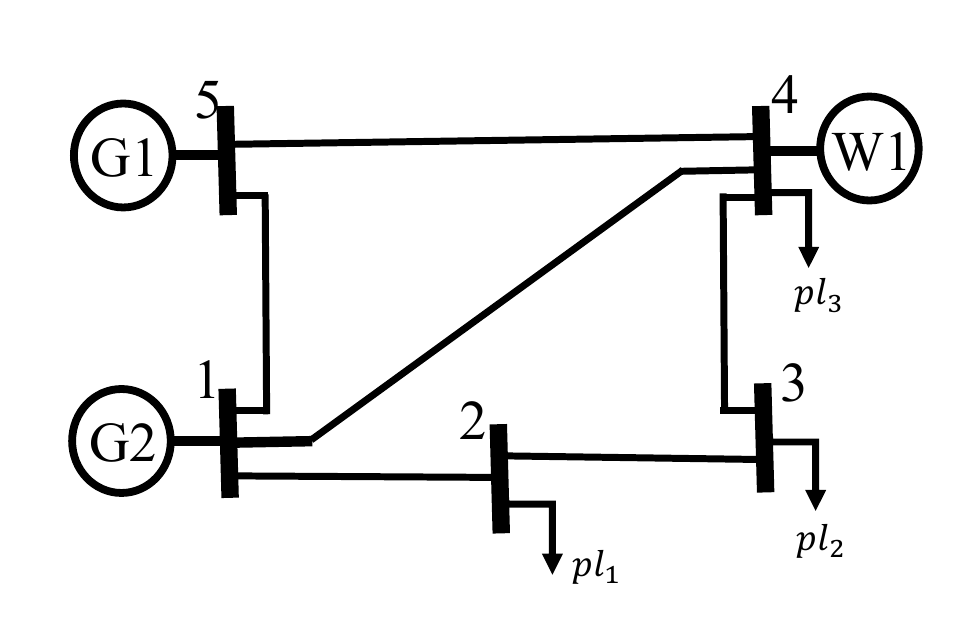} 
            \caption{Topology of PJM-5Bus Power Transmission System}
   \label{fig:AppS1}
    \end{figure}

\begin{table}[!hbtp]
     \label{tab:S1T1}
        \centering
        \caption{Parameters of Generators -- PJM--5Bus System}
        \includegraphics[width=16cm]{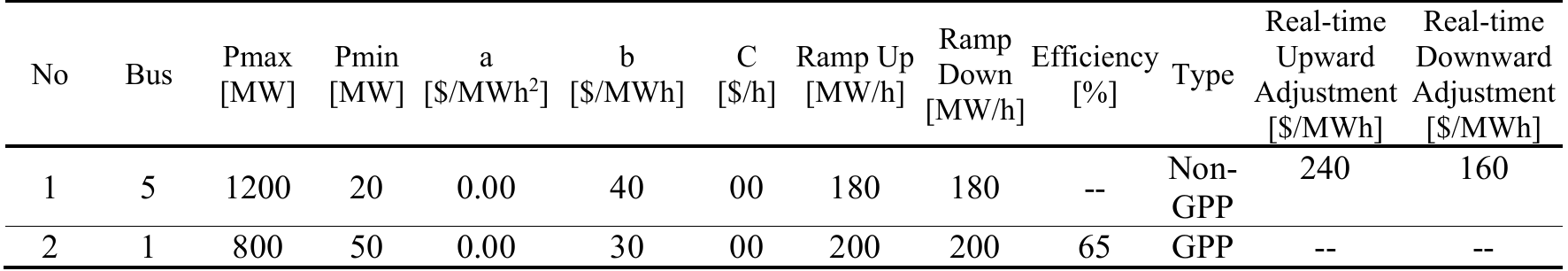}
    \end{table}

    \begin{table}[!hbtp]
     \label{tab:S1T2}
        \centering \caption{Parameters of P2G Facilities -- PJM--5Bus System}
        \includegraphics[width=16cm]{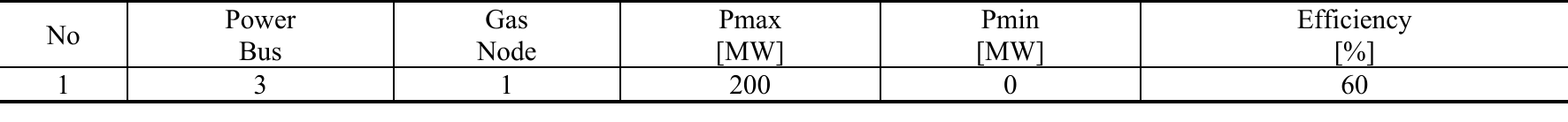}
    \end{table}

\begin{table}[!hbtp]
     \label{tab:S1T3}
        \centering  \caption{Parameters of Wind Power Generation [Pmin = 0, Pmax = 1000MW, wind curtailment penalty = 500\$/MWh] -- PJM--5Bus System}
        \includegraphics[width=16cm]{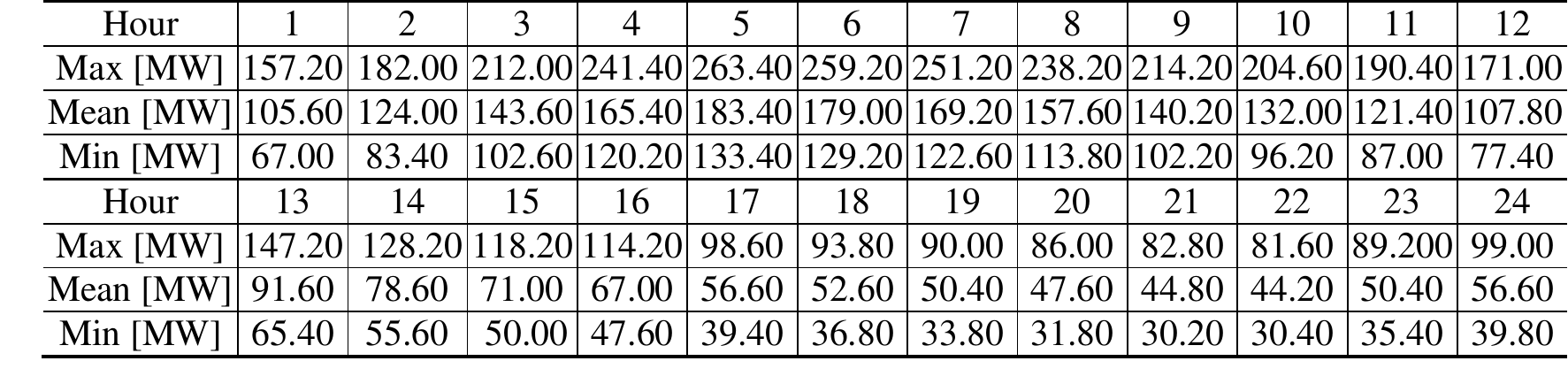}
    \end{table}

\begin{table}[!hbtp]
     \label{tab:S1T4}\centering
\caption{Parameters of Power Lines -- PJM--5Bus System}
        \includegraphics[width=16cm]{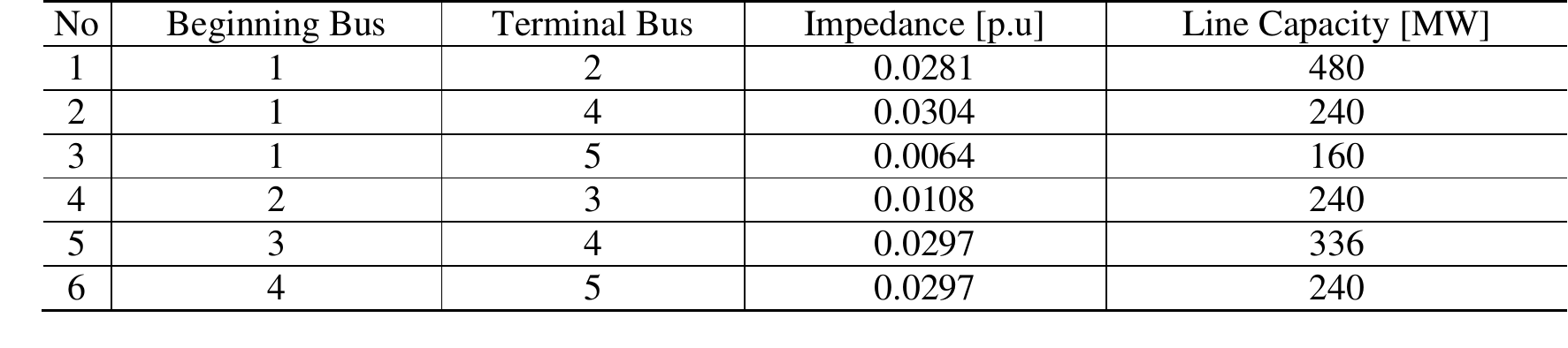}
    \end{table}

\begin{table}[!hbtp]
     \label{tab:S1T5}\centering
\caption{Unit Commitment -- PJM--5Bus System}
        \includegraphics[width=16cm]{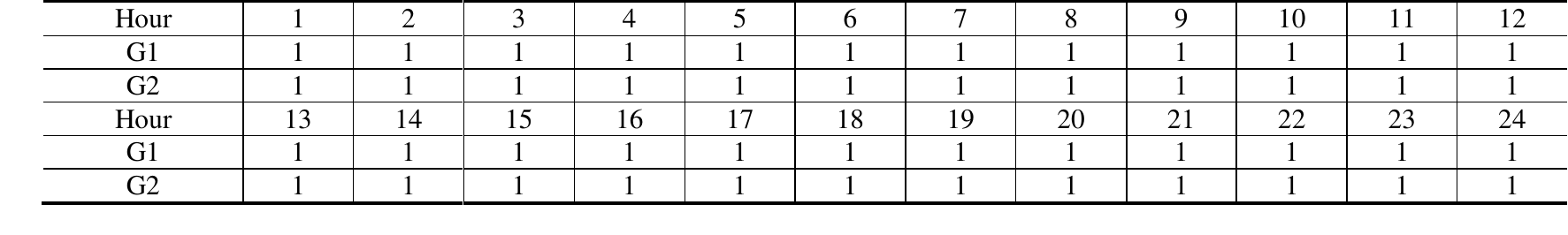}
    \end{table}

    \begin{table}[!hbtp]
     \label{tab:S1T6}
        \centering \caption{Parameters of Load Demand -- PJM--5Bus System}
        \includegraphics[width=16cm]{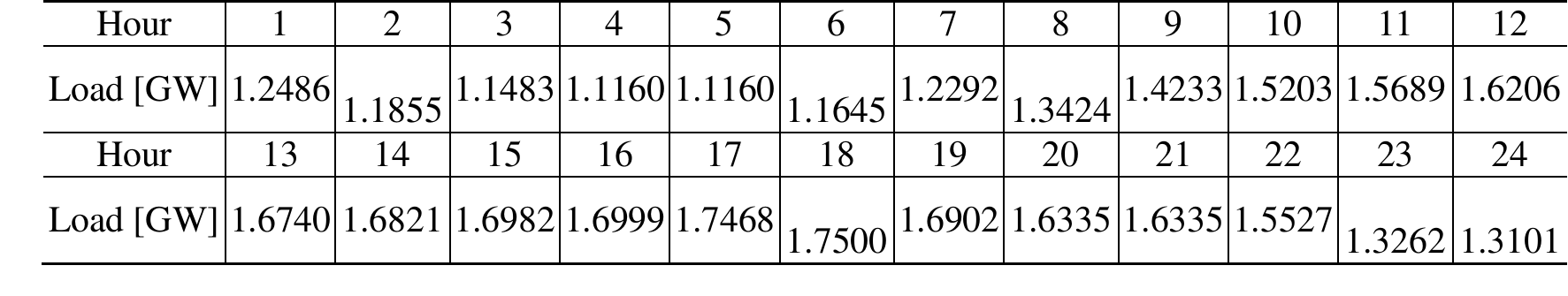}
    \end{table}

    \begin{table}[!hbtp]
     \label{tab:S1T7}
        \centering \caption{Load Portion-- PJM--5Bus System}
        \includegraphics[width=16cm]{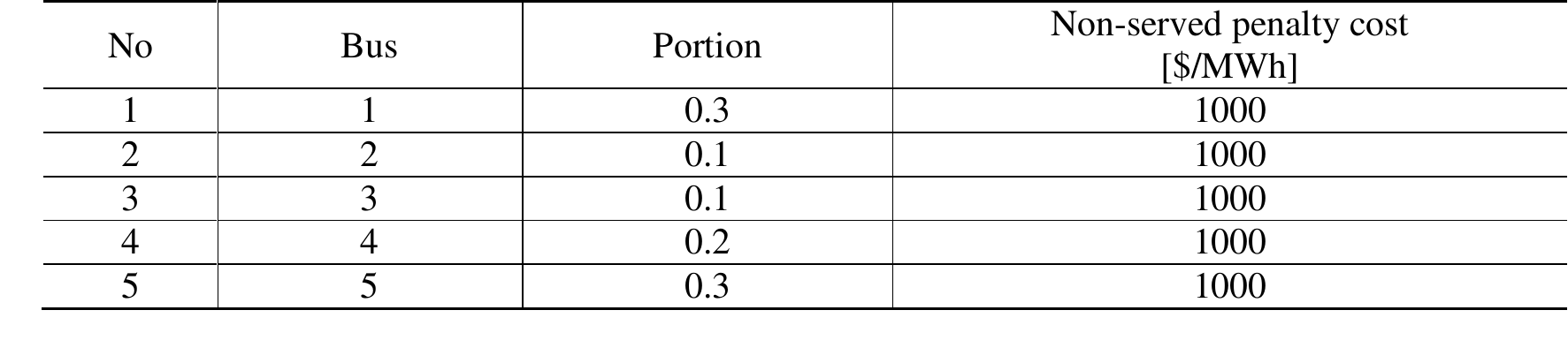}
    \end{table}

\subsection{IEEE--39Bus Power Transmission System} \label{App:39Bus}
    \begin{figure}[!ht]
        \centering
        \includegraphics[width=10cm]{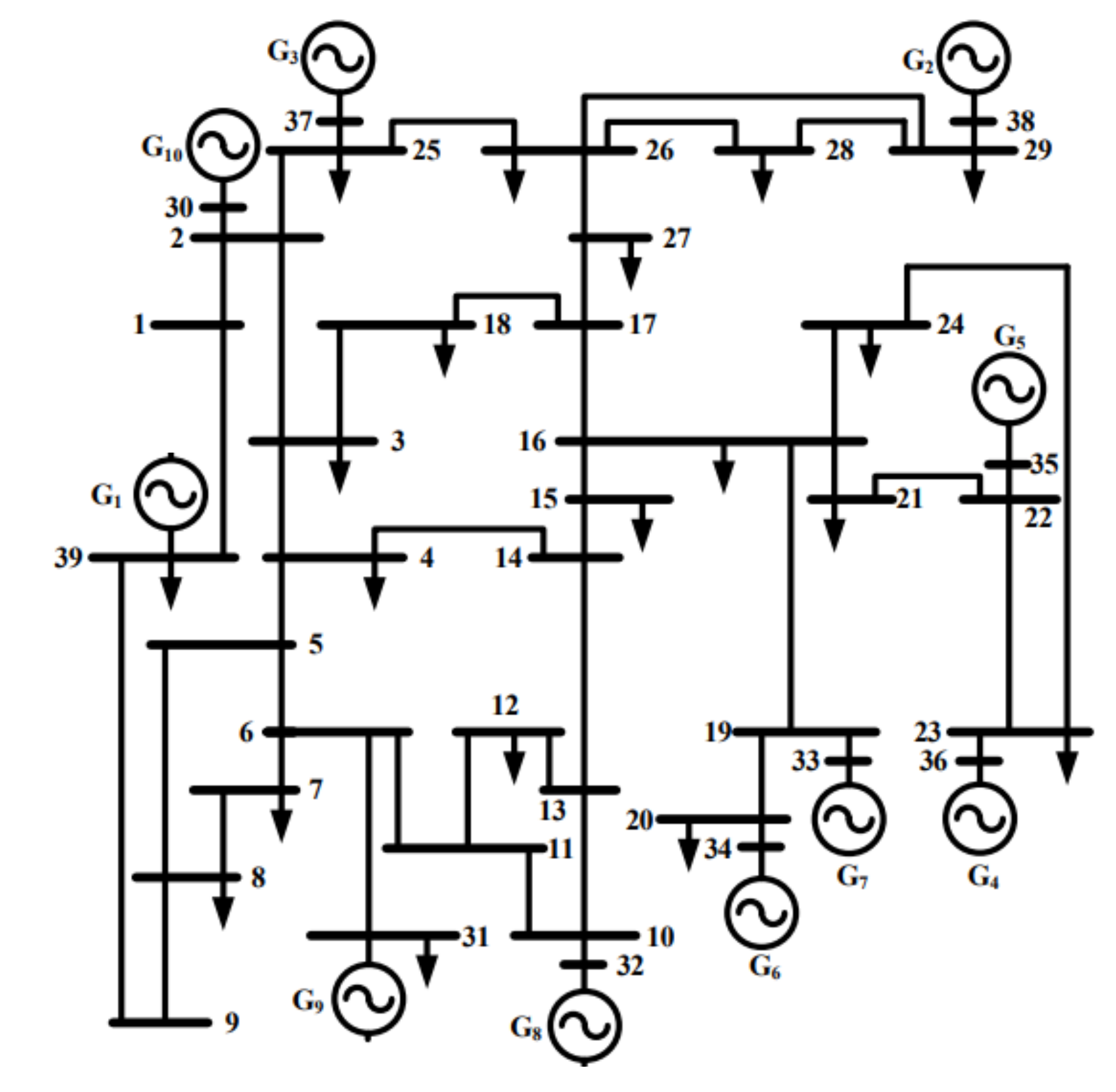}  
            \caption{Topology of IEEE--39Bus Power Transmission System}
   \label{fig:AppS3}
    \end{figure}
    We used the same parameters of power system which is utilized in paper “Robust Defense Strategy for Gas Electric Systems against Malicious Attacks”, and there is no particular mention. Please see the system parameters from https://sites.google.com/site/chengwang0617/home/data-sheet. The additional parameters are defined as follows.

    \begin{table}[!hbtp]
     \label{tab:S3T1}
        \centering \caption{Adjustment costs for non-GPUs and Efficiencies of GPUs-- IEEE--39Bus System}
        \includegraphics[width=16cm]{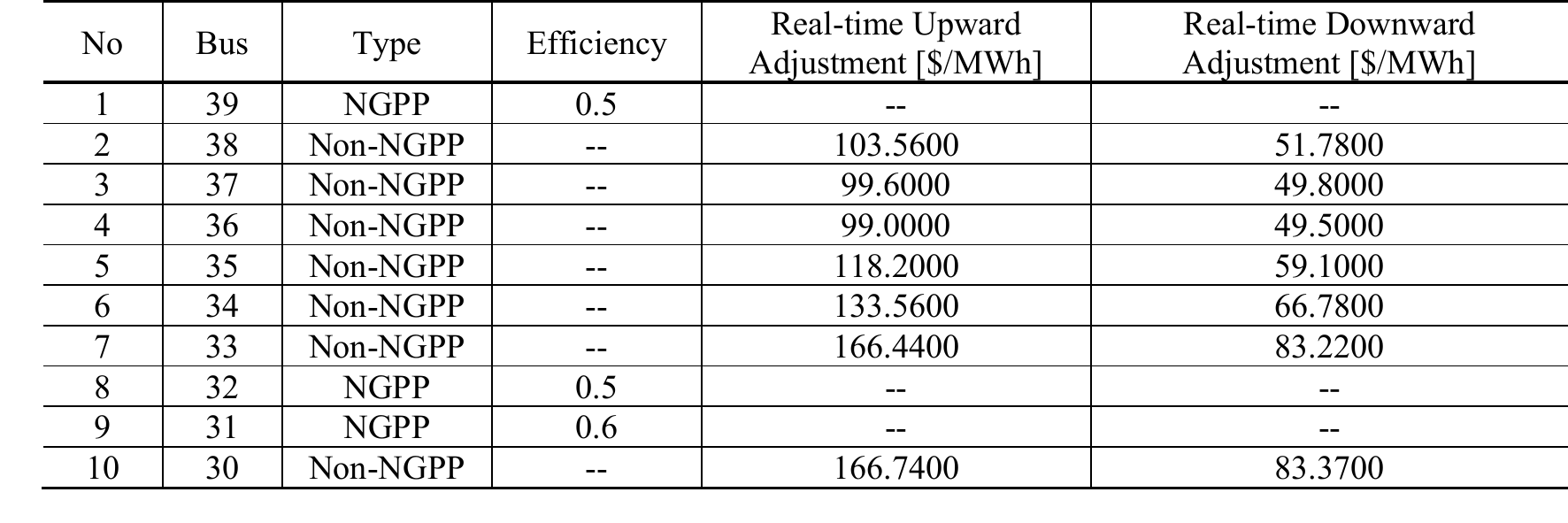}
    \end{table}
    \newpage
\subsection{IEEE--118Bus Power Transmission System}  \label{App:118Bus}
    \begin{figure}[!ht]
        \centering
        \includegraphics[width=14cm]{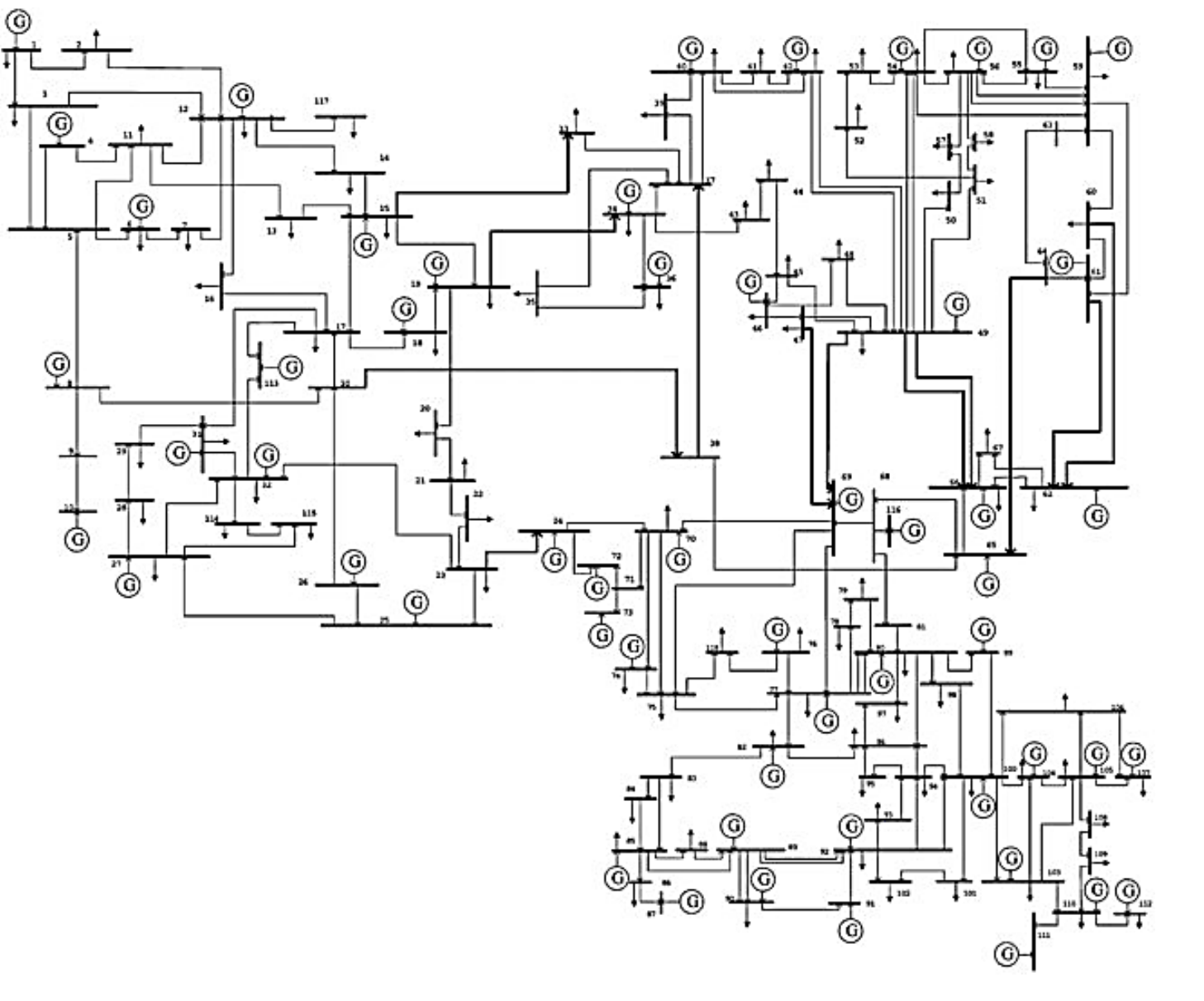} 
            \caption{Topology of IEEE--118Bus Power Transmission System}
   \label{fig:AppS3}
    \end{figure}

      \begin{table}[!hbtp]
     \label{tab:S1T20}
        \centering \caption{Parameters of Wind Power Generation [Pmin = 0, Pmax = 1000MW, wind curtailment penalty = 500\$/MWh] -- IEEE--118Bus  System}
        \includegraphics[width=16cm]{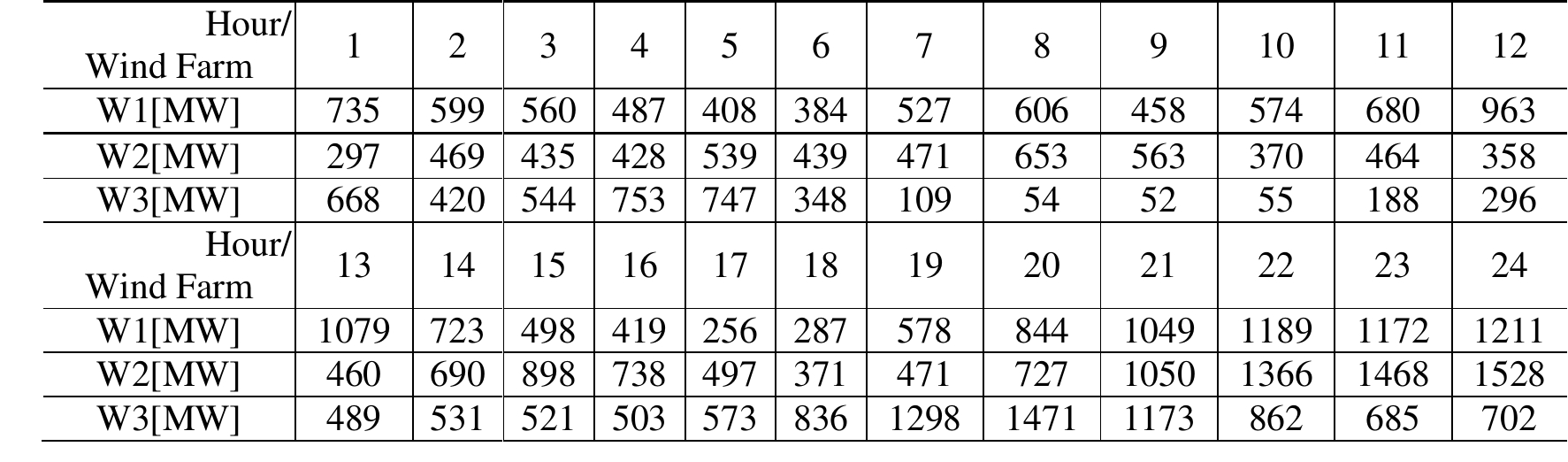}
    \end{table}

     \begin{table}[!hbtp]
     \label{tab:S1T19}
        \centering \caption{Parameters of P2G Facilities -- IEEE--118Bus  System}
        \includegraphics[width=16cm]{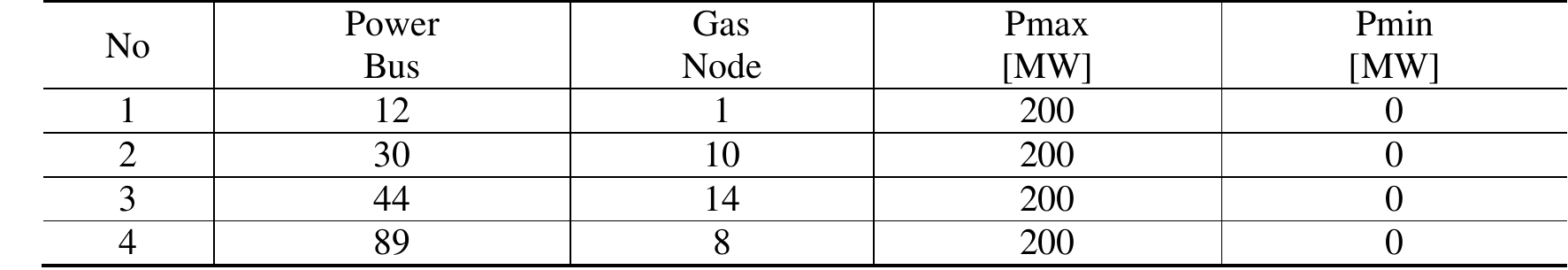}
    \end{table}

    \begin{table}[!hbtp]
     \label{tab:S1T17}
        \centering \caption{Parameters of non-GPUs Generators -- IEEE--118Bus  System}
        \includegraphics[width=16cm]{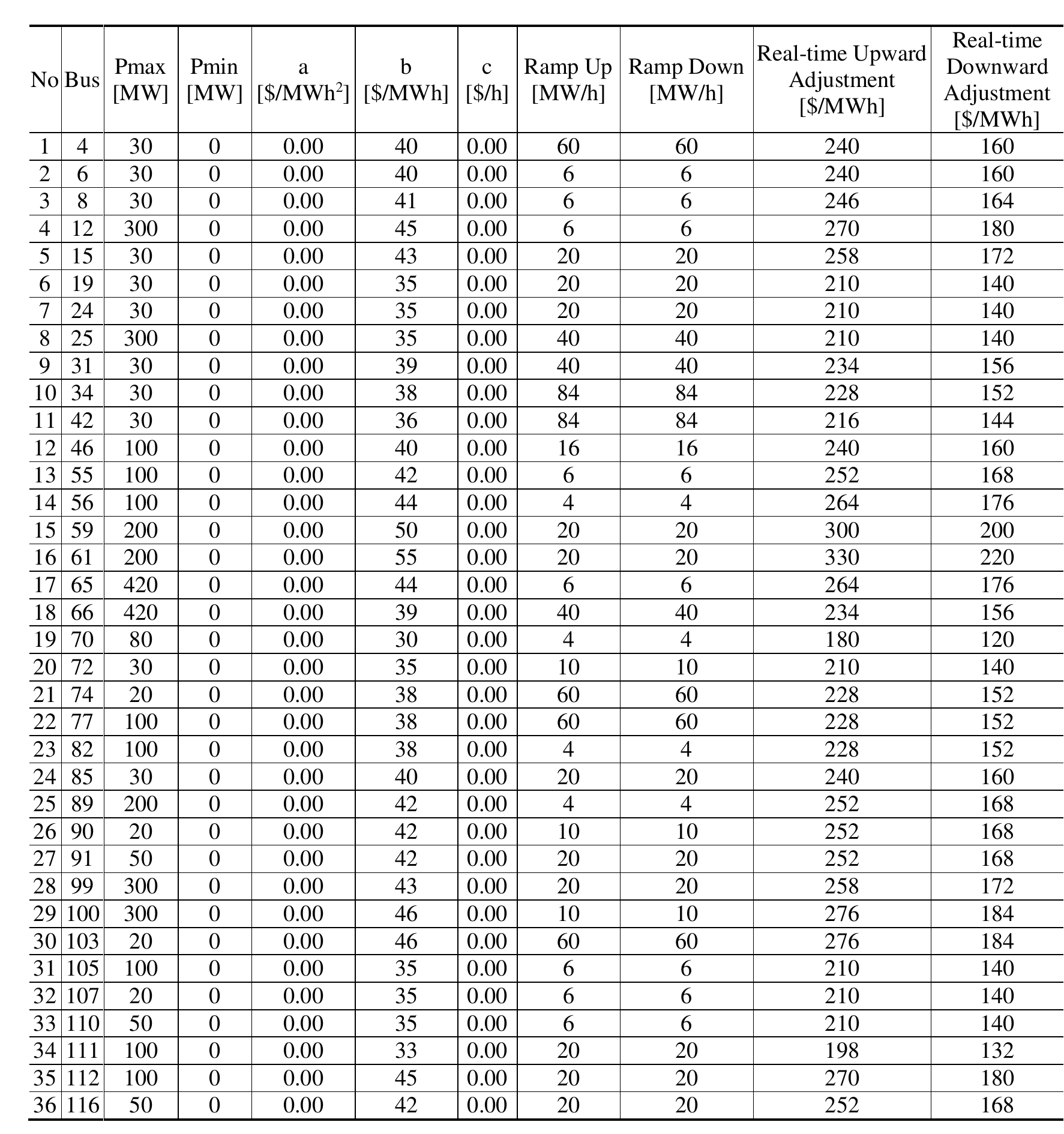}
    \end{table}

    \begin{table}[!hbtp]
     \label{tab:S1T18}
        \centering \caption{Parameters of GPUs Generators  -- IEEE--118Bus  System}
        \includegraphics[width=16cm]{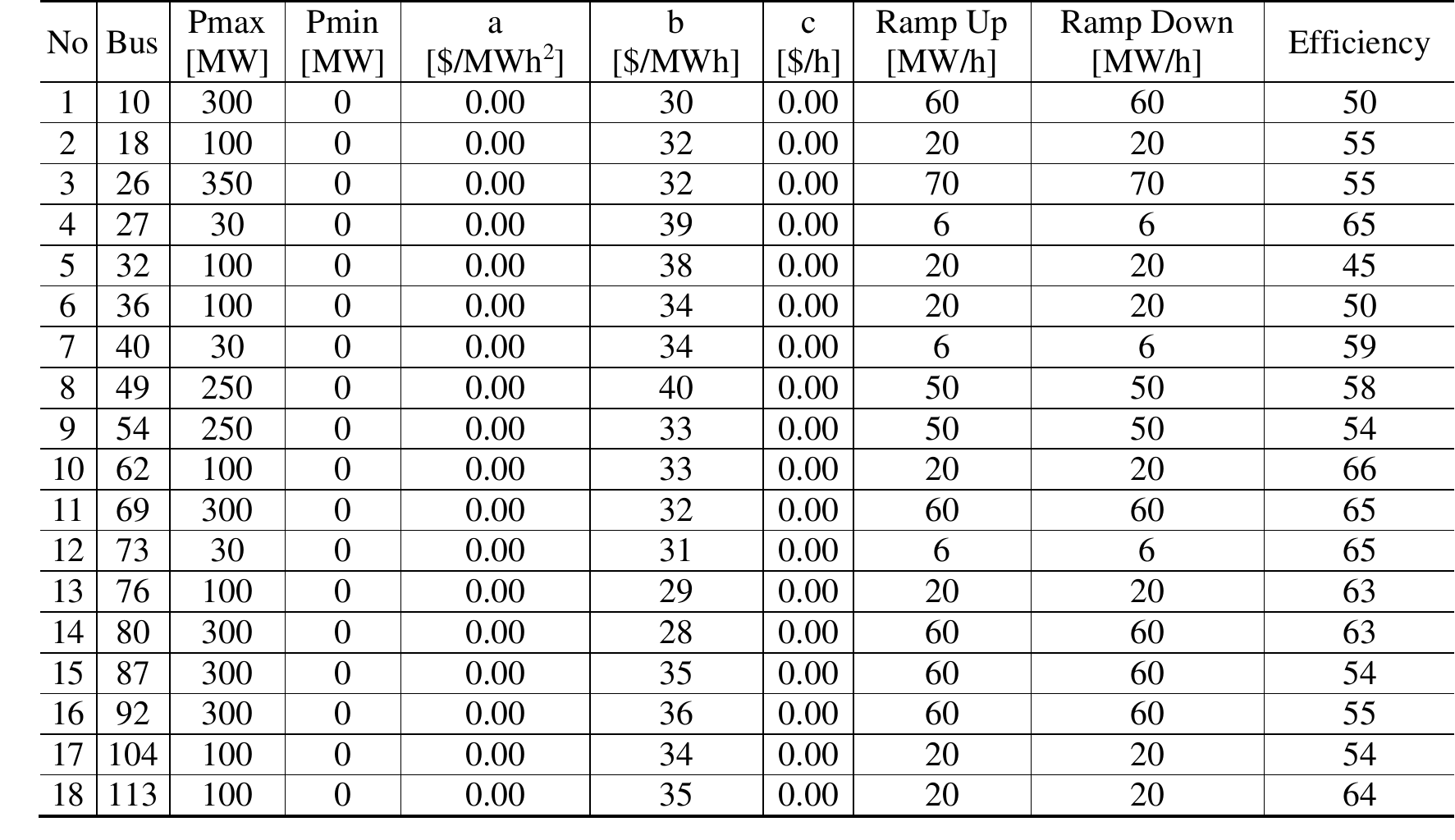}
    \end{table}

    \begin{table}[!hbtp]
     \label{tab:S1T21}
        \centering \caption{Unit Commitment(NGPPs==GPUs) -- IEEE--118Bus  System}
        \includegraphics[width=16cm]{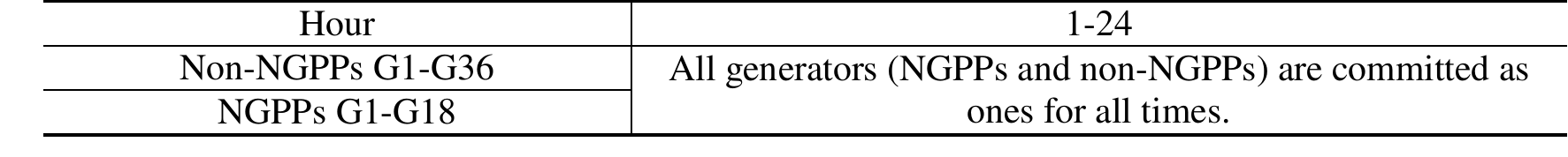}
    \end{table}

    \begin{table}[!hbtp]
     \label{tab:S1T22}	
        \centering \caption{Parameters of Load Demand -- IEEE--118Bus  System}
        \includegraphics[width=16cm]{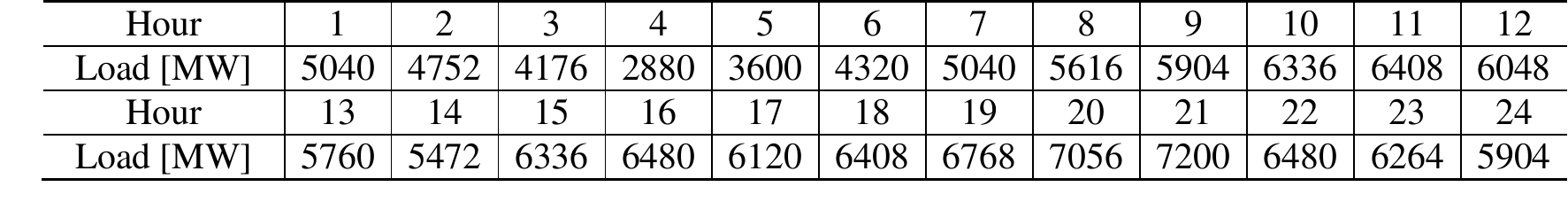}
    \end{table}

    \begin{table}[!hbtp]
     \label{tab:S1T23}	
        \centering \caption{Load Portion -- IEEE--118Bus  System}
        \includegraphics[width=16cm]{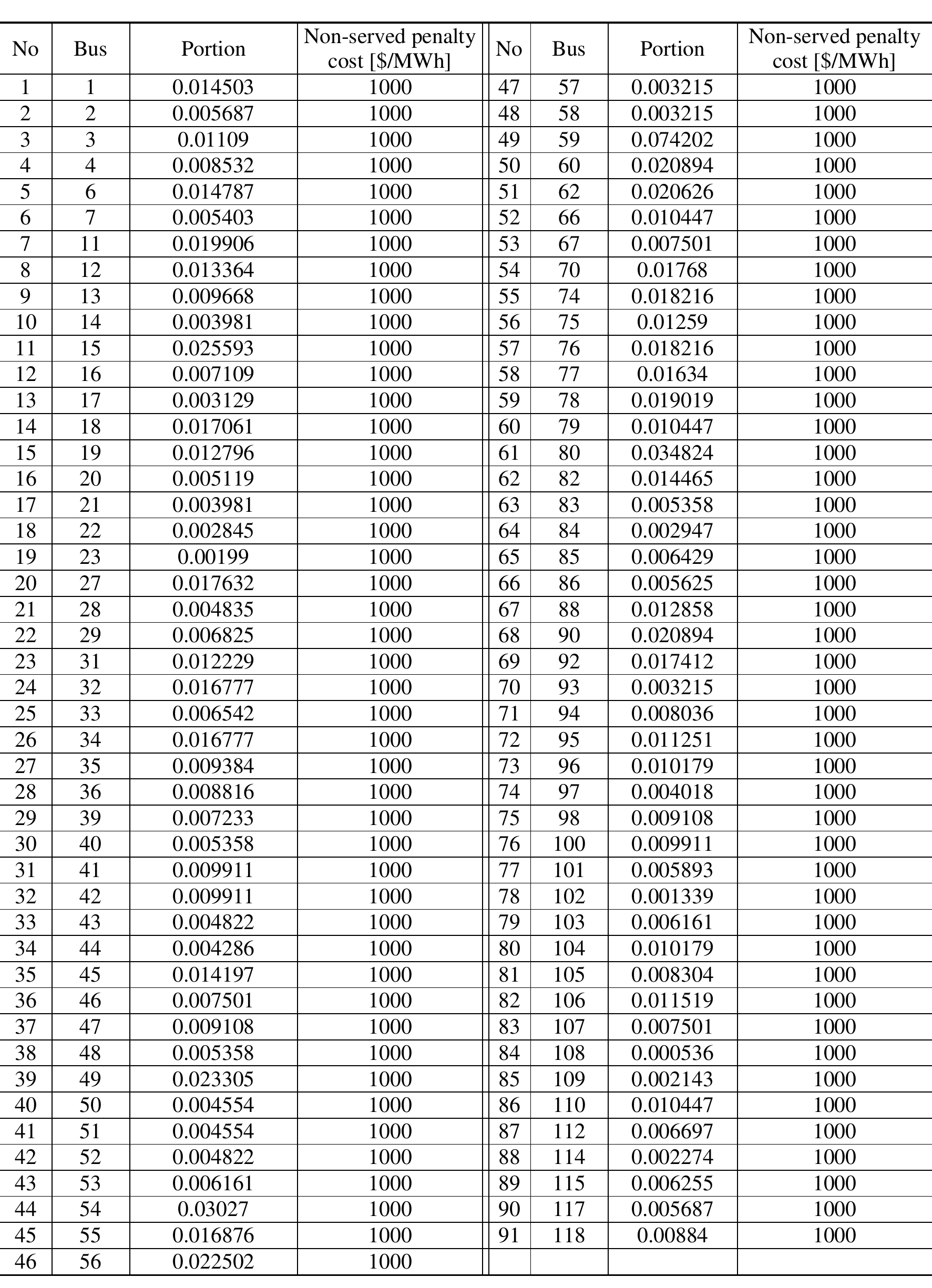}
    \end{table}

     \begin{table}[!hbtp]
     \label{tab:S118T1a}	
        \centering \caption{Parameters of Power Lines -- IEEE--118Bus  System}
        \includegraphics[width=16cm]{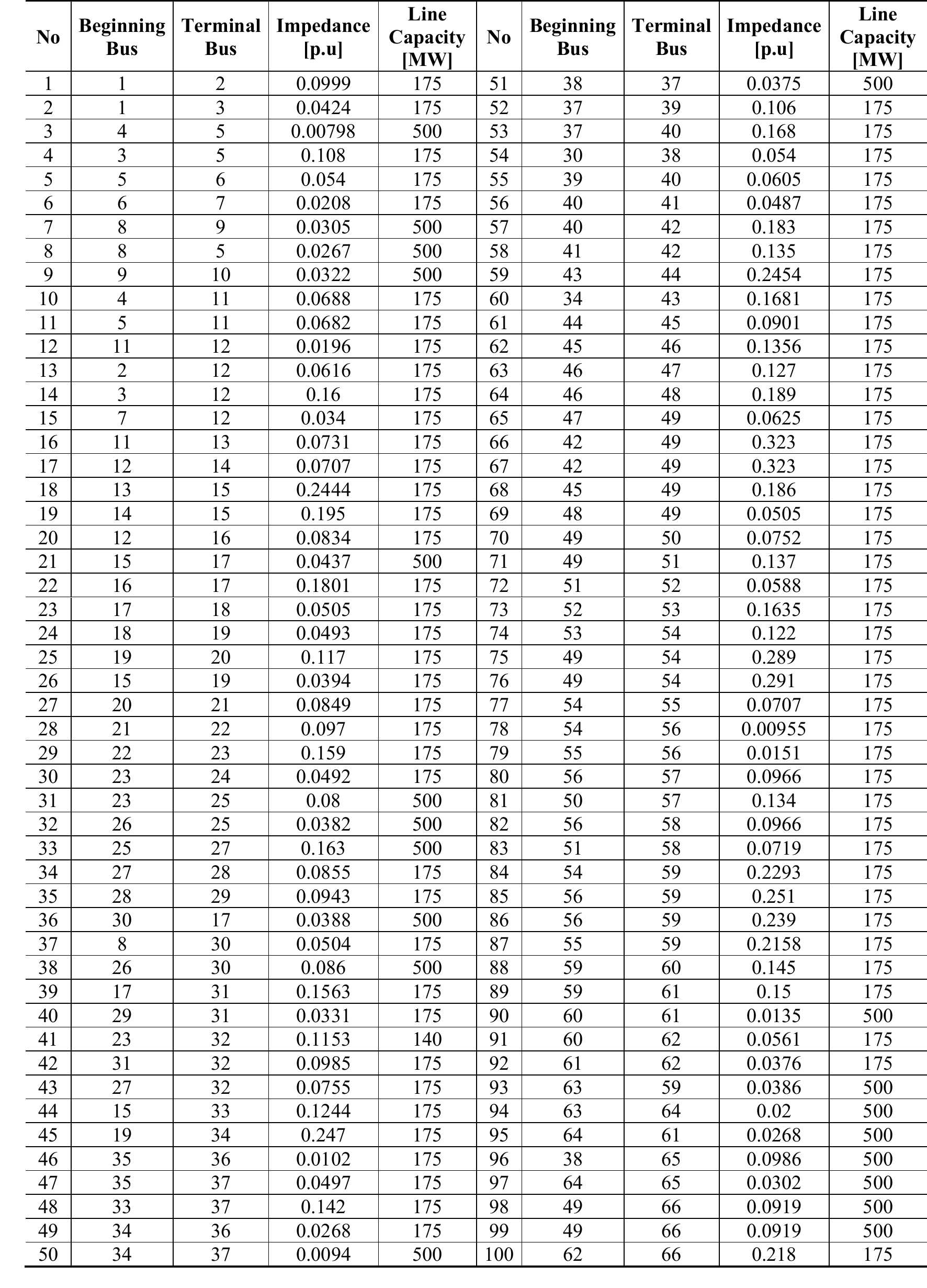}
    \end{table}
     \begin{table}[!hbtp]
     \label{tab:S118T1b}	
        \centering
        \includegraphics[width=16cm]{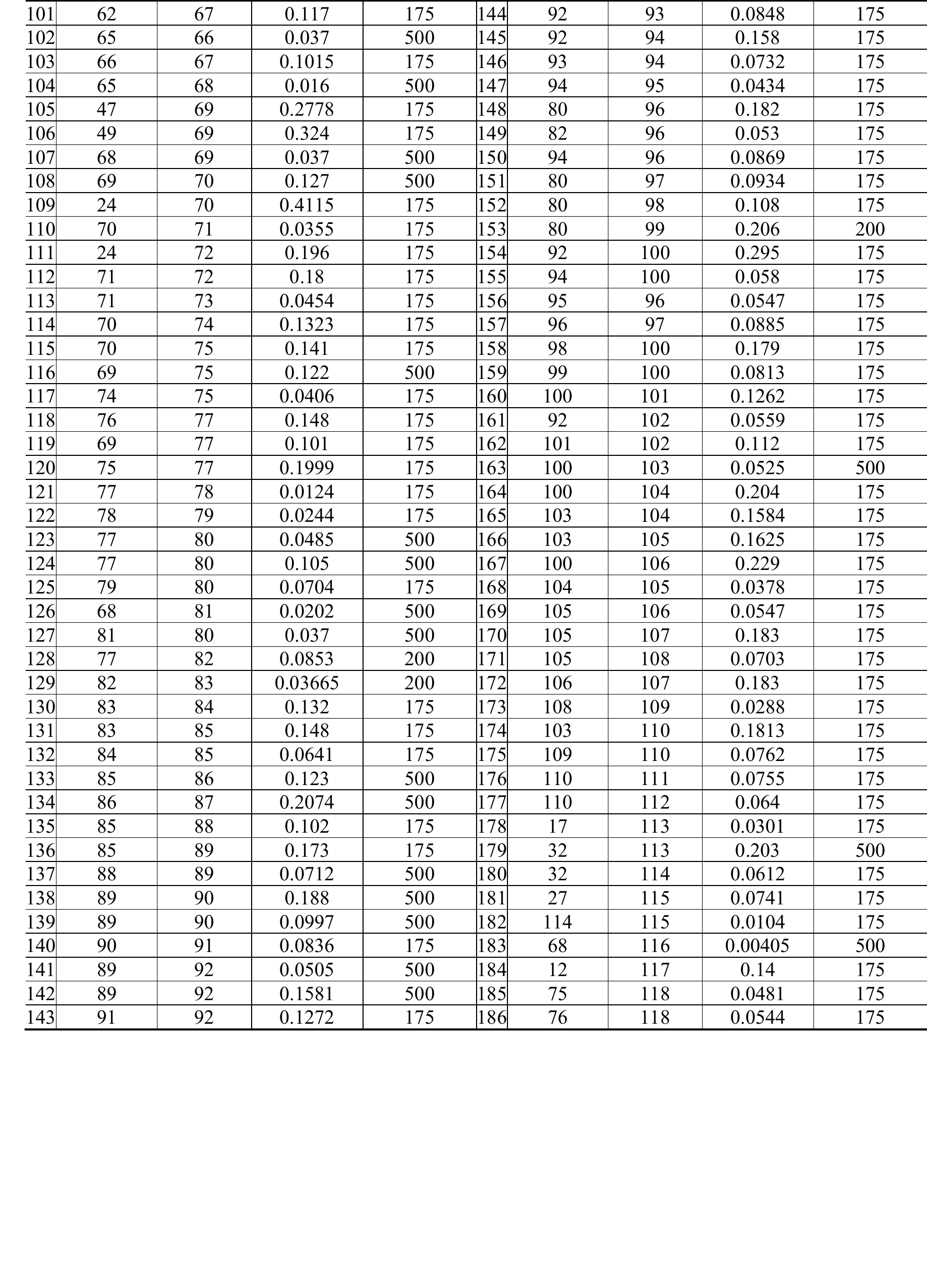}
    \end{table}

\newpage
\section{Power Distribution Networks}
\subsection{IEEE-13Bus Power Distribution Network} \label{App:13Bus}
\begin{figure}[!ht]
        \centering
            \includegraphics[width=12cm]{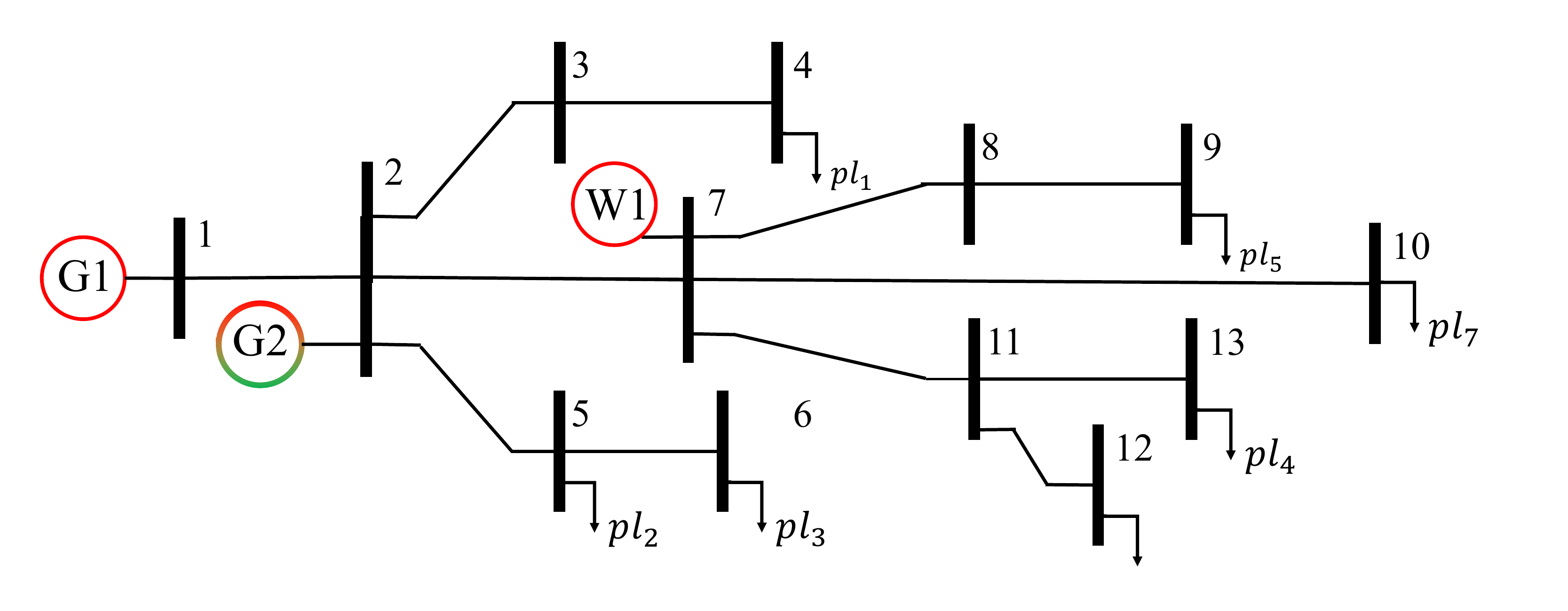} 
            \caption{Topology of IEEE--13Bus Power Distribution System}
   \label{fig:AppS5}
    \end{figure}

\begin{table}[!hbtp]
     \label{tab:S2T1}
        \centering
        \caption{Parameters of Generators -- IEEE--13Bus System}
        \includegraphics[width=16cm]{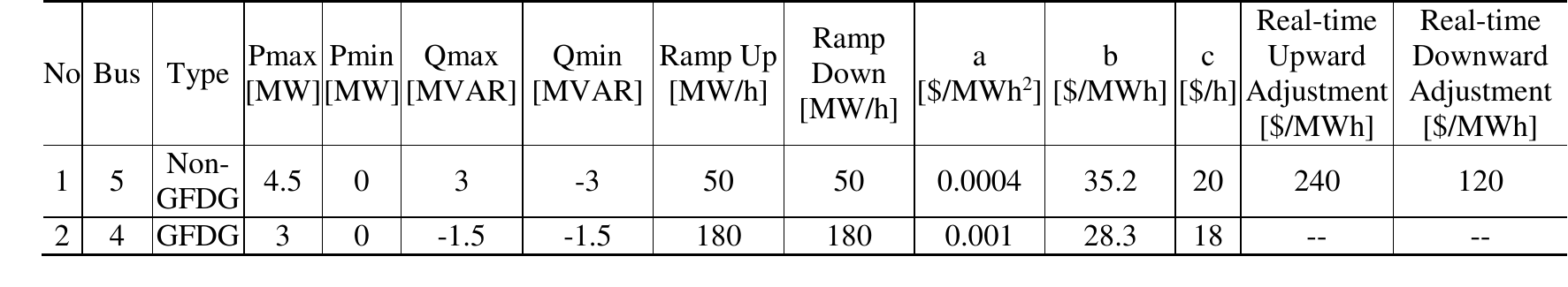}
    \end{table}

\begin{table}[!hbtp]
     \label{tab:S2T2}
        \centering
        \caption{Parameters of P2G Facilities -- IEEE--13Bus System}
        \includegraphics[width=16cm]{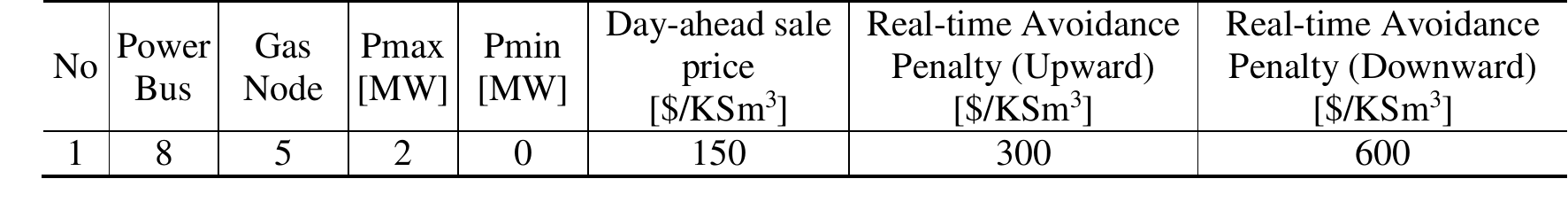}
    \end{table}

\begin{table}[!hbtp]
     \label{tab:S2T3}
        \centering
        \caption{Parameters of Wind Power Generation [Pmin = 0, Pmax = 2.5MW]  -- IEEE--13Bus System}
        \includegraphics[width=16cm]{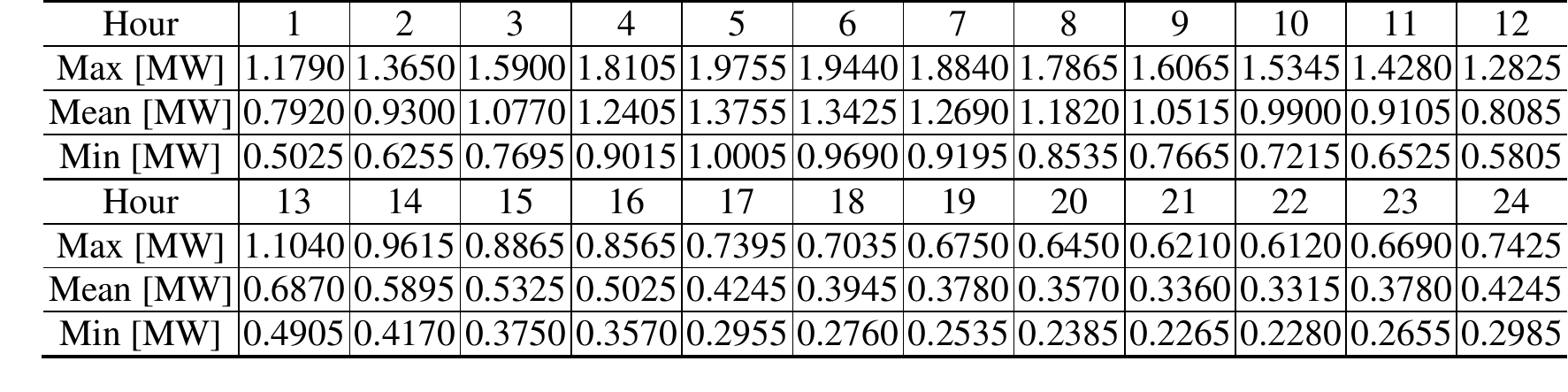}
    \end{table}

\begin{table}[!hbtp]
     \label{tab:S2T4}
        \centering
        \caption{Parameters of Power Feeders -- IEEE--13Bus System}
        \includegraphics[width=16cm]{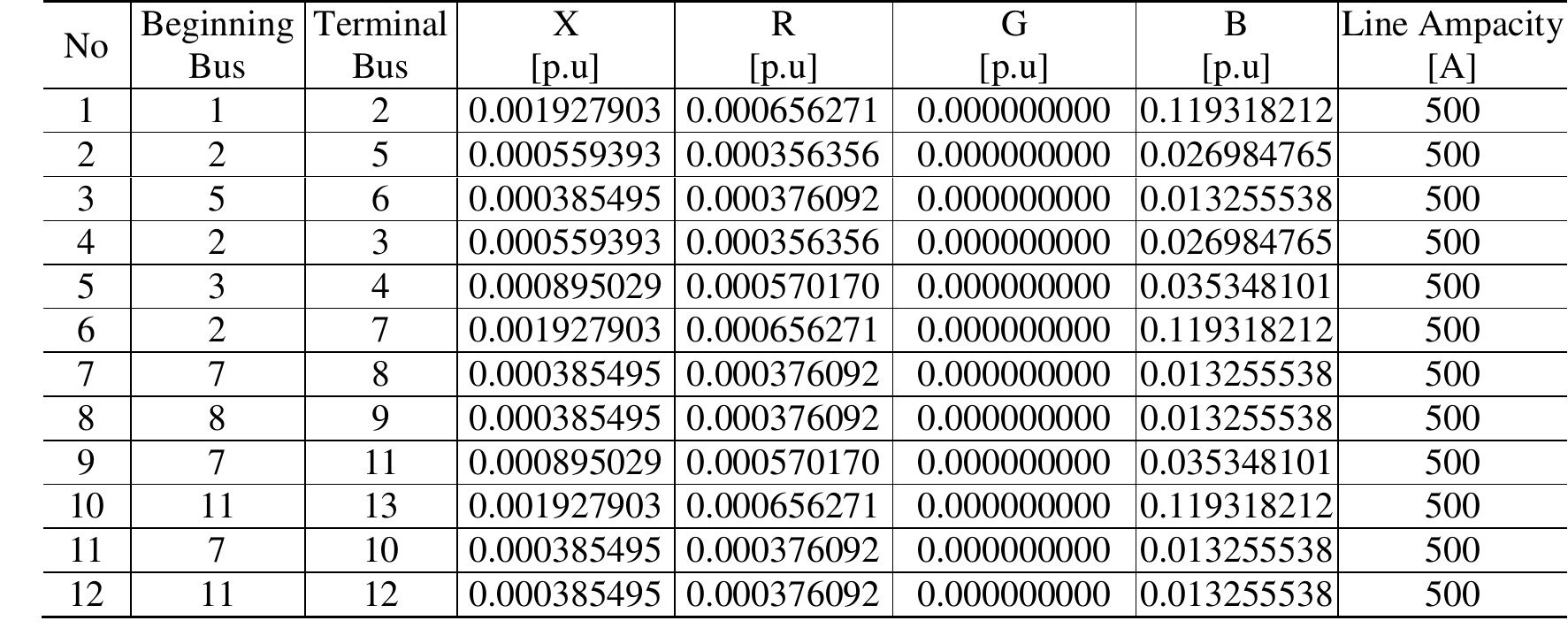}
    \end{table}

\begin{table}[!hbtp]
     \label{tab:S2T5}
        \centering
        \caption{Parameters of Power Buses -- IEEE--13Bus System}
        \includegraphics[width=16cm]{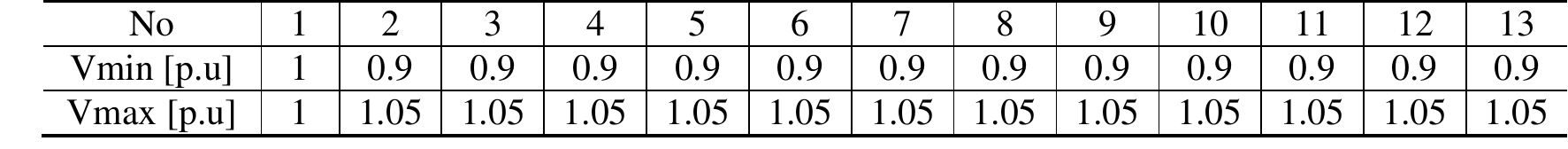}
    \end{table}

\begin{table}[!hbtp]
     \label{tab:S2T6}
        \centering
        \caption{Parameters of Load Demand -- IEEE--13Bus System}
        \includegraphics[width=16cm]{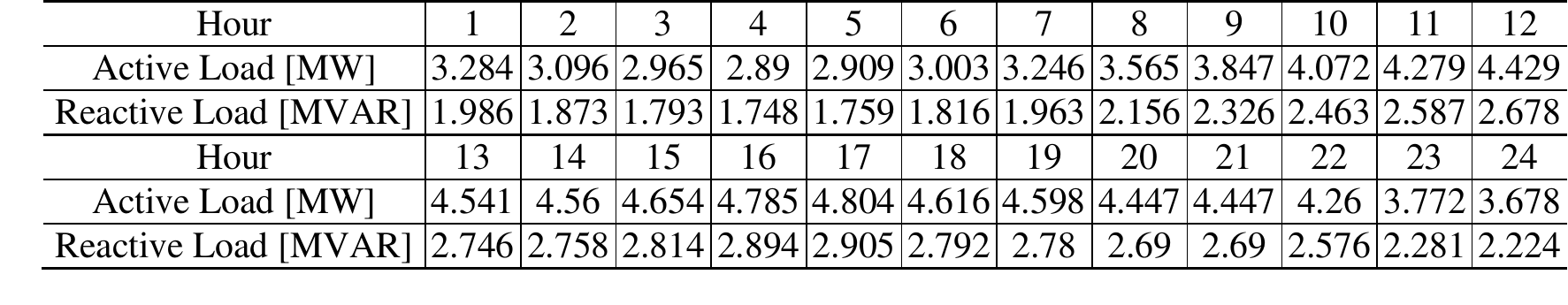}
    \end{table}

\begin{table}[!hbtp]
     \label{tab:S2T7}
        \centering
        \caption{Load Portion -- IEEE--13Bus System}
        \includegraphics[width=16cm]{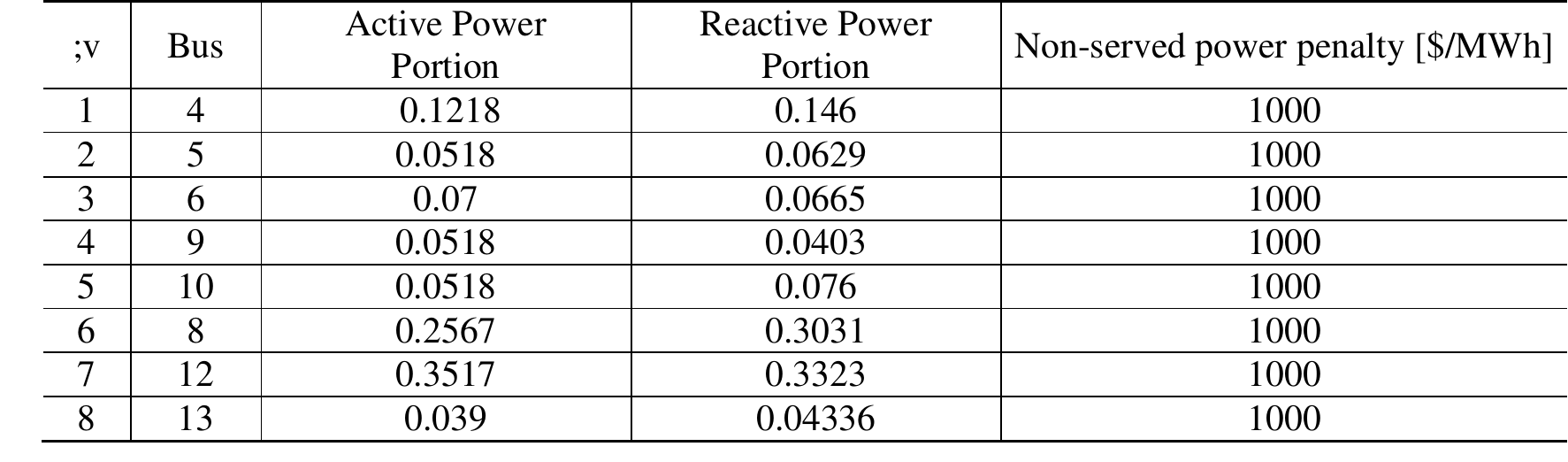}
    \end{table}

\newpage
\subsection{IEEE-123Bus Power Distribution Network}  \label{App:123Bus}
\begin{figure}[!ht]
        \centering
            \includegraphics[width=15cm]{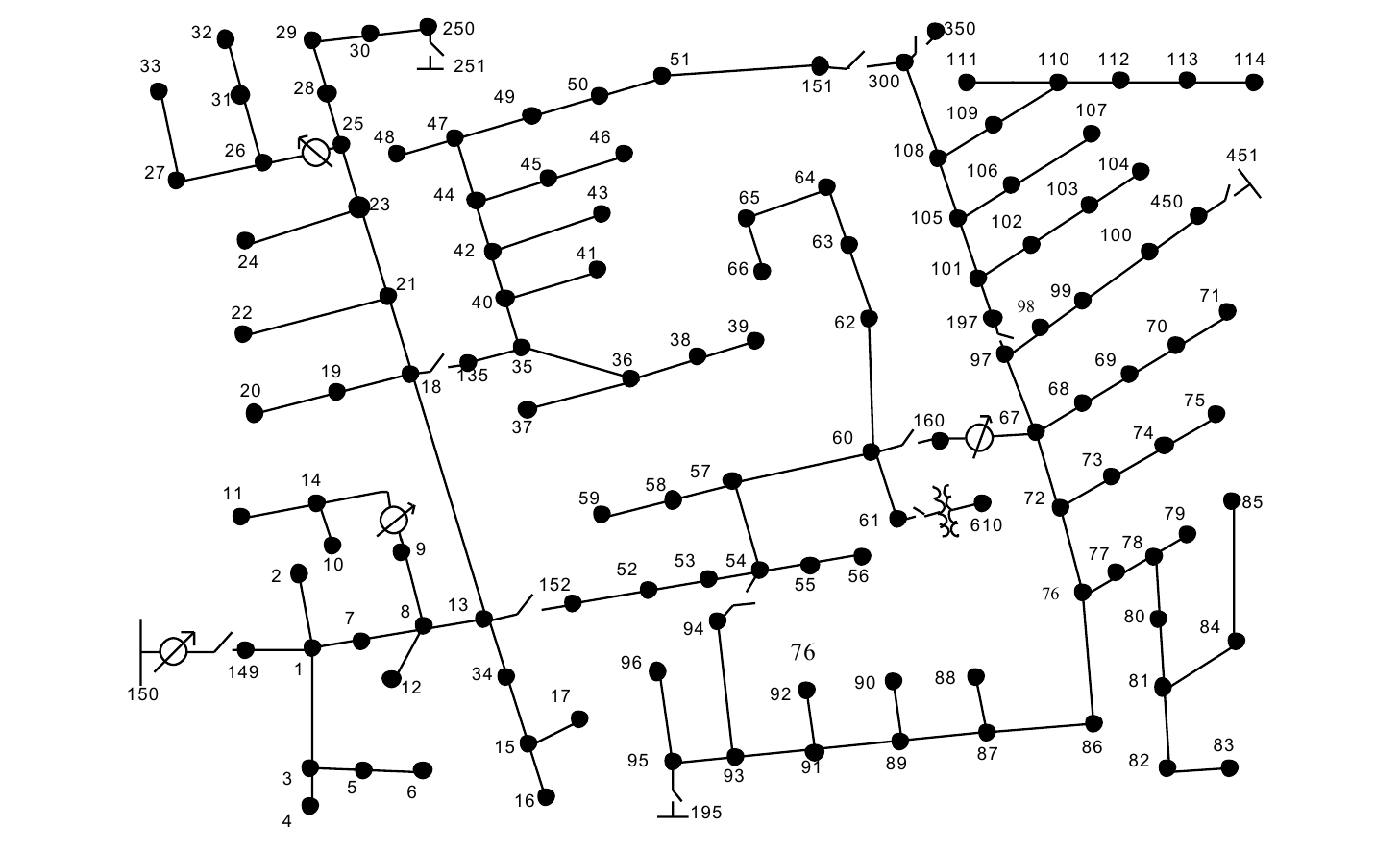} 
            \caption{Topology of IEEE--123Bus Power Distribution System}
   \label{fig:App123}
    \end{figure}
\begin{table}[!hbtp]
     \label{tab:S2T18}
        \centering
        \caption{Parameters of generators -- IEEE--123Bus System}
        \includegraphics[width=16cm]{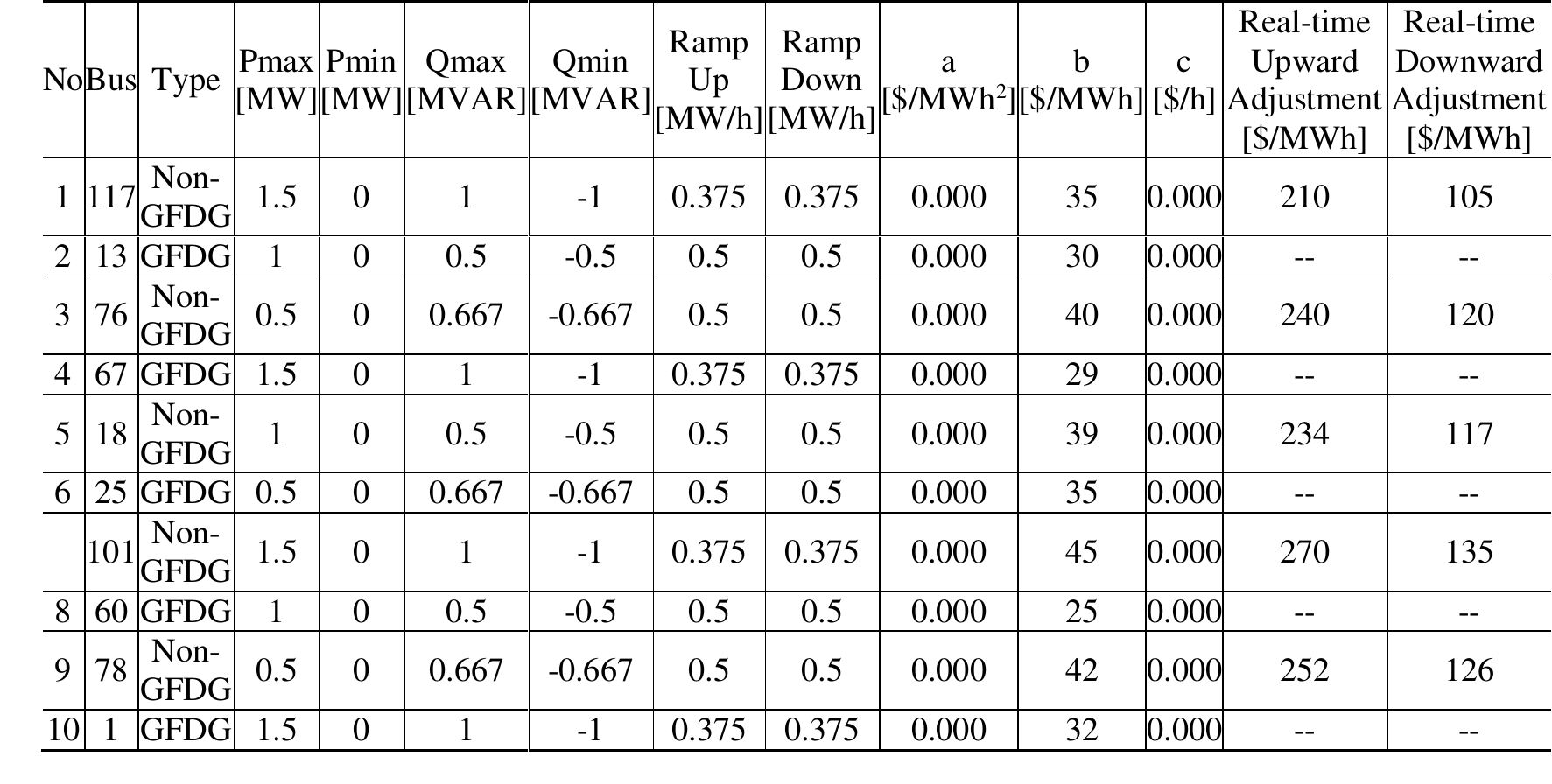}
    \end{table}

\begin{table}[!hbtp]
     \label{tab:S2T19}
        \centering
        \caption{Parameters of P2G Facilities -- IEEE--123Bus System}
        \includegraphics[width=16cm]{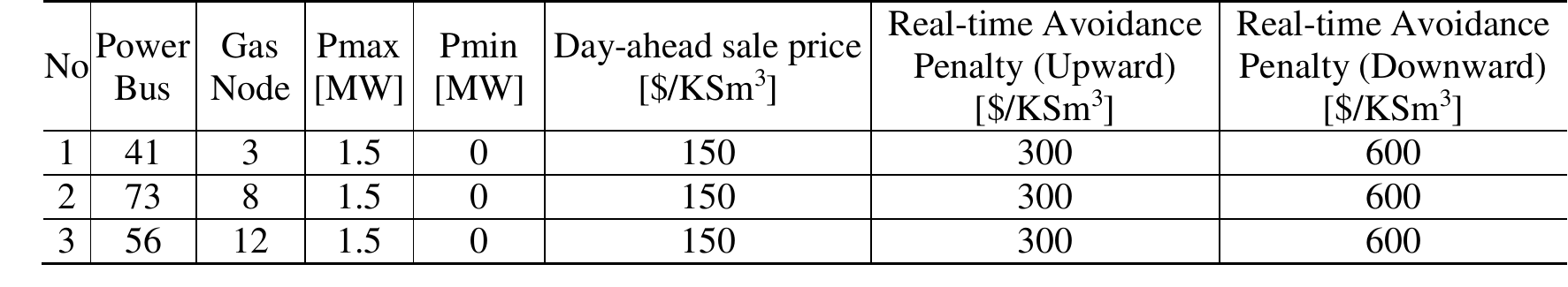}
    \end{table}

\begin{table}[!hbtp]
     \label{tab:S2T20}
        \centering
        \caption{Parameters of Wind Power Generation 1 [Pmin = 0, Pmax = 0.5 MW, Bus 54] -- IEEE--123Bus System}
        \includegraphics[width=16cm]{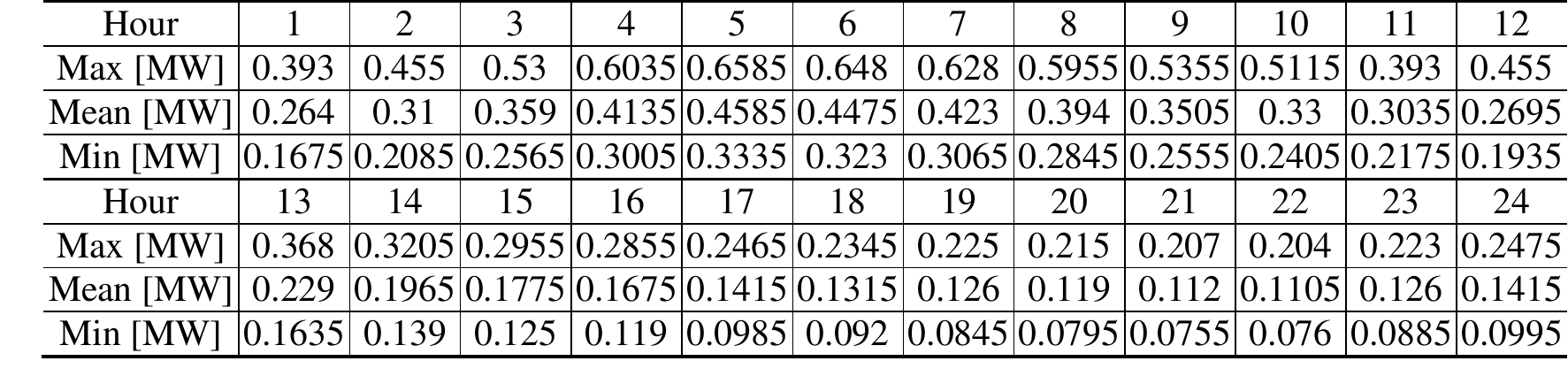}
    \end{table}
\begin{table}[!hbtp]
     \label{tab:S123T1}
        \centering
        \caption{Parameters of Power Feeders -- IEEE--123Bus System}
        \includegraphics[width=16cm]{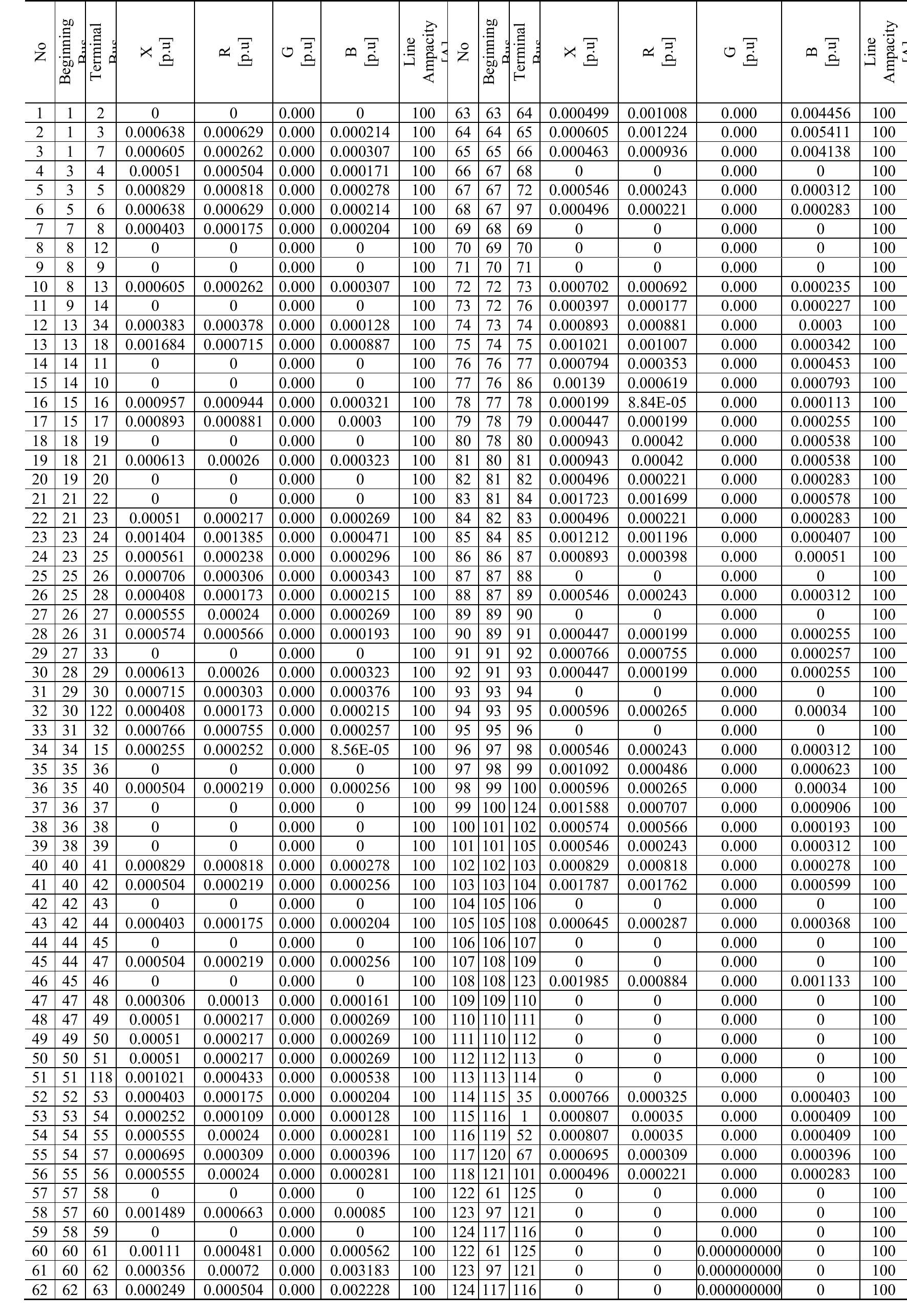}
    \end{table}
\begin{table}[!hbtp]
     \label{tab:S2T21}
        \centering
        \caption{Parameters of Wind Power Generation 2 [Pmin = 0, Pmax = 1 MW, Bus 72] -- IEEE--123Bus System}
        \includegraphics[width=16cm]{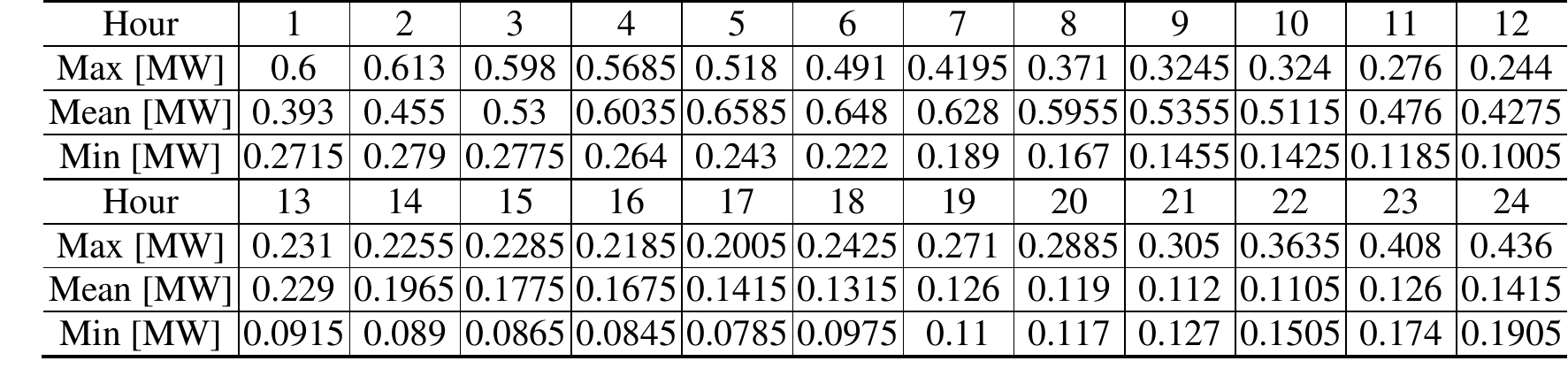}
    \end{table}

\begin{table}[!hbtp]
     \label{tab:S2T22}
        \centering
        \caption{Parameters of Wind Power Generation 3 [Pmin = 0, Pmax = 0.5MW, Bus 35]-- IEEE--123Bus System}
        \includegraphics[width=16cm]{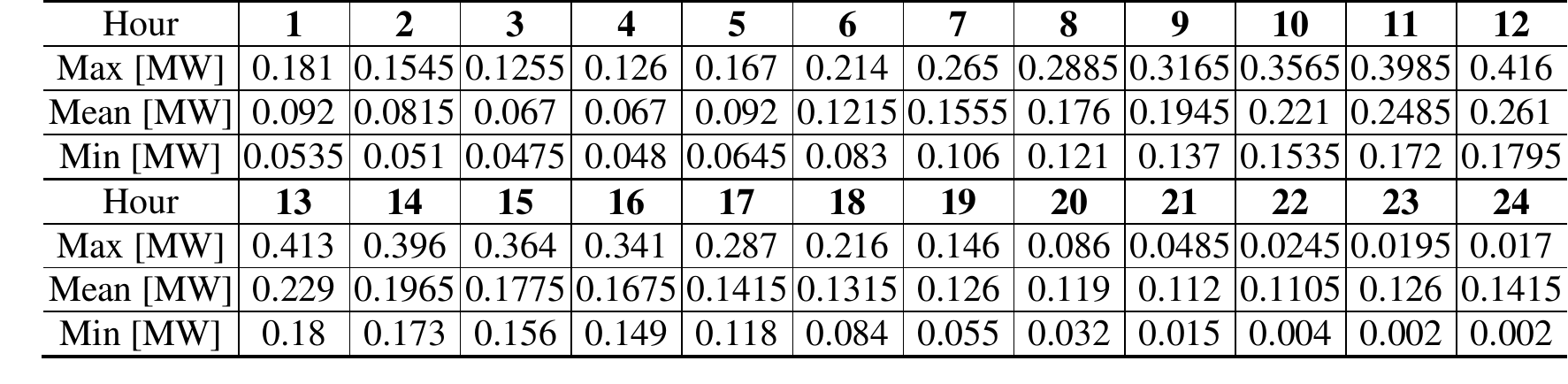}
    \end{table}

\begin{table}[!hbtp]
     \label{tab:S2T23}
        \centering
        \caption{Parameters of Power Buses -- IEEE--123Bus System}
        \includegraphics[width=16cm]{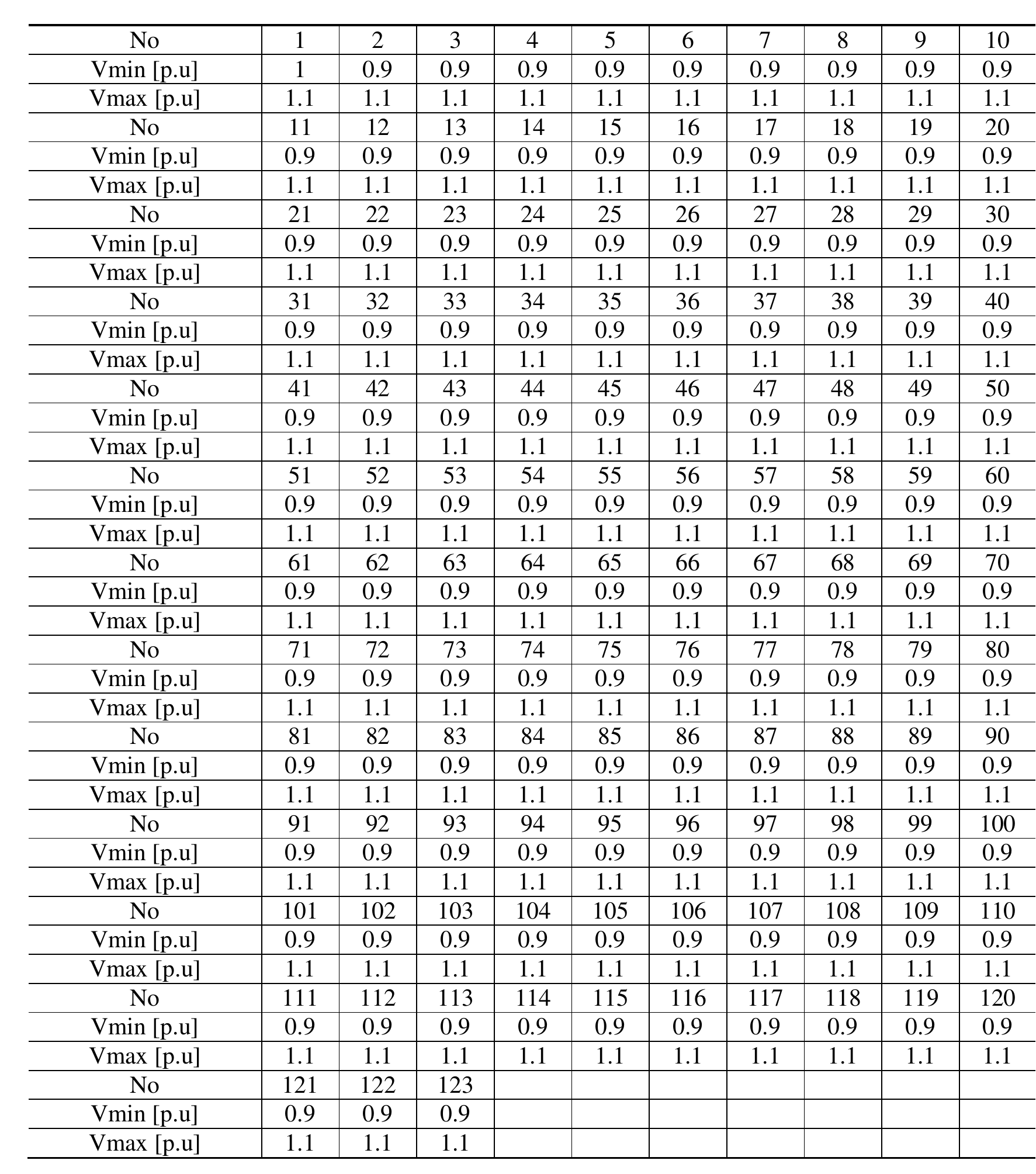}
    \end{table}

\begin{table}[!hbtp]
     \label{tab:S2T24}
        \centering
        \caption{Parameters of Load Demand -- IEEE--123Bus System}
        \includegraphics[width=16cm]{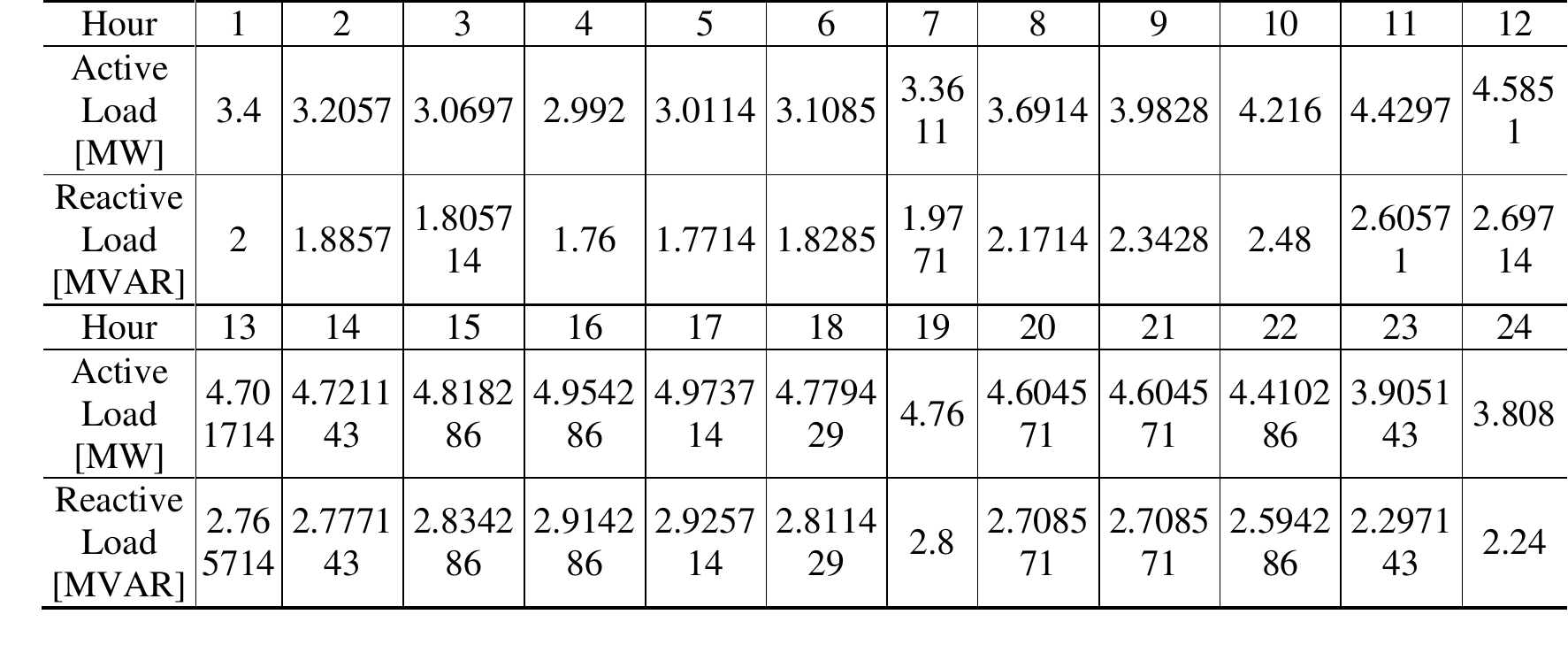}
    \end{table}

\newpage
\section{Gas Systems}
\subsection{7Nodes Gas System}  \label{App:sevenNode}
\begin{figure}[!ht]
        \centering
            \includegraphics[width=9cm]{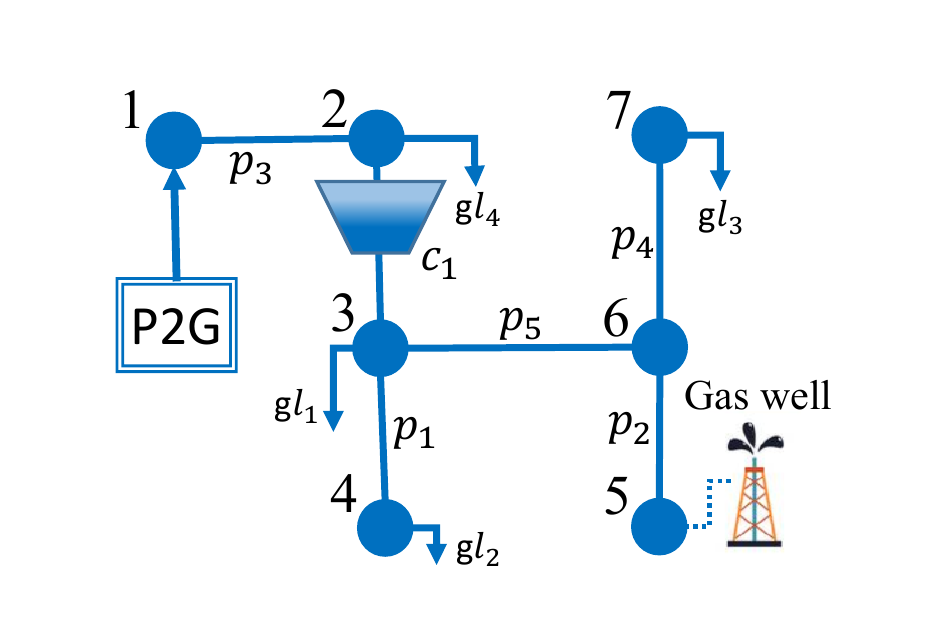} 
            \caption{Topology of 7Nodes Gas System}
   \label{fig:AppS2}
    \end{figure}
    \begin{table}[!hbtp]
     \label{tab:S1T10}
        \centering \caption{Parameters of Gas Pipelines -- 7Nodes Gas System}
        \includegraphics[width=16cm]{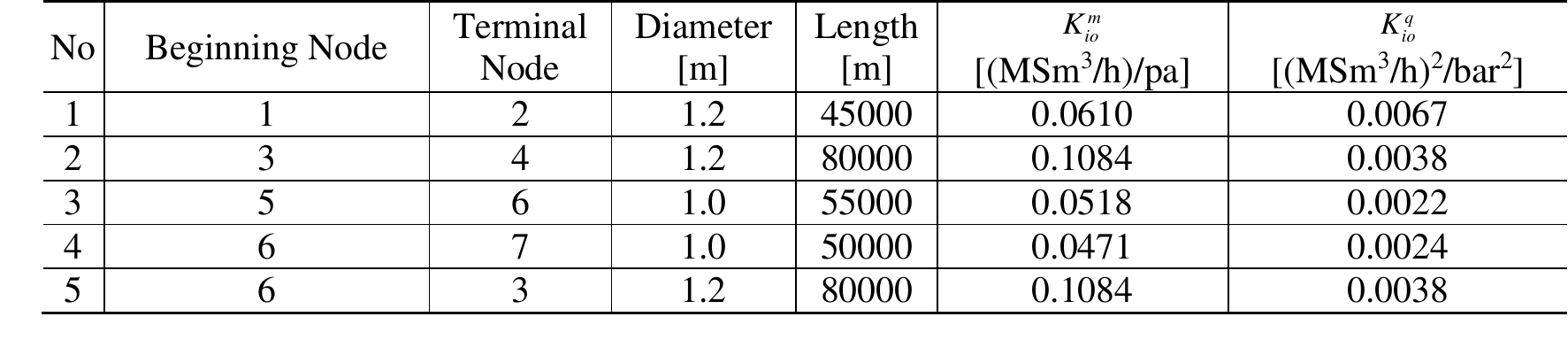}
    \end{table}
\begin{table}[!hbtp]
     \label{tab:S1T9}
        \centering \caption{Parameters of Gas Compressors -- 7Nodes Gas System}
        \includegraphics[width=16cm]{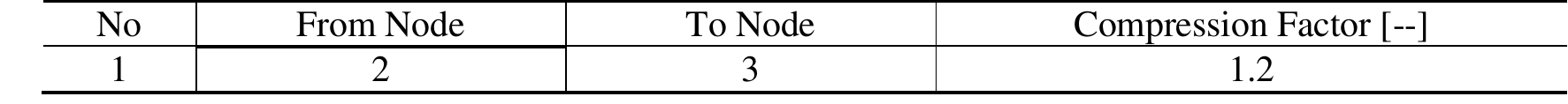}
    \end{table}

    \begin{table}[!hbtp]
     \label{tab:S1T8}
        \centering \caption{Parameters of Gas Wells -- 7Nodes Gas System}
        \includegraphics[width=16cm]{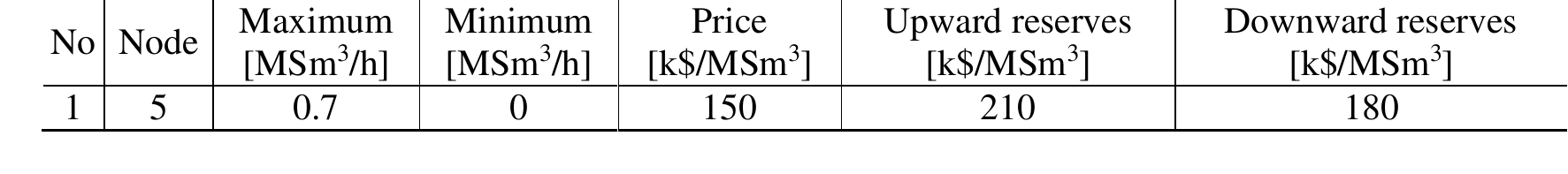}
    \end{table}

    \begin{table}[!hbtp]
     \label{tab:S1T11}
        \centering \caption{Parameters of Load Demand -- 7Nodes Gas System}
        \includegraphics[width=16cm]{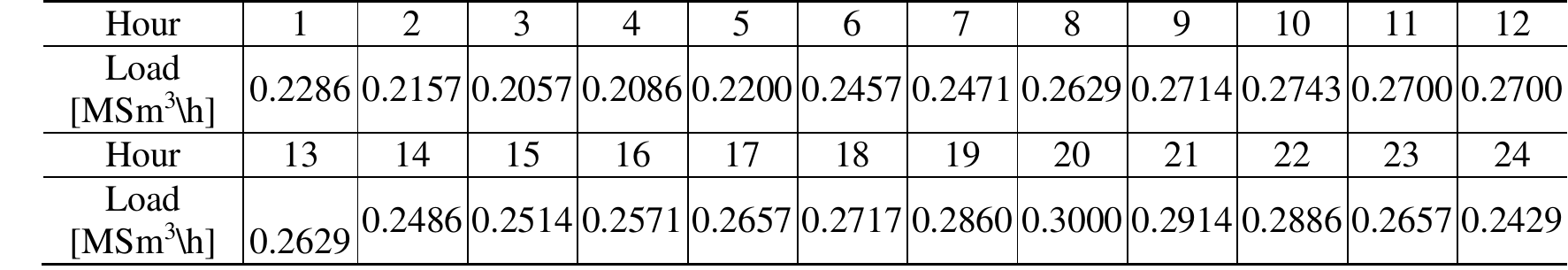}
    \end{table}

    \begin{table}[!hbtp]
     \label{tab:S1T12}
        \centering \caption{Parameters of Gas Nodes and their Load Portion -- 7Nodes Gas System}
        \includegraphics[width=16cm]{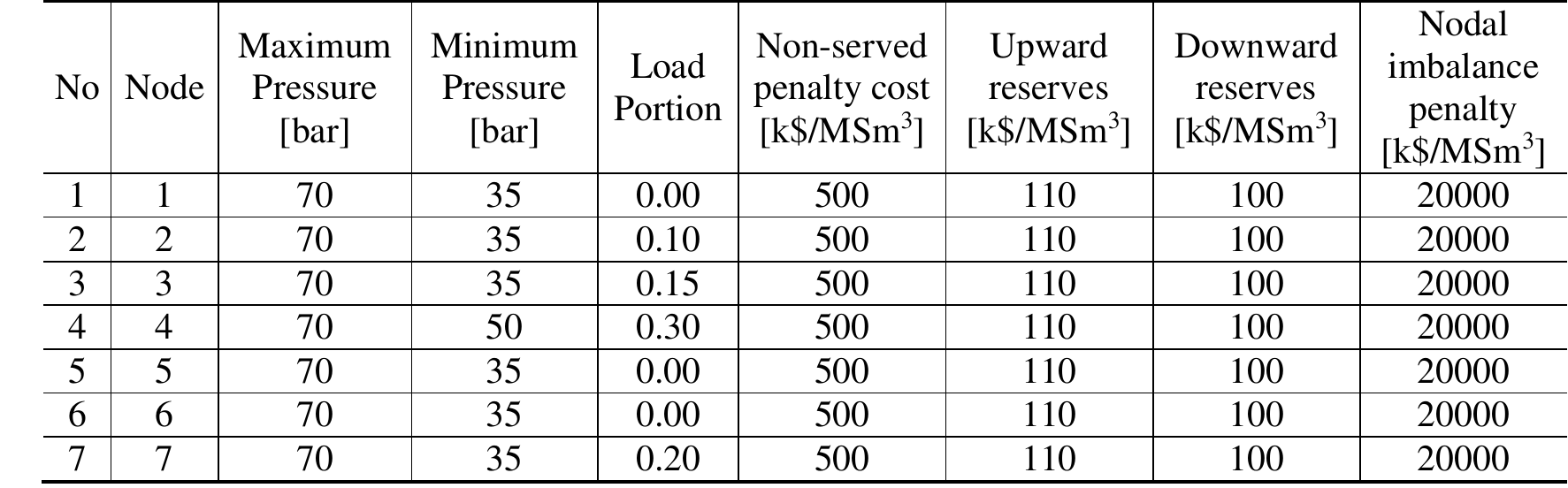}
    \end{table}

    \begin{table}[!hbtp]
     \label{tab:S1T13}
        \centering \caption{Connection Lines Between the PJM--5Bus System coupled and 7Nodes Gas System}
        \includegraphics[width=16cm]{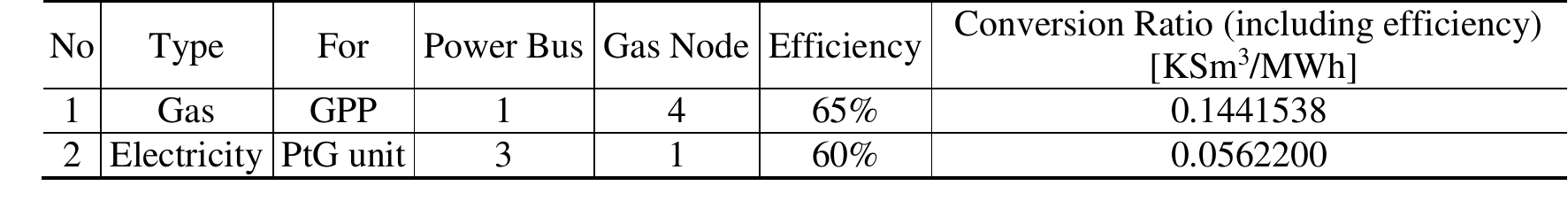}
    \end{table}
\newpage
\subsection{8Nodes Gas System}\label{App:8Node}

\begin{figure}[!ht]
        \centering
            \includegraphics[width=14cm]{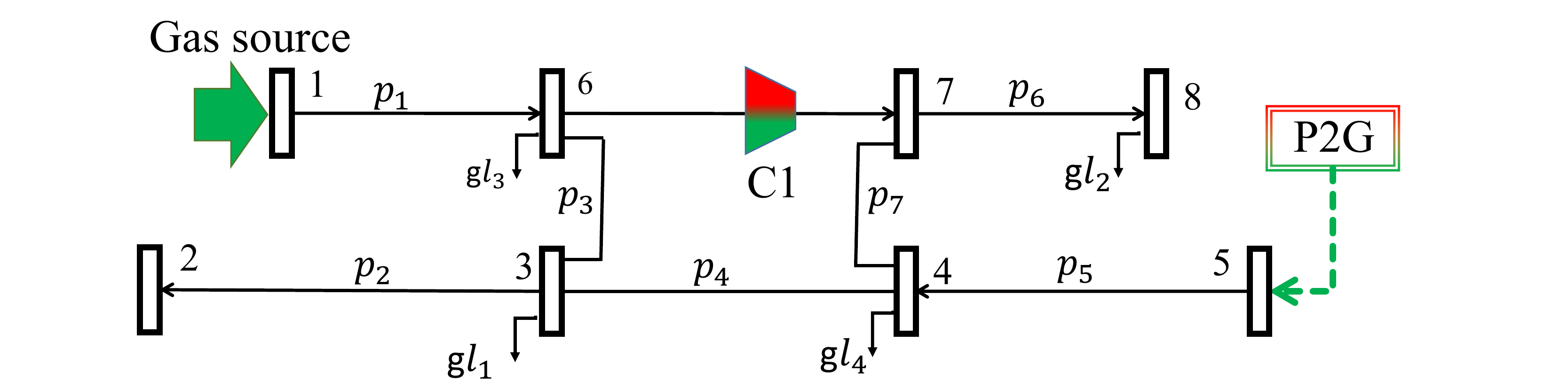}
            \caption{Topology of 8Nodes Gas System}
   \label{fig:AppS6}
    \end{figure}

    \begin{table}[!hbtp]
     \label{tab:S2T8}
        \centering
        \caption{Parameters of Generators -- 8Nodes Gas System}
        \includegraphics[width=16cm]{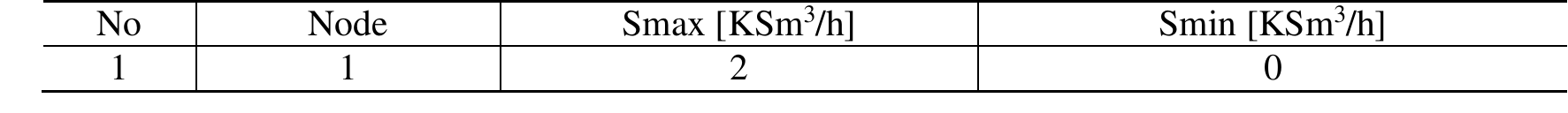}
    \end{table}

\begin{table}[!hbtp]
     \label{tab:S2T9}
        \centering
        \caption{Parameters of Gas sources -- 8Nodes Gas System}
        \includegraphics[width=16cm]{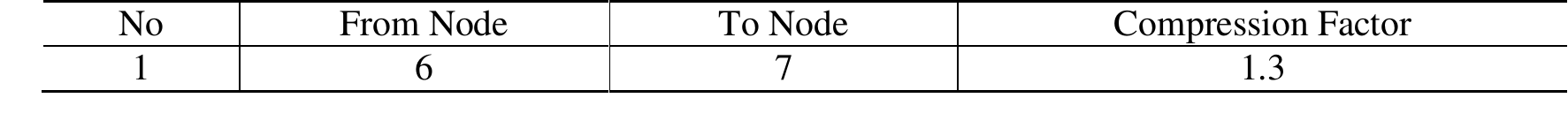}
    \end{table}

\begin{table}[!hbtp]
     \label{tab:S2T10}
        \centering
        \caption{Parameters of Gas Compressors -- 8Nodes Gas System}
        \includegraphics[width=16cm]{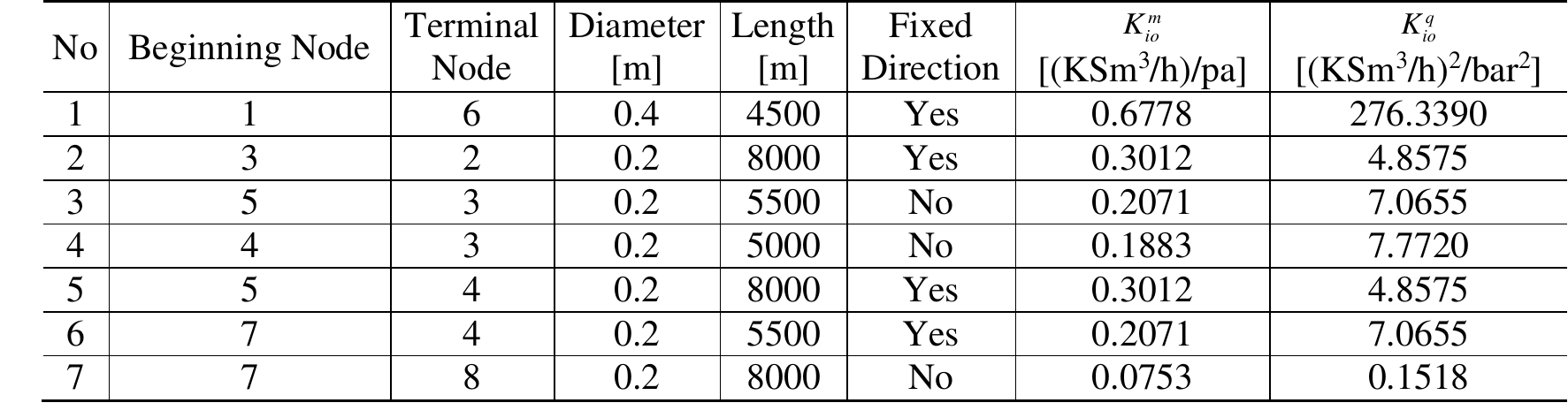}
    \end{table}

\begin{table}[!hbtp]
     \label{tab:S2T11}
        \centering
        \caption{Parameters of Load Demand -- 8Nodes Gas System}
        \includegraphics[width=16cm]{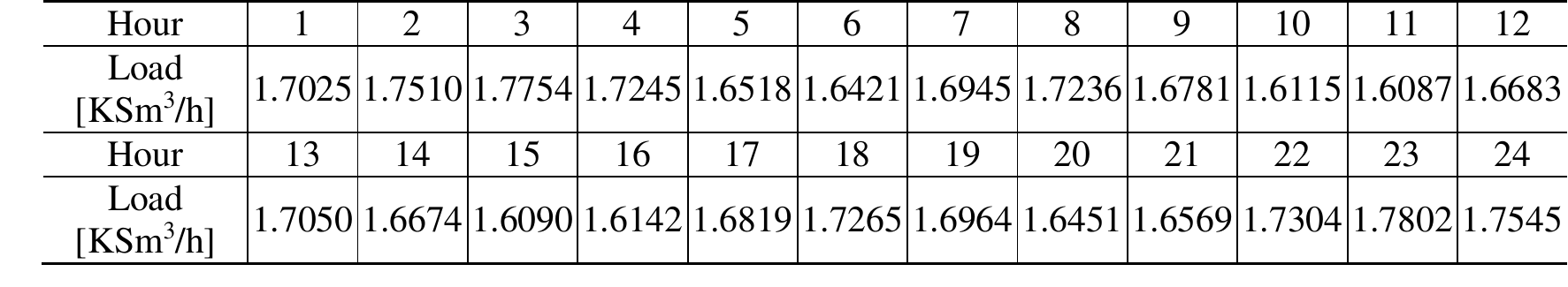}
    \end{table}
    \begin{table}[!hbtp]
     \label{tab:S2T12}
        \centering
        \caption{Parameters of Gas Nodes and their Load Portion -- 8Nodes Gas System}
        \includegraphics[width=16cm]{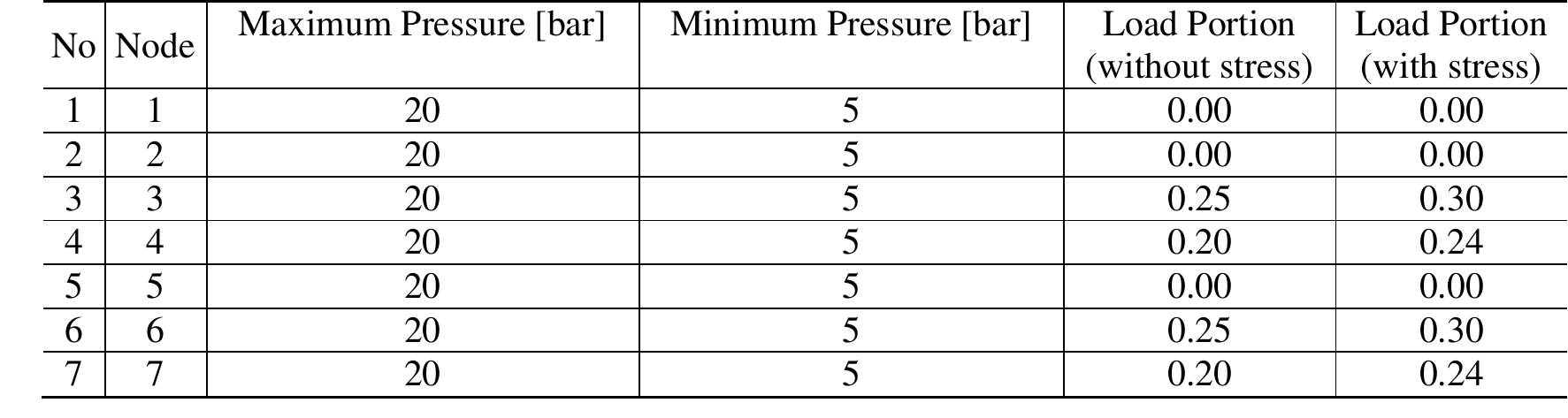}
    \end{table}
\begin{table}[!hbtp]
     \label{tab:S2T13}
        \centering
        \caption{Connection Lines Between IEEE--13Bus System and 8Nodes Gas System}
        \includegraphics[width=16cm]{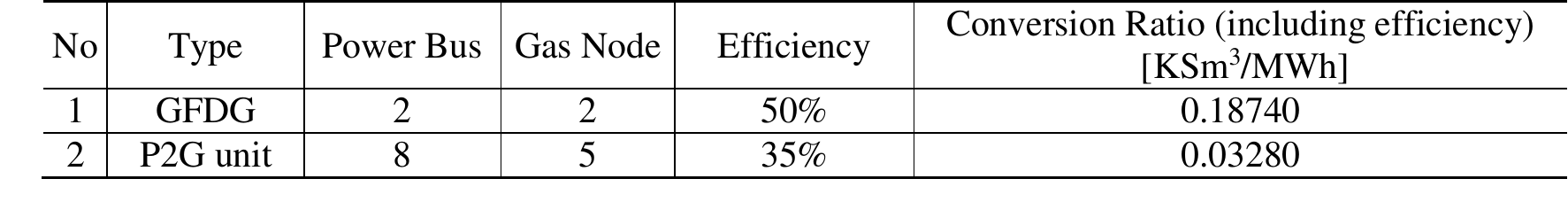}
    \end{table}

\begin{table}[!hbtp]
     \label{tab:S2T14}
        \centering
        \caption{G2P Gas Contracts Between IEEE--13Bus System and 8Nodes Gas System}
        \includegraphics[width=16cm]{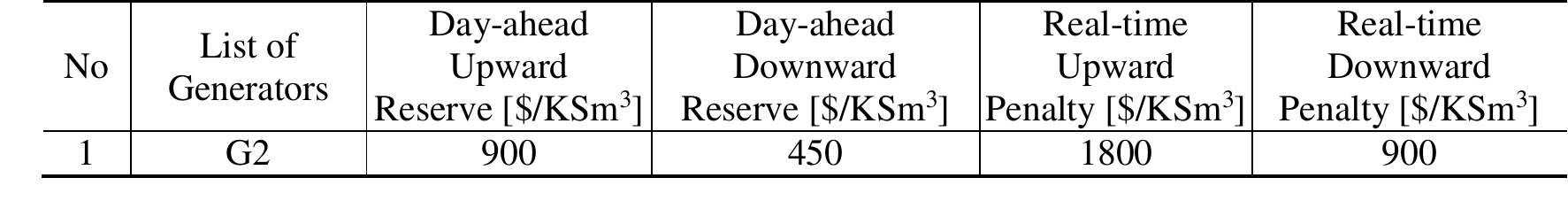}
    \end{table}

\begin{table}[!hbtp]
     \label{tab:S2T15}
        \centering
        \caption{P2G Gas Contracts Between IEEE--13Bus System and 8Nodes Gas System}
        \includegraphics[width=16cm]{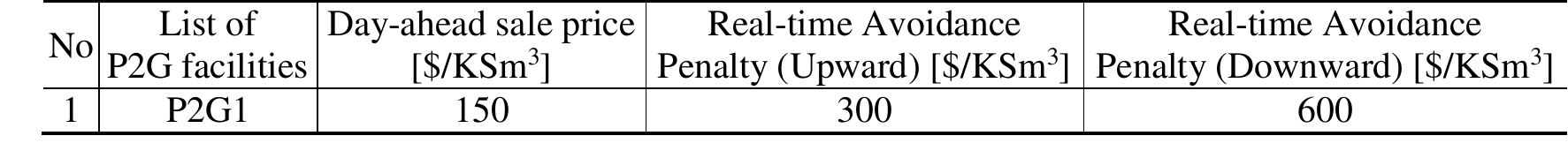}
    \end{table}
\newpage
\subsection{20Nodes Gas System} \label{App:20Node}
    \begin{figure}[!ht]
        \centering
            \includegraphics[width=9cm]{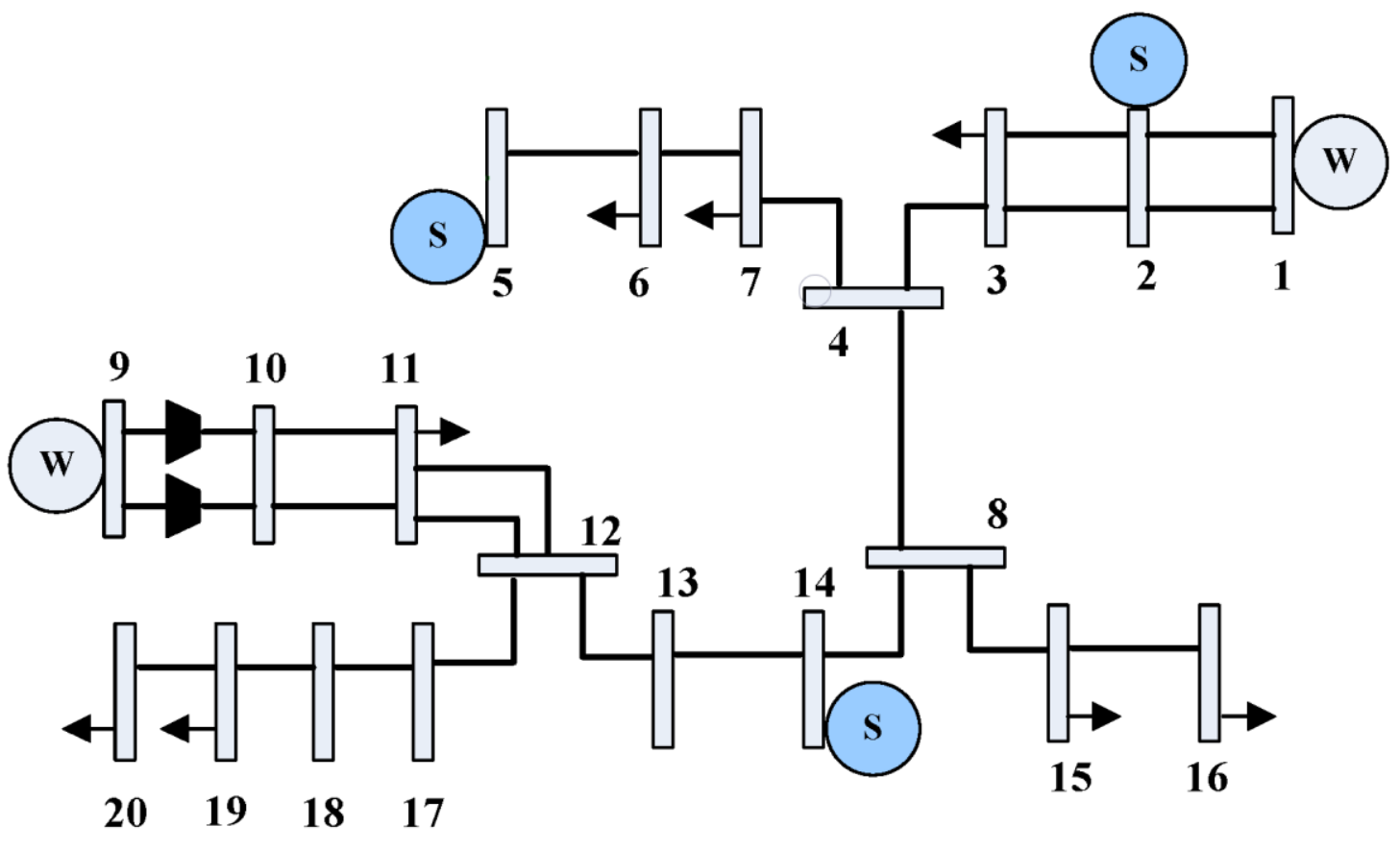}
            \caption{Topology of 20Nodes Gas System}
   \label{fig:AppS4}
    \end{figure}
    \begin{table}[!hbtp]
    \label{tab:S1T24}	
        \centering \caption{Parameters of Gas Wells -- 20Nodes Gas System}
        \includegraphics[width=16cm]{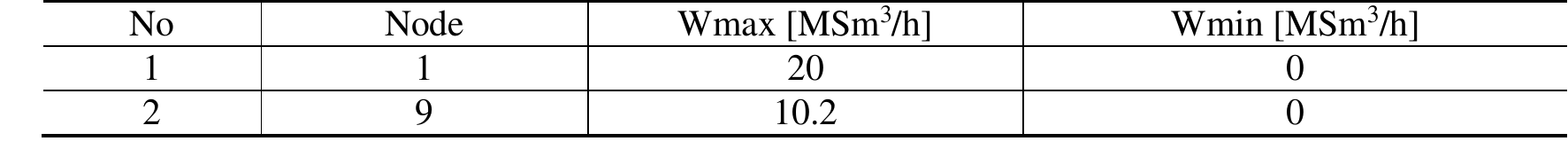}
    \end{table}

    \begin{table}[!hbtp]
     \label{tab:S1T25}	
        \centering \caption{Parameters of Gas Compressors -- 20Nodes Gas System}
        \includegraphics[width=16cm]{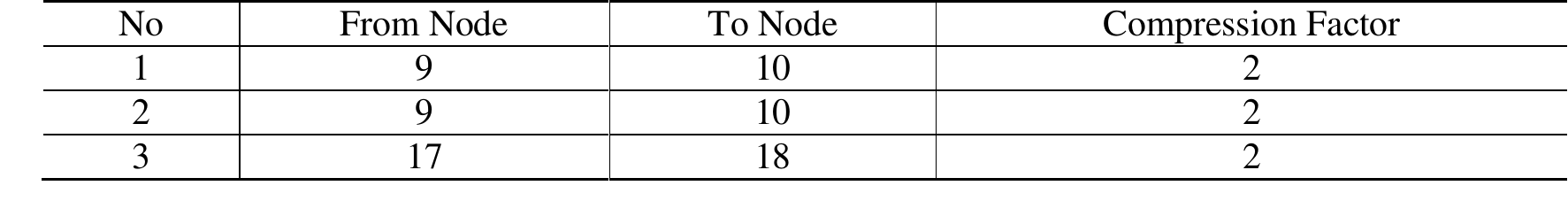}
    \end{table}
    \begin{table}[!hbtp]
     \label{tab:S2T25}
        \centering
        \caption{Connection Lines Between IEEE--123Bus System and 20Nodes Gas System}
        \includegraphics[width=16cm]{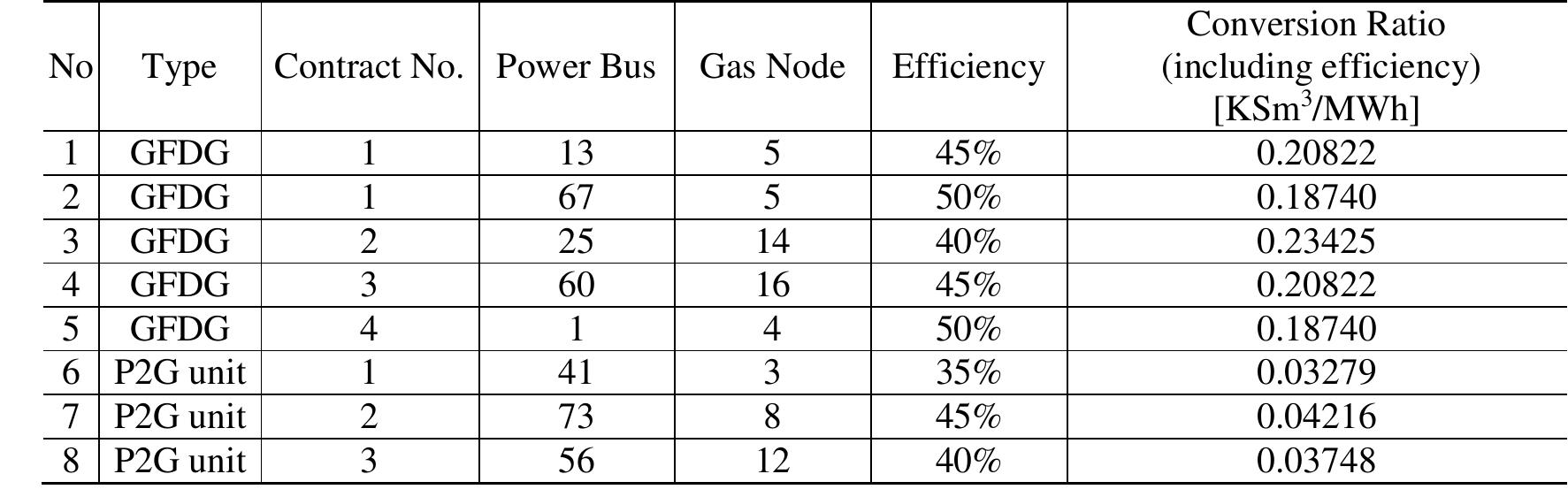}
    \end{table}

    \begin{table}[!hbtp]
     \label{tab:S1T26}	
        \centering \caption{Parameters of Gas Pipelines -- 20Nodes Gas System}
        \includegraphics[width=16cm]{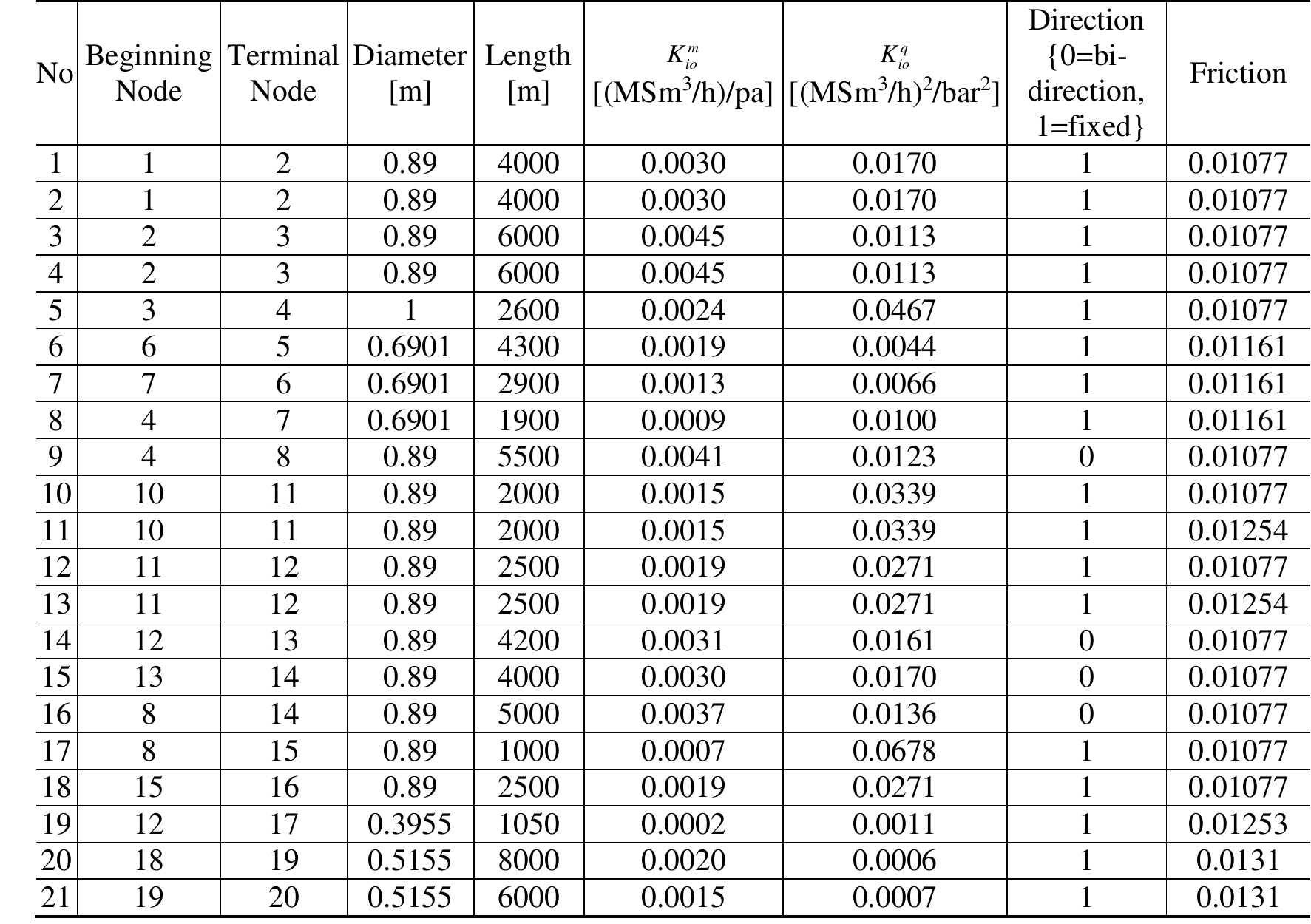}
    \end{table}
    \begin{table}[!hbtp]
     \label{tab:S2T26}
        \centering
        \caption{G2P Gas Contracts Between IEEE--123Bus System and 20Nodes Gas System}
        \includegraphics[width=16cm]{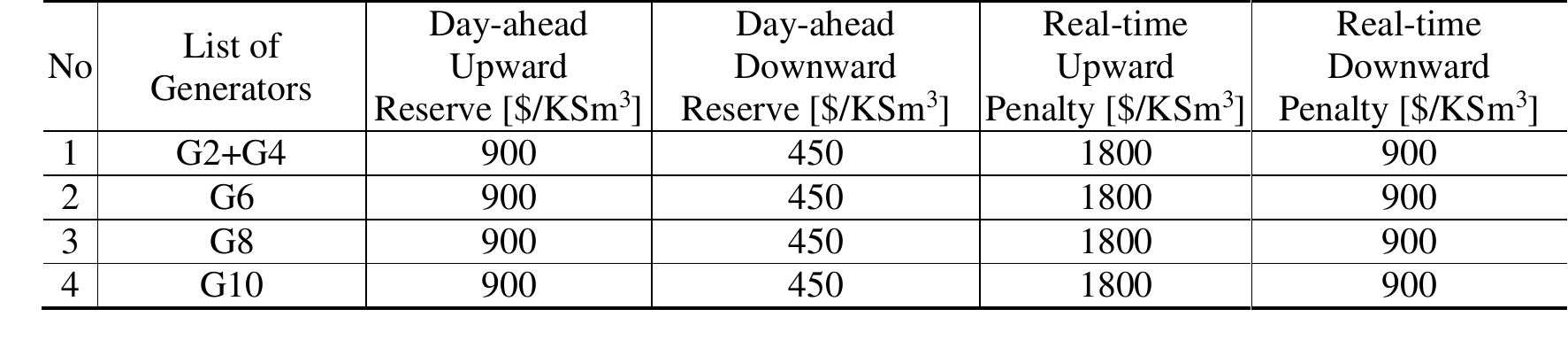}
    \end{table}
\begin{table}[!hbtp]
     \label{tab:S1T27}	
        \centering \caption{Parameters of Load Demand -- 20Nodes Gas System}
        \includegraphics[width=16cm]{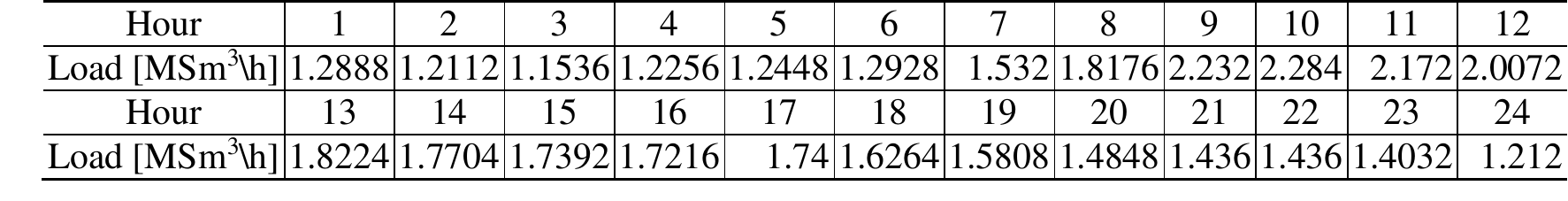}
    \end{table}

    \begin{table}[!hbtp]
     \label{tab:S1T28}	
        \centering \caption{Parameters of Gas Nodes and their Load Portion --  20Nodes Gas System}
        \includegraphics[width=16cm]{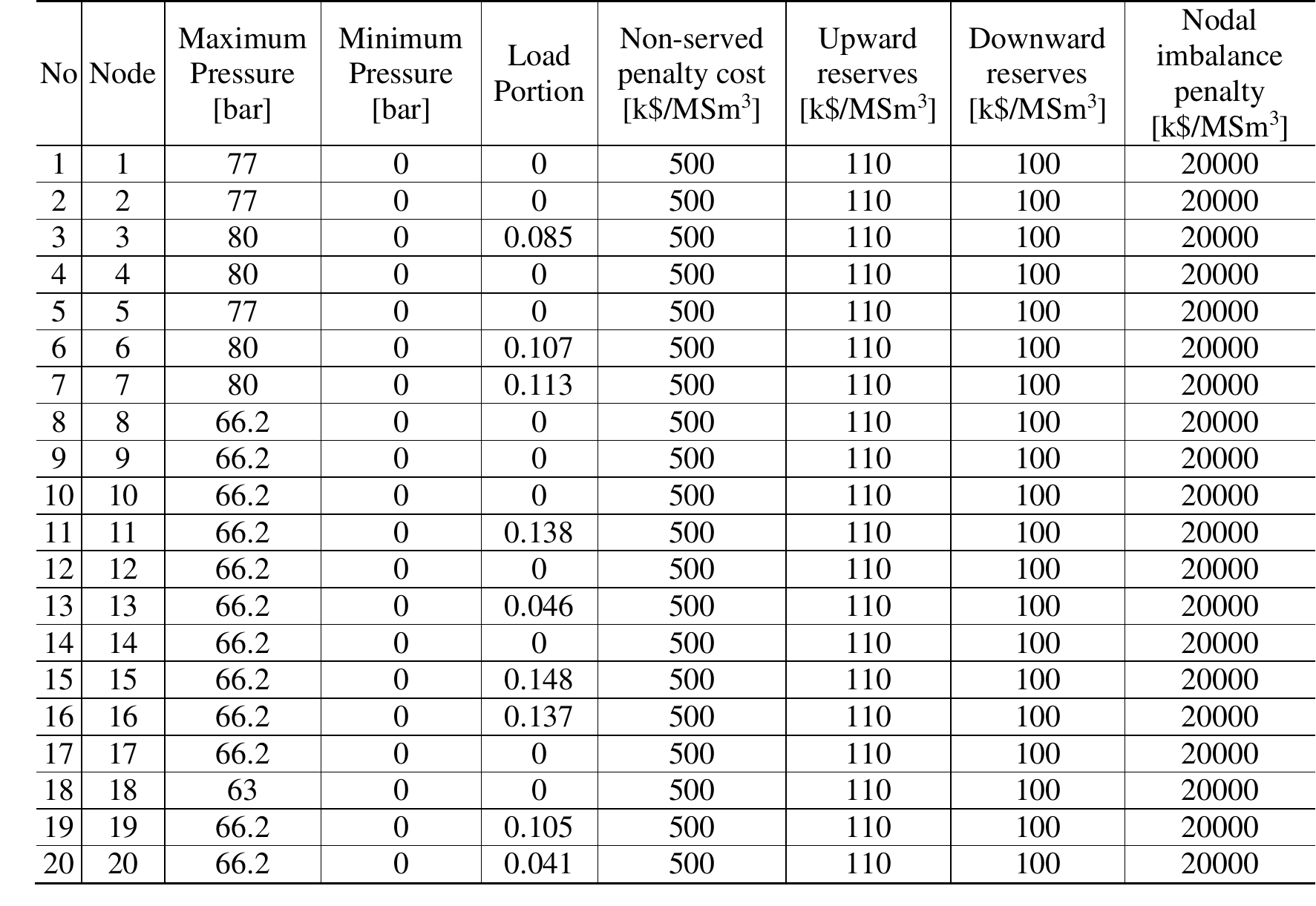}
    \end{table}

\begin{table}[!hbtp]
     \label{tab:S2T27}
        \centering
        \caption{P2G Gas Contracts Between IEEE--123Bus System and 20Nodes Gas System}
        \includegraphics[width=16cm]{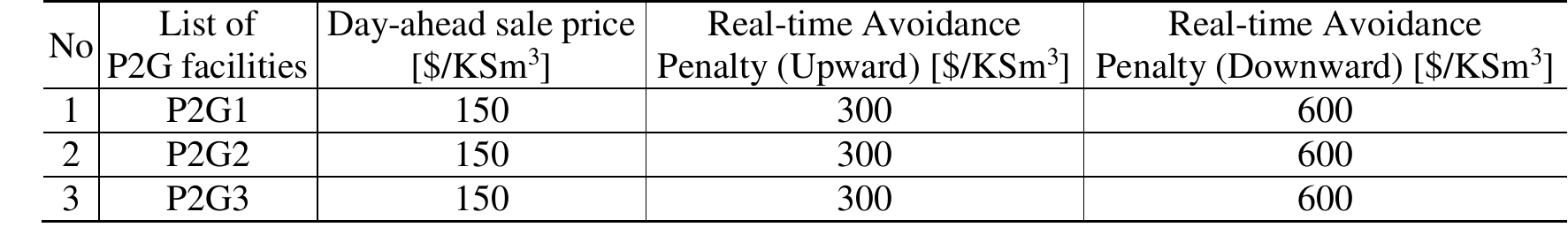}
    \end{table}

    \begin{table}[!hbtp]
     \label{tab:S1T29}	
        \centering \caption{Connection Lines Between IEEE--118Bus System and 20Nodes Gas System}
        \includegraphics[width=16cm]{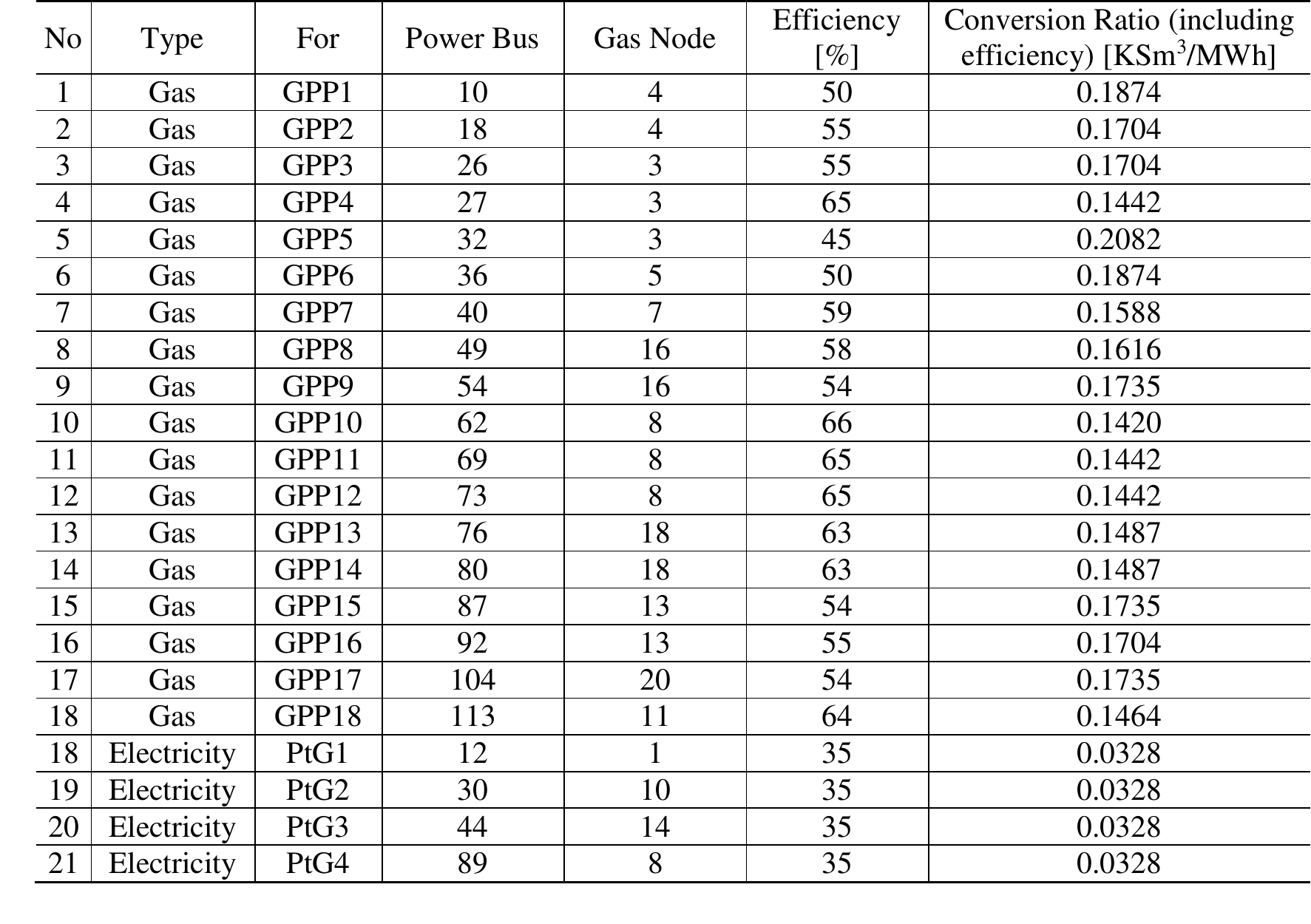}
    \end{table} 


\printbibliography[heading=bibintoc,title=References]

\backmatter
\begin{acknowledgements}
\addchaptertocentry{\acknowledgementname} 

First and foremost thanks to Allah. Without his help and blessing, I would not have been able to finish this work.

Then, I wish to express my sincere gratitude to my supervisors Prof. Tianshu Bi and Dr. Cheng Wang for encouraging me the guidance and unlimited support to me during my work. I am truly grateful to them for trusting my ability to complete the work. Their patience and kindness are greatly appreciated.

I would also like to thank all my professors and colleagues at the School of Electrical and Electronic Engineering for all the assistance when needed, especially Engineers; Arsalan Masood, Mohamed Abdelkarim, Mishkat, Sheeraz Iqbal, Sayed Zaki, and Haseeb.

I’m grateful to the examiners for constructive suggestions and valuable comments, which improve the quality of the thesis.

Last but not least, I am always indebted to all my family members, especially my parents, for their endless support and love. I greatly appreciate the sacrifices and understanding of my beloved wife during my struggling years.

%
%
%
%

\end{acknowledgements}


\end{document}